\documentclass[review]{elsarticle}

\usepackage{enumitem}
\usepackage[colorlinks,citecolor=blue,linktoc=all,linkcolor=cyan]{hyperref}
\usepackage{graphicx}
\usepackage{bm}
\usepackage[T1]{fontenc}
\usepackage{dsfont}               % use mathds instead of mathbb for outline fonts
\usepackage{mathrsfs}             % provides mathscr without overwriting mathcal
\usepackage{slashed}              % For Dirac slash notation.
\usepackage{amsmath}
\usepackage{amssymb}
\usepackage{amsbsy}
\usepackage{amsfonts}
\usepackage{amsfonts}
\usepackage{bbm}
\numberwithin{equation}{section}
\numberwithin{table}{section}
\numberwithin{figure}{section}

\usepackage{titlesec}
\usepackage{sectsty}
\titleformat{\section}{\normalfont\Large\bfseries}{\thesection}{1em}{}
\titleformat{\subsection}{\normalfont\large\bfseries}{\thesubsection}{1em}{}
\titleformat{\subsubsection}{\normalfont\normalsize\bfseries}{\thesubsubsection}{1em}{}
\graphicspath{{figures/}}

\usepackage{xcolor}

\biboptions{numbers,sort&compress} % make the refs more concise
\begin{document}
	\begin{frontmatter}
		%\title{Studying High Energy Nuclear Physics with Machine Learning}
            \title{Exploring QCD matter in extreme conditions with Machine Learning}
		%authors, affiliations, corresponding author mention 
	    \author[fias,cuhk]{Kai Zhou\corref{mycorrespondingauthor}}
            \cortext[mycorrespondingauthor]{Corresponding authors}
            \ead{zhou@fias.uni-frankfurt.de}
        
	    \author[fias]{Lingxiao Wang\corref{mycorrespondingauthor}}
            \ead{lwang@fias.uni-frankfurt.de}
        
        \author[ccnu]{Long-Gang Pang\corref{mycorrespondingauthor}}
            \ead{lgpang@ccnu.edu.cn}
        
        %\author[thu,sbu]{Shuzhe Shi\corref{mycorrespondingauthor}}
        \author[thu,sbu]{Shuzhe Shi\corref{mycorrespondingauthor}}
	       \ead{shuzhe.shi@stonybrook.edu}		
            
		\address[fias]{Frankfurt Institute for Advanced Studies (FIAS), D-60438 Frankfurt am Main, Germany.}
            \address[cuhk]{School of Science and Engineering, The Chinese University of Hong Kong, Shenzhen 518172, China}
  		\address[ccnu]{Institute of Particle Physics and Key Laboratory of Quark and Lepton Physics (MOE), Central China Normal University, Wuhan, 430079, China.}
        %\address[thu]{Department of Physics, Tsinghua University, Beijing 100084, China.}
		\address[sbu]{Center for Nuclear Physics, Department of Physics and Astronomy, Stony Brook University, Stony Brook, NY 11794-3800, USA.}

		\begin{abstract}
            In recent years, machine learning has emerged as a powerful computational tool and novel problem-solving perspective for physics, offering new avenues for studying strongly interacting QCD matter properties under extreme conditions. This review article aims to provide an overview of the current state of this intersection of fields, focusing on the application of machine learning to theoretical studies in high energy nuclear physics. It covers diverse aspects, including heavy ion collisions, lattice field theory, and neutron stars, and discuss how machine learning can be used to explore and facilitate the physics goals of understanding QCD matter. The review also provides a commonality overview from a methodology perspective, from data-driven perspective to physics-driven perspective. We conclude by discussing the challenges and future prospects of machine learning applications in high energy nuclear physics, also underscoring the importance of incorporating physics priors into the purely data-driven learning toolbox. This review highlights the critical role of machine learning as a valuable computational paradigm for advancing physics exploration in high energy nuclear physics.

		\end{abstract}
		
		
	\end{frontmatter}
 
	\newpage
	\thispagestyle{empty}
	\tableofcontents
	%to begin the line numbers: 
	%\linenumbers
	
	%beginning of the core of the manuscript
	\section{Introduction}\label{sec:intro}
High energy nuclear physics and machine learning, though being seemingly disparate, their interplay has already begun to emerge and been yielding promising results over the past decade. This review aims to introduce to the community the current status and report an overview of applying machine learning for high energy nuclear physics studies. From different aspects, we will present how scientific questions involved in this field can be tackled or assisted with the state-of-the-art techniques.

\subsection{Background}\label{subsec:bg}

The study of nuclear physics focuses on comprehending the nature of nuclear matter, its properties under various conditions, its building blocks, and the fundamental interactions that govern them. The behavior of nuclear matter is fundamentally governed by the strong interaction, described by quantum chromodynamics (QCD)~\cite{Gross:2022hyw}.
While nucleons serve as the main degrees of freedom for traditional low energy nuclear physics, high energy nuclear physics (HENP) cares more about QCD matter in extreme conditions, where the basic degrees of freedom are typically quarks and gluons.

A primary goal of HENP is to understand how nuclear matter behaves under extreme conditions, which remains an unresolved area of study~\cite{Yagi:2005yb, Wang:2016opj, Fukushima:2020yzx,Shuryak:2004pry}. Intuitively, when matter is subjected to extreme conditions, its basic constituents can be revealed. For example, when nuclear matter is heated to high temperatures, mesons and resonances are gradually excited out, and at some point causing hadrons or nucleons to \textit{overlap} (the same can happen when nuclear matter is compressed to high densities) thus their constituent quarks and gluons can move freely inside a larger area. The induced state of matter, known as quark-gluon plasma (QGP)~\cite{Rafelski:2019twp}, is a deconfined state in which quarks and gluons are the basic degrees of freedom and strongly interacting. The existence of such primordial state of matter is believed to have occurred in the early universe after the Big-Bang, and may also exist in the dense interior of neutron stars~\cite{Luo:2022mtp}.
The first-principle lattice QCD simulations predict that normal nuclear matter in the form of a dilute hadronic gas can undergo a crossover transition to the strongly-interacting, deconfined state of QGP at high temperature and low baryon chemical potentials, where chiral symmetry is also restored~\cite{Ding:2015ona, Karsch:2022opd, Aarts:2023vsf}. 

However, the situation at high baryon chemical potential is not well understood due to the fermionic sign problem that impedes direct lattice QCD simulations~\cite{Nagata:2021ugx}. The structures of the QCD phase diagram and the properties of matter in these regions have to be studied through effective model calculations and experimental searches. The formation and study of hot and dense QGP can be accomplished through relativistic heavy ion collision (HIC) experiments~\cite{Busza:2018rrf}, which act as a heating and compression machine to QCD matter to reach extreme conditions~\cite{Bzdak:2019pkr}. However, the collision of heavy nuclei creates a rapidly evolving and dynamic process with many entangled and not-fully understood physics factors~\cite{Luo:2022mtp}. Decoding the physics properties related to the early-formed QGP from the final measurements of heavy ion collisions therefore remains a challenging task.
In general, both theoretical calculations and experimental facilities are crucial for advancing the field of HENP, which is anyhow becoming increasingly intricate and producing ever-larger amounts of data from sources such as detectors in experiments e.g., Relativistic Heavy Ion Collider (RHIC) at BNL and Large Hadron Collider (LHC) at CERN, etc., and model simulations such as relativistic hydrodynamics, transport cascades, and lattice QCD simulations. Nevertheless, performing the involved calculations, analyzing the involved data, and decode underlying physics from confronting theory with data,i  are routinely demanding due to factors such as high-dimensionality, background contamination, and potentially entangled physical influences. 

Artificial Intelligence (AI) has grown rapidly in popularity with the goal of imparting intelligence to machines. The field of computer science, dating back to the 1960s, has experienced a revival in recent years with the development of modern learning algorithms, advanced computing power, and the vast amount of available data~\cite{Hogg:1987nn,AIreview:2021}. Machine learning (ML), an essential part of AI, is viewed as a modern computational paradigm that gives machines the ability to learn without being explicitly programmed. It enables computers to go beyond being instructed by humans and learn how to perform specific tasks on their own. This novel paradigm, particularly deep learning (DL)~\cite{lecun2015deep}, a subfield of ML that uses deep neural networks for hierarchical representation, has shown successful applications~\cite{sarker2021deep} in computer vision(CV), natural language processing(NLP), game AI, and drug discovery. It also demonstrates great promise in transforming scientific research~\cite{osti_1604756}. It allows for automatically recognizing patterns hidden in large, complex datasets, and has proven to be particularly effective in handling intricate structures and non-linear correlations that are beyond conventional analysis~\cite{osti_1604756}. 
%\lw{Remark 1 from referee - Perhaps a bit more context of the use of AI/ML methods in other fields would be helpful, with short examples (and references) of where it has been particularly useful. At the moment there is only brief general mention of some other fields in Sec. 1.1.}
%{\color{red}

To list several additional \textbf{AI for Science} examples on demonstrating how these computational methodologies have been employed in various scientific domains: 
\begin{itemize}
    \item Material Science: AI and ML have expedited the discovery of new materials by predicting properties and optimizing compositions, e.g., as shown in Ref.~\cite{PhysRevLett.120.145301}, ML can help to predict the quantum mechanical properties of a variety of materials, paving the way for faster discovery of new materials with desired properties.
    \item Neuroscience: AI/ML methods are applied for decoding neural activity and understanding brain functionality. Ref.~\cite{SUK2014569} employed deep learning to analyze large datasets of brain imaging (MRI and PET) to identify patterns associated with neurological disorders (Alzheimer's Disease and Mild Cognitive Impairment).
    \item Autonomous Systems: ML and DL are central to the development of autonomous systems and the next generation of industries. They enable real-time decision-making, navigation, and adaptation to changing environments for self-driving cars~\cite{2016arXiv160407316B}. They also provide real-time cavitation and turbulent detection for smart valves deploying just sensors to monitor the acoustic signals out of the running pipeline systems~\cite{SHA2022110897,SHA2022104904}.
    \item Geoscience: AI/ML tools are utilized for the prediction of natural resource locations, seismic event magnitude estimation~\cite{10.1029/2022JB024595}, earthquake detection and seismic phase identification~\cite{10.3389/feart.2022.953007}.
    \item Healthcare and Biology: AI/ML techniques have been instrumental in diagnostic imaging and disease mechanism investigation. For instance, DL algorithms have shown remarkable accuracy in identifying medical conditions from radiological images~\cite{LITJENS201760}. Moreover, ML approaches are being employed to decipher genetic data and understand the underlying mechanisms of various diseases, which aids in the development of personalized treatment plans~\cite{article_DL_genomics}.
    \item Environmental Science: AI/ML algorithms have been pivotal in monitoring and predicting environmental changes. They are utilized for climate modeling, pollution control, and natural disaster forecasting. For example, ML methods have demonstrated promise in predicting extreme weather events such as hurricanes and floods, which is crucial for early warning systems~\cite{article_dl_earth}.
    \item Agricultural Science: ML models have been applied to optimize crop yields, monitor soil health, and predict environmental impacts on agricultural productivity. AI/ML techniques are also utilized in precision agriculture for pest and disease detection~\cite{KAMILARIS201870}.
    \item Astronomy: the rapidly-developed field of astroinformatics employs ML algorithms to sift through vast amounts of astronomical data, to help in classifying celestial objects, detecting exoplanets, and understanding cosmic phenomena. Ref.~\cite{George:2016hay} applied convolutional neural networks to detect gravitational waves, a monumental achievement that opened new vistas in observational astronomy. Tidal properties have also shown to be capable of being extracted with DL in analyzing gravitational waves recently~\cite{Soma:2023rmq}.
    \item Lightning Analysis: ML algorithms have been used in analyzing intricate very-high-frequency(VHF) lightning data from LOFAR(LOw Frequency ARray). Highlighting the limitations of manual evaluations, Ref.~\cite{Wang:2023yul} champions an unsupervised ML solution by synergizing t-SNE and clustering techniques. This adeptly identifies and categorizes correlated structures for the lightning phenomenon, potentially enabling swift and automated analysis in lightning physics and LOFAR data processing. 
    \item Epidemiology Dynamics: the application of AI/ML in studying epidemiological dynamics has accelerated the understanding of the infectious disease's transmission and vaccination influence, significantly aiding global containment efforts. As shown in Ref.~\cite{Wang_2021}, ML algorithms could enable the identification of the spatio-temporal epidemic dynamics of COVID-19 from the Germany-reported infection data, and allow for further prediction of the prevalence of the multi-scale COVID-19.
    \item DeepMind: %AlphaFold, AlphaZero, AlphaCode, tokamak desgin...
    Predicting the three-dimensional structures of proteins from their amino acid sequences has been a persistent challenge for over 50 years. However, the AlphaFold AI system has successfully solved this "protein structure prediction problem" with remarkable accuracy, thereby advancing biological research significantly \cite{Jumper2021}.
    \item Nuclear Fusion: Nuclear fusion is one of the most daunting real-world challenges, primarily due to the difficulty in dynamically controlling the shape of hot plasma to prevent rapid instability growth. Nevertheless, deep reinforcement learning has been effectively utilized to design Tokamak magnetic controllers \cite{Degrave2022}, resulting in high performance and potentially accelerating the development of clean nuclear fusion energy.
    \item Deep Potential: The Deep Potential model  \cite{Han2018} employs deep neural networks to represent the energy surface of atomic and molecular systems accurately and efficiently. This approach has been widely adopted in molecular dynamics and material design research. 
\end{itemize}
Furthermore, there are numerous other applications of deep learning that have revolutionized the entire scientific research field. For more information on these applications, please refer to \href{https://www.anl.gov/ai-for-science-report}{``AI for Science, Energy, and Security Report''} for an overview.
%}

%\lw{The collowing content should be merged into the above paragraphs.}
Therefore, a variety of scientific tasks can be aided by these new computational methodologies, such as extracting knowledge from measurements or calculations, improving simulations, detecting anomalies, or discovering new phenomena. 
The deployment of machine learning in various scientific disciplines has become progressively ubiquitous.
It is believed that machine learning has the potential to greatly enhance physics research as well (see e.g., ~\cite{Mehta:2018dln} for a pedagogical introduction). With its flexibility in tackling general computational physics tasks, machine learning is actively being explored to support physics studies~\cite{Carleo:2019ptp}. Conversely, many concepts and foundations in machine learning can be traced back to physics, such as the Boltzmann machine and variational generative algorithms, which have close ties to statistical physics~\cite{Mezard:2009ipa}. The synergy between these fields will not only benefit physics but also deepen our understanding of the underlying working mechanisms of machine learning~\cite{Bahri:2020smd} or even intelligence.

%%%%%%%%%%%%%%%%%%%%%%%%%%%%%%%%%%%%%%%%%%%%%%%%%%%%%%%%
\begin{figure}[htbp!]
    \centering
    \includegraphics[width = 0.85\textwidth]{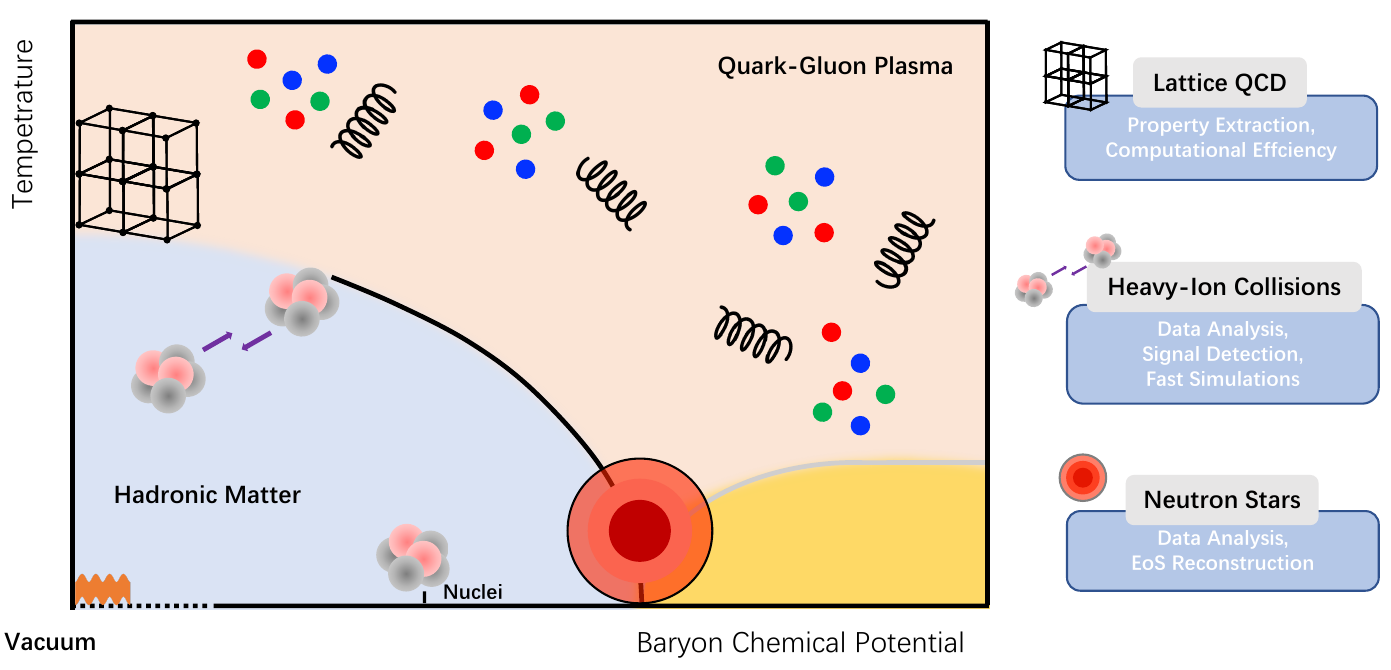}
    \caption{Schematic QCD phase diagram and the applications of machine learning in three domains, i.e., Lattice QCD, Heavy-Ion Collisions, and Neutron Stars.
    \label{fig:ML4HENP}}
\end{figure}
%%%%%%%%%%%%%%%%%%%%%%%%%%%%%%%%%%%%%%%%%%%%%%%%%%%%%%%%

High energy nuclear physics can greatly benefit from machine learning techniques due to its computational and data-intensive nature. In recent years, the intersection of machine learning and HENP has shown great potential and unconventional possibilities~\cite{Bedaque:2021bja,Boehnlein:2021eym}. HENP has long been at the forefront of big data analysis, using techniques such as neural networks and statistical learning~\cite{Humpert:1990rw,Gyulassy:1990et,Bass:1993vx,Bass:1996ez}, but recent advancements in deep learning have brought new developments to the crossing of these two fields. Deep learning has improved the ability to handle large amounts of high-dimensional data, extended the feasibility of going beyond human-crafted heuristics, and generated new possibilities such as fast and effective simulation, sensitive observable design, and new physics detection (see a living review~\cite{Feickert:2021ajf} and a collection of datasets in high energy physics~\cite{Benato:2021olt}). This growing and influential research area deserves dedicated recognition and further study to release the full potential of such novel computational paradigms for high energy nuclear physics.

The use of machine learning in high energy nuclear physics encompasses a wide range of topics and tasks, from experimental data analysis with normal discriminative algorithms to speeding up various simulations with generative models. We will review them systemically from three sectors, as shown in Fig.~\ref{fig:ML4HENP}, i.e., heavy-ion collisions in Sec.~\ref{sec:hic}, lattice QCD in Sec.~\ref{sec:lat} and neutron stars in Sec.~\ref{sec:astro}. Besides, the involved studies are quite often with multi-scale and high-dimensional computations and require customized ML methods that address the specific physics constraints and characteristics, e.g., physical differential equations, corresponding symmetries, and conservation laws. This leads to the development of unique techniques for science automation and physics exploration with ML paradigm. A refined summary of some new and advanced developments and their applications in high energy nuclear physics will be discussed in Sec.~\ref{sec:new}. This review, although unavoidably limited and with bias in terms of content selection, aims to offer a practical introduction to readers interested in exploring the rapidly developing field of applying machine learning to high energy nuclear physics. A general overview of the ongoing activities in this field will be covered. In the remainder of this section, we will briefly introduce the basics of machine learning for those unfamiliar with the terminology. This information and related techniques can be revisited when delving into the subsequent physics research discussions.

\subsection{Machine Learning in a Nutshell}\label{subsec:ML}
Before delving into the physics aspects of this review, we aim to present a concise and comprehensive introduction to the fundamental concepts of machine learning, with particular emphasis on the current advancements in deep learning. The concepts and techniques discussed here will be revisited in later sections of this review. Although it cannot cover all aspects of machine learning~\footnote{The missing parts include but are not limited to dimensionality reduction, clustering, linear/Logistic regressions, (support vector machines)SVMs, and ensemble learning.}, the main objective is to introduce the advanced deep learning approaches. For a comprehensive understanding of machine learning, we recommend perusing books~\cite{Bishop2006,Goodfellow2016}, or recent reviews for its practical applications in physics~\cite{Mehta:2018dln,Boehnlein:2021eym,Dawid:2022fga,Shanahan:2022ifi}.

\subsubsection{Bayesian Inference}\label{subsubsec:bi}
From a probabilistic perspective, one can infer the probability distribution of random variables (A) from limited trials/observations over its consequences (B), based on the Bayes' theorem~\cite{Murphy2012},
\begin{equation}
    P(A \mid B)=\frac{P(B \mid A) P(A)}{P(B)}, 
    \label{eq:bayesian}
\end{equation}
where in scientific applications, $A$ often represents parameters of theoretical models and $B$ refers to experimental observations. The Bayes formula is used to update or revise our belief/knowledge of a theoretical model in terms of $A$ using new experimental observations $B$. It involves the calculation of the posterior probability distribution $P(A \mid B)$, which takes into account the prior probability distribution $P(A)$ and the conditional probability $P(B \mid A)$, known as the likelihood. Uncertainties in the experimental observations $B$ will impact the posterior distribution of $A$ via likelihood, causing it to become wider. The posterior distribution $P(A \mid B)$ is a probability density function over A, which can be characterized by its mean, peak value, and variance. One way to determine the most likely value of $A$ is by finding the group of parameter combinations $A=a$ that maximizes the posterior distribution $P(A \mid B)$, known as the maximum a posteriori estimation (MAP). This process allows us to locate the peak of the posterior distribution.

The variance of posterior $P(A\mid B)$ can be used to determine the uncertainty or confidence level of the theoretical model. If $P(A\mid B)$ is a narrow distribution with a sharp peak, its variance is small, indicating that the data are decisive and useful in constraining the model with small uncertainty. The marginal distribution $P(B)=\sum_a P(B \mid A=a) P(A=a)$ is called evidence, which provides a normalization factor by traversing the space of the theoretical model. For an n-dimensional parameter space $A=\left(A_1, A_2, \cdots, A_n \right)$, if the random variable has $m$ possible choices in each dimension, one must repeat the theoretical calculations $m^n$ times to get the normalization factor $P(B)$. This procedure encounters the challenge of \textit{curse of dimensionality} for high-dimensional parameter spaces. Fortunately, $P(B)$ is not needed using Markov Chain Monte Carlo (MCMC) methods, which work for an unnormalized posterior distribution as follows,
\begin{equation}
    P(A \mid B) \propto P(B \mid A) P(A), 
    \label{eq:bayesian_unnorm}
\end{equation}
The way MCMC methods work for Eq.~\eqref{eq:bayesian_unnorm} is by importance sampling using random walk~\cite{garnett_bayesoptbook_2023}. Starting from any position $A = a_0$ in the parameter space, walk to the next position using a random step size $\epsilon$ sampled from a uniform or normal distribution, following the Markov Principle that each update only depends on its current position, $a_{n+1} = a_{n} + \epsilon$. The trajectory of one random walk in the parameter space forms a set $\{ a_n \}$ for $n=0, 1, 2\cdots$ whose density estimation produces a probability distribution. This probability distribution does not equal $P(A|B)$ without any constraints to the process of random walk. The MCMC algorithms set an acceptance--rejection probability to each update such that $A=a$ values in the parameter space are visited more frequently with larger $P(A=a|B)$ than smaller ones to form a stationary distribution that is close to $P(A|B)$. E.g., in Metropolis-Hastings algorithm which is one kind of MCMC algorithm, the acceptance--rejection ratio is set to be ${\rm min}(1, \alpha)$ where,
\begin{align}
\alpha = \frac{P(A=a_{n+1}|B) q(a_{n} | a_{n+1})}{P(A=a_{n}|B) q(a_{n+1-} | a_{n})} =  \frac{P(A=a_{n+1}|B)}{P(A=a_{n}|B)}
\label{eq:rejection_rate}
\end{align}
where $q(a_{n+1} | a_{n})$ is called a proposal function that can be a uniform distribution or a normal distribution centered at $a_n$. The proposal function is symmetric in Metropolis-Hastings algorithm, which can be eliminated in Eq.~\eqref{eq:rejection_rate}.
This is got from the detailed balance condition with $P(A=a_{n+1}|B) q(a_{n} | a_{n+1}) = \alpha P(A=a_{n}|B) q(a_{n+1} | a_{n})$. In practice, one can sample $a_{n+1}$ from a normal distribution $\mathcal{N}(a_n, \sigma)$ centered at $a_n$ with standard deviation $\sigma$. The parameter $\sigma$ can be adjusted adaptively for efficient sampling.

\textit{\textbf{Gaussian Processes}} --- Gaussian process(GP) is a probabilistic model that can be used for regression or classification~\cite{garnett_bayesoptbook_2023_update,NIPS1995_7cce53cf}. Formally, as a type of stochastic process, a GP dictates a probability distribution over functions, $f(\bm{x})\sim \mathcal{GP}(m(\bm{x}),k(\bm{x},\bm{x}'))$, which is fully specified by a mean function $m(\bm{x})$ (usually chosen to be zero) and a covariance function $k(\bm{x},\bm{x}')$ (measures the similarity between two input values). The GP is defined as a collection of random variables (e.g., the evaluations of its sample function at different inputs $\{f(\bm{x}_i)\}$), with its any finite set following a joint Gaussian (i.e., multivariate normal) distribution, $f(\bm{x}_1),...,f(\bm{x}_N)\sim\mathcal{N}(\bm{\mu},\bm{K})$ where $\bm{\mu}_i=m(\bm{x}_i)$ and $\bm{K}_{ij}=k(\bm{x}_i,\bm{x}_j)$. 

For a regression task, one can use GP to construct a non-parametric prior to the to-be-fitted function. Per observing a training set $\bm{X}=\{\bm{x}_i\}^N_{i=1}$ with its corresponding targets $Y=\{y_i\}^N_{i=1}$, one can update the prior by conditioning on the observations, which essentially restrict the functions from the GP to be in line with the training data. Specifically, consider the prediction $Y^{*}$ for new inputs $\bm{X}^{*}$, together with the training targets $Y$, one has the joint prior distribution
$$
\begin{bmatrix}
 Y\\
Y^{*}
\end{bmatrix} \sim \mathcal{N}(\bm{0},
\begin{bmatrix}
      \bm{K}(\bm{X},\bm{X}) & \bm{K}(\bm{X}, \bm{X}^{*}) \\
      \bm{K}(\bm{X}^{*},\bm{X}) & \bm{K}(\bm{X}^{*},\bm{X}^{*})
\end{bmatrix} ),
$$
and the conditioning of this joint Gaussian prior on the training data results in the posterior for the prediction,
$$\mathrm{p}(Y^{*}|\bm{X}^{*},\bm{X},Y)=\mathcal{N}(\bar{Y}^{*},\mathrm{cov}(Y^{*})),$$ with $\bar{Y}^{*}=\bm{K}(\bm{X}^{*},\bm{X})\bm{K}(\bm{X},\bm{X})^{-1}Y$ and $\mathrm{cov}(Y^{*})=\bm{K}(\bm{X}^{*},\bm{X}^{*})-\bm{K}(\bm{X}^{*},\bm{X})\bm{K}(\bm{X},\bm{X})^{-1}\bm{K}(\bm{X},\bm{X}^{*})$.

In practice, choosing the appropriate kernel function $K$ for a GP is an important step in the modeling process. It determines the shape of the covariance function and thus the overall structure of the model. The choice of kernel function depends on the specific problem at hand, as well as any prior knowledge or assumptions about the structure of the underlying function. Some common kernel functions contain: \textit{linear kernel}, $ k(x, x') = x^T x',$ where $x$ and $x'$ are vectors representing the input values; \textit{polynomial kernel}, $ k(x, x') = (x^T x' + c)^d$, where $c$ is a constant, and $d$ is the degree of the polynomial; \textit{squared exponential (SE)} or \textit{Radial Basis Function (RBF)} kernel~\cite{Genton:2002ker}. This is a popular choice for regression problems, as it can capture complex, non-linear relationships between inputs and outputs. The RBF kernel is defined as, 
$$ k(x, x') = \sigma_f^2 \exp\left(-\frac{1}{2l^2}(x - x')^2\right),$$
where $\sigma_f^2$ is the variance and $l$ is the length scale, which determines the smoothness of the function. Besides, there are many other choices such as \textit{Mat\'ern kernel}~\cite{Genton:2002ker} kernel for capturing a variety of functional shapes. Many other examples can be found in the book~\cite{Genton:2002ker}.  Once a kernel has been chosen, the parameters of the kernel (such as $\sigma_f^2$, $l$, and $d$ in the above examples) can be estimated from the data using a variety of optimization methods.

\subsubsection{Deep Learning}\label{subsubsec:dl}
%%%%%%%%%%%%%%%%%%%%%%%%%%%%%%%%%%%%%%%%%%%%%%%%%%%%%%%%
\begin{figure}[htbp!]
    \centering
    \includegraphics[width = 0.72\textwidth]{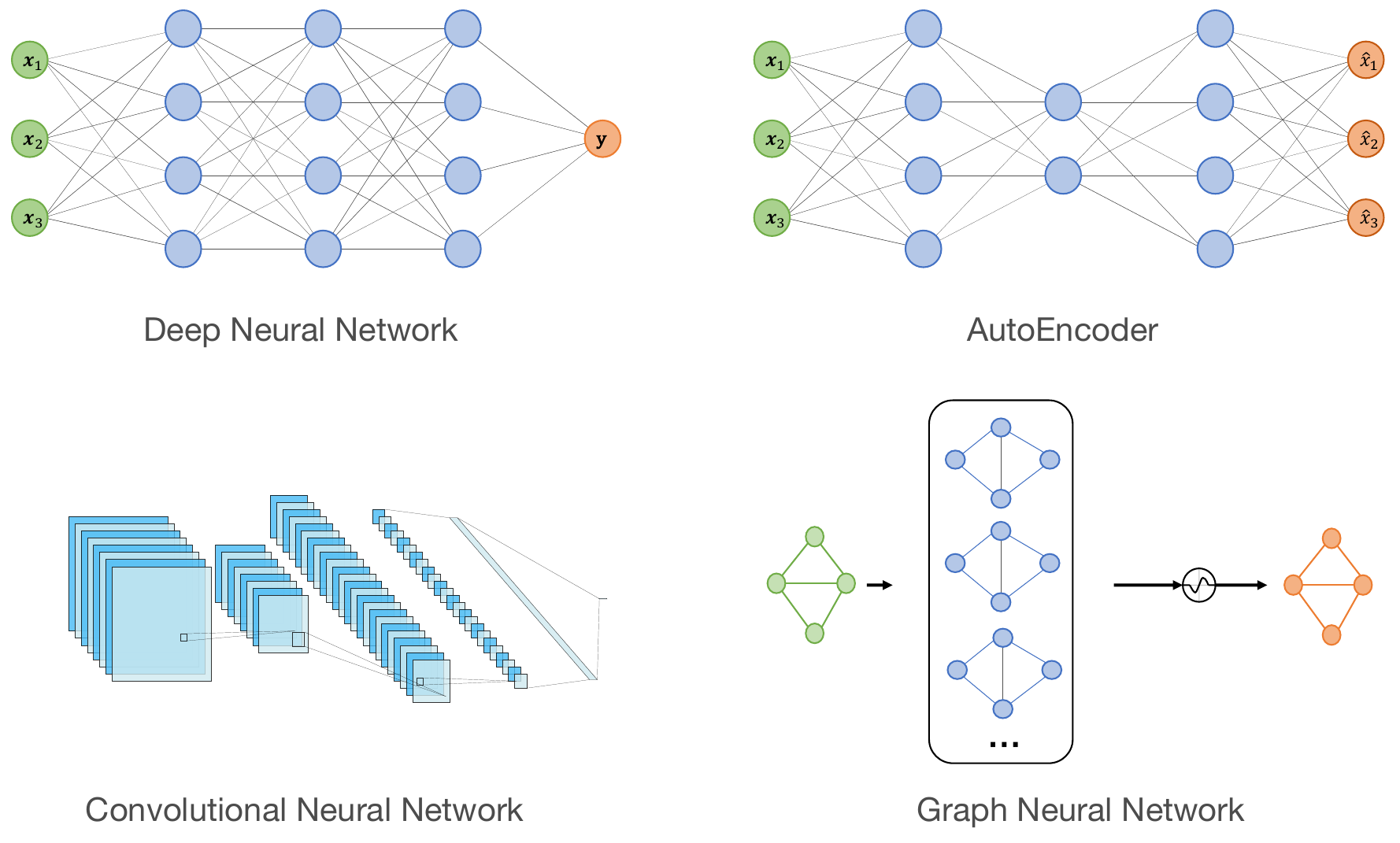}
    \caption{A demonstration of the common neural network structures that will be covered in this review. Deep Neural Networks (DNNs), AutoEncoders (AEs) with "bottleneck latent" layers, Convolutional Neural Networks (CNNs) utilizing convolutional operations, and Graph Neural Networks (GNNs) capable of processing graph-type inputs.
    \label{fig:NNs}}
\end{figure}
%%%%%%%%%%%%%%%%%%%%%%%%%%%%%%%%%%%%%%%%%%%%%%%%%%%%%%%%
Deep learning, as a modern branch of machine learning~\cite{lecun2015deep,schmidhuber2015deep}, usually involving the use of deep neural networks, provides tools for automatically extracting patterns from the massive multi-fidelity observational data currently available. The \textbf{enormous data} makes it possible to establish a reliable mapping between input features and desired targets with deep models. \textbf{Deep models} are typically constructed using artificial neural networks (ANNs), which were originally designed to mimic the functions of the human brain~\cite{mcculloch1943logical}. Subsequent studies have demonstrated their ability to represent a wide variety of continuous functions, usually expressed as the universal approximation theorem~\cite{cybenko1989approximation,hornik1991approximation}. In an oversimplified picture, deep neural networks (DNNs) are a type of composite function that has multi-layered nesting structures: $f_\theta(x) = z^l(\cdots z^2(z^1(x))), z^i(z^{i-1}) = \sigma(\omega^i z^{i-1} + b^i)$, where $x$ is input data defined in a set of samples $\mathbb{X}$, $l$ denotes the number of layers and $i$ is the corresponding index, and $z^0 \equiv x $. $\sigma(\cdot)$ is a nonlinear activation function and $\{\omega,b\}$ are weights and bias respectively. Weights and bias are all trainable parameters, they could be abbreviated as $\{\theta\}$. Some common neural network structures are shown in Figure.~\ref{fig:NNs}. They will be mentioned throughout the rest of this review.

Given a well-defined\textbf{loss function} $\mathcal{L}(f_\theta(x))$, these trainable parameters can be tuned in an optimization problem,
\begin{equation}
    \mathbf{\theta} = \mathop{argmin}\limits_{\theta}\mathcal{L}(f_\theta(\mathbb{X}))=\mathop{argmin}\limits_{\theta}\sum_j^m\mathcal{L}(f_\theta(x^{(j)})),
    \label{eq:1:op}
\end{equation}
where $j$ labels the index of a sample. If one can collect corresponding labels $y\in \mathbb{Y}$ for input $x$, the optimization process becomes a \textit{Supervised Learning} task. Common Loss functions contain Mean Squared Error(MSE), Mean Absolute Error(MAE), Cross Entropy, etc.

In fact, one can interpret the above optimization from a \textit{probabilistic perspective}:
\begin{equation}
    \mathbf{\theta} = \mathop{argmax}\limits_{\theta}p_\theta(\mathbb{X})=\mathop{argmax}\limits_{\theta}\prod_j^m p_\theta(x^{(j)}),
\end{equation}
which is the \textit{maximum likelihood estimation(MLE)} for parameters $\{\theta\}$ given the data-set $\mathbb{X}$. From this perspective, training a deep model involves approaching the conditional probability distribution $p_\text{data}(y|x)$. Furthermore, it can naturally expand to the task of Unsupervised Learning, which involves approaching the underlying distribution $p_\text{data}(x)$ of unlabeled data. A helpful technique is to maximize the log-likelihood, $\mathop{log}(p_\theta)$, converting the product to a summation (see Table.\ref{tab:metrics}). 
%%%%%%%%%%%%%%%%%%%%%%%%%%%%%%%%%%%%%%%%%%%%%%%%%%%%%%%%%%%%%%%%%%%%%
\begin{table}[!hbpt]
\centering
\begin{tabular}{l |l | l }
\hline\hline
    Metrics & Formula & Probabilistic Description \\
\hline
MSE & 
    $(f_\theta(x) - y)^2$&
    Gaussian\\
MAE & 
    $|f_\theta(x) - y|$&
    Laplace\\
Binary Cross-Entropy & 
    $-\left[ y \mathop{log} f_\theta(x) + (1 - y) \mathop{log} (1-f_\theta(x)) \right]$ &
    Bernoulli\\
\hline\hline
\end{tabular}
\caption{Common Loss functions and their probabilistic descriptions.}
\label{tab:metrics}
\end{table}
%%%%%%%%%%%%%%%%%%%%%%%%%%%%%%%%%%%%%%%%%%%%%%%%%%%%%%%%%%%%%%%%%%%%%

Before introducing the optimization algorithm, it is important to consider the generalization ability of deep models. In simple terms, this refers to the capacity of a model to perform well on data it has never encountered before. If there is too little training data to accurately represent the underlying distribution of the available data, or the model is over-represented in the available data, one may observe that performance metrics are better for training than for testing data. This situation is known as \textit{over-fitting}. In contrast, when the model is too simple to represent the available data, it is called \textit{under-fitting}. \textit{Under-fitting} can be easily overcome by increasing the complexity of deep models. However, \textit{over-fitting} is a challenge because it results from a \textbf{trade-off} between model flexibility and accuracy. Adding priors to the parameters is a conventional and efficient strategy to avoid possible \textit{over-fitting}. From a Bayesian perspective, it transforms MLE to the \textit{Maximum A Posteriori (MAP)} estimation,
\begin{equation}
    \mathbf{\theta}_\text{MAP} = \mathop{argmax}\limits_{\theta}p_\theta(\mathbb{X}|\theta)=\mathop{argmax}\limits_{\theta}\mathop{log}p(\mathbb{X}|\theta)+\mathop{log}p(\theta),
    \label{eq:1:op-map}
\end{equation}
where the last term $\mathop{log}p(\theta)$ is the prior distribution of parameters. The Gaussian and Laplace priors correspond to $L_2$ and $L_1$ regularization, respectively~\cite{Goodfellow2016}.

From an optimization perspective, once the loss function and model are given, one can choose many different algorithms to train the model. However, the \textbf{gradient-based optimizer} within the differentiable programming paradigm has been proven successful in deep models~\cite{Lecun1998,lecun2015deep,schmidhuber2015deep}. In fact, almost all current deep learning packages are adopting gradient-based methods, which are implemented in \textit{\textbf{automatic differentiation(AD)}} frameworks. AD is different from either the symbolic differentiation or the numerical differentiation~\cite{baydin2018automatic}. Its backbone is the chain rule, which can be programmed in a standard computation with the calculation of derivatives. Lots of AD libraries have been developed in the past several years, rendering easy access to AD usage for researchers, such as Tensorflow~\cite{AbaBar16Tensorflow}, PyTorch~\cite{NEURIPS2019_9015} and JAX~\cite{2019arXiv191204232S}.
%%%%%%%%%%%%%%%%%%%%%%%%%%%%%%%%%%%%%%%%%%%%%%%%%%%%%%%%
\begin{figure}[ht!]
    \centering
    \includegraphics[width = 0.45\textwidth]{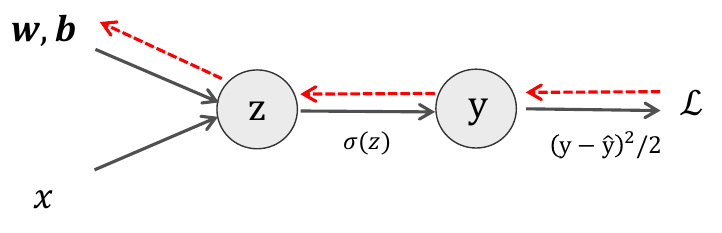}
    \caption{Computational graph of a logistic least squares regression in a vanilla net. 
    \label{fig:adsample}}
\end{figure}
%%%%%%%%%%%%%%%%%%%%%%%%%%%%%%%%%%%%%%%%%%%%%%%%%%%%%%%%

 In a basic example depicted in Fig.~\ref{fig:adsample}, the logistic least squares regression computation involves a sequence of differentiable operations. The forward mode is represented by black arrows that denote derivatives. To compute $\mathcal{L}$ from the input $x$, the appropriate derivatives can be computed simultaneously by utilizing the chain rule. With this example, readers can gain a clear understanding of the back-propagation (BP) algorithm~\cite{schmidhuber2015deep, Goodfellow2016}. The generalized BP algorithm corresponds to the reverse mode of automatic differentiation, where derivatives propagate backward from a given output. The adjoint is defined as $\bar{v}_i = \frac{\partial \mathcal{L}}{\partial v_i}$, which reflects how the output changes with respect to changes of intermediate variables $v_i$. The forward and reverse computations can be demonstrated as follows,
\begin{align}
\begin{split}
    &z=w x+b,\quad y=\sigma(z),\quad \mathcal{L}=\frac{1}{2}(y-\hat{y})^{2},\\
    &\overline{\mathcal{L}}=1,\quad \bar{y}=y-\hat{y},\quad \bar{z}=\bar{y} \sigma^{\prime}(z),\quad \bar{w}=\bar{z} x,\quad \bar{b}=\bar{z}.
\end{split}
\end{align}
The first line is the forward process and the second line indicates the reverse mode. Given a target $\hat{y}$, one can define a proper loss function $\mathcal{L}(y,\hat{y})$ and fine-tune the parameters $\mathbf{\theta} = (\omega,b)$ with the gradient $\partial_{\mathbf{\theta}}\mathcal{L}$, which is the gradient-based optimization introduced before. The BP is crucial for training a DNN because the derivatives can be used to optimize a high dimensional parameter set layer by layer. In most deep learning tasks, models are trained to approximate a $\mathbb{R}^N\rightarrow \mathbb{R}^1$ mapping. In such a case, the reverse mode of AD holds dominant advantages in the gradient-based optimization compared with the forward mode or the numerical differentiation, see more details in Ref~\cite{baydin2018automatic}.

In the optimization problem defined in \eqref{eq:1:op} or \eqref{eq:1:op-map}, a well-known \textbf{gradient-based optimizer} is the \textit{Stochastic Gradient Descent(SGD)}~\cite{Lecun1998}. This algorithm updates the parameters $\{\theta\}$ by subtracting the gradients derived from the BP algorithm,
\begin{equation}
    \mathbf{\theta}_{t+1} = \mathbf{\theta}_t - \eta\, \partial_{\mathbf{\theta}}\mathcal{L}_t,
\end{equation}
where $\eta$ is the \textit{learning rate} and $t$ indicates the $t$-th time repetition (\textit{epoch}) for the training data. In each epoch, the \textit{SGD} updates parameters on batches randomly selected from the training data set. It results in the stochastic updates of parameter $\mathbf{\theta}$  on the loss function landscape. This stochasticity has been shown necessary for escaping possible local minima~\cite{Bottou1999}. One representative and widely used \textit{SGD} algorithm is \textit{Adam}~\cite{Adam2015}. It is designed to combine adaptive estimates of lower-order moments with the stochastic optimization for improving the stability.
As a practical example, the optimizer implemented in our following discussions can be expressed as,
\begin{align}
    \theta_{t+1} &=  \theta_{t} - \eta \frac{\hat{m}}{\sqrt{\hat{v}} + \xi},\\
    \hat{m} &\equiv m_{t+1} = \frac{\beta_1}{1-\beta_1} m_{t} + \partial_\theta\mathcal{L}_{t},\\
    \hat{v} &\equiv v_{t+1} = \frac{\beta_2}{1-\beta_2} v_{t} + (\partial_\theta\mathcal{L}_{t})^2,
\end{align}
where $\xi$ is a small enough scalar for preventing divergence and $\beta_1, \beta_2$ are the forgetting factors($0.9, 0.99$ in default setting) for momentum term $\hat{m}$ and its weight $\hat{v}$.

Before entering the next section, it's helpful to introduce a specific architecture of neural networks, the AutoEncoder(AE)~\cite{tschannen2018recent}, shown in Figure~\ref{fig:NNs}. It consists of an encoder and a decoder. The encoder maps the input data $x$ to a typically smaller dimensional variable in a latent space, $z=f_{\phi}(x)$, while the decoder tries to convert the latent variable $z$ back into data, $\tilde{x}=g_{\theta}(z)$. The two components can be designed as two neural networks. The AE can learn a compressed representation of the input data that captures its most important features while minimizing the dissimilarity between input and output, for which the mean square error $L(x,\tilde{x})=\Sigma_i(x_i-\tilde{x}_i)^2$ serves heuristically well as the reconstruction loss. This architecture can be used to manifest the data in an interpretable latent space, which has proven to be a successful paradigm in physics~\cite{Iten:2020dpc}. Similarly, Principal Component Analysis (PCA) can be used as a linear method for dimensionality reduction and feature extraction~\cite{ladjal2019pca}. PCA computes the covariance matrix $C = X^T X $ and its eigenvectors and eigenvalues, where $X$ is the centered data matrix in a $d$-dimensional feature space. These eigenvectors are called the principal components. Sorting the principal components in descending order of eigenvalues, the first $k$ components with the largest eigenvalues are used to form the $k$-dimensional representation of a sample $x_i$ by projecting it onto the first $k$ principal components using $z_i = V_k^T (x_i - \mu)$. Where $V_k$ is the matrix of the first $k$ eigenvectors of $C$. Later, many works presented in this review involve applications of PCA, e.g., in sections~\ref{outlier}, \ref{flow_hic}, and \ref{sec_phases_obs}.

\subsubsection{Generative Models}\label{subsubsec:gm}

In general, deep learning algorithms can be categorized into two primary tasks: %two typical tasks can be cast for categorizing deep learning algorithms: 
discriminative modeling and generative modeling. From a probabilistic perspective, discriminative modeling aims to capture a conditional probability, $p(y|x)$, to enable the prediction over associated properties or class identities $y$ for a given object $x$, while generative modeling seeks to learn and sample from a joint probability (often situated in a high-dimensional space), $p(x,y)$, facilitating the generation of new data samples that follow the specified distribution or the same statistics represented by the training data. There are several excellent tutorials~\cite{Mehta:2018dln, Wang2018GenerativeMF} tailored to physics research that can be referred to for details. In physics, we often encounter similar generative tasks, such as in statistical many-body systems or lattice QFT simulations, we use Markov Chain Monte Carlo (MCMC) to sample new configurations and further estimate the physical observables of the system.

Various generative models have been developed by the ML community with deep inspiration from and into physics. Basically, most (but not all) of the generative models at present are with the maximum likelihood principle as common guidance. Specifically, generative models can be viewed as parametric models, $p_{\theta}(x)$, to approximate the probability distribution $p(x)$ of a system for which we may or may not have collected data. The Kullback-Leibler (KL) divergence (also called relative entropy), which measures the dissimilarity between the two distributions, can be used for this task, 
\begin{equation}
    \mathcal{D}_{KL}(p(x)||p_{\theta}(x))=\int p(x)\log\frac{p(x)}{p_{\theta}(x)}dx, 
    \label{eq:KL_divergence}
\end{equation}
the minimization of which under
given observational data for the system, $\mathcal{D}=\{x\}$, is equivalent to minimization of the negative log-likelihood (NLL),
\begin{equation}
    \mathcal{L}=\frac{1}{|\mathcal{D}|}\sum_{x\in\mathcal{D}}\log p_{\theta}(x), 
    \label{eq:NLL}
\end{equation}
thus the maximum likelihood principle. Note that for situations we do have collected data but with the distribution known up to its normalization factor, the reverse KL divergence, $\mathcal{D}_{KL}(p_{\theta}(x)||p(x))$, will be used instead, which corresponds to the variational free energy in essence.

In the following, we will briefly introduce several of the popular generative models with deep neural networks involved, including variational autoencoder (VAE), generative adversarial networks (GAN), autoregressive model, and normalizing flow (NF).

\begin{figure}[ht!]
    \centering
    \includegraphics[width = 0.55\textwidth]{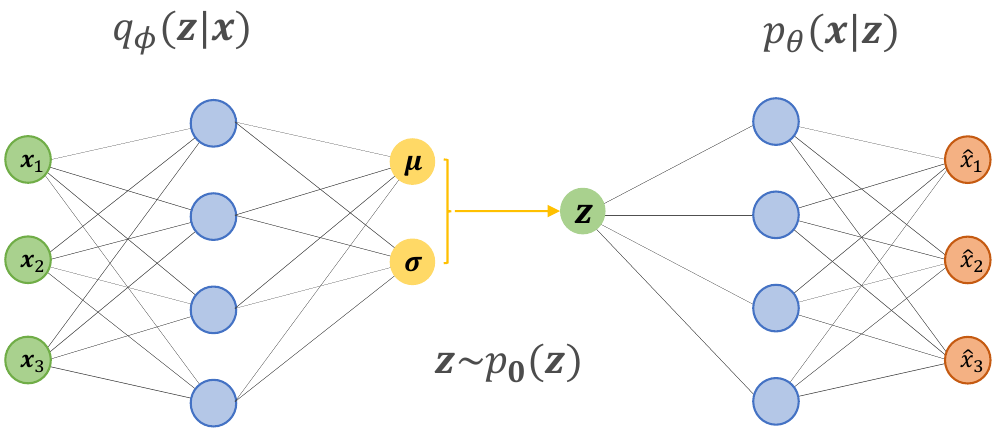}
    \caption{A demonstration of VAE whose latent layers output the parameters of prior $p_0(z)$.
    \label{fig:vae}}
\end{figure}

\textbf{Variational autoencoder (VAE)}~\cite{2013arXiv1312.6114K} added probabilistic latent to the classical autoencoder, which can perform variational inference under latent variable framework to the data distribution learning. For classical autoencoder (AE), the basic idea is to learn to reconstruct the input data going through a bottleneck structured pipeline, where the bottleneck hidden layer describes the \textit{code} to more efficiently represent the data also called latent variables usually with lower dimensional. The transformation from input data $x$ to the latent code $z$ is called encoder, and the left half of the network that reconstructs from the latent code to data manifold in output is called decoder or generator as shown in figure~\ref{fig:vae}.
%%%%%%%%%%%%%%%%%%%%%%%%%%%%%%%%%%%%%%%%%%%%%%%%%%%%%%%%
%%%%%%%%%%%%%%%%%%%%%%%%%%%%%%%%%%%%%%%%%%%%%%%%%%%%%%%%

In VAE, the latent variable per generation is conveniently assumed to follow a simple distribution called prior $p_0(z)$, such as a multivariate Gaussian distribution whose parameters, i.e., means $\mathbf{\mu}$ and standard deviations $\mathbf{\sigma}$, are taken from outputs of the encoder. The decoder thus provides a generative model by the trainable conditional probability $p_{\theta}(x|z)$. However, the introduction of the latent variable makes the entire data generation distribution intractable, since the potentially high-dimensional integration, $p_{\theta}(x)=\int p_{\theta}(x|z)p_0(z)dz$. This also leads to the intractable encoder probability or posterior distribution $p(z|x)=p_{\theta}(x|z)p_0(z)/p_{\theta}(x)$. To facilitate the MLE, or equivalently the minimization of the NLL on the training set, VAE uses a variational approach. Basically, VAE introduces a parameterized model $q_{\phi}(z|x)$ (modeled by a neural network, called an encoder) to approximate the posterior for the latent variable $p(z|x)$, which is, of course, the KL divergence between $q_{\phi}(z|x)$ and $p(z|x)$, $\mathcal{D}_{KL}(q_{\phi}(z|x)||p(z|x)$, can be invoked for the training objective, which derives the variational lower bound (also called evidence lower bound, ELBO) for the likelihood as the cornerstone for VAE,
\begin{equation}
    \mathcal{L}= \mathbb{E}_{q_{\phi}(z \vert x)}[\log p_{\theta}(x \vert z) + \log p_0(z) - \log q_{\phi}(z \vert x)]\le\log p_{\theta}(x), 
    \label{eq:ELBO}
\end{equation}
where
\begin{equation}
\mathbb{E}_{p(x)}[A(x)]=\int dx \,p(x) A(x), \quad \int dx\,p(x)=1\,,
\end{equation}
denotes the expectation value over the normalized probability distribution $p(x)$.

\begin{figure}[htbp!]
    \centering
    \includegraphics[width = 0.9\textwidth]{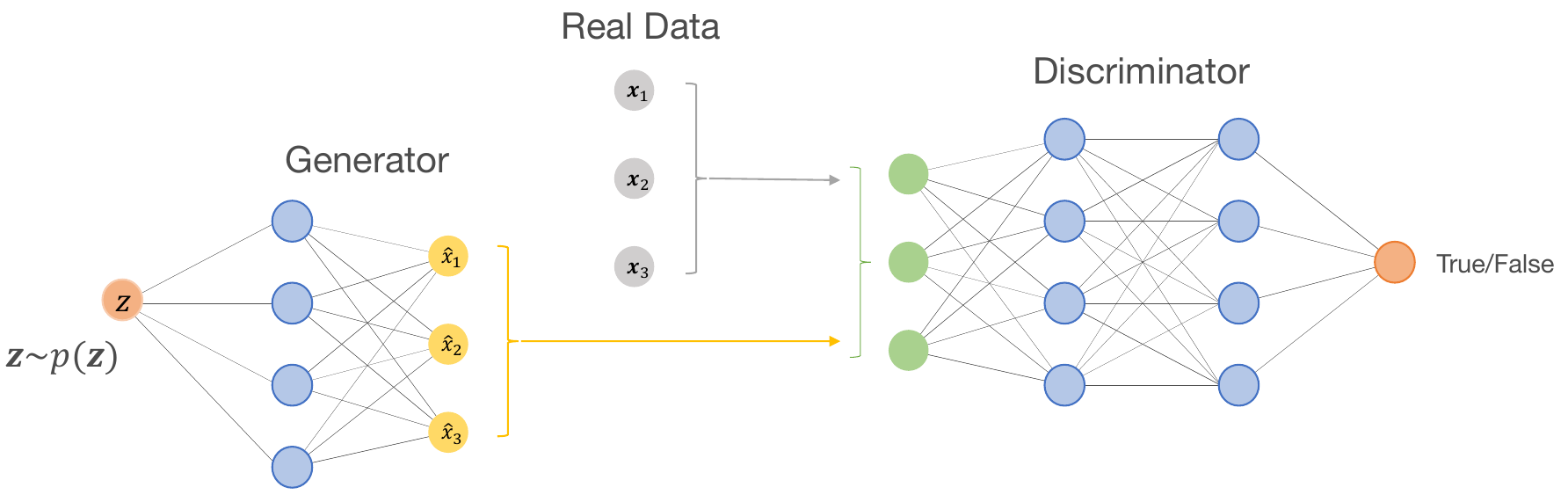}
    \caption{A demonstration of GAN, consisting of the generator and discriminator for generating and recognizing data.
    \label{fig:gan}}
\end{figure}

\textbf{Generative adversarial network (GAN)}~\cite{2014arXiv1406.2661G} is also a latent variable generative model, and introduces an adversarial training strategy to optimize the generator together with a discriminator (to be explained later). Basically, the GAN framework contains two non-linear differentiable functions, both represented by adaptive neural networks (see Fig.\ref{fig:gan}). The first is the generator $G$, which maps a random latent variable $z$ from a prior distribution $p(z)$ (usually multivariate uniform or Gaussian) to the target data space, $\tilde{x}=G(z)$, which induces an implicit synthetic distribution $p_G(x)$, which per training is to be pushed to the target distribution $p_{true}(x)$. The second one is called a discriminator $D(x)$ with a single scalar output for each data sample (fake or real), which tries to discriminate between real data $x$ and generated data $\hat{x}$ by training to output $D(x)=1$ and $D(\tilde{x})=0$. These two networks are trained in turn to improve their respective abilities against each other, mimicking a two-player min-max game (also called a zero-sum game) where their respective loss functions sum to zero, $\mathcal{L}_G + \mathcal{L}_D =0$. After optimization, the respective parameters $\theta_G$ and $\theta_D$ in the GAN will converge to the \textit{Nash equilibrium} state,
\begin{equation}
\theta_{G,D}^{*}=\arg \min\limits_{\theta_G} \max\limits_{\theta_D}(-\mathcal{L}_D(\theta_G,\theta_D))\,.
\label{nash-eq}
\end{equation}
where the generator excels in synthesizing samples that the discriminator cannot differentiate anymore from real ones, therefore the data distribution has been captured by the generator after training. The original GAN uses the loss function
\begin{equation}
\mathcal{L}_D=-\mathbb{E}_{x\sim p_{\rm true}}[\log D(x)] - \mathbb{E}_{z\sim p_{\rm prior}}[\log(1-D(G(z)))]\,,
\label{gan_l1}
\end{equation}
and we used
\begin{equation}
\mathbb{E}_{z\sim p_{\rm prior}}[A(G(z))] = \mathbb{E}_{\hat x \sim p_G} [A(\hat x)]\,.
\end{equation}
Mathematically it's proved that the training of GAN is equivalent to minimizing the Jensen–Shannon(JS) divergence generalized from the KL divergences, 
\begin{equation}
    \mathcal{D}_{JS}(p_{real}||p_G)=\frac{1}{2}(\mathcal{D}_{KL}(p_{real}||p_{mix})+\mathcal{D}_{KL}(p_G||p_{mix})), 
    \label{eq:JS}
\end{equation}
with $p_{mix}=(p_{real}+p_G)/2$. Thus, the GAN belongs to the implicit MLE-based generative model as well.
%%%%%%%%%%%%%%%%%%%%%%%%%%%%%%%%%%%%%%%%%%%%%%%%%%%%%%%%
%%%%%%%%%%%%%%%%%%%%%%%%%%%%%%%%%%%%%%%%%%%%%%%%%%%%%%%%
In order to add more flexible conditional control and improve the training stability, a variety of advanced development for GAN have been proposed, such as ACGAN~\cite{2016arXiv161009585O}, Wasserstein-GAN~\cite{2017arXiv170107875A}, improved Wasserstein-GAN~\cite{2017arXiv170400028G}. The main difference between Wasserstein GAN and the original GAN lies in the optimization objective, which turns the Wasserstein distance (also called Earth Mover distance) superior to the JS divergence in the Vanilla GAN.

\begin{figure}[ht!]
    \centering
    \includegraphics[width = 0.4\textwidth]{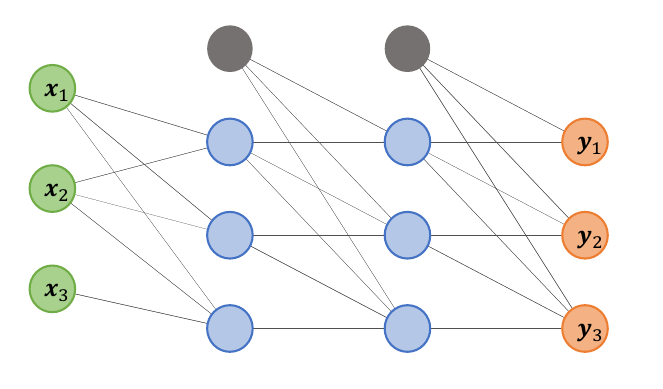}
    \caption{A neural network with autoregressive property, i.e., the value of outputs (red solid circle) at a given order only depends on its past inputs (green solid circles) and bias (black solid circle).
    \label{fig:autore}}
\end{figure}

\textbf{Autoregressive model}~\cite{2015arXiv150203509G,2016arXiv160106759V}, as an explicit MLE-based generative model, invokes the chain rule to decompose a joint probability into a series of conditionals,
\begin{equation}
    p_{\theta}(x)=\prod_{i}^{N}p_{\theta}(x_i|x_1,x_2,...,x_{i-1}),
\label{auto_prob}
\end{equation}
to model the data likelihood $p_{\theta}(x)$. The autoregressive model usually uses neural networks to represent each of the conditional probabilities involved, the collection of which as a whole can be viewed as a network (see Fig.\ref{fig:autore}) with a triangular weight parameter matrix for the simple fully connected network case (for cases using CNN or RNN, the integral weight matrix is masked accordingly), in order to respect the autoregressive properties specified by the probability decomposition (Eq.~\eqref{auto_prob}). In other words, it's designed so that each output element of the autoregressive network is independent of those input elements with a later index within a given order. Such a construction for the generative model with Eq.~\eqref{auto_prob} also allows efficient sampling from the joint distribution $p_{\theta}(x)$ by sequential sampling from each conditional probability.
%%%%%%%%%%%%%%%%%%%%%%%%%%%%%%%%%%%%%%%%%%%%%%%%%%%%%%%%
%%%%%%%%%%%%%%%%%%%%%%%%%%%%%%%%%%%%%%%%%%%%%%%%%%%%%%%%

It was initially developed for generating time-series data and, like recurrent neural networks (RNNs), also possesses autoregressive properties. For structured systems, it was extended to utilize convolutional layers, and accordingly, the PixelCNN structure~\cite{2016arXiv160605328V,2017arXiv170105517S} was constructed to fulfill the autoregressive transformation. If the network has been parameterized for Eq.~\eqref{auto_prob}, explicit MLE can be performed by minimizing the forward KL divergence $\mathcal{D}_{KL}(p_{real}||p_{\theta})$ if collected samples are available. Alternatively, if only the unnormalized distribution is known, minimizing the backward KL divergence $\mathcal{D}_{KL}(p_{\theta}||p_{real})$ can be achieved through sampling to estimate the associated entropy term.

\textbf{Normalizing flow (NF)}~\cite{2015arXiv150505770J, 2019arXiv190809257K} puts more effort into performing explicitly the maximum likelihood estimation (MLE), through introducing an invertible, bijective, and differentiable transformation $f_{\theta}$ (parameterized with networks) between a simple latent space $z$ and the complex data space $x=f_{\theta}(z)$ (see Fig.~\ref{fig:nflow}). The essential idea lies in the change of variable theorem which manifests the conservation of probability in connecting the latent space and target data space, 
\begin{equation}
    p_{\theta}(x)=p(z)|\det(\frac{\partial z}{\partial x})|=p(f_{\theta}^{-1}(x))|\det(\mathcal{J}_{f_{\theta}^{-1}})|,
    \label{eq:change-of-variable}
\end{equation}
through which the likelihood can be evaluated and maximized to optimize the transformation, thus the generator $f_{\theta}(z)=x$. One usually adopts special lattice structures that make Jacobian determinant evaluations accessible and practically computable, such as NICE (Non-linear Independent Components Estimation)~\cite{2014arXiv1410.8516D} or Real NVP (Real-valued Non-Volume Preserving)~\cite{2016arXiv160508803D}, both with a triangular Jacobian matrix. 

%%%%%%%%%%%%%%%%%%%%%%%%%%%%%%%%%%%%%%%%%%%%%%%%%%%%%%%%
\begin{figure}[htbp!]
    \centering
    \includegraphics[width = 0.75\textwidth]{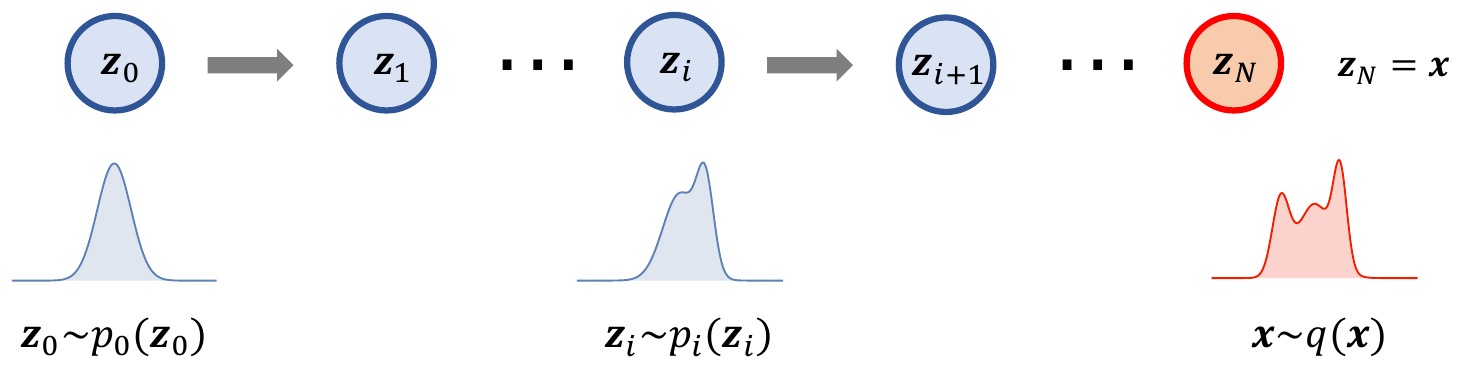}
    \caption{Sketch of a normalizing flow model, deforming a simple distribution $p(\mathbf{z}_0)$ to a complex one $q(\mathbf{x})$ step by step. 
    \label{nflow}}
\end{figure}
%%%%%%%%%%%%%%%%%%%%%%%%%%%%%%%%%%%%%%%%%%%%%%%%%%%%%%%%

Basically, NICE or Real NVP composes a sequence of simple affine coupling layers (represented by DNN) which split the sample (e.g., image, or field configuration as discussed in Sec.~\ref{sec:3:flow_based}) into two subsets and transform one subset variables conditioned on the other, thus the transformation is with triangular Jacobian matrix, whose determinant can be efficiently evaluated. Specifically, the $i^{\text{th}}$ affine coupling layer in NICE~\cite{2014arXiv1410.8516D} reads,
\begin{equation}
\left\{
\begin{aligned}
    z^{i}_{1:k} &= z^{i-1}_{1:k} \\
    z^{i}_{k+1:N} &= z^{i-1}_{k+1:N}  + m_{\theta}^i(z^{i-1}_{1:k}),
\end{aligned}
\right.
\label{flow_nice}
\end{equation}
with $z^i$ the output of the $i^{\text{th}}$ affine coupling layer (thus $z^0 = z\sim p(z)$ and final output $z^{L}=x\sim p(x)$), and $N$ the dimensionality or the number of variables in one sample. The above coupling layer operation can be easily inverted,
\begin{equation}
\left\{
\begin{aligned}
    z^{i-1}_{1:k} &= z^{i}_{1:k} \\
    z^{i-1}_{k+1:N} &= z^{i}_{k+1:N} - m_{\theta}^i(z^{i}_{1:k}),
\end{aligned}
\right.
\end{equation}
The neural network representation comes into play for the mapping construction of the translation function ($m_{\theta}: \mathbb{R}^{k}\rightarrow \mathbb{R}^{N-k}$), or in Real-NVP for the involved scaling and translation functions (as introduced in Sec.~\ref{sec:3:flow_based}). The nonlinearity or complexity introduced by the network would not affect the accessibility for the Jacobian determinant, which is the trace for the lower triangular matrix, and in the case of NICE, it is just unity.
Usually, several such affine coupling layers are combined to construct the normalizing flow transformation $f_{\theta}(z)$.

\subsubsection{Physics-motivated New Developments}
In the above, we reviewed the deep-learning techniques in detail.
Deep learning is a powerful tool that can be broadly applied to different aspects of high-energy nuclear physics. 
If physics knowledge is encoded in the set-up of the learning process, one can further improve the efficiency, even applicability, of deep learning. 
In the history of its development, several concepts have been raised following this philosophy in deep learning --- 
\textit{physics-inspired}, \textit{physics-informed}, and \textit{physics-driven} deep learning.

The concept of \textit{physics-inspired} machine learning is lacking of clear definition yet loosely summarized in Ref.~\cite{Ahmad:2020kdd}. It basically describes a paradigm that transfers ideas originating in physics into machine learning. Historically, many concepts from statistical physics deeply influence the foundations of machine learning, e.g., information entropy, information bottleneck, and energy-based model, see a more systematic introduction in Ref.~\cite{sompolinsky1988statistical,mezard2009information}. In many modern applications, such as Restricted Boltzmann Machines(RBM)~\footnote{It consists of two layers of nodes: a visible layer, which represents the input, and a hidden layer, which models the underlying interactions. The nodes in the two layers are linked by weighted connections, and the RBM defines a probability distribution over the visible units given the hidden units and vice versa. The RBMs have been used to construct many-body quantum states~\cite{2017Sci...355..602C}(mentioned in Sec.~\ref{sec_phases_obs}), also see more advanced research in Ref.~\cite{Carrasquilla:2020mas,Jia:2021qaa} about recent developments in Neural Network Quantum State (NNQS).}~\cite{ackley1985learning,fischer2012introduction}, Geometric Learning~\cite{Bronstein:2016thv,Bronstein:2021mdi} and Diffusion Models~\footnote{The Diffusion Model is a cutting-edge development in deep learning. It consists of a forward diffusion process and a reverse generation process. The forward diffusion process incrementally adds Gaussian noise to the samples in multiple small steps until it reaches a purely normal distribution. After training, generation can be initiated from a normal distribution by repeatedly applying the learned inverse denoising transformation until the sample is in the data space. See a recent review~\cite{Yang2022DiffusionMA}.}~\cite{ho:2020denoising,song2021scorebased}, the physics origins are explicit. The backbone of modern physics, symmetry, is also widely manifested in the design of neural network architecture~\cite{Goodfellow2016,Mattheakis:2019tyi,Kicki:2021so}, e.g., translation invariance in CNNs~\cite{zhang1988shift,Goodfellow2016}, 
Group Equivariant Convolutional Networks~\cite{pmlr-v48-cohenc16}, permutation invariance in Point Nets~\cite{qi2016pointnet}, and rotational invariance in E3NN~\cite{e3nn_paper}, and gauge invariance in the partition function of a lattice field theory~\cite{Albergo:2019eim, Kanwar:2020xzo, Albergo:2021vyo}.

Physics knowledge is implemented in \textit{physics-informed} deep learning~\cite{Raissi:2017zsi,2021NatRP...3..422K} in a different manner. Some physics properties cannot be encoded in the network architecture explicitly. Instead, one can include extra terms in the loss function to ``punish'' the violation of the desired properties. Let us take a dynamical system evolving in time as an example. One can train a network with the loss function, including the violation of the observation and the underlying equation of motion. A successfully trained network would be able to predict the observables at any given time without solving the differential evolution equation. Compared to \textit{physics-inspired} DL, \textit{physics-informed} ones take the physics knowledge as a soft constraint, as one can hardly guarantee the loss term to be exactly zero.
More introduction can be found in recent review~\cite{2021NatRP...3..422K}. 

Last but not least, \textit{physics-driven} deep learning is proposed to train differentiable problems. The Back-Propagation method discussed in Sec.~\ref{subsubsec:dl} is designed to optimize the parameters in a network so that its output matches the desired observables. In some problems, the observables are not naively the network output but some function (or functional) of it. If such a function or functional relation is differentiable, one may incorporate their derivatives with the Back-Propagation method such that parameters can be efficiently optimized using gradient-based methods. See also Ref.~\cite{2021arXiv210905237T} as a comprehensive introduction.

In general, physics-driven machine learning incorporates more physics information during training, resulting in a reduced data requirement compared to the other two methods. It is important to note that these three approaches are not mutually exclusive, and one can apply various techniques to the same problem from different perspectives. A specific instance aimed at reconstructing the equation of state of nuclear matter from measurements of neutron star mass and radius will be provided in Section~\ref{sec:new}.

\subsection{Outline}
\begin{figure}[!htbp]
\begin{center}
\includegraphics[width=1.0\textwidth]{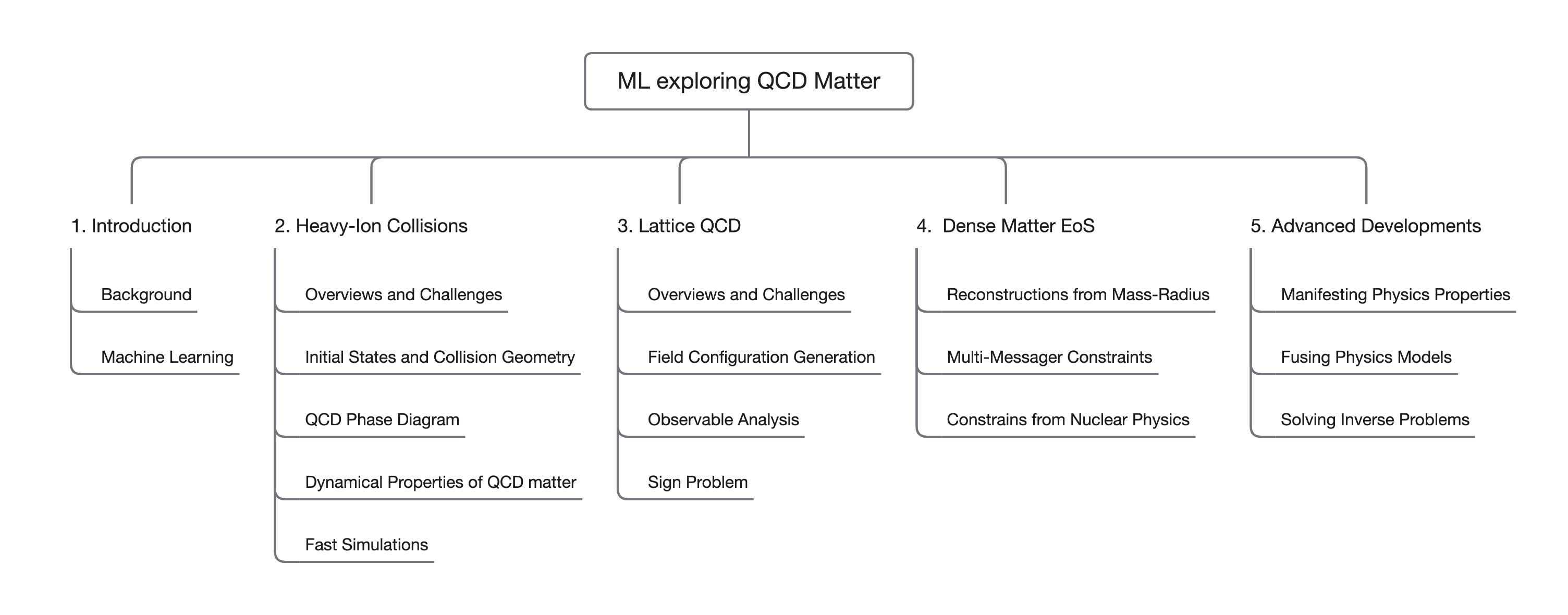}
\caption{Outline of this review.}\label{fig:1:Outline}
\end{center}
\end{figure}

The organization of this review is summarized in Fig.~\ref{fig:1:Outline}. We cover ML applications in Heavy-ion collisions (Sec.~\ref{sec:hic}),  Lattice QCD (Sec.~\ref{sec:lat}) and Neutron star equation of states (Sec.~\ref{sec:astro}). Finally, in Sec.~\ref{sec:new}, we also demonstrate some new advanced developments.
%%%%%%%%%%%%%%%%%%%%%%%%%%%%%%%%%%%%%%%%%%%%%%%%%%%%%%%%%%%%%%%%%%%%%%%%%%%%%%%%%%%%
%%%%%%%%%%%%%%%%%%%%%%%%%%%%%%%%%%%%%%%%%%%%%%%%%%%%%%%%%%%%%%%%%%%%%%%%%%%%%%%%%%%%
	\newpage
	\section{Heavy-Ion Collisions}\label{sec:hic}

\subsection{Overview and Challenges for HICs}\label{sub:hic_method}

Relativistic heavy-ion collisions (HICs) create an extreme environment for QCD matter in the laboratory~\cite{BRAHMS:2004adc, PHENIX:2004vcz, PHOBOS:2004zne, STAR:2005gfr, Schukraft:2011cz, Steinberg:2011qq, Wyslouch:2011zz}, e.g., at RHIC and LHC (also see Ref.~\cite{Baym:2001in} for a history background retrospect), with the highest temperature~\cite{Muller:2012zq, Braun-Munzinger:2015hba, Muller:2006ee}, the strongest magnetic field~\cite{STAR:2018gyt, Muller:2018ibh}, and fastest rotational angular velocity on Earth~\cite{STAR:2017ckg}. In this extreme environment, hadronic matter is expected to be deconfined to form a strongly interacting and free-roaming quark--gluon plasma (QGP), which is the same matter that was created in the early universe $10^{-6}$s after the \textbf{Big Bang}~\cite{Yagi:2005yb, Wang:2016opj, Fukushima:2020yzx}.

The goals of the HIC physics include (but not limited to),
\begin{itemize}
\item \textbf{Look for evidence of deconfinement.} E.g., the strong collective flow~\cite{Teaney:2000cw}, the enhancement of strange particles~\cite{Rafelski:1982pu}, the suppression of high-$p_T$ particles explained by the jet-medium interaction in QGP~\cite{Wang:1992qdg} and the suppression of heavy quarkonium~\cite{Matsui:1986dk}.
\item \textbf{Study the QCD phase structure.} For instance, what is the nature of the phase transition between QGP and a hadron resonance gas (HRG) with equal proportions of matter and antimatter? Moreover, what type of phase transition occurs between QGP and normal nuclear matter with a finite baryon number? If there are crossover and first-order phase transitions present in the diagram, does it indicate the presence of a critical endpoint in between? Furthermore, what experimental evidence supports the occurrence of critical phenomena? For further information on these topics, please refer to~\cite{Gupta:2011wh, Luo:2017faz}.
\item \textbf{Study the equation of state (EoS) of QGP}, e.g., the relations between local pressure, local energy density, entropy density, and local temperature. Will the EoS provided by lattice QCD lead to a reasonable dynamical evolution of QGP that can describe the momentum distribution of final state particles in HICs? For a more comprehensive overview, refer to~\cite{Shuryak:2004cy, Bazavov:2009zn, Borsanyi:2016ksw}. 
\item \textbf{Study the sensitivity of the physical observables to the physical parameters in the collisions.} For example, what are the effects of shear and bulk viscosity of QGP? What are the effects of the freeze-out temperature and the hadronic cascade? What are the effects of initial nuclear structure and gluon saturation? One may refer to~\cite{Bernhard:2018hnz, Bernhard:2019bmu, JETSCAPE:2020shq, Nijs:2020ors} and the references therein.
\item \textbf{Study the jet-medium interaction and in-medium effects for heavy bound states.} E.g., the shape and structure of the jet shower due to the jet-medium interaction, the sensitivity of the medium response to the underlying EoS of QGP, and the heavy quark potential inside the medium. Refer to~\cite{Wang:2004dn, Vitev:2002pf, JET:2013cls, Rapp:2018qla} for an overview.
\item \textbf{Search for CP violation and rotation/spin polarization (Chirality and Vorticity in QCD)} in the strong interaction. coupling between the classical, collective orbital angular momentum and the spin, an intrinsic quantum property, of a single-hadron. 
\end{itemize}

\subsubsection{``Standard Model'' of Simulating HICs}\label{sub:hic_method:standard_model}
To describe the whole process of a heavy ion collision, one has to construct hybrid models with different physics at different stages of the collision~\cite{Yagi:2005yb, Busza:2018rrf, Lappi:2016gmk, Elfner:2022iae}. Here we briefly summarize the state-of-the-art modeling for each stage. In the pre-collision state, a \textit{Monte Carlo model} is used to determine the positions of the nucleons inside a 3-dimensional nucleus, to determine the collision patterns between two nuclei, e.g., the impact parameter of a collision, the orientation of the deformed nucleus, which can lead to more complex tip-tip or body-body collisions. During the collision, \textit{color-glass condensate and saturation models} are used to account for the physics of special relativity and vacuum fluctuations in the nucleus to calculate the fluctuating local entropy deposition in the overlap region. After the formation of locally equilibrated QGPs, \textit{relativistic hydrodynamics} is used to describe the dynamical evolution of the local temperature and fluid velocity in the expanding QGP. Hadrons form at the boundaries of the QGP and will interact with each other via \textit{transport models}. In parallel to the dynamical evolution of the soft(low energy) particles in HIC, the hard(high energy) partons with extremely high energy and momentum will pass through the QGP and interact with the thermal partons in the QGP. The physics is described by QCD, where partons will split and collide. The differential cross-sections are usually provided by pQCD calculations in leading order. 
    
\emph{\textbf{Monte Carlo models for the initial nuclear structure}} ---
According to the charge distribution of the nucleons in the nucleus, the density distribution of the nucleons is modeled using a deformed Woods--Saxon function~\cite{Woods:1954zz, Kahana:1969zz},
\begin{align}
\rho(r, \theta, \phi) = \frac{\rho_0}{e^{(r - R_0(1 + \beta_2 Y_{20}(\theta) + \beta_4 Y_{40}(\theta)))/a} + 1},
\end{align}
where $\rho_0$ is the nucleon density inside the nucleus, $R_0$ is the radius, $a$ is the diffusiveness, $\beta_2$ and $\beta_4$ are two deformation factors whose values determine the shape of the nucleus, e.g., whether it is prolate or oblate.
For $^{208}$Pb, whose proton number $82$ and neutron number $126$ are both magic numbers, the deformation parameters $\beta_2=\beta_4=0$, its shape is a perfect sphere~\cite{10.2307/1758208,Loizides:2017ack}. Other heavy-ionic nuclei like Au, U, Cu, O, Xe, Ru, and Zr are deformed to varying degrees.

In Monte Carlo simulations, $A$ nucleons are first sampled from the above deformed Woods--Saxon distribution for each nucleus. For each pair of the sampled nuclei, the probability $P(b) = 2b / R^2$ is used to sample the impact parameter $b$, where $R$ is the maximum distance between two nucleons for overlap. However, in this procedure, each nucleon is independently sampled, without accounting for the nucleon-nucleon correlation, clustering effects, or differences in proton and neutron distributions. These factors may need to be considered in specific studies.

\emph{\textbf{Color Glass Condensate}} ---
Due to special relativity (Lorentz contraction and dilation effect, specifically), at extremely high energy with respect to the lab frame, the shape of the nucleus is compressed along the beam direction and the lifetime of quantum fluctuations in the nucleus is extended~\cite{Mueller:1989st, McLerran:1993ka, McLerran:1993ni, McLerran:1994vd, Lappi:2006fp, Gelis:2010nm}. Consequently, virtual quark-antiquark pairs and gluons live long enough to participate in high-energy collisions.
As the collision energy increases, the gluons of typical longitudinal momentum correspond to a small momentum fraction($x$) of the incoming nucleons. Noting the small-$x$ region of parton distribution e.g., from Deep Inelastic Scatterings(DISc), the dominant contribution is from gluons, and the number of gluons in the projectile seen by the target increases with the collision energy. To leading order, the energy-momentum after the collision is given by
\begin{align}
T^{\mu\nu} = {1\over 4} g^{\mu\nu} F^{\alpha\beta}F_{\alpha\beta} - F^{\mu\alpha}F^{\nu}_{\beta},
\end{align}
where $F^{\mu\nu}$ is the field strength of the classical retarded color field $A^{\mu}$ described by the classical Yang--Mills equation,
\begin{align}
[D_{\mu}, F^{\mu\nu}] = I^{\nu}
\end{align}
where $I^{\nu}= \delta^{\nu+}\rho_1 + \delta^{\nu -} \rho_2$ is the external current associated with the fast-moving partons in the projectile with density $\rho_1$ and in the target with density $\rho_2$.

The IPGlasma model is a successful attempt to describe the HIC initial condition by solving the classical Yang--Mills equation for gluons radiated from color sources~\cite{Schenke:2012wb, Schenke:2012hg}. Solving the field equation is, however, computationally expensive. One may adopt phenomenological models, such as the Trento Monte Carlo model~\cite{Moreland:2014oya}
\begin{align}
s({\bf x_T}) = \left( \frac{T_A^p({\bf x_T}) + T_B^p({\bf x_T})}{2}\right)^{1/p}.
\end{align}
which generates the initial condition of entropy density($s$) as a $p$-powered average of the thickness functions ($T_A$ and $T_B$), where $T_A$($T_B$) is the nuclear matter in projectile(target) with the longitudinal direction integrated. $p$ is a dimensionless parameter that can be tuned. Also note that such CGC inspired initial condition to HICs is also phenomenologically interesting by itself~\cite{Xu:2014ega,Stoecker:2015zea,Stocker:2015nka,Zhou:2017zql}.
From a Bayesian global fit~\cite{Bernhard:2016tnd}, it has been found that the anisotropy of entropy deposition in the transverse plane of IPGlasma can be approximated by choosing $p \approx 0$.

\emph{\textbf{Relativistic Hydrodynamics}} ---
\label{sec:hydro}
The dynamical evolution of created fireball (QGP and HRG in local equilibrium) from HICs can be described by relativistic hydrodynamic equations,
\begin{align}
\nabla_{\mu} T^{\mu\nu}  = 0, \quad\;
\nabla_{\mu} J^{\mu} = 0,
\end{align}   
where $\nabla_{\mu}$ represents the covariant derivatives, $T^{\mu\nu} = (\varepsilon + P + \Pi) u^{\mu}u^{\nu} - (P + \Pi) g^{\mu\nu} + \pi^{\mu\nu}$ is the energy-momentum tensor of hot nuclear matter, with $\varepsilon$ the local energy density,  $P$ the local pressure given by the equation of state $P = P(\varepsilon, \mu_B)$, $\Pi$ the bulk viscosity, $u^{\mu}$ the fluid four-velocity satisfying $u^2=1$, $\pi^{\mu\nu}$ the shear viscous tensor, $g^{\mu\nu}$ the metric tensor. 
$J^{\mu} = n u^{\mu} + v^{\mu}$ is the charge current, where $n$ is the net charge density,
$v^{\mu}$ is the diffusion of the net charge, e.g., the net baryon diffusion.  
This set of equations is solved together with the Israel--Steward equations~\cite{Israel:1979wp} for $\Pi$, $\pi^{\mu\nu}$ and the baryon diffusion current $v^{\mu}$. 

One merit of relativistic hydrodynamics is that it encodes the equation of state provided by lattice QCD calculations.  Meanwhile, relativistic hydrodynamics is a complex dynamical evolution that involves many fundamental properties of hot QCD matter, e.g., shear viscosity over entropy density $\eta/s$, bulk viscosity over entropy density $\zeta /s$, baryon diffusion parameter $k_B$, the initial time $\tau_0$, the freeze-out temperature $T_f$. With the recent experimental data on HIC, Bayesian global fitting analysis has been used to determine these fundamental properties in relativistic hydrodynamics. Recent studies such as those by Bernhard et al. ~\cite{Bernhard:2018hnz, Bernhard:2019bmu, JETSCAPE:2020shq, Nijs:2020ors} have made significant progress in this area, providing valuable insights into the nature of hot and dense QCD matter. Hydrodynamics provide the complete evolutionary history of the soft partons, e.g., the energy density, the pressure, and the temperature at any given space-time point $(t, x, y, z)$. This information is invaluable as it provides not only the spectra and momentum anisotropy of soft hadrons but also serves as a crucial background for understanding jet quenching and the production of direct photons and dileptons. Relativistic hydrodynamics thus presents multiple avenues to explore the properties of Quark--Gluon Plasma (QGP).

There are many different implementations of the relativistic hydrodynamics, either in 2+1D or in 3+1D~\cite{Kolb:2003dz, Muronga:2001zk, Hirano:2005xf, Chaudhuri:2006jd, Romatschke:2007jx, Dusling:2007gi, Song:2007ux, Du:2019obx, Inghirami:2016iru, Okamoto:2017ukz, Nijs:2021clz, Florkowski:2010cf, Strickland:2012bc, Schenke:2010nt, Ryu:2015vwa, Shen:2017bsr, Bazow:2013ifa, Shen:2014vra, Pierog:2013ria, Sakai:2020pjw, Karpenko:2013wva, Pang:2018zzo, Yin:2015fca, Hattori:2022hyo, Guo:2019mgh, Shi:2020htn}. 
2+1D hydrodynamics assumes that the fluid is boost-invariant along the beam direction, while 3+1D hydrodynamics does not. This difference has important implications for the study of high-energy nuclear collisions. 2+1D hydrodynamics provides a good approximation for mid-rapidity HIC phenomena, while 3+1D hydrodynamics is more suitable for understanding rapidity-dependent phenomena. However, simulating one HIC event in 3+1D hydrodynamics with finite shear viscosity can take up to 60 times longer than simulating the same event in 2+1D hydrodynamics. As a result, it is often more convenient to accumulate a large amount of data using 2+1D hydrodynamics, which can also be beneficial for machine learning studies. For example, in the Bayesian analysis~\cite{Bernhard:2016tnd} that requires millions of events, VISH2+1 hydrodynamic model is used to generate data with event-by-event fluctuating initial conditions. 

\emph{\textbf{Transport models for the hadronic cascade}} ---
The particle-yield ratios between different hadrons are well described by the statistical model, assuming that the particles emitted from the freeze-out hypersurface obey the Fermi--Dirac distribution for baryons and the Bose--Einstein distribution for mesons, with mass $m_i$ and chemical potential $\mu_i$,
\begin{equation}
{1 \over 2\pi}{d N \over dY p_T dp_T d\phi} = {\rm dof} \int {d^3 p \over (2\pi)^3}f(p\cdot u, T),
\label{eq:freezeout}
\end{equation}
where ${\rm dof} = 2 \times {\rm spin} + 1$ is the spin degeneracy, $f$ is the distribution function given by,
\begin{equation}
 f = {1 \over e^{p\cdot u/T} \pm 1},
\end{equation}
where $\pm$ stands for baryons and mesons respectively. In hybrid models, one can thus sample hadrons using the above distribution function in the comoving frame of each fluid cell with local temperature $T$ and fluid velocity $u^{\mu}$ on the freeze-out hyper surface. 
    
The hadrons emerging from the freeze-out hypersurface can undergo two distinct processes. In the immediate aftermath of particlization, the hadron density remains high, and the many-body interactions between hadrons are too strong to be accurately described by transport models. During this initial stage, relativistic hydrodynamics remains a reliable tool. As the hadronic system evolves and the hadron density decreases, the interactions between hadrons become more tractable and can be described by hadronic cascade models such as UrQMD~\cite{Bleicher:1999xi} and SMASH~\cite{Weil:2016zrk}. 
    
\emph{\textbf{Jet quenching in QGP}} --- There have been tremendous efforts in developing theoretical tools to model the interactions between energetic partons and thermal partons in QGP and simulate them in phenomenological studies ~\cite{Gyulassy:1993hr, Wang:1994fx, Baier:1996sk, Baier:1994bd, Baier:1996kr, Zakharov:1996fv, Wiedemann:2000za, Guo:2000nz, Wang:2001ifa, Zhang:2003yn, Schafer:2007xh, He:2015pra, Cao:2016gvr, Casalderrey-Solana:2014bpa, Shi:2018izg, Cao:2020wlm, Qin:2015srf, Majumder:2010qh, Armesto:2011ht, Qin:2009bk, Jeon:2003gi, Noronha-Hostler:2016eow, Andres:2016iys, Bianchi:2017wpt, Chien:2015vja, Andres:2019eus, Yazdi:2022bru, Shi:2022rja}. While these models take different assumptions in the derivation, here we take the linear Boltzmann transport equations~\cite{He:2015pra} as an example to demonstrate the theoretical framework,
\begin{equation}
p_i \cdot \partial f_i=\int \sum_{i,j,k} \prod_{a=i, j, k} \frac{1}{(2 \pi)^3 }\frac{d^3 p_a}{2 E_a}\left(f_k f_l-f_i f_j\right)\left|\mathcal{M}_{i j \rightarrow k l}\right|^2 \frac{\gamma_j}{2} S_2(\hat{s}, \hat{t}, \hat{u})(2 \pi)^4 \delta^4\left(p_i+p_j-p_k-p_l\right)+\text { inelastic },
\label{eq:LBT}
\end{equation}
where $f_i$ is the distribution function for the hard parton $i$ whose initial position is sampled from the distribution of binary collisions between nucleons and whose four-momentum is provided by the Pythia Monte Carlo model~\cite{Sjostrand:2019zhc}. The right-hand side includes the gain term $f_k f_l$ and the loss term $-f_i f_j$ due to the collisions of hard partons with thermal partons. The scattering amplitude $\mathcal{M}_{i j \rightarrow k l}$ is given by tree-level pQCD calculations. $\gamma_j$ is the spin and color degeneracy of the thermal parton $j$. $S_2$ is a control factor to eliminate the collinear divergence, which is given by
\begin{equation}
S_2(\hat{s}, \hat{t}, \hat{u}) = \theta(-\hat{s} + \mu_D^2 < \hat{t} < -\mu_D^2) \theta(\hat{s} > 2 \mu_D^2),
\label{eq:s2_collinear}
\end{equation}
where $\hat{s}, \hat{t}, \hat{u}$ are three Mandelstam variables and $m_D = {\sqrt{6} \over 2} g T$ is the Debye screening mass for gluons and light quarks. 

The last inelastic term accounts for the gluon radiation induced by elastic scattering. The gluon radiation rate $\Gamma_a^{\text {inel}}$ is taken from higher-twist calculations,
\begin{align}
 \frac{d \Gamma_a^{\text {inel}}}{d z d k_{\perp}^2}=\frac{6 \alpha_{\mathrm{s}} P_a(z) k_{\perp}^4}{\pi\left(k_{\perp}^2+z^2 m^2\right)^4} \frac{p \cdot u}{p_0} \hat{q}_a(x) \sin ^2 \frac{\tau-\tau_i}{2 \tau_f}
\label{eq:gluon_radiation}
\end{align}
where $z$ is the energy fraction of the emitted gluon with respect to the hard parton $a$, $k_{\perp}$ is the transverse momentum of the emitted gluon, $\alpha_{\mathrm{s}}=\frac{g^2}{4\pi}$ is the coupling constant, $P_a(z)$ is the parton splitting function, $\hat{q}_a(x)$ is the transverse momentum transfer per unit length due to elastic scattering. The $\sin$ function encodes the quantum interference between gluons emitted at different time. The $\tau_i$ is the production time of the parent parton $a$ and $\tau_f = 2 p_0 z(1-z) / (k_{\perp}^2+z^2 m^2)$ is the formation time of the emitted gluon.

All simulation techniques and packages mentioned above were developed by various research groups, making it essential to organize them systematically for a phenomenological study of high-energy nuclear collisions. To address this challenge, the JETSCAPE topical collaboration was formed to provide a comprehensive framework that implements the state-of-the-art simulation packages for every stage. This allows for consistent and coherent studies of both soft and hard physics in heavy ion collisions. In particular, the JETSCAPE framework constructs a multi-stage jet-medium interaction model that accounts for different energy and virtuality scales of hard partons in the jet shower compared to the medium. For instance, JETSCAPE uses MATTER~\cite{Kordell:2017hmi} to simulate the parton splitting at early times when the parton $a$ is highly virtual, while LBT~\cite{He:2015pra}, MARTINI~\cite{Schenke:2009gb}, or AdS/CFT~cite{Pablos:2017csi} are used to simulate interactions between medium and hard partons with low virtuality. By integrating these models into the JETSCAPE Monte Carlo model, the production of simulated data in heavy ion collisions can be achieved in a consistent and unified manner.
	
\subsubsection{HIC Challenges}
\label{hic_challenges}
Theoretical simulations of high-energy heavy ion collisions (HIC) and experiments at RHIC and LHC have generated a vast amount of data. Unlike $e^+ + e^-$ and $p+p$ collisions, HIC produces thousands of final-state hadrons in every single Au+Au or Pb+Pb collision event at the highest energies of RHIC and LHC. These produced hadrons consist of both soft and hard particles, with the former primarily originating from the freeze-out of the quark--gluon plasma (QGP) while the latter arising from jet fragmentation or heavy quarkonium decay.
    
\begin{figure}[htbp!]
    \centering
    \includegraphics[width = 0.6\textwidth]{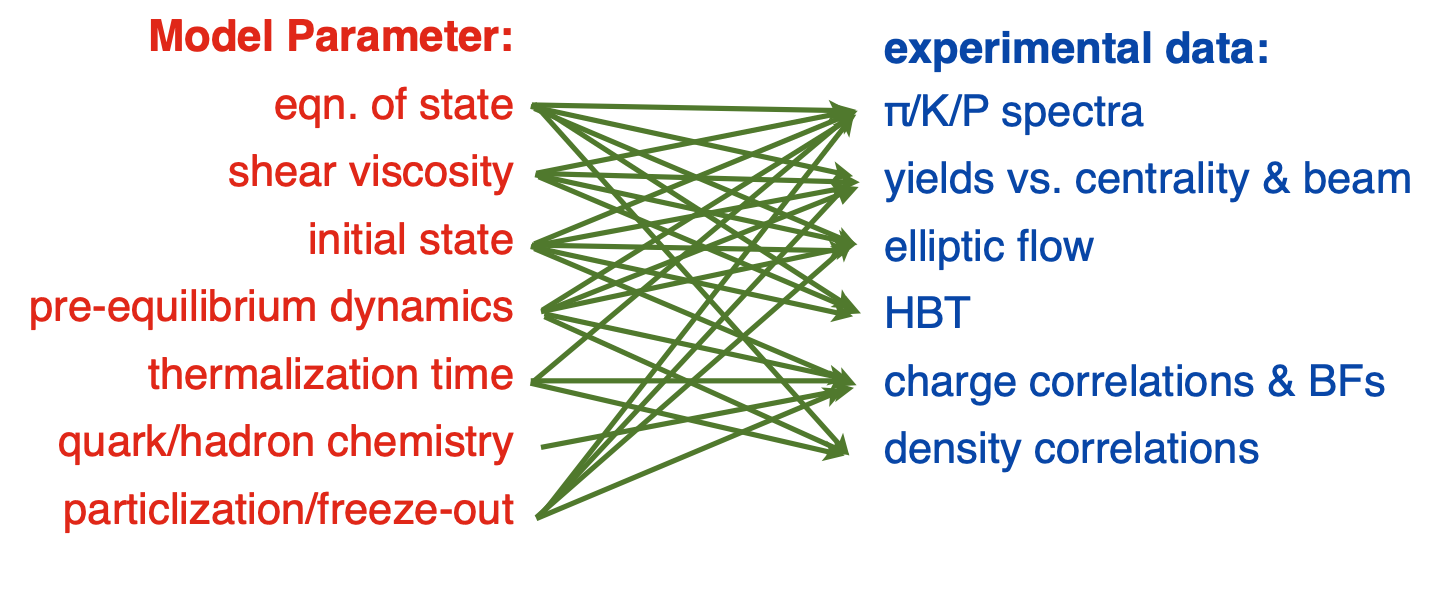}
    \caption{The entanglement between different model parameters and physical observables in heavy ion collisions. Taken from Ref.~\cite{Bass:2017zyn} with permission.
    \label{fig:Bass}}
\end{figure}

The data is routinely compressed to low-dimensional representations in physics space for data-model comparison, e.g., the charged multiplicity as a function of pseudo-rapidity, the $p_T$ spectra, the anisotropic flow coefficients, the di-hadron correlation, etc. However, the initial state, the QCD matter EoS, the QGP properties such as shear and bulk viscosity are all coupled to final state observables intricately. For example, both the shear viscosity and the freezing temperature will change the slope of the $p_T$ spectra. Traditional data analysis techniques encounter difficulties in determining a physical property using the final state observables, as its value will change adaptively with the values of other model parameters, as shown in Fig.~\ref{fig:Bass}.

%%%%%%%%%%%%%%%%%%%%%%%%%%%%%%%%%%%%%%%%%%%%%%%%%%%%%%%%
%%%%%%%%%%%%%%%%%%%%%%%%%%%%%%%%%%%%%%%%%%%%%%%%%%%%%%%%
    
Data analysis in high-energy HICs is a typical \textit{inverse problem}\footnote{See Sec.~\ref{sec:5:inverse} for more discussions on inverse problems.}: given all the different known physics factors, well-established standard computational models such as 3+1 D hydrodynamics can mimic the collision dynamics and obtain corresponding final state information, but given only limited and cross-impacted final state measurements, how to extract knowledge about the early time physics happened from the intricate entanglement influence, it is a very non-trivial inverse inference task. The lifetime of the formed QGP in high-energy HICs is only about $10^{-23}$ seconds, which is too short (also too small) to be resolved. What can be determined experimentally is the four-momentum of the final state hadrons or their decay products, but what we are interested in is the initial state and the properties of the QGP early in the collision evolution. It is unknown whether the physical information will survive the violent expansion and leave an imprint in the final state due to entropy production and memory decay induced by information loss. Additionally, it is unknown whether this is an ill-defined inverse problem, where different parameter combinations may lead to degenerate output (final-state information).

In recent years, two techniques have shown promise for extracting physical information from exotic final state particles of high-energy HICs. The first method is Bayesian analysis, which involves using all available experimental data to simultaneously determine multiple model parameters through global fitting~\cite{Bass:2017zyn, Bernhard:2016tnd, Bernhard:2018hnz, Bernhard:2019bmu, JETSCAPE:2020shq, Nijs:2020ors}. The second method is deep learning, which can search for observables that are sensitive to only one physical property \cite{Pang:2016vdc,Steinheimer:2019iso,Benato:2021olt}. Deep learning has been shown to be the best pattern recognition method for extracting features and feature combinations from high-dimensional data and mapping them to specific physical properties.
	
\subsection{Initial States and Collision Geometry}

Since HIC challenges can be viewed as an inverse problem, it is natural to ask whether the initial state of HIC can be extracted from the momentum distribution of the final state hadrons. Unfortunately, due to entropy production and information loss, there is no guarantee that the initial state can be recovered from the final state particles' information. However, we do know that some of the initial state information survives the complex dynamical evolution of strongly coupled matter and is present in the exotic particles of the final state. For instance, momentum anisotropies have a strong correlation with the geometric eccentricity of the initial state as well as with the impact parameter. It is also known that the deformation of the nuclear structure leads to intrinsic collision patterns related to the distributions of the final state charge multiplicity and momentum anisotropy. It would be interesting to explore whether other initial state fluctuations and correlations are transformed into correlations of final state particles in momentum space, such as neutron skin, nucleon-nucleon correlations, $alpha$ clusters in heavy nuclei~\cite{He:2021uko}, and gluon saturation at relativistic energies.

\subsubsection{Impact Parameter Determination}\label{hic_b}
If one wants to extract one number at the initial state from the final state particles using machine learning, the first one to try is the impact parameter, which is the transverse distance between two colliding nuclei. It is essential to know the impact parameter $b$ for determining the event geometry and further analysis, e.g., the volume estimation in fluctuation analysis. However, we have no direct control or measurement over $b$ in experiments. Usually, final state observables such as charged multiplicity are used to define centrality classes based upon models such as (Monte-Carlo)-Glauber simulation~\cite{Miller:2007ri}, through which one can get only a likely distribution of $b$ for a given centrality class. Here the centrality classes are usually specified from percentiles of those final state observables, and further guide the grouping of events but in a rough manner. For centrality estimation, the number of participant $N_{part}$, has been shown to be a strongly model dependent quantity~\cite{OmanaKuttan:2023cno}.
%This impact parameter is so important that recently RHIC spent several million on the detector to improve the precision of impact parameter determination~\cite{Kagamaster:2020oon}. 
The significance of the impact parameter is also underscored by the recent massive investment by RHIC in their detector enhancement specifically aimed at improving the precision of impact parameter determination~\cite{Kagamaster:2020oon}.
ML algorithms can provide a useful tool in discriminating initial conditions for HICs from the final state accessible information.

\emph{\textbf{Early Attempts with ML to Determine Impact Parameter}} ---
Many early attempts with the usage of ML techniques for determining the impact parameter mainly resort to simple algorithms, e.g., feed-forward neural network using conventional observables~\cite{David:1994qc,Bass:1996ez,Haddad:1996xw} or support vector machine (SVM)~\cite{DeSanctis:2009zzb} or Bayesian inference with also K-means clustering~\cite{Li:2022mni}. Later, working directly on the two-dimensional transverse momentum and rapidity spectra of final state particles, the DNN, Light Gradient Boosting Machine (LightGBM) and the Convolutional Neural Networks (CNN) algorithms were employed to estimate the impact parameter at intermediate energies~\cite{Li:2020qqn, Li:2021plq, Zhang:2021zxd} with UrQMD provide the simulated events, and then also on realistic cases including detector responses of the S$\pi$RIT Time Projection Chamber into the simulation events~\cite{Tsang:2021rku}. Compared to conventional means, these ML-based methods show better performance in estimating the impact parameter, especially giving rise to the ability in recognizing the central collision events. Such strategy for impact parameter estimation using DNN and CNN was also discussed for Au+Au collisions at $\sqrt{s_{NN}}=200$ GeV~\cite{Xiang:2021ssj} using final state energy spectrum in $(p_x, p_y)$ space as input. With data simulated from AMPT model, the trained CNN gives good prediction accuracy for impact parameter with a mean absolute error about 0.4 fm for $2<b<12.5 fm$, while for central and peripheral collisions the performance gets worse. For HICs at LHC energies, the Gradient Boosted Decision Trees (GBDTs) were used~\cite{Mallick:2021wop} for impact parameter regression in Pb+Pb collisions at $\sqrt{s_{NN}}=5.03$ TeV with charged-particle multiplicity ($\langle dN_{ch}/d\eta\rangle$, $\langle N_{ch}^{TS}\rangle$) and mean transverse momentum ($\langle p_T\rangle$) as the input features, meanwhile, the transverse spherocity is obtained which characterize in two limits the hard and soft events\footnote{Transverse spherocity is defined for unit vector $\hat{\mathbf{n}}$ which minimizes the ratio $S_0=\frac{\pi^2}{4}(\frac{\sum_i\vec{p_{Ti}}\times \hat{\mathbf{n}} }{\sum_i p_{Ti}})^2$.}. In Ref.~\cite{Saha:2022skj}, besides the determination of impact parameter, two other quantities in characterizing the initial geometry--eccentricity and participant eccentricity, were also included as targets within ML-based regressions, where k-NearestNeighbors (kNN), ExtraTrees Regressor(ET) and the Random Forest Regressor(RF) models were employed based on the transverse momentum spectra as input features, model dependencies and generalizability of the trained were also discussed.  

\emph{\textbf{End-to-End b-meter with PointCloud Network}} ---
The detector's record in HIC has an inherent point cloud structure (as will also be discussed in Sec.~\ref{sec:2:eos:pcn}), which is defined as a collection of points as an unordered list with their record attributes, e.g., the position, charge, or momentum of particles. Such point cloud format in principle should hold the permutation invariance. Being specially developed, the PointNet provides an appropriate structure handling point cloud dataset and is meanwhile invariant under the ordering of the points (i.e., particles). Therefore, for HICs study, the PointNet-based models open up the possibility to work directly on the detector readout for physics exploration using pattern recognition strategy in the big data sense. As introduced in above, an accurate estimation of impact parameters on an event-by-event basis is non-trivial, much less the demand in working with detector output (hits or tracks) directly even before the particle identification. This actually forms an inverse problem, where the task of determining the initial impact parameter given purely detector output for the final state individual event is implicit by itself. In Ref.~\cite{OmanaKuttan:2020brq,OmanaKuttan:2021axp}, an end-to-end\footnote{Here \textbf{end-to-end} means the inference is performed on direct detector output without much preprocessing for the data.} impact parameter meter is devised with PointNet-based deep learning models and demonstrated for the Compressed Baryonic Matter (CBM) experiment. 
    
\begin{figure}[htbp!]
    \centering
    \includegraphics[width = 0.8\textwidth]{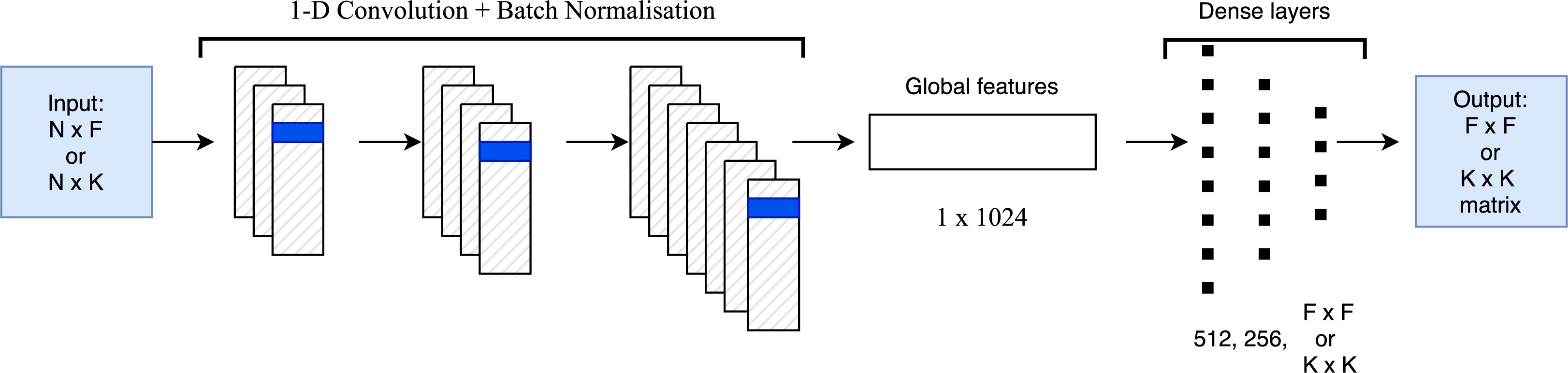}
    \caption{General structure of joint alignment network which induce transformations on input or features inside the PointNet. Taken from Ref.~\cite{OmanaKuttan:2020brq} with permission.
    \label{fig:pointnet}}
\end{figure}

As is under construction within the FAIR program at GSI, CBM aims at studying the properties of strongly compressed nuclear matter through heavy ion collisions with beam energies ranging from 2 to 10$A$ GeV. The key feature of CBM experiment is that it will have high event rate and trigger rate rendering rare particle detection and high statistic evaluation for some observables (e.g., higher order fluctuations or correlations), which however calls for fast real-time analysis in selecting events from the flooding stream of data produced from the collision experiment. To this end, thus being able to work with directly the detector output, Ref.~\cite{OmanaKuttan:2020brq, OmanaKuttan:2021axp} adopted a supervised training strategy to construct an end-to-end impact parameter meter using PointNet, where the training data is prepared from UrQMD followed by the CBM detector simulation using CbmRoot. As input to the PointNet-based impact parameter estimator, each event is represented with particle hits or tracks information in point cloud format. One PointNet-based model consisting of two joint alignment networks as shown in Fig.~\ref{fig:pointnet} is constructed to capture the inverse mapping from the detector output of CBM to impact parameter and was trained supervisedly with training data simulated from UrQMD+CbmRoot. 

%%%%%%%%%%%%%%%%%%%%%%%%%%%%%%%%%%%%%%%%%%%%%%%%%%%%%%%%
%%%%%%%%%%%%%%%%%%%%%%%%%%%%%%%%%%%%%%%%%%%%%%%%%%%%%%%%

 As seen from Fig.~\ref{fig:pointnet}, the adopted 1-D convolutions with kernels of size 1 across point features for individual points (particles) ensures the operations to be order invariant, after which symmetric function like global Average or Max pooling is used to aggregate all global features for further processing (usually use dense layers), thus preserves the order invariance for the model. Note that the used 1-D CNN is equivalent to applying a shared dense layer operation to lift the feature space of each particle to a high-dimensional transformed feature space.  As demonstrated in Ref.~\cite{OmanaKuttan:2020brq}, such PointNet-based model provides fast and accurate, end-to-end and event-by-event impact parameter determination, showing around $0.5 fm$ mean squared error. By using such a trained model, It is promising to access event-by-event centrality estimation from the hits record for CBM experiment.

\subsubsection{Unsupervised Centrality Outlier Detection}\label{outlier}
Though there is a well-established ``standard model'' for HICs simulations, due to the inaccessible first principle calculation to the collisional dynamics, new or not well-understood physics may be lacking in the models, e.g., the critical phenomenon. From the experimental point of view, in HICs the new and interesting physics might be hidden in rare events and/or rare particles and also their inter-correlations, such as higher-order cumulants of particle multiplicity distributions. In addressing such rare probes, \textit{large event rate} is scheduled especially for the new experiments, like CBM or PANDA at FAIR, the RHIC beam energy scan, and the ALICE experiment at CERN. Efficient online event selection is then in urgent need, which for example could pop out events potentially containing different characteristics or statistics as compared to the background. Experimentally the disability in this issue may induce artifacts for the interpretation of the observable, such as an imperfect centrality determination or detector malfunction's contamination induced different event types within a bulk background, manifested as two-bump distribution in the proton number distribution (which could also be induced by critical fluctuation physically), is discussed~\cite{Bzdak:2018uhv} to be able to explain the STAR observed deviation~\cite{Luo:2015ewa} for net-proton multiplicity distribution from simple binomial in Au+Au collisions at $\sqrt{s_{NN}}=7.7$ GeV. Otherwise, this deviation may also signal a critical endpoint in the QCD phase diagram.
%%%%%%%%%%%%%%%%%%%%%%%%%%%%%%%%%%%%%%%%%%%%%%%%%%%%%%%%%%%%%%%%%%%%%%%%%%%%
\begin{figure}[!hbtp]\centering
\includegraphics[width=0.43\textwidth]{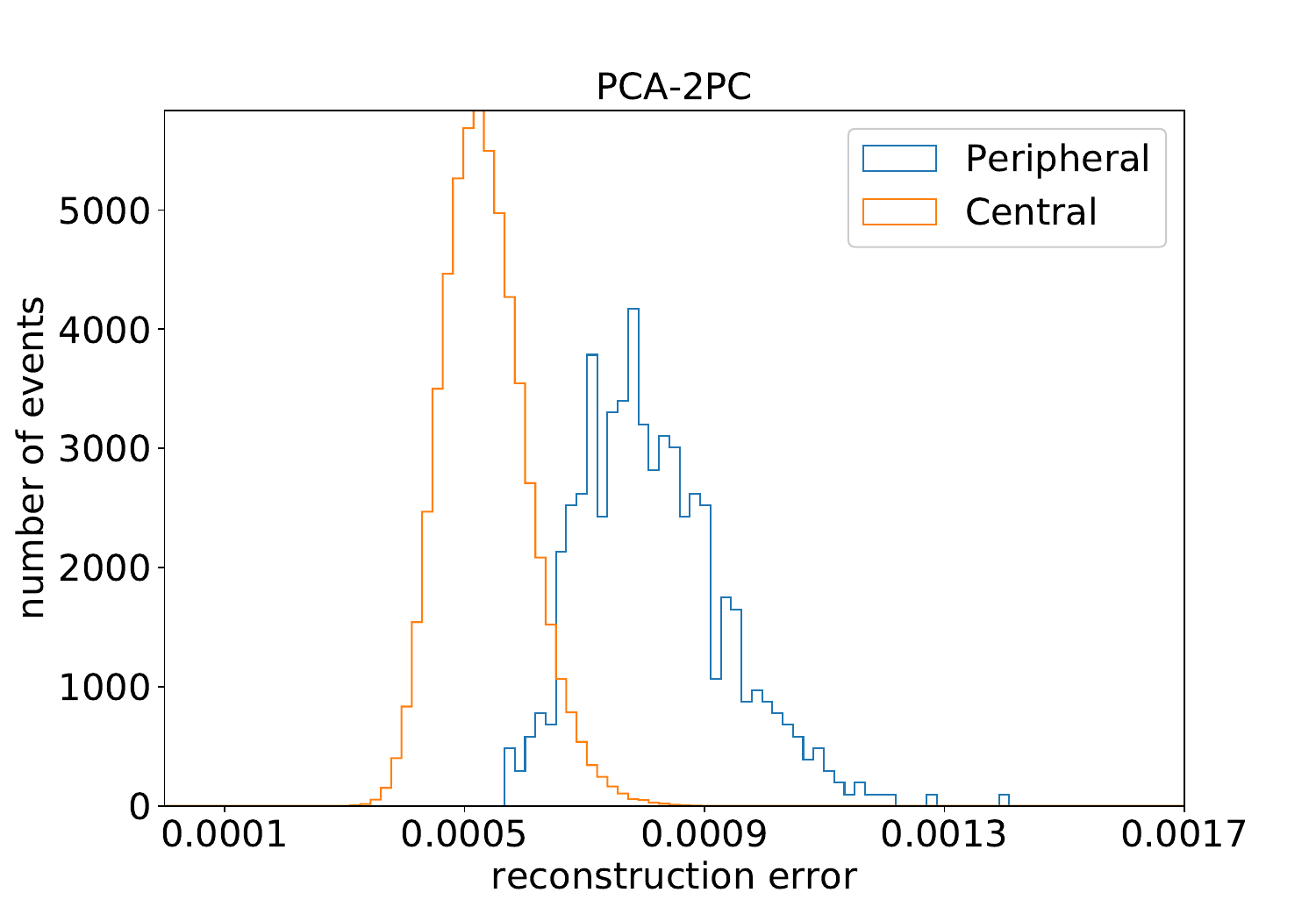}
\includegraphics[width=0.44\textwidth]{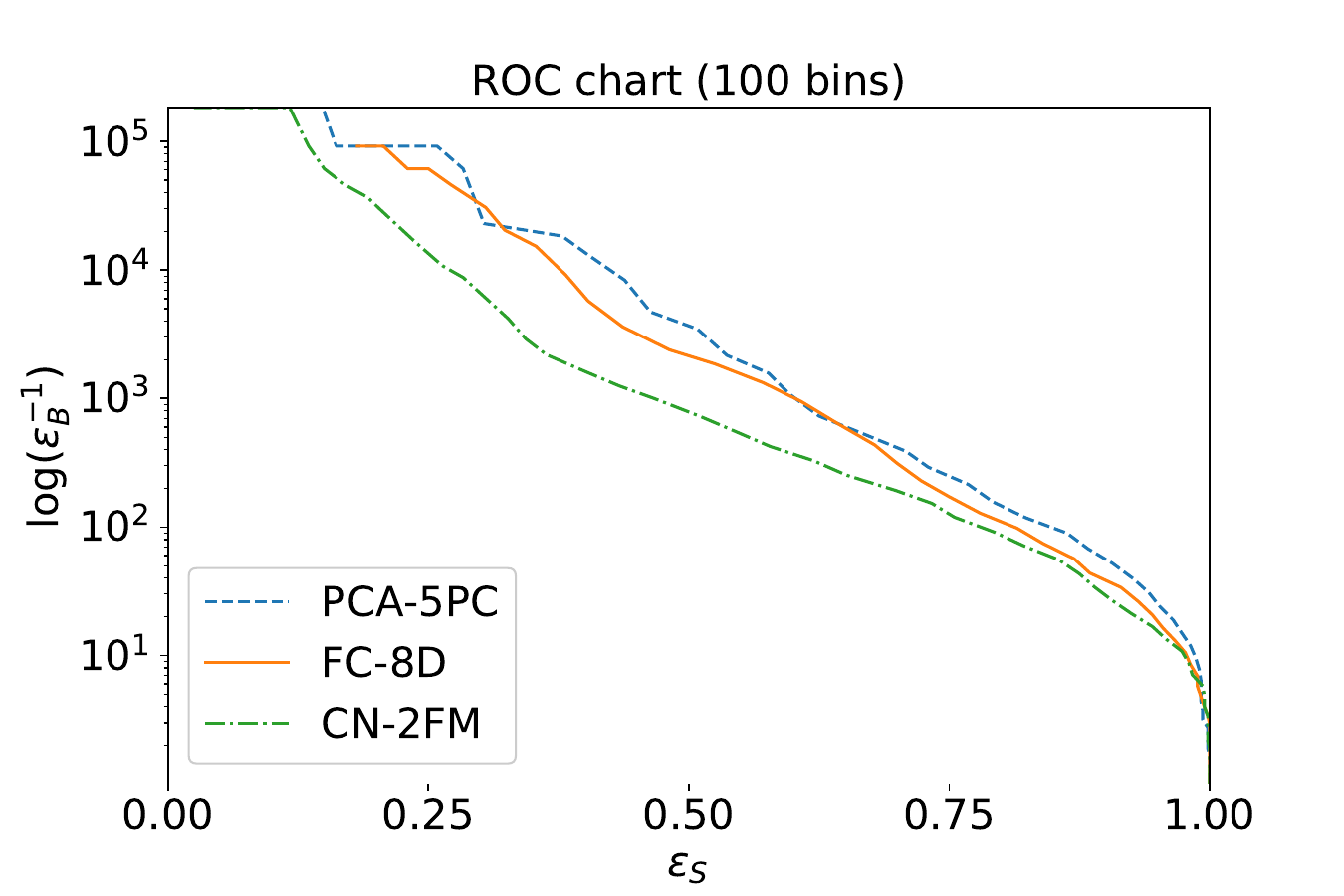}
\caption{Taken from Refs.~\cite{Thaprasop:2020mzp}. (left) the histogram of the RE in Eq.~\eqref{eq:re} for PCA using 2 PC; (right) the ROC curves for best models of each type: PAC, AEN-FC, AEN-CNN.\label{fig:outlier}}
\end{figure}
%%%%%%%%%%%%%%%%%%%%%%%%%%%%%%%%%%%%%%%%%%%%%%%%%%%%%%%%%%%%%%%%%%%%%%%%%%%%

In such a context, outlier detection which in ML community has been well-developed is proposed to detect anomaly event in HICs~\cite{Thaprasop:2020mzp}. As an exploratory study in testing the methodology, Ref.~\cite{Thaprasop:2020mzp} generated an ensemble of mixed events with the majority to be central collisions (with impact parameter $b=3$ fm) and a few portions to be peripheral ($b=7$ fm) using UrQMD transport model, with the ratio of the two centrality classes set to be able to give the anomalous proton number distribution observed from STAR. The task is unsupervisedly select the outlier--peripheral collision events out of the background--large number of central events. Each event is represented by the 2D-histograms of transverse momentum ($p_x, p_y$) with normalization performed to eliminate the easy characteristic of total number of particles per event in distinguishing the centrality classes. Two kinds of algorithms were discussed for this outlier detection, principal component analysis (PCA) and autoencoder network (AEN), both can achieve dimensionality reduction over the training data-set while the output (for PCA i.e., the inverse transformation) largely reproduce the input. In the reduced feature space (e.g., with just 2 or 3 dimensions), the feasible visualization provides the most simple way for data clustering and outlier identification, which however is limited by the expressibility of the used feature space. Another way to realize outlier detection with arbitrary latent feature space is utilizing the \textit{reconstruction error} (RE) of each event to guide the selection, e.g., the mean-squared error (see Fig.~\ref{fig:outlier} for its histograms and performance in identifying outlier), 
\begin{equation}
RE(x)=\frac{1}{N}\sqrt{\sum^N_{i=1} (x^{rec}_i-x_i)^2},
    \label{eq:re}
\end{equation}
where $x^{rec}_i$ is the $i^{th}$ component of the reconstructed event for $x$. Since the learning algorithms (PCA or AEN) is trained to reproduce the input event through reduced latent space on data-set with majority to be background types, the RE reflecting the reconstruction loss is expected to be different between two types of events if they possess distinct characteristics or statistics, thus provides also a promising indicator for anomaly identifier. It is found that the PCA with five principal components gives better outlier detection performance than complex AEN in the considered task~\cite{Thaprasop:2020mzp}.

\subsubsection{Nuclear Structure Inference}\label{hic_nuclear_structure}
The momentum distribution of final state hadrons produced in heavy ion collisions exhibits a clear dependence on the initial nuclear structure, providing an opportunity to determine the initial nuclear structure using the final state hadrons in heavy ion collisions~\cite{Bally:2022vgo}. The momentum anisotropy as a function of the charged particle multiplicity differs significantly between Pb+Pb and U+U collisions. $^{208}{\rm Pb}$ is a double magic nucleus with a shape close to a perfect sphere, while $^{238}{\rm U}$ has a shape close to a prolate watermelon, resulting in more complex collision geometries in U+U collisions compared to Pb+Pb collisions. However, nuclear structure is not limited to shape deformations. Other nuclear structures include the neutron skin ~\cite{PREX:2021umo}, which reveals the difference between the distribution of protons and neutrons in the nucleus, the $\alpha$ cluster, crucial for light nuclei, and the nucleon-nucleon correlation. A deep residual neural network was trained to predict the initial nuclear deformation parameters $\beta_2$ and $\beta_4$ ~\cite{Pang:2019aqb}. The network was found to accurately obtain the absolute values of $\beta_2$ and $\beta_4$ using distributions of geometric eccentricity and total entropy. Deep learning is also used to classify whether there is $\alpha$ cluster structure in $^{12}{\rm C}$ and $^{16}{\rm O}$ using AMPT simulations of collisions between light nuclei \cite{Bailey:2021bzo,He:2021uko}.
    
One difficulty in determining the nuclear structure using high-energy heavy ion collisions is caused by the effect of special relativity. On one hand, the nuclear structure information along the beam direction is strongly destroyed due to Lorentz contraction. On the other hand, fluctuations of sea quarks and gluons will participate in the collision due to time dilation. The size of the nucleons inside the nucleus may increase due to the parton cloud surrounding the nucleons. This will definitely change the initial fluctuations in the overlapped region of the colliding nucleus, e.g., the sizes of the hot spots in the initial entropy-density distribution. Fortunately, the Trento Monte Carlo model has taken this change into account for the initial condition.

Another difficulty in determining nuclear structure from high-energy heavy ion collisions is due to the evolutionary processes that link the initial state to the final observed particles. Such processes include pre-hydro evolution, hydrodynamic expansion, and hadron scattering, where many transport coefficients are undetermined. It has been proposed~\cite{Jia:2021oyt} that such a subtlety might be avoided in a contrast experiment for isobar systems --- a pair of nuclei that have the same mass number but different electric charge, and therefore different nucleon structures. The ratio of the final state observables between the isobar collisions is expected to be sensitive only to the nucleon structure and not to the transport parameters. The natural question then arises as to whether it is possible to determine the nucleon structures of the isobar pair by focusing only on the ratio of the final state observables. In Ref.~\cite{Cheng:2023ciy}, the authors attempted to answer this question by taking the Monte Carlo Glauber model as an ``emulator'' to map the nuclear structure onto the final state particle distribution, and performing Bayesian inference of the nuclear structure parameters from different combinations of ``mock data''. The authors found that it was not possible to constrain the structure for both nuclei simultaneously if only the ratios of the multiplicity distribution and the elliptical, triangular and radial flows were fitted. However, the authors found that simultaneous reconstruction is plausible if one includes the multiplicity distributions for both collision systems. Such a pioneering investigation paves the way for further nuclear structure studies in HICs using Bayesian inference based on more computationally expensive but realistic models for the final state particle distribution.

\subsection{QCD Phase Diagram}
\label{qcd_phase_hic}

\subsubsection{Bayesian Analysis of QCD EoS at $\mu_{B}=0$}
Although the Lattice QCD predicts that the transition between QGP and HRG is a smooth crossover in the high temperature and zero baryon chemical potential region, there is no clear evidence from experimental data produced at RHIC and LHC. Ref.~\cite{Pratt:2015zsa} uses Bayesian analysis to extract the speed of sound square $c_s^2$ as a function of temperature, with the help of relativistic hydrodynamic simulations and experimental data. The $c_s^2 = {dP / d\epsilon}$ is a direct measure of QCD EoS, comparing $c_s^2(T)$ from data with that given by Lattice QCD provides direct evidence of a smooth crossover.

 \label{sec:2:eos_bayes}     
	\begin{figure}[htbp!]
    \centering
    \includegraphics[width = 0.62\textwidth]{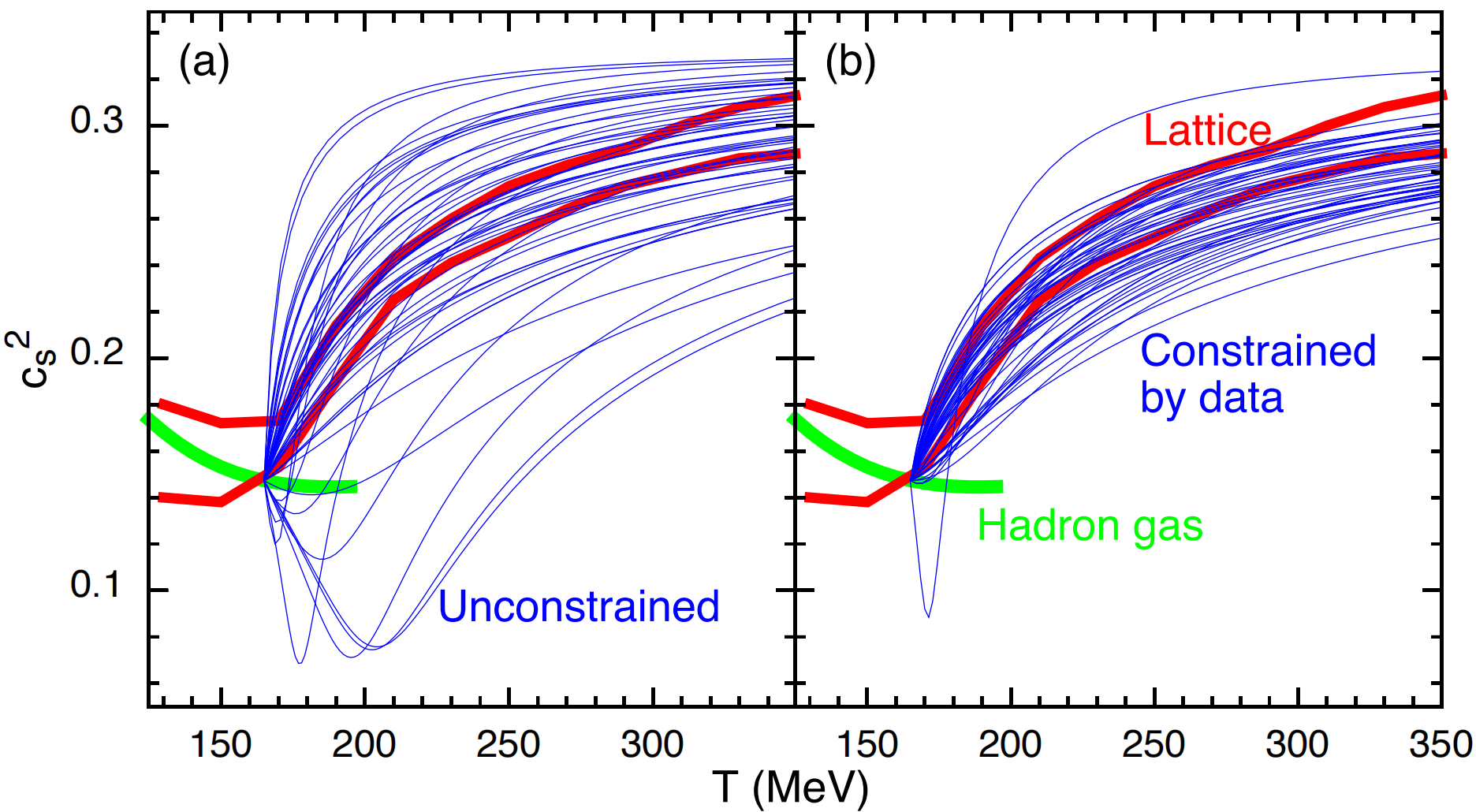}
    \caption{The prior and the posterior of the speed of sound square $c_s^2$ of hot nuclear matter using Bayesian analysis from Ref.~\cite{Pratt:2015zsa}.
    \label{fig:bayes_eos}}
\end{figure}

The $c_s^2$ is parameterized as a function of energy density in the following,
\begin{equation}
c_s^2(\epsilon) =  c_s^2(\epsilon_h) + \left({1\over 3} -  c_s^2(\epsilon_h) \right) {X_0 x + x^2 \over X_0 x + x^2 + X'^2 }
\label{eq:cs2_vs_T}
\end{equation}
where $X_0 = \sqrt{12} R X' c_s(\epsilon_h)$, $x \equiv \ln \frac{\epsilon}{\epsilon_h}$, $\epsilon_h$ is the energy density at $T=165$ MeV, $R$ and $X'$ are the two parameters in the EoS to be determined. Randomly choosing $R$ and $X'$ from the range $-0.9 < R < 2$ and $0.5<X'<5$ generate the unconstrained EoS that varies in a large region between $c_s^2=0.05$ and $c_s^2=0.33$, as shown in Fig.~\ref{fig:bayes_eos}-a. This corresponds to the 
a priori distribution of $c_s^2$ parameters together with other 12 parameters in the model $P(\theta)$. 

The likelihood between experimental data and relativistic hydrodynamic outputs is given by,
\begin{align}
    P(D | \theta) = \Pi_i \exp\left( -(z_i(\theta)-z_{ i,{\rm exp}})^2 / 2\right)
\end{align}
where $D$ represent the experimental observables, $z_i(\theta)$ and $z_{i, {\rm exp}}$ are the principle components of $D$ from model outputs and the experimental data correspondingly. In Bayesian analysis, MCMC is used to sample $\theta$ from the posterior distribution of parameters $P(\theta | D)\propto P(D | \theta) P(\theta)$ to generate a batch of $c_s^2(T)$ curves. The EoS constrained by data is in good agreement with Lattice QCD calculations for $T>150$ MeV. 

Note that recently in Ref.~\cite{OmanaKuttan:2022aml} the Bayesian method has been successfully applied in constraining the density dependence of the QCD EoS for dense nuclear matter(at high $\mu_{B}$) based on low energy HICs experimental data. Specifically, the mean transverse kinetic energy and integrated elliptic flow of protons from HICs in the beam energy range $\sqrt{s_{NN}}=2\sim 10$ GeV are taken as the evidence for the inference. Up to around 4 times nuclear saturation density ($4\rho_{0}$) the EoS is extracted from the analysis which describes other observables (the directed flow $v_1$ and the $p_T$ dependent elliptic flow $v_2$).

\subsubsection{Identify QCD Phase Transitions using CNN}

Lattice QCD predicts that the transition between QGP and hadron resonance gas (HRG) at high temperature and zero baryon chemical potential is a smooth crossover~\cite{Ding:2015ona}. It also produces the QCD EoS, which describes the pressure as a function of energy density. Using Bayesian analysis, the QCD EoS can be parameterized and used in relativistic hydrodynamic simulations of the HIC to extract the QCD EoS at the highest RHIC and LHC energies. The extracted EoS is in agreement with lattice QCD calculations~\cite{Pratt:2015zsa}.
It is conjectured that at high baryon chemical potentials, the transition between QGP and HRG is a first order phase transition. The endpoint of the first-order phase transition close to the crossover is called the critical endpoint. Tremendous efforts have been made to look for this critical endpoint (CEP) or region, including also the deployment of machine learning techniques. 

\begin{figure}[htbp!]
    \centering
    \includegraphics[width = 0.7\textwidth]{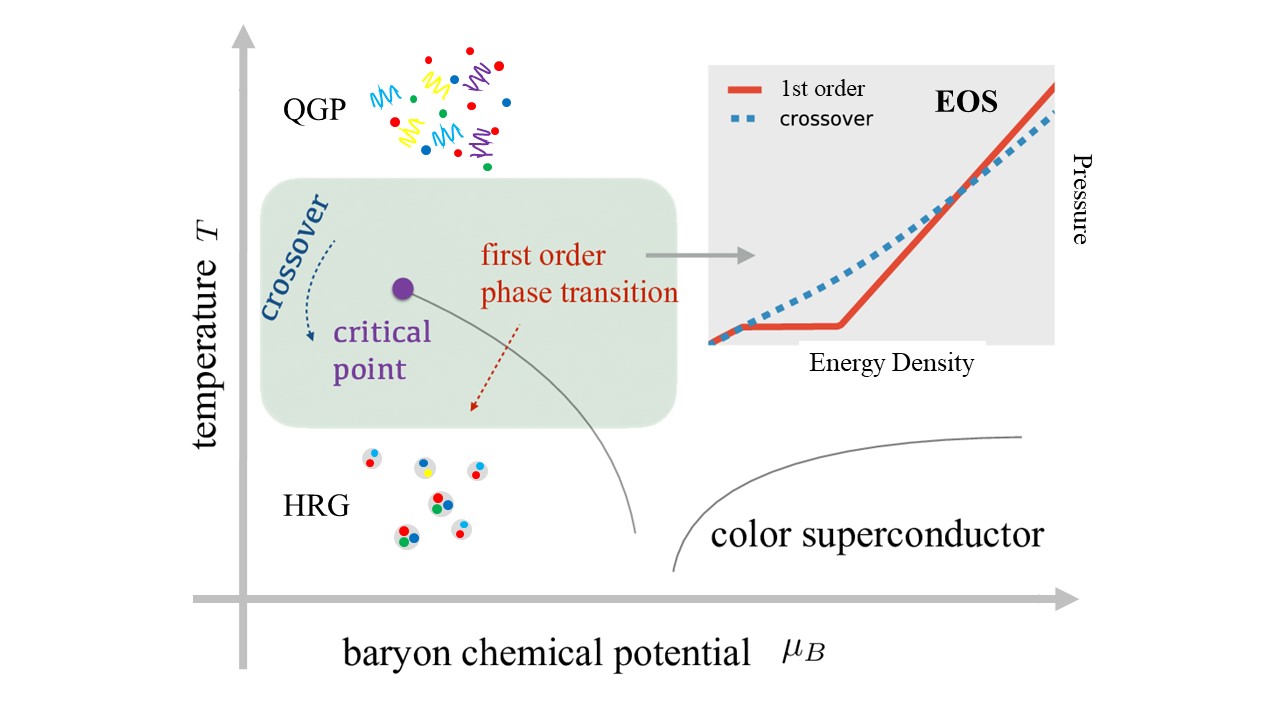}
    \caption{A schematic chart for QCD transition in the QCD phase diagram and the associated equation of state.
    \label{fig:phase_diag_eos}}
\end{figure}

Fig.~\ref{fig:phase_diag_eos} shows two different transition regions between QGP and hadron resonance gas (HRG) in the QCD phase diagram, the crossover region and the first order phase transition region. Different phase transitions lead to different equation of states  (EOS). 
For crossover EOS, the pressure as a function of energy density(the blue dashed curve) is smooth in the region between QGP phase at high energy density and HRG at low energy density. For first order phase transition (the red solid line), the pressure as a function of energy density has a plateau between QGP phase and HRG phase. As a result, the pressure gradient is zero in this region. Notice that the main driven force of fireball expansion is the pressure gradient. Difference of the pressure gradient in the phase transition regions between two EOS will lead to different evolution histories, which may encode the information of phase transition to the final state particles. 

%%%%%%%%%%%%%%%%%%%%%%%%%%%%%%%%%%%%%%%%%%%%%%%%%%%%%%%%

\begin{figure}[htbp!]
    \centering
    \includegraphics[width = \textwidth,trim=0 4cm 0 5cm, clip]{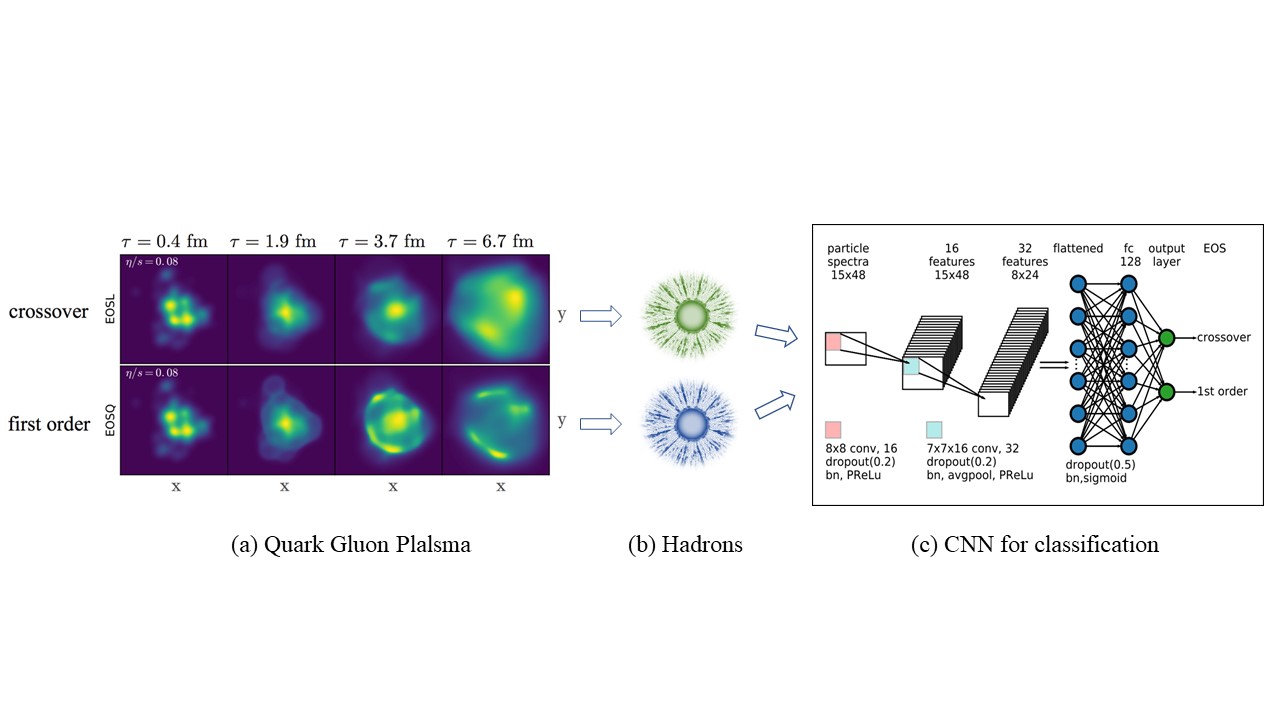}
    \caption{A schematic chart for QCD transition binary classification with CNN using final particle spectra from HIC as input.
    \label{fig:cnn_eos}}
\end{figure}

Fig.~\ref{fig:cnn_eos}-(a) compares the energy density distributions in the transverse plane for 4 time snapshots, for two different EOSs in CLVisc relativistic hydrodynamic simulations. The evolution histories are visually different and can be easily distinguished by human. The energy density distributions using a EOS with crossover as shown in the first row is much smoother than that shown in the second row for a first order phase transition. However, the QGP will convert to hadrons through particlization and hadronic cascade, as shown in Eq.~\ref{eq:freezeout}. 
What have been detected in experiment are the final state particles in momentum space. It is verified in ~\cite{Pang:2016vdc} that the event-by-event distributions of traditional observables for 2 different EOSs almost overlap. The shape of the event-by-event distribution are also sensitive to the initial condition, the shear viscosity and the freeze out temperature. It seems that there is no way to identify the used EOS from the spectra of a single collision event.

%%%%%%%%%%%%%%%%%%%%%%%%%%%%%%%%%%%%%%%%%%%%%%%%%%%%%%%%
%%%%%%%%%%%%%%%%%%%%%%%%%%%%%%%%%%%%%%%%%%%%%%%%%%%%%%%%
Ref.~\cite{Pang:2016vdc} gives the first exploratory study of using deep learning techniques to directly connect the raw final state information from the HIC experiment to the bulk properties of this QCD phase transition type encoded in the EoS. Inspired by the success of image recognition in computer vision, a deep convolutional neural network(CNN) is employed to learn and to directly represent the potentially existing inverse mapping from the final state information to the early time QCD matter bulk properties. Specifically, the CNN is used to classify two phase transition regions using simulated data from the parallelized relativistic hydrodynamics on GPU~\cite{Pang:2016vdc, Pang:2019int, Zhou:2018hsl}, with different EoS and phase transition types embedded in. 

Fig.~\ref{fig:cnn_eos}-(c) shows the schematic flow chart of the CNN. The input to the CNN is the the $(p_T, \phi)$ distribution of final state pions simulated by the CLVisc hydrodynamic model, with fluctuating initial conditions and different values of shear viscosity. The two dimensional spectra has 15 different $p_T$ and 48 different azimuthal angle $\phi$ which serves as input images with $15\times 48$ pixels. The output of the CNN have 2 neurons, representing the probability of the crossover and the first order phase transition, given by softmax activation as shown in Eq.~\ref{eq:eos_2class},
\begin{align}
    \hat{y} = \begin{bmatrix}
      p_{\rm crossover} \\
      p_{\rm 1st\ order}
    \end{bmatrix} = \begin{bmatrix}
      {e^{z_1} \over e^{z_1} + e^{z_2}} \\
      {e^{z_2} \over e^{z_1} + e^{z_2}}
    \end{bmatrix} 
    \label{eq:eos_2class}
\end{align}
where $\hat{y}$ is the discrete probability density distribution predicted by the network. 
$z_1$ and $z_2$ are the values of the last two neurons before the softmax activation. Multi-layer CNN and MLP are used to extracts correlations between particles in different momentum space in the input image, that can be used to make a final decision in the output layer.

The true label for crossover is $y=\left[y_1, y_2\right]^T=\left[1, 0\right]^T$, indicating that $p_{\rm crossover} = 1$ and $p_{\rm 1st\ order}=0$. The true label for 1st order phase transition is thus $y=\left[y_1, y_2\right]^T=\left[0, 1\right]^T$. The loss function contains a cross entropy loss between 2 distributions and the $l_2$ regularization term,
\begin{align}
    l(\theta) = - {1 \over m} \sum_{i=1}^m \left(y^i_1 \log p^i_{\rm crossover} + y^i_2 \log p^i_{\rm 1st\ order}\right) + {\lambda \over 2}||\theta||_2^2 
\end{align}
where $\theta$ represents all the trainable parameters such as the matrix elements in the convolution kernels, the weights and bias in the MLP. $m$ is the mini-batch size. The index $i$ represents the $i$-th sample in the mini-batch. The $||\theta||_2 = \sqrt{\sum_k \theta_k^2}$ is the $l_2$-norm used to constrain the magnitude of model parameters. The $\lambda$ is a small number set manually. After trained, the network generalizes well to data generated with another relativistic hydrodynamic model (iEBE-VISHNU~\cite{Shen:2014vra}), or to data generated with CLVisc with different initial conditions. In average, the prediction accuracy achieves $93\%$. 
 
It is worth noting that, the demonstration in Ref.~\cite{Pang:2016vdc} of using deep CNN to identify the QCD transition types from the final pion's spectra is only on the level of pure hydrodynamic evolution, with its superb classification accuracy in the testing stage clearly indicating that: the early time transition information especially its types within hydrodynamics (mimicking HICs) evolution can survive to the final state. The constructed inverse mapping by the trained deep CNN also shows robustness to different initial fluctuations and shear viscosities. In modeling heavy ion collision with more realistic consideration, hybrid simulations combining hydrodynamics together with after-burner hadronic cascade transport become more appropriate. Accordingly, for HICs this strategy of using supervised learning to capture the inverse mapping from final accessible information to early time desired physics could be further deepened.

\begin{figure}[hbpt!]
\centering
\includegraphics[width=0.48\textwidth]{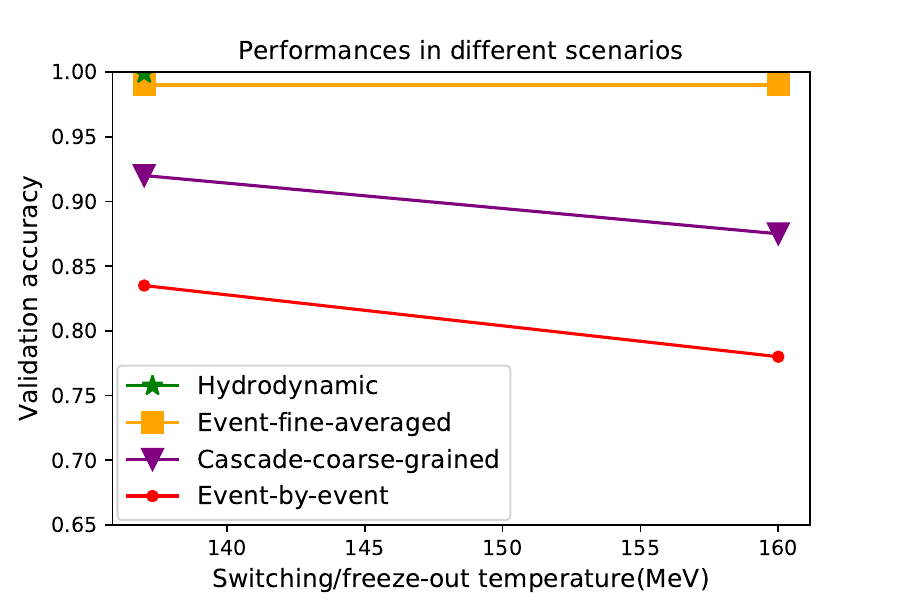} 
\caption{Taken from ~\cite{Du:2019civ}. Comparison between the validation accuracy in all the different sub-scenarios studied. The green star depicts the pure hydrodynamic result~\cite{Pang:2016vdc}. The orange square, the purple triangle and the red filled circle symbols depict the results for the 30-events-fine-averaged, cascade-coarse-grained and event-by-event spectra, respectively, in different switching temperatures.}
\label{hybrid_performance}
\end{figure}

Indeed, later, the same method is used to classify the EoS using final state hadrons sampled from the freeze-out hypersurface and passing through the hadronic cascade via UrQMD~\cite{Du:2019civ}. Both the stochastic particlization being followed by hadronic rescattering and the resonance decay effects are taken into account in producing the final state pion spectra $\rho(p_T,\phi)$. Initially, it was found that the accuracy for CNN with event-by-event spectra is significantly lower than that using smooth particle spectra in pure hydrodynamic case. This performance decline manifests the concealing over the fingerprint of QCD transition inside final state, by the fluctuations due to finite particles and resonance decay. A scenario with event-fine-averaged spectra as input is investigated, where the performance is showed to be greatly improved by feeding such averaged spectra with 30 events within the same fine centrality bin into the deep CNN. See Fig.~\ref{hybrid_performance} for a performance comparison from Refs.~\cite{Du:2019civ,Du:2020poe}.
%%%%%%%%%%%%%%%%%%%%%%%%%%%%%%%%%%%%%%%%%%%%%%%%%%%%%%%%%%%%%%%%%%%%%%%%%%%%%%%%%%%%%%%

%%%%%%%%%%%%%%%%%%%%%%%%%%%%%%%%%%%%%%%%%%%%%%%%%%%%%%%%%%%%%%%%%%%%%%%%%%%%%%%%%%%%%%%
\subsubsection{Learning Stochastic Process with QCD Phase Transition}
The aforementioned EoS identification assumes that the phase transition process involved in the entire HICs evolution take place in equilibrium states. However, the actual collisional evolution dynamics, especially the phase transition process involved, should be a non-equilibrium evolutionary behavior.
It is nontrivial to uncover the phase transition and involved dynamical information from a stochastically evolving dynamical system. Refs.~\cite{Jiang:2021gsw,Wang:2021yjw} generalized the idea of identifying transition type from HICs final state further, to recognize the phase order and extract the dynamical parameters in a stochastic dynamical process would happen in HICs. The general thermodynamics and phase behavior of QCD can be reasonably approximated by a linear sigma model~\cite{Nahrgang:2011mg}, the effective potential of which describes the crossover transition at small chemical potentials and the first-order phase transition at large chemical potentials. As simplified modelling to the phase transition processes in HICs, the Langevin equation is adopted to describe the semi-classical evolution of the long wavelength mode of the sigma field, 
\begin{equation}
\partial ^{\mu }\partial _{\mu }\sigma \left( t,x\right) +\eta \partial
_{t}\sigma \left( t,x\right) +\frac{\delta V_{eff}\left( \sigma \right) }{
\delta \sigma }=\xi \left( t,x\right),
\label{eq:langevin}
\end{equation}
with the effective potential $V_{eff}$ controlling the type of phase transition in the stochastic process. The damping coefficient $\eta$, and the noise term $\xi \left( t,x\right)$, follow the fluctuation-dissipation theorem. As a simplified description for HICs, Hubble-like decaying temperature field and constant baryon chemical potential are assumed for the heat bath. A Gaussian-type spatial noise with different overall strengths is adopted for $\xi$ to inject fluctuations associated with the phase transition.
%%%%%%%%%%%%%%%%%%%%%%%%%%%%%%%%%%%%%%%%%%%%%%%%%%%%%%%%%%%%%%%%%%%%%%%%%%%%
\begin{figure}[!hbtp]\centering
\includegraphics[width=0.33\textwidth]{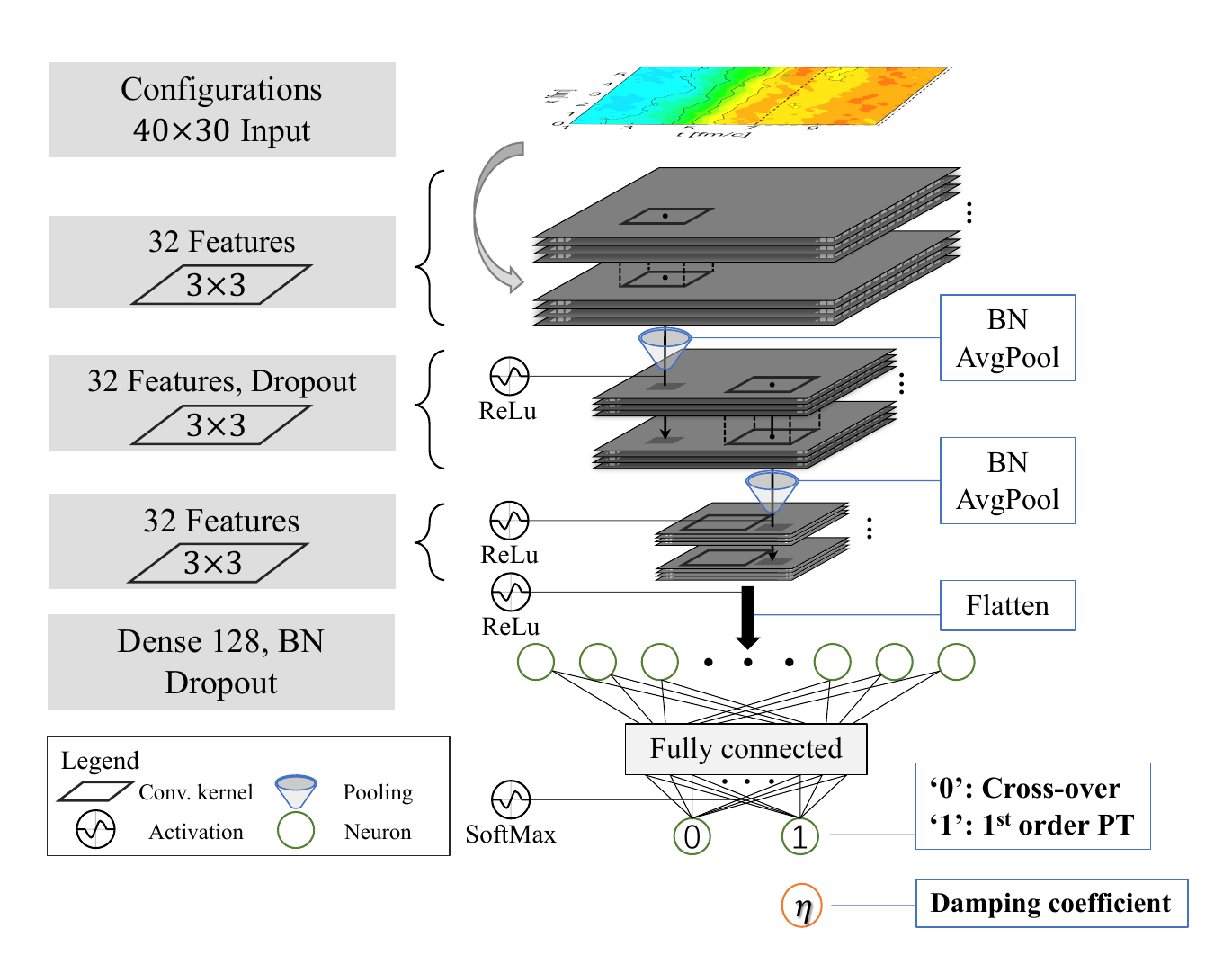}
\includegraphics[width=0.34\textwidth]{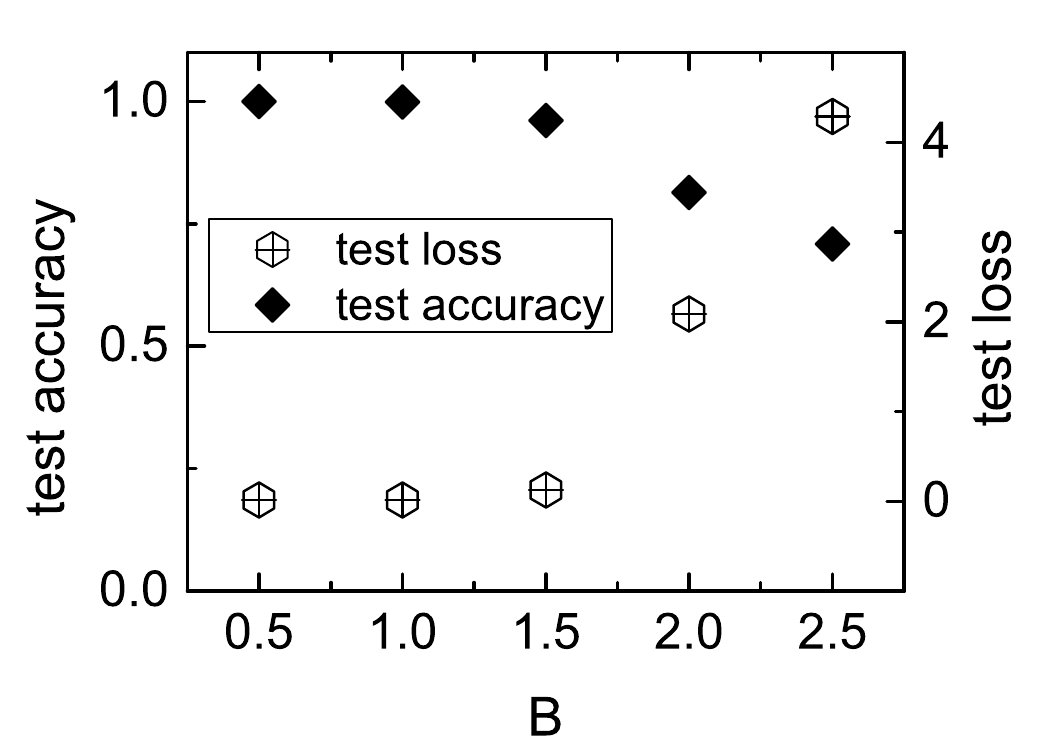}
\includegraphics[width=0.31\textwidth]{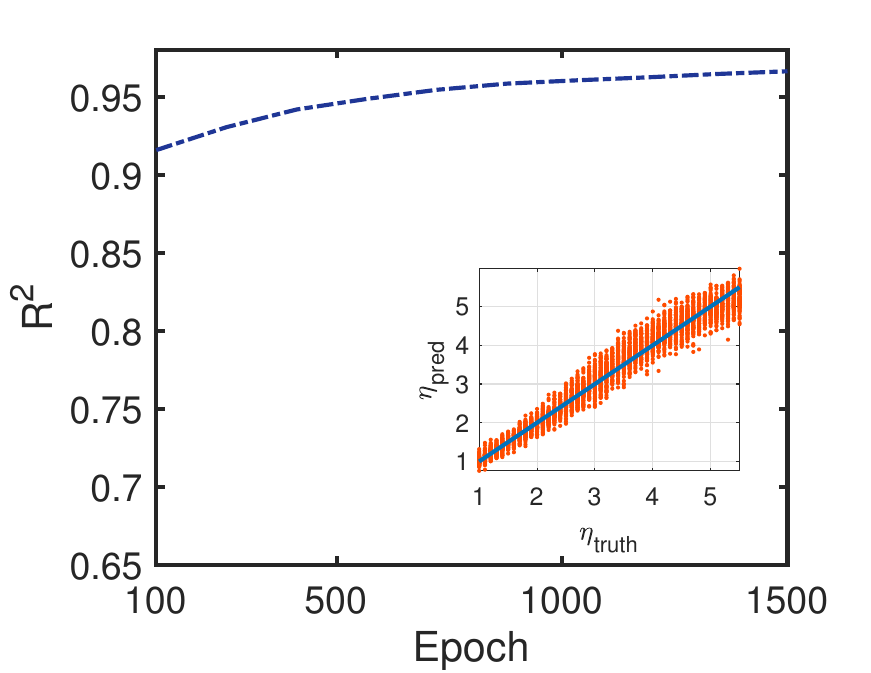}
\caption{Taken from Refs.~\cite{Jiang:2021gsw,Wang:2021yjw}. (left) the deep CNN for phase transition identification and damping coefficient regression; (middle) phase order classification accuracy in the testing stage with different spatial noise magnitude B; (right) the regression performance for damping coefficient during training and comparison to ground truth.\label{fig:langevin}}
\end{figure}
%%%%%%%%%%%%%%%%%%%%%%%%%%%%%%%%%%%%%%%%%%%%%%%%%%%%%%%%%%%%%%%%%%%%%%%%%%%%
Then a deep CNN, as shown in the left panel of Fig.~\ref{fig:langevin}, is devised to detect possible QCD phase transition order and predict the damping coefficient from the recorded $\sigma$ field spatio-temporal configurations in later stage of the non-equilibrium evolution. It is shown that the trained CNN is able to identify the encoded phase transition to be cross-over or first-order based on the final stage stochastic evolution of the field, which works even on different magnitude of noise, see middle panel of Fig.~\ref{fig:langevin}. For the damping coefficient prediction, the network gives a performance with on average $R^2=0.97$ in the scenario of first-order phase transition, as shown in the right panel of Fig.~\ref{fig:langevin}.

\subsubsection{Point Cloud Network for Spinodal Clumping Identification}
\label{sec:2:eos:pcn}
Several follow-up studies used convolutional neural networks (CNNs) to map final particle spectra to EoS types within hadronic transport models~\cite{Kvasiuk:2020izb,Sergeev:2020fir,Wang:2020tgb} or nuclear symmetry energy~\cite{Wang:2021xbb}. The CNN is the state-of-the-art for image recognition, so the momenta of the final state particles have to be converted into images. This is done by density estimation in $(p_x, p_y)$ space or $(p_t, \phi)$ space using 2-dimensional histograms. In both cases, there is a loss of information just from this pre-processing to the final state. As also been introduced in Sec.~\ref{hic_b}, the natural data structure of particles in momentum space is a \textit{point cloud}, where each particle is a point in momentum and feature space. The \textit{PointCloud network} is used to classify two first-order phase transitions with Maxwell and spinodal constructions~\cite{Steinheimer:2019iso}. With the latter construction, QGP evolves into many blobs that hadronize separately. However, it is rather  difficult  to distinguish the equation-of-state type from the final state hadrons in momentum space. Another point cloud neural network is trained using reconstructed tracks of final state hadrons given by the CBM detector simulations~\cite{OmanaKuttan:2020btb}. It was found that the performance decreases when considering realistic detector acceptance cuts and resolution. On the other hand, the performance improves when multiple events are combined. 

\begin{figure}[htbp!]
\centering
\includegraphics[width = 0.8\textwidth]{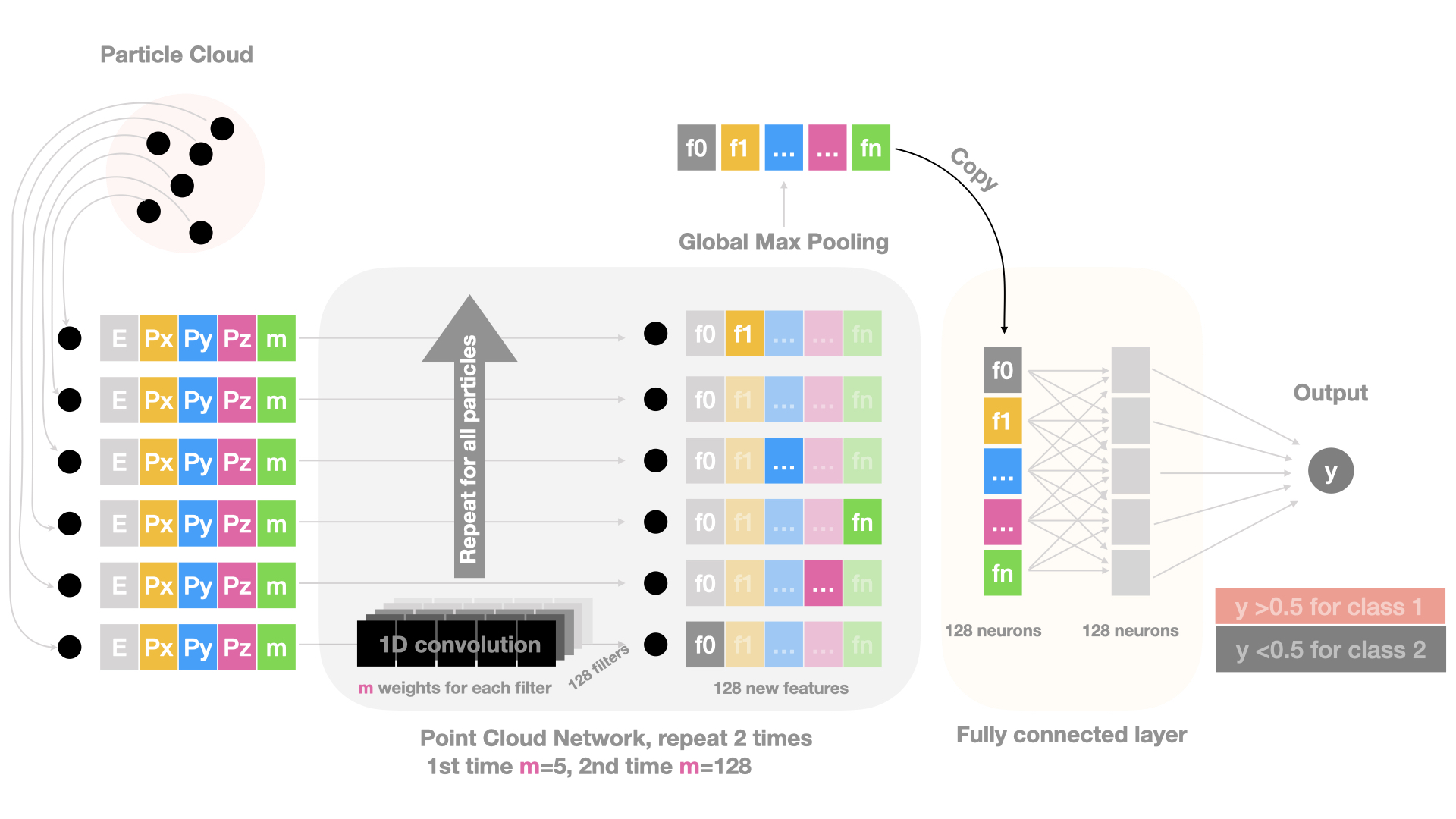}
\caption{A simple  point cloud network used  to  classify two kinds  of  first  order  phase  transitions.
\label{fig:pointcloud}}
\end{figure}

Shown in Fig.~\ref{fig:pointcloud} is a  simple point  cloud  network used classify  two  different  equation  of state. The point cloud network uses an MLP to transform the momentum and species information of each particle into a high-dimensional feature space (128 dimensions in this example). This MLP is shared by all particles in the cloud, so it is also called a 1-dimensional CNN.  Notice  that  CNN is not a specific technique  for image  processing,  any operations having  the  properties of local connection and weight sharing can be called CNN. Afterwards, a permutation-symmetric operations is used here to extract the multi-particle correlations hidden in the point cloud. A commonly used  permutation-symmetric operation  is the Global Max Pooling that is able to extract the boundaries of this point cloud in high-dimensional feature space, as shown below,
\begin{align}
f_i = max(\{f_{ij}\}, {\rm along}=j)
\end{align}
where $\{ f_{ij} \}$ is the feature matrix with $i$ denoting the i-th feature, $j$ representing the j-th particle and $f_{ij} = {\rm MLP}_i(\vec{p}_j)$, $f_i$ is the maximum value of the i-th feature among all the particles in the cloud,  extracted by the Global Max Pooling. Note that other permutation-symmetric operations can be used here to replace the global max-pooling. For example, the widely used Global Average Pooling is shown below, 
\begin{align}
f_i = {1 \over N}\sum_{j=1}^N f_{ij}
\end{align}
where $N$ is the number of particles in the cloud.

In principle, the point cloud network trained with 100 particles can be applied to a point cloud  with 1000 or more particles,  as the MLP is  shared by all particles and the permutation-symmetric operation is not sensitive to the number of particles. The number of particles may be different for different particle clouds produced in different HIC events. However, implementing a deep neural network with varying number of input particles during training meet technical difficulties, because a mini-batch of input data are required to have the same shape for GPU parallel acceleration.  Padding is usually used to produce the same number of particles in each cloud. There are different ways of padding, e.g., a list of virtual particles whose properties are all zeros, or particles from other events can be added which have no correlation with particles in the current event. To eliminate the effect of padding particles, another property can be added to each particle, called a "mask". The masks are equal to 1 for real particles in the cloud, but 0 for padding particles. Using these masks, the padding particles will not participate in the global max(average) pooling.

\subsubsection{Dynamical Edge Convolution Network for Critical Similarity}
There are rich critical phenomena near the critical endpoint, such as fluctuation enhancement and the emergence of self-similarity. The self-similarity in momentum space is modeled by Critical Monte Carlo (CMC) and encoded in the interparticle distances in momentum space. If self-similarity is present in only a few percent of all final state hadrons, the signal-to-background ratio is small. Traditionally, intermittency analysis is used to search for self-similarity in momentum space. However, this method fails at the current experimental $p_T$ resolution of about $0.1$ GeV/$c$. A point cloud network and a dynamical edge convolution neural network are used to identify events with interparticle similarity and to label correlated particles~\cite{Huang:2021iux}. 
The dynamical edge convolution network succeeds with a test accuracy of $92\%$ on events with $5\%$ signal particles. The method developed here appears to be powerful in searching for multi-particle correlations that are sensitive to specific physics.
%%%%%%%%%%%%%%%%%%%%%%%%%%%%%%%%%%%%%%%%%%%%%%%%%%%%%%%%%%%%%%%%%%%%%%%%%%%%%%%%%%%%
\begin{figure}[htbp!]
\centering
\includegraphics[width = 0.9\textwidth]{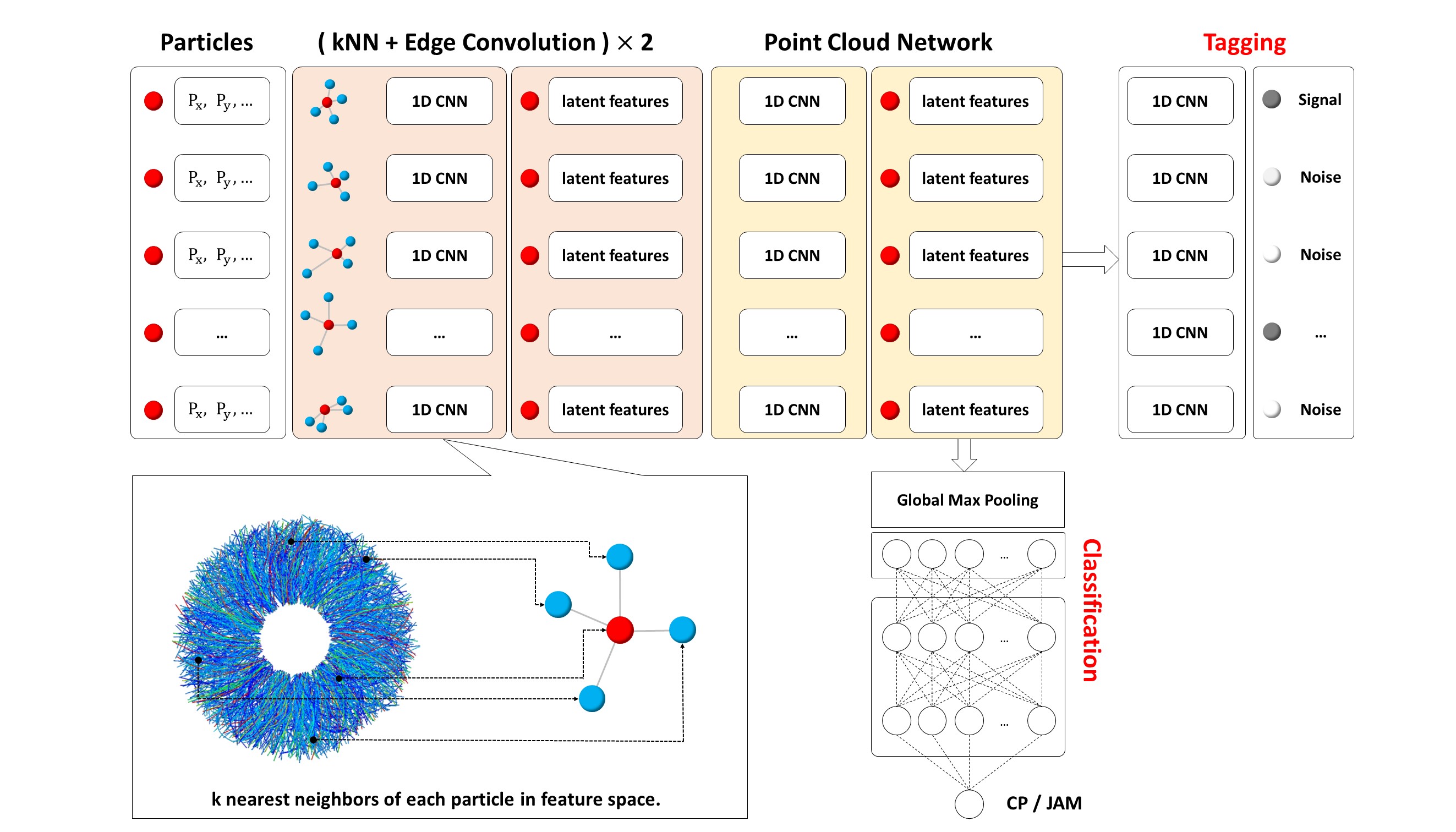}
\caption{A schematic diagram of critical self-similarity search with a dynamical edge convolutional neural network using the HIC particle cloud as input. Taken from~\cite{Huang:2021iux}.
\label{fig:decnn}}
\end{figure}
%%%%%%%%%%%%%%%%%%%%%%%%%%%%%%%%%%%%%%%%%%%%%%%%%%%%%%%%%%%%%%%%%%%%%%%%%%%%%%%%%%%%

Fig.~\ref{fig:decnn} shows the structure of the dynamical edge convolution neural network used to search for multi-particle correlations in the particle cloud.
The input to the network is a particle cloud produced in heavy ion collisions, with multiple features for each particle. The output of the network has two branches, one for classification and the other for tagging. The dynamic edge convolutional neural network extracts correlations between multiple particles to build a map between the input and output. Each input particle can have the four momenta, the charge, the baryon number, or the strange number as properties. In practice, the four momentum is usually normalized to the maximum energy of all particles in the cloud. Other quantum numbers can also be normalized to values between $-1$ and $1$.

In the dynamic edge convolution network, the k nearest neighbors of each particle are used to capture the multi-particle correlation.
In the first layer, the distances between different particles in coordinate or momentum space are computed to explicitly determine the neighbors. 
A fully connected neural network (MLP) is used to compute the correlation between the central particle and one of its neighbors, using the following formula
\begin{align}
    e_{ij} = {\rm MLP}({p}_i, p_j)
\end{align}
where $e_{ij}$ is the edge feature between particle $i$ and particle $j$, the $p_i$ represent the input features of particle $i$. 
This MLP network is shared by each pair of particles to produce edge features. 
The name "edge convolution" comes from the fact that the MLP is locally connected and its weights are shared among all edges.
After the first edge convolution, there is a point cloud network that transforms the input of each particle together with its neighbors into a high-dimensional latent space, whose input is a concatenation of the central particle and its k edge features,
\begin{align}
    f_i = [p_i, e_{i1}, e_{i2}, ..., e_{ik}].
\end{align}
Using the features $f_i$ in the high-dimensional latent space, the edge convolution can be applied again and again. As a result, the k nearest neighbors of each particle are determined by the distances between different particles in feature space afterwards. In other words, particles can be close in feature space but far apart in momentum space, allowing the network to capture long-range multi-particle correlation, which is critical for some specific classification or tagging tasks.

\subsubsection{Active Learning for Unstable Regions in QCD EoS}

\begin{figure}[htbp!]
    \centering
    \includegraphics[width = 0.65\textwidth]{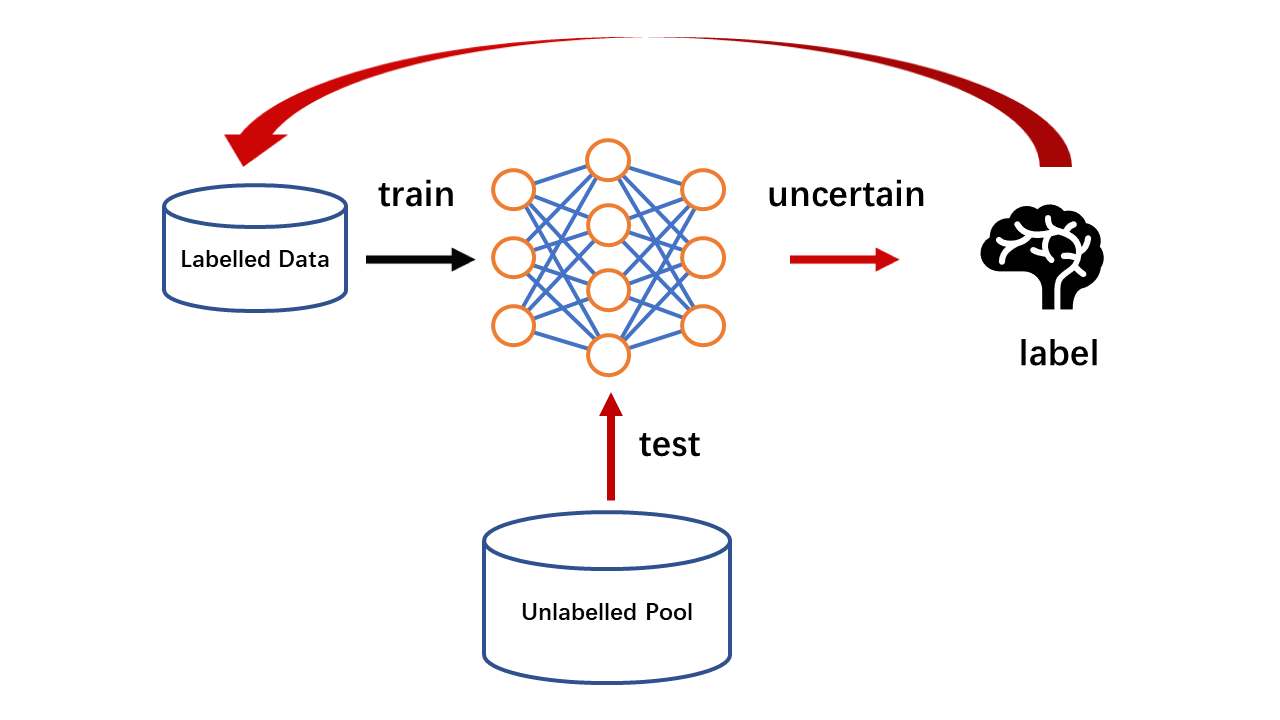}
    \caption{A schematic diagram of active learning.
    \label{fig:aclearn}}
\end{figure}

Active learning is a sub-field of machine learning which can get higher accuracy using fewer labels \cite{Burr2009}.
This is achieved by allowing the machine to choose data from which it learns.
The motivation of using active learning is that labels may be difficult to obtain in some tasks (expensive or time consuming) and
the machine can learn better on some instances than on others.  
Active learning starts with a small labeled data set for supervised training. The trained model is used to make predictions about samples from a large unlabeled pool, as Figure~\ref{fig:aclearn} shows. If the network is uncertain about a sample, e.g., the sample is predicted to be stable with probability $51\%$ and unstable with probability $49\%$, it is labeled and added to the labeled dataset for further supervised training. Labels can be made by human in the loop or by computer programs. The key of active learning is to propose samples for labeling, the first method is to use the uncertainty criterion as mentioned above, which is the entropy of the prediction,
\begin{align}
    s = - \sum_i p_i \log p_i
\end{align}
where $p_i$ is the predicted probability that the sample is in the $i$th category. 
The entropy $s$ is known to be maximized if the predicted distribution is uniform.
The entropy criterion is similar to the margin criterion $m = p_1 - p_2$,
where $p_1$ and $p_2$ are the first and the second most probable category labels for input $x$, predicted by the pre-trained network.
There are other criteria based on which the machines make their proposal.
For example, four neural networks can be trained in parallel and vote on samples from unsupervised pool.
If their judgments disagree the most on one sample, that sample will be voted for labeling.
Another criterion is the gradient which quantifies how much the machine will learn from this instance.
The negative gradients $-\partial l / \partial \theta$ have dependence on the input,
it is thus possible to choose input that leads to the largest change to the network for labeling.

%%%%%%%%%%%%%%%%%%%%%%%%%%%%%%%%%%%%%%%%%%%%%%%%%%%%%%%%%%%%%%%%%%%%%%%%%%%%%%%%%%%%%%
%%%%%%%%%%%%%%%%%%%%%%%%%%%%%%%%%%%%%%%%%%%%%%%%%%%%%%%%%%%%%%%%%%%%%%%%%%%%%%%%%%%%%%
Ref.~\cite{mroczek2022} uses active learning to detect thermodynamically unstable and acausal regions in QCD EoS. 
Lattice QCD fails to produce the QCD EoS around the critical endpoint region due to the sign problem (it will be discussed in the next section). The Beam Energy Scan Theory (BEST) collaboration~\cite{An:2021wof} constructs a QCD EoS by mapping the 3D Ising model and the lattice QCD EoS. However, the shape and position of the critical region depend on several parameters. Some parameter combinations lead to unstable and acausal QCD EoS that should be discarded in the future model to data comparisons, as the hybrid model of HIC is computationally expensive and sensitive to acausal EoS in the medium.

The pressure is a sum of the contributions from Lattice QCD and the 3D Ising model as demonstrated in~\cite{An:2021wof}, 
\begin{align}
P(T, \mu) = P_{\rm lattice}(T, \mu) + P_{\rm Ising}(T, \mu)
\end{align}
where the pressure from Lattice QCD calculations at non-zero baryon chemical potential $\mu$ is given by the Tylor expansion around $\mu=0$,
\begin{align}
{P_{\rm lattice}(T, \mu) \over T^4 } =  \sum_n c_n(T) ({\mu \over T})^n
\end{align}
where the coefficients are $c_n(T) = {1\over n!} \chi^B_n(T) = {1 \over n!}{\partial^n P/T^4 \over \partial (\mu/T)^n }$,
the $\chi_n^B$ is the $n$th order baryon susceptibility.
Using the pressure provided above, the complete thermal dynamic relations are given below,  
\begin{align}
{n_B(T, \mu) \over T^3} &= {1 \over T^3}\left({\partial P(T, \mu) \over \partial \mu}\right)_{T} \\
{s(T, \mu) \over T^3} &= {1 \over T^3}\left({\partial P(T, \mu) \over \partial T}\right)_{\mu} \\
{\epsilon(T, \mu) \over T^4} &= {s(T, \mu) \over T^3} - {P(T, \mu) \over T^4 } + {\mu \over T}{n_B \over T^3}
\end{align}

There are $4$ parameters in total, whose combinations determine the size and the shape of critical endpoint in QCD EoS,
as well as the stability and the causality of the derived EoS,
\begin{align}
\left(\mu, \alpha_{\text {diff }}, w, \rho\right) \mapsto P\left(T, \mu\right) \mapsto\{\rm acceptable, unstable, acausal \}
\end{align}
where stability requires the positivity of the energy density $\epsilon$, the pressure $P$, the entropy density $s$, the net baryon density $n_B$,
the second order baryon susceptibility $\chi_2^B$ and the heat capacity $\left({\partial s \over \partial T}\right)_{n_B}$. 
The causality requires that the speed of sound square satisfies $0 \le c_s^2 \le 1$.
Any parameter combinations that lead to unstable or acausal EoS will be discarded.

\subsection{Dynamical Properties of QCD Matter}

\subsubsection{Shear and Bulk Viscosity}
\label{transport}
 Relativistic hydrodynamic simulations have shown that the quark--gluon plasma (QGP) produced in high-energy heavy-ion collisions (HIC) is a strongly coupled plasma with a shear viscosity over the entropy density ratio $\eta/s$ close to the universal limit ${1\over 4\pi}$ by AdS/CFT calculations ~\cite{Heinz:2005bw,Romatschke:2007mq,Song:2010mg,Buchel:2003tz} . Both $\eta/s$ and $\zeta/s$ strongly influence the collective expansion of hot nuclear matter ~\cite{Teaney:2003kp}, leaving clear signals in the momentum anisotropy of final state hadrons produced in HIC. However, quantitatively determining the values of $\eta/s$ and $\zeta/s$ from the final state hadrons is challenging due to the difficulty in solving the inverse problem and the temperature dependence of $\eta/s(T)$ and $\zeta/s(T)$.
 Ref.~\cite{Niemi:2015qia} shows that relativistic hydrodynamic simulations with different fluctuating initial conditions can all describe data, with different effective shear viscosity. For example, using ``IPGlasma'' initial condition in MUSIC hydrodynamic model, the simulation describes the event by event fluctuations of $v_n$,
the mean $v_n$ and the $v_n(p_T)$ with constant $\eta/s=0.12$ at RHIC energy and $\eta/s=0.2$ at LHC energy~\cite{Gale:2012rq}. Using ``MC-KLN'' initial condition in VISHNU hydrodynamic model, the simulation describes the charged multiplicity, 
the $p_T$ spectra as well as the elliptic flow at both RHIC and LHC energy, with constant shear viscosity to entropy ratio $\eta/s=0.16$~\cite{Song:2013qma}. Using NLO improved EKRT initial condition in viscous hydrodynamics, it was observed that both the
 constant $\eta/s=0.2$ and a temperature dependent $\eta/s(T)$ give equal good overall fitting to RHIC and LHC data. The used linear parametrization is $0.12 < \eta/s < 0.12 + (0.18/320)(T /\text{MeV} - 180) $ for temperature above 180 MeV in the QGP phase and $\eta/s(T) = 0.12 - (0.20/80)(T /\text{MeV} - 180)$ for temperature below 180 MeV in the hadron resonance gas phase. A global analysis on all available RHIC and LHC data using Bayesian parameter estimation is required to quantitatively constrain the temperature-dependent shear and bulk viscosity~\cite{Novak:2013bqa,Pratt:2015zsa,Bernhard:2016tnd, Bernhard:2019bmu,JETSCAPE:2020shq, JETSCAPE:2020mzn,Nijs:2020ors, Nijs:2020roc}.

\begin{figure}[htbp!]
    \centering
    \includegraphics[width = 0.95\textwidth]{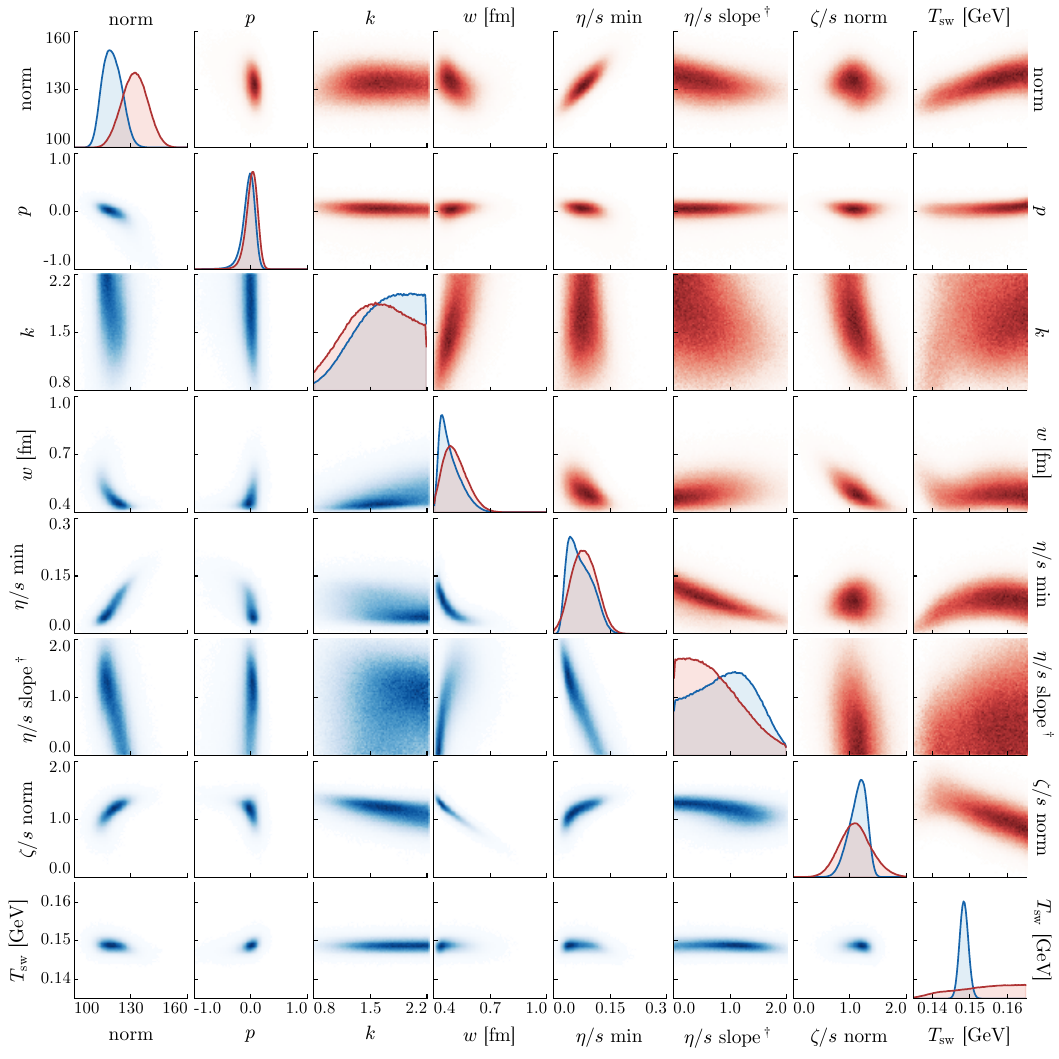}
    \caption{The posterior distribution of 9 model parameters for initial condition and transport coefficients using Bayesian analysis. Taken from~\cite{Bernhard:2016tnd}.
    \label{fig:bayes}}
\end{figure}

The studies by Bernhard et al. in Refs.~\cite{Bernhard:2016tnd, Bernhard:2019bmu} use a Trento + IEBE-VishNew + UrQMD hybrid model to perform a global fitting of the charged multiplicity, transverse momentum spectra, elliptic flow, and triangular flow as a function of centrality. This analysis helps to constrain both initial condition parameters and temperature-dependent shear and bulk viscosity. The results demonstrate that the extracted initial conditions are consistent with the gluon saturation model and indicate a clear signal of non-zero bulk viscosity.
A set of nine model parameters in total are constrained using Bayesian analysis, whose posterior distributions are given in Fig.~\ref{fig:bayes}. These parameters include:
\begin{enumerate}
    \item Norm, the overall normalization factor determining the initial total entropy that is responsible for the multiplicity of final state hadrons;
    \item $p$, the entropy deposition parameter, which is used to mimic different kinds of initial entropy production, from MC-Glauber to saturation-based models;
    \item $k$, multiplicity fluctuation shape parameter, which determines the multiplicity distribution in minimum-bias proton+proton collisions;
    \item $w$, Gaussian nuclear width, which determines the size of nucleons during collision;
    \item $(\eta/s)_{\rm hrg}$, a constant parameter for the value of $\eta/s$ in hadron resonance gas phase;
    \item $(\eta/s)_{\rm min}$, the minimum $\eta/s$ at the critical temperature $T_c=154$ MeV;
    \item $(\eta/s)_{\rm slope}$, the slope of $\eta/s(T)$ above the critical temperature $T_c$;
    \item $\zeta/s$ norm, one overall normalization factor for the given $\zeta/s(T)$ function;
    \item $T_{\rm switch}$, the particlization temperature.
\end{enumerate}

%%%%%%%%%%%%%%%%%%%%%%%%%%%%%%%%%%%%%%%%%%%%%%%%%%%%%%%%%%%%%%%%%%%%%%%%%%%%%%%%%%%%%%
%%%%%%%%%%%%%%%%%%%%%%%%%%%%%%%%%%%%%%%%%%%%%%%%%%%%%%%%%%%%%%%%%%%%%%%%%%%%%%%%%%%%%%
These parameters form a nine-dimensional parameter space, each point in this space representing a combination of model parameters. Without any constraints, the values of these parameters vary within a given range provided by physical a priori. According to Bayesian parameter estimation, the unnormalized posterior distributions of these parameters are given by $P(\theta|D)\propto P(D|\theta)P(\theta)$, where the experimental data $D$ used in the Bayes formula consists of differential yields per unit rapidity (d$N/dy$), mean transverse momentum $\langle p_T \rangle$ for $\pi^{\pm}$, $K^{\pm}$, and p$\bar{p}$ at mid-rapidity, as well as two-particle cumulants $v_n(2)$ with $n=2$, $3$, and $4$ for charged hadrons at various centrality classes. The likelihood $P(D|\theta)$ is computed using the model output and experimental data, while the prior $P(\theta)$ is usually chosen to be a uniform distribution. It is worth mentioning that the computationally expensive high-energy heavy-ion collision simulations are run for a finite number of parameter sets and then emulated by a Gaussian process (see Sec.~\ref{sec:hic:emulator} for more details).

The diagonal subplots in Fig.~\ref{fig:bayes} display the marginal distributions of these nine parameters. The posterior distribution in red color is constrained using charged particles, while the distribution in blue color is constrained using identified particles. For many parameters, the optimal values differ when using these two groups of experimental data. However, the marginal distributions of $p$ and $\zeta/s$ norm have narrow peaks whose locations agree with each other when using both groups of data. The results indicate that the initial condition is consistent with saturation-based models, and the bulk viscosity is clearly non-zero.

Follow up Bayesian Inferences of transport parameters are performed by the JETSCAPE collaboration~\cite{JETSCAPE:2020shq, JETSCAPE:2020mzn} and Nijs et al~\cite{Nijs:2020ors, Nijs:2020roc}. These independent researches take different experimental results as evidence and focus on different parameters. The former focused on $p_T$ averaged observables in both RHIC and LHC energies, whereas the latter looked into the $p_T$ differential ones in LHC collisions. A clear tension is observed in the bulk viscosity, which reveals the model dependence in the parameter extraction and calls for further investigation.

Bayesian Inference provides a statistically systematic way to make full use of the low-dimensional observables\footnote{i.e., preprocessed and projected from original detector raw record to expert-designed quantities like elliptic flow or $p_T$ spectra within rapidity cut and centrality bin.} provided by experiment as well as the physical prior, where the uncertainty of the inference can be properly estimated with the experimental error and theoretical modeling error being able to be incorporated. However, most of the Bayesian analysis in HICs use handcrafted parametrizations for many of the inference targets (e.g., EoS or shear viscosity's temperature dependence) thus the final results may be limited or dependent on the setup and also priors. The other concern lies in the probable information loss of using only low-dimensional processed ``observations'' instead of the original high-dimensional raw record. A naive pushing forward for Bayesian inference to, e.g., event-by-event quantity analysis would hugely increase the computational demanding and time. Direct mapping capturing by deep learning or combining deep generative models into traditional Bayesian Inference are thus necessary alternatives.

\subsubsection{Jet Energy Loss in Hot QGP}\label{hic_jet}
    
As energetic partons pass through the hot QGP, they lose energy through elastic scattering and inelastic gluon radiation. The energy loss of the jet thus serves as a good probe of the properties of the medium. Bayesian analysis is used to study the temperature and momentum dependence of the heavy quark diffusion parameter~\cite{Xu:2017obm}, the jet quenching or transverse diffusion coefficient $\hat{q}$~\cite{Soltz:2019aea} and the jet energy loss distributions~\cite{He:2018gks}. Bayesian analysis typically assumes a parameterized function whose parameters are determined by maximum a posteriori (MAP) estimation. Recently, the information field has been used to generate random functions that can be used as a model for Bayesian analysis to obtain the temperature-dependent $\hat{q}(T)/T^3$ using both RHIC and LHC data~\cite{Xie:2022ght}. The information field provides a non-parametric representation of the unknown function, which can eliminate the unnecessary long-range correlations in the prior. 
%%%%%%%%%%%%%%%%%%%%%%%%%%%%%%%%%%%%%%%%%%%%%%%%%%%%%%%%%%%%%%%%%%%%%%%%%%%%%%%%%%%%%%
\begin{figure}[htbp!]
    \centering
    \includegraphics[width = 0.6\textwidth]{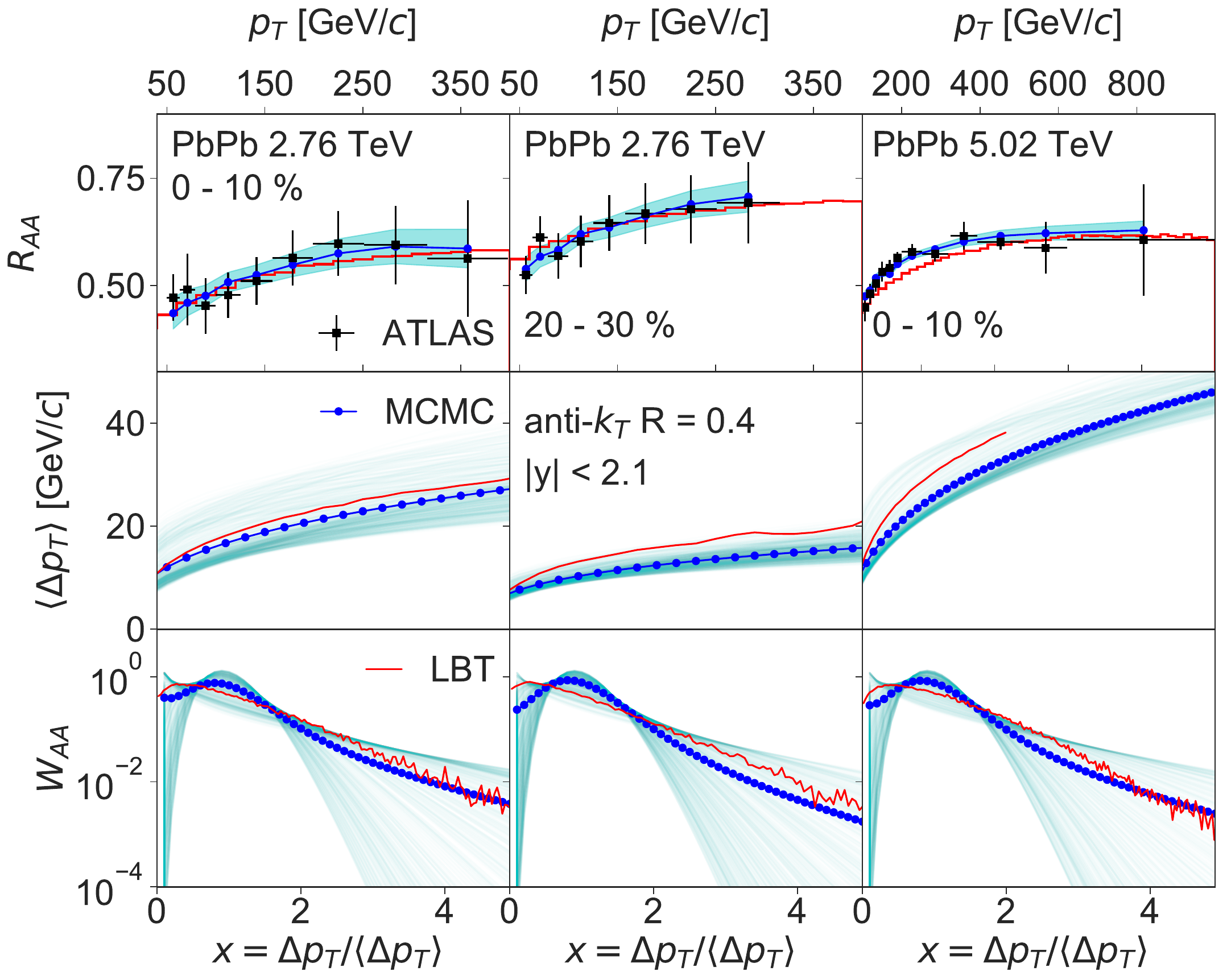}
    \caption{The extracted jet energy loss distribution and the computed $R_{AA}$ as compared with Linear Boltzmann Transport model calculations. Taken from~\cite{He:2018gks}.
    \label{fig:jet_bayes}}
\end{figure}
%%%%%%%%%%%%%%%%%%%%%%%%%%%%%%%%%%%%%%%%%%%%%%%%%%%%%%%%%%%%%%%%%%%%%%%%%%%%%%%%%%%%%%
Fig.~\ref{fig:jet_bayes} shows the application of Bayesian parameter estimation in determining the jet energy loss distribution~\cite{He:2018gks}. The jet cross-section is approximated by the convolution between p+p jet cross-section and the jet energy loss distribution function, where the latter is parameterized using a Gamma distribution function,
\begin{align}
    W_{AA}(x) = {\alpha^{\alpha} x^{\alpha - 1} e^{-\alpha x} \over \Gamma(\alpha)},
\end{align}
where $x= \Delta p_T / \langle \Delta p_T \rangle$, $\alpha$ can be interpreted as the average number of jet-medium scatterings that take energy out of the jet cone. The energy per scattering is thus $\langle p_T \rangle / \alpha$ with the mean transverse momentum loss as a function of $p_T$ defined as follows,
\begin{align}
   \langle \Delta p_T \rangle (p_T) = \beta\, p_T^{\gamma}\, \log p_T.
\end{align}

Note that the function form of the averaged energy loss is motivated by theoretical calculations~\cite{He:2015pra}. There are thus three parameters $(\alpha, \beta, \gamma)$ in the modification factor $R_{AA}$ to be constrained using the Bayesian analysis. The first row of Fig.~\ref{fig:jet_bayes} demonstrates that the $R_{AA}$ calculated from the constrained jet energy loss distribution is in good agreement with experimental data in Pb+Pb 2.76 and 5.02 TeV collisions. The second and the third row compare the mean  $p_T$ loss and the energy loss distribution $W_{AA}(x)$, as compared with Monte Carlo simulations using the Linear Boltzmann Transport model. The extracted value of $\alpha$ is small, indicating a large transverse momentum loss per scattering.
The results suggest that the observed jet quenching is mainly caused by a few out-of-cone scatterings.

\subsubsection{Jet Classification and Jet  Tomography}

Numerous studies have been conducted to investigate jets in high energy physics \cite{Larkoski:2017jix,Louppe:2017ipp,Dreyer:2021hhr,deLima:2021fwm,Romero:2021qlf,Konar:2021zdg,Karagiorgi:2021ngt,Nguyen:2021xnq,Luchmann:2022iih,Gong:2022lye,Bedeschi:2022rnj,Qu:2022mxj,CMS:2022wjj,Cal:2022fnm,Cranmer:2021gdt,Rossi:2023qvf,Chien:2018rgm,Du:2021qwv,Apolinario:2021olp,Du:2020pmp, Du:2021pqa,Yang:2022yfr,Feickert:2021ajf}. Machine learning-based jet-flavor and event classifications may find applications in future electron-ion colliders \cite{Lee:2022kdn}. Jet-flavor tagging using machine learning can be crucial for determining the collinear and transverse momentum dependent parton distribution functions (PDFs) \cite{Arratia2020CharmJA}. The classification of e+P and e+A collisions can help uncover observables sensitive to the cold nuclear effect. Additionally, the charm and anti-charm tagging in di-jet events can aid in constraining the gluon TMD and gluon Sivers function~\cite{Arrington:2021yeb}. Jet substructure techniques have the potential to constrain the gluon PDF \cite{Caletti:2021ysv}.

In Ref.\cite{Apolinario:2021olp}, a deep neural network was used to differentiate between vacuum-like jets from p+p collisions and jets in medium from Pb+Pb collisions. This approach helped identify the effect of jet quenching in the presence of QGP in Pb+Pb collisions. By utilizing the energy loss and charged particle distribution within a jet cone as inputs to a deep neural network, the initial jet energy and production positions can be predicted \cite{Du:2020pmp, Du:2021pqa, Yang:2022yfr}. Machine learning-assisted jet tomography can be employed to select jet events with similar initial energy or production locations for more detailed differential studies.

Machine learning assisted jet tomography is employed to study Mach cones produced in high energy heavy ion collisions \cite{Yang:2022yfr}. Mach cones are expected to form in the smallest nuclear liquid droplet when energetic partons traverse through QGP and deposit energy and momentum in QGP \cite{Baumgardt:1975qv,Rischke:1990jy,Casalderrey-Solana:2004fdk,Satarov:2005mv,Dremin:2005an,Koch:2005sx,Ma:2006mz,Gubser:2007ga,Betz:2008js,Neufeld:2008dx,Torrieri:2008aqg,Qin:2009uh,Roy:2009cc}. The speed of hard partons is close to the speed of the light, which is much larger than the speed of sound $c_S$ of QGP. There is a simple formula that relates the  half-angle $\theta$ of Mach cone to the speed of sound $c_s$ and the nuclear equation of state (pressure as a function of energy density),
\begin{align}
    {d P \over d \epsilon}  =  c_s^2  = \sin^2 \theta 
\end{align}
where $P$ is the local pressure and $\epsilon$ is the local energy density of hot nuclear matter. The Mach cone angle thus provides a direct probe of the QCD EoS, if its effects on the final state hadrons are well understood.

\begin{figure}[htbp!]
    \centering
    \includegraphics[width=13.0cm]{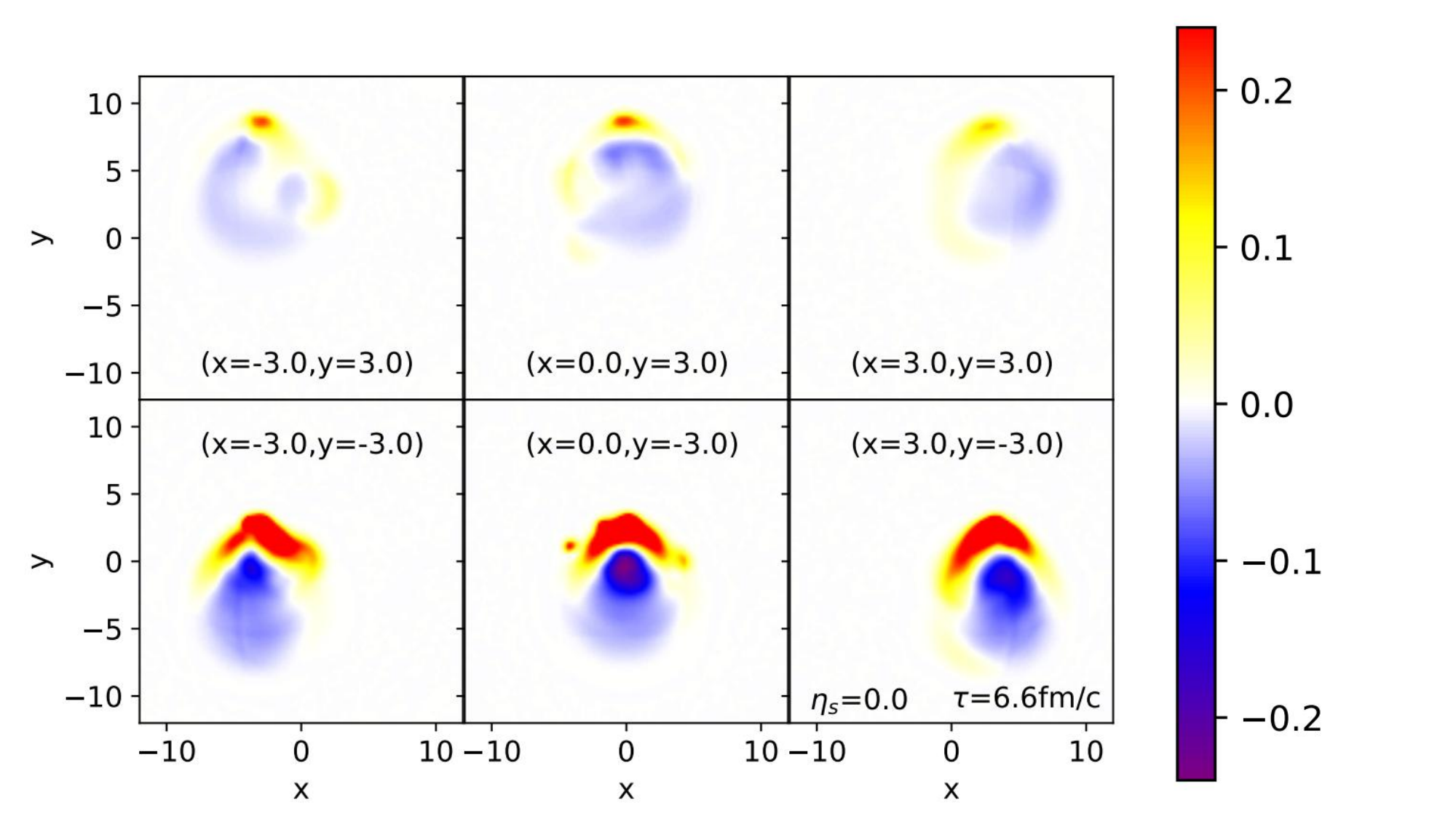}
    \setlength{\belowcaptionskip}{-0.cm}
    \caption{The shape of Mach cones for jets that are initiated at different positions and travel in the same positive-y direction. Reproduced from Ref.~\cite{Yang:2022yfr}.}
    \label{fig:machcones}
\end{figure}

In practice, Mach cones are challenging to detect due to the fact that jets are produced at different positions and travel in diverse directions \cite{Satarov:2005mv}. The shape of Mach cones can be influenced by both temperature gradients and strong collective flow \cite{Tachibana:2015qxa}. As illustrated in Fig.~\ref{fig:machcones}, the medium response is stronger for jets with longer path lengths (originated from $y=-3.0$) compared to those with shorter path lengths (originated from $y=3.0$) in QGP. However, the Mach cone shape and associated diffusion wake are deformed by the initial jet production position and collective flow on the jet trajectory. For jets initiated from $x=-3, y=-3$, the medium response is stronger on the right wave front than its left branch due to increased temperature and parton density in the central region of QGP, which leads to an enhanced number of scatterings between jet partons and thermal partons. Additionally, the diffusion wake is affected by strong radial flow, which pushes the medium response outwards. It is envisioned that if it becomes possible to determine the initial jet production positions for event selection, the effects of medium response, Mach cone signals, and diffusion wake will all be amplified, leading to improved sensitivity to the underlying dynamics of relativistic heavy-ion collisions.

In Ref.~\cite{Yang:2022yfr}, the authors employ a point cloud network to determine the initial jet production positions, which is then used to select jets with similar path lengths and directions that are shown to be able to amplify the signal of 3D mach cones. The network in this study utilizes two distinct types of inputs: one is the point cloud in momentum space for particles with $p_T>2$ GeV, while the other is the global information of $\gamma$-trigger and the full jet. These features are processed differently, with the particle cloud undergoing a point cloud network and global max pooling, while the global information is processed using a simple MLP. The two inputs are then concatenated to make the final decision, which are two real numbers to represent the initial jet production locations, resulting in a regression task solved using supervised learning. The training data is generated using LBT and CLVisc, and after being trained, the network is capable of predicting the initial jet production positions with reasonable accuracy.

\begin{figure}[htbp!]
    \centering
    \includegraphics[width=0.8\textwidth]{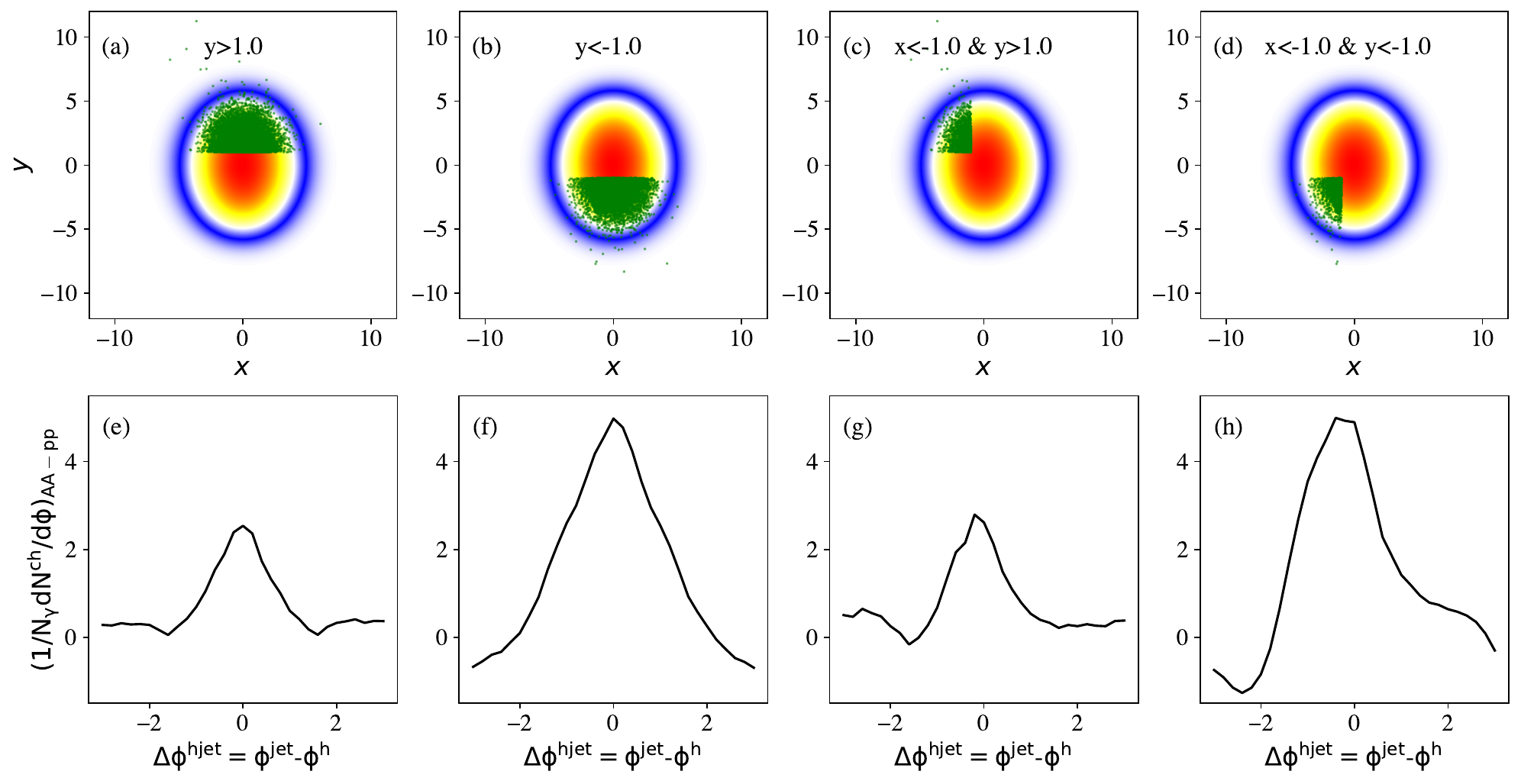}
    \setlength{\belowcaptionskip}{-0.cm}
    \caption{The hadron-jet angular correlation for selected jet events that are initiated from different regions in the transverse plane. Reproduced from Ref.~\cite{Yang:2022yfr}.}
    \label{fig:jet_vs_pos}
\end{figure}

The ML assisted jet tomography is used to select events for jets initiated from the same region. Fig.~\ref{fig:jet_vs_pos} shows the angular correlation between the final state hadrons and the jet, for jet partons initiated from 4 different regions. The magnitudes of the angular correlation in sub-figures (f) and (h) are stronger than (e) and (g), as the path lengths in these events are selected to be larger. The suppression of hadrons in the opposite direction of jets is caused by the diffusion wake. Constraining initial jets to the left half of the QGP fireball will introduce asymmetric angular correlation, which reflects the effect of radial flow to the diffusion wake. Observing these features in experimental data using ML assisted jet tomography will provide direct evidence of the existence of Mach cones and diffusion wake. 

\subsubsection{Chiral Magnetic Effect Detection}\label{hic_cme}

The chiral magnetic effect (CME) is a phenomenon that occurs in chiral matter, where the combination of a chiral anomaly and a magnetic field leads to the separation of electric charges along the direction of the magnetic field~\cite{Fukushima:2008xe}. The CME is important in the study of the quark--gluon plasma (QGP) created in HIC experiments~\cite{Kharzeev:2013ffa}. Recently, researchers have been trying to search for this effect in HICs~\cite{Kharzeev:2015znc, Zhao:2019hta, Li:2020dwr}, but the traditional methods (e.g., $\gamma$-correlator) have proven to introduce large amounts of background contamination, e.g., the elliptic flow, global polarization and indeterminate background noises. Zhao et al. have developed a \textit{CME-meter} using a deep convolutional neural network (CNN), to analyze the entire final-state hadronic spectrum as big data and reveal the distinctive signatures of CME.
%%%%%%%%%%%%%%%%%%%%%%%%%%%%%%%%%%%%%%%%%%%%%%%%%%%%%%%%
\begin{figure}[htbp!]
    \centering
    \includegraphics[width=12.0cm]{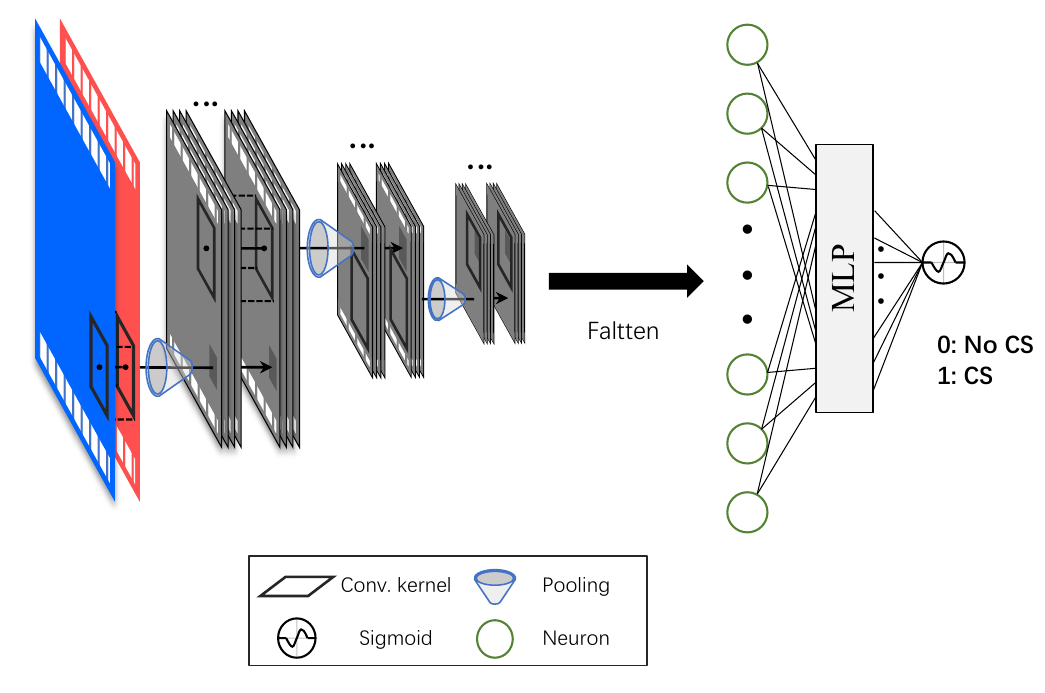}
    \setlength{\belowcaptionskip}{-0.cm}
    \caption{The convolutional neural network architecture with $\pi^+$ and $\pi^-$ spectra $\rho^{\pm}(p_T, \phi)$ as input. Reproduced from Ref.~\cite{Zhao:2021yjo}.}
    \label{fig:CNN}
\end{figure}
%%%%%%%%%%%%%%%%%%%%%%%%%%%%%%%%%%%%%%%%%%%%%%%%%%%%%%%%%%%

They trained the meter using data prepared from a multiphase transport model(AMPT)~\cite{Lin:2004en, Ma:2011uma, Jin:2018fwq} and found that it was accurate at identifying the CME-related charge separation in the final-state pion spectra. In data preparation, for CME events, the y-components of momenta of a fraction of downward moving light quarks and their corresponding anti-quarks were switched to upward moving direction, with the CS fraction, $f$, defining the events as ``no CS'' (labeled as ``0'') for those with $f=0$, and ``CS'' (labeled as ``1'') for those with $f>0$. Each event was represented as two-dimensional transverse momentum and azimuthal angle spectra of charged pions in the final state, $\rho_{\pi}(p_T,\phi)$. The training set includes multiple collision beam energies and centralities to ensure diversity in the dataset.  With two different charge separation fractions as super parameters of the meter (5\% and 10\%), the meter is trained to recognize the CME signal under supervision. The sketch of the framework is shown in Fig.~\ref{fig:CNN} and the detailed architecture can be found in Ref.~\cite{Zhao:2021yjo}.

%%%%%%%%%%%%%%%%%%%%%%%%%%%%%%%%%%%%%%%%%%%%%%%%%%%%%%%%%%%%%%%%%%%%%%%%%%%%
\begin{figure}[!hbtp]\centering
    \includegraphics[width=0.38\textwidth]{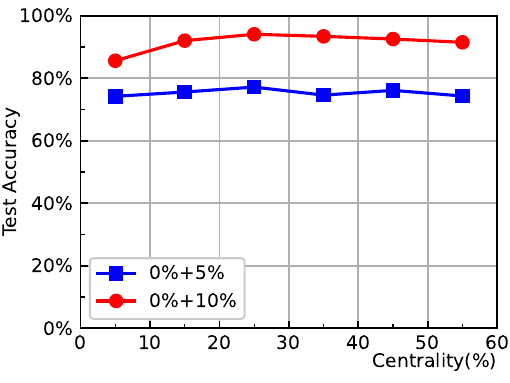}
    \includegraphics[width=0.38\textwidth]{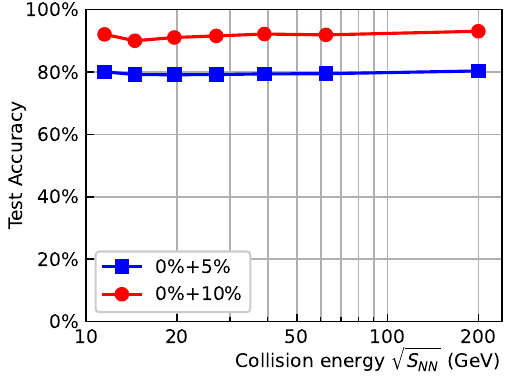}
\caption{Taken from Refs.~\cite{Zhao:2021yjo} with permission. (left) The test accuracy of models on different data-sets containing mixed collision energies along with centralities; (right) the test accuracy of models on different data-sets containing mixed centralities along with collision energies.\label{fig:acc}}
\end{figure}
%%%%%%%%%%%%%%%%%%%%%%%%%%%%%%%%%%%%%%%%%%%%%%%%%%%%%%%%%%%%%%%%%%%%%%%%%%
The well-trained CNN models are denoted as (0\%+5\%) and (0\%+10\%), respectively, in Fig.~\ref{fig:acc}. Their ability to recognize the CS signal is also shown in the figure. Although there is a discrepancy between the two models, they both display robust performance under varying collision conditions, such as different $\sqrt{s_{NN}}$ and centralities. This indicates that the CS signals are not completely lost or contaminated during the collision dynamics, and can still be detected by our network-based CME-meter. In addition to its accuracy and robustness in handling collision conditions, it is noteworthy that the well-trained machine, embedded with the knowledge of CME in the final state of HICs, can extrapolate this pattern to different charge separation fractions, collision systems, and even different simulation model, i.e., Anomalous-Viscous Fluid Dynamics(AVFD)~\cite{Shi:2017cpu, Shi:2018sah, Shi:2019wzi}. The well-trained machine also provides a powerful tool for quantifying the CME and is insensitive to the backgrounds dominated by elliptic flow ($v_2$) and local charge conservation(LCC), compared to the conventional $\gamma$-correlator. 

To implement the trained CME-meter in real experiments, the first step would be to reconstruct the reaction plane of each collision event and form averaged events as input for the meter. In general, reconstructing the reaction plane requires measuring correlations among final state particles, which is subject to finite resolution and background effects. However, the well-trained CME-meter is still capable of recognizing the CS signals even with restricted event plane reconstruction. As shown in Fig.~\ref{fig:CNN}, the network output consists of two nodes, which can be naturally interpreted as the probability of the meter recognizing a given input spectrum as a CME event ($P_1$) or a non-CME event ($P_0=1-P_1$). From a hypothesis test perspective, a characteristic distribution of $\max{P_1(\phi)}$ can be obtained, which can be used as a criterion for determining the existence of CME in HICs. Further details on deploying the CME-meter on single event measurements can be found in Ref.~\cite{Zhao:2021yjo}. Finally, DeepDream, a method used to visualize the patterns learned by convolutional neural networks (CNNs), is applied as a validation test using $P_1$ to detect the CME. It helps us uncover the hidden physical knowledge in the well-trained machine, including charge conservation and specific charge distributions.

\subsubsection{Anisotropic Flow Analysis}
\label{flow_hic}

One of the key signals for the production of a Quark--Gluon Plasma (QGP) in high-energy heavy-ion collisions is the strong collective flow~\cite{Shuryak:2004cy}. In semi-central and peripheral collisions, the initial geometric eccentricity of the system is transformed into momentum anisotropy in the final state hadrons through the process of strong collective expansion. This momentum anisotropy can be quantified by Fourier decomposing the azimuthal angular distribution in momentum space to obtain the well-known elliptic flow $v_2$, which represents the second-order harmonic coefficient in this decomposition. The scaling of $v_2$ with the number of constituent quarks for identified particles in HIC provides compelling evidence that the underlying degree of freedom is indeed partons. Interestingly, the discovery of triangular flow, which is sensitive to initial state fluctuations, was made much later than elliptic flow. Subsequently, higher orders of harmonic flow were also discovered, along with non-linear correlations between $v_4$ and $v_2$, as well as between $v_5$, $v_2$, and $v_3$. These correlations between different orders of harmonic flows at different transverse momenta have been used as new observables to constrain the properties of the QGP.

\begin{figure}[htbp!]
    \centering
    \includegraphics[width=\textwidth]{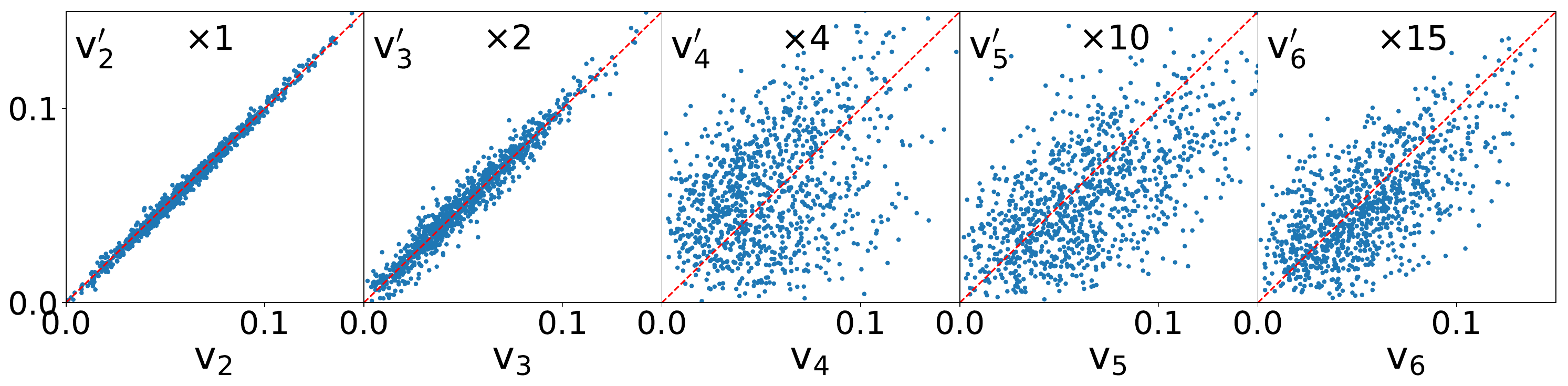}
    \setlength{\belowcaptionskip}{-0.cm}
    \caption{The anisotropic flow harmonics from principal component analysis as compared with traditional Fourier decomposition method. Taken from ~\cite{Liu:2019jxg} with permission.}
    \label{fig:pca_flow}
\end{figure}

Principal component analysis (PCA) is employed in Ref.~\cite{Liu:2019jxg} to obtain the first several principal components of the single-particle distribution. As illustrated in Fig.~\ref{fig:pca_flow}, the $v_2'$ and $v_3'$ obtained from PCA are consistent with $v_2$ and $v_3$ using Fourier decomposition, but higher-order harmonic flows with $n\geq4$ exhibit significant deviations from the Fourier decomposition method. The correlations between different harmonic flows decrease, while the Pearson correlation between $v_n'$ and the initial geometric eccentricity is enhanced as compared to traditional methods. These findings suggest that relativistic hydrodynamics may not be as highly non-linear as previously thought, as it does not couple different modes effectively. In PCA, anisotropic flows to various orders can be rediscovered, with triangular flow emerging as the next most important component. This highlights the significance of PCA in shedding light on the underlying physics of heavy ion collisions.

In Ref.~\cite{Mallick:2022alr}, a deep neural network is employed to learn the elliptic flow from the final-state particle kinetic information. This supervised learning technique utilizes elliptic flow calculated from the event-plane method as labels and is trained on minimum-bias data from Pb+Pb $5.02$ TeV collisions. After trained, the network succeeds in computing elliptic flows at various other beam energies and centrality classes. The successful completion of this task demonstrates the potential for machine learning tools to replace time-consuming data analysis tasks. Deep neural networks have demonstrated their capability in computing critical physical observables such as anisotropic flow in heavy ion collisions, suggesting its potential for revealing new physical observables that are sensitive to specific QGP properties within complex hybrid models. By training a deep neural network to map the physical property of interest to the final state hadrons, one can accomplish objectives such as correlations among different traditional observables or non-linear correlations between particles from different regions in the momentum space. For follow-up developments, see Ref.~\cite{Mallick:2023vgi} for their estimation of $v_2$ for light-flavor identified particles.

\subsection{Fast Simulations for HICs}
Simulations are a vital, albeit highly computationally demanding, element for interlinking theoretical strides with experimental findings. This is especially the case for the field of HENP, particularly in studying heavy ion collision physics.  As mentioned in Sec.~\ref{sec:hydro}, relativistic hydrodynamics simulations have been shown crucial in modeling the heavy ion collisions, rendering them an indispensable apparatus for studying the properties and evolution of the strongly interacting QCD matter under extreme conditions in the context of HICs. The quest to decode the physics propelling HICs necessitates a huge amount of collision event simulations to effectively confront the experimental measurements. However, the conventional numerical methods earmarked for HIC simulations have been computationally resources-intensive and time-consuming, especially when tasked with simulation for a large number of events.

\subsubsection{Efficient Emulator with Transfer Learning}\label{sec:hic:emulator}
As introduced in Sec.~\ref{transport}, when performing Bayesian inference to estimate the likely model parameters, the MCMC procedures that are needed to sample the posterior distribution for the to-be-inferred parameters would require huge model simulations to explore the parameter space. In practice, a model emulator or surrogate is necessary to reduce the computational demand in enabling the MCMC for inference. In general, a model emulator, usually constructed with machine learning, gives a fast map from any arbitrary point in multidimensional parameter space to the physical model predictions on the desired observables with uncertainties associated. In the context of QGP properties constraints in HICs or other parameter estimation for computationally intensive models~\cite{higdon2015bayesian, higdon2008computer}, Gaussian Process (GP) emulators have been shown successful and taken as the standard practice for a decade. Notwithstanding, the training of a reliable emulator including preparation of training data is still highly computationally intensive, especially when different collision systems need to be considered jointly for the inference, since each individual system needs a separate emulator for facilitating the global Bayesian inference.

By realizing that different collision systems share common physics are thus have related physical observables, Ref.~\cite{Liyanage:2022byj} applied transfer learning to build an efficient emulator (or target task, $f_T(\bold{x})$) from emulator (for source task, $f_S(\bold{x})$) trained from limited existing training data simulated for a different situation. Specifically, the Kennedy--O'Hagan (KO) model is applied to model the discrepancy function between different nuclear collision systems or dynamical simulations, i.e., the systematic difference between target and source systems is modeled with one independent GP prior,
\begin{equation}
f_T(\bold{x}) = \rho f_S(\bold{x}) + \delta(\bold{x}), \quad \delta(\bold{x})\sim GP\{\mu_{\delta}, k^{SE}_{\delta}(\cdot,\cdot)\},\quad f_S(\bold{x})\sim GP\{\mu_{S}, k^{SE}_S(\cdot, \cdot)\},
\label{eq:transfer}
\end{equation}
where the squared-exponential kernels are used with different variance and length-scale parameters for the source emulator and the discrepancy function.
With a collection of training data consisting of enough (suppose size to be $m$) model simulations for the source system ($\{\mathbf{X}_S, \mathbf{y_S} \}$) and a much smaller sized (suppose size to be $n\ll m$) simulations for the target system ($\{\mathbf{X}_T, \mathbf{y}_T \}$), the posterior distribution for the target system $f_T$ at any given different parameter $\mathbf{x}_{\rm new}$ can be derived as
\begin{equation}
[f_T(\mathbf{x}_{\rm new})|\mathbf{y}_S, \mathbf{y}_T] \sim \mathcal{N}(\mu^{*}_T(\mathbf{x}_{\rm new}),\sigma^{2*}_T(\mathbf{x}_{\rm new})),
\label{eq:transfer_posterior}
\end{equation}
\begin{equation}
\mu^{*}_T(\mathbf{x}_{\rm new})=\rho\mu_S + \mu_{\delta} + \mathbf{k}^{\top}\mathbf{\Sigma}^{-1}\left(
  \begin{bmatrix}
    \mathbf{y}_S\\
    \mathbf{y}_T
  \end{bmatrix} - 
  \begin{bmatrix}
    \mu_S\mathbf{1}_{m}\\
    (\rho\mu_S+\mu_\delta)\mathbf{1}_{n}
  \end{bmatrix} \right),
\label{eq:transfer_posterior_mean}
\end{equation}
\begin{equation}
{\sigma^2_T}^*(\mathbf{x}_{\rm  new}) = \rho^2\mathbf{k}_S(\mathbf{x}_{\rm new},\mathbf{x}_{\rm new})+\mathbf{k}_\delta(\mathbf{x}_{\rm new},\mathbf{x}_{\rm new}) - \mathbf{k}_{\rm new}^\top \mathbf{\Sigma}^{-1}\mathbf{k}_{\rm new},
\label{eq:transfer_posterior_variance}
\end{equation}
where $\mathbf{k}_{\rm new}=[\mathbf{k}_{\rm new}^S,\mathbf{k}_{\rm new}^T]=[[k(\mathbf{x}_{\rm new},\mathbf{x}^S_i)]_{i=1}^{m}, [k(\mathbf{x}_{\rm new},\mathbf{x}^T_j)]_{j=1}^{n}]$ and 
\begin{align*}
    \mathbf{\Sigma} &= 
    \begin{bmatrix}
        \mathbf{K}_S(\mathbf{X}_S)+\gamma^2_S\mathbf{I}_{m} & \rho\mathbf{K}_S(\mathbf{X}_S,\mathbf{X}_T)\\
        \rho\mathbf{K}_S(\mathbf{X}_S,\mathbf{X}_T)^T & \rho^2\mathbf{K}_S(\mathbf{X}_T)+\mathbf{K}_\delta(\mathbf{X}_T)+\gamma^2_T\mathbf{I}_{n}
    \end{bmatrix}.
\end{align*}\
These thus serve the transfer learning emulator model with $\mu^{*}_T$ providing the prediction and $\sigma^{2*}_T$ quantifying the uncertainty, which by construction and in spirit transfers the captured knowledge from an easily accessed or already existed source term emulator to a different target system.

\subsubsection{Accelerating Hydrodynamic Simulations}
Fast simulation via machine learning for heavy ion collision modeling with relativistic hydrodynamics is discussed in Ref.~\cite{Huang:2018fzn}. A stacked U-Net (sU-net) is designed and trained to capture the non-linear mapping from the initial state to final state profiles of the fluid field in the special case of being with ideal EoS ($p=e/3$), zero viscosity, zero charge densities and a fixed spatiotemporal discretization. The initial and final energy-momentum tensor profiles from hydrodynamics, $T^{\tau\tau}(x,y), T^{\tau x}(x,y), T^{\tau y}(x,y)$, are taken as the input and output for the deep neural network. 
%%%%%%%%%%%%%%%%%%%%%%%%%%%%%%%%%%%%%%%%%%%%%%%%%%%%%%%%
\begin{figure}[htbp!]
    \centering
    \includegraphics[width = 0.75\textwidth]{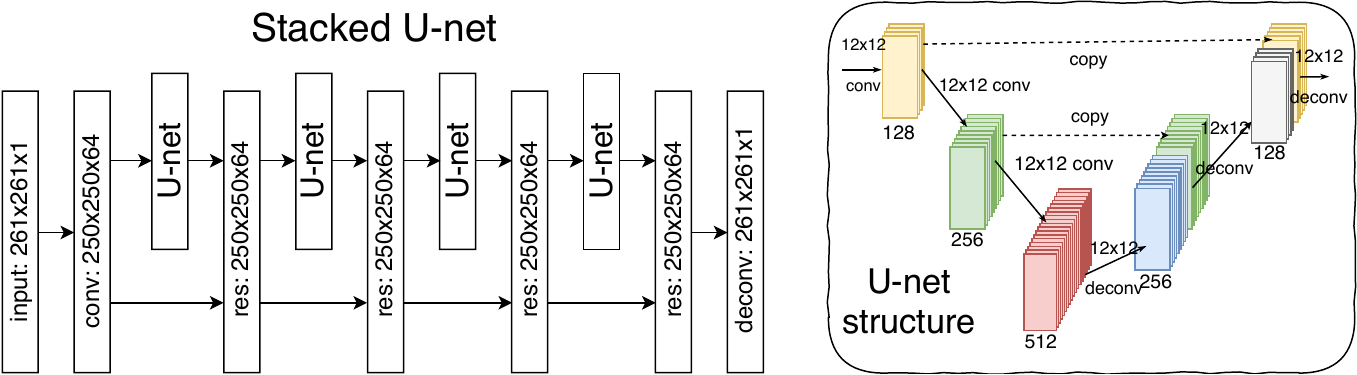}
    \caption{Taken from ~\cite{Huang:2018fzn} with permission. The stacked U-net structure for predicting flow field in the relativistic hydrodynamic simulation of heavy ion collisions.
    \label{fig:sunet}}
\end{figure}
%%%%%%%%%%%%%%%%%%%%%%%%%%%%%%%%%%%%%%%%%%%%%%%%%%%%%%%%

As a variation of the autoencoder structure with short-cut residual connections, the constructed sU-net architecture has been shown to be powerful for image synthesis or object segmentation, with the ability to identify local patterns. In the context of HICs, with simulation from hydrodynamic non-linear partial differential equations, sU-net, as shown in Fig.~\ref{fig:sunet}, was demonstrated to be able to predict well the final energy density and flow velocity profiles given any arbitrary initial profile from an initial condition model. Different initial conditional models were implemented for the generalizability test of the trained sU-Net, including MC-KLN, AMPT, and TRENTo with vanishing transverse flow-velocity, which were used for training the MC-Glauber initial condition. For physics interests in investigating the deformation and inhomogeneity of the created QGP medium, the eccentricity coefficients associated with the final energy density profiles from the sU-Net prediction were calculated and found to reproduce the results from VISH2+1 hydro simulation well.
Note in Ref.~\cite{Taradiy:2021pxd} demonstrated on simple non-relativistic hydrodynamics, one single DNN was shown to be able to predict the fluid dynamics in extrapolation manner on both initial hydro profile and simulation duration. Another promising strategy not fully explored for HICs yet may come from a physics-informed neural network, which has been demonstrated for fluid dynamics~\cite{2021AcMSn..37.1727C, RAO2020207, BAI2022114740} prediction in applied physics. It is also interesting to explore the inverse inference through such trained PINN for the relativistic Hydrodynamics simulations.

\subsection{Summary}\label{sub:hic_outlook}

In this chapter, the ``standard model'' of high energy heavy ion collisions is reviewed. It is a hybrid model of gluon saturation, relativistic fluid dynamics for QGP expansion, hadronic transport, and the jet-medium interaction. The model is used to generate a vast amount of data that has been used in Bayesian analysis to constrain the QCD EoS at high energy, the initial state entropy deposition, the temperature-dependent shear and bulk viscosity, and the jet energy loss coefficients/distributions. The data are also used in PCA or deep learning to rediscover physical observables, which verifies the representation power of machine learning.
It is worth noting that, besides the research mentioned above, the determination of other bulk thermodynamics, like the entropy production~\cite{Habashy:2021qku}, or related particle ratios~\cite{Rahman:2022tfq} and particles multiplicity~\cite{Habashy:2021poi} have been very recently investigated with DNN or Bayesian approach~\cite{Yousefnia:2021cup}.

The deep neural network is employed in supervised learning to establish connections between the initial nuclear structure, the QCD Equation of State (EoS), the Chiral Magnetic Effect (CME), jet flavors, and jet production positions to final state hadrons.
In conclusion, machine learning tools can serve as rapid simulators of high-energy collisions. The active learning algorithm can be utilized to identify important regions of the parameter space that should be simulated with hybrid models to enhance data generation efficiency. After training to create connections between data and model parameters, the network $f(x, \theta)$ can function as a novel physical observable, offering inspiration for both experimentalists and theorists in interpreting its results.
	\newpage
    \section{Lattice QCD}\label{sec:lat}
As the fundamental theory for describing the strong interaction, QCD is challenging to be solved due to its characteristic asymptotic freedom, which embarrasses perturbative treatment at low energy scales~\cite{Gross:2022hyw}. Therefore, non-perturbative methods are needed to study nuclear matter in the context of strong interactions. Lattice field theory plays an important role in ab-initio computation\footnote{Functional methods e.g., functional renormalization group (fRG) and Dyson-Schwinger (D--S) equations form the other first-principles approaches. See Ref.~\cite{Fischer:2018sdj} for a review.} of general many-body systems~\cite{Wilson:1974sk}. In this approach, the system's partition function or path integrals are discretized on a lattice of Euclidean spacetime, and field configurations or paths can be generated using importance sampling for further evaluation of physical observables.

Generally, algorithms related to Markov Chain Monte Carlo (MCMC), such as the Hybrid Monte Carlo (HMC) algorithm~\cite{Duane:1987de}, are employed to generate ensembles of configurations in lattice calculations~\cite{Knechtli:2017sna}, and have succeeded in providing important results, e.g., for the QCD phase structure~\cite{Fukushima:2010bq,Philipsen:2012nu,Ding:2015ona,Guenther:2020jwe, Karsch:2022opd}. See also Ref.~\cite{Ratti:2018ksb} for a recent review. However, the inherent sequential nature and diffusive update involved in these Monte Carlo algorithms seriously hamper the sampling efficiency for larger and finer lattices due to the phenomenon of critical slowing down(CSD)~\cite{Wolff:1989wq}. The computation becomes even more challenging for evaluating real-time dynamical properties. Recent strides in machine learning techniques may provide a promising way to either circumvent or alleviate these computational obstacles involved in lattice QFT and subsequent QCD studies~\cite{Boyda:2022nmh}. From a methodological point of view, many of the algorithmic advancements were initially explored on simple many-body physics or QFT systems. Yet, they hold the potential for broader application to QCD studies. Therefore, this chapter discusses the topic with a generalized perspective on lattice field theory.%the discussion in this chapter is with generality into lattice field theory.

\subsection{Overview and Challenges in Lattice Field Theory}
\label{sec:lattice_cha}
In general, Lattice QFT offers a non-perturbative approach to solving the path integral for the field system on a discretized Euclidean spacetime lattice. This allows the expectation value of an observable, $\mathcal{O}(\phi)$, to be represented by the discretized action,
\begin{equation}
    \langle\mathcal{O}\rangle=
    \frac{1}{Z}\int\mathcal{D}\phi\,\mathcal{O}(\phi)e^{-S(\phi)}.
    \label{eq:obs}
\end{equation}
Here $Z=\int\mathcal{D}\phi e^{-S(\phi)}$ is the partition function, while $\int\mathcal{D}\phi$ signifies the integration over all possible configurations of the discretized quantum field $\phi$. All the dynamical and interaction information of the fields are encoded in the action $S(\phi)$. The specific numerical evaluation of Eq.~\ref{eq:obs} hinges on sampling configurations following the action-dictated probability distribution 
\begin{equation}
    p(\phi)=\frac{1}{Z}e^{-S(\phi)}.
    \label{eq:target}
\end{equation}

We briefly introduce several challenges and computational tasks associated with lattice field theory calculations in the following. Subsequently, we will review the current advancements in leveraging machine learning techniques to address these challenges.

\begin{itemize}
\item{\textbf{Critical slowing down}}

Monte Carlo (MC) simulation can provide unbiased sampling for lattice field theory, which has been proven to be ergodic and asymptotically exact to approach the target distribution $p(\phi)$ under detailed balance condition for the involved proposal probability~\cite{krauth2006statistical}.
Within a classical MCMC framework, the process of configuration sampling typically involves proposing a new configuration based on the preceding one. This proposal then undergoes a subsequent judgment for acceptance or rejection, with the decision often made by evaluating a specific probability using both the proposed and the previous field configurations --- as exemplified by the Metropolis-Hastings algorithm~\cite{Metropolis:1953am,Hastings:1970mcs}. In these diffusive updates based sampling approaches, the proposal for new field configuration is usually made by local perturbation of the previous one or heat bath updates, which thus is inefficient in drawing independent configurations. Improvements can be realized by HMC which relies on evolving the configuration $\phi(x)$ jointly with a “conjugate momentum” $\pi(x)$ under classical Hamiltonian dynamics. Practically the computational costs involved are due to the strong autocorrelation of local updates, or the single local update itself is computationally expensive. The autocorrelation time is expected to scale as $\tau\sim\xi^z$, where $\xi$ is the correlation length that diverges around the critical point or approaching the continuum limit, and $z$ is the dynamical critical exponents which are $\approx2$ for standard local update algorithms~\cite{Wolff:1989wq}. Correspondingly, the induced severe inefficiency of sampling is called \textit{critical slowing down} (CSD). Being related, in the context of lattice field theories with well-defined topological behavior like QCD, \textit{topological freezing} can happen where topological observables usually have exponential scaling $\tau\sim e^{z\xi}$~\cite{DelDebbio:2004xh,Schaefer:2010hu}. CSD and topological freezing form the main barrier of lattice QFT computations.

\item{\textbf{Observables measurements and physics analysis}}

With ensembles of field configurations sampled from the desired distribution $p(\phi)$, one can evaluate correlation functions and further various physical observables. This process involves intensive operations on field configurations which are high-dimensional tensors. Furthermore, the physics underlying the studied system must be analyzed, including thermodynamics, phase transitions (characterized by the order parameter), % Physics is thus also to be analyzed such as thermodynamics or phases transition (order parameter) of the system, 
and also different real-time dynamical properties, e.g., the reconstruction of spectral function or parton distribution functions (PDF) represent a particularly challenging, ill-conditioned inverse problem. 

\item{\textbf{Sign problem}}

The sign problem in lattice QCD calculations refers to the fact that the functional integral defined in the partition function is not positive definite in many cases (e.g., at finite density or for Minkowski-time dynamical physics), which complicates the application of standard Monte Carlo methods~\cite{Troyer:2004ge}. In lattice QCD, the partition function is expressed as a functional integral over all possible configurations of quarks and gluons on a discrete four-dimensional lattice in the form~\cite{Aarts:2015tyj,Nagata:2021ugx},
\begin{equation}
    Z = \int D[A] D[\psi] D[\bar{\psi}] e^{-S[A,\psi,\bar{\psi}])},
\end{equation}
where $S$ is the QCD action, $A$ is the gauge field, and $\psi$ and $\bar{\psi}$ are the quark and antiquark fields, respectively. 
In finite-density QCD, the nonzero chemical potential($\mu$) renders the fermion determinant complex within the partition function, $Z = \int D[A] \text{exp}(-S_\text{YM})\text{det}M(\mu)$, where $S_\text{YM}$ is the Yang-Mills action. A primary concern arises since the exponent in the integrand, $-S$, is not always positive, which leads to the non-positive definite integral. It induces the ill-defined probability when sampling configurations in standard Monte Carlo techniques. Though certain methods might technically navigate around the predicament, they will inevitably encounter the highly fluctuating phase factors which result in an exponential growth of computational cost as the system's volume expands~\cite{Berger:2019odf,Alexandru:2020wrj}.

\end{itemize}

\subsection{Field Configuration Generation}

In lattice QFT research, generating field configurations is often the most computationally demanding task, primarily because of the typically required Monte Carlo simulation on Markov chains. These simulations tackle the sampling of high dimensional distribution distributions, especially as they strive to approach the continuum limit.
In fact, for general many-body system studies that use Monte Carlo simulation, considerable efforts have been invested in crafting intelligent proposals or global updates during the MCMC procedures. The goal here is to reduce the autocorrelation time, which, when increased, tends to impede the efficiency of MCMC sampling. Besides the critical slowing down as introduced in Sec.~\ref{sec:lattice_cha}, which is a crucial barrier when pushing the lattice calculation to the continuum limit, the other challenge faced by conventional MCMC simulations is sampling from multimodal distributions. Navigating between widely separated modes of the target distribution using update-based samplers is far from straightforward~\cite{DelDebbio:2004xh,Hasenbusch:2017fsd}. This challenge of high-dimensional sampling and the quest for efficient proposal design is also a prominent theme in the machine learning community. It has spurred the rapid advancement of generative models~\cite{Wang2018GenerativeMF}. Research in both general classical/quantum many-body statistical physics studies and lattice QFT has demonstrated that modern generative algorithmic development stemming from AI/ML can offer improved efficiency, and serve as a valuable complement to traditional Monte Carlo methods, especially in the context of generating configurations for physical systems.

\subsubsection{GAN-based Algorithms}
\begin{figure}[htbp!]
  \centering
  \includegraphics[width = 0.41\textwidth]{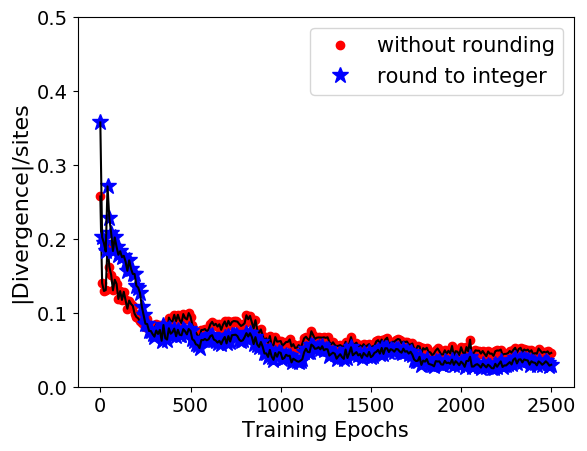}
  \includegraphics[width = 0.45\textwidth]{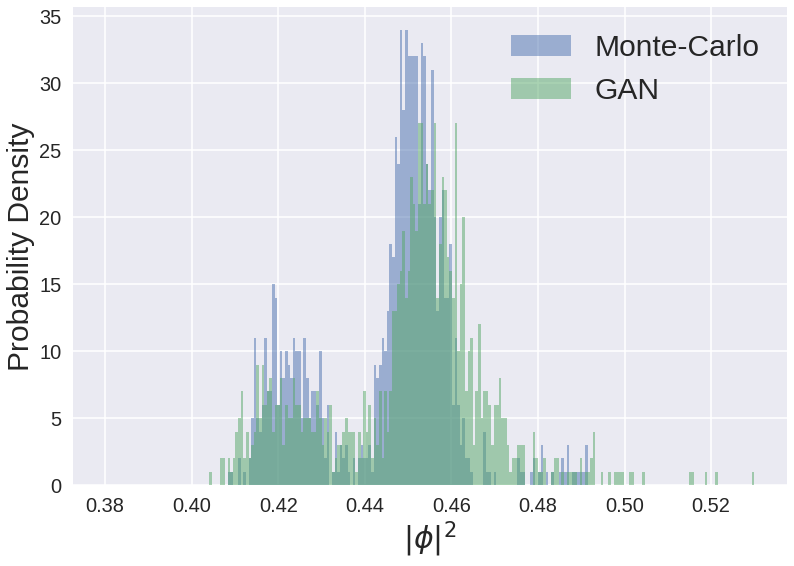}
  \caption{(left) The degree of divergence satisfaction for configurations from trained GAN generator; (right) the probability density distribution for the squared field $\phi^2$ from GAN and from Monte-Carlo simulations at a fixed chemical potential. Taken from Ref.~\cite{Zhou:2018ill}.}
  \label{fig:scalar_gan}
\end{figure}

As introduced in Sec.~\ref{subsubsec:gm}, the Generative Adversarial Network(GAN), is a deep generative model to realize implicit MLE for distribution learning through adversarial training. This generative model has recently been explored for many-body statistical systems (e.g., for Ising model in~\cite{2017arXiv171004987L}) and also for the configuration generation of quantum field systems.
In Refs.~\cite{Zhou:2018ill,Zhou:2021vza,Zhou:2020yna}, the GAN is used for the configuration generation of complex scalar $\phi^4$ field at non-zero chemical potential in the context of lattice QFT study. In Sec.~\ref{sec_phases_obs} about \textit{\textbf{Regression in QFT}} sector, this QFT system under worldline formalism was also introduced for regression exploration via deep learning. New configuration generation via Wasserstein-GAN scheme is also explored~\cite{Zhou:2018ill} in the task of generating uncorrelated configurations satisfying the physical distribution, with training ensembles of configurations prepared by the worm algorithm. Since the considered field system under the dualization approach should satisfy an important local divergence-type constraint reflecting flux conservation, the effectiveness of the generative algorithms, in terms of whether the generated configurations are physical or not, is verified by checking this divergence-type constraint condition for training. Surprisingly, as shown in the left part of Fig.~\ref{fig:scalar_gan}, it is found that this highly implicit physical constraint condition is realized in a converging manner (more and more together with training epochs) for configurations generated by the trained generator within GAN, without explicit guidance of this constraint. 
%%%%%%%%%%%%%%%%%%%%%%%%%%%%%%%%%%%%%%%%%%%%%%%%%%%%%%%%
%%%%%%%%%%%%%%%%%%%%%%%%%%%%%%%%%%%%%%%%%%%%%%%%%%%%%%%%
Also, the distribution with respect to normal physical observables from the generated samples was found to agree well with the MCMC evaluation. And here, for the considered complex scalar field system, it even agrees reasonably well over the multimodally distributed number density and field square (see the right side of Fig.~\ref{fig:scalar_gan}), which is not trivial for conventional sampling approaches without special treatment. Furthermore, a conditional generative network, cGAN, is proposed in~\cite{Zhou:2018ill} to generalize the configuration generation ability of the generator to be dependent/conditional on physical observables (specifically the number density $n$ was used for demonstration) and go beyond the distribution on which it was trained. Ref.~\cite{Singha:2021nht} later applied the conditional GAN in the lattice Gross Neveu model to mitigate the critical slowing down when approaching the critical region.

This strategy with the conditional GAN has also been applied to the spin configuration generation of the two-dimensional XY model~\cite{2021ScPP...11...43S}, with several different architectures proposed including one specific output distribution entropy maximization regulator to mitigate the mode-collapse problem. As the non-local defining feature for the involved topological phase transition, the vortices distribution is also especially scrutinized besides other relevant observables (e.g., magnetization and energy). With the trained GAN, Ref.~\cite{2021ScPP...11...43S} further proposed the GAN fidelity in terms of the discriminator network output which is shown to be able to detect the phase transition unsupervisedly.

The above demonstration, especially the consistency between the distribution from the GAN and the desired one, actually indicates that the trained GAN here can be taken as a good enough proposal on a Markov Chain when one wants to guarantee the ergodicity and detailed balance properties for the sampling process, thus make sure the correct physics can be estimated in converging manner. This indeed is further demonstrated in Ref.~\cite{Pawlowski:2018qxs} on scalar field theory together with the Hamiltonian Monte Carlo (HMC) approach. 
Specifically, the trained GAN is proposed to serve as overrelaxation (i.e., configuration space exploration that keeps the action unchanged, $\delta S=0$) procedure within the action-based MCMC sampling algorithm. Starting from some initial configuration $\phi$ after a number of HMC steps, the gradient flow is performed on the latent variable $z$ to achieve the GAN overrelaxation proposal $G(z')$, 
\begin{equation}
    z'(\tau+\epsilon)=z'(\tau)-\epsilon\frac{\partial (S[G(z)]-S[\phi])^2}{\partial z}, 
    \label{eq:gflow_gan}
\end{equation}
where $\epsilon$ is the learning rate and $\tau$ the training epochs. In Metropolis steps, proposal configuration with $\delta S=0$ (w.r.t. the last configuration) is accepted automatically given the transition probability for the proposal is symmetric. Heuristic arguments are provided in~\cite{Pawlowski:2018qxs} on this issue for GAN's selection probability $P(G(z')|\phi)$. With demonstration on 2d scalar field theory, under such GAN overrelaxation method, the autocorrelation time has been shown to be largely reduced. 

Note that the GAN-based approach in general requires a training data set to be prepared from conventional sampling means, which hampers its ability to assist the field configuration generation purely from the physically known action or Hamiltonian. Also, though it can be taken as a proposal in generating independent and physically promising configurations, the vanilla GAN has no evaluation of the likelihood and thus the sample probability. More potential can be unlocked for the GAN-based approach if the explicit likelihood estimation can be added inside the adversarial learning in the future. In contrast, the methods introduced in the following can render the self-training just from physical action or Hamiltonian, with also the sample probability to be accessible.

\subsubsection{Self-Learning through Effective Action}

\emph{\textbf{Self-Learning Monte Carlo}} --- 
The self-learning Monte Carlo (SLMC) method is a general-purpose numerical algorithm for configuration generation of many-body systems based on machine learning. The development of SLMC originally is within classical spin systems\cite{2017PhRvB..95d1101L} and shows clear improvement in curing the critical slowing down problem. It works also with other general quantum models that are concerned with condensed matter physics and quantum chemistry.~\cite{PhysRevB.102.041124}. The basic idea is to learn a tunable and effective Hamiltonian or Action that can be associated with a more efficient update algorithm (such as cluster-update or other global moving ways to propose uncorrelated configurations), with which, the Metropolis-Hasting test using the real action can turn it to be an exact sampler for the system.

\begin{figure}[!hbtp]
    \centering
    \includegraphics[width=0.6\textwidth]{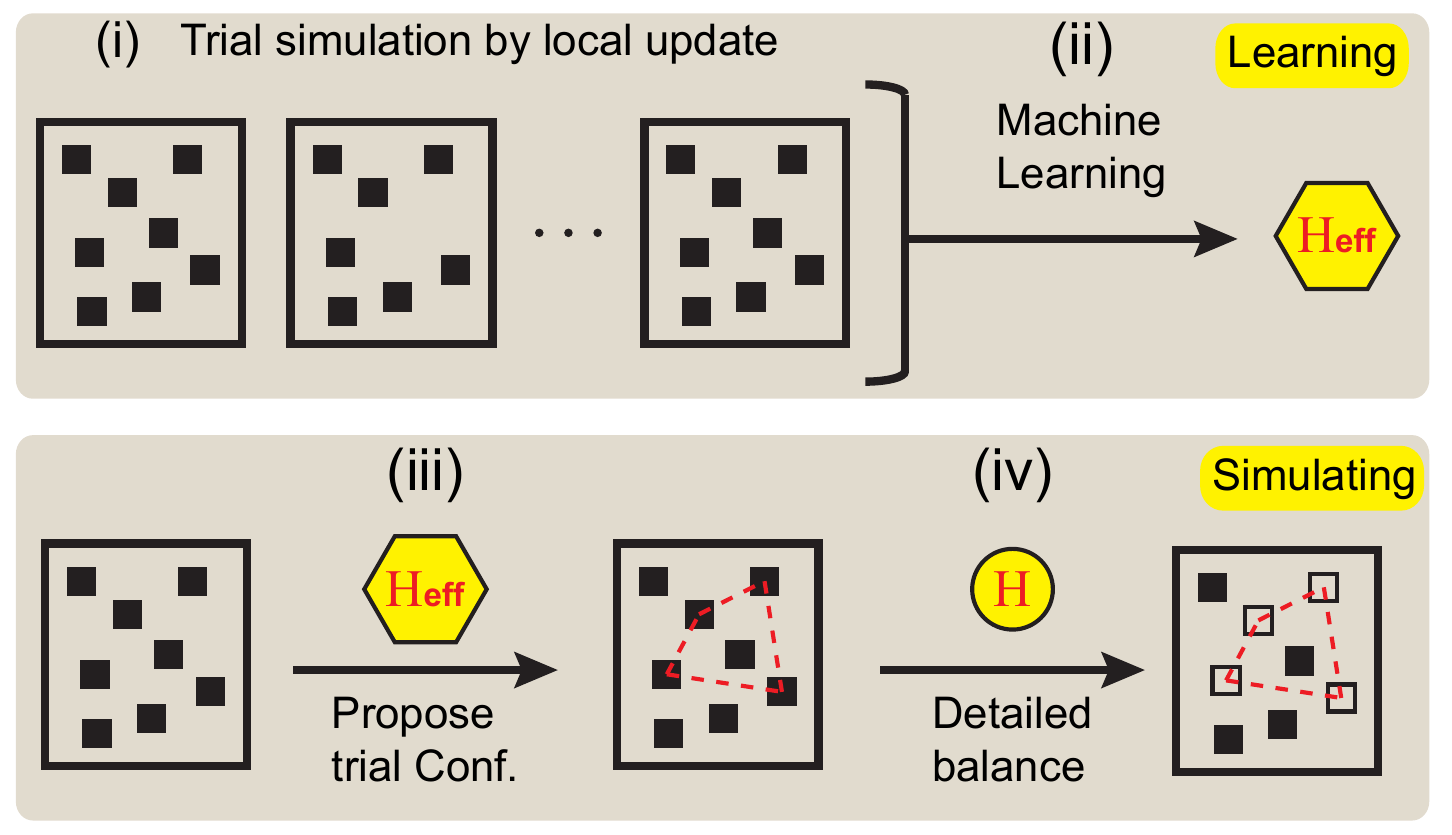}
    \caption{(From Ref.~\cite{2017PhRvB..95d1101L})Schematic illustration of learning process (top panel) and simulating process (bottom panel) in SLMC. \label{fig:SLMC}}
\end{figure}

As detailed in Ref.~\cite{2017PhRvB..95d1101L} on the example of a statistical spin model, the SLMC consists of a learning phase and exploration phase within actual MC simulation in general: (1) a conventional MC simulation under local update can be performed to produce a series of configurations with its weight (energy or action evaluation) also being known; (2) based on these training data an effective Hamiltonian $H_{eff}$ can be learned, to which one can use a global update for a faster simulation (e.g. when only two-particle interactions are contained in $H_{eff}$); (3) the learned $H_{eff}$ can be used to make proposal in actual MC simulation and (4) perform Metropolis Hasting test with the original Hamiltonian to correct (reject or accept). See Fig.~\ref{fig:SLMC} for a schematic illustration of SLMC in a simple spin model.
The training over $H_{eff}$ can also be self-improved by iteratively reinforcing training $H_{eff}$ with generated configuration from the last step self-learning update with $H_{eff}$. Basically, a more efficient MC update model can be obtained from the SLMC. It's demonstrated that the SLMC indeed largely reduced the autocorrelation time $\tau$ especially near the phase transition, and gives around 10 to 20-fold speedup on the considered 2D generalized Ising model.
It's worth noting that similar ideas by training and simulating Restricted Boltzmann Machines (RBM)~\cite{10.1162/089976602760128018} were adopted in Refs.~\cite{PhysRevB.95.035105,2016PhRvB..94p5134T,Wang:2017mzw,2019PhRvE.100d3301P} for proposing efficient Monte Carlo updates, which is not limited to Trotter decomposition for fermionic system and renders faster proposal since the used Gibbs sampling in RBM.

This self-learning MC strategy was further pushed forward into more realistic fermionic systems where Trotter decomposition is employed~\cite{Liu:2016zmg, 2017PhRvB..96d1119X} and for continuous time Monte Carlo simulation~\cite{2017PhRvB..96p1102N}.  Deep neural network techniques are also additionally introduced into the effective action or Hamiltonian's construction in Ref.~\cite{2018PhRvB..97t5140S, Nagai:2018sav,2019PhRvB.100d5153S}, which can further increase the flexibility of the effective action and improve the following acceptance rate in the Monte Carlo sampling stage. 

The SLMC for non-abelian SU(2) gauge theory with dynamical staggered fermions at zero and finite temperature was developed in Ref.~\cite{Nagai:2020jar}.
Ref.~\cite{Tomiya:2021ywc} devised a gauge covariant neural network for 4-dimensional non-abelian gauge field theory, and adopted such network to construct effective action within HMC to achieve self-learning HMC scheme, with demonstration on the case of two color QCD including un-rooted staggered fermion.

\emph{\textbf{Action parameter regression}} ---
In tackling the issue of CSD for lattice field theory study, with promise, multiscale methods are proposed to overcome the CSD by refining ensemble of configurations at a coarse scale. Such methods require \textbf{action matching} across different scales via renormalization group (RG), then the sampling at coarse-scale level can already render the approach to continuum limit physics evaluation together with a cheap re-thermalization with the original fine action. The key challenge involved is parametric regression for identifying the proper action parameters that best describe physics at coarse scale from an ensemble of configurations generated at a finer scale. It's thus proposed to use deep neural networks to tackle this action-matching regression task\cite{Shanahan:2018vcv}, where the coarsened ensemble of SU(2) gauge field configurations are taken as input and the required action parameters are the output. It's worth noting that the mismatch from the regression can be corrected by re-thermalization steps on the finer scale with the corresponding action.

The simple fully-connected neural network was tried first and found to appear successful in validation, which however fails to generalize to different parameter cases or even new Hybrid Monte Carlo (HMC) simulation streams of the same parameters as of the training set. The failure of such naive neural network is argued in Ref.\cite{Shanahan:2018vcv} to be induced by the lack of symmetries of the gauge field configurations, and further proposed a customized symmetry-preserving network to reduce effective degrees of freedom for the task. The embedding of the symmetries is designed by featuring an initial preprocessing layer to yield possible symmetry-invariant quantities as input to the following fully-connected layers.  This gives accurate parameter regression and successful generalization for even ensembles not distinguishable from principal component analysis (PCA). It thus provides a solution to the action matching.

\subsubsection{Variational Autoregressive Network}
\label{van}

The key object for a general many-body system is the free energy, which contains all the information about the system in principle. From a probabilistic point of view, the usual MCMC approach basically uses importance sampling to implicitly approach the free energy, which is an intractable high dimensional integration. However, the direct evaluation of the free energy is not possible for the naive classical MCMC approach (note that there are variants of MCMC algorithms developed to be able to approximately assess the free energy, which is computationally expensive). As another alternative strategy based on variational point of view, the \textit{mean field approach} or related information transfer algorithms can be adopted for the free energy estimation, which then basically performs the minimization over the variational free energy. In probabilistic language this is equivalent to the minimization over the reverse (or backward) KL divergence between the variational distribution $q_{\theta}(\phi)$ and the target distribution in Eq.~\ref{eq:target} or in statistical physics $p(\phi)=e^{-\beta H(\phi)}/Z$ (where $\beta$ is the inverse temperature, $H(\phi)$ the Hamiltonian of the system, and $Z$ is the partition function),
\begin{equation}
    \mathcal{D}_{KL}(q_{\theta}(\phi)||p(\phi))=\int\mathcal{D}[\phi] q_{\theta}(\phi)\log\frac{q_{\theta}(\phi)}{p(\phi)}=\beta(F_{q}-F), 
    \label{eq:KL_van}
\end{equation}
with the variational free energy defined as
\begin{equation}
    \beta F_q=\mathbb{E}_{\phi\sim q_{\theta}}[\beta H(\phi)+\log q_{\theta}(\phi)], 
    \label{eq:var_free}
\end{equation}
Because of the non-negativity for KL divergence ($\mathcal{D}_{KL}\ge0$, also known as \textit{Gibbs-Bogoliubov-Feynman inequality}~\cite{Peierls:1938zz, 10.1063/1.1704383, PhysRevLett.22.631}), the true free energy $F$ is upper bounded by the variational free energy $F_q$, and the equality happens when the variational distribution $q_{\theta}$ really reaches the target distribution $p$ exactly.  Note that in machine learning community as proposed in the beginning as general density estimation method, the autoregressive network usually uses the (forward) KL divergence between data empirical distribution (e.g., constructed with training data) and the variational distribution, $\mathcal{D}_{KL}(p_{data}||q_{\theta})$. For many-body physics or QFT study, the proposal to use the reverse KL divergence together with the easily sampleable variational ansatz makes it possible for \textbf{self-training} starting solely from the unnormalized target distribution\footnote{i.e., the training is performed with sampled generated from the variational distribution, instead of collected samples from the true distribution in advance}, e.g., knowing the action or Hamiltonian for the physical system in equilibrium.

Despite its popularity and success, the \textit{mean field calculation} is quite often limited by the assumed \textit{variational ansatz} (e.g., factorized), especially when the system is with strong correlations between their degrees of freedom, thus most of the time it is valid only in high temperature cases or when the system's topological structure fulfills the requirement from mean field approximation. Here the autoregressive model naturally stands out, because both the direct sampling and tractable likelihood evaluation (which is desired in evaluating the variational free energy) can be simultaneously realized. Meanwhile, when the autoregressive model is constructed as variational ansatz, the variational free energy (Eq.~\ref{eq:var_free}), which serves as the loss function, can be estimated unbiasedly and stochastically by drawing ensembles of samples from the sequential stochastic process specified by the autoregressive model (see Eq.~\ref{auto_prob} as introduced in Sec.~\ref{subsubsec:gm}). Thus, in every optimization iteration, using data sampled from the autoregressive model it suffices to perform \textbf{self-training} on the model.

Ref.~\cite{2019PhRvL.122h0602W} for the first time proposed to introduce the neural network to give a more powerful yet tractable variational ansatz, taking advantage of the strong representational ability of neural networks due to the universal approximation theorem. To keep the evaluation of the variational free energy tractable, one needs to design the network such that the variational distribution represented by the network is accessible and efficiently computable. Accordingly, the autoregressive networks were adopted by the authors to decompose the joint probability over all lattices to a product of conditional probabilities, and parametrized each conditional with neural networks\footnote{Note that to represent the conditional probability, the output of neural networks are interpreted as essential parameters of the distribution. More explanation about autoregressive networks can refer to Sec.~\ref{subsubsec:gm}}. The idea was directly demonstrated on a simple 2D ferromagnetic Ising system, also generalized to the PixelCNN structure with convolutional layers included to respect the locality and the translational symmetry of the system. Compared to conventional MCMC evaluation, this variational autoregressive network can also give evaluation of the free energy by giving its well-minimized upper bound. This strategy of using autoregressive probabilistic models was later combined with quantum circuits, and extended the variational quantum eigensolver (VQE) to investigate thermal properties and excitations of quantum lattice model, termed as $\beta$-VQE~\cite{Liu:2021bst}. This quantum-classical hybrid algorithm was also applied to the Schwinger model at finite temperature and density~\cite{Tomiya:2022chr}, with a large volume limit evaluated and a continuum limit took in obtaining the phase diagram.
%%%%%%%%%%%%%%%%%%%%%%%%%%%%%%%%%%%%%%%%%%%%%%%%%%%%%%%%%%%%%%%%%%%%%%%%%%%%%%%%%%%%
\begin{figure}[htbp!]
    \centering
    \begin{minipage}[t]{0.48\textwidth}
        \centering
        \includegraphics[width=7.cm]{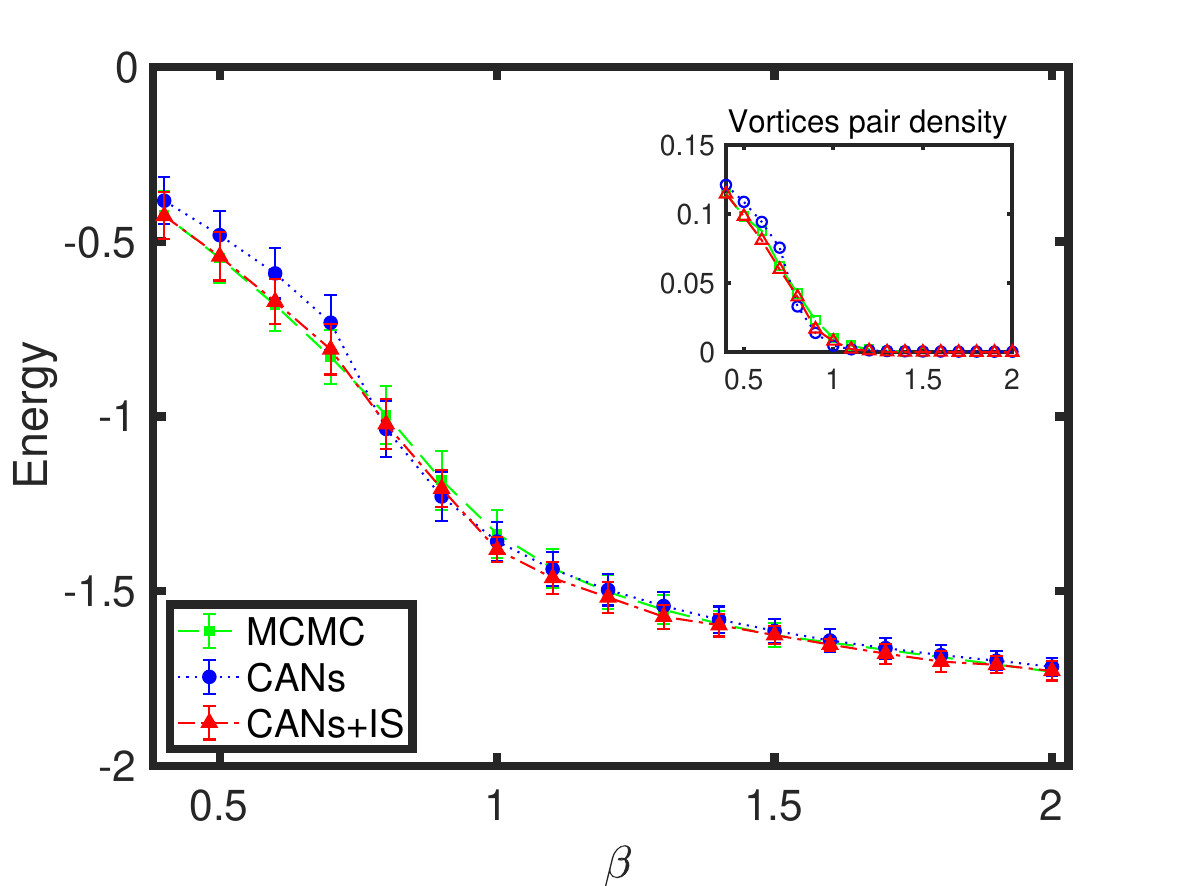}
        \caption{The energy per site of 2D XY model ($L=16$) from CANs, CANs+IS and MCMC. The upper inset shows the number of vortex pairs density with inverse temperature. Taken from ~\cite{Wang:2020hji}.}
        \label{fig:xy_en_can}
    \end{minipage}
    \hspace{0.5cm}
    \begin{minipage}[t]{0.48\textwidth}
        \centering
        \includegraphics[width=9cm]{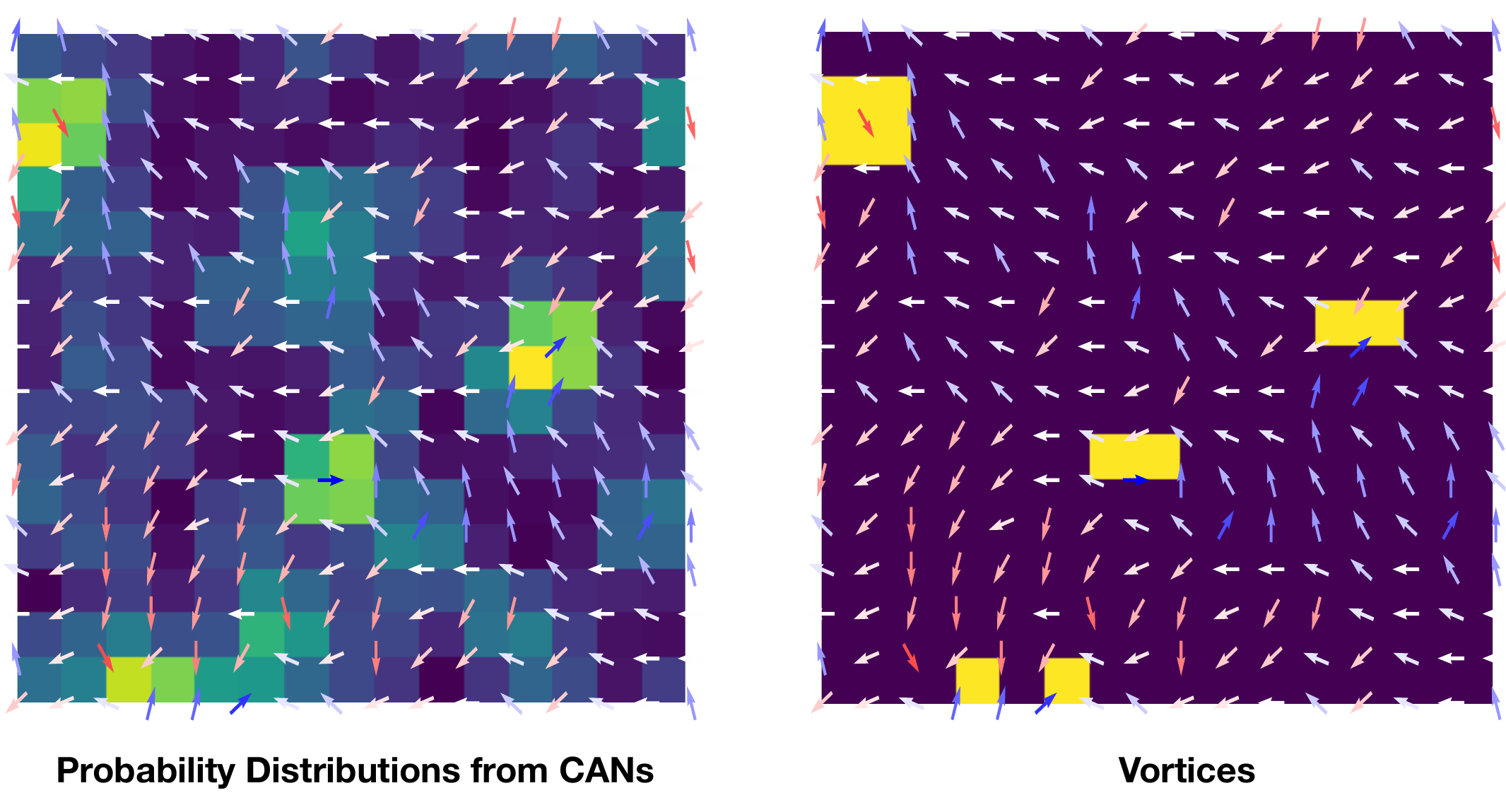}
        \caption{Probabilities analysis with CANs and corresponding vortices for a random 2D XY model configuration sampled from well-trained CANs at $\beta = 1.0$. Taken from ~\cite{Wang:2020hji}.}
        \label{fig:xy_vor_can}
    \end{minipage}
\end{figure}
%%%%%%%%%%%%%%%%%%%%%%%%%%%%%%%%%%%%%%%%%%%%%%%%%%%%%%%%%%%%%%%%%%%%%%%%%%%%%%%%%%%%

Authors in Ref.~\cite{Wang:2020hji} further generalized the above method to a general many-body system with continuous variables, where the probability interpretation of the introduced autoregressive is devised to be a mixture beta distribution, instead of a Bernoulli distribution for Ising spins. This newly extended continuous-mixture autoregressive network (denoted as CAN in the paper) is well demonstrated on the 2D XY model which exhibits non-trivial topological KT phase transition. The thermodynamics of the systems were shown to be captured successfully (see Fig.~\ref{fig:xy_en_can}), and the underlying emergent degree of freedom--vertex--is also found to be rediscovered by this CAN method. Furthermore, it is found that the trained CAN network can automatically give rise to the vortices' distribution for any random given XY spin configurations with its conditional probability components output from plaquette (see Fig.~\ref{fig:xy_vor_can}), directly indicating that the network captured the underlying emergent physics about this many-body system. Note that this CAN method can capture the O(2) symmetry for the 2D XY model, which is equivalent to global U(1).

Ref.~\cite{Wu:2021tfb} proposed symmetry-enforcing updates within MCMC with autoregressive neural network as global update proposer, since the system action or Hamiltonian remain invariant under specific symmetry operations (e.g., translation and reflections as considered in Ref.~\cite{Wu:2021tfb}). This introduced symmetry operation largely reduced the ergodicity problem from those exponentially suppressed configurations (i.e., those with exponentially smaller $q_{\theta}(\phi)$ compared to $p(\phi)$, however, will hardly influence the variational free energy evaluation). Additionally, a neural cluster update scheme is devised in Ref.~\cite{Wu:2021tfb} utilizing the decomposition structure of autoregressive model, which can lower the autocorrelation time by setting only a subset of the lattice to be changed instead of the whole in each Monte Carlo step.

\subsubsection{Flow-based Variational Learning}
\label{sec:3:flow_based}
Analogous to the above-mentioned variational autoregressive network models (refer to Sec.~\ref{van}), flow-based models employ normalizing flows (NF) to construct the variational ansatz for the target distribution, with the primary objective of approximating and learning the desired distribution by minimizing the variational free energy. Again, technically, this process involves minimizing the reverse mode of Kullback-Leibler (KL) divergence between the variational distribution constructed by the flow-based model and the target distribution.

%%%%%%%%%%%%%%%%%%%%%%%%%%%%%%%%%%%%%%%%%%%%%%%%%%%%%%%%%%%%%%%%%%%%%%%%%%%%%%%%%%%%
\begin{figure}[htbp!]
  \centering
  \includegraphics[width = 0.6\textwidth]{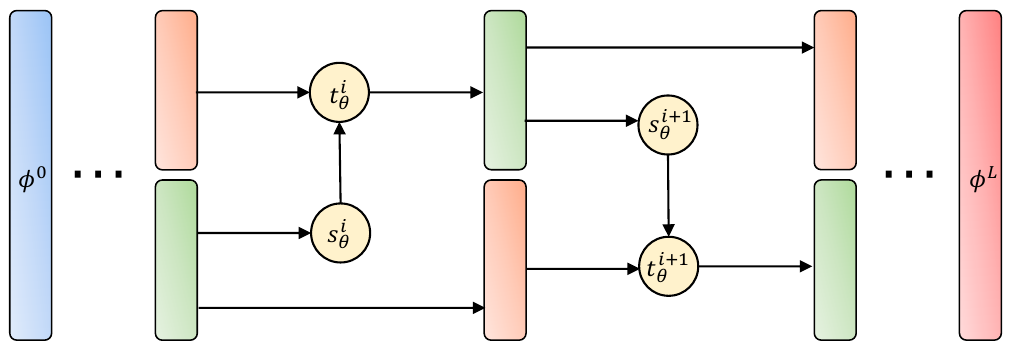}
  \caption{A schematic demonstration for the normalizing flow of the Real NVP, in which the yellow circles represent affine transformations represented by the neural networks.}
  \label{fig:3:normflow}
\end{figure}
%%%%%%%%%%%%%%%%%%%%%%%%%%%%%%%%%%%%%%%%%%%%%%%%%%%%%%%%%%%%%%%%%%%%%%%%%%%%%%%%%%%%

Though previously introduced in Sec.~\ref{subsubsec:gm}, we provide a brief explanation of the normalizing flow (NF) applications in the context of field configuration generation for the sake of terminology consistency in the following discussion. Generally, NF establishes a flexible and tractable probability distribution by facilitating a generative bijective transformation between a naive prior distribution $p_u(u)$ (e.g., multivariate Gaussians) and the variational distribution $q_{\theta}(\phi)$ (aimed at approaching the target distribution), $\phi=f_{\theta}(u)\sim q_{\theta}(\phi)$. This transformation is parametrized by neural networks with trainable parameters denoted as $\{\theta\}$, and is specifically designed to possess invertibility. Through the principle of probability conservation (i.e., the change of variable theorem for the probability distribution), one can easily derive the connection between the variational distribution and prior as $q_{\theta}(\phi)=p_u(u)|\det\frac{\partial f_{\theta}(u)}{\partial u}|^{-1}$. Given that the prior probability is easily evaluable and the Jacobian determinant $J_f$ is computable, one can calculate the variational probability $q_{\theta}(\phi)$ on any samples, allowing for further minimization of the variational free energy. Therefore, the key recipe in constructing a viable flow model is ensuring that the network represented transformation $f_{\theta}$ is invertible, differentiable, and composable. As the most common flow-based model, Real NVP structure (see Sec.~\ref{subsubsec:gm} for details) is frequently employed, which constructs the transformation $f_{\theta}$ by composing a series of affine coupling layers (see its schematic diagram in Fig.~\ref{fig:3:normflow}) each structured as follows:
\begin{equation}
\left\{
\begin{aligned}
    \phi^{i}_{1:k} &= \phi^{i-1}_{1:k} \\
    \phi^{i}_{k+1:N} &= \phi^{i-1}_{k+1:N} \odot e^{s_{\theta}^i(\phi^{i-1}_{1:k})} + t_{\theta}^i(\phi^{i-1}_{1:k}),
\end{aligned}
\right.
\end{equation}
with $\phi^i$ the output of the $i^{\text{th}}$ affine coupling layer (thus $\phi^0 = u\sim p_u(u)$ and final output $\phi^{L}=\phi\sim p(\phi)$), $k$ the separation point of the configuration variables into two subsets, and $N$ the number of variables in one configuration. The scaling and translation functions can be parameterized by DNNs as $s_{\theta}: \mathbb{R}^{k}\rightarrow \mathbb{R}^{N-k}$ and $t_{\theta}: \mathbb{R}^{k}\rightarrow \mathbb{R}^{N-k}$. The Jacobian determinant can also be directly evaluated to be $(\det J_{T}^i) = \Pi_j^{N-k}e^{s_{\theta}^i(X_{1:k})_j}$. Additionally, due to the splitting, the above coupling layer can be easily inverted,
\begin{equation}
\left\{
\begin{aligned}
    \phi^{i-1}_{1:k} &= \phi^{i}_{1:k} \\
    \phi^{i-1}_{k+1:N} &= (\phi^{i}_{k+1:N} - t_{\theta}^i(\phi^{i}_{1:k}))\odot e^{-s_{\theta}^i(\phi^{i}_{1:k})}.
\end{aligned}
\right.
\end{equation}

With the above-mentioned flows, i.e., a series of bijective transformation layers, one effectively attains a parametric model with an explicit probability for each sample, $q_{\theta}(\phi)$, and with tunable parameters denoted as $\theta$. Analogous to autoregressive models, the reverse KL divergence expressed in Eq.~\ref{eq:KL_van} can be employed to steer the optimization of $q_{\theta}(\phi)$ towards approaching the target distribution $p(\phi)$ in Eq.~\ref{eq:target},
\begin{equation}   \theta^{*}=\arg\min_{\theta}\mathcal{D}_{KL}(q_{\theta}(\phi)||p(\phi))=\arg\min_{\theta}[\int\mathcal{D}\phi q_{\theta}(\phi)(S(\phi)+\log q_{\theta}(\phi)) + \log Z ], 
    \label{eq:trainingKL}
\end{equation}

Ref.~\cite{Albergo:2019eim} presented the first application of the flow-based model to lattice quantum field theory simulations\footnote{An earlier similar flow-based model was developed and demonstrated on the two-dimensional Ising model in its continuous dual version, with the novel concept of neural network based variational Renormalization Group approach proposed\cite{2018PhRvL.121z0601L} proposed.}. Taking the two-dimensional $\phi^4$ field theory as an example, this study showed that the flow-based sampler confers substantial advantages over conventional sampling algorithms (with local Metropolis sampling and HMC considered for benchmarking), as evidenced by the systematic reduction in autocorrelation times when deploying the flow sampler as a proposal within the Markov Chain. The incorporation of the trained flow sampler with MCMC (e.g., Metropolis-Hastings) guarantees the asymptotic exactness of the sampling, as the Metropolis-Hastings acceptance/rejection step serves as a corrector to ensure that the sampling distribution precisely converges to the target distribution.
This method is commonly referred to as \textit{flow-based MCMC} in the literature. One of its notable attributes is the alleviation of the critical slowing down issue associated with the Markov Chain sampling, as the samples generated from the flow model are uncorrelated. Consequently, the associated cost is transitioned to the up-front training expenditure for the flow generative model.
It should be noted that in calculating observables ($\langle\mathcal{O}(\phi) \rangle$), besides the stochastic MCMC correction method, reweighting (i.e., importance sampling) can also be employed, which assigns a weight $w(\phi)=p(\phi)/q_{\theta}(\phi)$ to each sample $\phi$ when computing the expectation of the observables. A detailed introduction to \textit{flow-based MCMC} for the scalar field with code written in \textit{PyTorch} can be found in Ref.~\cite{Albergo:2021vyo}. We also summarize in Tab.~\ref{tab:flow} some related applications of this strategy to different many-body physics systems.
In Ref.~\cite{Nicoli:2020njz}, a simpler normalizing flow construction with Non-linear Independent Component Estimation (NICE, see Eq.~\ref{flow_nice}) is adopted for lattice simulation of 2-D real scalar field theory. In this study, the $\mathbb{Z}_2$-invariance is explicitly introduced by constraining the network to be with \textit{tanh} non-linearity and vanishing biases, which ensures that the transformation derived from the flow is an odd function, $f_{\theta}(-u)=-f_{\theta}(u)$. Interestingly, this work also proposed a direct estimator for the \textit{free energy} ($\hat{F}=-T\ln\hat{Z}$)\footnote{The temperature $T=\frac{1}{a N_T}$ where $a$ is the lattice spacing and $N_T$ is the number of lattice points in the temporal direction.} from the trained flow model via a Monte-Carlo approximation on the partition function,
\begin{equation}
\hat{Z}=\frac{1}{N}\sum_{\phi_i\sim q_{\theta}}[e^{-S(\phi_i)}/q_{\theta}(\phi_i)] .
\label{eq:eff_free}
\end{equation}
This approach is conceptually appealing as it circumvents the cumbersome integration error accumulation inherent in conventional MCMC-based estimations for the partition function. Once the free energy is well estimated, other thermodynamic observables can be naturally obtained by taking the derivative of the free energy, and can also be further refined by the importance sampling~\cite{Muller:2019nis}. The capability exhibited by flow-based models primarily stems from their explicit likelihood estimation ability, a feature also shared by variational autoregressive models as introduced in Sec.~\ref{van}.

In Ref.~\cite{Singha:2023cql}, a conditional normalizing flow (c-NF) model was trained on samples pre-generated from HMC in the non-critical region of the theory\footnote{Note that being different from the variational strategy of flow-based MCMC method, this work trains the flow on existing samples, thus using forward KL divergence as the loss function.}. Such a trained c-NF model is able to interpolate or extrapolate the dependency of configuration generation on the parameter, thereby being applicable to the critical regions (i.e., phase transition situation or continuum limit seeking). Although the interpolation or extrapolation of the trained c-NF model in a critical region for configuration generation may exhibit bias, it can be efficiently corrected with a few e.g., Metroplolis-Hasting steps due to the tractable probability evaluation for each generated (or ``proposed'') configurations. Since the flow model is shared across the entire parameter space except for the conditioning on them, it presents a cost-effective means to reduce the expenditure involved in employing a flow-based model for the phase diagram exploration of the theory.

%%%%%%%%%%%%%%%%%%%%%%%%%%%%%%%%%%%%%%%%%%%%%%%%%%%%%%%%%%%%%%%%%%%%%%%%%%%%%%%%%%%
\begin{table}[htbp!]
\caption{Normalizing Flows applied to different physical systems in terms of lattice study}
\label{tab:flow}
\centering
\begin{tabular}{ccccccc}
\hline
systems\quad & Ising \quad & Scalar \quad & U(1) gauge \quad & SU(N) gauge \quad & Yukawa  \quad & SU(3) gauge \\ \hline
dimensionality\quad & 2D \quad & 2D \quad & 2D \quad & 2D \quad & 2D  \quad & 2D \\ \hline
fermions\quad & no \quad & no \quad & no \quad & no \quad & Staggered  \quad & 2 flavors \\ \hline
Ref.  \quad & \cite{2018PhRvL.121z0601L} \quad & \cite{Albergo:2019eim,DelDebbio:2021qwf,Hu:2019nea} \quad & \cite{Kanwar:2020xzo,Foreman:2021ixr} \quad & \cite{Boyda:2020hsi} \quad & \cite{Albergo:2021bna} \quad & \cite{Abbott:2022zhs} \\ \hline
%$\sigma_{R_{\rm iso}}$ \quad & 2.52\% \quad & 2.05\% \quad & 2.03\% \quad & 1.87\% \quad & 2.25\% \quad & 2.64\% \\ \hline
\end{tabular}
\end{table}
%%%%%%%%%%%%%%%%%%%%%%%%%%%%%%%%%%%%%%%%%%%

It is worth noting that, such NF-build (with networks) diffeomorphic transformations $f_{\theta}$ between a naive simply distributed ``prior field'', $u$, and the physical field configurations, $\phi$, resembles the \textit{trivializing maps} approach suggested by L\"uscher in Ref.~\cite{Luscher:2009eq}, under which one actually is constructing for the system an effective action,
\begin{equation} 
S_{\rm{eff}}[u]=S[f_{\theta}(u)] - \log\det J_{f_{\theta}}(u). 
    \label{eq:flow_eff_action}
\end{equation}
that decouples the field variables by using the transformation whose Jacobian hopefully cancels out the interaction part in the original action, such that easier sampling can be realized for the (inversely) transformed new system $u=f_{\theta}^{-1}(\phi)\sim e^{-S_{\rm{eff}}(u)}$. Thus, one can in general view NF as neural network parametrizations for trivializing maps. By noting such perspective and taking the trained flow model inverse as an approximation of the trivializing map, Ref.~\cite{Foreman:2021ljl} and Ref.~\cite{Albandea:2023wgd} performed HMC on the flow transformed system $u$ for a 2-dimensional U(1) gauge theory and $\phi^4$ scalar field theory, respectively. Then to the HMC induced Markov chain $\{u_i\}^N_{i=1}\sim e^{-S_{\rm{eff}}(u)}$ which is with shorter autocorrelation times, the application of the inverse flow transformation just gives the recovered field configuration samples, $\{f_{\theta}(u_i)\}^N_{i=1}=\{\phi_i\}^N_{i=1}\sim e^{-S(\phi)}$. It was also demonstrated for the 2-dimensional $U(1)$ theory that such deep learning assisted HMC (DLHMC)~\cite{Foreman:2021ixr, Foreman:2021ljl} allows mix between modes of different topological charge sectors. Note that performing HMC on the inversely transformed system (can also be called the latent space for the original system) was proposed also by Ref.~\cite{2018PhRvL.121z0601L} on Ising system and showed enhanced efficiency, and later Ref.~\cite{Hu:2019nea} adopted similar approach of the neural RG to complex $\phi^4$ field theory with interpretation as automatic construction of exact holographic mapping of AdS-CFT.

The splitting operation in Real NVP (i.e., partitioning the field configuration into two parts before entering each flow coupling layer, as Fig.~\ref{fig:3:normflow} shown.) may constrain the expressibility of the flow-induced field transformation $f_{\theta}$, such as in terms of the symmetry embedding. Ref.~\cite{Gerdes:2022eve} introduced continuous normalizing flow to define this invertible map, $f_{\theta}: u\to\phi$, as the solution to neural ODE~\cite{2018arXiv180607366C} with a fixed time $\mathcal{T}$ ($x$ indicate the lattice cites):
\begin{equation}
\frac{d\phi(t)_x}{dt}=g_{\theta}(\phi(t),t)_x,\quad \text{with} \quad u\equiv\phi(0), \quad \phi\equiv\phi(\mathcal{T}),
\label{eq:flow_ode}
\end{equation}
where $g_{\theta}$ is a neural network represented vector field, to which symmetries can be built in more easily. The probability of the generated configuration $\phi$ follows the solution from a second ODE~\cite{2018arXiv180607366C}:
\begin{equation}
\frac{d\log p(\phi(t))}{dt}=-(\nabla_{\phi}\cdot g_{\theta})(\phi(t),t), \quad \text{with} \quad p(\phi(0))\equiv p_u(u), \quad p(\phi(\mathcal{T}))\equiv q_{\theta}(\phi),
\label{eq:flow_ode_prob}
\end{equation}
Again, taking the two-dimensional scalar field theory as the testing ground, authors in Ref.~\cite{Gerdes:2022eve} proposed a vector field for the neural ODE inspired by Fourier expansion, $g_{\theta}(\phi(t),t)_x=\sum_{y,a,f}W_{xyaf}K(t)_a\sin(\omega_f\phi(t)_y)$, with $\omega_f$ the learnable frequencies and $W$ the learnable weight tensor and $K(t)_a$ the first several terms of Fourier expansion on the interval $[0,\mathcal{T}]$. With such a vector field, the required symmetry and also the internal Z2 ($\phi\to -\phi$) symmetry of the scalar $\phi^4$ theory can be easily satisfied. Compared to Real NVP, such continuous flow method shows quite enhanced sampling efficiency: the effective sample size (ESS)\footnote{The effective sample size, ESS, can be computed as $\text{ESS}=\frac{[N^{-1}\sum^N_{i=1}p(\phi_i)/q_{\theta}(\phi_i)]^2}{N^{-1}\sum^N_{i=1}[p(\phi_i)/q_{\theta}(\phi_i)]^2}$ with $p(\phi)$ the target distribution and $q_{\theta}(\phi)$ the flow-sampler induced distribution} increases from $1\%$ to $91\%$ for $32\times 32$ lattice size.

For target distributions with multimodal structure, it is well known that the usual (local) update-based sampling methods, such as MCMC, face the challenge of traversing regions between different modes, e.g., Higgs modes/double-well potentials. ``Freezing'' may probably happen when modes are so widely separated that the sampler tends to collapse to only one or few modes~\cite{DelDebbio:2004xh,Hasenbusch:2017fsd}. It was pointed out~\cite{Hackett:2021idh} that the flow-based MCMC encounters such difficulties as well, due to the tendency of ``mode collapse'' from the original version of the flow model. Then several trials were performed in Ref.~\cite{Hackett:2021idh} to construct and train flow models in sampling multimodal distributional in lattice field theory with the example of the $\mathbb{Z}_2$-broken phase of real scalar field theory, including data preparation in mixture models and data-augmented forwards KL divergence training, or adiabatic retraining with flow-distance regularization. But these methods either need prior knowledge for the mode structure to be provided, or are difficult to really control the multimodal sampling with action parameters adjustments. Recently one interesting development~\cite{Chen:2022ytr} related is to introduce the Fourier transformation layer into the flow construction and showed promising in solving multimodal distribution sampling problem, details can refer to Sec.~\ref{sec_phy_manifest} (Note that later very recently, similar ideas named as \textit{power spectral density layer} for flow models is introduced in Ref.~\cite{Komijani:2023fzy} and applied to scalar $\phi^4$ field theory).

For lattice gauge fields, new architectures of flow-based models have recently been developed to preserve the relevant local symmetries. A gauge invariant flow model is designed for sampling configurations for a $\mathrm{U}(1)$ gauge theory in Ref.~\cite{Kanwar:2020xzo}. See Sec.~\ref{flow_symmetry} for more details. Soon this kind of flow-based scheme was developed further to system with $SU(N)$ links~\cite{Boyda:2020hsi} and also to fermionic system as shown in Ref.~\cite{Albergo:2021bna,Abbott:2022zhs}. In the lattice community, it is now under progress in translating these developments into real QCD simulations, as reported in the recent proceeding~\cite{Abbott:2022hkm}. Though promising, note however that, such flow-based methods currently are with training costs growing very fast as approaching the continuum limit, as pointed out recently~\cite{DelDebbio:2021qwf, Abbott:2022zsh, Komijani:2023fzy}. Thus, further efforts are needed to improve the approach, e.g., incorporating a transfer learning strategy.

\subsection{Observables Analysis for QFT}
In lattice QFT simulations, after the field configuration generation, usually the other big portion of the computations lie in the estimation of physical observables from the generated ensembles of field configurations or the accessible correlation functions, for example, the thermodynamics evaluation, and real-time physics extraction. Some problems especially for real-time physics reconstruction will often encounter ill-posedness from an inverse problem perspective, such as spectral functions extraction, the parton distribution functions computation, and other related transport coefficient analysis based upon Euclidean correlators from lattice calculation. ML and DL techniques have now been explored for tackling these problems in recent years.

\subsubsection{Thermodynamics and Phases}
\label{sec_phases_obs}

%Thermodynamic states of matter, especially the accompanying phase transition, is one very important and extensively observed phenomenon in various many-body physics systems. 
The study of thermodynamic states of matter, particularly the associated phase transitions, constitutes a critical and extensively researched phenomenon in various many-body physics systems.
Viewed from both theoretical (analytical or numerical) and experimental perspectives, the exploration of phase diagrams has long been a focal point in physics.
Typically, estimators for physical quantities calculated on numerically sampled configurations (e.g., from MCMC) are constructed with close relation to phase indicative parameters such as order parameters. However, identifying certain physically important states, like topological phases, using these estimators is not always straightforward.
%It is however not always easy to identify some physically important states with such estimators, like for the topological phase. 
In this context, data-driven machine learning is exceptionally well-suited for phase identification, which is largely due to its relevant strengths in discerning hidden patterns and correlations from complex datasets. Both supervised and unsupervised learning techniques have been explored for this specific task~\cite{Dawid:2022fga}.

\begin{figure}[htbp!]
  \centering
  \includegraphics[width = 0.8\textwidth]{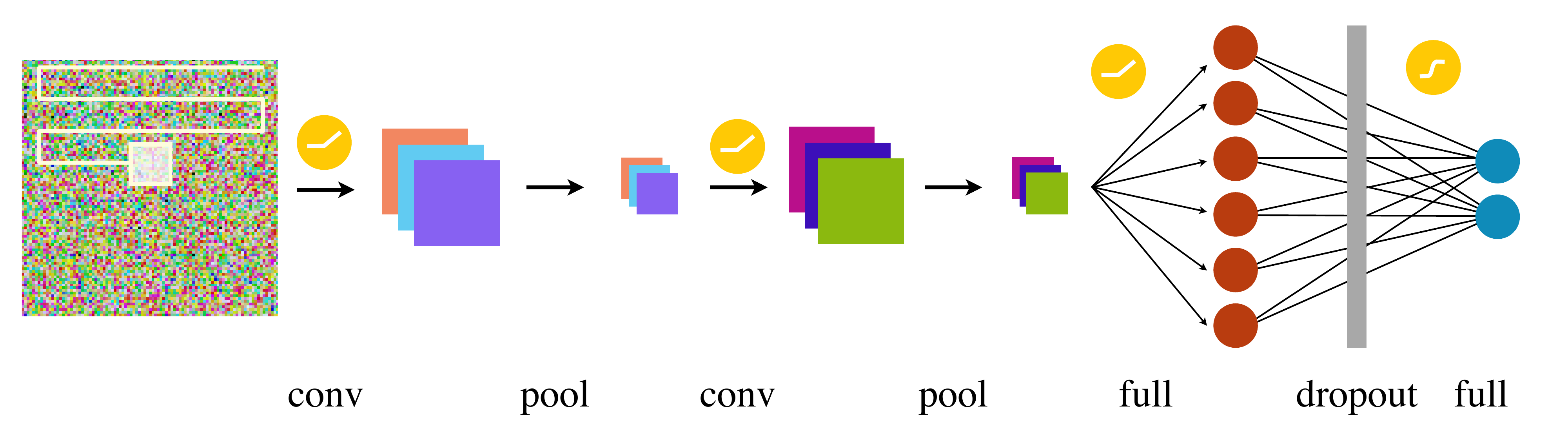}
  \caption{A typical convolutional neural network for phase binary classification. Taken from Ref.~\cite{2017NatSR...7.8823B} with permission.}
  \label{fig:cnn_phase}
\end{figure}

\emph{\textbf{Supervised Phase Classification}} ---
The initial incorporation of deep learning into thermodynamics and phase identification within general many-body systems can be traced back to Ref~\cite{2017NatPh..13..431C}. In this foundational work, it was convincingly demonstrated, using simple Ising systems, 
%The very beginning of introducing deep learning methods into thermodynamics/phase identification in general many-body systems can be dated back to Ref~\cite{2017NatPh..13..431C}, where it was firmly demonstrated on simple Ising systems 
that deep neural networks could be trained to recognize phases and phase transitions solely based on raw configurations sampled from Monte Carlo simulations\footnote{This strategy is dubbed as ``supervised'' since it relies on labeled configurations from domain or prior knowledge}. This pivotal work attests to the potential for leveraging machine learning algorithms to identify order parameters of the many-body system through supervised training. Subsequent research has expanded upon this strategy, including applications in the identification of quantum phases in fermionic system~\cite{2017NatSR...7.8823B}. See Fig.~\ref{fig:cnn_phase} for a general schematic view of such a supervised phase classification network.
%%%%%%%%%%%%%%%%%%%%%%%%%%%%%%%%%%%%%%%%%%%%%%%%%%%%%%%%
%%%%%%%%%%%%%%%%%%%%%%%%%%%%%%%%%%%%%%%%%%%%%%%%%%%%%%%%

By training a convolutional neural network (CNN) to capture the correlation between the inverse temperature and the 2D Ising spin configurations, Ref.~\cite{Tanaka:2016rtu} found that the trained CNN can isolate characteristic features associated with the phase transition of the system with scrutinization of the filters (weight parameters $W$) of the network. Accordingly, a neural network assisted order parameter is defined to give rise to critical temperature estimation in this work. Ref.~\cite{Li:2021yet} generalized the phase identification with deep learning on the Ising system to 3-dimensional case (employed 3D CNN on 3D Ising model), also discussed the average magnetization and energy regression, and the second-order and first-order phase transition learning. Being of special interest, it is expected that the QCD critical region has similar critical behaviors as in the 3D Ising model because of their shared universality.

\emph{\textbf{Unsupervised Phase Clustering}} ---
Focusing on the task of phase classification, which holds significant importance in many-body physics studies, numerous studies grounded in machine learning methods hinge on prior knowledge of the system's order parameters, along with the preparation of ensembles of microscopic configurations of the physical system. In other words, a supervised training strategy necessitating a substantial dataset of correctly labeled configurations-phase pairs is often required, as previously mentioned.% i.e., supervised training with correctly labeled ``configuration-phase'' large data set is needed as mentioned in above.
However, labels of phase classes for ensembles of field configuration are routinely a hurdle, especially in the investigation of newly studied systems. On a promising note, the domain of machine learning offers also unsupervised learning strategies. These strategies are capable of autonomously discerning crucial patterns within the amassed data, which could potentially correlate with phase information in the context of phase identification. Such an unsupervised approach to phase identification is particularly advantageous, as it does not require prior knowledge regarding the phases under scrutiny.% without knowing prior knowledge about the phase's information under consideration. 

\begin{figure}[htbp!]
  \centering
  \includegraphics[width = 0.42\textwidth]{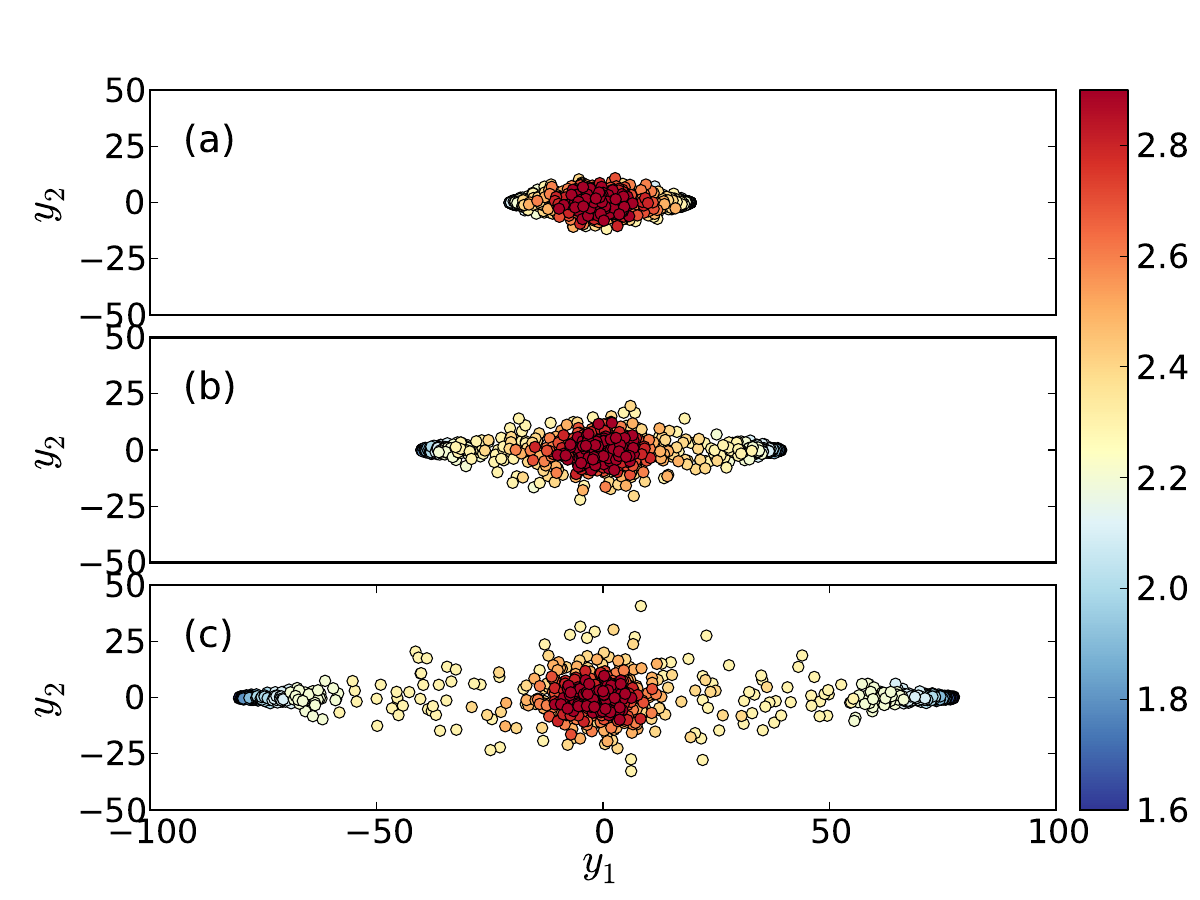}
  \hspace{1cm}
  \includegraphics[width = 0.3\textwidth]{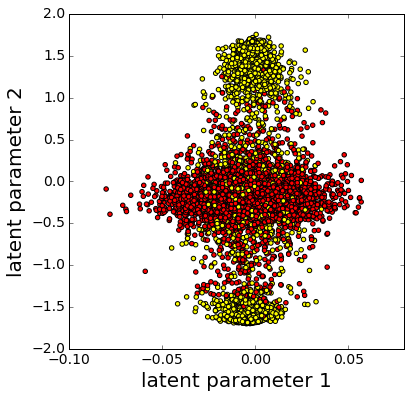}
  \caption{Taken from Ref.~\cite{2016PhRvB..94s5105W} (left) and ~\cite{2017PhRvE..96b2140W} (right).}
  \label{fig:pca_ising}
\end{figure}

The pioneering application of unsupervised learning to phase transition recognition was introduced in Ref.~\cite{2016PhRvB..94s5105W}, where neither the knowledge of the order parameter, indicating the presence of a phase transition, nor the location of the critical point was necessitated. Specifically illustrated through the 2D classical Ising model, Principal Component Analysis (PCA) was employed to extract the most significant components for the collected configurations of the system, and then the projection of the spin configurations onto the first two principal components just automatically split into clusters matching well with the corresponding physical phases (see left of Fig.~\ref{fig:pca_ising}). The proposed simple PCA phase exploration approach also exhibited success in analyzing the Ising model with a conserved order parameter. Extending this rudimentary unsupervised learning phase identification approach, Ref.~\cite{2017PhRvE..95f2122H} applied it to various physical models, including the square and triangular-lattice Ising models, the Blume--Capel model, the biquadratic-exchange spin-1 Ising (BSI) model and the 2D XY model. The study affirmed that the extracted principal components could facilitate the exploration of symmetry-breaking induced phase structure, besides aiding in identifying the phase transition type and locating the transition points. 

However, the naive PCA analysis has inherent limitations due to the involved linear transformations, rendering it unsuitable for deciphering more intricate transitions characterized by non-linear patterns. As illustrated in Ref.~\cite{2017PhRvE..95f2122H}, the vorticity structure in the BSI model and XY model could not be captured from raw spin configurations by the PCA, thus motivating the exploration of nonlinear unsupervised machine learning algorithms.
Ref.~\cite{2017NatPh..13..435V} proposed a \textit{confusion scheme}, which does not depend on labeled data and therefore can be taken as a generic tool to investigate unexplored phases of matter. Basically, the neural network is trained on deliberately by hand -“labeled” data for confusion purposes, then the performance of the trained network was found to give a universal \textit{W-shape} as a function of the guessed critical point for the parameter (e.g., chemical potential or temperature). This interesting idea was successfully demonstrated on Kitaev chain for topological phase transition, the classical Ising system for thermal phase transition, and also the quantum many-body-localization transition in a disordered Random-field Heisenberg chain. However, this framework failed when applied to 2D XY model (with original unprocessed spin configurations as input) which shows unconventional topological phase transition, as demonstrated in Ref.~\cite{2018PhRvB..97d5207B}. It is found that significant feature engineering on the spin configurations is needed for the correct phase classification, which is closely related to the underlying vortex patterns of the KT transition.
%%%%%%%%%%%%%%%%%%%%%%%%%%%%%%%%%%%%%%%%%%%%%%%%%%%%%%%%
%%%%%%%%%%%%%%%%%%%%%%%%%%%%%%%%%%%%%%%%%%%%%%%%%%%%%%%%

As another popular unsupervised learning method, deep generative models such as the variational autoencoder (VAE) and generative adversarial networks (GAN), were also applied to investigate phase transition detection of physical systems by learning latent parameters of the original configurations. In Ref.~\cite{2017PhRvE..96b2140W} the authors studied the phase transition with PCA and VAE on states of the 2D Ising model (see right of Fig.~\ref{fig:pca_ising}) and 3D XY model. It was found that the unsupervisedly learned latent representations of the spin configurations are clustered automatically and correspond to the known order parameters of the system under investigation. Also, the reconstruction loss from the training can serve as an universal identifier for phase transition. In Ref~\cite{2020NatSR..1013047W} this strategy was further extended to explore crossover region identification to reveal a deeper understanding of the latent space, which is achieved by studying the response of the learned latent variable mappings of the Ising configurations along with external non-vanishing magnetic fields and temperatures. 

\emph{\textbf{Unsupervised Anomaly Detection}} ---
With a different strategy, DNN autoencoder (AE) is also deployed by performing anomaly detection to explore phase diagrams of quantum many-body systems~\cite{2020PhRvL.125q0603K} in an automated and completely unsupervised manner. Intuitively, the anomaly detection can be realized by utilizing the reconstruction loss from the well-trained AE, as introduced in Sec.~\ref{subsubsec:dl} and also applied for outlier detection in HICs in Sec.~\ref{outlier}, to single out newly confronted data those are showing larger reconstruction error compared to training classes. Thus, by investigating the loss map of the trained autoencoder on different points in the parameters space of a physical system (with input could be full state vector, entanglement spectrum, or correlations for the system), one can possibly map out the phase diagram without physical a priori knowledge for example of the order parameter. This anomaly detection scheme was tested on the extended Bose-Hubbard Model, which shows a rich phase diagram. Aside from the several standard phases, the method also reveals a new phase showing unexpected properties on the phase diagram. 

Under a similar scheme, the generative adversarial network (GAN) is trained in Ref.~\cite{Contessi:2021mrn} as the anomaly detector to identify gapless-to-gapped phase transition in different one-dimensional models. Specifically, the detection of the elusive Berezinskii--Kosterlitz--Thouless (BKT) phase transitions in the XYZ spin chain, the Bose--Hubbard model and the generalized two-component Bose-Hubbard model (all at zero temperature) is demonstrated~\cite{Contessi:2021mrn} with entanglement spectrum (measuring the degree of quantum correlation among sub-portions of the system) as dataset.

\emph{\textbf{Interpretable Learning}} --- 
Though phases of matter are shown to be detectable through supervised or unsupervised learning strategies, it's not clear yet in physics interpretation what the learning algorithm captured from such classification tasks. Some early works explored under the supervised kernel framework of support vector machine (SVM)\footnote{Briefly speaking, in the course of binary classification given training set $(\mathbf{x}^{(k)},y^{(k)}\in\{\pm 1 \})$, the SVM aims at determining a decision boundary, $\mathbf{\omega} \cdot \mathbf{x} +b=0$, a hyperplane with parameters $\mathbf{\omega}$ and $b$ to separate data into two classes. To clearly separate the data, a margin without any training data contained is inserted, which is with the boundaries defined as $\mathbf{\omega}\cdot\mathbf{x}+b=\pm 1$, and the margin width $2/||\omega||$ is expected to be maximized in generating the best separation of data. $\mathbf{\omega}\cdot\mathbf{x}+b=d(\mathbf{x})$ provide the \textbf{decision function}.} to give interpretable decision functions and further physical discriminators from the trained machines~\cite{2017PhRvB..96t5146P,2019PhRvB..99j4410L,2019PhRvB..99f0404G}.
Ref.~\cite{Wetzel:2017ooo} proposed a \textit{correlation probing neural network} to reveal the fact that the learned decision functions originate from physical quantities. It is also shown that a full explicit expression of the learned decision function can be reconstructed, from which one can further extract the quantities to facilitate the network's decision-making in classifying phases. The proposed procedure is demonstrated first on Ising model, where it dug out the magnetization and energy per spin as the decision support of the trained network. Then on SU(2) lattice gauge theory, a whole ML pipeline combining PCA and the \textit{correlation probing neural network} is constructed to examine the deconfinement phase transition related. The PCA with the average reconstruction loss serving as a universal phase transition identifier is shown to be able to capture the phase indicator, though the involved Polyakov loop is a non-linear order parameter in SU(2) gauge theory. This enables the awareness of an existing phase transition unsupervisedly. Then the \textit{correlation probing network} is trained to correctly predict phases and further construct the explicit expression of the decision function, to which it is found that the decision is based upon the Polyakov loop as a non-local and non-linear order parameter.

In many of the ML-based explorations for physics study, the more complicated algorithms though might be with better performance yet quite often lacking transparency and interpretability, especially when one seeks for new physical insight or comprehension of the learned representations from the data. In  
Ref.\cite{Blucher:2020mjt} it was proposed to adopt ``explainable AI'' techniques--specifically the layer-wise relevance propagation (LRP) method--to identify relevant features that influence the trained algorithms towards or against the particular recognition result. The work takes the (2+1) dimensional scalar Yukawa theory as a demonstration context, which displays an interesting phase structure with two broken phases (ferromagnetic denoted as FM, and antiferromagnetic denoted as AFM) separated by a symmetric paramagnetic phase (PM). These can be indicated by the normal magnetization and the staggered magnetization. With both the field configurations and preprocessed physical observables prepared, several machine learning models were trained to infer the action parameters (the hopping parameter $\kappa$ is taken in this work) from the known observables (labeled as approach A) or solely from the raw field configurations (labeled as approach B). It should be noted that the action parameter learning in this work is taken as just a pretext task to reveal the underlying phase structure or related physical insights. This is achieved by the adopted LRP to propagate the initially assigned relevance on the output to input layer by layer.

\emph{\textbf{Observables' Regression in QFT}} ---
The application of deep neural networks for regressive tasks in lattice quantum field theoretical setting was explored in
Ref~\cite{Zhou:2018ill} to unravel the dynamical information related to phase transition and physical observables. In Ref~\cite{Zhou:2018ill} the authors considered a complex massive scalar field with quartic coupling $\lambda$ in $(1+1)$-dimensional Euclidean space-time at nonzero temperature, and a finite chemical potential $\mu$ is introduced to control the charge density fluctuation, which also makes the action complex. The worldline formalism is taken for simulating the field configurations and circumventing the sign problem involved. Correspondingly, this $1+1$-d charged $\phi^4$ field is fully represented by 4 integer-valued dual variables: $k_1$, $k_2$ and $l_1$, $l_2$. Observables like number density and field square can also be calculated with the re-expressed partition function under the dualization approach. 
The ``silver blaze'' behavior is expected for this system at low temperature and low chemical potential $\mu$, that the particle density gets suppressed (with a mass gap) until some threshold value $\mu>\mu_{th}$ then enter the condensate region and increase considerably. Though for 2-dimensional systems there is no real phase transition related to symmetry breaking, one can refer to this pronounced change in density as a transition to condensation as treated in Ref.~\cite{Zhou:2018ill}. With the configurations training set prepared, two regressive tasks were attempted: phase identification and physical observable regression. 

\begin{figure}[htbp!]
  \centering
  \includegraphics[width = 0.41\textwidth]{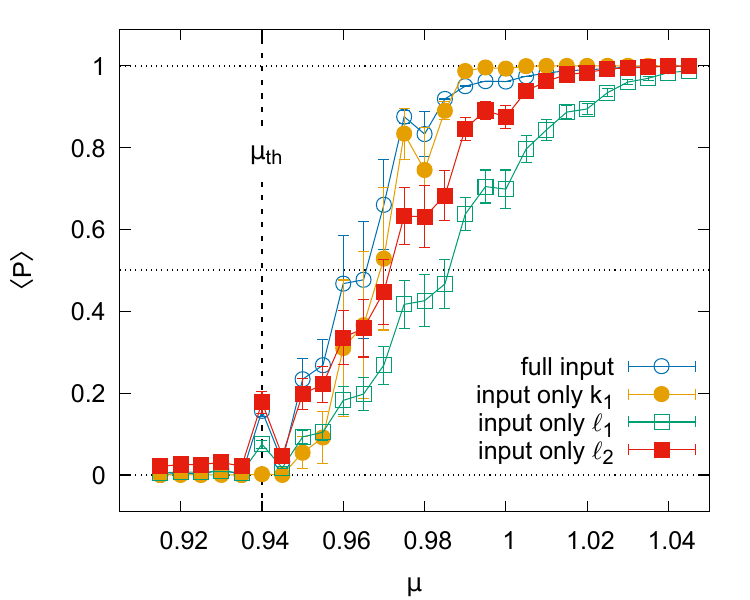}
  \includegraphics[width = 0.42\textwidth]{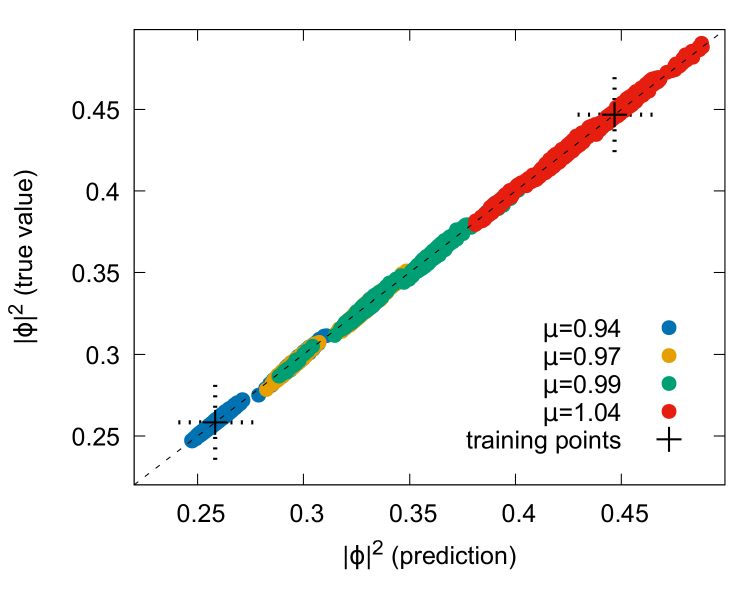}
  \caption{(left) Ensemble average of condensation probability in testing stage along with chemical potential; (right) Comparison of the network predicted squared field $\phi^2$ and the true values at different chemical potentials. Taken from Ref.~\cite{Zhou:2018ill} with permission.}
  \label{fig:scalar_regress}
\end{figure}

For the phase identification,  a convolutional neural network (CNN) was devised to perform the phase binary classification task using the field configurations as input. Purposely, the training set is prepared to consist of only two ensembles of configurations with one well above and one well below the transition point ($\mu_{th}$), while the testing set is constructed with many ensembles of configurations at different chemical potential values sit in
between the two chemical potential values of training ensembles. The network output is interpreted as \textit{condensation probability}, $P(\phi)$, for each inputted configuration. Strong correlations were observed between the network output and number density or field square, without any specific supervision on the role of these observables in distinguishing the phases. The ensemble average of the network predicted condensation probability, $\langle P(\phi) \rangle$ serves as an accurate \textit{phase classifier}, with its first non-vanishing point well indicating the transition threshold value. It is worthy to note that such concept of interpreting the network output $P$ as an observable was also extended to undergo histogram reweighting \cite{Bachtis:2020dmf}\footnote{Specifically, the reweighting in terms of inverse temperature is performed for 2-dimensional Ising system as $\langle P\rangle_{\beta}=\frac{\sum^N_i P_{\sigma_{i=1}}\exp(-(\beta-\beta_0)H(\sigma_i))}{\sum^N_{i=1}\exp(-(\beta-\beta_0)H(\sigma_i))}$, with $\beta_0$ the place of inverse temperature where MCMC measurements for $P$ is given.} to construct an effective order parameter and perform scaling analysis (see also combination with transfer learning in studying unknown phase transitions \cite{Bachtis:2020ajb}). Ref.~\cite{Zhou:2018ill} further tried to reduce the input features of this phase classification task and found that with even restricted training input e.g.,~only $l$ or even only $k_1$ variables, the network still can well distinguish the two phases. Note that the ``order parameter'' --- number density of the system, $n$, is given by the sum of $k_2$ variables for this field system (See left panel of Fig.~\ref{fig:scalar_regress}). This thus indicates that the network has captured hidden structures in $k_2$ variables, though it ($k_2$) conventionally is not able to distinguish the low-density and high-density phases.
%%%%%%%%%%%%%%%%%%%%%%%%%%%%%%%%%%%%%%%%%%%%%%%%%%%%%%%%
%%%%%%%%%%%%%%%%%%%%%%%%%%%%%%%%%%%%%%%%%%%%%%%%%%%%%%%%

The same supervised training paradigm was used for the observables (number density $n$ and field square $\phi^2$)\footnote{Specifically the number density is given by $n=\frac{1}{N_1 N_2}\sum_x k_2(x)$ and the squared field is given by $|\phi|^2=\frac{1}{N_1 N_2}\sum_x\frac{W[s(k,l;x)+2]}{W[s(k,l;x)]}$ where the weight $W[s]=\int_0^{\infty}r^{s+1}e^{-(4+m^2)^2-\lambda r^4dr}$ and $s(k,l;x)=\sum_{\nu}[|k_{\nu}(x)| + |k_{\nu}(x-\hat{\nu})| + 2(l_{\nu}(x)+l_{\nu}(x-\hat{\nu}))]$.} regression task, with only the network output layer adapted to linear activation, and the loss function for training changed from cross entropy (for classification) to MSE. The same two ensembles of configurations were taken as the training set, with which the regression network is trained and then tested on previously unseen configurations at also different chemical potentials showing different number densities and field square. The regression performance was found to be successful over a broad range of chemical potentials, as shown in the right panel of Fig.~\ref{fig:scalar_regress}. While for the number density regression, the needed pattern may seem simple, the field square is a highly non-linear function of the high-dimensional input (i.e., field configurations with $4\times N_1 \times N_2$=8000 entries). Thus, the good predictive ability of the network in regressing field squares is non-trivial and impressive. Similar findings in the context of lattice Yang-Mills theories (SU(2) and SU(3)) about the transferability of the neural network learned regression function to a different parameter space are also reported in Ref.~\cite{Boyda:2020nfh}, where the Polyakov loop as gauge invariant deconfinement order parameter is the prediction target. 
For similar tasks investigated in Ref.~\cite{Zhou:2018ill}, authors in Ref.~\cite{Bulusu:2021rqz} further explored the influences of translational equivariance satisfaction of the used network structure on the regression performance and generalization capabilities.
Note also there's earlier trial~\cite{Yoon:2018krb} with traditional machine learning technique, specifically a boosted decision tree (BDT) regression algorithm, to reduce the computational cost of evaluating lattice QCD observables, by means of predicting observables from simpler and less compute-intensive observables' evaluation those are correlated with the target observable. 

\emph{\textbf{Enhanced Regression with Symmetries Embedded Networks}} --- 
Interactions for physics systems always respect some symmetries, which possess fundamental importance to physics theories across all scales nowadays. The incorporation of the symmetries for the system into the analyzing procedures like machine learning architectures has been proven to be beneficial in improving the performance of the algorithms. One such popular simple example is the convolutional neural network (CNN), which is good at pattern recognition for image-like data structure because of the satisfied global translational equivariance (as manifested in the sharing of convolutional kernels). This concept has now been extended to yield up group equivariant CNNs (G-CNN)~\cite{2016arXiv160207576C}, where more general symmetries including rotations and reflections are discussed, with also local symmetry e.g., for data on curved manifolds~\cite{2019arXiv190204615C}. See also a recent snowmass white paper~\cite{Bogatskiy:2022hub} for a report about symmetry group equivariant architectures across physics studies.
In the context of quantum field theory, symmetries provide important constraints on the action and thus are essential for lattice field theory study, their proper consideration is also crucial in obtaining meaningful results in lattice simulations.

As the fundamental theory for strong interactions that guide high-energy nuclear physics phenomenon, QCD is a (non-abelian) gauge theory that the Lagrangian should be invariant under local symmetry transformations that form symmetry group SU(3). Being relevant, there are lattice gauge equivariant (LGE) CNNs being proposed recently~\cite{Favoni:2020reg}. Consider a SU($N_c$) Yang-Mills theory on a lattice $\Lambda$ and discretized in terms of links variables (parallel transporters) $U_{\mu}(x)=\exp[-igA^{\mu}(x+a\hat{\mu}/2)]$,  the gauge links are transformed by group elements $\Omega_x$ as 
\begin{equation}
U_\mu(x) \to \tilde{U}_\mu(x) = \Omega(x) {U}_\mu(x) \Omega^\dagger(x+\hat{\mu}),
\label{eq:u_gauge}
\end{equation}
with $\Omega:\mathbb{R}^4\to$SU($N_c$) gauge transformations of the gauge fields. Taking the Wilson action as an approximation for the Yang-Mills theory with coupling $g$,
\begin{equation}
S_W[U]=\frac{2}{g^2}\sum_{x\in\Lambda}\sum_{\mu<\nu} \mathrm{Re} \mathrm{Tr}[\mathbbm{1}-U_{\mu\nu}(x)],
\label{eq:wilson_action}
\end{equation}
with $U_{\mu\nu}(x)$ the plaquette ($1\times1$ Wilson loop),
\begin{equation}
U_{\mu\nu}(x)=U_{\mu}(x)U_{\nu}(x+\mu)U_{-\mu}(x+\mu+\nu)U_{-\nu}(x+\nu),
\label{eq:plaquette}
\end{equation}
and transform under gauge transformation locally as $U_{\mu\nu}(x)\to\Omega(x)U_{\mu\nu}(x)\Omega^{\dagger}(x)$.To construct lattice gauge equivariant network architectures, Ref.~\cite{Favoni:2020reg} devised several elementary layers to explicitly respect the gauge symmetry. One essential starting point is processing the input fields to be tuples $(\mathcal{U,W})$, where $\mathcal{U}=\{U_{\mu}(x) \}$ the set of links of the configuration and $\mathcal{W}=\{ W_i(x) \}$ with $W_i(x)\in\mathbb{C}^{N_c\times N_c}$ a set of locally transforming complex matrices like the plaquettes is used as example ($W_i(x)\to\Omega(x)W_i(x)\Omega^{\dagger}(x)$, note that Polyakov loops can also be included as stated in Ref.~\cite{Favoni:2020reg}). Then two gauge equivariant operations are introduced to act on the tuple data, $(\mathcal{U,W})$, one is performing convolutions named as LGE convolution (L-Convs) and the other is LGE bilinear layer shortly named as L-Bilin, both leaving the gauge links variables unchanged while modifying only the $\mathcal{W}$ part in a covariant manner. Specifically, the L-Convs generalizes the normal convolutional operation to account for the parallel transport under geodesics to meet the requirement of gauge equivariance,
\begin{equation}
W^{'}_i(x)=\sum_{j,\mu,k}\omega_{i,j,\mu,k}U_{k\cdot \mu}(x)W_j(x+k\cdot \mu)U^{\dagger}_{k\cdot \mu}(x),
\label{eq:l_convs}
\end{equation}
with the kernel weights $\omega_{i,j,\mu,k}\in\mathbb{C}$ and $1\le i\le N_{ch,out}, 1\le j\le N_{ch,in}, 0\le\mu\le D, -K\le k\le K$ where $K$ specifies the kernel size and $N_{ch}$ the feature map channel number. Such L-Convs operation combines data at different lattice sites with parallel transport well taken into account. The L-Bilin layer on the other hand is acting on a single lattice site, which combines two input tuples with bilinear product (note again the $\mathcal{U}$ part are the same after operation):
\begin{equation}
W^{''}_i(x)=\sum_{j,k}\alpha_{i,j,k}W_j(x)W^{'}_j(x),
\label{eq:l_bilin}
\end{equation}
%%%%%%%%%%%%%%%%%%%%%%%%%%%%%%%%%%%%%%%%%%%%%%%%%%%%%%%%
\begin{figure}[htbp!]
  \centering
  \includegraphics[width = 0.8\textwidth]{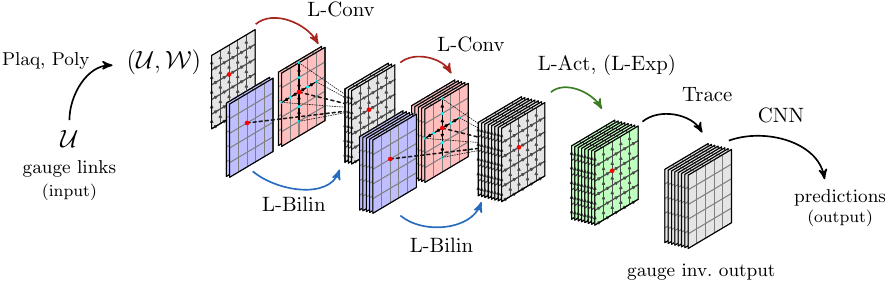}
  \caption{A generic lattice gauge equivariant CNN as from Ref.~\cite{Favoni:2020reg} with permission.}
  \label{fig:lge_cnn}
\end{figure}
%%%%%%%%%%%%%%%%%%%%%%%%%%%%%%%%%%%%%%%%%%%%%%%%%%%%%%%%
with the trainable weights $\alpha_{i,j,k}\in\mathbb{C}$ and $1\le i\le N_{out}, 1\le j\le N_{in,1}, 1\le k\le N_{in,2}$. Besides L-Convs and L-Bilin, LGE activation function and LGE Trace layer, plaquettes and Polyakov loops calculation layers (pre-processing layer to prepare the $\mathcal{W}$ information) are also proposed as elementary components to construct the gauge equivariant L-CNNs for lattice gauge field configuration treatment. See Fig.~\ref{fig:lge_cnn} for typical L-CNNs comprising the above-proposed LGE layers as from Ref.~\cite{Favoni:2020reg}. It is further demonstrated that such specially devised L-CNNs surpass traditional CNN models in the regression of gauge invariant observables. Recently, the L-CNN architecture has been extended to include additional global symmetries such
as rotations and reflections~\cite{Aronsson:2023rli}.

\subsubsection{Variational Neural-Network Quantum States}
\label{vnqs}
Many of the previous studies shown above rely on existing or pre-prepared ensembles of configurations for the physical system, up until Ref.~\cite{2017Sci...355..602C} pioneered the utilization of artificial neural networks to represent the wave function and presented a stochastic reinforcement learning scheme for solving the many-body problem without prior knowledge of exact samples. This gives a state-of-the-art accurate description of both the ground-state and time-dependent quantum states for a given Hamiltonian $\mathcal{H}$ across several prototypical spin systems, including 1 and 2-dimensional transverse-field Ising (IFI) and antiferromagnetic Heisenberg (AFT) models. The corresponding wave function, represented by a neural network, is basically a mapping from the N discrete-valued degrees of freedom set $\mathcal{S}= (\mathcal{S}_1,\mathcal{S}_2, ..., \mathcal{S}_N)$ to the complex phase and amplitude information, $\Psi(\mathcal{S})$ (taking the Restricted Boltzmann machine, RBM\footnote{Note in Ref.~\cite{2018PhRvB..97h5104C} it was demonstrated that the RBM possess holds an equivalence to tensor network states, which are widely used in quantum many-body physics}, with M hidden spin variables $h_i=\{\pm 1\}$ for example), 
\begin{equation}
\Psi_M(\mathcal{S};,\mathcal{W})=\Sigma_{h_i}e^{\Sigma_j a_j \mathcal{S}_j + \Sigma_i b_i h_i + \Sigma_{ij}W_{ij}h_i\mathcal{S}_j},
    \label{eq:nqs}
\end{equation}
is termed as neural-network quantum states (NQS) with the trainable parameters $\mathcal{W}=\{a_i, b_i, W_{ij}\}$. Such RBM representation is formally equivalent to a two-layer feed-forward neural network with special activation functions, i.e., $z^1(x)=\log\cosh(x)$,$z^2(x)=\exp(x)$.
Through the minimization of the energy expectation $E(\mathcal{W})=\langle\Psi_M|\mathcal{H}|\Psi_M\rangle/\langle \Psi_M|\Psi_M \rangle$, the network parameters $\mathcal{W}$ can be optimized via variational Monte Carlo (VMC) sampling. It has been shown that this proposed scheme can accurately evaluate the ground state energy in the TFI and AFH examples. Later, an extension of this approach to calculate excited states was also introduced, with both RBM and deeper fully connected neural networks~\cite{2018PhRvL.121p7204C}.

For the dynamical properties of the many-body state, which are elucidated upon solving the time-dependent Schr\"odinger equation, the NQS framework remains efficacious with extension of the network parameters to be complex-valued and time-dependent, $\mathcal{W}(t)$~\cite{2017Sci...355..602C}. Accordingly, per the Dirac-Frenkel time-dependent variational principle, the network parameter at each time $t$ can be trained, where the variational residuals are taken as the objective function,
\begin{equation}
R(t:\dot{\mathcal{W}}(t))=dist(\partial_t\Psi(\mathcal{W}(t)),-i\mathcal{H}\Psi),
    \label{eq:nqs_t}
\end{equation}
a feat attainable stochastically via a time-dependent VMC method. On both TFI and AFH models, this time-dependent NQS scheme has been demonstrated to capture with high precision the unitary dynamics induced by quantum quenches. There are further developments in using NQS for many-body physics studies, see Refs.~\cite{Noormandipour:2020dqp,2020PhRvL.124b0503S,Wu:2021tfb} and Refs.~\cite{Carrasquilla:2021zlj,2021arXiv210111099C}.

\subsubsection{Real-Time Dynamics Analysis}\label{subsubsec:realt}

Within modern theoretical physics, the dynamics of strongly correlated systems hold the central role for many pressing research problems, e.g., the hadronic spectrum/behaviors at zero temperature or immersed inside a thermal medium~\cite{Asakawa:2000tr,Rothkopf:2022fyo}, the non-equilibrium evolution and transport properties for the created QGP in heavy ion collisions~\cite{Rothkopf:2019ipj,Zhao:2020jqu}, the understanding for parton distribution functions of nucleons and nuclei~\cite{Ji:2020ect,Constantinou:2020pek,Candido:2023nnb}. The computation of these real-time physics is often noncompliant with perturbative analysis, thus calling for nonperturbative treatment such as lattice QFT simulations. These first principle Monte-Carlo-based simulations are usually performed in Euclidean space-time (after a Wick rotation $t\to i t\equiv\tau$) and provide only Euclidean correlators.
Accessing real-time physics from imaginary-time correlation's “measurements” in quantum Monte Carlo or lattice QFT simulation generally forms ill-conditioned inverse problems. Spectral representation forms a bridge to approach the real-time information of the dynamics from the Euclidean correlators. Also, quite often the relevant physics can be decoded directly from the spectral functions, like transport coefficients or in-medium hadronic behaviors~\cite{Asakawa:2000tr}.

\emph{\textbf{Spectral Function Reconstruction}} ---
The involved spectral reconstruction problem can, in general, be cast from a Fredholm equation of the first kind, $g(t)=\int_a^b K(t, s)\rho(s)ds$, with the aim of rebuilding the function $\rho(s)$ given the kernel function $K(t, s)$ and limited numerical evaluation on $g(t)$. Once only a finite set of evaluation data with non-vanishing uncertainties are possible for $g(t)$, the inverse transformation of the above convolution becomes ill-conditioned (see Ref.~\cite{Shi:2022yqw} and Sec.~\ref{sec:5:inverse} for more details). Basically, one can expand the convolution kernel (as a linear operator) by basis functions in a Hilbert space, within which it's shown in Refs.~\cite{J_G_McWhirter_1978} and \cite{Shi:2022yqw} respectively that the Laplace transformation kernel, $K(t,s)=e^{-st}$, and K\"allen--Lehmann (KL) kernel, $K(t,s)=s(s^2+t^2)/\pi$, possess arbitrarily small eigenvalues thus correspond to eigenfunctions being able to induce negligible changes for the integral result of $g(t)$. Consequently, for the inversion operator these eigenfunctions, termed null-modes, are related to arbitrarily large eigenvalues and will bring about numerically unstable inversion from noisy $g(t)$ to $\rho(s)$.
%%%%%%%%%%%%%%%%%%%%%%%%%%%%%%%%%%%%%%%%%%%%%%%%%%%%%%%%
\begin{figure}[htbp!]
  \centering
  \includegraphics[width = 0.7\textwidth]{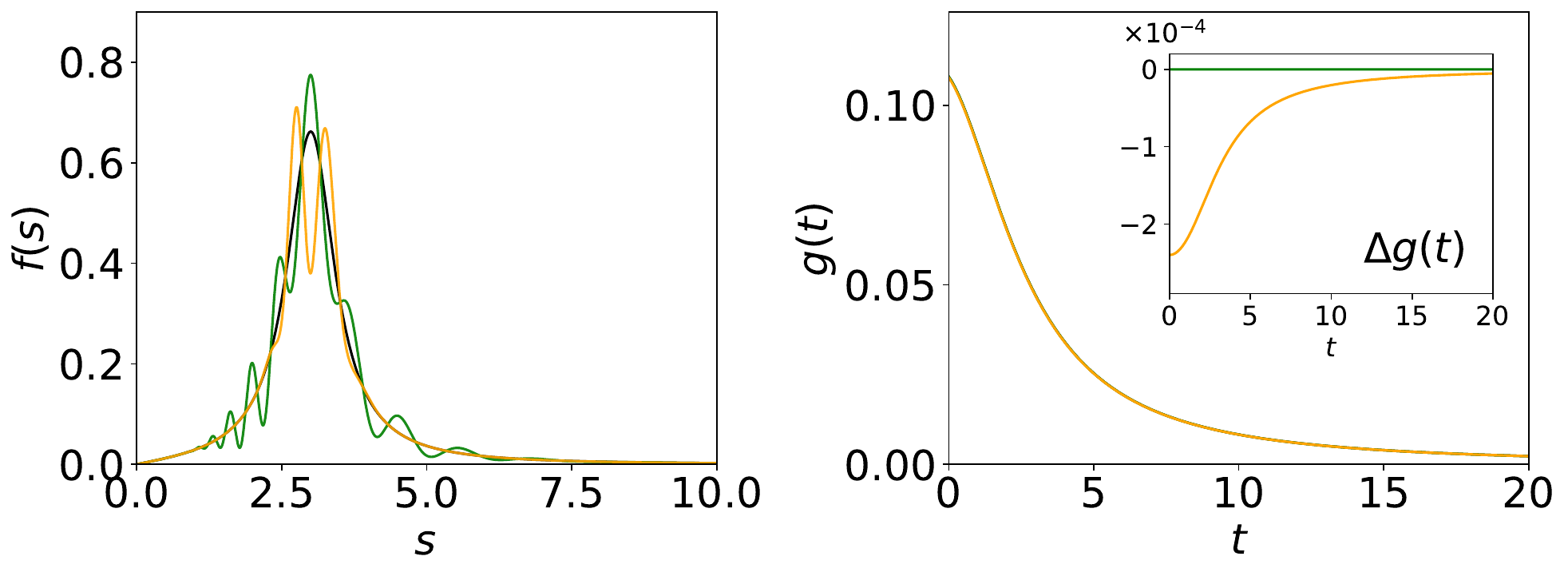}
  \caption{Spectral functions differed by null-modes (left) and their corresponding K\"allen--Lehmann correlation functions (right). The inset figure shows the differences-in-propagator caused by null modes. Taken from~\cite{Wang:2021jou} with permission.}
  \label{fig:spectral_samples}
\end{figure}
%%%%%%%%%%%%%%%%%%%%%%%%%%%%%%%%%%%%%%%%%%%%%%%%%%%%%%%%
In the context of QFTs the involved target function, $\rho(s)$, is the spectral function, and the integral $g(t)$ corresponds to the correlator which can be measured from lattice calculations. To break the degeneracy for facilitating the inversion related, different regulator terms indicating prior domain knowledge have been proposed over the past several decades, such as Tikhonov regularization(L2 regularization)~\cite{Tikhonov1943OnTS,tikhonov1995numerical}, sparse modeling approach(L1 regularization)~\cite{Otsuki:2017sma,Itou:2020azb}, maximum entropy method (MEM)~\cite{Jarrell:1996rrw,Asakawa:2000tr} and related Bayesian Reconstruction (BR) method~\cite{Burnier:2013nla}.

Recent endeavors have integrated machine learning (ML) techniques to unravel this ill-posed problem, predominantly through a supervised, data-driven paradigm, albeit with some ventures into unsupervised learning realms. 
As the early stage ML application, in Ref.~\cite{2016arXiv161204895A}, both the kernel ridge regression (KRR) and kernel quantile regression (KQR) models were harnessed to invert the Fredholm integral of the first kind. Through the preparation of the training database, coupled with the restriction on basis functions and kernels involved, a regularization is naturally provided by such projected regression treatment, thus taming the ill-conditioned inversion.

Ref.~\cite{2018PhRvB..98x5101Y} firstly introduced deep convolutional neural network (CNN) and variants of stochastic gradient descent optimizer into the spectral reconstruction from imaginary time Green's function, which is domain-knowledge-free as distinct from Ref.~\cite{2016arXiv161204895A}. Being demonstrated on a Mott-Hubbard insulator and metallic spectrum, the deep CNN gives good reconstruction performance superior to the classical MEM method. It's also found that the usage of CNN structure achieved better reconstruction than a fully connected neural network structure. A comparable strategy, delineated in Ref.~\cite{PhysRevLett.124.056401}, incorporated the principal component analysis (PCA) to reduce the dimensionality of the QMC simulated imaginary time correlation function. Within a prototypical problem of quantum harmonic oscillator linearly coupled to an ideal heat bath--which presents a more physically pertinent scenario--Ref.~\cite{PhysRevLett.124.056401} showcased that a deep neural network, with PCA processed input, outperforms the maximum entropy method (MEM) in reconstructing the spectral function from the single particle fermionic Green's function, particularly as the data noise level increases. Ref.~\cite{Kades:2019wtd} extended such a data-driven supervised approach into QFT domain, with the K\"allen--Lehmann (KL) spectral representation considered. The database is prepared in the form of a combination of Breit--Wigner peaks, $\rho^{BW}(\omega)=4A\Gamma\omega/((M^2+\Gamma^2-\omega^2)^2+4\Gamma^2\omega^2)$. In rendering the network output -- spectral function, two schemes were examined: one is with parameters of the Breit--Wigner peaks inside the spectral function and the other is with a list of discretized data points of the spectral directly. The performance of reconstruction was found to be at least on par and in numerous instances surpassing classical methods particularly in large noise cases. 

Drawing inspiration from the adeptly crafted Shannon--Jaynes entropy term in regularizing this ill-posed inverse problem, Ref.~\cite{Chen:2021giw} proposed a novel framework dubbed SVAE based on the variational autoencoder (VAE) together with an incorporated entropy term “S” within the loss function during reconstructing spectral functions from Euclidean correlators. A Gaussian mixture model was employed to construct the spectral function database, while physically motivated spectral corresponded correlators were curated for testing. Realistic noise levels of lattice QCD data were infused both in training and testing datasets. It was discerned that the trained SVAE, in most of the cases, offered reconstruction quality comparable to MEM, and in cases with sharp spectral peaks with fewer data points for the correlator, SVAE demonstrated superior results in comparison to MEM.

The aforementioned studies predominantly hinge on training data set preparation to regularize the inverse problem, and, as a corollary, may exhibit dependency on the specific kinds of training data utilized. Diverging from this, certain studies have embraced unsupervised learning strategies to execute the inversion directly. For instance, Ref.~\cite{Zhou:2021bvw} employed the radial basis function network (RBFN) to represent the spectral, essentially approximating the spectral as a linear combination of radial basis functions (RBF),
\begin{equation}
  \rho(\omega)=\sum_{j=1}^N w_{j}\phi(\omega-m_j),\label{eq:linear_summation}
\end{equation}
with $\phi$ the active RBF unit at an adjustable center $m_j$, and $w_j$ the trainable weight. Upon discretization, Eq.~\ref{eq:linear_summation} can be rewritten into a matrix format $\rho=\Phi\,W$, whereby the K\"allen--Lehmann spectral representation integral becomes
\begin{equation}\label{eq:RBFMatForm}
  G_i=\sum_{j=1}^{M} \sum_{k=1}^N K_{ij}\Phi_{jk}w_k\equiv \sum_{k=1}^{M} \tilde{K}_{ik}w_k, \ \ \ i=1,\cdots,\widehat{N}
\end{equation}
where $\tilde{K}$ manifests as an irreversible $\hat{N}\times M$ matrix. By equating $M$ and $N$, the truncated singular value decomposition (TSVD) method can be seamlessly applied to deduce $w_j$. Contrasted with supervised learning endeavors, this methodology is fast in training and also free of over-fitting issues. Compared to traditional methods, RBFN resulted in better spectral reconstruction, especially for the low-frequency part which is pivotal for the extraction of transport coefficients in the Kubo formula.

As an alternate representation, Gaussian Processes (GP) are incorporated within the Bayesian inference procedure to reconstruct the 2+1 flavor QCD ghost and gluon spectral functions~\cite{Horak:2021syv}. Generally, a GP defines a probability distribution over functions, characterized by a selected kernel function, as expressed in,
\begin{align}
\rho(\omega)\sim \mathcal{GP}(\mu(\omega),C(\omega, \omega')),
\end{align}
where $\mu(\omega)$ is the mean function, usually set as zeros, and $C(\omega,\omega')$ denotes the covariance dictated by the kernel function. Ref.~\cite{Horak:2021syv} used the radial basis function (RBF) kernel. Actually, it has been proven that a GP is equivalent to an infinitely wide neural network. In this sense, the spectral representation in Ref.~\cite{Horak:2021syv} provides a nuanced expansion upon the one in Ref.~\cite{Zhou:2021bvw} since the utilization of the RBF activation.A distinctive facet of Ref.~\cite{Horak:2021syv}is the integration of GP-represented spectral priors within the Bayesian framework to construct the likelihood of the ghost and gluon spectral. The corresponding reconstruction of the spectral function for ghosts and gluons shows a similar peak structure as to the fRG reconstruction of the Yang-Mills propagator, highlighting a convergence in understanding across different methodological approaches.

In Refs.~\cite{Wang:2021jou,Wang:2021cqw, Shi:2022yqw} a novel approach pivoted on automatic differentiation and a general deep neural network representation [$\rho(\omega)=NN(\omega)$] has been devised (refer to Fig.~\ref{fig:5:inverse:ad} for the flow chart), categorizing it within the unsupervised learning domain as well, thus circumventing the over-fitting issue and negating the necessity for training data preparation in advance. Given its general strategy towards addressing inverse problems, we summarized them in Sec.~\ref{sec:5:inverse}, refer there for an elaborated discourse on the technical intricacies and insights garnered from the corresponding results.

\emph{\textbf{In-medium Heavy Quark Potential}} ---
Another intriguing and important real-time physics in the context of high-energy nuclear physics delves into the in-medium effects on hard probes, for example, the jets or heavy quarkonium (the bound states of heavy quark and its anti-quark). Being regarded as a smoking gun indicative of Quark--Gluon Plasma (QGP) formation, heavy quarkonium has been intensively studied both theoretically~\cite{Chen:2012gg, Zhao:2010nk, Zhou:2014kka, Zhao:2020jqu, Rothkopf:2019ipj} and experimentally~\cite{CMS:2011all, CMS:2012gvv}. A cardinal endeavor lies in comprehending the in-medium heavy quark interaction, the computation of which represents a big challenge for non-perturbative strong interaction calculations. Due to the large mass and small relative velocities for the inter quarks inside the bound state, a non-relativistic treatment of them is permissible, and also the color electric interactions inside will be dominant. It has been long expected that color screening will attenuate the heavy quark interaction, analogous to the Debye screening phenomenon observed in Quantum Electrodynamics (QED). Additionally, studies from both the hard thermal loop (HTL) calculations~\cite{Laine:2006ns, Beraudo:2007ky} and the recent effective field theory approach, e.g., pNRQCD calculations~\cite{Brambilla:2008cx, Brambilla:2010vq}, underscore the emergence of a non-zero imaginary part for the heavy quark interaction beyond mere screening effect within the QCD medium. To attain a comprehensive understanding, a non-perturbative framework like lattice QCD is warranted. Over the past decade, examinations centered on the real-time heavy quark interaction based upon lattice QCD calculations have been undertaken, mainly employing a Bayesian reconstruction technique~\cite{Rothkopf:2011db, Burnier:2014ssa, Burnier:2015tda} for spectral functions of the thermal Wilson loop. On the other hand, quantification of the in-medium Bottomonium masses and thermal widths have been released through the very recent lattice QCD studies~\cite{Larsen:2019bwy, Larsen:2019zqv, Larsen:2020rjk}. Empirically, it proffers an intriguing inquiry to discern whether an in-medium heavy quark potential $V(T,r)$ under a Quantum Mechanical potential picture can accommodate these in-medium properties unveiled in lattice studies, a query yet to be addressed from a field-theoretic point of view.
%%%%%%%%%%%%%%%%%%%%%%%%%%%%%%%%%%%%%%%%%%%%%%%%%%%%%%%%%%%%%%%%%%%%%%%%%%%%
\begin{figure}[!hbtp]
    \centering
    \includegraphics[width=0.75\textwidth]{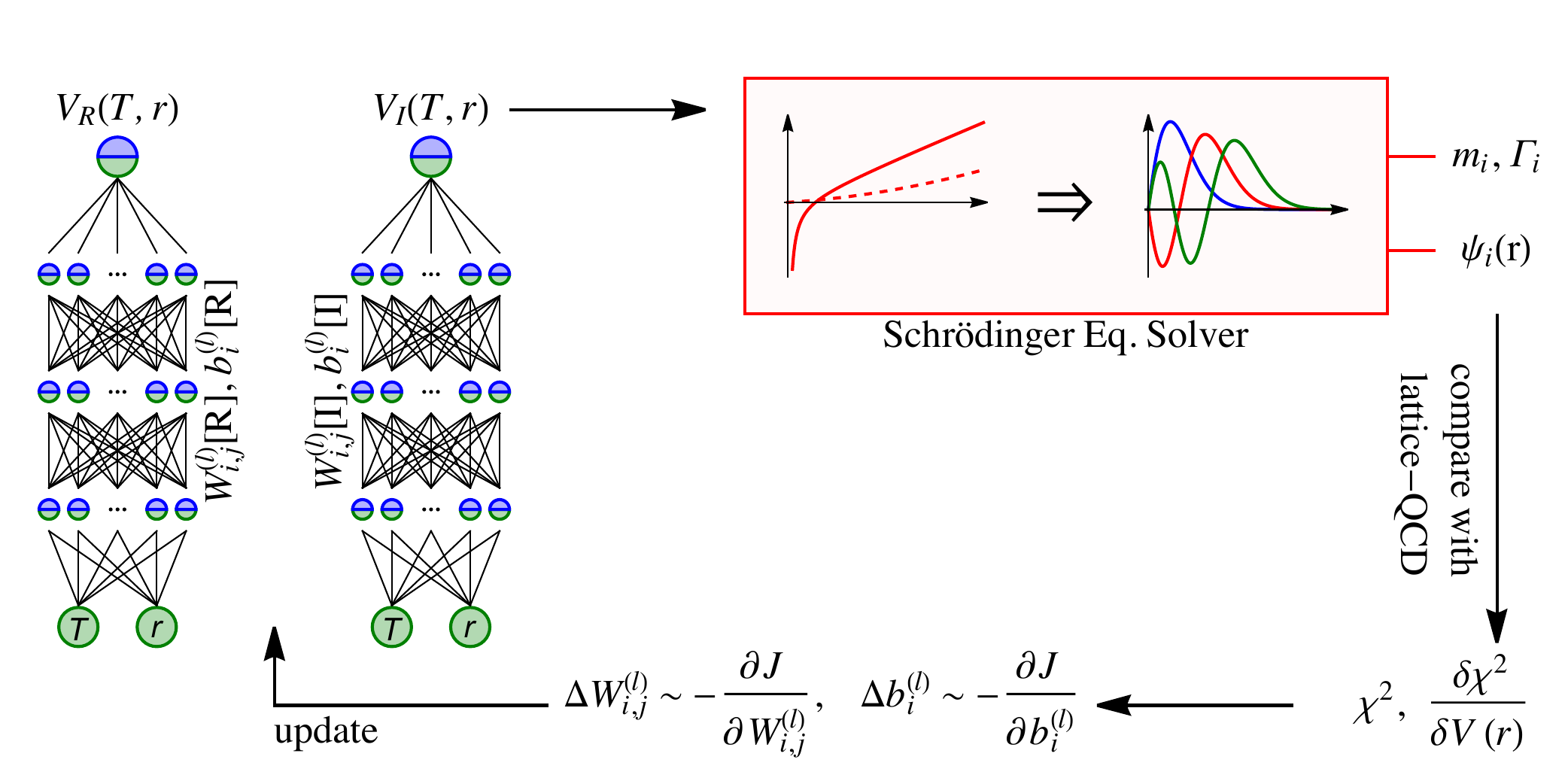}
    \caption{Flow chart of in-medium heavy quark empirical potential reconstruction from LQCD measurements of mass and thermal width. Taken from Ref.~\cite{Shi:2021qri} with permission.\label{fig:hq_flow_chart}}
\end{figure}
%%%%%%%%%%%%%%%%%%%%%%%%%%%%%%%%%%%%%%%%%%%%%%%%%%%%%%%%%%%%%%%%%%%%%%%%%%%%

Ref.~\cite{Shi:2021qri} adeptly devised a DNN-based method to infer the in-medium heavy quark interaction, starting from the lattice QCD unveiled in-medium properties for Bottomonium. The DNN is introduced to parametrize the temperature and inter-quark distance-dependent complex potential $V(T,r)$ in a model-independent fashion, and is coupled to the Schr\:odinger equation to give rise to the bound state mass and thermal width. By comparing to the lattice QCD data, the corresponding $\chi^2$ can be evaluated and serve as the loss function to optimize the DNN representation. Perturbation response analysis is performed to derive the gradient of the loss function with respect to network parameters which involves naturally the Feynmann--Hellmann theorem, with which the optimization can be done using stochastic gradient descent algorithms. Fig.~\ref{fig:hq_flow_chart} displays the flow chart of this DNN-based automatic differentiation inference for heavy quark potential. The uncertainties of the reconstruction can be properly evaluated via Bayesian inference, which takes into account both the aleatoric and epistemic uncertainty per evaluation of the posterior distribution. For a more in-depth discourse, see Sec.~\ref{sec:5:inverse}.

\emph{\textbf{Parton Distribution Function Reconstruction}} ---
The Parton Distribution Function (PDF) is a fundamental property that elucidates the internal structure of hadrons, which specifically portrays the probability distribution of the momentum fraction carried by the constituent quarks and gluons inside the hadron. It can be measured in high energy deep inelastic scattering experiments and also computed in theoretical calculations, such as lattice field theory. One may refer to~\cite{Lin:2017snn, Forte:2020yip} for an overall review of recent developments in the study of PDF. There have been some pioneer efforts to employ deep learning techniques to help extract the parton distribution function in a nucleon, where the NNPDF approach~\cite{Forte:2002fg,Ball:2009mk,Ball:2010de,Ball:2012cx,NNPDF:2014otw,Bertone:2017tyb} which conducts global QCD analyses has been systematically applied in determining the PDF in proton and also the fragmentation functions (FFs) in the light-hadron. With neural network parametrization for the PDF, NNPDF performs the global QCD fit based on high-energy collider experimental data and higher-order perturbative calculation in QCD and QED/Electroweak theory. This adept methodology has also been extended to the determination of the nuclear parton distribution function (nPDF)~\cite{AbdulKhalek:2019mzd,AbdulKhalek:2020yuc,AbdulKhalek:2022fyi} with the $\chi^2$ minimization achieved via stochastic gradient descent. 

In the practice of lattice QCD calculation, a useful intermediate quantity is the Ioffe-time distribution --- the Fourier transformation of the PDF, $Q_\mathrm{Ioffe}(\lambda;\mu) = \int_{-1}^{1} dx\, e^{ix\lambda}\, q_\mathrm{PDF}(x;\mu)$, where $\mu$ is the energy scale, $x$ denotes the momentum fraction, and $\lambda$ represents the Ioffe-time. Meanwhile, the observables that can be computed in lattice QCD calculation can be expressed as a convolution of the Ioffe-time distribution. Therefore, if $Q_\mathrm{Ioffe}(\lambda;\mu)$ is obtained from lattice QCD calculation, one can then perform the inverse Fourier transformation and compute the PDF.
Ref.~\cite{Karpie:2019eiq} reconstructed the Ioffe time distribution using two approaches, i) a Bayesian Inference reconstruction and ii) a DNN representation. In the latter, the network parameters are updated according to a generic algorithm, which takes random walks in the network parameter space and selects the configurations that reduce the difference between the desired data and the reconstructed ones.
The study in Ref.~\cite{Gao:2022iex} further accentuates this framework by employing a DNN to represent $Q_\mathrm{Ioffe}(\lambda;\mu)$ and executing the gradient-driven update method (elucidated in Sec.~\ref{sec:5:inverse}) to optimize the network parameters. The new method significantly increases both the efficiency and accuracy of the Ioffe-time distribution reconstruction.
More applications can be found in Refs.~\cite{Forte:2002fg, Forte:2002us, Zhang:2019qiq, DelDebbio:2020rgv, DelDebbio:2021whr}.

\subsection{Sign Problem}
\label{sec:4:sign}
Many efforts have been made to find ways to surmount the sign problem in lattice QCD, yet it persists as a challenging and active area of research (see recent reviews in Ref.~\cite{Berger:2019odf,Alexandru:2020wrj,Nagata:2021ugx}). The most direct method can be summarized as the ``statistical approach", which attempts to enhance the statistics directly in tackling the problem. \textit{Reweighting} the observable with phase factors~\cite{Ferrenberg:1988yz} and representing the actions with \textit{density-of-states}~\cite{Wang:2000fzi} are two practical attempts. Moreover, direct methods such as Taylor expansions to $\mu/T$ around zero chemical potential~\cite{Allton:2002zi,Borsanyi:2015axp}, and analytic continuation from imaginary chemical potential~\cite{deForcrand:2002hgr,deForcrand:2009zkb}, are other viable strategies.

The alternative approach to solving the problem is through the ``new variable" method. This entails redefining a new set of variables to reformulate the complex action. One successful example is \textit{Dualization}, which uses dual variables to represent the partition function in terms of positive quantities~\cite{Rossi:1984cv,Berger:2019odf}, thereby bypassing the complexities associated with the sign problem. To properly handle the action on the complex plane, researchers have developed complex Langevin methods and integration contour deformations. The former approach originates from stochastic quantization~\cite{Parisi:1980ys} and involves processing the complex action with two coupled Langevin dynamics~\cite{Aarts:2013uxa,Attanasio:2020spv}. The latter approach should rely on the thimble method, with the latest advancements in machine learning techniques focusing on this approach~\cite{Alexandru:2020wrj}. Both the complex Langevin and Lefschetz thimbles methods utilize complexification of the field degrees of freedom to shift the path integration into the complex plane.

The key idea of the \textit{thimble method} is to continuously deform the integration contour for the path integral from the original real fields ($\in\mathbb{R}^{n}$) into an N-dimensional real manifold $\mathcal{M}$ embedded in the complexified field space ($\in\mathbb{C}^n$), on which the dramatic phase fluctuations induced by complex actions can be mitigated or even removed~\cite{1997CPL...270..382R}. Earlier attempts designated the manifold $\mathcal{M}$ as the set of Lefshetz thimbles, resembling the high-dimensional generalization of the ``steepest descent direction" or ``stationary phase path". Consequently, the resulting integrand turns out to be real and positive up to a total phase over the thimble~\cite{Cristoforetti:2012su,Cristoforetti:2013wha}. The reason is that the imaginary part of the action, $S_I[\phi]$, morphs into a locally constant entity, and the real part, $S_R[\phi]$, is as close as possible to the ``saddle point", thus constructing the best landscape to perform stochastic evaluations of the path integrals. This conceptual framework further inspired the ``generalized thimble method" ~\cite{Alexandru:2015sua,Nishimura:2017vav}, wherein the integration contour is deformed to a manifold $\mathcal{M}_T$ chosen as the evolution results of the so-called holomorphic gradient flow equation over a fixed flow time T, starting from the original integration domain,
\begin{equation}
    \frac{d\phi}{dt} = \frac{\overline{\partial S}}{\partial \phi},
\end{equation}
where $S$ is a generic Euclidean action, and the bar indicates the complex conjugation. The time $t$ is an auxiliary variable denoting the evolution of the equation. Each flowed configuration $\phi(T)$ (collectively constituting the manifold $\mathcal{M}_T$) uniquely corresponds to an original configuration, $\phi(t=0)=\zeta\in\mathbb{R}^n$, thereby establishing a one-to-one mapping $\tilde{\phi}(\zeta)=\phi(T)$. Starting from the initial point $\phi(0)\equiv \zeta$, as the flow time increases, $\tilde{\phi}(\zeta)\rightarrow\phi(T)$, the ensemble of fields eventually approaches the right combination of thimbles, aligning with the original integral.

\subsubsection{NN-based Manifold}
To address the identification of the appropriate thimbles for alleviating the sign problem in larger systems, a significant flow time is required, which would lead to substantial computational consumption, particularly when assessing the required Jacobian. In Ref.~\cite{Alexandru:2017czx}, the authors proposed using a feed-forward neural network to approximate the thimble (or the generalized manifold) instead of directly solving the gradient flow equation and Jacobian evaluation. They referred to this network as "learnifold" since it predicts the imaginary part of the flowed manifold based on the real configuration input $\phi_R$,
\begin{equation}
    \tilde{\phi}(\phi_R) = \phi_R + i \tilde{f}_\theta(\phi_R),
\end{equation}

where the function $\tilde{f}_\theta(\cdot)$ is represented by a neural network. The \textit{learnifold} method, in contrast to the standard generalized thimble method, uses inputs from the real part manifold, resulting in a large Jacobian in practice. Additionally, the gaps between the integral contributing regions (thimbles) are smaller, making it easier to explore all relevant regions of integration objectively. This method can effectively address the issues of time-consuming flow evolution and multi-modal search in Monte Carlo sampling. The \textit{learnifold} network can incorporate translational symmetries. The authors tested the approach on sizable lattices in a 1+1 dimensional Thirring model with Wilson fermions and validated its effectiveness. 

This approach has been extended to the Hubbard model by Rodekamp et al.~\cite{Wynen:2020uzx}. Nevertheless, conventional (real-valued) neural networks continue to face computational challenges, primarily due to the extensive volume scaling of the Jacobian determinant. Recently, researchers have developed advanced neural networks that can learn the mapping from the integration manifold to the target manifold directly, utilizing complex values~\cite{Rodekamp:2022xpf, Rodekamp:2022ylw}, $\tilde{\phi}(\phi) = \tilde{f}_\theta(\phi)$. In this novel architecture, the Jacobian can be evaluated with high efficiency, due to the use of affine coupling layers. This results in a reduction of the scaling of the Jacobian determinant from a general cubic scaling to a linear scaling in volume. The method has been demonstrated in systems of different sizes.

\subsubsection{Normalizing Flow for Complex Actions}
Recalling the flow-based model introduced in Sec.~\ref{sec:3:flow_based}, one can observe similarities with contour deformation. The main idea of a normalizing flow is to construct an isomorphic deformation on smooth manifolds. While it is incapable of naturally dealing with complex actions, a generalization can be conceived by defining an integration contour to help alleviate the sign problem. In their works, Lawrence and Yamauchi~\cite{Lawrence:2021izu,Yamauchi:2021kpo} discuss the necessary conditions for the existence of a manifold that can solve the sign problem exactly, which establishes the requirement for constructing complex normalizing flows. The authors demonstrate the effectiveness of the manifold numerically over a range of couplings for the Schwinger--Keldysh sign problem associated with a real scalar field in 1+1 dimensions. Manifolds that can approximately solve the sign problem may be found in various physical systems, as indicated by Lawrence et al. in Ref.~\cite{Lawrence:2022afv}. 
In addition, in a recent paper~\cite{Pawlowski:2022rdn}, Pawlowski and Urban proposed to compute the density directly with the normalizing flow, which is the core of the \textit{density-of-states} approach for tackling sign problems. They validated the method in a two-component complex scalar field theory in which an imaginary external field explicitly breaks O(2) symmetry.\footnote{Note that in condensed matter physics, there are earlier trials using automatic differentiation to optimize a sufficiently general Hubbard--Stratonovich transformation on fermionic system to mitigate the sign problem~\cite{Wan:2020lff}, aligning with flow-based strategies in terms of field transformation optimization.}

\subsubsection{Path Optimization Method}

In Ref.~\cite{Mori:2017pne,Mori:2017nwj}, the authors first proposed the path optimization method. This new approach addresses the sign problem as an optimization problem of the integration path. They utilized a cost function to specify the path in the complex plane and adjusted it to minimize the cost function that represents the degree of weight cancellation,
\begin{equation}
    F[\phi(t)] = \frac{1}{2}\int dt |e^{i\theta(t)} - e^{i\theta_0}|^2 |J(\phi(t))e^{-S[\phi(t)]}|,
\end{equation}
where $\theta(t)$ is the complex phase of the parameterized integrand $J(\phi(t))e^{-S[\phi(t)]}$, and $\theta_0$ is the complex phase of the original integrand. The original partition function becomes $Z= \int_\mathcal{C}\mathcal{D}t J[\phi(t)]\text{exp}\{-S[\phi(t)]\}$. This method eliminates the need for solving the gradient flow found in the Lefschetz-thimble method. Instead, the construction of the integration-path contour becomes an optimization problem that can be solved using various efficient methods, e.g., gradient-based algorithms. This method has been successfully extended to e.g., 2D complex $\lambda\,\phi^4$ theory~\cite{Mori:2017nwj}, the Polyakov-loop extended Nambu--Jona--Lasinio model~\cite{Kashiwa:2019lkv,Kashiwa:2018vxr}, the 0+1 dimensional Bose gas~\cite{Bursa:2018ykf}, the 0+1 dimensional QCD~\cite{Mori:2019tux}, as well as SU(N) lattice gauge theory~\cite{Detmold:2021ulb}.

Simultaneously, in Ref.~\cite{Alexandru:2018fqp,Alexandru:2018ddf}, Alexandru et al. also investigated the path optimization method by parameterizing the manifold with neural networks $\mathcal{M}_\theta$. They first showed the results for the 1+1 dimensional Thirring model with Wilson fermions on lattice sizes up to $40\times10$. Then, they demonstrated the performance in the 2+1 dimensional Thirring model~\cite{Alexandru:2018ddf}. The recent progress of the path optimization method can be found in the review~\cite{Alexandru:2020wrj}.

Namekawa et al. further investigated the efficiency of gauge-invariant inputs~\cite{Namekawa:2021nzu} and gauge-covariant neural networks approximating the integral path~\cite{Namekawa:2022liz} for the path optimization method. The motivation is that the path optimization method with a completely gauge-fixed link-variable input can tame the sign problem in a simple gauge theory, but does not work well when the gauge degrees of freedom remain. To overcome this problem, the authors proposed employing a gauge-invariant input, such as a plaquette, or a gauge-covariant neural network, which is composed of the Stout-like smearing for representing the modified integral path. The efficiency is evaluated in the two-dimensional U(1) gauge theory with a complex coupling. The average phase factor is significantly enhanced by the path optimization with the plaquette or gauge-covariant neural network, indicating good control of the sign problem. Furthermore, another improvement is dropping the Jacobian during the learning process, which reduces the numerical cost of the Jacobian calculation from $O(N^3)$ to $O(1)$, where $N$ is the number of degrees of freedom of the theory.  Although a slight increase in the statistical error will emerge with the approximation, this practical strategy with invariant/covariant designs will push the path optimization toward solving complicated gauge theories.

\subsection{Summary}
The intricate landscape of lattice field theory presents a unique set of computational challenges, 
%This chapter first introduces the unique set of computational challenges encountered in lattice field theory, 
notably in the endeavor to undertake non-perturbative QCD calculations pivotal for understanding the properties of extreme nuclear matter. This chapter delineates these challenges and encapsulates significant explorations in three paramount sectors: %We then summarize relevant explorations from three sectors: 
field configuration generation, lattice data and physics analysis, and the sign problems, with an emphasis on the burgeoning role of the machine and deep learning techniques in propelling lattice field studies forward. %Around these three facets, we present the current state of using machine/deep-learning techniques to facilitate lattice field study. 
Many recent advanced developments are also discussed in this chapter, such as the symmetry embedding into the learning algorithms, the integration of physics priors through automatic differentiation in tackling inverse problems involved in real-time physics extraction, flow-based quantum field configuration generation, and a series of novel trials in applying deep neural networks for mitigating the sign problems in lattice simulation.
	\newpage
    \section{Dense Matter Equation of State}\label{sec:astro}
 In terrestrial laboratories, heavy-ion collisions compress nuclear matter to such high densities but inevitably involve high temperatures~\cite{Dexheimer:2020zzs, Fukushima:2020yzx}. In numerical calculations which serve as virtual laboratories, non-perturbative lattice QCD calculations can explore the finite-temperature region of the QCD phase diagram, yet leaves a long-standing challenge in finite-density part due to the inevitable sign-problem (see introductions in Sec.~\ref{sec:4:sign}). Nevertheless, studies dedicated to probe the cold dense nuclear matter properties have benefited from astronomical observations in past decades~\cite{Watts:2016uzu}. Neutron stars~(NSs) serve as cosmic laboratories for the study of neutron-rich nuclear matter, with densities far greater than the nuclear saturation density~($\rho_0\sim0.16~\text{fm}^{-3}$). NS structures~(e.g., mass and radius) are connected with their bulk properties~[e.g., equation of state~(EoS)]. Therefore, to understand QCD matter in such an ultra-high density, low temperature and large proton-neutron imbalance environment, one can infer the physical properties from NS observables, as an inverse problem (see, e.g., Refs.~\cite{Ozel:2016oaf, Baym:2017whm, Baiotti:2019sew, Kojo:2020krb, Lattimer:2021emm} for recent reviews).
 
 The EoS, a relationship of pressure ($p$) and energy density ($\varepsilon$), can be employed to deduce many-body interactions in nuclear matter or the presence of de-confined quarks at high densities~\cite{Fukushima:2013rx,Burgio:2021vgk}. For densities within the range of $n\simeq 1\text{--}2\, n_0$, one can utilize a combination of \textit{ab initio} techniques and the nuclear force derived from Chiral Effective Theory($\chi$EFT)~\cite{Drischler:2021kxf}. 
In the extremely high density region of $n\geq 50\,n_0$, perturbative QCD calculations provide a reliable understanding~\cite{Ghiglieri:2020dpq}. Accordingly, the neutron stars, with densities of up to a few times of the nuclear saturation density, covers the intermediate region($n\simeq 2\text{--}10\,n_0$). Now, rapidly cumulating neutron star observations have opened a new window for extracting the EoS. The conventional measurements like the Shapiro delay provide observations of \textit{massive pulsars}($M>2M_{\odot}$), extending the mass upper limit of the neutron star realized before~\cite{Ozel:2016oaf}. Quiescent low-mass X-ray binary systems(qLMXBs) and thermonuclear burst sources can determine the radius~\cite{Miller:2013tca,Miller:2016pom}. The latest Neutron Star Interior Composition Explorer (NICER; see~\cite{10.1117/12.2056811,gendreau2017searching}) collaboration provides more accurate X-ray spectral-timing measurements of pulsars' \textit{masses}($M$) and \textit{radii}($R$)~\cite{Yunes:2022ldq}. In addition, there are also \textit{compactness} ($M/R$), \textit{moment of inertia}($I$), \textit{quadrupole moment}($Q$) and \textit{tidal deformability}($\lambda$) can be measured. The last three observables can also be connected in $I$-Love-$Q$ formula without knowledge of the inner matter~\cite{Yagi:2013bca}. Observations of long-lived and colliding NSs are growing since the advent of gravitational waves (GWs) and multi-messenger astrophysics~\cite{Yunes:2022ldq}. The GW measurements from LIGO/VIRGO/KAGRA collaborations reveal an increasing number of events~\cite{LIGOScientific:2018mvr,LIGOScientific:2020ibl}, e.g., GW170817~\cite{LIGOScientific:2017vwq,LIGOScientific:2018cki}, GW190425~\cite{LIGOScientific:2020aai}, and GW200105/GW200115~\cite{LIGOScientific:2021qlt}. The former two have been analyzed sufficiently, which sets a tidal deformability boundary for undetermined EoSs~\cite{Yunes:2022ldq,Annala:2021gom}. One can expect that more well-analyzed events will constrain EoSs to a more narrow area with statistical approaches and machine learning techniques. Meanwhile, the astrophysics observations themselves are beneficial from the rapid development of machine learning, but it is not our main topic in this review (one can read related references in Ref.~\cite{2020WDMKD..10.1349F}).

\subsection{Reconstructions from Neutron Star Mass--Radius}\label{subsec:mr}
As mentioned before, the mass-radius($M$-$R$) curve of NSs is strongly dependent on the EoS of its internal dense matter. Its underlying physical law comes from the Einstein equation~\cite{Baym:2017whm,Schaffner-Bielich:2020psc}, which relates the geometric structure of space-time with the distribution of matter within it. In a spherically symmetric body of isotropic matter, the Tolman--Oppenheimer--Volkoff (TOV) equation~\cite{Tolman:1939jz,Oppenheimer:1939ne} gives a hydro-static description of the balance between pressure ($p$) and gravity (mass $m$) included in a neutron star with radius ($r$) from the center
\begin{align}
\begin{split}
\frac{\mathrm{d}p}{\mathrm{d}r} =\;&
   -G \frac{m(r) \varepsilon(r)}{r^2}\left(1+\frac{p(r)}{\varepsilon(r)}\right)\left(1+\frac{4 \pi r^3 p(r)}{m(r)}\right)\left(1-\frac{2 G m(r)}{r}\right)^{-1}
   \,, \\
\frac{\mathrm{d}m}{\mathrm{d}r} =\;&
    4\pi r^2 \varepsilon \,,
\end{split}
\label{eq:4:tov}
\end{align}
where $\varepsilon$ is the energy density. The observed mass and radius are set at the surface of neutron stars, as $M$ and $ R$ where the pressure $p(r=R)\simeq 0$. The similar equals mean the boundary of a neutron star will depend on how to define the vacuum pressure. To solve the TOV equations, one needs EoS in each shell of NSs. In principle, EoS depends on two independent variables, i.e., temperature and baryon chemical potential. However, compared with its Fermi temperature (>$10^{12}$K), the dense matter in the NSs is actually cold enough($10^8-10^{10}$K) to be treated as in zero-temperature environments~\cite{Yunes:2022ldq}. Then, there is only one independent thermodynamic degree of freedom, and one may represent the EoS as the relation, $\varepsilon = \varepsilon(p)$, between total energy density $\varepsilon$ and pressure $p$. 

%%%%%%%%%%%%%%%%%%%%%%%%%%%%%%%%%%%%%%%%%%%%%%%%%%%%%%%%%%%%%%%%%%%%%%%%%%%%%%%%%%%%
\begin{figure}[!htbp]
\begin{center}
\includegraphics[width=0.8\textwidth]{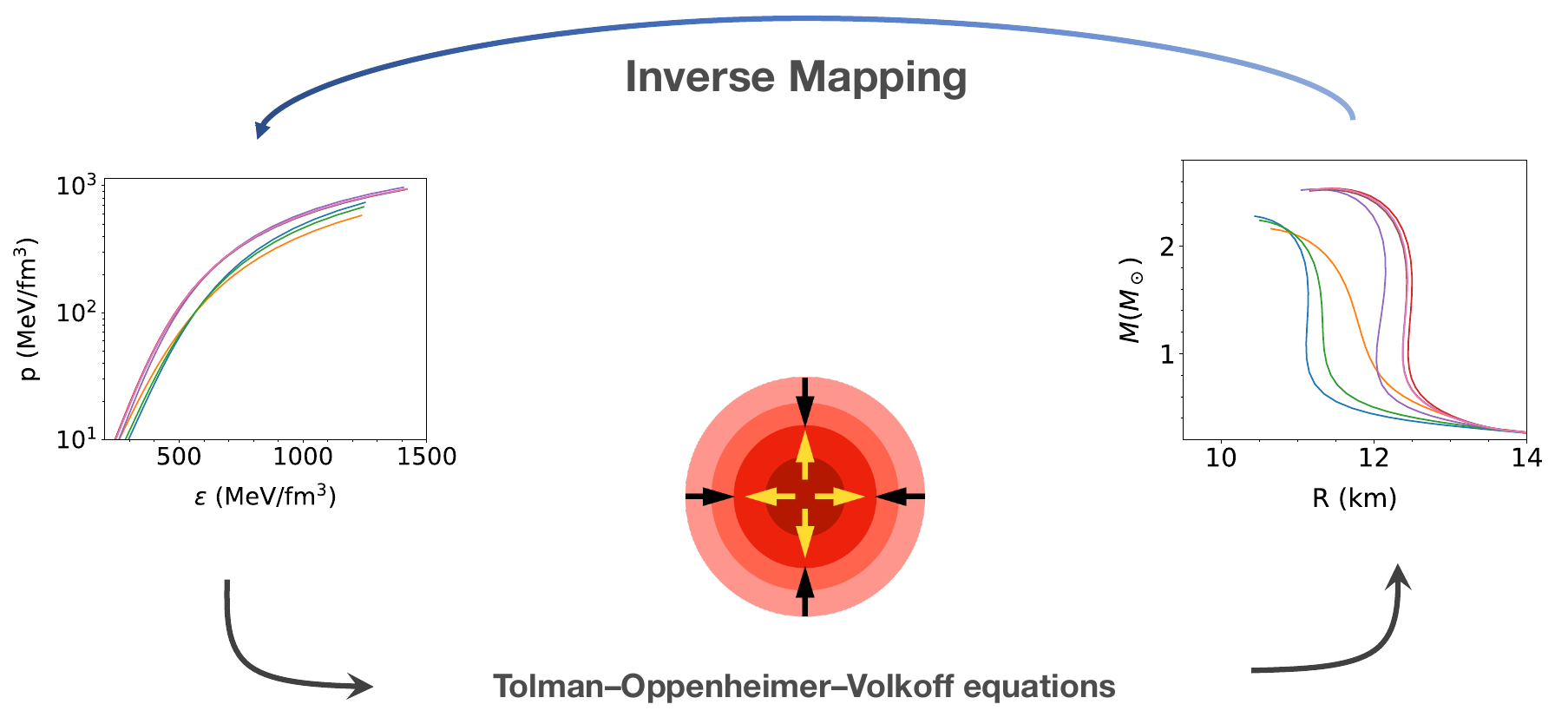}
\caption{A flow chart of TOV and its inverse mapping. The red circle represents a compact star, whereas the black arrows represent the gravity and the yellow arrows indicate the pressure inside it.}\label{fig:4:TOV}
\end{center}
\end{figure}
%%%%%%%%%%%%%%%%%%%%%%%%%%%%%%%%%%%%%%%%%%%%%%%%%%%%%%%%%%%%%%%%%%%%%%%%%%%%%%%%%%%%%

\subsubsection{Statistical Inference}\label{subsubsec:infer}
Before inferring the equation of state, it is important to specify the parameterization schemes and the necessary physical constraints. For a realistic EoS which can reflect properties of the dense matter, it should satisfy~\cite{Kojo:2020krb,Krastev:2021reh,Han:2021kjx},
\begin{enumerate}[label= \arabic*)]
	\item the microscopically stable condition, i.e.,$(dp/d\varepsilon)\geq0$,
	\item the causality condition, i.e., the speed of sound $c_s$ obeys,
        \begin{equation}
            \frac{dp}{d\varepsilon} = \frac{c_s^2}{c^2}<1,
        \end{equation}
        where $c$ is the speed of light in vacuum.
	\item the experimental constraints, e.g., the massive star observations forcing the EoS to produce a neutron star with mass at least $M\sim2M_\odot$, nuclear physics experiments suggesting the low-density part of the EoS.
\end{enumerate}
The first two constraints should serve as a ``hard'' condition for reconstructed EoSs, while the fact such as the massive mass neutron stars(NSs) would be introduced as priors of the Bayesian Inference. In fact, observations of $M>2M_\odot$ NSs, such as measurement of PSR J1614-2230~\cite{Demorest:2010bx} effectively has ruled out too ``soft'' equation of states. A ``soft'' EoS, for which the pressure increases slowly as the energy density increases, leads to smaller maximum masses. In contrast, the pressure-energy density curve of the ``stiff'' EoS has a larger slope, making for a larger maximum mass of NSs. Although the cores of the massive mass NSs could be composed of quark--gluon matter, most non-baryonic EOS models(e.g., hyperons, kaons, pure quark stars, etc.~\cite{Baym:2017whm}) have been ruled out.

Proper parametrization of EoSs means introducing as small as possible bias that is irrelevant to the physical priors~\footnote{There is a fascinating work to explore correlations among different physics model-motivated EoSs  by means of dimensionality reduction algorithms~\cite{Lobato:2022ajs}.}. The common schemes, such as the spectral representation~\cite{Lindblom:2010bb, Lindblom:2022mkr} and the piece-wise polytropic expansion~\cite{Read:2008iy, Ozel:2009da, Steiner:2010fz, Steiner:2012xt,Raithel:2016bux}, have been proved to be useful in inferring EoS. In Ref~\cite{Raithel:2017ity}, Raithel et al. manifest the feasibility of inferring pressures at five density segments from mock NS masses and radii. Although the authors demonstrate the five-polytropic model can infer possible phase transitions within $30\%$ error, it is still limited by the coarse representation. Thus, non-parametric methods, e.g, the Gaussian process(GP)~\cite{Landry:2018prl} and neural networks~\cite{Han:2021kjx, Soma:2022qnv, Soma:2022vbb} have been proposed to avoid the biased outcome due to misspecification~\cite{Han:2022sxt}. In addition to parametric forms, the self-consistent introduction of physical prior knowledge can also improve reconstructions. For instance, the meta-modeling EoS intuitively continue the Taylor expansions beyond the saturation density of symmetric nuclear matter~\cite{Margueron:2017eqc,Margueron:2017lup}. The well-measured nuclear empirical parameters can be used as priors~\cite{Xie:2019sqb}. Besides, parametrizing the speed of sound, $c_s^2$, rather than $\varepsilon$ itself, is beneficial for detecting phase transitions or crossover in dense matter~\cite{Brandes:2022nxa,Altiparmak:2022bke,Ecker:2022xxj,Jiang:2022tps}. Inspired by a similar form in spectral approach~\cite{Lindblom:2010bb}, a useful expression defines an auxiliary variable $\phi = \mathrm{log}(c^2/c^2_s - 1)$, in which the stability and causality conditions can be naturally satisfied~\cite{Landry:2018prl}.

With the parametric EoS, $\varepsilon_\theta(p)$[or speed of sound $c_{s,\theta}(p)$, auxiliary variable $\phi_\theta(p)$], one can infer parameters $\{\theta\}$ from observations following the Bayesian approach described in Sec.~\ref{subsubsec:bi}. The posterior is $P(\theta \mid \text{data})$ which describes the probability of obtaining a particular parametric EoS from a data set. Using Eq.~\eqref{eq:bayesian}, one can rewrite it as,
\begin{equation}
    P(\theta\mid\text{data}) \propto P(\text{data}|\theta)\frac{P(\theta)}{P(\text{data})},
\end{equation}
where $P(\theta) $ and $ P(\text{data})$ are priors on the parameters and observations, respectively. Given a set of EoS parameters $\{\theta\}$, one can derive the likelihood from data sets of observables $O$ as,
\begin{equation}
    P(\text{data}\mid \theta) = \prod_{i=1}^N P(O_i\mid\theta),\label{eq:4:likelihood}
\end{equation}
where $N$ counts the number of observations and the observable $O$ can be total mass-radius observations $(M,R)$ or the observations from multi-messenger measurements. To combine uncorrelated observables from different sources, one can multiply the corresponding likelihood at the right-hand side of Eq.~\eqref{eq:4:likelihood}. Eventually the likelihood can be estimated through $\chi^2$ fitting~\cite{d2003bayesian}
\begin{equation}
    \chi^2(O \mid \theta) = \sum_{i=1}^N\frac{(O_i - \tilde{O}_i(\theta))^2}{\sigma_i^2},
\end{equation}
where $\sigma_i$ is the measurement uncertainty associated with the observable $O_i$, and the prediction $\tilde{O}_i$ is from calculations based on the corresponding $\epsilon_\theta(p)$. Serving as an efficient alternative of maximizing likelihood, minimizing $\chi^2$ can provide an approximation to the posterior $P(\text{data}|\theta)$~\cite{berkson1980minimum}. The MCMC algorithm introduced in Sec.~\ref{subsubsec:bi} is widely applied for optimizing parameters $\{\theta\}$.

\begin{figure}[!htbp]
\begin{center}
\includegraphics[width=0.8\textwidth]{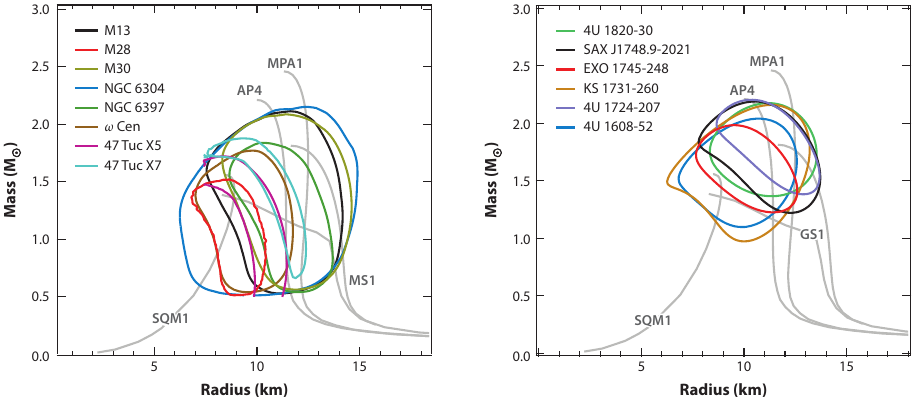}
\caption{The neutron-star mass and radius constraints (68\% Confidence Level), (left panel) low-mass X-ray binary neutron stars in a quiescent state, and (right panel) neutron stars exhibiting thermonuclear bursts. The mass relations corresponding to various EoSs are depicted using light gray lines. Figures are taken from Ref.~\cite{Ozel:2016oaf}. }\label{fig:4:mrobs}
\end{center}
\end{figure}

In the early works, although observations were limited, the Bayesian Inference of parametric EoSs anchored our basic understanding of dense nuclear matter. In Ref.~\cite{Steiner:2010fz}, Steiner et al. estimated eight parameters of EoSs in four energy density regimes from six NS masses and radii. The determined EoS and symmetry energy around the saturation density are ``soft'', leading to $R_{1.4M_{\odot}}=11-12$ km. The predicted EoS is ``stiff'' at higher densities, leading to a maximum mass of about $1.9-2.2 M_{\odot}$, consistent with the massive pulsar observation announced later~\cite{Demorest:2010bx}. In the following work~\cite{Steiner:2012xt}, the authors extended the data set to 8 NSs, and attempted to combine more constraints from both HICs and quantum MC calculations at relatively low densities. In Figure~\ref{fig:4:mrobs}, \:Ozel and Freire summarized NS observations can be used in Bayesian techniques at that time~\cite{Ozel:2016oaf}.

%%%%%%%%%%%%%%%%%%%%%%%%%%%%%%%%%%%%%%%%%%%%%%%%%%%%%%%%%%%%%%%%%%%%%%%%%%%%%%%%%%%%
%%%%%%%%%%%%%%%%%%%%%%%%%%%%%%%%%%%%%%%%%%%%%%%%%%%%%%%%%%%%%%%%%%%%%%%%%%%%%%%%%%%%%

Since 2017, GW observations have played a profound role in the study of dense matter EoSs~\cite{LIGOScientific:2017vwq, LIGOScientific:2018cki, LIGOScientific:2018mvr, LIGOScientific:2020aai, LIGOScientific:2020ibl, LIGOScientific:2021qlt, Bogdanov:2022faf}. In Bayesian inference, the measurements can be conveniently considered after marginalization with multi-source observations. They are discussed in the next section. In addition, the massive mass NSs (PSR J1614-2230~\cite{Demorest:2010bx}, PSR J0348+0432~\cite{2013Sci...340..448A}, and PSR J0740+6620~\cite{NANOGrav:2019jur}) set a baseline in inference when marginalizing over the mass measurement by considering the measurement uncertainty. The measurements of the NS moment of inertia, e.g. PSR J0737-3039~\cite{Landry:2018jyg, Miao:2021gmf}, can also serve as a constraint in inference. Since mass and radius are jointly determined by the inner matter EoSs of NSs, more accurate $M$-$R$ measurements can provide more precise constraints in inference. NASA's X-ray timing mission (NICER), which is currently in operation, has made $R$ and $M$ measurements of some of the radio millisecond pulsars that produce thermal radiation (J0030+0451~\cite{Miller:2019cac, Riley:2019yda}, J0740+6620~\cite{Miller:2021qha, Riley:2021pdl}).

Besides these observations, for improving the inference, another recent development is the application of machine learning in representing EoSs. For instance, Gaussian process is a non-parametric method~\cite{Landry:2018prl,Essick:2019ldf} can be used to represent the auxiliary variable $\phi$ at each pressure $p$ as a multivariate normal distribution, $\phi\sim\mathcal{N}(\mu(p_i),K(p_i,p_j))$, where $K$ is a kernel function for approximating the covariance. In priors composed of seven well-established EoS models, the GP process is implemented to estimate the posterior $P(\text{EoS}|\text{data})$ using the Markov Chain Monte Carlo (MCMC) algorithm. The alternative method introduced in Ref.~\cite{Han:2021kjx} is a shallow neural network representation of EoS that can also handle uncertainties from observations with MCMC while preserving flexibility. To alleviate the difficulty of sampling in a high dimensional space of parameters, the authors develop a variational auto-encoder(VAE) assisted framework for reducing the number of parameters in representing EoS~\cite{Han:2022sxt}.

\subsubsection{Deep Learning Inverse Mapping}\label{subsubsec:dlim}
As shown in Fig.~\ref{fig:4:TOV}, the mapping from the $M$-$R$ curve to the EoS relation in the TOV equations is a one-to-one correspondence~\cite{1992ApJ...398..569L}, which sets up a well-defined inverse problem when the observations are sufficient, i.e., constructing the EoS from a continuous $M$-$R$ curve. The task becomes natural to build deep neural networks to construct the complicated mapping from observed data to indeterminate physical variables~\cite{PANG:2020lda, Boehnlein:2021eym}. 

%%%%%%%%%%%%%%%%%%%%%%%%%%%%%%%%%%%%%%%%%%%%%%%%%%%%%%%%
\begin{figure}[!htbp]
\begin{center}
\includegraphics[width=0.7\textwidth]{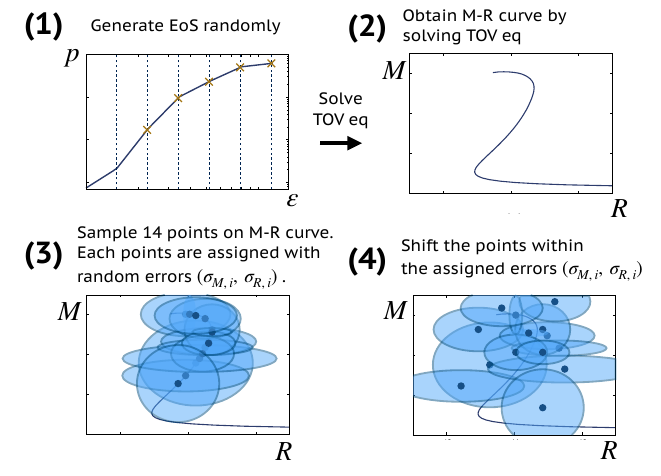}
\caption{Schematic flowchart of data generation procedure for deep learning the inverse mapping, in which the step(3) introduces the data augmentation. Figures from Ref.~\cite{Fujimoto:2021zas} with permission.}\label{fig:data_gen_dl}
\end{center}
\end{figure}
%%%%%%%%%%%%%%%%%%%%%%%%%%%%%%%%%%%%%%%%%%%%%%%%%%%%%%%%%%%

Fujimoto et al.~\cite{Fujimoto:2017cdo,Fujimoto:2019hxv} develop a supervised learning method to constrain the nuclear matter EoS, where a piecewise polytropic expansion is used to represent the EoS. The authors take the squared sound speed $c^2_s$ at the corresponding pressure as the output of the network, and the mass, radius, and their variances as inputs, ($M_i,R_i;\sigma_{M,i},\sigma_{R,i}$). Thus, one can use the NS observations to obtain the parameters of the nuclear matter EoS via the trained network. After a mock observation validation in the preliminary proof-of-concept study~\cite{Fujimoto:2017cdo}, the authors then use the mapping to construct the EoS from $M$-$R$ distributions of 14 observed neutron stars~\cite{Fujimoto:2019hxv}. In a more comprehensive work~\cite{Fujimoto:2021zas}, the authors introduce data augmentation to improve the performance and compare the deep learning approach with polynomial regression, proving the robustness of their DNN method. The training data generation procedures are shown in Figure~\ref{fig:data_gen_dl}, the new data augmentation strategy was introduced by assigning random uncertainty parameters($\sigma_{M,i},\sigma_{R,i}$) for sampled 14 points from the solved $M$-$R$ curve. To build the uncertainty for the reconstruction, they further proposed the validation method and the ensemble method, but it is still a statistical estimation to the real uncertainties.

In Ref.~\cite{Morawski:2020izm}, Morawski and Bejger further use a 4 hidden layer deep neural network to reconstruct EoSs from masses, radii, and tidal deformabilities. This model can also construct EoS and produce a $M$-$R$ curve that closely matches the observations within a range of about $1$--$7$ times the nuclear saturation density. They demonstrate the effectiveness of this method by applying it to mock data generated from a randomly selected polytropic EoS, achieving reasonable accuracy with only 11 mock $M$-$R$ pairs, similar to the current number of actual observations. They also validate the approach on mock data containing realistic EoSs. Unlike Ref.~\cite{Fujimoto:2019hxv}, the authors also investigate how the NS radii can be determined using only the GW observations of tidal deformability. In their recent work~\cite{Morawski:2022aud}, the problem of phase transition detection is transformed into an anomaly detection task. The normalizing flow, trained on samples of observations without phase transition signatures, interprets a phase transition sample as an anomaly. Although the results are inconclusive due to limited observations and large errors, it offers an alternative way to detect phase transitions that may occur in dense matter.

In Ref.~\cite{Traversi:2020dho}, Traversi and Char compare two methods for estimating the quark matter EoS: Bayesian inference and deep learning. These methods are applied to study the constant speed of sound EoS and the structure of quark stars in the two-family scenario. The observations include mass and radius estimates from various X-ray sources, and mass and tidal deformability measurements from gravitational wave events. The results from both methods are consistent, and the predicted speed of sound is in agreement with the conformal limit. 

All of the work discussed in this subsection falls into the category of supervised learning, where the goal is to learn the inverse mapping from a large ensemble of training data. Well-trained deep neural networks can successfully approximate this mapping if the generated dataset is diverse and of sufficient size. This approach provides a new perspective for constructing EoS from mass-radius pairs and can be extended to other observations. Meanwhile, it inevitably requires the preparation of training data and faces difficulties in obtaining uncertainty from finite noisy observations. In the next section, a gradient-based approach is presented to address these challenges.

\subsubsection{Gradient-Based Inference}\label{subsubsec:ad}
%%%%%%%%%%%%%%%%%%%%%%%%%%%%%%%%%%%%%%%%%%%%%%%%%%%%%%%%
\begin{figure}[!htbp]
\begin{center}
\includegraphics[width=0.8\textwidth]{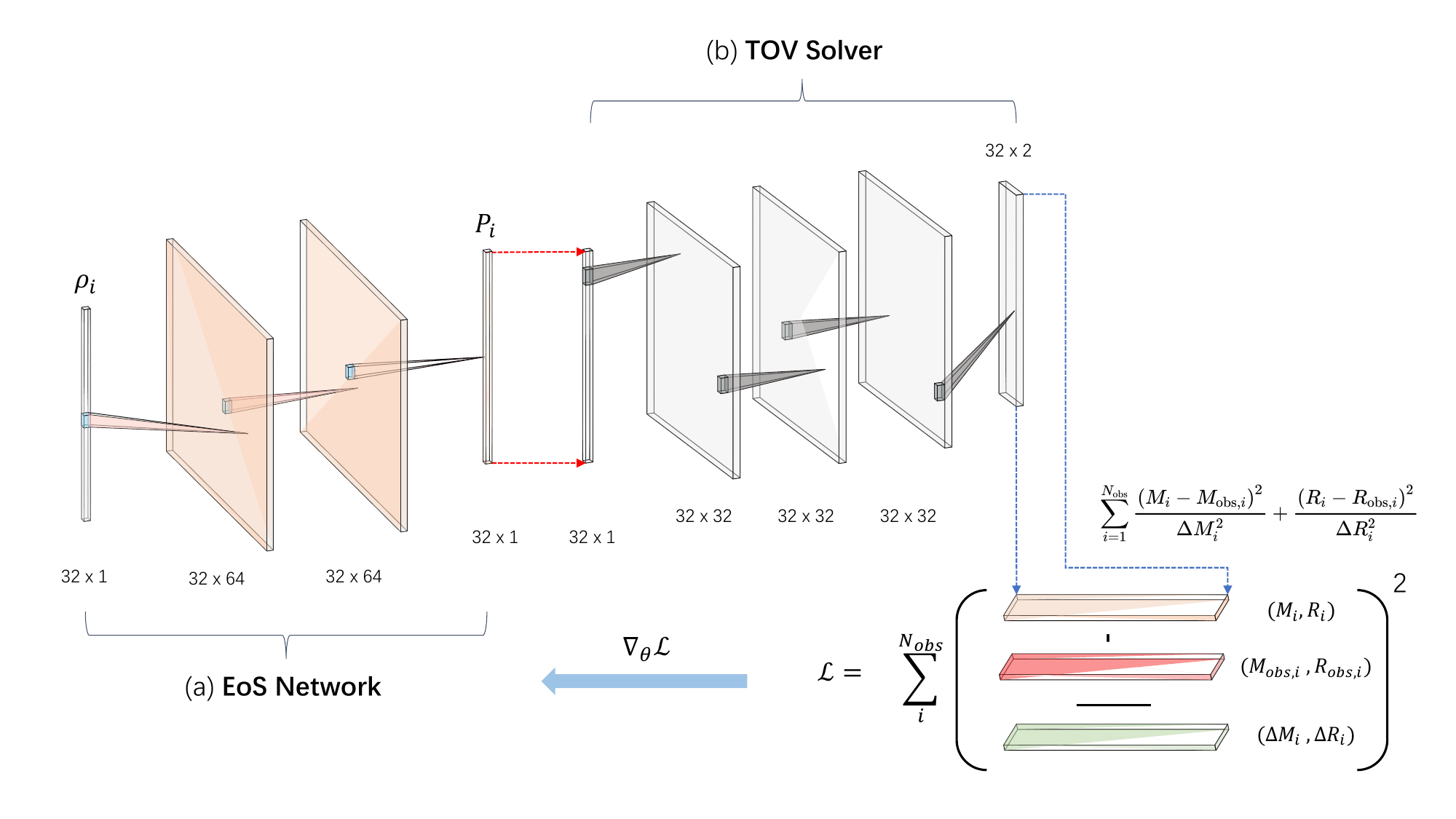}
\caption{A flow chart of AD methods, with (a) the EoS represented by neural networks named as \texttt{EoS Network}. Note that in (b) the \texttt{TOV-Solver} is a well-trained and static network.}\label{fig:framework}
\end{center}
\end{figure}
%%%%%%%%%%%%%%%%%%%%%%%%%%%%%%%%%%%%%%%%%%%%%%%%%%%%%%%%%%%
In recent work~\cite{Soma:2022qnv,Soma:2022vbb}, Shriya et al. propose an automatic differentiation (AD) framework to reconstruct the EoS from finite observations. A sketch of the framework is shown in Fig.~\ref{fig:framework}. It consists of two differentiable modules: (a) the \texttt{EoS Network}, $p_{\theta}(\mathbf{\rho})$, an unbiased and flexible DNN parameterization of the pressure as a function of the baryon number density; and (b) the \texttt{TOV Solver}, a DNN for translating any given EoS into its corresponding $M$-$R$ curve. The latter is an emulator for solving the TOV equations. With a well-trained \texttt{TOV-Solver} network, the \texttt{EoS Network} can be optimized in an unsupervised manner. Given $N_{\text{obs}}$ number of NS observations, we train the \texttt{EoS Network} to fit the pairwise ($M$, $R$) predictions from the pipeline (\texttt{EoS Network} $+$ \texttt{TOV-Solver}) to the observations.

A gradient-based algorithm within the AD framework is deployed to minimize the loss function, $\mathcal{L}\equiv\chi^2$, which is expressed as
\begin{equation}
\chi^2 = \sum_{i=1}^{N_{\text{obs}}} \frac{(M_{i} - M_{\text{obs},i})^2}{\Delta M_i^2}
+    \frac{(R_{i} - R_{\text{obs},i})^2}{\Delta R_i^2}. \label{eq:chi2}
\end{equation}
Here ($M_{i},R_i$) represents the output of the \texttt{TOV-Solver}, and ($M_{\text{obs},i},R_{\text{obs},i}$) are observations which have an uncertainty ($\Delta M_{i},\Delta R_i$). With a static well-trained \texttt{TOV-Solver} network, the gradients of the loss with respect to parameters of the \texttt{EoS Network} are
\begin{equation}
\frac{\partial\chi^2}{\partial \theta} = \sum_{i=1}^{N_{\text{obs}}}\int
\bigg[\frac{\partial\chi^2}{\partial M_i} \frac{\delta M_i}{\delta p_{\theta}(\rho)}
+\frac{\partial\chi^2}{\partial R_i} \frac{\delta R_i}{\delta p_{\theta}(\rho)}\bigg]
\frac{\partial p_{\theta}(\rho)}{\partial \theta} \mathrm{d}\rho,
\label{eq:ad}
\end{equation}
where the \texttt{TOV-Solver} is a functional mapping $f: p_{\theta}(\rho) \rightarrow {(M_i, R_i)}$. The partial/functional derivatives ${\partial p_{\theta}}/{\partial \theta}$, ${\delta M_i}/{\delta p_{\theta}(\rho)}$, and ${\delta R_i}/{\delta p_{\theta}(\rho)}$ can be computed directly with a backpropagation algorithm~\cite{baydin2018automatic} within the AD framework. By optimizing the parameters of the \texttt{EoS Network}, the best fit to the finite and noisy observational $M$-$R$ data can be obtained from the well-prepared \texttt{TOV-Solver}~\cite{Soma:2022qnv}.

\textbf{Uncertainty estimation}. To evaluate the reconstruction uncertainty, one can adopt a Bayesian perspective and focus on the posterior distribution of EoSs given the astrophysical observations, $\text{Posterior}(\boldsymbol{\theta}_{\text{EoS}}|\text{data})$. In the computations, the authors first draw an ensemble of $M$-$R$ samples from the fitted Gaussian distribution for real observations (see details in Ref.~\cite{Soma:2022vbb}). From this ensemble, one can infer the corresponding EoS deterministically with maximum likelihood estimation. Given the ensemble of reconstructed EoSs, one can then apply the importance sampling technique to estimate the uncertainty associated with the desired posterior distribution, where an appropriate weight is assigned to each EoS. In general, a physical variable $\hat{O}$ can be estimated as,
\begin{equation}
    \bar{O} = \langle \hat{O} \rangle = \sum_j^{N} w^{(j)} O^{(j)},
\end{equation}
with the standard deviation can also be estimated as $(\Delta O)^2 = \langle \hat{O}^2 \rangle - \bar{O}^2$. The weights are
\begin{equation}
   w^{(j)} = \frac{\text{Posterior}(\boldsymbol{\theta}^{(j)}_{\text{EoS}}|\text{data})}{\text{Proposal}(\boldsymbol{\theta}^{(j)}_{\text{EoS}})}  \propto \frac{P(\text{data}|\boldsymbol{\theta}^{(j)}_{\text{EoS}})\; \text{Prior}(\boldsymbol{\theta}^{(j)}_{\text{EoS}})}{P(\boldsymbol{\theta}^{(j)}_{\text{EoS}}|\text{samples}^{(j)})\; P(\text{samples}^{(j)}|\text{data})\; \text{Prior}(\text{data})},
\end{equation}
where $j$ indicates index of samples and $i$ indicates index of $M$-$R$ observables. $P(\boldsymbol{\theta}^{(j)}_{\text{EoS}}|\text{samples}^{(j)})=1$, because it is deterministic. $P(\text{data}|\boldsymbol{\theta}^{(j)}_{\text{EoS}})\propto \exp{(-\chi^2(M_{\boldsymbol{\theta}^{(j)}_{\text{EoS}}},R_{\boldsymbol{\theta}^{(j)}_{\text{EoS}}})/2})$ and $P(\text{samples}|\text{data})=\mathcal{N}(M_{\text{obs}},\Delta{M}^2)\mathcal{N}(R_{\text{obs}},\Delta{R}^2)$ calculated from the corresponding Gaussian distribution. In practical, weights should be normalized so that $1 = \sum_j w^{(j)}$ and cut off to avoid outliers in samples, which will also remove the prior terms.
%%%%%%%%%%%%%%%%%%%%%%%%%%%%%%%%%%%%%%%%%%%%%%%%%%%%%%%%%%%%%%%%%%%%%%%%%%%%%%%%%%%%
\begin{figure}[htbp!]
    \centering
    \begin{minipage}[t]{0.48\textwidth}
        \centering
        \includegraphics[width=8.cm]{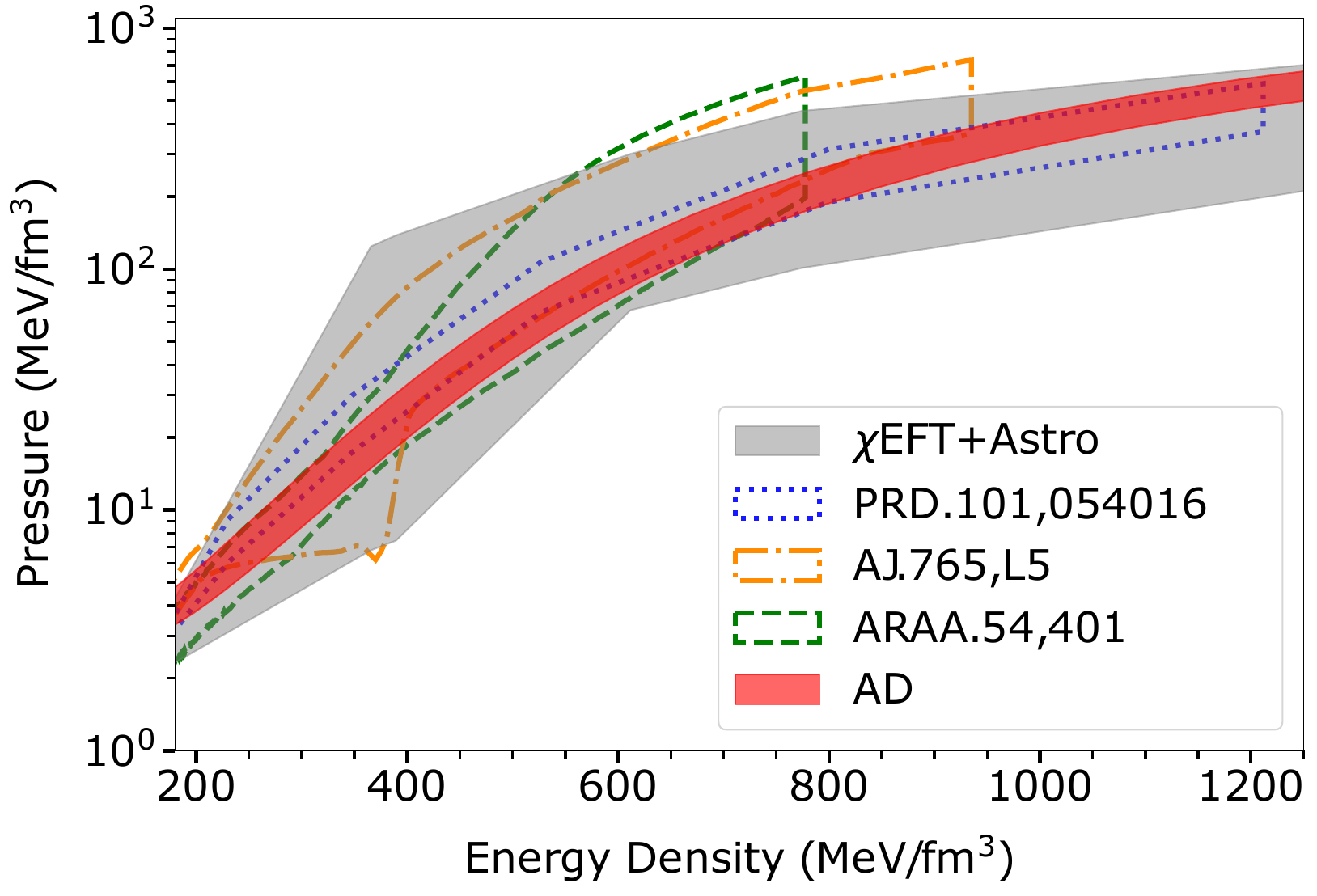}
        \caption{The EoS reconstructed from observational data. The gray band locates the $\chi$EFT prediction. The red shaded area represents the automatic differentiation results (AD~\cite{Soma:2022vbb}). Other results are derived from Bayesian methods (AJ.765,L5~\cite{Steiner:2012xt} and ARAA.54,401~\cite{Ozel:2016oaf}) and the direct inverse mapping (PRD.101,054016~\cite{Fujimoto:2019hxv}) are also included.}
        \label{fig:rec-eos}
    \end{minipage}
    \hspace{0.5cm}
    \begin{minipage}[t]{0.48\textwidth}
        \centering
        \includegraphics[width=8.1cm]{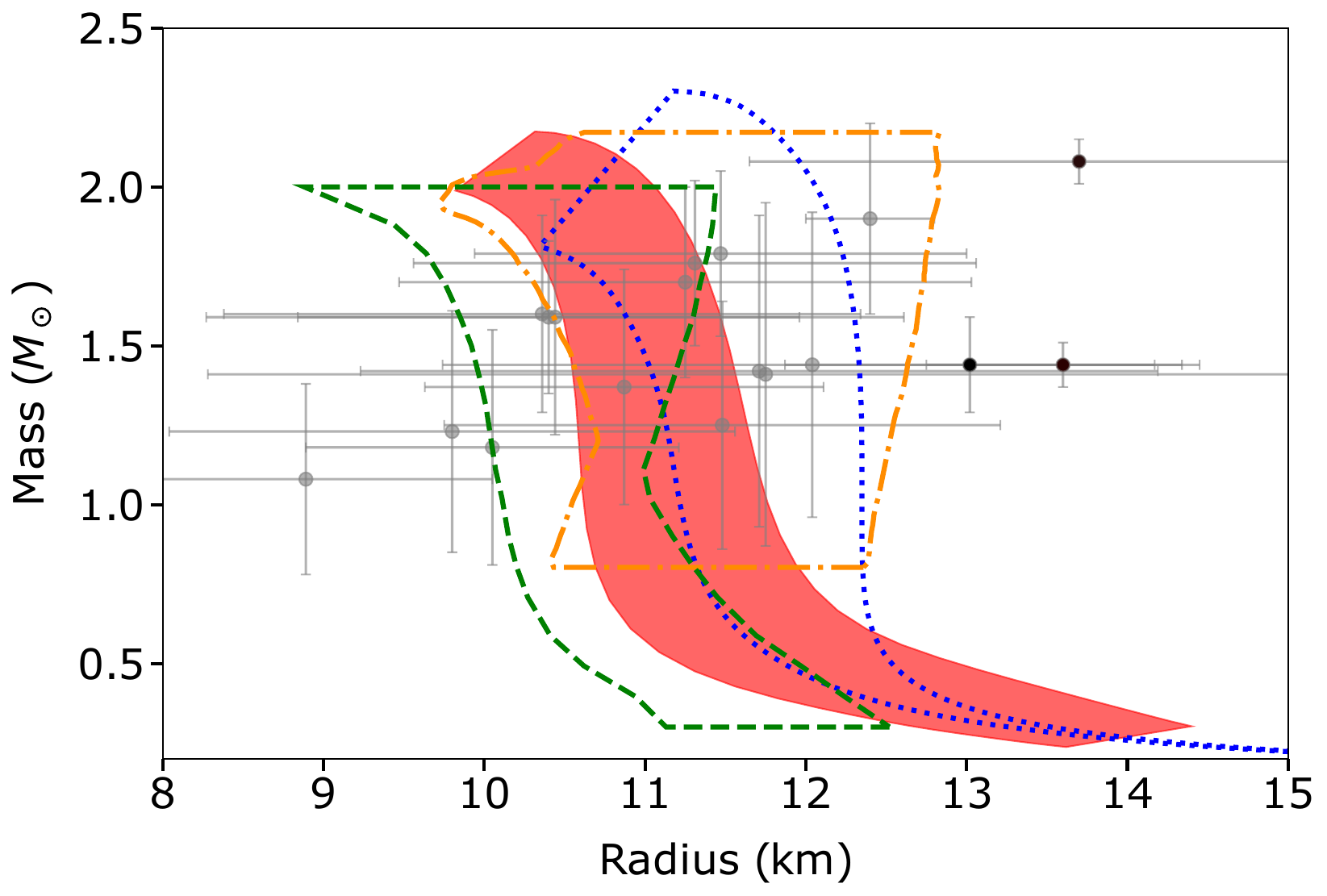}
        \caption{$M$-$R$ contour corresponding to the reconstructed EoSs in the left panel. The dots with uncertainties are fitting observations summarized in Ref.~\cite{Soma:2022vbb}, in which black dots are from NICER observations.}
        \label{fig:rec-mr}
    \end{minipage}
\end{figure}
%%%%%%%%%%%%%%%%%%%%%%%%%%%%%%%%%%%%%%%%%%%%%%%%%%%%%%%%%%%%%%%%%%%%%%%%%%%%%%%%%%%%

\textbf{Reconstructions}. In Fig.~\ref{fig:rec-eos} reconstructed EoSs from different works are plotted for comparison. The $\chi$EFT calculation combined with the polytropic extrapolation and the two-solar-mass pulsar constraint is labeled ``$\chi$EFT+Astro'' and shown as a gray band. It is expected that all reasonable predictions should fall within this gray band, since this first-principles calculation sets a theoretical baseline. In addition, the Bayesian analyses (labeled as``ARAA,54,401'' by \:Ozel et al.~\cite{Ozel:2016oaf} and ``AJ.765,L5'' by Steiner et al.~\cite{Steiner:2012xt}) and the supervised learning result (labeled as``PRD.101,054016'' by Fujimoto et al.~\cite{Fujimoto:2019hxv}) are also shown in the figure. It is worth noting that \cite{Ozel:2016oaf} and \cite{Fujimoto:2019hxv} use the same data set, including eight neutron stars in quiescent low-mass X-ray binaries (qLMXBs) and six thermonuclear bursters, while \cite{Steiner:2012xt} uses a subset of the data, i.e. eight of the X-ray sources. The AD results (labeled as ``AD'' by Shriya et al.~\cite{Soma:2022vbb}) are shown as a red band, which includes four additional observations: 4U 1702-429~\cite{Nattila:2017wtj}, PSR J0437-4715~\cite{Bogdanov:2019ixe}, J0030+0451~\cite{Miller:2019cac,Riley:2019yda}, and J0740+6620~\cite{Miller:2021qha,Riley:2021pdl}, the three pulsars from the latest NICER analysis. Figure~\ref{fig:rec-mr} shows the corresponding $M$-$R$ curves derived from the EoSs in Fig.~\ref{fig:rec-eos}. The latest reconstructions are certainly consistent with massive neutron stars, i.e. $M\geq2M_\odot$.

\subsection{Constraints from Multimessenger Observations}\label{subsec:gw}
Multimessenger measurements have provided new insights into ultrahigh-density matter, including gravitational waves (GWs) and electromagnetic (EM) signals. This new era began with the discovery of a binary neutron star merger by the Advanced LIGO and Virgo gravitational wave detectors in 2017~\cite{LIGOScientific:2017vwq}.  Because the properties of neutron stars are sensitive to the microscopic interactions that govern the EoS, gravitational waves from the mergers of neutron stars can carry information about the inner cold dense matter. The EoS of dense matter can further be strongly constrained by gravitational wave observations of neutron stars.

\textbf{Tidal deformability.} The finite size of the stars gives rise to tidal interactions that affect the evolution of the binary system during the late stages of the inspiral of coalescing NSs~\cite{Schaffner-Bielich:2020psc}. The tidal field produced by each binary component induces a quadrupole moment on the companion star, resulting in enhanced emission of gravitational radiation and a slight increase in their relative acceleration, accelerating the inspiral phase. The magnitude of the induced quadrupole moment is related to the internal structure of the NS. Larger stars are less compact and thus more easily deformable under the influence of an external field of a given amplitude. The tidal deformability $\lambda$ is a parameter for the quantification of the effects~\cite{Flanagan:2007ix,Hinderer:2007mb}. The dimensionless tidal deformability $\Lambda$ is defined as the ratio of the induced quadrupole moment to the external perturbing tidal field, $\Lambda = \lambda/M^5 = {2k_2}/{(3\beta^5)}$, where $k_2$ is the tidal Love number which is determined as,
\begin{align}
     k_2\left(\beta, y_R\right) &=\frac{8}{5} \beta^5(1-2 \beta)^2\left[2-y_R+2 \beta\left(y_R-1\right)\right] \times\left\{2 \beta\left[6-3 y_R+3 \beta\left(5 y_R-8\right)\right]\right. \nonumber \\
     &+4 \beta^3\left[13-11 y_R+\beta\left(3 y_R-2\right)+2 \beta^2\left(1+y_R\right)\right.] + 3(1-2 \beta)^2\left[2-y_R+2 \beta\left(y_R-1\right)\right] \ln (1-2 \beta)\}^{-1},
\end{align}
where $\beta\equiv M/R$ is dimensionless compactness parameter. $y_R \equiv y(R)$ is the boundary value of $y(r)$, which follows the differential equation
\begin{equation}
    \frac{d y(r)}{d r}=-\frac{y(r)^2}{r}-\frac{y(r)}{r} F(r)-r Q(r),
\end{equation}
with
\begin{align}
    F(r) &=\left\{1-4 \pi r^2[\varepsilon(r)-p(r)]\right\}\left[1-\frac{2 m(r)}{r}\right]^{-1}, \\ 
    Q(r) &=4 \pi\left[5 \varepsilon(r)+9 p(r)+\frac{\varepsilon(r)+p(r)}{c_s^2(r)}-\frac{6}{r^2}\right]\left[1-\frac{2 m(r)}{r}\right]^{-1} - \frac{4 m^2(r)}{r^4}\left[1+\frac{4 \pi r^3 p(r)}{m(r)}\right]^2\left[1-\frac{2 m(r)}{r}\right]^{-2}.
\end{align}
The total tidal effect of two compact stars in an inspiraling binary system is given by the mass-weighted (dimensionless) tidal deformability,
\begin{equation}
    \tilde{\Lambda}=\frac{16}{13}\frac{(M_1 + 12M_2)M_1^4\Lambda_1+(M_2+12M_1)M_2^4\Lambda_2}{(M_1+M_2)^5},
\end{equation}
where the subscripts for $\Lambda$ and $M$ indicate different stars. For single neutron stars, one can calculate their $\Lambda$, predicting the tidal phase contribution for a given binary system from each EOS, for the universality of the nuclear matter EoS. The weighted (dimensionless) deformability $\tilde{\Lambda}$ is usually plotted as a function of the chirp mass $\mathcal{M}=(M_1M_2)^{3/5}/(M_1+M_2)^{1/5}$ for asymmetric mass binary systems. See a more systematic introduction in Ref.~\cite{Schaffner-Bielich:2020psc}.

%%%%%%%%%%%%%%%%%%%%%%%%%%%%%%%%%%%%%%%%%%%%%%%%%%%%%%%%
\begin{figure}[!htbp]
\begin{center}
\includegraphics[width=0.9\textwidth]{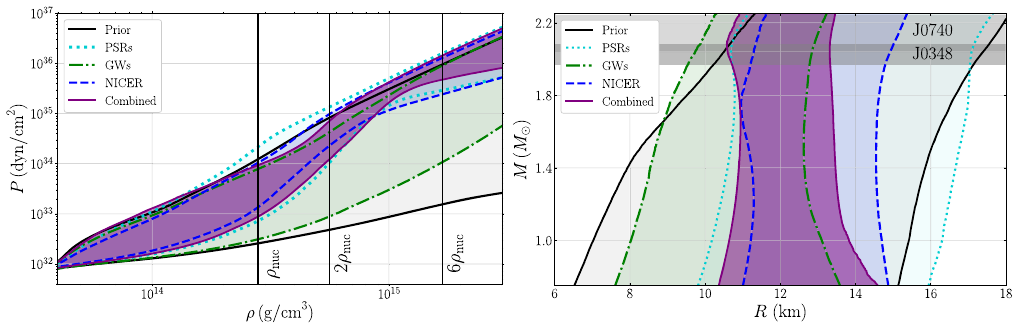}
\caption{In Ref.~\cite{Landry:2020vaw}, the authors presented individual-event constraints from various observation classes. The 90\% credible intervals for the EoS are displayed in the pressure-density plane (left panel) and the mass-radius plane (right panel). The prior range is indicated by black lines, and the posterior using only heavy pulsar measurements is represented by turquoise lines. The posterior based on only GW data is depicted by green shaded regions, and the posterior using only NICER data is depicted by blue shaded regions. The left panel includes vertical lines marking multiples of the nuclear saturation density, while the right panel includes horizontal shaded regions showing the 68\% credible mass estimate for the two heaviest known pulsars. Figures are taken from Ref.~\cite{Landry:2020vaw}.}\label{fig:5:gp}
\end{center}
\end{figure}
%%%%%%%%%%%%%%%%%%%%%%%%%%%%%%%%%%%%%%%%%%%%%%%%%%%%%%%%%%%
In terms of Bayesian Inference (BI) methods, there are many works using or combining the current GW observations to reconstruct the dense matter EoS~\cite{Brandes:2022nxa,Ecker:2022xxj,Miao:2021nuq,Golomb:2021tll,Chimanski:2022wzi,Chimanski:2022wzi}. As a modern variant of BI methods, the Gaussian Process (GP) has been used to represent the EoS by inferring it from GW observations in the work of Landry et al.~\cite{Landry:2018prl}. The authors first build a mapping from binary observables $(M_1,M_2,\Lambda_1,\Lambda_2)$ to the GP-represented EoSs and then use MCMC to infer their corresponding posteriors directly. These synthetic EOSs maintain high flexibility, i.e., they cover a sufficiently large interval of stiffnesses and core pressures while satisfying other properties introduced in Sec.~\ref{subsubsec:infer}. The authors validate the inference approach on simulated GW170817-like signals using real detector noise, with reasonable efficiency in recovering a known EOS. Eventually, the authors demonstrate the inferred EOSs and corresponding observables under posterior constraints from the single GW170817 event. In two prior set-ups, the deformability of $\tilde{\Lambda}=210^{+383}_{-113}(631^{+164}_{-122})$ and $\Lambda_{1.4}=160^{+448}_{-113}(556^{+163}_{-172})$, maximum mass of $M_{\text{max}}=2.09^{+0.37}_{-0.16}(2.04^{+0.22}_{-0.002}) M_{\odot}$ are consistent with previous analyses~\cite{LIGOScientific:2018cki,De:2018uhw}. In the following works, Essick et al.~\cite{Essick:2019ldf} extend the analysis, obtaining a more reliable conclusion that is the GW170817 favors a ``softer'' EoS. Through combining more information from other observations, i.e., maximum masses of NSs, more GW events shown in Table~\ref{tab:4:GW} and $M$-$R$ in the newest X-ray observation (J0030+0451). Landry et al.~\cite{Landry:2020vaw} further improve the inference by combining different observations. As a demonstration, individual constraints from different observations can be found in Fig.~\ref{fig:5:gp}. Through combining them, the authors find the radius of the $1.4 M_{\odot}$ NS is $R_{1.4}=12.32^{+1.09}_{-1.47}$~km.

%%%%%%%%%%%%%%%%%%%%%%%%%%%%%%%%%%%%%%%%%%%%%%%%%%%%%%%%%%%%%%%%%%%%%
\begin{table}[!hbpt]
\centering
\begin{tabular}{llll}
\hline\hline
    Events & $\mathcal{M}[M_\odot]$ & $\tilde{\Lambda}$ & $\Lambda_{1.4}$ \\
\hline
    GW170817 & 
    $1.186^{+0.001}_{-0.001}$~\cite{LIGOScientific:2017vwq}&
    $300^{+500}_{-190}$~\cite{LIGOScientific:2018hze}&
    $190^{+390}_{-120}$~\cite{LIGOScientific:2018cki}\\
    GW190425~\cite{LIGOScientific:2020aai} & 
    $1.44^{+0.02}_{-0.02}$&
    $\lesssim 600$&
    ---\\
\hline\hline
\end{tabular}
\caption{Summary of the GW events and their astrophysical observations. It 
 contains the median and uncertainties (90\% credible level) of  the chirp mass M, and the tidal parameters $\tilde{\Lambda}$ and its corresponding value $\Lambda_{1.4}$ at 1.4 solar mass.}
\label{tab:4:GW}
\end{table}

%%%%%%%%%%%%%%%%%%%%%%%%%%%%%%%%%%%%%%%%%%%%%%%%%%%%%%%%%%%%%%%%%%%%%

As introduced in Sec.~\ref{subsubsec:dlim}, Morawsk et al. implement deep learning to rebuild EoSs from GW observations~\cite{Morawski:2020izm, Morawski:2022aud}, in which the learnable inverse mapping consists of mass-radius and tidal deformability as inputs. In Ref~\cite{Han:2021kjx, Han:2022sxt}, the mass-tidal-deformability data of GW170817 event has also contributed to the inference of neural network represented EoSs. In the work of Ferreira et al.~\cite{Ferreira:2019bny}, the authors prepare training data from a set of metamodeling EoSs. The inverse mapping is learned to predict the parameters of the nuclear matter from the tidal deformability and the radius of the NS. In addition to DNNs, they adopt a classical machine learning method of support vector machine regression (SVM-R). Although the DNNs show a high level of accuracy than the SVM-R in their tests, it provides an alternative way to explore the possible inverse mapping. For classical machine learning algorithms, in Refs.~\cite{PhysRevD.100.103009, HernandezVivanco:2020cyp}, Hernandez et al. develop a random forest regressor algorithm to interpolate the marginalized likelihood for each gravitational-wave observation. It can be utilized in the hierarchical Bayesian Inference for constraining the EoS from GW170817 and GW190425 events directly. They provide a new constraint for the 1.4~$M_\odot$ neutron star with radius $R=11.6_{-0.9}^{+1.6}$~km. In addition, the AD framework introduced in Sec.~\ref{subsubsec:ad} can conceivably be implemented with the above differentiable formula to combine the observation of tidal deformability with the stellar structures to improve the reconstruction.

\textbf{Multi-messenger measurements.} It is also feasible to explore the nuclear matter EoS from multi-messenger measurements directly with modern deep learning techniques. In Ref.~\cite{Goncalves:2022smd}, Goncalves et al. investigate the Audio Spectrogram Transformer (AST) model for the analysis of gravitational wave data. The AST machine-learning model is a convolution-free classifier that captures long-range global dependencies through a purely attention-based mechanism. In this work, the model is applied to a simulated dataset of inspiral GW signals from binary neutron star coalescences constructed from five EoSs. It is shown that the model can correctly classify the EOS purely from the gravitational-wave spectra. Additionally, the generalization ability of the machine is investigated in testing data set. Overall, the results suggest that the well-trained attention-based model can infer the cold nuclear matter EOS directly from the GW signals using the simplified setup of noise-free waveforms. In work of Farrell et al.~\cite{Farrell:2022lfd}, the Transformer model is implemented to deduce EoSs directly from X-ray spectra of neutron stars. The approach can determine the EoS by analyzing the high-dimensional spectra of observed stars, bypassing the need for intermediate mass-radius calculations. They show that the end-to-end machine performs slightly but consistently better than the two-step method using the intermediate $M$-$R$ information. The network used in this approach takes into account the sources of uncertainty for each star, enabling a comprehensive propagation of uncertainties to the EOS. These attempts enlighten us that in the future, using deep learning techniques to process more measurements~\cite{Bogdanov:2022faf}(e.g., short gamma-ray burst~\cite{DAvanzo:2015kdp} and subsequent kilonova~\cite{Cowperthwaite:2017dyu}) will help us to eventually obtain a concrete understanding of the dense matter EoS~\cite{Huang:2022mqp,Fujimoto:2022xhv}.

\subsection{Constraints from Nuclear Physics}

\subsubsection{Connections with Nuclear Symmetry Energy}
One important aspect of the nuclear matter EoS is the symmetry energy, $E_\text{sym}(\rho)$, which represents the energy cost of converting neutrons into protons or vice versa. This quantity can affect many properties of nuclei, including the stability of nuclei and nuclear processes~\cite{Baldo:2016jhp}. The nucleonic component of the nuclear matter EoS can be expressed in terms of the energy per nucleon $\rho$ as,
\begin{equation}
    E(\rho, A) =  E(\rho,0) + E_\text{sym}(\rho)A^2,
\end{equation}
where $ E(\rho,0)\equiv E_{SNM}(\rho)$ is the energy per nucleon of symmetric nuclear matter(SNM) and $A = (\rho_n - \rho_p)/\rho$ is the isospin asymmetry with $\rho_{n/p}$ being the neutron (proton) density. Thus, as a part of the EoS, the symmetry energy $E_\text{sym}(\rho)$ can also be constrained by studying neutron star observations~\cite{Xie:2019sqb,Li:2021thg}. Around saturation density $\rho_0$, $E_{SNM}(\rho)$ can be expanded as $E_0 + K_0 x^2/2 +  J_0 x^3/6$ with $x\equiv (\rho - \rho_0)/(3\rho_0)$. $E_0, K_0$ and $ J_0$ are the binding energy, incompressibility, and skewness of symmetric nuclear matter; the symmetry energy can be expanded as $E_\text{sym}(\rho) = S_0 + L x + K_\text{sym} x^2/2 +  J_\text{sym} x^3/6$, where $S_0, L, K_\text{sym}$ and $ J_\text{sym}$ are the magnitude, slope, curvature, and skewness of the symmetry energy at $\rho_0$. These physical parameters can be determined from NS observables~\cite{Zhang:2018vrx}. This information can be further used to improve our understanding of how nuclear matter behaves under extreme conditions and to test nuclear models~\cite{Li:2019xxz}. In Ref.~\cite{Anil:2020lch}, the authors apply a machine learning approach of support vector machines(SVMs) to predict binding energies of nuclei which are highly related to the outer crust of neutron stars in the density range $\lesssim 10^{-3} \rho_0$. In a series of works from Li et al.~\cite{Xie:2019sqb, Zhang:2018vrx, Xie:2020rwg}, they have developed the Bayesian Inference approach to infer, e.g., EoS, phase transitions, symmetry energy and empirical parameters, from observations of neutron stars together with nuclear experiments. See a recent review in Ref.~\cite{Li:2021thg}.

In Ref.~\cite{Krastev:2021reh}, Krastev trains a deep neural network(DNN) to determine the nuclear symmetry energy as a function of density directly from observational neutron star data. The author demonstrates that DNNs can accurately reconstruct the nuclear symmetry energy from a set of available observables of the neutron star, i.e., the masses and the tidal deformabilities. In Ref.~\cite{Ferreira:2022nwh}, Ferreira et al. also use DNNs to analyze the relationship between the cold $\beta$-equilibrium matter of NSs and the properties of nuclear matter, which is a non-trivial mapping~\cite{Mondal:2021vzt}. Using a Taylor expansion of the energy per particle of homogeneous nuclear matter, they generate a data set of different $\beta$-equilibrium neutron star matter scenarios and their corresponding nuclear matter properties. The neural network is trained and achieves high accuracy on the test set. In a real case scenario, they test the neural network on 33 nuclear models and are able to accurately recover the nuclear matter parameters with reasonable standard deviations for both the symmetry energy slope and the nuclear matter incompressibility at saturation. In the work of Thete et al.~\cite{Thete:2022eif}, a DNN is trained to predict EoSs from nuclear matter saturation parameters, then used as a static module to infer a set of seven nuclear parameters from astrophysical observations.

\subsubsection{Constraints from $\chi$EFT, HICs, and pQCD}

In addition to using observations from astrophysics, one can improve the performance of the reconstruction by combining more physical priors from other domains of nuclear physics, i.e., relativistic mean-field(RMF) calculations at low-density region~\cite{Serot:1984ey,Oertel:2016bki}, heavy-ion collisions (HICs) at mid-density region~\cite{Fukushima:2020yzx,Dexheimer:2020zzs,Huth:2021bsp,Sorensen:2023zkk} and perturbative calculations at nuclear matter and quark matter regions respectively~\cite{Drischler:2021kxf,Ghiglieri:2020dpq}. They have been treated as straightforward physics constraints for reducing the dense matter EoS ~\cite{Annala:2021gom,Demircik:2021zll,Shirke:2022tta}.

Chiral effective field theory ($\chi$EFT) is a widely used framework for studying the properties of nuclear matter at moderate densities (see recent reviews ~\cite{Sammarruca:2019ncy,Drischler:2019xuo,Drischler:2021kxf}). Recent advances in $\chi$EFT have led to a computationally efficient tool for determining the properties of nuclear matter at densities up to $\sim 2\,n_0$. In $\chi$EFT, the effective strong interaction \textit{degree of freedom}s are nucleons and pions (and delta isobars), presenting throughout most of the neutron star interior.  The effective Lagrangian is formed in an expansion in powers of the hadron momenta and the quark masses, which are the small parameters in the theory. A new framework is introduced for quantifying the correlated uncertainties of the nuclear matter EoS that is derived from $\chi$EFT~\cite{Drischler:2020hwi,Drischler:2020yad}. This framework uses Gaussian processes with physics-based \textit{multitask} design to efficiently identify theoretical uncertainties from $\chi$EFT truncation errors to derived quantities, e.g., pressure, sound velocity, symmetry energy and its slope. This approach has been applied to the calculations with nucleon-nucleon and three-nucleon interactions up to fourth order in the expansion. At nuclear saturation density, the predicted symmetry energy and its slope are in agreement with various experimental constraints. This work has provided a statistically robust uncertainty estimate in the low density region, which can contribute as a reliable constraint.

The RMF approach uses a relativistic Lagrangian written in terms of baryon and meson fields with simplicity and flexibility~\cite{Oertel:2016bki}. The in-medium coupling constants are chosen to reproduce nuclear physics measurements around $n\simeq n_0$.
To determine the EOS based on a relativistic approach with minimal constraints, Malik et al.~\cite{Malik:2022zol} use the Bayesian approach on a set of models based on the RMF framework with density-dependent coupling parameters and no nonlinear mesonic terms. These models are constrained by the EoS derived from $\chi$EFT and four saturation properties of nuclear matter: density, binding energy per particle, incompressibility, and symmetry energy. The authors have verified that the derived posterior distribution of NS maximum masses, radii, and tidal deformabilities are consistent with recent NS observations. In related works from Li et al.~\cite{Zhu:2022ibs,Sun:2022yor}, the parametric interactions of RMF models are inferred from astrophysical observations. This is also a promising attempt to understand the nuclear matter from hadron degree of freedom.

Heavy-ion collisions involve the collision of heavy atomic nuclei at high energies, creating a state of matter known as the quark--gluon plasma. This state of matter exists in the cores of neutron stars. The results of heavy-ion collision experiments can be used to study the equation of state of dense matter under these extreme conditions~\cite{Huth:2021bsp,Lovato:2022vgq}. Bayesian methods have been employed to constrain the density dependence of the QCD EoS for dense nuclear matter using the mean transverse kinetic energy and elliptic flow of protons from HICs~\cite{OmanaKuttan:2022aml}, in the beam energy range $\sqrt{s_\text{NN}}=2$--$10$~GeV. The analysis results in tight constraints on EoS at density $n>4\,n_0$. The extracted EoS is found to be consistent with other observables measured in HIC experiments and with constraints from astrophysical observations, which were not used in the analysis. 

At extremely high densities, perturbative QCD (pQCD) can vary the order of renormalization scales, which allows one to derive thermodynamical contributions of higher-order corrections~\cite{Kurkela:2009gj}. In recent works of Gorda et al.~\cite{Gorda:2022jvk,Gorda:2022lsk}, the authors demonstrate that the pQCD calculations provide significant and non-trivial information about the nuclear matter EoS, going beyond what can be obtained from current astrophysical observations. The EoS is extrapolated with a Gaussian process and conditioned with observations and the QCD input. They find that imposing the QCD input does not require extrapolation to a large-$n$ density, while providing strong additional constraints at high densities. In practice, the reliable interpolation between low- and high-density parts can produce an ensemble of EoSs that can be further constrained by astrophysical observations~\cite{Annala:2021gom}.

\subsection{Summary}
This chapter provides an overview of recent research on nuclear matter EoS inference using statistical learning algorithms (e.g., Bayesian inference) and advanced deep learning methods (e.g., deep neural networks). Recent measurements, including pulsar masses and radii, are discussed. The neural network EoS and the automatic differentiation framework for solving the inverse problem are also presented. Multimessenger observations and nuclear constraints are discussed as ways to improve the reconstruction. As more data are accumulated, such as from GW observations~\cite{LIGOScientific:2019lzm,LIGOScientific:2023vdi}, and as prior knowledge (e.g., physical constraints) is incorporated into deep models, reconstruction performance will continue to improve.

	\newpage
    \section{Advanced Developments --- Physics Meets ML}\label{sec:new}

In the previous sections, we have reviewed how deep learning can be broadly applied to diverse aspects of high-energy nuclear physics, serving as a powerful tool for addressing specific problems abstracted from physical domains.
As we also noted, both accumulating large amounts of data rapidly and collecting small amounts of physical observations slowly pose challenges to readily usable deep learning techniques. Indeed, one needs to further enhance the efficiency, if not the applicability, of deep learning by incorporating physics knowledge into the learning process. In the development of deep learning, several attempts have been made to introduce physics knowledge, including:
\textit{Physics-inspired}~\cite{Ahmad:2020kdd, sompolinsky1988statistical, mezard2009information} deep learning, where physics properties such as symmetry or conservation laws are implemented in the architecture of the network and strictly enforced. For properties or rules that cannot be directly encoded in the network setup, one can adopt the \textit{physics-informed}~\cite{2021NatRP...3..422K} approach and include a loss function to penalize their violation. Finally, the \textit{physics-driven}~\cite{2021arXiv210905237T} approach extends the backpropagation procedure to train network parameters when the output of a network is related to observables through a differentiable function or functional mapping.

The EOS reconstruction from Neutron Star mass and radius observations that was discussed in Sec.~\ref{sec:astro} is a good example of how these three approaches can be simultaneously implemented in different aspects of a particular problem. 
One may reconstruct the nuclear matter equation of state, i.e., energy density as a function of pressure $\varepsilon(P)$, from the observation of mass and radius for neutron stars. One may adopt a neural network to represent the inverse of the speed of sound, $c_s^{-2}(P)\equiv \mathrm{d}\varepsilon(P)/\mathrm{d}P$. Then the causality requirement can be fulfilled by setting the activation function of the output layer to be $\sigma(z) = 1+e^z$. Such procedure encodes the physics constraint in the setup, and therefore it is a \textit{physics-inspired} procedure. Then prior knowledge from chiral Effective Field Theory(perturbative QCD calculation) can be implemented as a term in the loss function to constrain the low(high)-pressure sector of the equation of state, which fits the definition of \textit{physics-informed} deep learning. Finally, \textit{physics-driven} deep learning would be reflected by the variation analysis of the TOV equation ---  the functional mapping between $\varepsilon(P)$ and the mass-radius relation --- in which the functional derivative of mass and radius with respect to arbitrary change of the equation of state can be computed to guide the optimization of the network parameter in gradient-based training.

In this section, we focus on some works that have been briefly mentioned earlier in the previous sections and expand on their advanced developments of method which implement physics knowledge and/or constraints into the machine learning task. 

\subsection{Manifesting Physics Properties in NNs}
\label{sec_phy_manifest}
An illustration of physics-inspired deep learning is the integration of physics properties, such as symmetry, into the architecture of a neural network. For further information on this topic, readers can refer to references~\cite{Goodfellow2016, Mattheakis:2019tyi, Kicki:2021so}.
Many special neural network architectures have been designed to incorporate specific physics properties. For example, Convolutional Neural Networks (CNNs) are known to be invariant under coordinate translation~\cite{zhang1988shift,Goodfellow2016}. The E3NN architecture~\cite{e3nn_paper} is designed to preserve rotational invariance, while Group Equivariant Convolutional Networks~\cite{pmlr-v48-cohenc16}, Point Nets~\cite{qi2016pointnet} with permutation invariance, and Lorentz group equivariant Neural Networks~\cite{Bogatskiy:2020} are other examples.

In the field of nuclear and particle physics, implementing symmetry properties has been applied to reduce the computational complexity of machine learning tasks. This includes point cloud representations for collider event classifications~\cite{Onyisi:2022hdh}, embedding guage symmetry in normalizing flow for lattice guage field calculations~\cite{Kanwar:2020xzo}, Feynman path generation using Fourier-flow-based models~\cite{Chen:2022ytr} and the use of CNNs to approximate the inversion of the renormalization group transformation in quantum field theory~\cite{Bachtis:2021eww}. The last three examples will be discussed in more detail in the remainder of the current subsection.

\subsubsection{Gauge Symmetry in Normalizing Flow}
\label{flow_symmetry}
As introduced in Section~\ref{sec:3:flow_based}, the flow-based generative machine learning methods offer a promising solution to addressing the critical slowing down(CSD) in lattice simulations. Specifically, in the context of high-energy nuclear physics (HENP) studies of quantum chromodynamics (QCD), lattice gauge field simulations require consideration of special local symmetries. In recent years, learning architectures that encode the relevant symmetries in lattice field theory have been developed~\footnote{See also the end of Section~\ref{sec_phases_obs} where the incorporation of symmetries into neural networks is discussed for a regression task in lattice gauge theory studies.}.

In Ref.~\cite{Kanwar:2020xzo}, a gauge invariant flow model was developed to sample configurations in a $\mathrm{U}(1)$ gauge theory. On a lattice, the field configuration can be represented by the gauge variable $U_\mu(x)$, which connects the neighboring sites $x$ and $x+\hat{\mu}$. For a theory with $N_d$ spacetime dimensions (i.e. $\mu = 1, 2, \cdots, N_d$), the variable $U(x)$ has $G$ degrees of freedom (e.g. color). The authors defined the lattice volume as $V$, and the field configurations have dimensions of $G^{N_d V}$. The theory is invariant under a gauge transformation,
\begin{align}
    U_\mu(x) \to \tilde{U}_\mu(x) = \Omega(x) {U}_\mu(x) \Omega^\dagger(x+\hat{\mu}),
\end{align}
where the space-time-dependent $G\times G$ matrices, $\Omega$, act on the intrinsic degrees of freedom.
Therefore, the physical distribution shall also be invariant, i.e., 
\begin{align}
    p(U) = p(\tilde{U}).
    \label{eq:5:gauge_invariance}
\end{align}
In comparison to learning symmetries across a training dataset, the gauge invariant flow model efficiently encodes the invariance of distribution. One may use the normalizing flow method introduced in Sec.~\ref{subsubsec:gm}, in which the invertible coupling layers ($g$) are bijective mappings between a $G^{N_d V}$-dimensional manifold and itself. Equation~\eqref{eq:5:gauge_invariance} will be fulfilled if two conditions are met: 1) the prior distribution ($\mathcal{P}$) is symmetrical and 2) each coupling layer commutes with the gauge transformations. While the first criterion can be easily satisfied by setting $\mathcal{P}$ as a uniform distribution, the second requires a specific design for the coupling layer.
%%%%%%%%%%%%%%%%%%%%%%%%%%%%%%%%%%%%%%%%%%%%%%%%%%%%%%%%%%%%%%%%%%%%%%%%%%
\begin{figure}[!hbtp]
    \centering
    \includegraphics[width=0.6\textwidth]{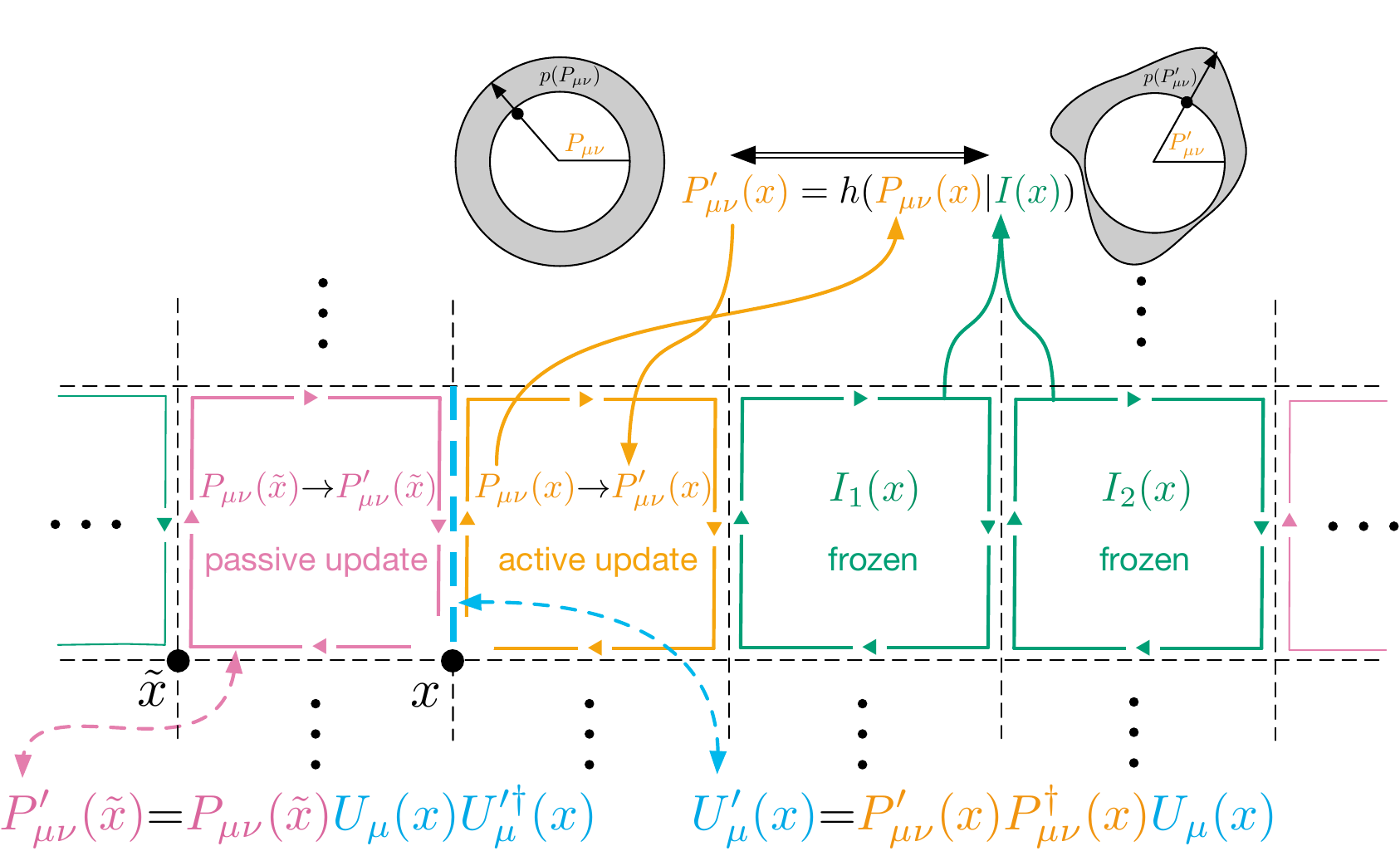}
    \caption{Demonstration of gauge invariant normalizing flow. Taken from~\protect{\cite{Kanwar:2020xzo}} with permission. \label{fig:5:gauge_inv}}
\end{figure}
%%%%%%%%%%%%%%%%%%%%%%%%%%%%%%%%%%%%%%%%%%%%%%%%%%%%%%%%%%%%%%%%%%%%%%%%%%
In~\cite{Kanwar:2020xzo}, each $g$ only acts on some of the gauge links, and one may divide the $G^{N_d V}$-dimensional manifold into two subsets, $U^A$ and $U^B$. All elements in the latter are invariant under $g$: $\forall U\in U^B$, $g(U)=U$; whereas a link in the former ($U^i \in U^A$) is mapped to
\begin{align}
    U'^i \equiv g(U^i) = h(U^i S^i|\boldsymbol{\Pi}_i) (S^{i})^{\dagger},
\end{align} 
in which $S^i$ is a product of links such that $U^i S^i$ forms a loop ending at the starting point. It belongs to $U^B$ to ensure invertibility, and its explicit form can be determined when a theory is specified. $h$ is a $G$-to-$G$ dimensional invertible kernel explicitly parametrized by a set of gauge invariant quantities ($\boldsymbol{\Pi}_i$) constructed from the elements of $U^B$, and satisfies,
%%%%%%%%%%%%%%%%%%%%%%%%%%%%%%%%%%%%%%%%%%%%%%%%%%%%%%%%%%%%%%%%%%%%%%%%%%%%%%%%%%%%%%
\begin{align}
    h(W X W^\dagger) = W h(X) W^\dagger, \qquad\forall X, W \in G,
\end{align}
so that $g$ commutes with the gauge transformation:
\begin{align}
\begin{split}
    U'^i \to \tilde{U}'^i 
=\;&
    h\Big(\Omega(x) U^i S^i \Omega^\dagger(x)\Big) \; \Big(\Omega(x+\hat{\mu}) S^{i} \Omega^\dagger(x)\Big)^{\dagger} 
\\=\;&
    \Omega(x)  h(U^i S^i) \Omega^\dagger(x)\; \Omega(x) (S^{i})^{\dagger} \Omega^\dagger(x+\hat{\mu})
\\=\;&
    \Omega(x) U'^i \Omega^\dagger(x+\hat{\mu}).
\end{split}
\end{align}
%%%%%%%%%%%%%%%%%%%%%%%%%%%%%%%%%%%%%%%%%%%%%%%%%%%%%%%%%%%%%%%%%%%%%%%%%%%%%%%%%%%%%%

Taking a $\mathrm{U}(1)$ gauge field in $1+1$ dimension as an application example, to which the lattice discretization for this field can be approximated through the Wilson action, defined as:
\begin{align}
S(U) = -\beta \sum_x \mathrm{Re} P(x),
\end{align}
where $P(x)=U_0(x) U_1(x+\hat{0}) U_0^\dagger(x+\hat{1}) U_1^\dagger(x)$ represents the plaquette at $x$. The choice of $S_i$ is depicted in Fig.~\ref{fig:5:gauge_inv}, with $U_i S_i$ representing $1\times1$ loops adjacent to each $U_i$. The elements of $U^A$ are sparse enough such that updates do not overlap with one another. With the aforementioned construction of gauge-equivariant coupling layers, a flow-based MCMC scheme was employed on this 1+1D $\mathrm{U}(1)$ gauge theory at a fixed lattice size ($L=16$). This resulted in more efficient calculations for topological quantities compared to traditional HMC and Heat Bath algorithms~\cite{Kanwar:2020xzo}.

\subsubsection{Inverse Renormalization Group Transformation} 
In addition to Normalizing Flow method~\cite{Albergo:2019eim, Kanwar:2020xzo, Albergo:2021vyo} which independently samples field configurations in each round, another approach to addressing the CSD using physics-inspired deep learning is the implementation of Convolutional Neural Networks (CNNs) in solving inversion of renormalization group (RG) transformations. This has been independently studied in the context of quantum field theory~\cite{Bachtis:2021eww} and spin systems~\cite{Shiina:2021pqe}.

%%%%%%%%%%%%%%%%%%%%%%%%%%%%%%%%%%%%%%%%%%%%%%%%%%%%%%%%%%%%%%%%%%%%%%%%%%%%
\begin{figure}[!hbtp]
    \centering
    \includegraphics[width=0.6\textwidth]{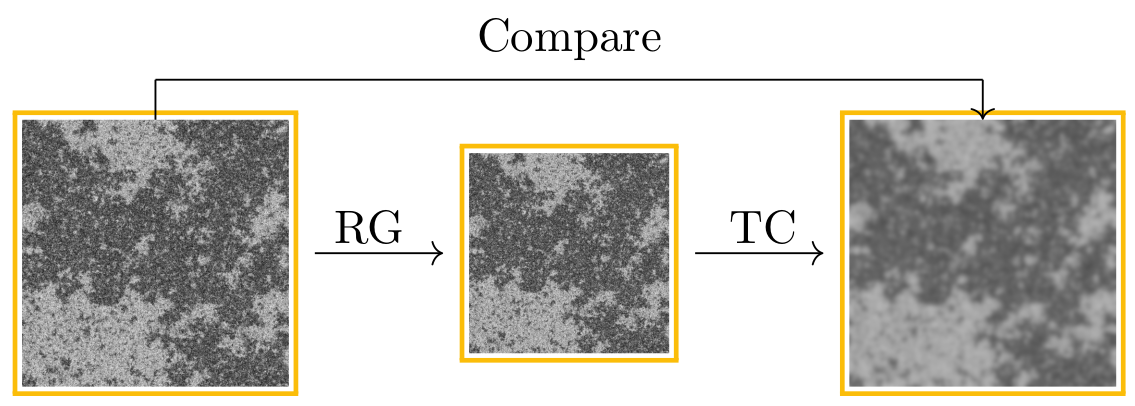}
    \caption{Taken from Ref.~\cite{Bachtis:2021eww} with permission. Demonstration of the inverse RG training with transposed convolution (TC).\label{fig:5:rg}}
\end{figure}
%%%%%%%%%%%%%%%%%%%%%%%%%%%%%%%%%%%%%%%%%%%%%%%%%%%%%%%%%%%%%%%%%%%%%%%%%%%%

At the critical point, a thermal system exhibits divergence of the correlation length and  therefore there is no length scale remaining in it. As a result, the expectation values of intensive observables are invariant under arbitrary scale transformations. This property is attributed to the theory of RG, which is widely used to study thermodynamics near the critical point. In the typical application of RG, one reduces the size of a system by coarse-graining the configurations of a thermal ensemble, comparing the thermal quantities before and after the operation. The number of RG steps is limited by the original finite size of the system, as the operations always result in a reduction in size. However, one needs to approach the critical behaviors in a large-size limit, which is routinely hindered by computational resources. It has been noted that the coarse-graining operation in real space can be treated as a convolution operation with step $2$ and filter size $2 \times 2$. The authors of~\cite{Bachtis:2021eww} and~\cite{Shiina:2021pqe} supervisedly train a transposed convolution or upsampling layer with filter size $2 \times 2$ to approximate the inverse operation of RG\footnote{See Ref.~\cite{2016arXiv160307285D} for details on the involved different convolution arithmetics}. Fig.~\ref{fig:5:rg} gives an illustration of such training using data generated from standard RG procedures. With this tool, it is able to generate large-size configurations from small-size ones with an infinite number of inverse RG steps. One can then accurately determine the location of the critical point and study the critical exponents in its vicinity.

\subsubsection{Fourier-flow Model Generating Feynman Paths}
\begin{figure}[!hbtp]
    \centering
    \includegraphics[width=0.75\textwidth]{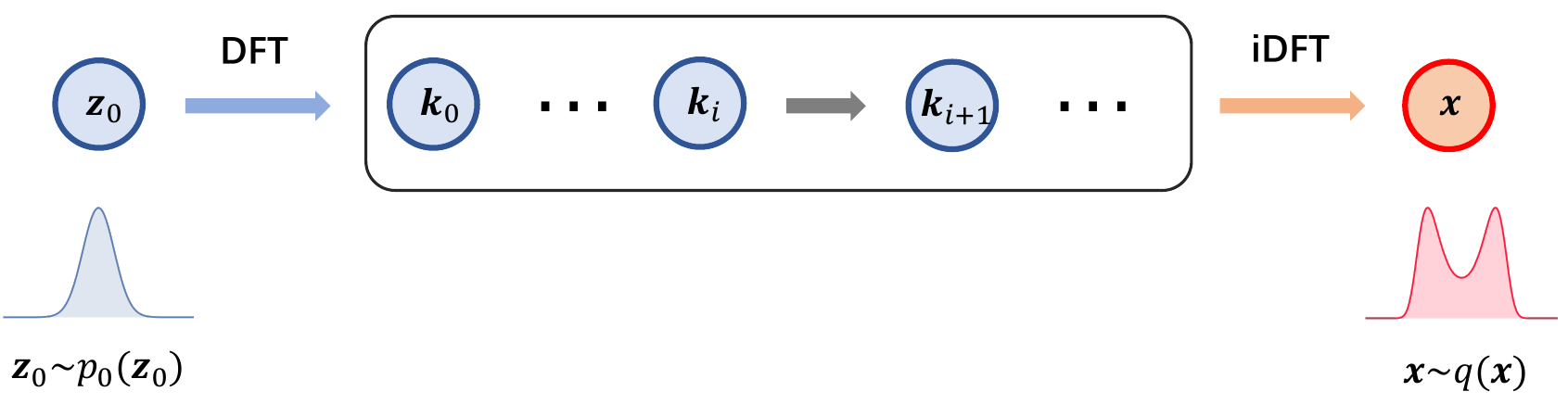}
    \caption{Demonstration of the Fourier flow model. The discrete Fourier transformation(DFT) and inverse discrete Fourier transformation(iDFT) are inserted before the input and after the output of a normalizing flow model respectively.\label{fig:5:ffm}}
\end{figure}

Besides embedding symmetries and approximating physical processes with neural networks, it is helpful to deploy the machine learning algorithms in a more efficient physics representation. In Ref.~\cite{Chen:2022ytr}, the authors proposed a Fourier flow model(FFM) to simulate the Feynman propagator and generate paths for quantum systems (see Fig.~\ref{fig:5:ffm}). The Fourier transformation is introduced to approach a Matsubara representation in order to preserve the physics condition for the system. The Euclidean action defined in a discrete time, $S_\text{E}[x_n]$, is converted into the Fourier space as, $S_\text{E}[X_k]$. The Fourier modes, $X_k$, represent multi-level correlations in coordinate space. Time-reversal symmetry requires that certain boundary conditions be imposed on the discretized paths ${x_n}$. They include invariance under translation (${n\rightarrow n+1}$), inversion (${n\rightarrow -n}$), and periodicity (${n\to n+N}$), where $n$ is the index of the site. These conditions are naturally satisfied in Fourier space. The path generator of FFM is validated on the harmonic and anharmonic oscillators as a demonstration. The latter is set as a multimode system in a double-well potential without an analytic solution. The ground-state wave function and low-lying energy levels are accurately estimated.

%%%%%%%%%%%%%%%%%%%%%%%%%%%%%%%%%%%%%%%%%%%%%%%%%%%%%%%%%%%%%%%%%%%%%%%%%%%%
%%%%%%%%%%%%%%%%%%%%%%%%%%%%%%%%%%%%%%%%%%%%%%%%%%%%%%%%%%%%%%%%%%%%%%%%%%%%

It should be noted that the kinetic term of action, $m(\Delta x/\Delta t)^2$ which correlates in coordinate space, will disentangle in frequency space as, $m|X_k|^2$(see details in Ref.~\cite{Chen:2022ytr}). As a matter of fact, the Feynman path describes the field in 0+1 dimensional field theory. The Fourier transformation converts the dynamical fields to the inverse of the two-point correlation function in frequency space (also known as the inverse power spectral density as discussed in Ref.~\cite{Komijani:2023fzy}). It can preserve the dynamical information at least at mean-field level~\cite{Georges:1996zz}. This method provides a novel way to manifest physics properties, such as time-reversal symmetry and dynamical mean-field theory, in the design of flow-based models.

\subsection{Fusing Physics Models into NNs}\label{sec:5:phy_mod}

\begin{figure}[!hbtp]
    \centering
    \includegraphics[width=0.5\textwidth]{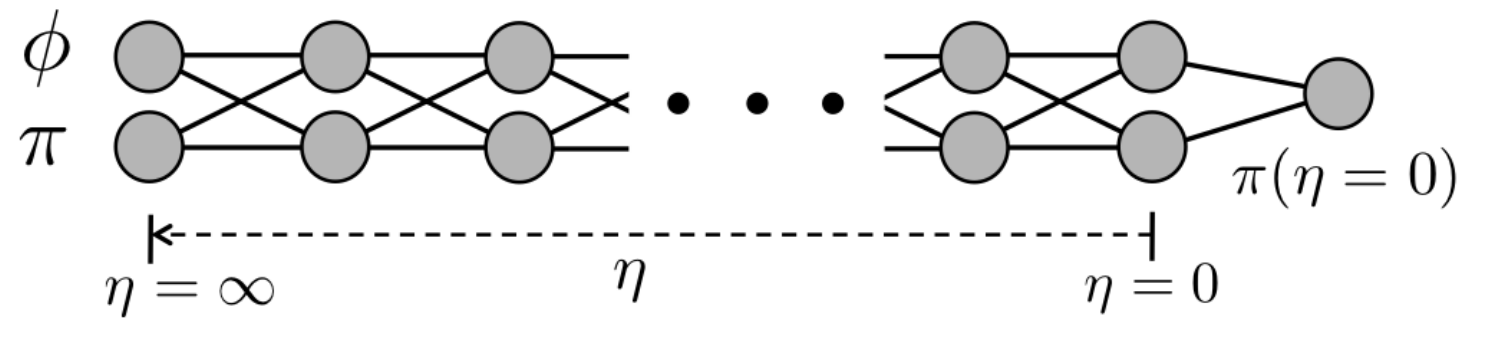}
    \caption{Demonstration of solving Eq.~\protect{\eqref{eq:5:adscft}} with DNN. Taken from~\protect{\cite{Hashimoto:2018ftp}} with permission. \label{fig:5:adscft}}
\end{figure}

Another type of physics-inspired machine learning leverages the mathematical similarity between a specific physics model and a neural network architecture, enabling the application of existing deep learning frameworks to solve physics problems with ease. Some pioneer attempts~\cite{Hashimoto:2018ftp,Hashimoto:2019bih,Hu:2019nea,Hashimoto:2021ihd} have been made in the direction of anti-de Sitter/conformal field theory(AdS-CFT) correspondence. The AdS-CFT correspondence~\cite{Maldacena:1997re,Gubser:1998bc,Witten:1998qj} is a renowned holographic relation between $d$-dimensional Quantum Field Theories and $(d+1)$-dimensional gravity. It has been widely applied in solving problems for strong coupling quantum fields. In the limit of large number of colors ($N_c$), the generating functional of the CFT boundary and the action of the gravity bulk are related by the Gubser--Klebanov--Polyakov--Witten relation,
\begin{align}
    Z_\mathrm{QFT}[J] = \exp(-S_\mathrm{gravity}[\phi]),
\end{align}
where $\phi(x,\eta)$ is the bulk field that satisfies the boundary condition $\phi(x,\eta=0) = J(x)$. It has been argued~\cite{Hashimoto:2019bih} that AdS/CFT correspondence as a deep Boltzmann machine~\footnote{It has also been found that the existence of the exact mapping between the deep Boltzmann machine and the renormalization group~\cite{mehta2014exact}.}. The Boltzmann machines are network models that give a probabilistic distribution of variables $v_i$ that defined as,
\begin{align}
    P(v_i) = \exp\Big(-\sum_i a_i v_i - \sum_{i,j} w_{ij} v_i v_j\Big),
\end{align}
with $a_i$ and $w_{ij}$ being network parameters. The deep Boltzmann machines further include hidden variables to enhance the expression ability, 
\begin{align}
    P(v_i) = \sum_{h_i^{(k)}} \exp\Big(-\sum_{i,j} w_{ij}^{(0)} v_i h^{(1)}_j - \sum_{k=1}^{N-1}\sum_{i,j} w_{i,j}^{(k)} h^{(k)}_i h^{(k+1)}_j\Big),
\end{align}
When comparing with AdS/CFT correspondence, the variables of interest are the quantum fields $J$, represented as $v_i$, and the bulk field $\phi$ corresponds to the hidden variables $h^{(k)}_{j}$. The bulk field is governed by its equation of motion, which can be modeled through the setup of a deep neural network in the deep Boltzmann machine, as claimed in~\cite{Hashimoto:2018ftp}. For instance, in the $(d+1)$-dimensional space-time with the metric,
\begin{align}
    \mathrm{d}s^2 = -f(\eta) \mathrm{d}t^2 +\mathrm{d}\eta^2
    + g(\eta)(\mathrm{d}x_1^2 + \cdots + \mathrm{d}x_{d-1}^2),
\end{align}
with $\eta$ being the holographic direction, a scalar field theory is defined by the action,
\begin{align}
    S = \frac{1}{2}\int \mathrm{d}^{d+1} x \, \sqrt{|g|} \Big(g^{\mu\nu}(\partial_\mu \phi) (\partial_\nu \phi) + m^2 \phi^2 + \frac{\lambda}{2} \phi^4 \Big),
\end{align}
where $\sqrt{|g|} = \sqrt{f(\eta) g^{d-1}(\eta)}$ is the spacetime volume factor. The classical equation of motion for a homogeneous, constant field ($\phi$) and its canonical momentum ($\pi$) reads
\begin{align}
\begin{split}
    \partial_\eta \pi + h(\eta) \pi - m^2 \phi - \lambda \,\phi^3 = 0\,,
\qquad
    \partial_\eta \phi = \pi\,.
\end{split}
\label{eq:5:adscft}
\end{align}
where $h(\eta) \equiv \partial_\eta \ln\sqrt{|g|}$. It has been claimed in~\cite{Hashimoto:2018ftp} that numerically solving \eqref{eq:5:adscft} with Euler's method is equivalent to a DNN with width $w=2$ at each layer, see Fig.~\ref{fig:5:adscft}. Each layer corresponds to an Euler iteration step. The weights at the $n^\mathrm{th}$ layer are set to be 
\begin{align}
W^{(n)} = 
\left(\begin{array}{cc}
    m^2\delta\eta & 1-h(\eta^{(n)})\Delta\eta\\
    1 & \Delta\eta
\end{array}\right),
\end{align}
whereas biases are set to zero. Linear activation is set for $\phi$ and non-linear one is set for $\pi$ to implement the interaction term ($\lambda \phi^3$).
%%%%%%%%%%%%%%%%%%%%%%%%%%%%%%%%%%%%%%%%%%%%%%%%%%%%%%%%%%%%%%%%%%%%%%%%%%%%
%%%%%%%%%%%%%%%%%%%%%%%%%%%%%%%%%%%%%%%%%%%%%%%%%%%%%%%%%%%%%%%%%%%%%%%%%%%%
In Ref.~\cite{Hashimoto:2018ftp}, the authors aim to determine the function $h(\eta)$ given the values of $\pi$ and $\phi$ at the $\eta=\infty$ and $\eta=0$ limits. As a mock test, the authors prepared a set of data by solving Eq.~\eqref{eq:5:adscft} exactly with a known $h(\eta)$. The parameters of the network [$h(\eta^{(n)})$] were then trained to match the relationship between the large and small boundary values in the holographic direction. The results showed that the reconstructed metric agreed well with the true values, with a deviation of about $30\%$ in the near-horizon (small $\eta$) region.

The procedure has been further improved by parameterizing $h(\eta)$ by polynomials~\cite{Hashimoto:2020jug}. The loss function and parameter training procedure of the neural ODE~\cite{2018arXiv180607366C} are borrowed to learn the polynomial coefficients in $h(\eta)$. Then, the authors trained the machine from lattice QCD results of the quark mass condensate and obtained results that are qualitatively consistent with the temperature dependence of the confinement and the Debye-screening behavior.

\subsection{Solving Complex Inverse Problems}\label{sec:5:inverse}
%%%%%%%%%%%%%%%%%%%%%%%%%%%%%%%%%%%%%%%%%%%%%%%%%%%%%%%%%%%%%%%%%%%%
\begin{table}[!hbpt]
\centering
\begin{tabular}{l |l | l | l}
\hline\hline
    problem & quantity of interest $\mathcal{Q}(x)$ & observable $\mathcal{O}_y$ & relation \\
\hline
Sec.~\ref{subsubsec:realt} & 
    in-medium potential $V(r)$&
    energy spectrum $\{E_n\}$&
    Schr\"odinger equation~\eqref{eq:5:schroedinger}\\
Sec.~\ref{subsec:mr} & 
    equation of state $\varepsilon(P)$&
    mass and radius $\{M_i, R_i\}$&
    TOV equation~\eqref{eq:4:tov}\\
Sec.~\ref{subsubsec:realt} & 
    spectral function $\rho(\omega)$&
    Euclidean correlator $D(k)$&
    K\"allen--Lehmann convolution~\eqref{eq:5:corr_D}\\
\hline\hline
\end{tabular}
\caption{Complex inverse problems: quantities of interest, observables, and the relation between them.}
\label{tab:inverse_example}
\end{table}
%%%%%%%%%%%%%%%%%%%%%%%%%%%%%%%%%%%%%%%%%%%%%%%%%%%%%%%%%%%%%%%%%%%%

We finally review how deep learning helps in solving inverse problems~\cite{Zhou:2023tvv}.
Nuclear physics, and more broadly, physics, presents numerous challenging inverse problems. In these problems, the forward problem is straightforward, but its inversion is not. For instance, consider a quantum system described by the  Schr\"odinger equation with a potential model, if the interaction potential is known, the microscopic properties such as the energy levels and corresponding wave functions can easily be predicted. However, for certain systems like mesons (bound states of quarks and anti-quarks), the energy spectrum can be experimentally measured, but the effective interaction potential remains unclear. In this case, extracting the interaction potential from the given energy spectrum, which is the inverse problem of solving the Schr\"odinger equation, poses a significant challenge, yet remains an important practical issue. Deep learning has been found suitable for solving such inverse problems, examples include reconstructions of heavy quark interaction potential~\cite{Shi:2021qri}, spectral function~\cite{Kades:2019wtd,Chen:2021giw,Zhou:2021bvw,Wang:2021jou,Wang:2021cqw,Shi:2022yqw,Horak:2021syv}, parton distribution function~\cite{DelDebbio:2007ee,Ball:2010de,Ball:2011gg,Ball:2011uy,Nocera:2014gqa,Gao:2022iex}, nuclear matter equation of state~\cite{Fujimoto:2017cdo,Fujimoto:2019hxv,Fujimoto:2021zas,Soma:2022qnv,Soma:2022vbb}, and effective parton mass in a finite-temperature QCD medium~\cite{Li:2022ozl}.

Such inverse problems share similar characteristics. The unknown functions of interest, denoted as $\mathcal{Q}(x)$, are continuous, while the observables which can be either continuous functions or discrete variables, are functionals of $\mathcal{Q}(x)$, i.e. $\mathcal{O}_y = \mathcal{F}_y[\mathcal{Q}(x)]$, where the subscript $y$ labels the continuous argument or discrete index of the observables. A summary of the quantities of interest, observables, and their relationship for the inverse problems discussed in this subsection can be found in Table~\ref{tab:inverse_example}.

%%%%%%%%%%%%%%%%%%%%%%%%%%%%%%%%%%%%%%%%%%%%%%%%%%%%%%%%%%%%%%%%%%%%
\begin{figure}[!hbpt]
    \centering
    \includegraphics[width=0.9\textwidth]{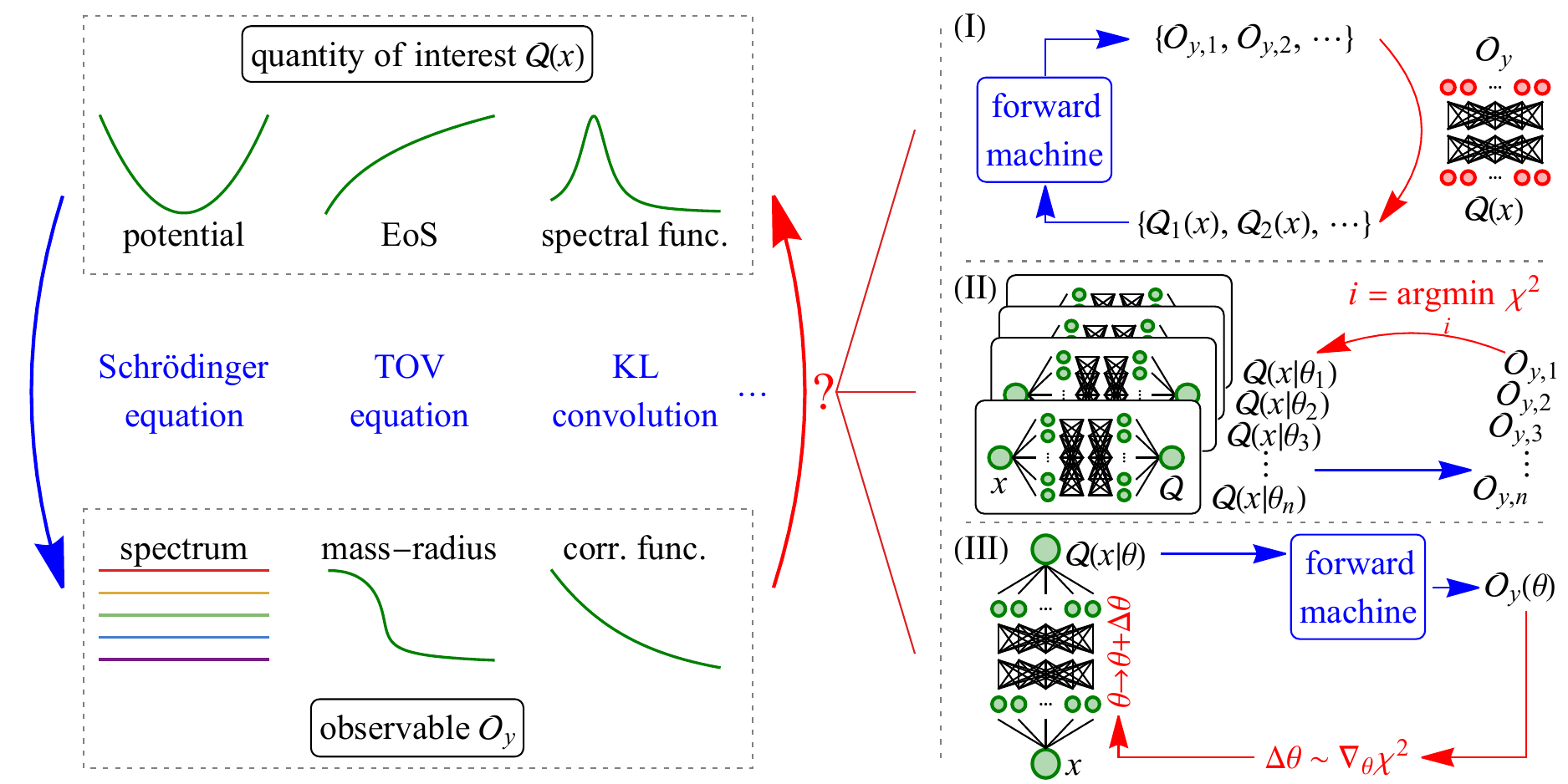}
    \caption{Different approaches of solving inverse problems.}
    \label{fig:5:inverse}
\end{figure}
%%%%%%%%%%%%%%%%%%%%%%%%%%%%%%%%%%%%%%%%%%%%%%%%%%%%%%%%%%%%%%%%%%%%

Generally speaking, there are three different approaches to solving such inverse problems. They are summarized in Fig.~\ref{fig:5:inverse} and listed as follows.
%%%%%%%%%%%%%%%%%%%%%%%%%%%%%%%%%%%%%%%%%%%%%%%%%%%%%%%%%%%%%%%%%%%%%%%%%%%%%%
\begin{itemize}
\item[(i)]
    The first approach is to parameterize or discretize $\mathcal{Q}(x)$, prepare a sufficiently large ensemble of different $\mathcal{Q}$'s, compute their corresponding $\mathcal{O}$'s from the feasible forward modelling and exploit machine learning techniques, such as regression or classification algorithms e.g., with deep neural networks(DNNs) to learn an inverse mapping from the set of $\{\mathcal{O},\mathcal{Q}\}$ pairs. Such an approach falls in the category of direct supervised learning. Physics priors are implicitly manifested inside the training data collection. This approach has been demonstrated in studies discussed in, e.g., Sections~\ref{hic_b}, \ref{hic_cme}, \ref{qcd_phase_hic}, \ref{sec_phases_obs}, and \ref{subsubsec:dlim}.
\item[(ii)]
    The second approach is to parameterize the quantity of interest by polynomials, neural networks or Gaussian Process(GP), $\mathcal{Q}(x)=\mathcal{Q}(x|\boldsymbol{\theta})$, update its parameters according to statistical approach (such as Markov-Chain Monte Carlo within Bayesian Inference) or Heuristic Algorithm (Generic Algorithm), in order to minimize the uncertainty-weighted difference between the target observable and the ones corresponding to $\mathcal{Q}(x|\boldsymbol{\theta})$,
\begin{align}
    \chi^2 = \sum_y \Big(\frac{\mathcal{F}_y[\mathcal{Q}_\text{NN}(x|\boldsymbol{\theta})] - \mathcal{O}_y}{\Delta\mathcal{O}_y} \Big)^2,
    \label{eq:5:chisq}
\end{align}
where $\Delta\mathcal{O}_y$ is the uncertainty of $\mathcal{O}_y$ and $\sum_y \to \int \mathrm{d}y$ for continuous arguments. Then being integrated into the Bayesian formula one can evaluate the posterior distribution for the target. See Sections~\ref{hic_nuclear_structure}, \ref{hic_jet}, and \ref{subsubsec:infer} for typical examples adopted this approach.
\item[(iii)]
    The third approach is similar to the second one, but to update parameters of neural networks according to a gradient-based method,
\begin{align}
    \frac{1}{2}\nabla_{\boldsymbol{\theta}}\chi^2 = \sum_y \frac{\mathcal{F}_y[\mathcal{Q}_\text{NN}(x|\boldsymbol{\theta})] - \mathcal{O}_y}{(\Delta\mathcal{O}_y)^2} 
    \int \mathrm{d}x \frac{\delta\mathcal{F}_y[\mathcal{Q}(x)]}{\delta \mathcal{Q}(x)}\bigg|_{\mathcal{Q}(x)=\mathcal{Q}_\text{NN}(x|\boldsymbol{\theta})} \nabla_{\boldsymbol{\theta}}\mathcal{Q}_\text{NN}(x|\boldsymbol{\theta}),
    \label{eq:5:chisq_grad}
\end{align}
with $\frac{\delta\mathcal{F}_y[\mathcal{Q}(x)]}{\delta \mathcal{Q}(x)}$ being the functional derivative of $\mathcal{F}_y$ with respect to $\mathcal{Q}(x)$. Intuitively, when the chi-square function reaches its minimum, the gradients vanish by definition, $\frac{1}{2}\nabla_{\boldsymbol{\theta}}\chi^2=0$, the parameters stop being updated and achieve the optimal point. This approach also allows the fusion of physical model or simulation into the optimization procedure with differentiable programming strategy. See Sections~\ref{subsubsec:realt}, \ref{subsubsec:ad}, and \ref{sec:5:phy_mod} for examples using this approach, and in the following we will have details explanations for three typical cases.
\end{itemize}
%%%%%%%%%%%%%%%%%%%%%%%%%%%%%%%%%%%%%%%%%%%%%%%%%%%%%%%%%%%%%%%%%%%%%%%%%%%%%%

Compared to Approach(ii), the physics-driven Approach(iii) directs the parameters to the minimum of $\chi^2$, rather than random walks. Therefore, it is more efficient for networks with numerous parameters. To apply the third method, the differentiability of $\mathcal{F}_y[\mathcal{Q}(x)]$ is a prerequisite, as shown in Eq.~\eqref{eq:5:chisq_grad}. In most cases, $\frac{\delta\mathcal{F}_y[\mathcal{Q}(x)]}{\delta \mathcal{Q}(x)}$ can be obtained either through analytical variation analysis or the modern Auto Differentiation framework(e.g., PyTorch, Tensorflow and Paddle); however, in some specific cases, the mapping from $\mathcal{Q}(x)$ to $\mathcal{O}_y$ may be implicit, requiring non-trivial physics knowledge and/or mathematical derivation to compute the functional derivative. In the remainder of this subsection, we provide several examples to highlight the differences between these approaches.

In Sec.~\ref{sec:astro}, we reviewed a few examples of inferring nuclear matter equation of state(EOS) from the neutron star mass-radius relation. In Refs.~\cite{Soma:2022qnv, Soma:2022vbb}, the authors follow Approach (iii) to implement a Neural Network to represent the EOS and unsupervisedly optimize the network parameter in order to fit the observation data, whereas Refs.~\cite{Fujimoto:2017cdo, Fujimoto:2019hxv, Fujimoto:2021zas} supervisedly train a network to represent the inverse mapping from mass-radius observations to the EOS.
The latter fit into Approach (i) and are natural applications of DL. However, one should carefully prepare the training data set to avoid bias. In this particular case, one should also be careful in ensuring that the sequence of different observations does not alter the result EOS. Also, reliable uncertainty quantification is challenging~\cite{Fujimoto:2021zas}. The aforementioned challenges can be addressed through unsupervised training. In this approach, a DNN is used as an unbiased parametrization of the EoS and fed into a Bayesian analysis. This enables the extraction of not only the mean value, but also the associated uncertainty.

\emph{\textbf{Inverse TOV equation}} ---
The success of unsupervised training lies in designing a physics-driven training process that guides the iteration of parameters towards maximizing likelihood. This can be achieved through the use of automatic differentiation (AD) as discussed in Sec.~\ref{subsec:mr}, or through analytical derivation. Despite the highly non-linear nature of the TOV equation~\eqref{eq:4:tov} with respect to the EoS, it is still possible to analyze the linear response of observables out of TOV equations to small changes in the EoS.

For the sake of convenience, we reformulate Eq.~\eqref{eq:4:tov} by changing the independent variable from the radius ($r$) to the logarithmic of pressure, $\ln(P/P_\text{bnd})$. $P_\text{bnd}$ denotes the boundary pressure, and it is small enough such that $M$ and $R$ are insensitive to it. Also, we consider the argument of $v\equiv r^3$ rather than $r$ to avoid numerical divergence. The TOV equation thus becomes,
\begin{align}
\begin{split}
\frac{\mathrm{d}v}{\mathrm{d}\ln \frac{P}{P_\text{bnd}}} =\;&
   - \mathcal{K}_v(P,\varepsilon,v,m)  \,, \\
\frac{\mathrm{d}m}{\mathrm{d}\ln \frac{P}{P_\text{bnd}}} =\;&
   - \mathcal{K}_m(P,\varepsilon,v,m)\,,
\end{split}\label{eq:5:tov}
\end{align}
where 
\begin{align}
\begin{split}
\mathcal{K}_v(P,\varepsilon,v,m) \equiv \;&
    \frac{3(v^{\frac{1}{3}} - 2\,m)}{(m/v+4\pi P)(1+\varepsilon/P)} \,,\\
\mathcal{K}_m(P,\varepsilon,v,m) \equiv \;&
    \frac{4\pi(v^{\frac{1}{3}} - 2\,m)\varepsilon}{(m/v+4\pi P)(1+\varepsilon/P)} \,.
\end{split}
\end{align}
For later convenience, we define $\mathcal{K}_{X,Y} \equiv
\frac{\partial \mathcal{K}_X}{\partial Y}$ for $X\in\{v,m\}$ and $Y\in\{P,\varepsilon,v,m\}$. Given a small, Dirac-$\delta$ function-like perturbation in the EoS, $\varepsilon(P) \to \varepsilon(P) + \lambda\, \delta(P-P')$, the change of $v$ and $m$, denoted as $\Delta v(P|P')$ and $\Delta m(P|P')$, are given by the linear variation of Eq.~\eqref{eq:5:tov},
\begin{align}
\begin{split}
0 =\;& \frac{\mathrm{d}\Delta X(P|P')}{\mathrm{d} \ln  \frac{P}{P_\text{bnd}}}  
      + \mathcal{K}_{X,v} \Delta v(P|P')
      + \mathcal{K}_{X,m} \Delta m(P|P')
      + \lambda\, \mathcal{K}_{X,\varepsilon} \delta(P-P')\,.
\end{split}\label{eq:5:tov_derivative}
\end{align}
It is obvious that both $\Delta v(P|P')$ and $\Delta m(P|P')$ are linearly depending on the perturbation parameter $\lambda$.
By solving Eq.~\eqref{eq:5:tov_derivative} with parameter $\lambda=1$ and argument $\ln \frac{P}{P_\text{bnd}}$ going from $\ln \frac{P_c}{P_\text{bnd}}$ to unity, the functional derivative can be computed as
\begin{align}
\begin{split}
    \frac{\delta M(P_c)}{\delta \varepsilon(P')} =\;& \Delta m(P|P'=P_\text{bnd})\,,\\
    \frac{\delta R(P_c)}{\delta \varepsilon(P')} =\;& \Big(\Delta v(P|P'=P_\text{bnd})\Big)^{1/3}\,.
\end{split}
\end{align}
Consequently, we demonstrate that it is possible to calculate the linear response of a Neutron Star's mass and radius to an infinitesimal perturbation in energy density, thereby supporting the numerical automatic differentiation method outlined in Sec.~\ref{sec:astro}.
%%%%%%%%%%%%%%%%%%%%%%%%%%%%%%%%%%%%%%%%%%%%%%%%%%%%%%%%%%%%%%%%%%%%%%%%%%%%%%%%%%%%
\begin{figure}[!hbt]
    \centering
    \includegraphics[width=0.35\textwidth]{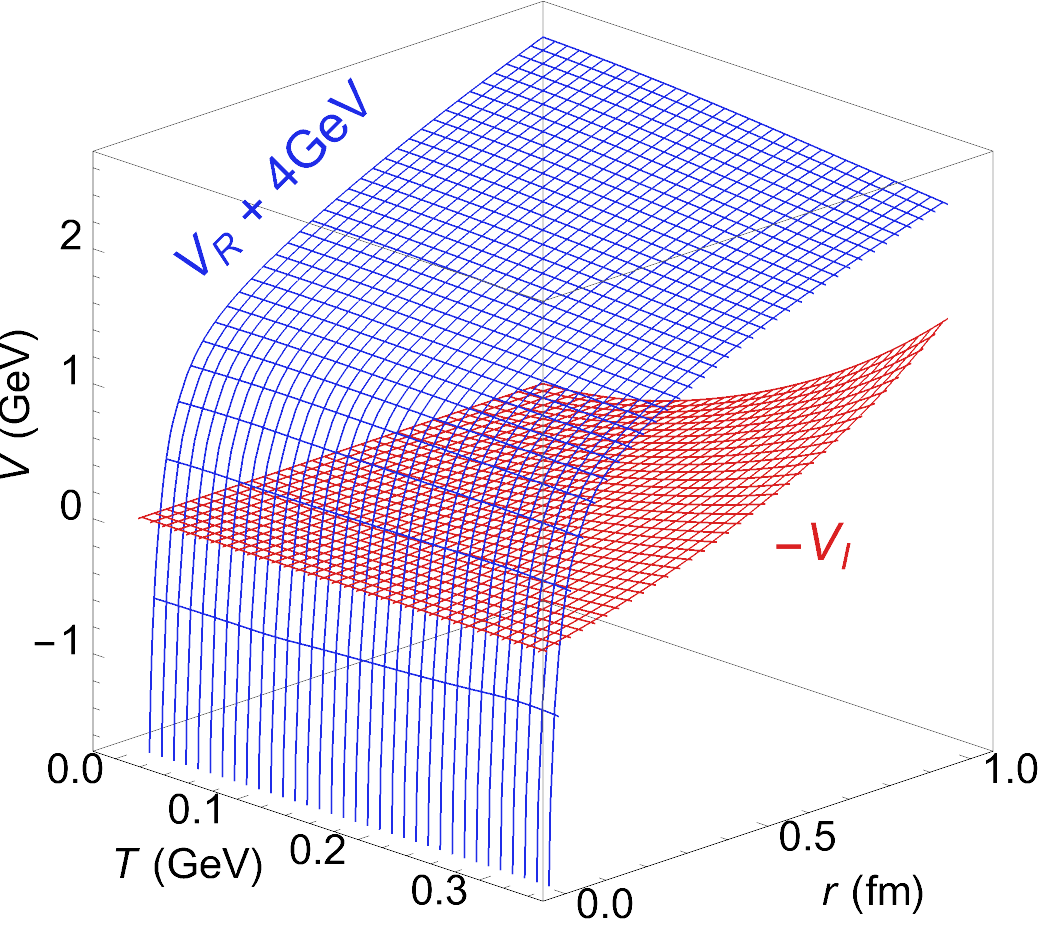}
    \includegraphics[width=0.3\textwidth]{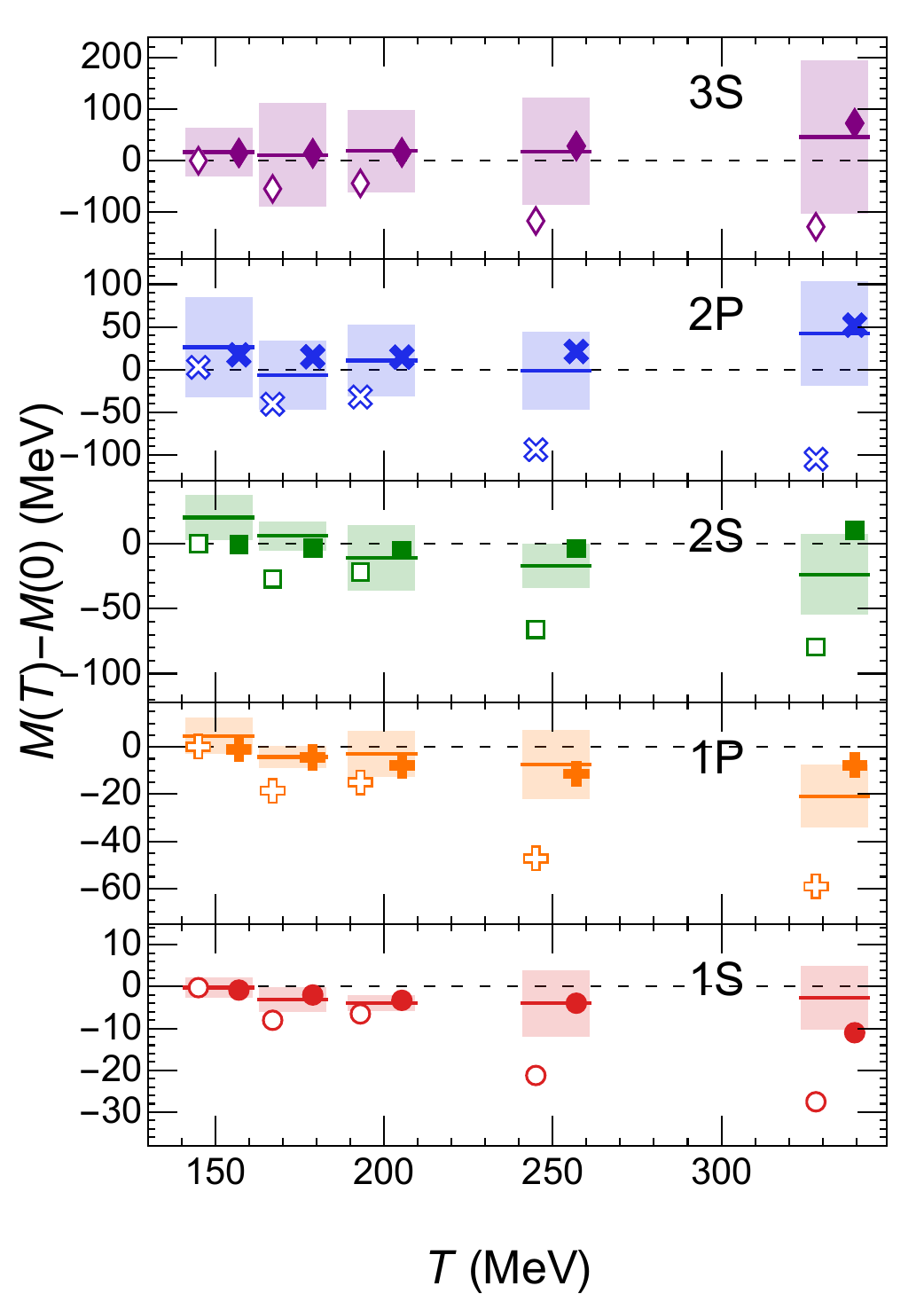}
    \includegraphics[width=0.3\textwidth]{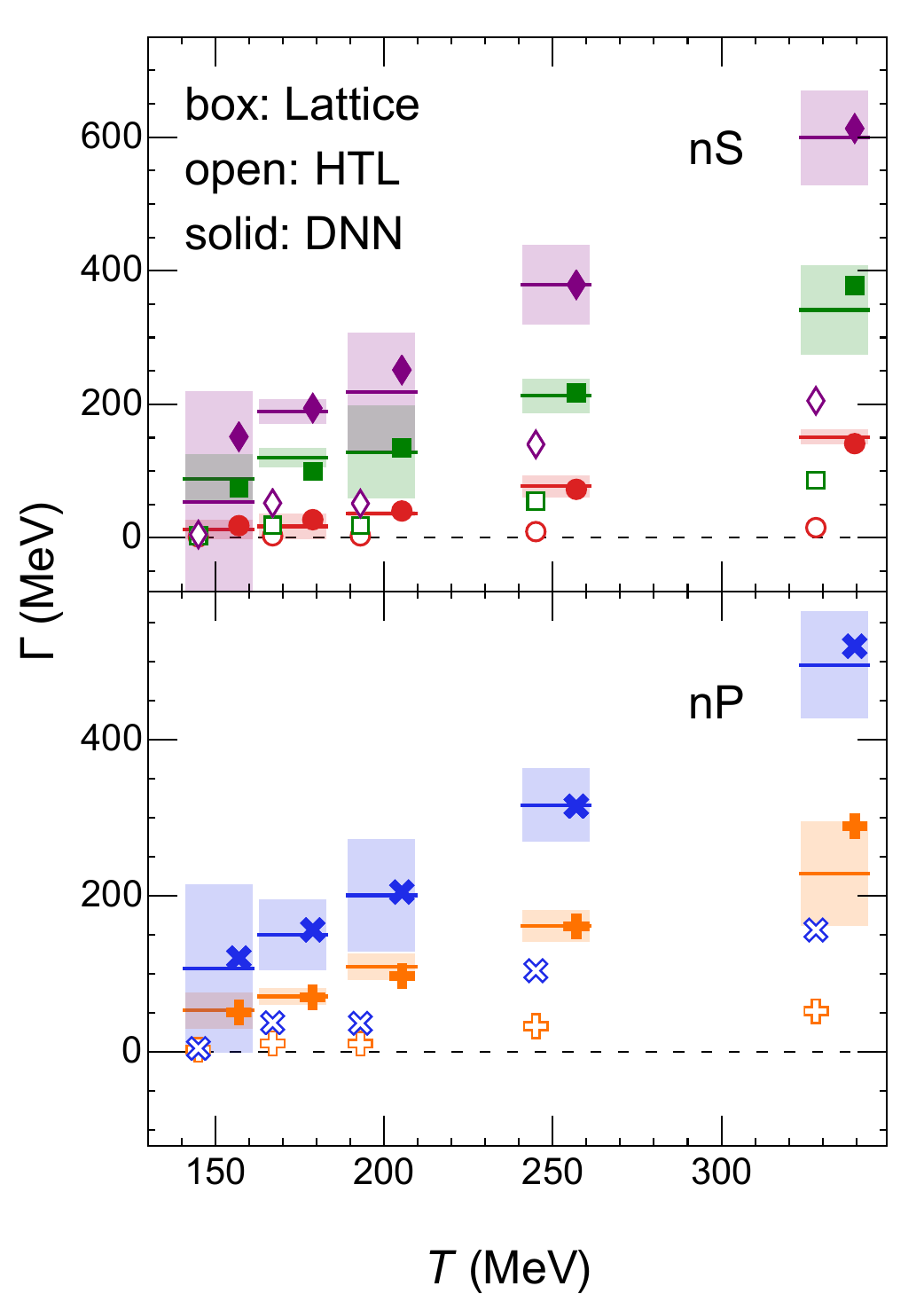}
    \caption{(Left)Real (blue) and imaginary (red) part of interaction potentials versus temperature and quark-antiquark distance extracted via DNNs.
    (Middle and Right) In-medium mass shifts with respect to the vacuum mass (middle) and the thermal widths (right) of different bottomonium states obtained from fits to lattice QCD results of Ref.~\cite{Larsen:2019zqv} (lines and shaded bands) using weak-coupling motivated functional forms~\cite{Lafferty:2019jpr} (open symbols) and DNN based optimization (solid symbols). The points are shifted horizontally for better visualization. $\Upsilon(1S)$, $\chi_{b_0}(1P)$, $\Upsilon(2S)$, $\chi_{b_0}(2P)$ and $\Upsilon(3S)$ states are represented by red circles, orange pluses, green squares, blue crosses and purple diamonds, respectively. Figures from Ref.~\cite{Shi:2021qri} with permission.}
    \label{fig:5:inverse:schroedinger}
\end{figure}
%%%%%%%%%%%%%%%%%%%%%%%%%%%%%%%%%%%%%%%%%%%%%%%%%%%%%%%%%%%%%%%%%%%%%%%%%%%%%%%%%%%%

\emph{\textbf{Inverse Schr\"odinger equation}}---
Another example of solving inverse problem with physics-driven learning is the reconstruction of finite-temperature interaction potential from the lattice QCD simulation of Bottomonia masses and thermal widths in the Quark-Gluon Plasma(QGP). 
The QGP is a new state of matter formed in high energy nuclear collisions and is composed of color-deconfined quarks and gluons, which is in contrast to the low-temperature phase where quarks and gluons are confined within hadrons. The suppression of heavy quarkonium production rates, which are the bound states of a heavy quark and its anti-quark, is evidence of QGP formation in high energy nuclear collisions~\cite{Matsui:1986dk}. Due to the very large mass and small velocity of heavy quarks, one is allowed to employ the Schr\"odinger equation with an effective finite-temperature potential and study the properties of bottomonium in medium~\cite{Satz:2005hx,Guo:2012hx},
\begin{equation}
-\frac{\nabla^2}{2m_\mu} \psi_n(r) + V(r) \psi_n(r) = E_n \psi_n(r) \,,
\label{eq:5:schroedinger}
\end{equation}
where the reduced mass is half of the $b$-quark mass, $m_\mu=m_b/2$, $\psi_n$ the relative wavefuntion, and $E_n$ the energy. At finite temperature, the interaction potential becomes complex~\cite{Laine:2006ns},  $V(T,r) = V_R(T,r) + i \cdot V_I(T,r)$, where the imaginary part emerges due to the Landau damping effect and transition between color-singlet and octet and vanishing in the vacuum. Accordingly, the energy eigenvalues are complex, with the real and imaginary parts correspond to the mass, $\mathrm{Re}[E_n] = m-2m_b$, and width, $\mathrm{Im}[E_n] = -\Gamma$, respectively.

In Ref.~\cite{Shi:2021qri}, it was found that the recent lattice QCD calculation of Bottomonia mass and width (Refs.~\cite{Larsen:2019bwy,Larsen:2019zqv,Larsen:2020rjk}) contradicts the weak-coupling motivated functional forms~\cite{Lafferty:2019jpr} (shown as open symbols in Fig.~\ref{fig:5:inverse:schroedinger}). To address this, the authors developed a model-independent approach for extracting the finite-temperature complex potential directly from the lattice QCD calculation. The complex-valued potential, $\mathcal{Q} \equiv {V_R(r,T), V_I(r,T)}$, is the quantity of interest, while the observables, the masses and widths for various bound states at different temperatures, $\mathcal{O}_y \equiv {m_n(T_j), \Gamma_n(T_j)}$, are obtained by solving the eigenvalue problem of a differential equation. This makes the application of general techniques in physics-driven deep learning challenging in computing the parameter gradients in Eq.~\eqref{eq:5:chisq_grad}.

To compute the functional derivative, one needs to know how the complex-valued energy eigenvalues response to an arbitrary perturbation in potential. Since we only consider small perturbations, the changes of energy eigenvalues are given by the Hellmann--Feynmann theorem in quantum mechanics, $\delta E_n = \int |\psi_n(r)|^2 \delta V(r) \mathrm{d}r$. In Ref.~\cite{Shi:2021qri}, the temperature and distance dependence of the real and imaginary potentials are expressed by the DNNs,
\begin{align}
    V_R(r,T) = V_{R,\mathrm{DNN}}(r,T|\boldsymbol{\theta}_R),
\quad
    V_I(r,T) = V_{R,\mathrm{DNN}}(r,T|\boldsymbol{\theta}_I),
\end{align}
with the network parameters being trained according to the parameter gradients~\eqref{eq:5:chisq_grad} with functional derivatives,
\begin{align}
\begin{split}
 \frac{\delta m_n}{\delta V_R(r)} =&\; -\frac{\delta \Gamma_n}{\delta V_I(r)} = |\psi_n(r)|^2 \,, \\
 \frac{\delta m_n}{\delta V_I(r)} =&\; \frac{\delta \Gamma_n}{\delta V_R(r)} = 0\,.
\end{split}
\end{align}

The wavefunctions $\psi_n(r)$ are obtained by solving the eigenvalue problem. The validity of this method has been confirmed through a closure test with known potentials. The authors began by using a given formula for complex-valued potentials and solving the Schr\"odinger equation at six different temperatures to generate a set of pseudo-data, which consists of the mass and width for five eigenstates. This choice of pseudo-data was consistent with the data obtained from lattice QCD calculations~\cite{Larsen:2019zqv}. The set of pseudo-data was then used to train the parameters of a DNN, and the resulting $V_{R,\mathrm{DNN}}$ and $V_{I,\mathrm{DNN}}$ were found to be in consistent with the ground-truth. The validated method is then applied to the masses and thermal widths computed from lattice QCD calculation. With the optimized parameter sets for $\boldsymbol{\theta}_R$ and $\boldsymbol{\theta}_I$, the real and imaginary potentials are shown in Fig.~\ref{fig:5:inverse:schroedinger} (left), and the corresponding mass shifts and thermal widths are shown by the solid symbols in Fig.~\ref{fig:5:inverse:schroedinger} (left and middle).

\emph{\textbf{Tackling ill-posed inverse problems}} --- In practice, when solving ill-posed inverse problems, one must introduce additional rules to better control the input function and eliminate degeneracy. This is demonstrated through the example of reconstructing spectral functions, which cannot be directly computed in non-perturbative Monte Carlo calculations such as lattice QCD. Instead, they must be inferred from limited sets of correlation data~\cite{Asakawa:2000tr}. The K"allen--Lehmann (KL) correlation functions are among the commonly studied observables,
\begin{align}
D(k) = \;& 
    \int_0^\infty \frac{1}{\pi}
    \frac{\omega \,\mathrm{d}\omega}{\omega^2 + k^2} 
    \rho(\omega),
    \label{eq:5:corr_D}
\end{align}
which is a linear transformation mapping a continuous real function to another continuous real function, and the arguments of both the input and output functions are defined in the real axis. Ref.~\cite{Shi:2022yqw} analytically solved the eigenvalue problem and found the eigenvalues to be $1/(2\cosh\frac{\pi s}{2})$. Here, $s \in \mathbb{R}$ is a continuous parameter that labels the eigenstates. The eigenvalues can be arbitrarily close to zero for large $s$, and the eigenvalues of the inverse operation, which are the inverses of those of the KL convolution, can be arbitrarily large. In the inverse operation, a small noise in the realistic numerical calculation of $D$ will be magnified to a large deviation in $\rho$ as shown in Fig.~\ref{fig:spectral_samples} of Sec.~\ref{subsubsec:realt}. Therefore, inverse KL convolution is ill-posed.

\begin{figure}[!hbt]
    \centering
    \includegraphics[width=0.85\textwidth]{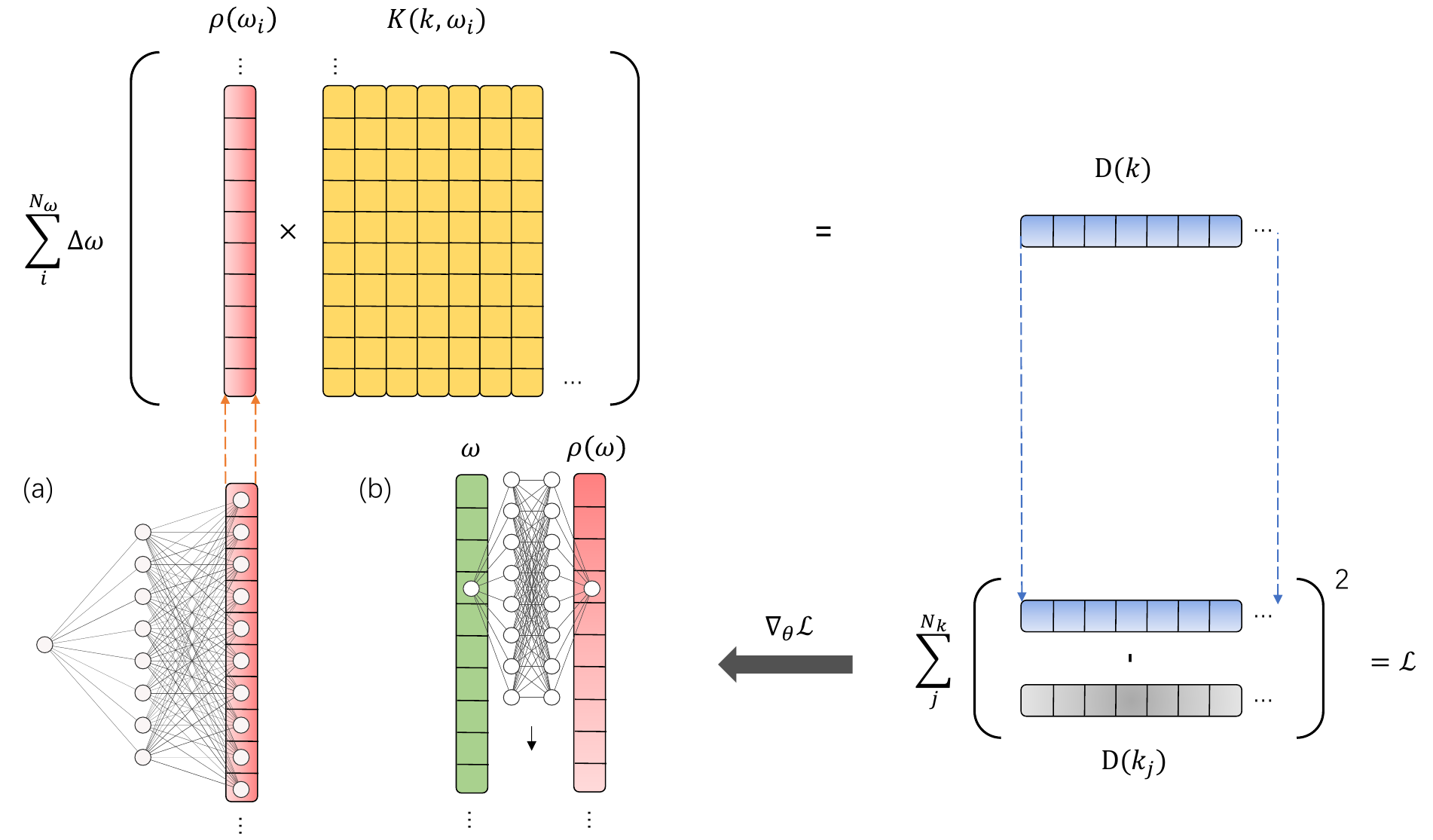}
    \caption{Automatic differential framework to reconstruct spectral from observations. (a) \texttt{NN}. Neural networks have outputs as a list representation of spectrum $\rho_i(\omega_i)$. (b) \texttt{NN-P2P}. Neural networks have input and output nodes as $(\omega_i,\rho_i)$ pairwise.}
    \label{fig:5:inverse:ad}
\end{figure}

Common methods for reconstructing spectral functions include the Tikhonov (TK) regulator~\cite{Tikhonov1943OnTS,tikhonov1995numerical}, the Maximum Entropy Method (MEM) that employs the Shannon--Jaynes entropy~\cite{Narayan:1986wj,Jarrell:1996rrw}, and the Bayesian Reconstruction (BR) method~\cite{Burnier:2013nla}. In references~\cite{Wang:2021jou,Wang:2021cqw,Shi:2022yqw}, the spectral function $\rho(\omega)$ is either formulated as a one-dimensional input and one-dimensional output deep neural network (DNN) that approximates the function (\texttt{NN-P2P} in Figure~\ref{fig:5:inverse:ad}), or as a unity-input and $N_\omega$-dimensional output DNN that represents the value of $\rho$ at $N_\omega$ points in $\omega$ (\texttt{NN} in Figure~\ref{fig:5:inverse:ad}). It has been shown~\cite{Shi:2022yqw} that both representations provide non-local regulators and lead to a unique solution of $\rho(\omega)$. To optimize the parameters of network representations $\{\boldsymbol{\theta}\}$ with loss function, the authors implemented gradient-based algorithms. It derives as,
\begin{align}
    \nabla_{\boldsymbol{\theta}} \mathcal{L} =
    \sum_{j,i}
    K(k_j,\omega_i)
    \frac{\partial \mathcal{L}}{\partial D(k_j)}
    \nabla_{\boldsymbol{\theta}} \rho_i,
\end{align}
where $K(k,\omega) = \frac{1}{\pi}\frac{\omega}{\omega^2 + k^2}$, and $\nabla_{\boldsymbol{\theta}} \rho_i$ is computed by the standard back-propagation  method The reconstruction error is propagated to parameters of the neural network, combined with gradients derived from automatic differentiation.

%%%%%%%%%%%%%%%%%%%%%%%%%%%%%%%%%%%%%%%%%%%%%%%%%%%%%%%%%%%%%%%%%%%%%%%%%%%%%%%%%%%
%%%%%%%%%%%%%%%%%%%%%%%%%%%%%%%%%%%%%%%%%%%%%%%%%%%%%%%%%%%%%%%%%%%%%%%%%%%%%%%%%%%

As a numerical demonstration, the spectral function formed with two Breit--Wigner peaks is choesen as the ground truth,
\begin{align}\label{eq.breit_wigner}
    \rho(\omega) = \sum_{n=1}^{2} \frac{4 A_n \Gamma_n \omega}{\left(M_n^{2}+\Gamma_n^{2}-\omega^{2}\right)^{2}+4 \Gamma_n^{2} \omega^{2}},
\end{align}
with $A_1= 0.8$, $A_2 = 1.0$, $\Gamma_1 = \Gamma_2 = 0.5$~GeV, $M_1= 2.0$~GeV, $M_2 =5.0$~GeV, and compute the corresponding KL correlation functions $D(k_i)$ at $k_i = i\times\Delta k$, with $i=1,2,\cdots,100$, and  $\Delta k=0.2$~GeV. To investigate the effects of noise in a realistic situation, mock data were prepared with random noise on the correlation function, i.e. $\mathrm{D}_i^\text{noisy} = D(k_i) + n_{i}$. See Refs.~\cite{Asakawa:2000tr,Shi:2022yqw} for the detailed setup. Then, for points $\omega_a = a\times \Delta\omega$, with $a=1, 2, \cdots, 500$, and $\Delta\omega=0.04$~GeV, the spectral functions $\rho(\omega)$ were reconstructed. 

%%%%%%%%%%%%%%%%%%%%%%%%%%%%%%%%%%%%%%%%%%%%%%%%%%%%%%%%%%%%%%%%%%%%%%%%%%%%
\begin{figure}[!hbtp]\centering
\includegraphics[width=0.32\textwidth]{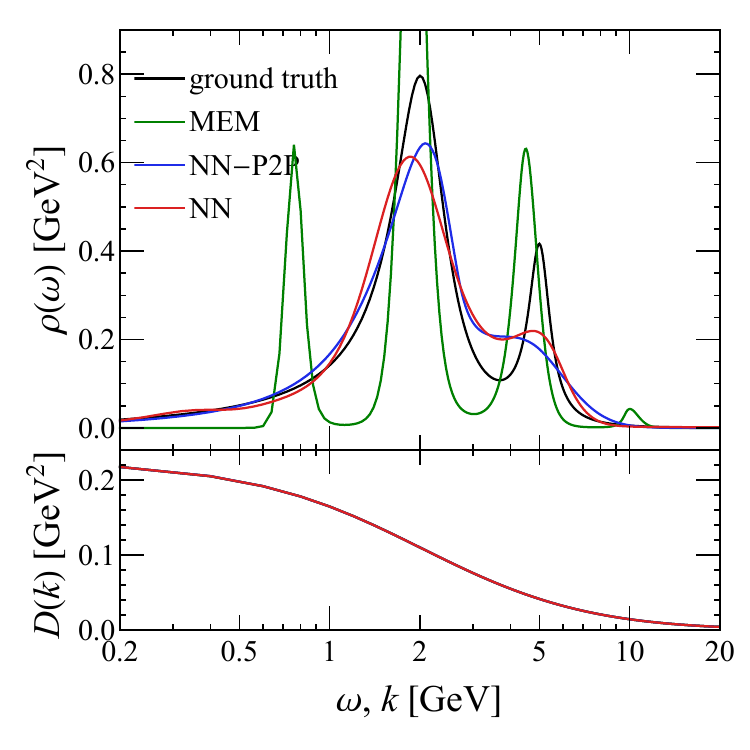}
\includegraphics[width=0.32\textwidth]{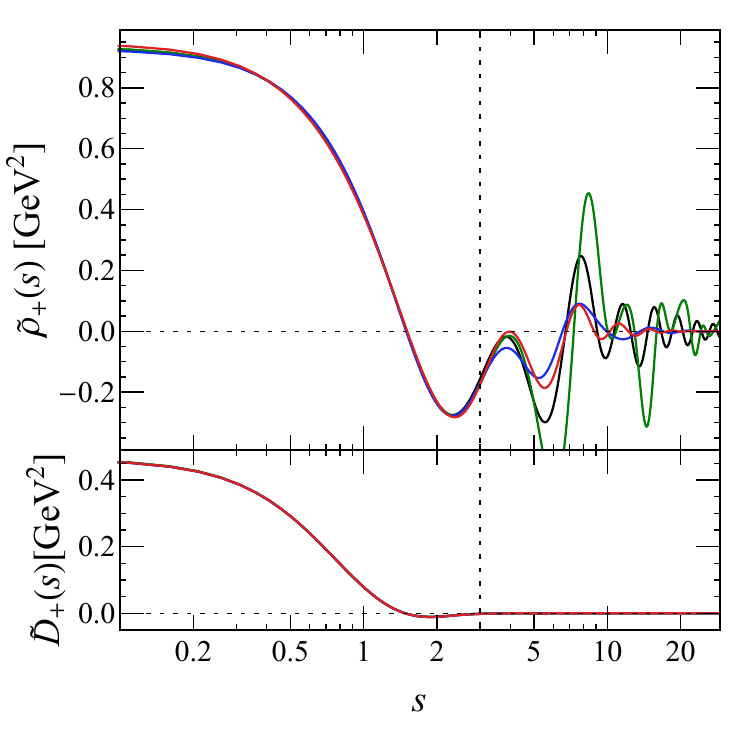}
\includegraphics[width=0.32\textwidth]{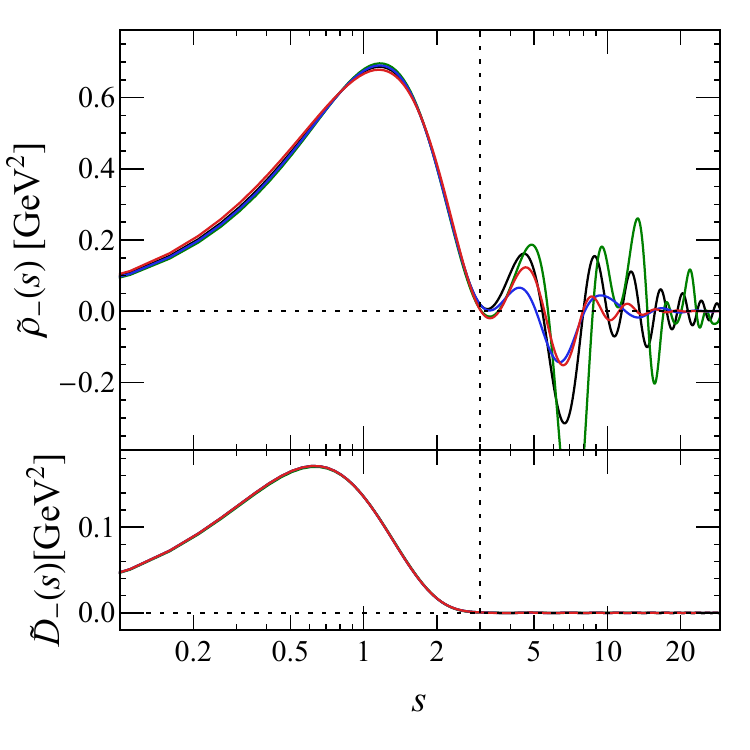}
\caption{Spectral functions using different reconstruction methods (upper panels) and their corresponding KL correlation functions (lower panels) in the generalized coordinate space (left) and generalized momentum space (middle and right). 
Black curves are for the ground truth using Breit--Wigner spectral functions. Numerically reconstructed functions using \texttt{NN}, \texttt{NN-P2P}, and MEM using $N_\text{basis}=100$ basis are represented by red, blue, and green curves, respectively. Figures reproduced from Ref.~\cite{Wang:2021cqw, Shi:2022yqw} with permission.\label{fig:5:inverse:comp}}
\end{figure}
%%%%%%%%%%%%%%%%%%%%%%%%%%%%%%%%%%%%%%%%%%%%%%%%%%%%%%%%%%%%%%%%%%%%%%%%%%%%

In Figure~\ref{fig:5:inverse:comp}, the spectral functions and their respective correlation functions in both the generalized coordinate and momentum spaces are displayed for three methods: \texttt{NN} and \texttt{NN-P2P} representations, and the widely used Maximum Entropy Method (MEM). To ensure the stability of the MEM results, the number of basis functions, $N_\text{basis}=N_k$, has been set to 100. The \texttt{NN} architecture includes three hidden layers with a width of 64, with the input layer consisting of a single constant node set to unity. The output layer contains $N_\omega$ nodes. For the \texttt{NN-P2P} architecture, the input and hidden layers are unchanged, but the output layer has only one node. All activation functions before the output have been selected as \texttt{ELU}. Although the behavior of $\rho(\omega)$ may vary significantly in the generalized coordinate space, the generalized momentum $\widetilde{\rho}_\pm(s)$ exhibits similar behavior for values of $s$ less than or equal to three, as indicated by the vertical dashed lines in Fig.~\ref{fig:5:inverse:comp}. This highlights the fundamental challenge in the reconstruction problem. Regardless of the reconstruction method used, one can always accurately recover the low-frequency modes of $\widetilde{\rho}$, but the high-frequency modes are prone to being polluted by noise or numerical inaccuracies in the correlation functions, making them nearly impossible to achieve. On the other hand, all $D$ and $\widetilde{D}$ values are nearly equal, which is ensured through the $\chi^2$-fitting.

Thus, it should be noted that all these methods are not guaranteed to provide the correct inversion function when $D$ is of finite precision. Meanwhile, the ``biases'' introduced in the supervised training data set become helpful in solving ill-posed inverse problems if unwanted biases are carefully avoided. A natural way to implement the physical regulators can be provided by Approach (i). This is demonstrated in studies that employ DNNs to learn the inverse mapping, such as~\cite{Kades:2019wtd, 2018PhRvB..98x5101Y, 2020PhRvL.124e6401F, PhysRevLett.124.056401, Chen:2021giw}. In these works, an ensemble of $\rho$'s that follow the physical properties is generated and the corresponding $D$'s are computed. DNNs are then trained to represent the inverse mapping from $D$ to $\rho$. The trained networks become automatically regularized inverse functions with embedded prior knowledge, but one should be careful about the risk of introducing unwanted biases in the training data.

\subsection{Summary}
In this section, we reviewed recent advancements in embedding more physical prior knowledge into machine learning methods to enhance the efficiency and applicability of physics exploration. We illustrated these physics-inspired, informed, and driven deep learning approaches with concrete examples of lattice field configuration sampling, inverse renormalization group transformation, Feynman path generation, and the network realization of AdS-CFT calculation. Additionally, we presented complex inverse problems such as nuclear equation of state reconstruction from neutron star masses and radii, interaction potential extraction from the Schr\"odinger equation energy spectrum, and spectral function reconstruction from Euclidean correlators. The examples presented in this section demonstrated significant advantages of using these advanced developments in ML/DL for solving physics problems. We hope that these examples will inspire further work in the near future.
    \newpage
    \section{Summary, Limitations, and Outlook}\label{sec:sum}

\emph{\textbf{Summary}} --- 
The wave of \textit{AI for science} has witnessed a marked surge in popularity and adoption across the broader scientific spectrum~\cite{osti_1604756}. Particularly, paradigms with modern machine- and deep-learning techniques are transforming across various scientific domains -- primarily those overwhelmed by large-scale intensive computations and/or large amounts of high-dimensional data, including high-energy nuclear physics (HENP)~\cite{Boehnlein:2021eym, He:2023zin, Calafiura:2022ges}. The integration of machine learning techniques has led to remarkable advances and a host of results in the field, opening a new horizon for exploration and discovery. These ML-based methods have revolutionized the way of analyzing data, improved the ability to discover new phenomena, and fostered the development of more efficient simulations. HENP is an extremely fruitful area in this sense, and many advances have been made in the past decade~\cite{Bzdak:2019pkr, Fukushima:2020yzx, Bogdanov:2022faf}. We are in the right and exciting era to work in this direction and to amplify our understanding of QCD matter in extreme conditions. 

This review aims to provide an overview of the current applications of ML in HENP theoretical studies, while spotlighting some of the recent developments in this rapidly evolving crossing field. Several aspects are reviewed, revolving around ML for the theoretical study of extreme QCD matter exploration, ranging from data analysis in high energy heavy-ion collisions(HICs) sector, advancing lattice QCD/QFT simulations, to the inference of Neutron Star(NS) interior matter properties, thereby encapsulating the ongoing endeavors in the study of strongly interacting nuclear matter. Subsequently, a refined advanced development summary and discussion are presented in the last chapter, trying to form a common ground from a methodology perspective in inspiring further exploration.

In terms of big \textbf{data analysis}, HENP unequivocally presents a golden playground to us. Copious amounts of multifarious data can be gleaned from a variety of sources including HIC experiment detectors, astrophysical observatories, and lattice QCD simulations. Many of the well-established physics models and software/packages further facilitate the generation of large-scale simulation data, to which the disentangled physics understanding and correlation analysis are yet daunting with conventional methods. Thus, as a modern numerical method to process complex data for hidden pattern decoding, ML and DL techniques have provided a powerful tool for exploring physics across different disciplines. 

Beyond being data-rich, numerous areas in HENP are notably \textbf{computation-intensive}, with many of the accessible measurements such as billions of collision events from the detectors requiring comprehension derived from theory simulations, yet remain not fully calculable in a first principle manner. For the purpose of either evolving or confirming our theory understandings, or exploring new physics for discoveries, the ability to perform efficient and prompt simulations is critical across many facets of HENP. ML and DL have also charted significant advancements in HENP computations, e.g., various algorithms have been developed to enhance the simulation speed and accuracy~\cite{Shanahan:2022ifi}. Advanced calculations with faster and optimized models in synergy with ML methodologies pave the way for better prediction and comprehension in confronting data.

\emph{\textbf{Limitations}} --- 
In addition to those head-on applications of ML/DL in HENP, there are still many challenges and questions that need to be further addressed. Probably the first-line concern from physicists is the \textbf{interpretability} of the ML approaches utilized in HENP research~\cite{Neubauer:2022zpn}. Detailed understandings of why and under what conditions the used methods work are desired in physics. More efforts need to be devoted to uncovering the inner workings of ML algorithms to rectify their “Black Box” characteristics, especially those with huge amounts of parameters. Techniques to reveal the patterns and validate the computations with ML are called for as well. Then per the purpose of physics exploration, those practical computational results from the ML methods need to be transferred or related to physics knowledge or inspirations, which better conform to the well-established language of physics in a controlled way, such as with uncertainties properly given. The incorporation of physics priors into those ML paradigms to specific physics studies deserves further development in addressing this concern~\cite{2021NatRP...3..422K}.

Another challenge is how to reliably \textbf{connect experimental measurements and physical theory} using ML/DL techniques, with the associated \textbf{uncertainties} also properly evaluated. As emphasized, in HENP, to understand the nature of strongly interacting matter, experiments, and observations from big scientific infrastructures play a crucial role, which needs to be transformed into physics knowledge per their mission. This however is with unprecedented scales and complexities, due to the high dimensionality and greatly correlated data stream from measurement, the multiscale and intricate physics simulations for the dynamical process involved in the measurement. Statistical learning methods such as Bayesian inference~\cite{Cranmer:2019eaq} and alternative ML/DL approaches provide a number of great demonstrations to analyze data for inferring physics knowledge. Principled uncertainty estimation for results inferred from naive ML/DL applications is not yet fully developed. These methods still have great potential to be unleashed in the exploration of QCD matter.

Moreover, the size of the training dataset imposes additional constraints. On one hand, high-energy nuclear collision experiments usually generate billions($\sim 10^9$) of events. Processing such a massive dataset necessitates a strong demand for high computational efficiency to avert excessive training time. On the other hand, typical training requires an adequate amount of well-labeled training data. The nature of the optimization procedure makes it challenging to apply it to rare physical processes with a limited number of events or phenomena that cannot be labeled.

Finally, as discussed in various examples throughout this review, the specific physics problem can impose further limitations in terms of machine learning techniques. Here are a few examples: one cannot address an ill-defined problem without introducing prior knowledge, whether using conventional or ML methods; the properties learned by the network are often tailored to the training dataset, rendering it unreliable to generalize to datasets governed by other different physics models; how ML methods can help when the physical model used\footnote{used e.g., in generating the training dataset, or used directly inside the ML methodology.} is missing some unknown but essential physics component in confronting particular phenomenon?

\emph{\textbf{Outlooks}} ---
As future prospects, with ML/DL assisted, important physics properties and hopefully new physics phenomena are expected to be explored from the growing measurements accumulation in heavy ion collision experiments or astrophysics observations, to inform also physics discovery. Specifically, ever \textbf{faster and more efficient analysis} to large amounts of data could be anticipated to better bridge experiment to physics theory; \textbf{smart, controllable and rapid simulation} with ML methods such as generative modeling may largely reduce the demand in computational resource for the field, and also mitigate the enormous complexities in disentangling different physics involved in the simulation; in the course of neutron star, it'd be intriguing to explore the possible potential of ML methods in identifying e.g., presence of exotic phases in NS interiors and phase transitions for dense QCD matter from observations, and also develop more reliable ways in \textbf{combining evidence from multi-source data} such as those from HICs together with those from multi-messenger astrophysics. Furthermore, in the lattice QCD sector, while efforts to \textbf{improve the efficiency and accuracy of lattice sampling and simulations} will continue, it may also be possible to enhance our understanding of QCD in terms of effective theory construction from an ML algorithmic perspective. The dynamic interplay between HENP and ML is rapidly advancing, and with the continued progress and enthusiasm in ML, we can expect even more exciting developments and remarkable achievements in the near future.

As introduced in this review, there has been a growing intersection between generative models and nuclear physics. This convergence is anticipated to catalyze further explorations employing advanced deep learning tools, such as Large Language Models(LLMs), in nuclear physics research. A prime representative example would be the Generative Pre-trained Transformer (GPT) models~\cite{openai2023gpt4}.

GPT Models have demonstrated immense promise across various domains due to their ability to generate coherent and contextually pertinent responses to queries. However, their potential extends beyond mere question answering. As the integration of generative models within nuclear physics intensifies, they are poised to assume pivotal roles in a multitude of facets.
\begin{itemize}
    \item Integration of GPT into traditional workflows:
    
    \begin{itemize}
    \item Aiding Experimental Design and Simulations: GPT models can be used to refine experimental designs and simulations by leveraging existing domain knowledge.
    \item Enhancing Existing Software Packages: GPT models can be integrated into existing software packages to enhance their performance. For instance, they can be used to identify and rectify inefficiencies or errors in current codes, leading to more accurate and efficient software solutions.
    \item Data Analysis. GPT models can propose different methods to analyze large datasets, extracting meaningful insights and patterns that might be missed by traditional analysis methods.
    \item Easing Framework Transition Between Different Programming Languages. GPT models can assist in translating code from one programming language to another, facilitating the transition between different frameworks.
    \end{itemize}
    
    \item Symbolic computation. Language models are well-suited for symbolic computations, as mathematical logic embodies a learnable, language-like set of rules. Besides direct application in formula derivations, GPT models can be used to compute specific physics processes following Feynman rules~\cite{Alnuqaydan_2023}. Recent developments highlight the potential of GPT models in proving mathematical theorems~\cite{yang2023leandojo}. Consequently, they may play a crucial role in advancing our understanding of fundamental physical laws.

    \item Education. After fine-tuning GPT models based on the nuclear physics domain knowledge, GPT models can serve as educational tools, providing hands-on instructions for applying machine learning in the exploration of QCD matter.
\end{itemize}

	\newpage
	\section*{Acknowledgements}  
	We express our sincere gratitude to all collaborators who have made invaluable contributions to our research projects. \\   
	This work was supported by the Germany BMBF funded KISS consortium (05D23RI1) in the ErUM-Data action plan (KZ), the AI grant of SAMSON AG, Frankfurt (LW and KZ), Xidian-FIAS International Joint Research Center (LW), Huawei Technologies Co., Ltd and the National Natural Science Foundation of China under contract No. 12075098 (LP), Tsinghua University under grant No. 53330500923 (SS), and U.S. Department of Energy, Office of Science, Office of Nuclear Physics, grant No. DE-FG88ER40388 (SS).
	
    \bibliography{reviews}

\begin{thebibliography}{100}
\expandafter\ifx\csname url\endcsname\relax
  \def\url#1{\texttt{#1}}\fi
\expandafter\ifx\csname urlprefix\endcsname\relax\def\urlprefix{URL }\fi
\expandafter\ifx\csname href\endcsname\relax
  \def\href#1#2{#2} \def\path#1{#1}\fi

\bibitem{Gross:2022hyw}
F.~Gross, et~al., {50 Years of Quantum Chromodynamics} (12 2022).
\newblock \href {http://arxiv.org/abs/2212.11107} {\path{arXiv:2212.11107}}.

\bibitem{Yagi:2005yb}
K.~Yagi, T.~Hatsuda, Y.~Miake, {Quark-gluon plasma: From big bang to little bang}, Vol.~23, 2005.

\bibitem{Wang:2016opj}
X.-N. Wang (Ed.), {Quark-Gluon Plasma 5}, World Scientific, New Jersey, 2016.
\newblock \href {https://doi.org/10.1142/9533} {\path{doi:10.1142/9533}}.

\bibitem{Fukushima:2020yzx}
K.~Fukushima, B.~Mohanty, N.~Xu, {Little-Bang and Femto-Nova in Nucleus-Nucleus Collisions}, AAPPS Bull. 31 (2021) 1.
\newblock \href {http://arxiv.org/abs/2009.03006} {\path{arXiv:2009.03006}}, \href {https://doi.org/10.1007/s43673-021-00002-7} {\path{doi:10.1007/s43673-021-00002-7}}.

\bibitem{Shuryak:2004pry}
E.~V. Shuryak, {The QCD vacuum, hadrons and the superdense matter}, Vol.~71, 2004.
\newblock \href {https://doi.org/10.1142/5367} {\path{doi:10.1142/5367}}.

\bibitem{Rafelski:2019twp}
J.~Rafelski, {Discovery of Quark-Gluon-Plasma: Strangeness Diaries}, Eur. Phys. J. ST 229~(1) (2020) 1--140.
\newblock \href {http://arxiv.org/abs/1911.00831} {\path{arXiv:1911.00831}}, \href {https://doi.org/10.1140/epjst/e2019-900263-x} {\path{doi:10.1140/epjst/e2019-900263-x}}.

\bibitem{Luo:2022mtp}
X.~Luo, Q.~Wang, N.~Xu, P.~Zhuang (Eds.), {Properties of QCD Matter at High Baryon Density}, Springer, 2022.
\newblock \href {https://doi.org/10.1007/978-981-19-4441-3} {\path{doi:10.1007/978-981-19-4441-3}}.

\bibitem{Ding:2015ona}
H.-T. Ding, F.~Karsch, S.~Mukherjee, {Thermodynamics of strong-interaction matter from Lattice QCD}, Int. J. Mod. Phys. E 24~(10) (2015) 1530007.
\newblock \href {http://arxiv.org/abs/1504.05274} {\path{arXiv:1504.05274}}, \href {https://doi.org/10.1142/S0218301315300076} {\path{doi:10.1142/S0218301315300076}}.

\bibitem{Karsch:2022opd}
F.~Karsch, {Lattice QCD at non-zero temperature and density} (12 2022).
\newblock \href {http://arxiv.org/abs/2212.03015} {\path{arXiv:2212.03015}}.

\bibitem{Aarts:2023vsf}
G.~Aarts, et~al., {Phase Transitions in Particle Physics}: {Results and Perspectives from Lattice Quantum Chromo-Dynamics}, Prog. Part. Nucl. Phys. 133 (2023) 104070.
\newblock \href {http://arxiv.org/abs/2301.04382} {\path{arXiv:2301.04382}}, \href {https://doi.org/10.1016/j.ppnp.2023.104070} {\path{doi:10.1016/j.ppnp.2023.104070}}.

\bibitem{Nagata:2021ugx}
K.~Nagata, {Finite-density lattice QCD and sign problem: Current status and open problems}, Prog. Part. Nucl. Phys. 127 (2022) 103991.
\newblock \href {http://arxiv.org/abs/2108.12423} {\path{arXiv:2108.12423}}, \href {https://doi.org/10.1016/j.ppnp.2022.103991} {\path{doi:10.1016/j.ppnp.2022.103991}}.

\bibitem{Busza:2018rrf}
W.~Busza, K.~Rajagopal, W.~van~der Schee, {Heavy Ion Collisions: The Big Picture, and the Big Questions}, Ann. Rev. Nucl. Part. Sci. 68 (2018) 339--376.
\newblock \href {http://arxiv.org/abs/1802.04801} {\path{arXiv:1802.04801}}, \href {https://doi.org/10.1146/annurev-nucl-101917-020852} {\path{doi:10.1146/annurev-nucl-101917-020852}}.

\bibitem{Bzdak:2019pkr}
A.~Bzdak, S.~Esumi, V.~Koch, J.~Liao, M.~Stephanov, N.~Xu, {Mapping the Phases of Quantum Chromodynamics with Beam Energy Scan}, Phys. Rept. 853 (2020) 1--87.
\newblock \href {http://arxiv.org/abs/1906.00936} {\path{arXiv:1906.00936}}, \href {https://doi.org/10.1016/j.physrep.2020.01.005} {\path{doi:10.1016/j.physrep.2020.01.005}}.

\bibitem{Hogg:1987nn}
T.~Hogg, B.~A. Huberman, {ARTIFICIAL INTELLIGENCE AND LARGE SCALE COMPUTATION: A PHYSICS PERSPECTIVE}, Phys. Rept. 156 (1987) 227.
\newblock \href {https://doi.org/10.1016/0370-1573(87)90096-2} {\path{doi:10.1016/0370-1573(87)90096-2}}.

\bibitem{AIreview:2021}
Y.~K. Dwivedi, L.~Hughes, E.~Ismagilova, G.~Aarts, C.~Coombs, T.~Crick, Y.~Duan, R.~Dwivedi, J.~Edwards, A.~Eirug, V.~Galanos, P.~V. Ilavarasan, M.~Janssen, P.~Jones, A.~K. Kar, H.~Kizgin, B.~Kronemann, B.~Lal, B.~Lucini, R.~Medaglia, K.~{Le Meunier-FitzHugh}, L.~C. {Le Meunier-FitzHugh}, S.~Misra, E.~Mogaji, S.~K. Sharma, J.~B. Singh, V.~Raghavan, R.~Raman, N.~P. Rana, S.~Samothrakis, J.~Spencer, K.~Tamilmani, A.~Tubadji, P.~Walton, M.~D. Williams, \href{https://www.sciencedirect.com/science/article/pii/S026840121930917X}{Artificial intelligence (ai): Multidisciplinary perspectives on emerging challenges, opportunities, and agenda for research, practice and policy}, International Journal of Information Management 57 (2021) 101994.
\newblock \href {https://doi.org/https://doi.org/10.1016/j.ijinfomgt.2019.08.002} {\path{doi:https://doi.org/10.1016/j.ijinfomgt.2019.08.002}}.
\newline\urlprefix\url{https://www.sciencedirect.com/science/article/pii/S026840121930917X}

\bibitem{lecun2015deep}
Y.~LeCun, Y.~Bengio, G.~Hinton, Deep learning, nature 521~(7553) (2015) 436--444.

\bibitem{sarker2021deep}
I.~H. Sarker, Deep learning: a comprehensive overview on techniques, taxonomy, applications and research directions, SN Computer Science 2~(6) (2021) 420.

\bibitem{osti_1604756}
R.~Stevens, V.~Taylor, J.~Nichols, A.~B. Maccabe, K.~Yelick, D.~Brown, \href{https://www.osti.gov/biblio/1604756}{Ai for science: Report on the department of energy (doe) town halls on artificial intelligence (ai) for science} (2 2020).
\newblock \href {https://doi.org/10.2172/1604756} {\path{doi:10.2172/1604756}}.
\newline\urlprefix\url{https://www.osti.gov/biblio/1604756}

\bibitem{PhysRevLett.120.145301}
T.~Xie, J.~C. Grossman, Crystal graph convolutional neural networks for an accurate and interpretable prediction of material properties, Phys. Rev. Lett. 120 (2018) 145301.
\newblock \href {https://doi.org/10.1103/PhysRevLett.120.145301} {\path{doi:10.1103/PhysRevLett.120.145301}}.

\bibitem{SUK2014569}
H.-I. Suk, S.-W. Lee, D.~Shen, Hierarchical feature representation and multimodal fusion with deep learning for ad/mci diagnosis, NeuroImage 101 (2014) 569--582.
\newblock \href {https://doi.org/https://doi.org/10.1016/j.neuroimage.2014.06.077} {\path{doi:https://doi.org/10.1016/j.neuroimage.2014.06.077}}.

\bibitem{2016arXiv160407316B}
M.~{Bojarski}, D.~{Del Testa}, D.~{Dworakowski}, B.~{Firner}, B.~{Flepp}, P.~{Goyal}, L.~D. {Jackel}, M.~{Monfort}, U.~{Muller}, J.~{Zhang}, X.~{Zhang}, J.~{Zhao}, K.~{Zieba}, {End to End Learning for Self-Driving Cars}, arXiv e-prints (2016) arXiv:1604.07316\href {https://doi.org/10.48550/arXiv.1604.07316} {\path{doi:10.48550/arXiv.1604.07316}}.

\bibitem{SHA2022110897}
Y.~Sha, J.~Faber, S.~Gou, B.~Liu, W.~Li, S.~Schramm, H.~Stoecker, T.~Steckenreiter, D.~Vnucec, N.~Wetzstein, A.~Widl, K.~Zhou, An acoustic signal cavitation detection framework based on xgboost with adaptive selection feature engineering, Measurement 192 (2022) 110897.
\newblock \href {https://doi.org/https://doi.org/10.1016/j.measurement.2022.110897} {\path{doi:https://doi.org/10.1016/j.measurement.2022.110897}}.

\bibitem{SHA2022104904}
Y.~Sha, J.~Faber, S.~Gou, B.~Liu, W.~Li, S.~Schramm, H.~Stoecker, T.~Steckenreiter, D.~Vnucec, N.~Wetzstein, A.~Widl, K.~Zhou, A multi-task learning for cavitation detection and cavitation intensity recognition of valve acoustic signals, Engineering Applications of Artificial Intelligence 113 (2022) 104904.
\newblock \href {https://doi.org/https://doi.org/10.1016/j.engappai.2022.104904} {\path{doi:https://doi.org/10.1016/j.engappai.2022.104904}}.

\bibitem{10.1029/2022JB024595}
M.~Chakraborty, D.~Fenner, W.~Li, J.~Faber, K.~Zhou, G.~Rümpker, H.~Stoecker, N.~Srivastava, Creime—a convolutional recurrent model for earthquake identification and magnitude estimation, Journal of Geophysical Research: Solid Earth 127~(7)  e2022JB024595.
\newblock \href {https://doi.org/https://doi.org/10.1029/2022JB024595} {\path{doi:https://doi.org/10.1029/2022JB024595}}.

\bibitem{10.3389/feart.2022.953007}
W.~Li, M.~Chakraborty, D.~Fenner, J.~Faber, K.~Zhou, G.~Rümpker, H.~Stöcker, N.~Srivastava, Epick: Attention-based multi-scale unet for earthquake detection and seismic phase picking, Frontiers in Earth Science 10 (2022).
\newblock \href {https://doi.org/10.3389/feart.2022.953007} {\path{doi:10.3389/feart.2022.953007}}.

\bibitem{LITJENS201760}
G.~Litjens, T.~Kooi, B.~E. Bejnordi, A.~A.~A. Setio, F.~Ciompi, M.~Ghafoorian, J.~A. {van der Laak}, B.~{van Ginneken}, C.~I. Sánchez, A survey on deep learning in medical image analysis, Medical Image Analysis 42 (2017) 60--88.
\newblock \href {https://doi.org/https://doi.org/10.1016/j.media.2017.07.005} {\path{doi:https://doi.org/10.1016/j.media.2017.07.005}}.

\bibitem{article_DL_genomics}
G.~Eraslan, Z.~Avsec, J.~Gagneur, F.~Theis, Deep learning: new computational modelling techniques for genomics, Nature Reviews Genetics 20 (2019) 1.
\newblock \href {https://doi.org/10.1038/s41576-019-0122-6} {\path{doi:10.1038/s41576-019-0122-6}}.

\bibitem{article_dl_earth}
M.~Reichstein, G.~Camps-Valls, B.~Stevens, M.~Jung, J.~Denzler, N.~Carvalhais, M.~Prabhat, Deep learning and process understanding for data-driven earth system science, Nature 566 (2019) 195.
\newblock \href {https://doi.org/10.1038/s41586-019-0912-1} {\path{doi:10.1038/s41586-019-0912-1}}.

\bibitem{KAMILARIS201870}
A.~Kamilaris, F.~X. Prenafeta-Boldú, Deep learning in agriculture: A survey, Computers and Electronics in Agriculture 147 (2018) 70--90.
\newblock \href {https://doi.org/https://doi.org/10.1016/j.compag.2018.02.016} {\path{doi:https://doi.org/10.1016/j.compag.2018.02.016}}.

\bibitem{George:2016hay}
D.~George, E.~A. Huerta, {Deep Neural Networks to Enable Real-time Multimessenger Astrophysics}, Phys. Rev. D 97~(4) (2018) 044039.
\newblock \href {http://arxiv.org/abs/1701.00008} {\path{arXiv:1701.00008}}, \href {https://doi.org/10.1103/PhysRevD.97.044039} {\path{doi:10.1103/PhysRevD.97.044039}}.

\bibitem{Soma:2023rmq}
S.~Soma, H.~St\"ocker, K.~Zhou, {Mass and tidal parameter extraction from gravitational waves of binary neutron stars mergers using deep learning} (6 2023).
\newblock \href {http://arxiv.org/abs/2306.17488} {\path{arXiv:2306.17488}}.

\bibitem{Wang:2023yul}
L.~Wang, B.~M. Hare, K.~Zhou, H.~St\"ocker, O.~Scholten, {Identifying lightning structures via machine learning}, Chaos Solitons and Fractals: the interdisciplinary journal of Nonlinear Science and Nonequilibrium and Complex Phenomena 170 (2023) 113346.
\newblock \href {https://doi.org/10.1016/j.chaos.2023.113346} {\path{doi:10.1016/j.chaos.2023.113346}}.

\bibitem{Wang_2021}
L.~Wang, T.~Xu, T.~Stoecker, H.~Stoecker, Y.~Jiang, K.~Zhou, \href{https://dx.doi.org/10.1088/2632-2153/ac0314}{Machine learning spatio-temporal epidemiological model to evaluate germany-county-level covid-19 risk}, Machine Learning: Science and Technology 2~(3) (2021) 035031.
\newblock \href {https://doi.org/10.1088/2632-2153/ac0314} {\path{doi:10.1088/2632-2153/ac0314}}.
\newline\urlprefix\url{https://dx.doi.org/10.1088/2632-2153/ac0314}

\bibitem{Jumper2021}
J.~Jumper, R.~Evans, A.~Pritzel, T.~Green, M.~Figurnov, O.~Ronneberger, K.~Tunyasuvunakool, R.~Bates, A.~{\v{Z}}{\'{\i}}dek, A.~Potapenko, A.~Bridgland, C.~Meyer, S.~A.~A. Kohl, A.~J. Ballard, A.~Cowie, B.~Romera-Paredes, S.~Nikolov, R.~Jain, J.~Adler, T.~Back, S.~Petersen, D.~Reiman, E.~Clancy, M.~Zielinski, M.~Steinegger, M.~Pacholska, T.~Berghammer, S.~Bodenstein, D.~Silver, O.~Vinyals, A.~W. Senior, K.~Kavukcuoglu, P.~Kohli, D.~Hassabis, \href{https://doi.org/10.1038/s41586-021-03819-2}{Highly accurate protein structure prediction with {AlphaFold}}, Nature 596~(7873) (2021) 583--589.
\newblock \href {https://doi.org/10.1038/s41586-021-03819-2} {\path{doi:10.1038/s41586-021-03819-2}}.
\newline\urlprefix\url{https://doi.org/10.1038/s41586-021-03819-2}

\bibitem{Degrave2022}
J.~Degrave, F.~Felici, J.~Buchli, M.~Neunert, B.~Tracey, F.~Carpanese, T.~Ewalds, R.~Hafner, A.~Abdolmaleki, D.~de~las Casas, C.~Donner, L.~Fritz, C.~Galperti, A.~Huber, J.~Keeling, M.~Tsimpoukelli, J.~Kay, A.~Merle, J.-M. Moret, S.~Noury, F.~Pesamosca, D.~Pfau, O.~Sauter, C.~Sommariva, S.~Coda, B.~Duval, A.~Fasoli, P.~Kohli, K.~Kavukcuoglu, D.~Hassabis, M.~Riedmiller, \href{https://doi.org/10.1038/s41586-021-04301-9}{Magnetic control of tokamak plasmas through deep reinforcement learning}, Nature 602~(7897) (2022) 414--419.
\newblock \href {https://doi.org/10.1038/s41586-021-04301-9} {\path{doi:10.1038/s41586-021-04301-9}}.
\newline\urlprefix\url{https://doi.org/10.1038/s41586-021-04301-9}

\bibitem{Han2018}
J.~Han, L.~Zhang, R.~Car, W.~E, \href{https://doi.org/10.4208/cicp.oa-2017-0213}{Deep potential: A general representation of a many-body potential energy surface}, Communications in Computational Physics 23~(3) (2018).
\newblock \href {https://doi.org/10.4208/cicp.oa-2017-0213} {\path{doi:10.4208/cicp.oa-2017-0213}}.
\newline\urlprefix\url{https://doi.org/10.4208/cicp.oa-2017-0213}

\bibitem{Mehta:2018dln}
P.~Mehta, M.~Bukov, C.-H. Wang, A.~G.~R. Day, C.~Richardson, C.~K. Fisher, D.~J. Schwab, {A high-bias, low-variance introduction to Machine Learning for physicists}, Phys. Rept. 810 (2019) 1--124.
\newblock \href {http://arxiv.org/abs/1803.08823} {\path{arXiv:1803.08823}}, \href {https://doi.org/10.1016/j.physrep.2019.03.001} {\path{doi:10.1016/j.physrep.2019.03.001}}.

\bibitem{Carleo:2019ptp}
G.~Carleo, I.~Cirac, K.~Cranmer, L.~Daudet, M.~Schuld, N.~Tishby, L.~Vogt-Maranto, L.~Zdeborov\'a, {Machine learning and the physical sciences}, Rev. Mod. Phys. 91~(4) (2019) 045002.
\newblock \href {http://arxiv.org/abs/1903.10563} {\path{arXiv:1903.10563}}, \href {https://doi.org/10.1103/RevModPhys.91.045002} {\path{doi:10.1103/RevModPhys.91.045002}}.

\bibitem{Mezard:2009ipa}
M.~Mézard, A.~Montanari, \href{https://doi.org/10.1093/acprof:oso/9780198570837.001.0001}{{Information, Physics, and Computation}}, Oxford University Press, 2009.
\newblock \href {https://doi.org/10.1093/acprof:oso/9780198570837.001.0001} {\path{doi:10.1093/acprof:oso/9780198570837.001.0001}}.
\newline\urlprefix\url{https://doi.org/10.1093/acprof:oso/9780198570837.001.0001}

\bibitem{Bahri:2020smd}
Y.~Bahri, J.~Kadmon, J.~Pennington, S.~S. Schoenholz, J.~Sohl-Dickstein, S.~Ganguli, \href{https://doi.org/10.1146/annurev-conmatphys-031119-050745}{Statistical mechanics of deep learning}, Annual Review of Condensed Matter Physics 11~(1) (2020) 501--528.
\newblock \href {https://doi.org/10.1146/annurev-conmatphys-031119-050745} {\path{doi:10.1146/annurev-conmatphys-031119-050745}}.
\newline\urlprefix\url{https://doi.org/10.1146/annurev-conmatphys-031119-050745}

\bibitem{Bedaque:2021bja}
P.~Bedaque, et~al., {A.I. for nuclear physics}, Eur. Phys. J. A 57~(3) (2021) 100.
\newblock \href {https://doi.org/10.1140/epja/s10050-020-00290-x} {\path{doi:10.1140/epja/s10050-020-00290-x}}.

\bibitem{Boehnlein:2021eym}
A.~Boehnlein, et~al., {Colloquium: Machine learning in nuclear physics}, Rev. Mod. Phys. 94~(3) (2022) 031003.
\newblock \href {http://arxiv.org/abs/2112.02309} {\path{arXiv:2112.02309}}, \href {https://doi.org/10.1103/RevModPhys.94.031003} {\path{doi:10.1103/RevModPhys.94.031003}}.

\bibitem{Humpert:1990rw}
B.~Humpert, {On the use of neural networks in high-energy physics experiments}, Comput. Phys. Commun. 56 (1990) 299--311.
\newblock \href {https://doi.org/10.1016/0010-4655(90)90016-T} {\path{doi:10.1016/0010-4655(90)90016-T}}.

\bibitem{Gyulassy:1990et}
M.~Gyulassy, M.~Harlander, {Elastic tracking and neural network algorithms for complex pattern recognition}, Comput. Phys. Commun. 66 (1991) 31--46.
\newblock \href {https://doi.org/10.1016/0010-4655(91)90005-6} {\path{doi:10.1016/0010-4655(91)90005-6}}.

\bibitem{Bass:1993vx}
S.~A. Bass, A.~Bischoff, C.~Hartnack, J.~A. Maruhn, J.~Reinhardt, H.~Stoecker, W.~Greiner, {Neural networks for impact parameter determination}, J. Phys. G 20 (1994) L21--L26.
\newblock \href {https://doi.org/10.1088/0954-3899/20/1/004} {\path{doi:10.1088/0954-3899/20/1/004}}.

\bibitem{Bass:1996ez}
S.~A. Bass, A.~Bischoff, J.~A. Maruhn, H.~Stoecker, W.~Greiner, {Neural networks for impact parameter determination}, Phys. Rev. C 53 (1996) 2358--2363.
\newblock \href {http://arxiv.org/abs/nucl-th/9601024} {\path{arXiv:nucl-th/9601024}}, \href {https://doi.org/10.1103/PhysRevC.53.2358} {\path{doi:10.1103/PhysRevC.53.2358}}.

\bibitem{Feickert:2021ajf}
M.~Feickert, B.~Nachman, {A Living Review of Machine Learning for Particle Physics} (2 2021).
\newblock \href {http://arxiv.org/abs/2102.02770} {\path{arXiv:2102.02770}}.

\bibitem{Benato:2021olt}
L.~Benato, et~al., {Shared Data and Algorithms for Deep Learning in Fundamental Physics}, Comput. Softw. Big Sci. 6~(1) (2022) 9.
\newblock \href {http://arxiv.org/abs/2107.00656} {\path{arXiv:2107.00656}}, \href {https://doi.org/10.1007/s41781-022-00082-6} {\path{doi:10.1007/s41781-022-00082-6}}.

\bibitem{Bishop2006}
C.~M. Bishop, Pattern Recognition and Machine Learning, Springer Verlag, Berlin, 2006.

\bibitem{Goodfellow2016}
I.~Goodfellow, Y.~Bengio, A.~Courville, Deep Learning, The MIT Press, Cambridge, Massachusetts, 2016.

\bibitem{Dawid:2022fga}
A.~Dawid, et~al., {Modern applications of machine learning in quantum sciences}, 2022.
\newblock \href {http://arxiv.org/abs/2204.04198} {\path{arXiv:2204.04198}}.

\bibitem{Shanahan:2022ifi}
P.~Shanahan, et~al., {Snowmass 2021 Computational Frontier CompF03 Topical Group Report: Machine Learning} (9 2022).
\newblock \href {http://arxiv.org/abs/2209.07559} {\path{arXiv:2209.07559}}.

\bibitem{Murphy2012}
K.~P. Murphy, Machine Learning: A Probabilistic Perspective, The MIT Press, Cambdridge, Massachusetts, 2012.

\bibitem{garnett_bayesoptbook_2023}
R.~Garnett, {Bayesian Optimization}, Cambridge University Press, 2023, to appear.

\bibitem{garnett_bayesoptbook_2023_update}
R.~Garnett, {Bayesian Optimization}, Cambridge University Press, 2023.

\bibitem{NIPS1995_7cce53cf}
C.~Williams, C.~Rasmussen, \href{https://proceedings.neurips.cc/paper/1995/file/7cce53cf90577442771720a370c3c723-Paper.pdf}{Gaussian processes for regression}, in: D.~Touretzky, M.~C. Mozer, M.~Hasselmo (Eds.), Adv. Neural Inf. Process. Syst., Vol.~8, MIT Press, 1996.
\newline\urlprefix\url{https://proceedings.neurips.cc/paper/1995/file/7cce53cf90577442771720a370c3c723-Paper.pdf}

\bibitem{Genton:2002ker}
M.~G. Genton, Classes of kernels for machine learning: A statistics perspective 2 (2002) 299–312.

\bibitem{schmidhuber2015deep}
J.~Schmidhuber, Deep learning in neural networks: An overview, Neural networks 61 (2015) 85--117.

\bibitem{mcculloch1943logical}
W.~S. McCulloch, W.~Pitts, A logical calculus of the ideas immanent in nervous activity, The bulletin of mathematical biophysics 5 (1943) 115--133.

\bibitem{cybenko1989approximation}
G.~Cybenko, Approximation by superpositions of a sigmoidal function, Mathematics of control, signals and systems 2~(4) (1989) 303--314.

\bibitem{hornik1991approximation}
K.~Hornik, Approximation capabilities of multilayer feedforward networks, Neural networks 4~(2) (1991) 251--257.

\bibitem{Lecun1998}
Y.~Lecun, L.~Bottou, Y.~Bengio, P.~Haffner, Gradient-based learning applied to document recognition, Proceedings of the IEEE 86~(11) (1998) 2278--2324.
\newblock \href {https://doi.org/10.1109/5.726791} {\path{doi:10.1109/5.726791}}.

\bibitem{baydin2018automatic}
A.~G. Baydin, B.~A. Pearlmutter, A.~A. Radul, J.~M. Siskind, Automatic differentiation in machine learning: a survey, Journal of Marchine Learning Research 18 (2018) 1--43.

\bibitem{AbaBar16Tensorflow}
M.~Abadi, P.~Barham, J.~Chen, Z.~Chen, A.~Davis, J.~Dean, M.~Devin, S.~Ghemawat, G.~Irving, M.~Isard, et~al., Tensorflow: A system for large-scale machine learning., in: OSDI, Vol.~16, 2016, pp. 265--283.

\bibitem{NEURIPS2019_9015}
A.~Paszke, S.~Gross, F.~Massa, A.~Lerer, J.~Bradbury, G.~Chanan, T.~Killeen, Z.~Lin, N.~Gimelshein, L.~Antiga, A.~Desmaison, A.~K\"{o}pf, E.~Yang, Z.~DeVito, M.~Raison, A.~Tejani, S.~Chilamkurthy, B.~Steiner, L.~Fang, J.~Bai, S.~Chintala, Pytorch: An imperative style, high-performance deep learning library, in: Proceedings of the 33rd International Conference on Neural Information Processing Systems, Curran Associates Inc., Red Hook, NY, USA, 2019.

\bibitem{2019arXiv191204232S}
S.~S. {Schoenholz}, E.~D. {Cubuk}, {JAX, M.D.: A Framework for Differentiable Physics}, arXiv e-prints (2019) arXiv:1912.04232\href {http://arxiv.org/abs/1912.04232} {\path{arXiv:1912.04232}}, \href {https://doi.org/10.48550/arXiv.1912.04232} {\path{doi:10.48550/arXiv.1912.04232}}.

\bibitem{Bottou1999}
L.~Bottou, On-Line Learning and Stochastic Approximations, Cambridge University Press, USA, 1999.

\bibitem{Adam2015}
D.~P. Kingma, J.~Ba, \href{http://arxiv.org/abs/1412.6980}{Adam: {A} method for stochastic optimization}, in: Y.~Bengio, Y.~LeCun (Eds.), 3rd International Conference on Learning Representations, {ICLR} 2015, San Diego, CA, USA, May 7-9, 2015, Conference Track Proceedings, 2015.
\newline\urlprefix\url{http://arxiv.org/abs/1412.6980}

\bibitem{tschannen2018recent}
M.~Tschannen, O.~Bachem, M.~Lucic, Recent advances in autoencoder-based representation learning, arXiv preprint arXiv:1812.05069 (2018).

\bibitem{Iten:2020dpc}
R.~Iten, T.~Metger, H.~Wilming, L.~del Rio, R.~Renner, \href{https://link.aps.org/doi/10.1103/PhysRevLett.124.010508}{Discovering physical concepts with neural networks}, Phys. Rev. Lett. 124 (2020) 010508.
\newblock \href {https://doi.org/10.1103/PhysRevLett.124.010508} {\path{doi:10.1103/PhysRevLett.124.010508}}.
\newline\urlprefix\url{https://link.aps.org/doi/10.1103/PhysRevLett.124.010508}

\bibitem{ladjal2019pca}
S.~Ladjal, A.~Newson, C.-H. Pham, A pca-like autoencoder, arXiv preprint arXiv:1904.01277 (2019).

\bibitem{Wang2018GenerativeMF}
L.~Wang, \href{"https://wangleiphy.github.io/lectures/GenerativeModels_PIL.pdf"}{Generative models for physicists}, 2018.
\newline\urlprefix\url{"https://wangleiphy.github.io/lectures/GenerativeModels_PIL.pdf"}

\bibitem{2013arXiv1312.6114K}
D.~P. {Kingma}, M.~{Welling}, {Auto-Encoding Variational Bayes}, arXiv e-prints (2013) arXiv:1312.6114\href {http://arxiv.org/abs/1312.6114} {\path{arXiv:1312.6114}}, \href {https://doi.org/10.48550/arXiv.1312.6114} {\path{doi:10.48550/arXiv.1312.6114}}.

\bibitem{2014arXiv1406.2661G}
I.~J. {Goodfellow}, J.~{Pouget-Abadie}, M.~{Mirza}, B.~{Xu}, D.~{Warde-Farley}, S.~{Ozair}, A.~{Courville}, Y.~{Bengio}, {Generative Adversarial Networks}, arXiv e-prints (2014) arXiv:1406.2661\href {http://arxiv.org/abs/1406.2661} {\path{arXiv:1406.2661}}, \href {https://doi.org/10.48550/arXiv.1406.2661} {\path{doi:10.48550/arXiv.1406.2661}}.

\bibitem{2016arXiv161009585O}
A.~{Odena}, C.~{Olah}, J.~{Shlens}, {Conditional Image Synthesis With Auxiliary Classifier GANs}, arXiv e-prints (2016) arXiv:1610.09585\href {http://arxiv.org/abs/1610.09585} {\path{arXiv:1610.09585}}, \href {https://doi.org/10.48550/arXiv.1610.09585} {\path{doi:10.48550/arXiv.1610.09585}}.

\bibitem{2017arXiv170107875A}
M.~{Arjovsky}, S.~{Chintala}, L.~{Bottou}, {Wasserstein GAN}, arXiv e-prints (2017) arXiv:1701.07875\href {http://arxiv.org/abs/1701.07875} {\path{arXiv:1701.07875}}, \href {https://doi.org/10.48550/arXiv.1701.07875} {\path{doi:10.48550/arXiv.1701.07875}}.

\bibitem{2017arXiv170400028G}
I.~{Gulrajani}, F.~{Ahmed}, M.~{Arjovsky}, V.~{Dumoulin}, A.~{Courville}, {Improved Training of Wasserstein GANs}, arXiv e-prints (2017) arXiv:1704.00028\href {http://arxiv.org/abs/1704.00028} {\path{arXiv:1704.00028}}, \href {https://doi.org/10.48550/arXiv.1704.00028} {\path{doi:10.48550/arXiv.1704.00028}}.

\bibitem{2015arXiv150203509G}
M.~{Germain}, K.~{Gregor}, I.~{Murray}, H.~{Larochelle}, {MADE: Masked Autoencoder for Distribution Estimation}, arXiv e-prints (2015) arXiv:1502.03509\href {http://arxiv.org/abs/1502.03509} {\path{arXiv:1502.03509}}, \href {https://doi.org/10.48550/arXiv.1502.03509} {\path{doi:10.48550/arXiv.1502.03509}}.

\bibitem{2016arXiv160106759V}
A.~{van den Oord}, N.~{Kalchbrenner}, K.~{Kavukcuoglu}, {Pixel Recurrent Neural Networks}, arXiv e-prints (2016) arXiv:1601.06759\href {http://arxiv.org/abs/1601.06759} {\path{arXiv:1601.06759}}, \href {https://doi.org/10.48550/arXiv.1601.06759} {\path{doi:10.48550/arXiv.1601.06759}}.

\bibitem{2016arXiv160605328V}
A.~{van den Oord}, N.~{Kalchbrenner}, O.~{Vinyals}, L.~{Espeholt}, A.~{Graves}, K.~{Kavukcuoglu}, {Conditional Image Generation with PixelCNN Decoders}, arXiv e-prints (2016) arXiv:1606.05328\href {http://arxiv.org/abs/1606.05328} {\path{arXiv:1606.05328}}, \href {https://doi.org/10.48550/arXiv.1606.05328} {\path{doi:10.48550/arXiv.1606.05328}}.

\bibitem{2017arXiv170105517S}
T.~{Salimans}, A.~{Karpathy}, X.~{Chen}, D.~P. {Kingma}, {PixelCNN++: Improving the PixelCNN with Discretized Logistic Mixture Likelihood and Other Modifications}, arXiv e-prints (2017) arXiv:1701.05517\href {http://arxiv.org/abs/1701.05517} {\path{arXiv:1701.05517}}, \href {https://doi.org/10.48550/arXiv.1701.05517} {\path{doi:10.48550/arXiv.1701.05517}}.

\bibitem{2015arXiv150505770J}
D.~{Jimenez Rezende}, S.~{Mohamed}, {Variational Inference with Normalizing Flows}, arXiv e-prints (2015) arXiv:1505.05770\href {http://arxiv.org/abs/1505.05770} {\path{arXiv:1505.05770}}, \href {https://doi.org/10.48550/arXiv.1505.05770} {\path{doi:10.48550/arXiv.1505.05770}}.

\bibitem{2019arXiv190809257K}
I.~{Kobyzev}, S.~J.~D. {Prince}, M.~A. {Brubaker}, {Normalizing Flows: An Introduction and Review of Current Methods}, arXiv e-prints (2019) arXiv:1908.09257\href {http://arxiv.org/abs/1908.09257} {\path{arXiv:1908.09257}}, \href {https://doi.org/10.48550/arXiv.1908.09257} {\path{doi:10.48550/arXiv.1908.09257}}.

\bibitem{2014arXiv1410.8516D}
L.~{Dinh}, D.~{Krueger}, Y.~{Bengio}, {NICE: Non-linear Independent Components Estimation}, arXiv e-prints (2014) arXiv:1410.8516\href {http://arxiv.org/abs/1410.8516} {\path{arXiv:1410.8516}}, \href {https://doi.org/10.48550/arXiv.1410.8516} {\path{doi:10.48550/arXiv.1410.8516}}.

\bibitem{2016arXiv160508803D}
L.~{Dinh}, J.~{Sohl-Dickstein}, S.~{Bengio}, {Density estimation using Real NVP}, arXiv e-prints (2016) arXiv:1605.08803\href {http://arxiv.org/abs/1605.08803} {\path{arXiv:1605.08803}}, \href {https://doi.org/10.48550/arXiv.1605.08803} {\path{doi:10.48550/arXiv.1605.08803}}.

\bibitem{Ahmad:2020kdd}
M.~A. Ahmad, c.~\"{O}z\"{o}nder, \href{https://doi.org/10.1145/3394486.3406464}{Physics inspired models in artificial intelligence}, in: Proceedings of the 26th ACM SIGKDD International Conference on Knowledge Discovery and Data Mining, KDD '20, Association for Computing Machinery, New York, NY, USA, 2020, p. 3535–3536.
\newblock \href {https://doi.org/10.1145/3394486.3406464} {\path{doi:10.1145/3394486.3406464}}.
\newline\urlprefix\url{https://doi.org/10.1145/3394486.3406464}

\bibitem{sompolinsky1988statistical}
H.~Sompolinsky, et~al., Statistical mechanics of neural networks, Physics Today 41~(21) (1988) 70--80.

\bibitem{mezard2009information}
M.~Mezard, A.~Montanari, Information, physics, and computation, Oxford University Press, 2009.

\bibitem{2017Sci...355..602C}
G.~{Carleo}, M.~{Troyer}, {Solving the quantum many-body problem with artificial neural networks}, Science 355~(6325) (2017) 602--606.
\newblock \href {http://arxiv.org/abs/1606.02318} {\path{arXiv:1606.02318}}, \href {https://doi.org/10.1126/science.aag2302} {\path{doi:10.1126/science.aag2302}}.

\bibitem{Carrasquilla:2020mas}
J.~Carrasquilla, {Machine Learning for Quantum Matter}, Adv. Phys. X 5~(1) (2020) 1797528.
\newblock \href {http://arxiv.org/abs/2003.11040} {\path{arXiv:2003.11040}}, \href {https://doi.org/10.1080/23746149.2020.1797528} {\path{doi:10.1080/23746149.2020.1797528}}.

\bibitem{Jia:2021qaa}
Z.-A. Jia, B.~Yi, R.~Zhai, Y.-C. Wu, G.-C. Guo, G.-P. Guo, Quantum neural network states: A brief review of methods and applications, Advanced Quantum Technologies 2~(7-8) (2019) 1800077.
\newblock \href {https://doi.org/https://doi.org/10.1002/qute.201800077} {\path{doi:https://doi.org/10.1002/qute.201800077}}.

\bibitem{ackley1985learning}
D.~H. Ackley, G.~E. Hinton, T.~J. Sejnowski, A learning algorithm for boltzmann machines, Cognitive science 9~(1) (1985) 147--169.

\bibitem{fischer2012introduction}
A.~Fischer, C.~Igel, An introduction to restricted boltzmann machines, in: Iberoamerican congress on pattern recognition, Springer, 2012, pp. 14--36.

\bibitem{Bronstein:2016thv}
M.~M. Bronstein, J.~Bruna, Y.~LeCun, A.~Szlam, P.~Vandergheynst, {Geometric Deep Learning: Going beyond Euclidean data}, IEEE Sig. Proc. Mag. 34~(4) (2017) 18--42.
\newblock \href {http://arxiv.org/abs/1611.08097} {\path{arXiv:1611.08097}}, \href {https://doi.org/10.1109/MSP.2017.2693418} {\path{doi:10.1109/MSP.2017.2693418}}.

\bibitem{Bronstein:2021mdi}
M.~M. Bronstein, J.~Bruna, T.~Cohen, P.~Veli\v{c}kovi\'c, {Geometric Deep Learning: Grids, Groups, Graphs, Geodesics, and Gauges} (4 2021).
\newblock \href {http://arxiv.org/abs/2104.13478} {\path{arXiv:2104.13478}}.

\bibitem{Yang2022DiffusionMA}
L.~Yang, Z.~Zhang, Y.~Song, S.~Hong, R.~Xu, Y.~Zhao, Y.~Shao, W.~Zhang, B.~Cui, M.-H. Yang, Diffusion models: A comprehensive survey of methods and applications, arXiv preprint arXiv:2209.00796 (2022).

\bibitem{ho:2020denoising}
J.~Ho, A.~Jain, P.~Abbeel, Denoising diffusion probabilistic models, in: Proc. 34th {{Int}}. {{Conf}}. {{Neural Inf}}. {{Process}}. {{Syst}}., {{NIPS}}'20, {Curran Associates Inc.}, pp. 6840--6851.

\bibitem{song2021scorebased}
Y.~Song, J.~Sohl-Dickstein, D.~P. Kingma, A.~Kumar, S.~Ermon, B.~Poole, \href{https://openreview.net/forum?id=PxTIG12RRHS}{Score-based generative modeling through stochastic differential equations}, in: Int. {{Conf}}. {{Learn}}. {{Represent}}.
\newline\urlprefix\url{https://openreview.net/forum?id=PxTIG12RRHS}

\bibitem{Mattheakis:2019tyi}
M.~Mattheakis, P.~Protopapas, D.~Sondak, M.~Di~Giovanni, E.~Kaxiras, {Physical Symmetries Embedded in Neural Networks} (4 2019).
\newblock \href {http://arxiv.org/abs/1904.08991} {\path{arXiv:1904.08991}}.

\bibitem{Kicki:2021so}
P.~Kicki, P.~Skrzypczyński, M.~Ozay, A new approach to design symmetry invariant neural networks, in: 2021 International Joint Conference on Neural Networks (IJCNN), 2021, pp. 1--8.
\newblock \href {https://doi.org/10.1109/IJCNN52387.2021.9533541} {\path{doi:10.1109/IJCNN52387.2021.9533541}}.

\bibitem{zhang1988shift}
W.~Zhang, J.~Tanida, K.~Itoh, Y.~Ichioka, Shift-invariant pattern recognition neural network and its optical architecture, in: Proceedings of annual conference of the Japan Society of Applied Physics, 1988, pp. 2147--2151.

\bibitem{pmlr-v48-cohenc16}
T.~Cohen, M.~Welling, Group equivariant convolutional networks, in: M.~F. Balcan, K.~Q. Weinberger (Eds.), Proceedings of The 33rd International Conference on Machine Learning, Vol.~48 of Proceedings of Machine Learning Research, PMLR, New York, New York, USA, 2016, pp. 2990--2999.

\bibitem{qi2016pointnet}
C.~R. Qi, H.~Su, K.~Mo, L.~J. Guibas, Pointnet: Deep learning on point sets for 3d classification and segmentation, arXiv preprint arXiv:1612.00593 (2016).

\bibitem{e3nn_paper}
M.~Geiger, T.~Smidt, \href{https://arxiv.org/abs/2207.09453}{e3nn: Euclidean neural networks} (2022).
\newblock \href {https://doi.org/10.48550/ARXIV.2207.09453} {\path{doi:10.48550/ARXIV.2207.09453}}.
\newline\urlprefix\url{https://arxiv.org/abs/2207.09453}

\bibitem{Albergo:2019eim}
M.~S. Albergo, G.~Kanwar, P.~E. Shanahan, {Flow-based generative models for Markov chain Monte Carlo in lattice field theory}, Phys. Rev. D 100~(3) (2019) 034515.
\newblock \href {http://arxiv.org/abs/1904.12072} {\path{arXiv:1904.12072}}, \href {https://doi.org/10.1103/PhysRevD.100.034515} {\path{doi:10.1103/PhysRevD.100.034515}}.

\bibitem{Kanwar:2020xzo}
G.~Kanwar, M.~S. Albergo, D.~Boyda, K.~Cranmer, D.~C. Hackett, S.~Racani\`ere, D.~J. Rezende, P.~E. Shanahan, {Equivariant flow-based sampling for lattice gauge theory}, Phys. Rev. Lett. 125~(12) (2020) 121601.
\newblock \href {http://arxiv.org/abs/2003.06413} {\path{arXiv:2003.06413}}, \href {https://doi.org/10.1103/PhysRevLett.125.121601} {\path{doi:10.1103/PhysRevLett.125.121601}}.

\bibitem{Albergo:2021vyo}
M.~S. Albergo, D.~Boyda, D.~C. Hackett, G.~Kanwar, K.~Cranmer, S.~Racani\`ere, D.~J. Rezende, P.~E. Shanahan, {Introduction to Normalizing Flows for Lattice Field Theory} (1 2021).
\newblock \href {http://arxiv.org/abs/2101.08176} {\path{arXiv:2101.08176}}.

\bibitem{Raissi:2017zsi}
M.~Raissi, P.~Perdikaris, G.~E. Karniadakis, {Physics Informed Deep Learning (Part I): Data-driven Solutions of Nonlinear Partial Differential Equations}, J. Comput. Phys. 378 (2019) 686--707.
\newblock \href {http://arxiv.org/abs/1711.10561} {\path{arXiv:1711.10561}}, \href {https://doi.org/10.1016/j.jcp.2018.10.045} {\path{doi:10.1016/j.jcp.2018.10.045}}.

\bibitem{2021NatRP...3..422K}
G.~E. {Karniadakis}, I.~G. {Kevrekidis}, L.~{Lu}, P.~{Perdikaris}, S.~{Wang}, L.~{Yang}, {Physics-informed machine learning}, Nature Reviews Physics 3~(6) (2021) 422--440.
\newblock \href {https://doi.org/10.1038/s42254-021-00314-5} {\path{doi:10.1038/s42254-021-00314-5}}.

\bibitem{2021arXiv210905237T}
N.~{Thuerey}, P.~{Holl}, M.~{Mueller}, P.~{Schnell}, F.~{Trost}, K.~{Um}, {Physics-based Deep Learning}, arXiv e-prints (2021) arXiv:2109.05237\href {http://arxiv.org/abs/2109.05237} {\path{arXiv:2109.05237}}.

\bibitem{BRAHMS:2004adc}
I.~Arsene, et~al., {Quark gluon plasma and color glass condensate at RHIC? The Perspective from the BRAHMS experiment}, Nucl. Phys. A 757 (2005) 1--27.
\newblock \href {http://arxiv.org/abs/nucl-ex/0410020} {\path{arXiv:nucl-ex/0410020}}, \href {https://doi.org/10.1016/j.nuclphysa.2005.02.130} {\path{doi:10.1016/j.nuclphysa.2005.02.130}}.

\bibitem{PHENIX:2004vcz}
K.~Adcox, et~al., {Formation of dense partonic matter in relativistic nucleus-nucleus collisions at RHIC: Experimental evaluation by the PHENIX collaboration}, Nucl. Phys. A 757 (2005) 184--283.
\newblock \href {http://arxiv.org/abs/nucl-ex/0410003} {\path{arXiv:nucl-ex/0410003}}, \href {https://doi.org/10.1016/j.nuclphysa.2005.03.086} {\path{doi:10.1016/j.nuclphysa.2005.03.086}}.

\bibitem{PHOBOS:2004zne}
B.~B. Back, et~al., {The PHOBOS perspective on discoveries at RHIC}, Nucl. Phys. A 757 (2005) 28--101.
\newblock \href {http://arxiv.org/abs/nucl-ex/0410022} {\path{arXiv:nucl-ex/0410022}}, \href {https://doi.org/10.1016/j.nuclphysa.2005.03.084} {\path{doi:10.1016/j.nuclphysa.2005.03.084}}.

\bibitem{STAR:2005gfr}
J.~Adams, et~al., {Experimental and theoretical challenges in the search for the quark gluon plasma: The STAR Collaboration's critical assessment of the evidence from RHIC collisions}, Nucl. Phys. A 757 (2005) 102--183.
\newblock \href {http://arxiv.org/abs/nucl-ex/0501009} {\path{arXiv:nucl-ex/0501009}}, \href {https://doi.org/10.1016/j.nuclphysa.2005.03.085} {\path{doi:10.1016/j.nuclphysa.2005.03.085}}.

\bibitem{Schukraft:2011cz}
J.~Schukraft, {ALICE results from the first Pb-Pb run at the CERN LHC}, J. Phys. G 38 (2011) 124003.
\newblock \href {http://arxiv.org/abs/1106.5620} {\path{arXiv:1106.5620}}, \href {https://doi.org/10.1088/0954-3899/38/12/124003} {\path{doi:10.1088/0954-3899/38/12/124003}}.

\bibitem{Steinberg:2011qq}
P.~Steinberg, {Recent Heavy Ion Results with the ATLAS Detector at the LHC}, in: {Meeting of the APS Division of Particles and Fields}, 2011.
\newblock \href {http://arxiv.org/abs/1110.3352} {\path{arXiv:1110.3352}}.

\bibitem{Wyslouch:2011zz}
B.~Wyslouch, {Overview of experimental results in PbPb collisions at $\sqrt{s_{NN}}=2.76$ TeV by the CMS Collaboration}, J. Phys. G 38 (2011) 124005.
\newblock \href {http://arxiv.org/abs/1107.2895} {\path{arXiv:1107.2895}}, \href {https://doi.org/10.1088/0954-3899/38/12/124005} {\path{doi:10.1088/0954-3899/38/12/124005}}.

\bibitem{Baym:2001in}
G.~Baym, {RHIC: From dreams to beams in two decades}, Nucl. Phys. A 698 (2002) XXIII--XXXII.
\newblock \href {http://arxiv.org/abs/hep-ph/0104138} {\path{arXiv:hep-ph/0104138}}, \href {https://doi.org/10.1016/S0375-9474(01)01342-2} {\path{doi:10.1016/S0375-9474(01)01342-2}}.

\bibitem{Muller:2012zq}
B.~Muller, J.~Schukraft, B.~Wyslouch, {First Results from Pb+Pb collisions at the LHC}, Ann. Rev. Nucl. Part. Sci. 62 (2012) 361--386.
\newblock \href {http://arxiv.org/abs/1202.3233} {\path{arXiv:1202.3233}}, \href {https://doi.org/10.1146/annurev-nucl-102711-094910} {\path{doi:10.1146/annurev-nucl-102711-094910}}.

\bibitem{Braun-Munzinger:2015hba}
P.~Braun-Munzinger, V.~Koch, T.~Sch\"afer, J.~Stachel, {Properties of hot and dense matter from relativistic heavy ion collisions}, Phys. Rept. 621 (2016) 76--126.
\newblock \href {http://arxiv.org/abs/1510.00442} {\path{arXiv:1510.00442}}, \href {https://doi.org/10.1016/j.physrep.2015.12.003} {\path{doi:10.1016/j.physrep.2015.12.003}}.

\bibitem{Muller:2006ee}
B.~Muller, J.~L. Nagle, {Results from the relativistic heavy ion collider}, Ann. Rev. Nucl. Part. Sci. 56 (2006) 93--135.
\newblock \href {http://arxiv.org/abs/nucl-th/0602029} {\path{arXiv:nucl-th/0602029}}, \href {https://doi.org/10.1146/annurev.nucl.56.080805.140556} {\path{doi:10.1146/annurev.nucl.56.080805.140556}}.

\bibitem{STAR:2018gyt}
J.~Adam, et~al., {Global polarization of $\Lambda$ hyperons in Au+Au collisions at $\sqrt{s_{_{NN}}}$ = 200 GeV}, Phys. Rev. C 98 (2018) 014910.
\newblock \href {http://arxiv.org/abs/1805.04400} {\path{arXiv:1805.04400}}, \href {https://doi.org/10.1103/PhysRevC.98.014910} {\path{doi:10.1103/PhysRevC.98.014910}}.

\bibitem{Muller:2018ibh}
B.~M\"uller, A.~Sch\"afer, {Chiral magnetic effect and an experimental bound on the late time magnetic field strength}, Phys. Rev. D 98~(7) (2018) 071902.
\newblock \href {http://arxiv.org/abs/1806.10907} {\path{arXiv:1806.10907}}, \href {https://doi.org/10.1103/PhysRevD.98.071902} {\path{doi:10.1103/PhysRevD.98.071902}}.

\bibitem{STAR:2017ckg}
L.~Adamczyk, et~al., {Global $\Lambda$ hyperon polarization in nuclear collisions: evidence for the most vortical fluid}, Nature 548 (2017) 62--65.
\newblock \href {http://arxiv.org/abs/1701.06657} {\path{arXiv:1701.06657}}, \href {https://doi.org/10.1038/nature23004} {\path{doi:10.1038/nature23004}}.

\bibitem{Teaney:2000cw}
D.~Teaney, J.~Lauret, E.~V. Shuryak, {Flow at the SPS and RHIC as a quark gluon plasma signature}, Phys. Rev. Lett. 86 (2001) 4783--4786.
\newblock \href {http://arxiv.org/abs/nucl-th/0011058} {\path{arXiv:nucl-th/0011058}}, \href {https://doi.org/10.1103/PhysRevLett.86.4783} {\path{doi:10.1103/PhysRevLett.86.4783}}.

\bibitem{Rafelski:1982pu}
J.~Rafelski, B.~Muller, {Strangeness Production in the Quark - Gluon Plasma}, Phys. Rev. Lett. 48 (1982) 1066, [Erratum: Phys.Rev.Lett. 56, 2334 (1986)].
\newblock \href {https://doi.org/10.1103/PhysRevLett.48.1066} {\path{doi:10.1103/PhysRevLett.48.1066}}.

\bibitem{Wang:1992qdg}
X.-N. Wang, M.~Gyulassy, {Gluon shadowing and jet quenching in A + A collisions at s**(1/2) = 200-GeV}, Phys. Rev. Lett. 68 (1992) 1480--1483.
\newblock \href {https://doi.org/10.1103/PhysRevLett.68.1480} {\path{doi:10.1103/PhysRevLett.68.1480}}.

\bibitem{Matsui:1986dk}
T.~Matsui, H.~Satz, {$J/\psi$ Suppression by Quark-Gluon Plasma Formation}, Phys. Lett. B 178 (1986) 416--422.
\newblock \href {https://doi.org/10.1016/0370-2693(86)91404-8} {\path{doi:10.1016/0370-2693(86)91404-8}}.

\bibitem{Gupta:2011wh}
S.~Gupta, X.~Luo, B.~Mohanty, H.~G. Ritter, N.~Xu, {Scale for the Phase Diagram of Quantum Chromodynamics}, Science 332 (2011) 1525--1528.
\newblock \href {http://arxiv.org/abs/1105.3934} {\path{arXiv:1105.3934}}, \href {https://doi.org/10.1126/science.1204621} {\path{doi:10.1126/science.1204621}}.

\bibitem{Luo:2017faz}
X.~Luo, N.~Xu, {Search for the QCD Critical Point with Fluctuations of Conserved Quantities in Relativistic Heavy-Ion Collisions at RHIC : An Overview}, Nucl. Sci. Tech. 28~(8) (2017) 112.
\newblock \href {http://arxiv.org/abs/1701.02105} {\path{arXiv:1701.02105}}, \href {https://doi.org/10.1007/s41365-017-0257-0} {\path{doi:10.1007/s41365-017-0257-0}}.

\bibitem{Shuryak:2004cy}
E.~V. Shuryak, {What RHIC experiments and theory tell us about properties of quark-gluon plasma?}, Nucl. Phys. A 750 (2005) 64--83.
\newblock \href {http://arxiv.org/abs/hep-ph/0405066} {\path{arXiv:hep-ph/0405066}}, \href {https://doi.org/10.1016/j.nuclphysa.2004.10.022} {\path{doi:10.1016/j.nuclphysa.2004.10.022}}.

\bibitem{Bazavov:2009zn}
A.~Bazavov, et~al., {Equation of state and QCD transition at finite temperature}, Phys. Rev. D 80 (2009) 014504.
\newblock \href {http://arxiv.org/abs/0903.4379} {\path{arXiv:0903.4379}}, \href {https://doi.org/10.1103/PhysRevD.80.014504} {\path{doi:10.1103/PhysRevD.80.014504}}.

\bibitem{Borsanyi:2016ksw}
S.~Borsanyi, et~al., {Calculation of the axion mass based on high-temperature lattice quantum chromodynamics}, Nature 539~(7627) (2016) 69--71.
\newblock \href {http://arxiv.org/abs/1606.07494} {\path{arXiv:1606.07494}}, \href {https://doi.org/10.1038/nature20115} {\path{doi:10.1038/nature20115}}.

\bibitem{Bernhard:2018hnz}
J.~E. Bernhard, {Bayesian parameter estimation for relativistic heavy-ion collisions}, Ph.D. thesis, Duke U. (4 2018).
\newblock \href {http://arxiv.org/abs/1804.06469} {\path{arXiv:1804.06469}}.

\bibitem{Bernhard:2019bmu}
J.~E. Bernhard, J.~S. Moreland, S.~A. Bass, {Bayesian estimation of the specific shear and bulk viscosity of quark\textendash{}gluon plasma}, Nature Phys. 15~(11) (2019) 1113--1117.
\newblock \href {https://doi.org/10.1038/s41567-019-0611-8} {\path{doi:10.1038/s41567-019-0611-8}}.

\bibitem{JETSCAPE:2020shq}
D.~Everett, et~al., {Phenomenological constraints on the transport properties of QCD matter with data-driven model averaging}, Phys. Rev. Lett. 126~(24) (2021) 242301.
\newblock \href {http://arxiv.org/abs/2010.03928} {\path{arXiv:2010.03928}}, \href {https://doi.org/10.1103/PhysRevLett.126.242301} {\path{doi:10.1103/PhysRevLett.126.242301}}.

\bibitem{Nijs:2020ors}
G.~Nijs, W.~van~der Schee, U.~G\"ursoy, R.~Snellings, {Transverse Momentum Differential Global Analysis of Heavy-Ion Collisions}, Phys. Rev. Lett. 126~(20) (2021) 202301.
\newblock \href {http://arxiv.org/abs/2010.15130} {\path{arXiv:2010.15130}}, \href {https://doi.org/10.1103/PhysRevLett.126.202301} {\path{doi:10.1103/PhysRevLett.126.202301}}.

\bibitem{Wang:2004dn}
X.-N. Wang, {Discovery of jet quenching and beyond}, Nucl. Phys. A 750 (2005) 98--120.
\newblock \href {http://arxiv.org/abs/nucl-th/0405017} {\path{arXiv:nucl-th/0405017}}, \href {https://doi.org/10.1016/j.nuclphysa.2004.12.037} {\path{doi:10.1016/j.nuclphysa.2004.12.037}}.

\bibitem{Vitev:2002pf}
I.~Vitev, M.~Gyulassy, {High $p_{T}$ tomography of $d$ + Au and Au+Au at SPS, RHIC, and LHC}, Phys. Rev. Lett. 89 (2002) 252301.
\newblock \href {http://arxiv.org/abs/hep-ph/0209161} {\path{arXiv:hep-ph/0209161}}, \href {https://doi.org/10.1103/PhysRevLett.89.252301} {\path{doi:10.1103/PhysRevLett.89.252301}}.

\bibitem{JET:2013cls}
K.~M. Burke, et~al., {Extracting the jet transport coefficient from jet quenching in high-energy heavy-ion collisions}, Phys. Rev. C 90~(1) (2014) 014909.
\newblock \href {http://arxiv.org/abs/1312.5003} {\path{arXiv:1312.5003}}, \href {https://doi.org/10.1103/PhysRevC.90.014909} {\path{doi:10.1103/PhysRevC.90.014909}}.

\bibitem{Rapp:2018qla}
A.~Beraudo, et~al., {Extraction of Heavy-Flavor Transport Coefficients in QCD Matter}, Nucl. Phys. A 979 (2018) 21--86.
\newblock \href {http://arxiv.org/abs/1803.03824} {\path{arXiv:1803.03824}}, \href {https://doi.org/10.1016/j.nuclphysa.2018.09.002} {\path{doi:10.1016/j.nuclphysa.2018.09.002}}.

\bibitem{Lappi:2016gmk}
T.~Lappi, {Theory overview of Heavy Ion collisions}, PoS LHCP2016 (2016) 016.
\newblock \href {http://arxiv.org/abs/1609.04917} {\path{arXiv:1609.04917}}, \href {https://doi.org/10.22323/1.276.0016} {\path{doi:10.22323/1.276.0016}}.

\bibitem{Elfner:2022iae}
H.~Elfner, B.~M\"uller, {The exploration of hot and dense nuclear matter: introduction to relativistic heavy-ion physics}, J. Phys. G 50~(10) (2023) 103001.
\newblock \href {http://arxiv.org/abs/2210.12056} {\path{arXiv:2210.12056}}, \href {https://doi.org/10.1088/1361-6471/ace824} {\path{doi:10.1088/1361-6471/ace824}}.

\bibitem{Woods:1954zz}
R.~D. Woods, D.~S. Saxon, {Diffuse Surface Optical Model for Nucleon-Nuclei Scattering}, Phys. Rev. 95 (1954) 577--578.
\newblock \href {https://doi.org/10.1103/PhysRev.95.577} {\path{doi:10.1103/PhysRev.95.577}}.

\bibitem{Kahana:1969zz}
S.~Kahana, H.~C. Lee, C.~K. Scott, {Effect of Woods-Saxon Wave Functions on the Calculation of A=18, 206, 210 Spectra with a Realistic Interaction}, Phys. Rev. 180 (1969) 956--966.
\newblock \href {https://doi.org/10.1103/PhysRev.180.956} {\path{doi:10.1103/PhysRev.180.956}}.

\bibitem{10.2307/1758208}
L.~Wilets, \href{http://www.jstor.org/stable/1758208}{Shape of the nucleus}, Science 129~(3346) (1959) 361--367.
\newline\urlprefix\url{http://www.jstor.org/stable/1758208}

\bibitem{Loizides:2017ack}
C.~Loizides, J.~Kamin, D.~d'Enterria, {Improved Monte Carlo Glauber predictions at present and future nuclear colliders}, Phys. Rev. C 97~(5) (2018) 054910, [Erratum: Phys.Rev.C 99, 019901 (2019)].
\newblock \href {http://arxiv.org/abs/1710.07098} {\path{arXiv:1710.07098}}, \href {https://doi.org/10.1103/PhysRevC.97.054910} {\path{doi:10.1103/PhysRevC.97.054910}}.

\bibitem{Mueller:1989st}
A.~H. Mueller, {Small x Behavior and Parton Saturation: A QCD Model}, Nucl. Phys. B 335 (1990) 115--137.
\newblock \href {https://doi.org/10.1016/0550-3213(90)90173-B} {\path{doi:10.1016/0550-3213(90)90173-B}}.

\bibitem{McLerran:1993ka}
L.~D. McLerran, R.~Venugopalan, {Gluon distribution functions for very large nuclei at small transverse momentum}, Phys. Rev. D 49 (1994) 3352--3355.
\newblock \href {http://arxiv.org/abs/hep-ph/9311205} {\path{arXiv:hep-ph/9311205}}, \href {https://doi.org/10.1103/PhysRevD.49.3352} {\path{doi:10.1103/PhysRevD.49.3352}}.

\bibitem{McLerran:1993ni}
L.~D. McLerran, R.~Venugopalan, {Computing quark and gluon distribution functions for very large nuclei}, Phys. Rev. D 49 (1994) 2233--2241.
\newblock \href {http://arxiv.org/abs/hep-ph/9309289} {\path{arXiv:hep-ph/9309289}}, \href {https://doi.org/10.1103/PhysRevD.49.2233} {\path{doi:10.1103/PhysRevD.49.2233}}.

\bibitem{McLerran:1994vd}
L.~D. McLerran, R.~Venugopalan, {Green's functions in the color field of a large nucleus}, Phys. Rev. D 50 (1994) 2225--2233.
\newblock \href {http://arxiv.org/abs/hep-ph/9402335} {\path{arXiv:hep-ph/9402335}}, \href {https://doi.org/10.1103/PhysRevD.50.2225} {\path{doi:10.1103/PhysRevD.50.2225}}.

\bibitem{Lappi:2006fp}
T.~Lappi, L.~McLerran, {Some features of the glasma}, Nucl. Phys. A 772 (2006) 200--212.
\newblock \href {http://arxiv.org/abs/hep-ph/0602189} {\path{arXiv:hep-ph/0602189}}, \href {https://doi.org/10.1016/j.nuclphysa.2006.04.001} {\path{doi:10.1016/j.nuclphysa.2006.04.001}}.

\bibitem{Gelis:2010nm}
F.~Gelis, E.~Iancu, J.~Jalilian-Marian, R.~Venugopalan, {The Color Glass Condensate}, Ann. Rev. Nucl. Part. Sci. 60 (2010) 463--489.
\newblock \href {http://arxiv.org/abs/1002.0333} {\path{arXiv:1002.0333}}, \href {https://doi.org/10.1146/annurev.nucl.010909.083629} {\path{doi:10.1146/annurev.nucl.010909.083629}}.

\bibitem{Schenke:2012wb}
B.~Schenke, P.~Tribedy, R.~Venugopalan, {Fluctuating Glasma initial conditions and flow in heavy ion collisions}, Phys. Rev. Lett. 108 (2012) 252301.
\newblock \href {http://arxiv.org/abs/1202.6646} {\path{arXiv:1202.6646}}, \href {https://doi.org/10.1103/PhysRevLett.108.252301} {\path{doi:10.1103/PhysRevLett.108.252301}}.

\bibitem{Schenke:2012hg}
B.~Schenke, P.~Tribedy, R.~Venugopalan, {Event-by-event gluon multiplicity, energy density, and eccentricities in ultrarelativistic heavy-ion collisions}, Phys. Rev. C 86 (2012) 034908.
\newblock \href {http://arxiv.org/abs/1206.6805} {\path{arXiv:1206.6805}}, \href {https://doi.org/10.1103/PhysRevC.86.034908} {\path{doi:10.1103/PhysRevC.86.034908}}.

\bibitem{Moreland:2014oya}
J.~S. Moreland, J.~E. Bernhard, S.~A. Bass, {Alternative ansatz to wounded nucleon and binary collision scaling in high-energy nuclear collisions}, Phys. Rev. C 92~(1) (2015) 011901.
\newblock \href {http://arxiv.org/abs/1412.4708} {\path{arXiv:1412.4708}}, \href {https://doi.org/10.1103/PhysRevC.92.011901} {\path{doi:10.1103/PhysRevC.92.011901}}.

\bibitem{Xu:2014ega}
Z.~Xu, K.~Zhou, P.~Zhuang, C.~Greiner, {Thermalization of gluons with Bose-Einstein condensation}, Phys. Rev. Lett. 114~(18) (2015) 182301.
\newblock \href {http://arxiv.org/abs/1410.5616} {\path{arXiv:1410.5616}}, \href {https://doi.org/10.1103/PhysRevLett.114.182301} {\path{doi:10.1103/PhysRevLett.114.182301}}.

\bibitem{Stoecker:2015zea}
H.~Stoecker, et~al., {Glueballs amass at RHIC and LHC Colliders! - The early quarkless 1st order phase transition at $T=270$ MeV - from pure Yang-Mills glue plasma to GlueBall-Hagedorn states}, J. Phys. G 43~(1) (2016) 015105.
\newblock \href {http://arxiv.org/abs/1509.00160} {\path{arXiv:1509.00160}}, \href {https://doi.org/10.1088/0954-3899/43/1/015105} {\path{doi:10.1088/0954-3899/43/1/015105}}.

\bibitem{Stocker:2015nka}
H.~Stocker, et~al., {Under-saturation of quarks at early stages of relativistic nuclear collisions: The hot glue initial scenario and its observable signatures}, Astron. Nachr. 336~(8/9) (2015) 744--748.
\newblock \href {http://arxiv.org/abs/1509.07682} {\path{arXiv:1509.07682}}, \href {https://doi.org/10.1002/asna.201512252} {\path{doi:10.1002/asna.201512252}}.

\bibitem{Zhou:2017zql}
K.~Zhou, Z.~Xu, P.~Zhuang, C.~Greiner, {Kinetic description of Bose-Einstein condensation with test particle simulations}, Phys. Rev. D 96~(1) (2017) 014020.
\newblock \href {http://arxiv.org/abs/1703.02495} {\path{arXiv:1703.02495}}, \href {https://doi.org/10.1103/PhysRevD.96.014020} {\path{doi:10.1103/PhysRevD.96.014020}}.

\bibitem{Bernhard:2016tnd}
J.~E. Bernhard, J.~S. Moreland, S.~A. Bass, J.~Liu, U.~Heinz, {Applying Bayesian parameter estimation to relativistic heavy-ion collisions: simultaneous characterization of the initial state and quark-gluon plasma medium}, Phys. Rev. C 94~(2) (2016) 024907.
\newblock \href {http://arxiv.org/abs/1605.03954} {\path{arXiv:1605.03954}}, \href {https://doi.org/10.1103/PhysRevC.94.024907} {\path{doi:10.1103/PhysRevC.94.024907}}.

\bibitem{Israel:1979wp}
W.~Israel, J.~M. Stewart, {Transient relativistic thermodynamics and kinetic theory}, Annals Phys. 118 (1979) 341--372.
\newblock \href {https://doi.org/10.1016/0003-4916(79)90130-1} {\path{doi:10.1016/0003-4916(79)90130-1}}.

\bibitem{Kolb:2003dz}
P.~F. Kolb, U.~W. Heinz, {Hydrodynamic description of ultrarelativistic heavy ion collisions} (2003) 634--714\href {http://arxiv.org/abs/nucl-th/0305084} {\path{arXiv:nucl-th/0305084}}.

\bibitem{Muronga:2001zk}
A.~Muronga, {Second order dissipative fluid dynamics for ultrarelativistic nuclear collisions}, Phys. Rev. Lett. 88 (2002) 062302, [Erratum: Phys.Rev.Lett. 89, 159901 (2002)].
\newblock \href {http://arxiv.org/abs/nucl-th/0104064} {\path{arXiv:nucl-th/0104064}}, \href {https://doi.org/10.1103/PhysRevLett.88.062302} {\path{doi:10.1103/PhysRevLett.88.062302}}.

\bibitem{Hirano:2005xf}
T.~Hirano, U.~W. Heinz, D.~Kharzeev, R.~Lacey, Y.~Nara, {Hadronic dissipative effects on elliptic flow in ultrarelativistic heavy-ion collisions}, Phys. Lett. B 636 (2006) 299--304.
\newblock \href {http://arxiv.org/abs/nucl-th/0511046} {\path{arXiv:nucl-th/0511046}}, \href {https://doi.org/10.1016/j.physletb.2006.03.060} {\path{doi:10.1016/j.physletb.2006.03.060}}.

\bibitem{Chaudhuri:2006jd}
A.~K. Chaudhuri, {Dissipative hydrodynamics in 2+1 dimension}, Phys. Rev. C 74 (2006) 044904.
\newblock \href {http://arxiv.org/abs/nucl-th/0604014} {\path{arXiv:nucl-th/0604014}}, \href {https://doi.org/10.1103/PhysRevC.74.044904} {\path{doi:10.1103/PhysRevC.74.044904}}.

\bibitem{Romatschke:2007jx}
P.~Romatschke, {Causal viscous hydrodynamics for central heavy-ion collisions. II. Meson spectra and HBT radii}, Eur. Phys. J. C 52 (2007) 203--209.
\newblock \href {http://arxiv.org/abs/nucl-th/0701032} {\path{arXiv:nucl-th/0701032}}, \href {https://doi.org/10.1140/epjc/s10052-007-0354-z} {\path{doi:10.1140/epjc/s10052-007-0354-z}}.

\bibitem{Dusling:2007gi}
K.~Dusling, D.~Teaney, {Simulating elliptic flow with viscous hydrodynamics}, Phys. Rev. C 77 (2008) 034905.
\newblock \href {http://arxiv.org/abs/0710.5932} {\path{arXiv:0710.5932}}, \href {https://doi.org/10.1103/PhysRevC.77.034905} {\path{doi:10.1103/PhysRevC.77.034905}}.

\bibitem{Song:2007ux}
H.~Song, U.~W. Heinz, {Causal viscous hydrodynamics in 2+1 dimensions for relativistic heavy-ion collisions}, Phys. Rev. C 77 (2008) 064901.
\newblock \href {http://arxiv.org/abs/0712.3715} {\path{arXiv:0712.3715}}, \href {https://doi.org/10.1103/PhysRevC.77.064901} {\path{doi:10.1103/PhysRevC.77.064901}}.

\bibitem{Du:2019obx}
L.~Du, U.~Heinz, {(3+1)-dimensional dissipative relativistic fluid dynamics at non-zero net baryon density}, Comput. Phys. Commun. 251 (2020) 107090.
\newblock \href {http://arxiv.org/abs/1906.11181} {\path{arXiv:1906.11181}}, \href {https://doi.org/10.1016/j.cpc.2019.107090} {\path{doi:10.1016/j.cpc.2019.107090}}.

\bibitem{Inghirami:2016iru}
G.~Inghirami, L.~Del~Zanna, A.~Beraudo, M.~H. Moghaddam, F.~Becattini, M.~Bleicher, {Numerical magneto-hydrodynamics for relativistic nuclear collisions}, Eur. Phys. J. C 76~(12) (2016) 659.
\newblock \href {http://arxiv.org/abs/1609.03042} {\path{arXiv:1609.03042}}, \href {https://doi.org/10.1140/epjc/s10052-016-4516-8} {\path{doi:10.1140/epjc/s10052-016-4516-8}}.

\bibitem{Okamoto:2017ukz}
K.~Okamoto, C.~Nonaka, {A new relativistic viscous hydrodynamics code and its application to the Kelvin\textendash{}Helmholtz instability in high-energy heavy-ion collisions}, Eur. Phys. J. C 77~(6) (2017) 383.
\newblock \href {http://arxiv.org/abs/1703.01473} {\path{arXiv:1703.01473}}, \href {https://doi.org/10.1140/epjc/s10052-017-4944-0} {\path{doi:10.1140/epjc/s10052-017-4944-0}}.

\bibitem{Nijs:2021clz}
G.~Nijs, W.~van~der Schee, {Predictions and postdictions for relativistic lead and oxygen collisions with the computational simulation code Trajectum}, Phys. Rev. C 106~(4) (2022) 044903.
\newblock \href {http://arxiv.org/abs/2110.13153} {\path{arXiv:2110.13153}}, \href {https://doi.org/10.1103/PhysRevC.106.044903} {\path{doi:10.1103/PhysRevC.106.044903}}.

\bibitem{Florkowski:2010cf}
W.~Florkowski, R.~Ryblewski, {Highly-anisotropic and strongly-dissipative hydrodynamics for early stages of relativistic heavy-ion collisions}, Phys. Rev. C 83 (2011) 034907.
\newblock \href {http://arxiv.org/abs/1007.0130} {\path{arXiv:1007.0130}}, \href {https://doi.org/10.1103/PhysRevC.83.034907} {\path{doi:10.1103/PhysRevC.83.034907}}.

\bibitem{Strickland:2012bc}
M.~Strickland, {Highly anisotropic dissipative hydrodynamics}, AIP Conf. Proc. 1560~(1) (2013) 658--662.
\newblock \href {http://arxiv.org/abs/1208.2626} {\path{arXiv:1208.2626}}, \href {https://doi.org/10.1063/1.4826864} {\path{doi:10.1063/1.4826864}}.

\bibitem{Schenke:2010nt}
B.~Schenke, S.~Jeon, C.~Gale, {(3+1)D hydrodynamic simulation of relativistic heavy-ion collisions}, Phys. Rev. C 82 (2010) 014903.
\newblock \href {http://arxiv.org/abs/1004.1408} {\path{arXiv:1004.1408}}, \href {https://doi.org/10.1103/PhysRevC.82.014903} {\path{doi:10.1103/PhysRevC.82.014903}}.

\bibitem{Ryu:2015vwa}
S.~Ryu, J.~F. Paquet, C.~Shen, G.~S. Denicol, B.~Schenke, S.~Jeon, C.~Gale, {Importance of the Bulk Viscosity of QCD in Ultrarelativistic Heavy-Ion Collisions}, Phys. Rev. Lett. 115~(13) (2015) 132301.
\newblock \href {http://arxiv.org/abs/1502.01675} {\path{arXiv:1502.01675}}, \href {https://doi.org/10.1103/PhysRevLett.115.132301} {\path{doi:10.1103/PhysRevLett.115.132301}}.

\bibitem{Shen:2017bsr}
C.~Shen, B.~Schenke, {Dynamical initial state model for relativistic heavy-ion collisions}, Phys. Rev. C 97~(2) (2018) 024907.
\newblock \href {http://arxiv.org/abs/1710.00881} {\path{arXiv:1710.00881}}, \href {https://doi.org/10.1103/PhysRevC.97.024907} {\path{doi:10.1103/PhysRevC.97.024907}}.

\bibitem{Bazow:2013ifa}
D.~Bazow, U.~W. Heinz, M.~Strickland, {Second-order (2+1)-dimensional anisotropic hydrodynamics}, Phys. Rev. C 90~(5) (2014) 054910.
\newblock \href {http://arxiv.org/abs/1311.6720} {\path{arXiv:1311.6720}}, \href {https://doi.org/10.1103/PhysRevC.90.054910} {\path{doi:10.1103/PhysRevC.90.054910}}.

\bibitem{Shen:2014vra}
C.~Shen, Z.~Qiu, H.~Song, J.~Bernhard, S.~Bass, U.~Heinz, {The iEBE-VISHNU code package for relativistic heavy-ion collisions}, Comput. Phys. Commun. 199 (2016) 61--85.
\newblock \href {http://arxiv.org/abs/1409.8164} {\path{arXiv:1409.8164}}, \href {https://doi.org/10.1016/j.cpc.2015.08.039} {\path{doi:10.1016/j.cpc.2015.08.039}}.

\bibitem{Pierog:2013ria}
T.~Pierog, I.~Karpenko, J.~M. Katzy, E.~Yatsenko, K.~Werner, {EPOS LHC: Test of collective hadronization with data measured at the CERN Large Hadron Collider}, Phys. Rev. C 92~(3) (2015) 034906.
\newblock \href {http://arxiv.org/abs/1306.0121} {\path{arXiv:1306.0121}}, \href {https://doi.org/10.1103/PhysRevC.92.034906} {\path{doi:10.1103/PhysRevC.92.034906}}.

\bibitem{Sakai:2020pjw}
A.~Sakai, K.~Murase, T.~Hirano, {Rapidity decorrelation of anisotropic flow caused by hydrodynamic fluctuations}, Phys. Rev. C 102~(6) (2020) 064903.
\newblock \href {http://arxiv.org/abs/2003.13496} {\path{arXiv:2003.13496}}, \href {https://doi.org/10.1103/PhysRevC.102.064903} {\path{doi:10.1103/PhysRevC.102.064903}}.

\bibitem{Karpenko:2013wva}
I.~Karpenko, P.~Huovinen, M.~Bleicher, {A 3+1 dimensional viscous hydrodynamic code for relativistic heavy ion collisions}, Comput. Phys. Commun. 185 (2014) 3016--3027.
\newblock \href {http://arxiv.org/abs/1312.4160} {\path{arXiv:1312.4160}}, \href {https://doi.org/10.1016/j.cpc.2014.07.010} {\path{doi:10.1016/j.cpc.2014.07.010}}.

\bibitem{Pang:2018zzo}
L.-G. Pang, H.~Petersen, X.-N. Wang, {Pseudorapidity distribution and decorrelation of anisotropic flow within the open-computing-language implementation CLVisc hydrodynamics}, Phys. Rev. C 97~(6) (2018) 064918.
\newblock \href {http://arxiv.org/abs/1802.04449} {\path{arXiv:1802.04449}}, \href {https://doi.org/10.1103/PhysRevC.97.064918} {\path{doi:10.1103/PhysRevC.97.064918}}.

\bibitem{Yin:2015fca}
Y.~Yin, J.~Liao, {Hydrodynamics with chiral anomaly and charge separation in relativistic heavy ion collisions}, Phys. Lett. B 756 (2016) 42--46.
\newblock \href {http://arxiv.org/abs/1504.06906} {\path{arXiv:1504.06906}}, \href {https://doi.org/10.1016/j.physletb.2016.02.065} {\path{doi:10.1016/j.physletb.2016.02.065}}.

\bibitem{Hattori:2022hyo}
K.~Hattori, M.~Hongo, X.-G. Huang, {New Developments in Relativistic Magnetohydrodynamics}, Symmetry 14~(9) (2022) 1851.
\newblock \href {http://arxiv.org/abs/2207.12794} {\path{arXiv:2207.12794}}, \href {https://doi.org/10.3390/sym14091851} {\path{doi:10.3390/sym14091851}}.

\bibitem{Guo:2019mgh}
X.~Guo, J.~Liao, E.~Wang, {Spin Hydrodynamic Generation in the Charged Subatomic Swirl}, Sci. Rep. 10~(1) (2020) 2196.
\newblock \href {http://arxiv.org/abs/1904.04704} {\path{arXiv:1904.04704}}, \href {https://doi.org/10.1038/s41598-020-59129-6} {\path{doi:10.1038/s41598-020-59129-6}}.

\bibitem{Shi:2020htn}
S.~Shi, C.~Gale, S.~Jeon, {From chiral kinetic theory to relativistic viscous spin hydrodynamics}, Phys. Rev. C 103~(4) (2021) 044906.
\newblock \href {http://arxiv.org/abs/2008.08618} {\path{arXiv:2008.08618}}, \href {https://doi.org/10.1103/PhysRevC.103.044906} {\path{doi:10.1103/PhysRevC.103.044906}}.

\bibitem{Bleicher:1999xi}
M.~Bleicher, et~al., {Relativistic hadron hadron collisions in the ultrarelativistic quantum molecular dynamics model}, J. Phys. G 25 (1999) 1859--1896.
\newblock \href {http://arxiv.org/abs/hep-ph/9909407} {\path{arXiv:hep-ph/9909407}}, \href {https://doi.org/10.1088/0954-3899/25/9/308} {\path{doi:10.1088/0954-3899/25/9/308}}.

\bibitem{Weil:2016zrk}
J.~Weil, et~al., {Particle production and equilibrium properties within a new hadron transport approach for heavy-ion collisions}, Phys. Rev. C 94~(5) (2016) 054905.
\newblock \href {http://arxiv.org/abs/1606.06642} {\path{arXiv:1606.06642}}, \href {https://doi.org/10.1103/PhysRevC.94.054905} {\path{doi:10.1103/PhysRevC.94.054905}}.

\bibitem{Gyulassy:1993hr}
M.~Gyulassy, X.-n. Wang, {Multiple collisions and induced gluon Bremsstrahlung in QCD}, Nucl. Phys. B 420 (1994) 583--614.
\newblock \href {http://arxiv.org/abs/nucl-th/9306003} {\path{arXiv:nucl-th/9306003}}, \href {https://doi.org/10.1016/0550-3213(94)90079-5} {\path{doi:10.1016/0550-3213(94)90079-5}}.

\bibitem{Wang:1994fx}
X.-N. Wang, M.~Gyulassy, M.~Plumer, {The LPM effect in QCD and radiative energy loss in a quark gluon plasma}, Phys. Rev. D 51 (1995) 3436--3446.
\newblock \href {http://arxiv.org/abs/hep-ph/9408344} {\path{arXiv:hep-ph/9408344}}, \href {https://doi.org/10.1103/PhysRevD.51.3436} {\path{doi:10.1103/PhysRevD.51.3436}}.

\bibitem{Baier:1996sk}
R.~Baier, Y.~L. Dokshitzer, A.~H. Mueller, S.~Peigne, D.~Schiff, {Radiative energy loss and p(T) broadening of high-energy partons in nuclei}, Nucl. Phys. B 484 (1997) 265--282.
\newblock \href {http://arxiv.org/abs/hep-ph/9608322} {\path{arXiv:hep-ph/9608322}}, \href {https://doi.org/10.1016/S0550-3213(96)00581-0} {\path{doi:10.1016/S0550-3213(96)00581-0}}.

\bibitem{Baier:1994bd}
R.~Baier, Y.~L. Dokshitzer, S.~Peigne, D.~Schiff, {Induced gluon radiation in a QCD medium}, Phys. Lett. B 345 (1995) 277--286.
\newblock \href {http://arxiv.org/abs/hep-ph/9411409} {\path{arXiv:hep-ph/9411409}}, \href {https://doi.org/10.1016/0370-2693(94)01617-L} {\path{doi:10.1016/0370-2693(94)01617-L}}.

\bibitem{Baier:1996kr}
R.~Baier, Y.~L. Dokshitzer, A.~H. Mueller, S.~Peigne, D.~Schiff, {Radiative energy loss of high-energy quarks and gluons in a finite volume quark - gluon plasma}, Nucl. Phys. B 483 (1997) 291--320.
\newblock \href {http://arxiv.org/abs/hep-ph/9607355} {\path{arXiv:hep-ph/9607355}}, \href {https://doi.org/10.1016/S0550-3213(96)00553-6} {\path{doi:10.1016/S0550-3213(96)00553-6}}.

\bibitem{Zakharov:1996fv}
B.~G. Zakharov, {Fully quantum treatment of the Landau-Pomeranchuk-Migdal effect in QED and QCD}, JETP Lett. 63 (1996) 952--957.
\newblock \href {http://arxiv.org/abs/hep-ph/9607440} {\path{arXiv:hep-ph/9607440}}, \href {https://doi.org/10.1134/1.567126} {\path{doi:10.1134/1.567126}}.

\bibitem{Wiedemann:2000za}
U.~A. Wiedemann, {Gluon radiation off hard quarks in a nuclear environment: Opacity expansion}, Nucl. Phys. B 588 (2000) 303--344.
\newblock \href {http://arxiv.org/abs/hep-ph/0005129} {\path{arXiv:hep-ph/0005129}}, \href {https://doi.org/10.1016/S0550-3213(00)00457-0} {\path{doi:10.1016/S0550-3213(00)00457-0}}.

\bibitem{Guo:2000nz}
X.-f. Guo, X.-N. Wang, {Multiple scattering, parton energy loss and modified fragmentation functions in deeply inelastic e A scattering}, Phys. Rev. Lett. 85 (2000) 3591--3594.
\newblock \href {http://arxiv.org/abs/hep-ph/0005044} {\path{arXiv:hep-ph/0005044}}, \href {https://doi.org/10.1103/PhysRevLett.85.3591} {\path{doi:10.1103/PhysRevLett.85.3591}}.

\bibitem{Wang:2001ifa}
X.-N. Wang, X.-f. Guo, {Multiple parton scattering in nuclei: Parton energy loss}, Nucl. Phys. A 696 (2001) 788--832.
\newblock \href {http://arxiv.org/abs/hep-ph/0102230} {\path{arXiv:hep-ph/0102230}}, \href {https://doi.org/10.1016/S0375-9474(01)01130-7} {\path{doi:10.1016/S0375-9474(01)01130-7}}.

\bibitem{Zhang:2003yn}
B.-W. Zhang, X.-N. Wang, {Multiple parton scattering in nuclei: Beyond helicity amplitude approximation}, Nucl. Phys. A 720 (2003) 429--451.
\newblock \href {http://arxiv.org/abs/hep-ph/0301195} {\path{arXiv:hep-ph/0301195}}, \href {https://doi.org/10.1016/S0375-9474(03)01003-0} {\path{doi:10.1016/S0375-9474(03)01003-0}}.

\bibitem{Schafer:2007xh}
A.~Schafer, X.-N. Wang, B.-W. Zhang, {Multiple Parton Scattering in Nuclei: Quark-quark Scattering}, Nucl. Phys. A 793 (2007) 128--170.
\newblock \href {http://arxiv.org/abs/0704.0106} {\path{arXiv:0704.0106}}, \href {https://doi.org/10.1016/j.nuclphysa.2007.06.009} {\path{doi:10.1016/j.nuclphysa.2007.06.009}}.

\bibitem{He:2015pra}
Y.~He, T.~Luo, X.-N. Wang, Y.~Zhu, {Linear Boltzmann Transport for Jet Propagation in the Quark-Gluon Plasma: Elastic Processes and Medium Recoil}, Phys. Rev. C 91 (2015) 054908, [Erratum: Phys.Rev.C 97, 019902 (2018)].
\newblock \href {http://arxiv.org/abs/1503.03313} {\path{arXiv:1503.03313}}, \href {https://doi.org/10.1103/PhysRevC.91.054908} {\path{doi:10.1103/PhysRevC.91.054908}}.

\bibitem{Cao:2016gvr}
S.~Cao, T.~Luo, G.-Y. Qin, X.-N. Wang, {Linearized Boltzmann transport model for jet propagation in the quark-gluon plasma: Heavy quark evolution}, Phys. Rev. C 94~(1) (2016) 014909.
\newblock \href {http://arxiv.org/abs/1605.06447} {\path{arXiv:1605.06447}}, \href {https://doi.org/10.1103/PhysRevC.94.014909} {\path{doi:10.1103/PhysRevC.94.014909}}.

\bibitem{Casalderrey-Solana:2014bpa}
J.~Casalderrey-Solana, D.~C. Gulhan, J.~G. Milhano, D.~Pablos, K.~Rajagopal, {A Hybrid Strong/Weak Coupling Approach to Jet Quenching}, JHEP 10 (2014) 019, [Erratum: JHEP 09, 175 (2015)].
\newblock \href {http://arxiv.org/abs/1405.3864} {\path{arXiv:1405.3864}}, \href {https://doi.org/10.1007/JHEP09(2015)175} {\path{doi:10.1007/JHEP09(2015)175}}.

\bibitem{Shi:2018izg}
S.~Shi, J.~Liao, M.~Gyulassy, {Global constraints from RHIC and LHC on transport properties of QCD fluids in CUJET/CIBJET framework}, Chin. Phys. C 43~(4) (2019) 044101.
\newblock \href {http://arxiv.org/abs/1808.05461} {\path{arXiv:1808.05461}}, \href {https://doi.org/10.1088/1674-1137/43/4/044101} {\path{doi:10.1088/1674-1137/43/4/044101}}.

\bibitem{Cao:2020wlm}
S.~Cao, X.-N. Wang, {Jet quenching and medium response in high-energy heavy-ion collisions: a review}, Rept. Prog. Phys. 84~(2) (2021) 024301.
\newblock \href {http://arxiv.org/abs/2002.04028} {\path{arXiv:2002.04028}}, \href {https://doi.org/10.1088/1361-6633/abc22b} {\path{doi:10.1088/1361-6633/abc22b}}.

\bibitem{Qin:2015srf}
G.-Y. Qin, X.-N. Wang, {Jet quenching in high-energy heavy-ion collisions}, Int. J. Mod. Phys. E 24~(11) (2015) 1530014.
\newblock \href {http://arxiv.org/abs/1511.00790} {\path{arXiv:1511.00790}}, \href {https://doi.org/10.1142/S0218301315300143} {\path{doi:10.1142/S0218301315300143}}.

\bibitem{Majumder:2010qh}
A.~Majumder, M.~Van~Leeuwen, {The Theory and Phenomenology of Perturbative QCD Based Jet Quenching}, Prog. Part. Nucl. Phys. 66 (2011) 41--92.
\newblock \href {http://arxiv.org/abs/1002.2206} {\path{arXiv:1002.2206}}, \href {https://doi.org/10.1016/j.ppnp.2010.09.001} {\path{doi:10.1016/j.ppnp.2010.09.001}}.

\bibitem{Armesto:2011ht}
N.~Armesto, et~al., {Comparison of Jet Quenching Formalisms for a Quark-Gluon Plasma 'Brick'}, Phys. Rev. C 86 (2012) 064904.
\newblock \href {http://arxiv.org/abs/1106.1106} {\path{arXiv:1106.1106}}, \href {https://doi.org/10.1103/PhysRevC.86.064904} {\path{doi:10.1103/PhysRevC.86.064904}}.

\bibitem{Qin:2009bk}
G.-Y. Qin, J.~Ruppert, C.~Gale, S.~Jeon, G.~D. Moore, {Jet energy loss, photon production, and photon-hadron correlations at RHIC}, Phys. Rev. C 80 (2009) 054909.
\newblock \href {http://arxiv.org/abs/0906.3280} {\path{arXiv:0906.3280}}, \href {https://doi.org/10.1103/PhysRevC.80.054909} {\path{doi:10.1103/PhysRevC.80.054909}}.

\bibitem{Jeon:2003gi}
S.~Jeon, G.~D. Moore, {Energy loss of leading partons in a thermal QCD medium}, Phys. Rev. C 71 (2005) 034901.
\newblock \href {http://arxiv.org/abs/hep-ph/0309332} {\path{arXiv:hep-ph/0309332}}, \href {https://doi.org/10.1103/PhysRevC.71.034901} {\path{doi:10.1103/PhysRevC.71.034901}}.

\bibitem{Noronha-Hostler:2016eow}
J.~Noronha-Hostler, B.~Betz, J.~Noronha, M.~Gyulassy, {Event-by-event hydrodynamics $+$ jet energy loss: A solution to the $R_{AA} \otimes v_2$ puzzle}, Phys. Rev. Lett. 116~(25) (2016) 252301.
\newblock \href {http://arxiv.org/abs/1602.03788} {\path{arXiv:1602.03788}}, \href {https://doi.org/10.1103/PhysRevLett.116.252301} {\path{doi:10.1103/PhysRevLett.116.252301}}.

\bibitem{Andres:2016iys}
C.~Andr\'es, N.~Armesto, M.~Luzum, C.~A. Salgado, P.~Zurita, {Energy versus centrality dependence of the jet quenching parameter $\hat{q}$ at RHIC and LHC: a new puzzle?}, Eur. Phys. J. C 76~(9) (2016) 475.
\newblock \href {http://arxiv.org/abs/1606.04837} {\path{arXiv:1606.04837}}, \href {https://doi.org/10.1140/epjc/s10052-016-4320-5} {\path{doi:10.1140/epjc/s10052-016-4320-5}}.

\bibitem{Bianchi:2017wpt}
E.~Bianchi, J.~Elledge, A.~Kumar, A.~Majumder, G.-Y. Qin, C.~Shen, {The $x$ and $Q^2$ dependence of $\hat{q}$, quasi-particles and the JET puzzle} (2 2017).
\newblock \href {http://arxiv.org/abs/1702.00481} {\path{arXiv:1702.00481}}.

\bibitem{Chien:2015vja}
Y.-T. Chien, A.~Emerman, Z.-B. Kang, G.~Ovanesyan, I.~Vitev, {Jet Quenching from QCD Evolution}, Phys. Rev. D 93~(7) (2016) 074030.
\newblock \href {http://arxiv.org/abs/1509.02936} {\path{arXiv:1509.02936}}, \href {https://doi.org/10.1103/PhysRevD.93.074030} {\path{doi:10.1103/PhysRevD.93.074030}}.

\bibitem{Andres:2019eus}
C.~Andres, N.~Armesto, H.~Niemi, R.~Paatelainen, C.~A. Salgado, {Jet quenching as a probe of the initial stages in heavy-ion collisions}, Phys. Lett. B 803 (2020) 135318.
\newblock \href {http://arxiv.org/abs/1902.03231} {\path{arXiv:1902.03231}}, \href {https://doi.org/10.1016/j.physletb.2020.135318} {\path{doi:10.1016/j.physletb.2020.135318}}.

\bibitem{Yazdi:2022bru}
R.~M. Yazdi, S.~Shi, C.~Gale, S.~Jeon, {Leading order, next-to-leading order, and nonperturbative parton collision kernels: Effects in static and evolving media}, Phys. Rev. C 106~(6) (2022) 064902.
\newblock \href {http://arxiv.org/abs/2206.05855} {\path{arXiv:2206.05855}}, \href {https://doi.org/10.1103/PhysRevC.106.064902} {\path{doi:10.1103/PhysRevC.106.064902}}.

\bibitem{Shi:2022rja}
S.~Shi, R.~Modarresi~Yazdi, C.~Gale, S.~Jeon, {Comparing the martini and cujet models for jet quenching: Medium modification of jets and jet substructure}, Phys. Rev. C 107~(3) (2023) 034908.
\newblock \href {http://arxiv.org/abs/2212.05944} {\path{arXiv:2212.05944}}, \href {https://doi.org/10.1103/PhysRevC.107.034908} {\path{doi:10.1103/PhysRevC.107.034908}}.

\bibitem{Sjostrand:2019zhc}
T.~Sj\"ostrand, {The PYTHIA Event Generator: Past, Present and Future}, Comput. Phys. Commun. 246 (2020) 106910.
\newblock \href {http://arxiv.org/abs/1907.09874} {\path{arXiv:1907.09874}}, \href {https://doi.org/10.1016/j.cpc.2019.106910} {\path{doi:10.1016/j.cpc.2019.106910}}.

\bibitem{Kordell:2017hmi}
M.~Kordell, A.~Majumder, {Event-by-Event Simulations of Jet Modification Using the MATTER Event Generator}, in: {8th International Conference on Hard and Electromagnetic Probes of High-energy Nuclear Collisions}: {Hard Probes 2016}, 2017.
\newblock \href {http://arxiv.org/abs/1702.05862} {\path{arXiv:1702.05862}}.

\bibitem{Schenke:2009gb}
B.~Schenke, C.~Gale, S.~Jeon, {MARTINI: An Event generator for relativistic heavy-ion collisions}, Phys. Rev. C 80 (2009) 054913.
\newblock \href {http://arxiv.org/abs/0909.2037} {\path{arXiv:0909.2037}}, \href {https://doi.org/10.1103/PhysRevC.80.054913} {\path{doi:10.1103/PhysRevC.80.054913}}.

\bibitem{Bass:2017zyn}
S.~A. Bass, J.~E. Bernhard, J.~S. Moreland, {Determination of Quark-Gluon-Plasma Parameters from a Global Bayesian Analysis}, Nucl. Phys. A 967 (2017) 67--73.
\newblock \href {http://arxiv.org/abs/1704.07671} {\path{arXiv:1704.07671}}, \href {https://doi.org/10.1016/j.nuclphysa.2017.05.052} {\path{doi:10.1016/j.nuclphysa.2017.05.052}}.

\bibitem{Pang:2016vdc}
L.-G. Pang, K.~Zhou, N.~Su, H.~Petersen, H.~St\"ocker, X.-N. Wang, {An equation-of-state-meter of quantum chromodynamics transition from deep learning}, Nature Commun. 9~(1) (2018) 210.
\newblock \href {http://arxiv.org/abs/1612.04262} {\path{arXiv:1612.04262}}, \href {https://doi.org/10.1038/s41467-017-02726-3} {\path{doi:10.1038/s41467-017-02726-3}}.

\bibitem{Steinheimer:2019iso}
J.~Steinheimer, L.~Pang, K.~Zhou, V.~Koch, J.~Randrup, H.~Stoecker, {A machine learning study to identify spinodal clumping in high energy nuclear collisions}, JHEP 12 (2019) 122.
\newblock \href {http://arxiv.org/abs/1906.06562} {\path{arXiv:1906.06562}}, \href {https://doi.org/10.1007/JHEP12(2019)122} {\path{doi:10.1007/JHEP12(2019)122}}.

\bibitem{He:2021uko}
J.~He, W.-B. He, Y.-G. Ma, S.~Zhang, {Machine-learning-based identification for initial clustering structure in relativistic heavy-ion collisions}, Phys. Rev. C 104~(4) (2021) 044902.
\newblock \href {http://arxiv.org/abs/2109.06277} {\path{arXiv:2109.06277}}, \href {https://doi.org/10.1103/PhysRevC.104.044902} {\path{doi:10.1103/PhysRevC.104.044902}}.

\bibitem{Miller:2007ri}
M.~L. Miller, K.~Reygers, S.~J. Sanders, P.~Steinberg, {Glauber modeling in high energy nuclear collisions}, Ann. Rev. Nucl. Part. Sci. 57 (2007) 205--243.
\newblock \href {http://arxiv.org/abs/nucl-ex/0701025} {\path{arXiv:nucl-ex/0701025}}, \href {https://doi.org/10.1146/annurev.nucl.57.090506.123020} {\path{doi:10.1146/annurev.nucl.57.090506.123020}}.

\bibitem{OmanaKuttan:2023cno}
M.~Omana~Kuttan, J.~Steinheimer, K.~Zhou, M.~Bleicher, H.~Stoecker, {Model dependence of the number of participant nucleons and observable consequences in heavy-ion collisions}, Eur. Phys. J. C 83~(9) (2023) 792.
\newblock \href {http://arxiv.org/abs/2303.07919} {\path{arXiv:2303.07919}}, \href {https://doi.org/10.1140/epjc/s10052-023-11968-z} {\path{doi:10.1140/epjc/s10052-023-11968-z}}.

\bibitem{Kagamaster:2020oon}
S.~Kagamaster, R.~Reed, M.~Lisa, {Centrality determination with a forward detector in the RHIC Beam Energy Scan}, Phys. Rev. C 103~(4) (2021) 044902.
\newblock \href {http://arxiv.org/abs/2009.01483} {\path{arXiv:2009.01483}}, \href {https://doi.org/10.1103/PhysRevC.103.044902} {\path{doi:10.1103/PhysRevC.103.044902}}.

\bibitem{David:1994qc}
C.~David, M.~Freslier, J.~Aichelin, {Impact parameter determination for heavy-ion collisions by use of a neural network}, Phys. Rev. C 51 (1995) 1453--1459.
\newblock \href {https://doi.org/10.1103/PhysRevC.51.1453} {\path{doi:10.1103/PhysRevC.51.1453}}.

\bibitem{Haddad:1996xw}
F.~Haddad, K.~Hagel, J.~Li, N.~Mdeiwayeh, J.~B. Natowitz, R.~Wada, B.~Xiao, C.~David, M.~Freslier, J.~Aichelin, {Impact parameter determination in experimental analysis using neural network}, Phys. Rev. C 55 (1997) 1371--1375.
\newblock \href {https://doi.org/10.1103/PhysRevC.55.1371} {\path{doi:10.1103/PhysRevC.55.1371}}.

\bibitem{DeSanctis:2009zzb}
J.~De~Sanctis, M.~Masotti, M.~Bruno, M.~D'Agostino, E.~Geraci, G.~Vannini, A.~Bonasera, {Classification of the impact parameter in nucleus-nucleus collisions by a support vector machine method}, J. Phys. G 36 (2009) 015101.
\newblock \href {https://doi.org/10.1088/0954-3899/36/1/015101} {\path{doi:10.1088/0954-3899/36/1/015101}}.

\bibitem{Li:2022mni}
X.~Chen, L.~Li, Y.~Cui, J.~Yang, Z.~Li, Y.~Zhang, {Bayesian reconstruction of impact parameter distributions from two observables for intermediate energy heavy ion collisions}, Phys. Rev. C 108~(3) (2023) 034613.
\newblock \href {http://arxiv.org/abs/2201.12586} {\path{arXiv:2201.12586}}, \href {https://doi.org/10.1103/PhysRevC.108.034613} {\path{doi:10.1103/PhysRevC.108.034613}}.

\bibitem{Li:2020qqn}
F.~Li, Y.~Wang, H.~L\"u, P.~Li, Q.~Li, F.~Liu, {Application of artificial intelligence in the determination of impact parameter in heavy-ion collisions at intermediate energies}, J. Phys. G 47~(11) (2020) 115104.
\newblock \href {http://arxiv.org/abs/2008.11540} {\path{arXiv:2008.11540}}, \href {https://doi.org/10.1088/1361-6471/abb1f9} {\path{doi:10.1088/1361-6471/abb1f9}}.

\bibitem{Li:2021plq}
F.~Li, Y.~Wang, Z.~Gao, P.~Li, H.~L\"u, H.~Lv, Q.~Li, C.~Y. Tsang, M.~B. Tsang, {Application of machine learning in the determination of impact parameter in the $^{132}$Sn+$^{124}$Sn system}, Phys. Rev. C 104~(3) (2021) 034608.
\newblock \href {http://arxiv.org/abs/2105.08912} {\path{arXiv:2105.08912}}, \href {https://doi.org/10.1103/PhysRevC.104.034608} {\path{doi:10.1103/PhysRevC.104.034608}}.

\bibitem{Zhang:2021zxd}
X.~Zhang, et~al., {Determining impact parameters of heavy-ion collisions at low-intermediate incident energies using deep learning with convolutional neural networks}, Phys. Rev. C 105~(3) (2022) 034611.
\newblock \href {http://arxiv.org/abs/2111.06597} {\path{arXiv:2111.06597}}, \href {https://doi.org/10.1103/PhysRevC.105.034611} {\path{doi:10.1103/PhysRevC.105.034611}}.

\bibitem{Tsang:2021rku}
C.~Y. Tsang, et~al., {Applying machine learning to determine impact parameter in nuclear physics experiments} (7 2021).
\newblock \href {http://arxiv.org/abs/2107.13985} {\path{arXiv:2107.13985}}.

\bibitem{Xiang:2021ssj}
P.~Xiang, Y.-S. Zhao, X.-G. Huang, {Determination of the impact parameter in high-energy heavy-ion collisions via deep learning *}, Chin. Phys. C 46~(7) (2022) 074110.
\newblock \href {http://arxiv.org/abs/2112.03824} {\path{arXiv:2112.03824}}, \href {https://doi.org/10.1088/1674-1137/ac6490} {\path{doi:10.1088/1674-1137/ac6490}}.

\bibitem{Mallick:2021wop}
N.~Mallick, S.~Tripathy, A.~N. Mishra, S.~Deb, R.~Sahoo, {Estimation of Impact Parameter and Transverse Spherocity in heavy-ion collisions at the LHC energies using Machine Learning}, Phys. Rev. D 103~(9) (2021) 094031.
\newblock \href {http://arxiv.org/abs/2103.01736} {\path{arXiv:2103.01736}}, \href {https://doi.org/10.1103/PhysRevD.103.094031} {\path{doi:10.1103/PhysRevD.103.094031}}.

\bibitem{Saha:2022skj}
A.~Saha, D.~Dan, S.~Sanyal, {Machine-learning model-driven prediction of the initial geometry in heavy-ion collision experiments}, Phys. Rev. C 106~(1) (2022) 014901.
\newblock \href {http://arxiv.org/abs/2203.15433} {\path{arXiv:2203.15433}}, \href {https://doi.org/10.1103/PhysRevC.106.014901} {\path{doi:10.1103/PhysRevC.106.014901}}.

\bibitem{OmanaKuttan:2020brq}
M.~Omana~Kuttan, J.~Steinheimer, K.~Zhou, A.~Redelbach, H.~Stoecker, {A fast centrality-meter for heavy-ion collisions at the CBM experiment}, Phys. Lett. B 811 (2020) 135872.
\newblock \href {http://arxiv.org/abs/2009.01584} {\path{arXiv:2009.01584}}, \href {https://doi.org/10.1016/j.physletb.2020.135872} {\path{doi:10.1016/j.physletb.2020.135872}}.

\bibitem{OmanaKuttan:2021axp}
M.~Omana~Kuttan, J.~Steinheimer, K.~Zhou, A.~Redelbach, H.~Stoecker, {Deep Learning Based Impact Parameter Determination for the CBM Experiment}, Particles 4~(1) (2021) 47--52.
\newblock \href {https://doi.org/10.3390/particles4010006} {\path{doi:10.3390/particles4010006}}.

\bibitem{Bzdak:2018uhv}
A.~Bzdak, V.~Koch, D.~Oliinychenko, J.~Steinheimer, {Large proton cumulants from the superposition of ordinary multiplicity distributions}, Phys. Rev. C 98~(5) (2018) 054901.
\newblock \href {http://arxiv.org/abs/1804.04463} {\path{arXiv:1804.04463}}, \href {https://doi.org/10.1103/PhysRevC.98.054901} {\path{doi:10.1103/PhysRevC.98.054901}}.

\bibitem{Luo:2015ewa}
X.~Luo, {Energy Dependence of Moments of Net-Proton and Net-Charge Multiplicity Distributions at STAR}, PoS CPOD2014 (2015) 019.
\newblock \href {http://arxiv.org/abs/1503.02558} {\path{arXiv:1503.02558}}, \href {https://doi.org/10.22323/1.217.0019} {\path{doi:10.22323/1.217.0019}}.

\bibitem{Thaprasop:2020mzp}
P.~Thaprasop, K.~Zhou, J.~Steinheimer, C.~Herold, {Unsupervised Outlier Detection in Heavy-Ion Collisions}, Phys. Scripta 96~(6) (2021) 064003.
\newblock \href {http://arxiv.org/abs/2007.15830} {\path{arXiv:2007.15830}}, \href {https://doi.org/10.1088/1402-4896/abf214} {\path{doi:10.1088/1402-4896/abf214}}.

\bibitem{Bally:2022vgo}
B.~Bally, et~al., {Imaging the initial condition of heavy-ion collisions and nuclear structure across the nuclide chart} (9 2022).
\newblock \href {http://arxiv.org/abs/2209.11042} {\path{arXiv:2209.11042}}.

\bibitem{PREX:2021umo}
D.~Adhikari, et~al., {Accurate Determination of the Neutron Skin Thickness of $^{208}$Pb through Parity-Violation in Electron Scattering}, Phys. Rev. Lett. 126~(17) (2021) 172502.
\newblock \href {http://arxiv.org/abs/2102.10767} {\path{arXiv:2102.10767}}, \href {https://doi.org/10.1103/PhysRevLett.126.172502} {\path{doi:10.1103/PhysRevLett.126.172502}}.

\bibitem{Pang:2019aqb}
L.-G. Pang, K.~Zhou, X.-N. Wang, {Interpretable deep learning for nuclear deformation in heavy ion collisions} (6 2019).
\newblock \href {http://arxiv.org/abs/1906.06429} {\path{arXiv:1906.06429}}.

\bibitem{Bailey:2021bzo}
S.~Bailey, et~al., {The identification of $\alpha$-clustered doorway states in $^{44,48,52}$Ti using machine learning}, Eur. Phys. J. A 57~(3) (2021) 108.
\newblock \href {https://doi.org/10.1140/epja/s10050-021-00357-3} {\path{doi:10.1140/epja/s10050-021-00357-3}}.

\bibitem{Jia:2021oyt}
J.~Jia, C.~Zhang, {Scaling approach to nuclear structure in high-energy heavy-ion collisions}, Phys. Rev. C 107~(2) (2023) L021901.
\newblock \href {http://arxiv.org/abs/2111.15559} {\path{arXiv:2111.15559}}, \href {https://doi.org/10.1103/PhysRevC.107.L021901} {\path{doi:10.1103/PhysRevC.107.L021901}}.

\bibitem{Cheng:2023ciy}
Y.-L. Cheng, S.~Shi, Y.-G. Ma, H.~St\"ocker, K.~Zhou, {Examination of nucleon distribution with Bayesian imaging for isobar collisions}, Phys. Rev. C 107~(6) (2023) 064909.
\newblock \href {http://arxiv.org/abs/2301.03910} {\path{arXiv:2301.03910}}, \href {https://doi.org/10.1103/PhysRevC.107.064909} {\path{doi:10.1103/PhysRevC.107.064909}}.

\bibitem{Pratt:2015zsa}
S.~Pratt, E.~Sangaline, P.~Sorensen, H.~Wang, {Constraining the Eq. of State of Super-Hadronic Matter from Heavy-Ion Collisions}, Phys. Rev. Lett. 114 (2015) 202301.
\newblock \href {http://arxiv.org/abs/1501.04042} {\path{arXiv:1501.04042}}, \href {https://doi.org/10.1103/PhysRevLett.114.202301} {\path{doi:10.1103/PhysRevLett.114.202301}}.

\bibitem{OmanaKuttan:2022aml}
M.~Omana~Kuttan, J.~Steinheimer, K.~Zhou, H.~St\"ocker, {The QCD EoS of dense nuclear matter from Bayesian analysis of heavy ion collision data} (11 2022).
\newblock \href {http://arxiv.org/abs/2211.11670} {\path{arXiv:2211.11670}}.

\bibitem{Pang:2019int}
L.-G. Pang, K.~Zhou, N.~Su, H.~Petersen, H.~St\"ocker, X.-N. Wang, {Classify QCD phase transition with deep learning}, Nucl. Phys. A 982 (2019) 867--870.
\newblock \href {https://doi.org/10.1016/j.nuclphysa.2018.10.077} {\path{doi:10.1016/j.nuclphysa.2018.10.077}}.

\bibitem{Zhou:2018hsl}
K.~Zhou, L.-G. Pang, N.~Su, H.~Petersen, H.~Stoecker, X.-N. Wang, {Identifying QCD Transition Using Deep Learning}, EPJ Web Conf. 171 (2018) 16005.
\newblock \href {https://doi.org/10.1051/epjconf/201817116005} {\path{doi:10.1051/epjconf/201817116005}}.

\bibitem{Du:2019civ}
Y.-L. Du, K.~Zhou, J.~Steinheimer, L.-G. Pang, A.~Motornenko, H.-S. Zong, X.-N. Wang, H.~St\"ocker, {Identifying the nature of the QCD transition in relativistic collision of heavy nuclei with deep learning}, Eur. Phys. J. C 80~(6) (2020) 516.
\newblock \href {http://arxiv.org/abs/1910.11530} {\path{arXiv:1910.11530}}, \href {https://doi.org/10.1140/epjc/s10052-020-8030-7} {\path{doi:10.1140/epjc/s10052-020-8030-7}}.

\bibitem{Du:2020poe}
Y.-L. Du, K.~Zhou, J.~Steinheimer, L.-G. Pang, A.~Motornenko, H.-S. Zong, X.-N. Wang, H.~St\"ocker, {Identifying the nature of the QCD transition in heavy-ion collisions with deep learning}, Nucl. Phys. A 1005 (2021) 121891.
\newblock \href {http://arxiv.org/abs/2009.03059} {\path{arXiv:2009.03059}}, \href {https://doi.org/10.1016/j.nuclphysa.2020.121891} {\path{doi:10.1016/j.nuclphysa.2020.121891}}.

\bibitem{Jiang:2021gsw}
L.~Jiang, L.~Wang, K.~Zhou, {Deep learning stochastic processes with QCD phase transition}, Phys. Rev. D 103~(11) (2021) 116023.
\newblock \href {http://arxiv.org/abs/2103.04090} {\path{arXiv:2103.04090}}, \href {https://doi.org/10.1103/PhysRevD.103.116023} {\path{doi:10.1103/PhysRevD.103.116023}}.

\bibitem{Wang:2021yjw}
L.~Wang, L.~Jiang, K.~Zhou, {Learning Langevin dynamics with QCD phase transition}, EPJ Web Conf. 259 (2022) 10017.
\newblock \href {http://arxiv.org/abs/2108.03987} {\path{arXiv:2108.03987}}, \href {https://doi.org/10.1051/epjconf/202225910017} {\path{doi:10.1051/epjconf/202225910017}}.

\bibitem{Nahrgang:2011mg}
M.~Nahrgang, S.~Leupold, C.~Herold, M.~Bleicher, {Nonequilibrium chiral fluid dynamics including dissipation and noise}, Phys. Rev. C 84 (2011) 024912.
\newblock \href {http://arxiv.org/abs/1105.0622} {\path{arXiv:1105.0622}}, \href {https://doi.org/10.1103/PhysRevC.84.024912} {\path{doi:10.1103/PhysRevC.84.024912}}.

\bibitem{Kvasiuk:2020izb}
Y.~Kvasiuk, E.~Zabrodin, L.~Bravina, I.~Didur, M.~Frolov, {Classification of Equation of State in Relativistic Heavy-Ion Collisions Using Deep Learning}, JHEP 07 (2020) 133.
\newblock \href {http://arxiv.org/abs/2004.14409} {\path{arXiv:2004.14409}}, \href {https://doi.org/10.1007/JHEP07(2020)133} {\path{doi:10.1007/JHEP07(2020)133}}.

\bibitem{Sergeev:2020fir}
F.~Sergeev, E.~Bratkovskaya, I.~Kisel, I.~Vassiliev, {Deep learning for quark\textendash{}gluon plasma detection in the CBM experiment}, Int. J. Mod. Phys. A 35~(33) (2020) 2043002.
\newblock \href {https://doi.org/10.1142/S0217751X20430022} {\path{doi:10.1142/S0217751X20430022}}.

\bibitem{Wang:2020tgb}
R.~Wang, Y.-G. Ma, R.~Wada, L.-W. Chen, W.-B. He, H.-L. Liu, K.-J. Sun, {Nuclear liquid-gas phase transition with machine learning}, Phys. Rev. Res. 2~(4) (2020) 043202.
\newblock \href {http://arxiv.org/abs/2010.15043} {\path{arXiv:2010.15043}}, \href {https://doi.org/10.1103/PhysRevResearch.2.043202} {\path{doi:10.1103/PhysRevResearch.2.043202}}.

\bibitem{Wang:2021xbb}
Y.~Wang, F.~Li, Q.~Li, H.~L\"u, K.~Zhou, {Finding signatures of the nuclear symmetry energy in heavy-ion collisions with deep learning}, Phys. Lett. B 822 (2021) 136669.
\newblock \href {http://arxiv.org/abs/2107.11012} {\path{arXiv:2107.11012}}, \href {https://doi.org/10.1016/j.physletb.2021.136669} {\path{doi:10.1016/j.physletb.2021.136669}}.

\bibitem{OmanaKuttan:2020btb}
M.~Omana~Kuttan, K.~Zhou, J.~Steinheimer, A.~Redelbach, H.~Stoecker, {An equation-of-state-meter for CBM using PointNet}, JHEP 21 (2020) 184.
\newblock \href {http://arxiv.org/abs/2107.05590} {\path{arXiv:2107.05590}}, \href {https://doi.org/10.1007/JHEP10(2021)184} {\path{doi:10.1007/JHEP10(2021)184}}.

\bibitem{Huang:2021iux}
Y.~Huang, L.-G. Pang, X.~Luo, X.-N. Wang, {Probing criticality with deep learning in relativistic heavy-ion collisions}, Phys. Lett. B 827 (2022) 137001.
\newblock \href {http://arxiv.org/abs/2107.11828} {\path{arXiv:2107.11828}}, \href {https://doi.org/10.1016/j.physletb.2022.137001} {\path{doi:10.1016/j.physletb.2022.137001}}.

\bibitem{Burr2009}
B.~Settles, {Active learning literature survey}, 2009.

\bibitem{mroczek2022}
D.~Mroczek, M.~Hjorth-Jensen, J.~Noronha-Hostler, P.~Parotto, C.~Ratti, R.~Vilalta, \href{https://arxiv.org/abs/2203.13876}{Mapping out the thermodynamic stability of a qcd equation of state with a critical point using active learning} (2022).
\newblock \href {https://doi.org/10.48550/ARXIV.2203.13876} {\path{doi:10.48550/ARXIV.2203.13876}}.
\newline\urlprefix\url{https://arxiv.org/abs/2203.13876}

\bibitem{An:2021wof}
X.~An, et~al., {The BEST framework for the search for the QCD critical point and the chiral magnetic effect}, Nucl. Phys. A 1017 (2022) 122343.
\newblock \href {http://arxiv.org/abs/2108.13867} {\path{arXiv:2108.13867}}, \href {https://doi.org/10.1016/j.nuclphysa.2021.122343} {\path{doi:10.1016/j.nuclphysa.2021.122343}}.

\bibitem{Heinz:2005bw}
U.~W. Heinz, H.~Song, A.~K. Chaudhuri, {Dissipative hydrodynamics for viscous relativistic fluids}, Phys. Rev. C 73 (2006) 034904.
\newblock \href {http://arxiv.org/abs/nucl-th/0510014} {\path{arXiv:nucl-th/0510014}}, \href {https://doi.org/10.1103/PhysRevC.73.034904} {\path{doi:10.1103/PhysRevC.73.034904}}.

\bibitem{Romatschke:2007mq}
P.~Romatschke, U.~Romatschke, {Viscosity Information from Relativistic Nuclear Collisions: How Perfect is the Fluid Observed at RHIC?}, Phys. Rev. Lett. 99 (2007) 172301.
\newblock \href {http://arxiv.org/abs/0706.1522} {\path{arXiv:0706.1522}}, \href {https://doi.org/10.1103/PhysRevLett.99.172301} {\path{doi:10.1103/PhysRevLett.99.172301}}.

\bibitem{Song:2010mg}
H.~Song, S.~A. Bass, U.~Heinz, T.~Hirano, C.~Shen, {200 A GeV Au+Au collisions serve a nearly perfect quark-gluon liquid}, Phys. Rev. Lett. 106 (2011) 192301, [Erratum: Phys.Rev.Lett. 109, 139904 (2012)].
\newblock \href {http://arxiv.org/abs/1011.2783} {\path{arXiv:1011.2783}}, \href {https://doi.org/10.1103/PhysRevLett.106.192301} {\path{doi:10.1103/PhysRevLett.106.192301}}.

\bibitem{Buchel:2003tz}
A.~Buchel, J.~T. Liu, {Universality of the shear viscosity in supergravity}, Phys. Rev. Lett. 93 (2004) 090602.
\newblock \href {http://arxiv.org/abs/hep-th/0311175} {\path{arXiv:hep-th/0311175}}, \href {https://doi.org/10.1103/PhysRevLett.93.090602} {\path{doi:10.1103/PhysRevLett.93.090602}}.

\bibitem{Teaney:2003kp}
D.~Teaney, {The Effects of viscosity on spectra, elliptic flow, and HBT radii}, Phys. Rev. C 68 (2003) 034913.
\newblock \href {http://arxiv.org/abs/nucl-th/0301099} {\path{arXiv:nucl-th/0301099}}, \href {https://doi.org/10.1103/PhysRevC.68.034913} {\path{doi:10.1103/PhysRevC.68.034913}}.

\bibitem{Niemi:2015qia}
H.~Niemi, K.~J. Eskola, R.~Paatelainen, {Event-by-event fluctuations in a perturbative QCD + saturation + hydrodynamics model: Determining QCD matter shear viscosity in ultrarelativistic heavy-ion collisions}, Phys. Rev. C 93~(2) (2016) 024907.
\newblock \href {http://arxiv.org/abs/1505.02677} {\path{arXiv:1505.02677}}, \href {https://doi.org/10.1103/PhysRevC.93.024907} {\path{doi:10.1103/PhysRevC.93.024907}}.

\bibitem{Gale:2012rq}
C.~Gale, S.~Jeon, B.~Schenke, P.~Tribedy, R.~Venugopalan, {Event-by-event anisotropic flow in heavy-ion collisions from combined Yang-Mills and viscous fluid dynamics}, Phys. Rev. Lett. 110~(1) (2013) 012302.
\newblock \href {http://arxiv.org/abs/1209.6330} {\path{arXiv:1209.6330}}, \href {https://doi.org/10.1103/PhysRevLett.110.012302} {\path{doi:10.1103/PhysRevLett.110.012302}}.

\bibitem{Song:2013qma}
H.~Song, S.~Bass, U.~W. Heinz, {Spectra and elliptic flow for identified hadrons in 2.76A TeV Pb + Pb collisions}, Phys. Rev. C 89~(3) (2014) 034919.
\newblock \href {http://arxiv.org/abs/1311.0157} {\path{arXiv:1311.0157}}, \href {https://doi.org/10.1103/PhysRevC.89.034919} {\path{doi:10.1103/PhysRevC.89.034919}}.

\bibitem{Novak:2013bqa}
J.~Novak, K.~Novak, S.~Pratt, J.~Vredevoogd, C.~Coleman-Smith, R.~Wolpert, {Determining Fundamental Properties of Matter Created in Ultrarelativistic Heavy-Ion Collisions}, Phys. Rev. C 89~(3) (2014) 034917.
\newblock \href {http://arxiv.org/abs/1303.5769} {\path{arXiv:1303.5769}}, \href {https://doi.org/10.1103/PhysRevC.89.034917} {\path{doi:10.1103/PhysRevC.89.034917}}.

\bibitem{JETSCAPE:2020mzn}
D.~Everett, et~al., {Multisystem Bayesian constraints on the transport coefficients of QCD matter}, Phys. Rev. C 103~(5) (2021) 054904.
\newblock \href {http://arxiv.org/abs/2011.01430} {\path{arXiv:2011.01430}}, \href {https://doi.org/10.1103/PhysRevC.103.054904} {\path{doi:10.1103/PhysRevC.103.054904}}.

\bibitem{Nijs:2020roc}
G.~Nijs, W.~van~der Schee, U.~G\"ursoy, R.~Snellings, {Bayesian analysis of heavy ion collisions with the heavy ion computational framework Trajectum}, Phys. Rev. C 103~(5) (2021) 054909.
\newblock \href {http://arxiv.org/abs/2010.15134} {\path{arXiv:2010.15134}}, \href {https://doi.org/10.1103/PhysRevC.103.054909} {\path{doi:10.1103/PhysRevC.103.054909}}.

\bibitem{Xu:2017obm}
Y.~Xu, J.~E. Bernhard, S.~A. Bass, M.~Nahrgang, S.~Cao, {Data-driven analysis for the temperature and momentum dependence of the heavy-quark diffusion coefficient in relativistic heavy-ion collisions}, Phys. Rev. C 97~(1) (2018) 014907.
\newblock \href {http://arxiv.org/abs/1710.00807} {\path{arXiv:1710.00807}}, \href {https://doi.org/10.1103/PhysRevC.97.014907} {\path{doi:10.1103/PhysRevC.97.014907}}.

\bibitem{Soltz:2019aea}
R.~Soltz, {Bayesian extraction of $\hat{q}$ with multi-stage jet evolution approach}, PoS HardProbes2018 (2019) 048.
\newblock \href {https://doi.org/10.22323/1.345.0048} {\path{doi:10.22323/1.345.0048}}.

\bibitem{He:2018gks}
Y.~He, L.-G. Pang, X.-N. Wang, {Bayesian extraction of jet energy loss distributions in heavy-ion collisions}, Phys. Rev. Lett. 122~(25) (2019) 252302.
\newblock \href {http://arxiv.org/abs/1808.05310} {\path{arXiv:1808.05310}}, \href {https://doi.org/10.1103/PhysRevLett.122.252302} {\path{doi:10.1103/PhysRevLett.122.252302}}.

\bibitem{Xie:2022ght}
M.~Xie, W.~Ke, H.~Zhang, X.-N. Wang, {Information-field-based global Bayesian inference of the jet transport coefficient}, Phys. Rev. C 108~(1) (2023) L011901.
\newblock \href {http://arxiv.org/abs/2206.01340} {\path{arXiv:2206.01340}}, \href {https://doi.org/10.1103/PhysRevC.108.L011901} {\path{doi:10.1103/PhysRevC.108.L011901}}.

\bibitem{Larkoski:2017jix}
A.~J. Larkoski, I.~Moult, B.~Nachman, {Jet Substructure at the Large Hadron Collider: A Review of Recent Advances in Theory and Machine Learning}, Phys. Rept. 841 (2020) 1--63.
\newblock \href {http://arxiv.org/abs/1709.04464} {\path{arXiv:1709.04464}}, \href {https://doi.org/10.1016/j.physrep.2019.11.001} {\path{doi:10.1016/j.physrep.2019.11.001}}.

\bibitem{Louppe:2017ipp}
G.~Louppe, K.~Cho, C.~Becot, K.~Cranmer, {QCD-Aware Recursive Neural Networks for Jet Physics}, JHEP 01 (2019) 057.
\newblock \href {http://arxiv.org/abs/1702.00748} {\path{arXiv:1702.00748}}, \href {https://doi.org/10.1007/JHEP01(2019)057} {\path{doi:10.1007/JHEP01(2019)057}}.

\bibitem{Dreyer:2021hhr}
F.~A. Dreyer, G.~Soyez, A.~Takacs, {Quarks and gluons in the Lund plane}, JHEP 08 (2022) 177.
\newblock \href {http://arxiv.org/abs/2112.09140} {\path{arXiv:2112.09140}}, \href {https://doi.org/10.1007/JHEP08(2022)177} {\path{doi:10.1007/JHEP08(2022)177}}.

\bibitem{deLima:2021fwm}
R.~T. de~Lima, {Sequence-based Machine Learning Models in Jet Physics} (2 2021).
\newblock \href {http://arxiv.org/abs/2102.06128} {\path{arXiv:2102.06128}}.

\bibitem{Romero:2021qlf}
A.~Romero, D.~Whiteson, M.~Fenton, J.~Collado, P.~Baldi, {Safety of Quark/Gluon Jet Classification} (3 2021).
\newblock \href {http://arxiv.org/abs/2103.09103} {\path{arXiv:2103.09103}}.

\bibitem{Konar:2021zdg}
P.~Konar, V.~S. Ngairangbam, M.~Spannowsky, {Energy-weighted message passing: an infra-red and collinear safe graph neural network algorithm}, JHEP 02 (2022) 060.
\newblock \href {http://arxiv.org/abs/2109.14636} {\path{arXiv:2109.14636}}, \href {https://doi.org/10.1007/JHEP02(2022)060} {\path{doi:10.1007/JHEP02(2022)060}}.

\bibitem{Karagiorgi:2021ngt}
G.~Karagiorgi, G.~Kasieczka, S.~Kravitz, B.~Nachman, D.~Shih, {Machine Learning in the Search for New Fundamental Physics} (12 2021).
\newblock \href {http://arxiv.org/abs/2112.03769} {\path{arXiv:2112.03769}}.

\bibitem{Nguyen:2021xnq}
T.~Q. Nguyen, {Searches for Nonresonant Higgs Boson Pair Production and Long-Lived Particles at the LHC and Machine-Learning Solutions for the High-Luminosity LHC Era}, Ph.D. thesis, Caltech (2021).
\newblock \href {https://doi.org/10.7907/knfz-q495} {\path{doi:10.7907/knfz-q495}}.

\bibitem{Luchmann:2022iih}
M.~A. Luchmann, {Making the most of LHC data - Bayesian neural networks and SMEFT global analysis}, Ph.D. thesis, U. Heidelberg (main) (2022).

\bibitem{Gong:2022lye}
S.~Gong, Q.~Meng, J.~Zhang, H.~Qu, C.~Li, S.~Qian, W.~Du, Z.-M. Ma, T.-Y. Liu, {An efficient Lorentz equivariant graph neural network for jet tagging}, JHEP 07 (2022) 030.
\newblock \href {http://arxiv.org/abs/2201.08187} {\path{arXiv:2201.08187}}, \href {https://doi.org/10.1007/JHEP07(2022)030} {\path{doi:10.1007/JHEP07(2022)030}}.

\bibitem{Bedeschi:2022rnj}
F.~Bedeschi, L.~Gouskos, M.~Selvaggi, {Jet flavour tagging for future colliders with fast simulation}, Eur. Phys. J. C 82~(7) (2022) 646.
\newblock \href {http://arxiv.org/abs/2202.03285} {\path{arXiv:2202.03285}}, \href {https://doi.org/10.1140/epjc/s10052-022-10609-1} {\path{doi:10.1140/epjc/s10052-022-10609-1}}.

\bibitem{Qu:2022mxj}
H.~Qu, C.~Li, S.~Qian, {Particle Transformer for Jet Tagging} (2 2022).
\newblock \href {http://arxiv.org/abs/2202.03772} {\path{arXiv:2202.03772}}.

\bibitem{CMS:2022wjj}
A.~Tumasyan, et~al., {Reconstruction of decays to merged photons using end-to-end deep learning with domain continuation in the CMS detector}, Phys. Rev. D 108~(5) (2023) 052002.
\newblock \href {http://arxiv.org/abs/2204.12313} {\path{arXiv:2204.12313}}, \href {https://doi.org/10.1103/PhysRevD.108.052002} {\path{doi:10.1103/PhysRevD.108.052002}}.

\bibitem{Cal:2022fnm}
P.~Cal, J.~Thaler, W.~J. Waalewijn, {Power counting energy flow polynomials}, JHEP 09 (2022) 021.
\newblock \href {http://arxiv.org/abs/2205.06818} {\path{arXiv:2205.06818}}, \href {https://doi.org/10.1007/JHEP09(2022)021} {\path{doi:10.1007/JHEP09(2022)021}}.

\bibitem{Cranmer:2021gdt}
K.~Cranmer, M.~Drnevich, S.~Macaluso, D.~Pappadopulo, {Reframing Jet Physics with New Computational Methods}, EPJ Web Conf. 251 (2021) 03059.
\newblock \href {http://arxiv.org/abs/2105.10512} {\path{arXiv:2105.10512}}, \href {https://doi.org/10.1051/epjconf/202125103059} {\path{doi:10.1051/epjconf/202125103059}}.

\bibitem{Rossi:2023qvf}
M.~Rossi, {Deep Learning Applications to Particle Physics: From Monte Carlo Simulation Acceleration to Protodune Reconstruction}, Ph.D. thesis, Milan U., Milan U. (2023).
\newblock \href {http://arxiv.org/abs/2302.03343} {\path{arXiv:2302.03343}}.

\bibitem{Chien:2018rgm}
Y.-T. Chien, {Probing heavy ion collisions using quark and gluon jet substructure with machine learning}, Nucl. Phys. A 982 (2019) 619--622.
\newblock \href {http://arxiv.org/abs/1808.04708} {\path{arXiv:1808.04708}}, \href {https://doi.org/10.1016/j.nuclphysa.2018.11.009} {\path{doi:10.1016/j.nuclphysa.2018.11.009}}.

\bibitem{Du:2021qwv}
Y.-L. Du, D.~Pablos, K.~Tywoniuk, {Classification of quark and gluon jets in hot QCD medium with deep learning}, PoS PANIC2021 (2022) 224.
\newblock \href {http://arxiv.org/abs/2112.00681} {\path{arXiv:2112.00681}}, \href {https://doi.org/10.22323/1.380.0224} {\path{doi:10.22323/1.380.0224}}.

\bibitem{Apolinario:2021olp}
L.~Apolin\'ario, N.~F. Castro, M.~Crispim Rom\~ao, J.~G. Milhano, R.~Pedro, F.~C.~R. Peres, {Deep Learning for the classification of quenched jets}, JHEP 11 (2021) 219.
\newblock \href {http://arxiv.org/abs/2106.08869} {\path{arXiv:2106.08869}}, \href {https://doi.org/10.1007/JHEP11(2021)219} {\path{doi:10.1007/JHEP11(2021)219}}.

\bibitem{Du:2020pmp}
Y.-L. Du, D.~Pablos, K.~Tywoniuk, {Deep learning jet modifications in heavy-ion collisions}, JHEP 21 (2020) 206.
\newblock \href {http://arxiv.org/abs/2012.07797} {\path{arXiv:2012.07797}}, \href {https://doi.org/10.1007/JHEP03(2021)206} {\path{doi:10.1007/JHEP03(2021)206}}.

\bibitem{Du:2021pqa}
Y.-L. Du, D.~Pablos, K.~Tywoniuk, {Jet Tomography in Heavy-Ion Collisions with Deep Learning}, Phys. Rev. Lett. 128~(1) (2022) 012301.
\newblock \href {http://arxiv.org/abs/2106.11271} {\path{arXiv:2106.11271}}, \href {https://doi.org/10.1103/PhysRevLett.128.012301} {\path{doi:10.1103/PhysRevLett.128.012301}}.

\bibitem{Yang:2022yfr}
Z.~Yang, Y.~He, W.~Chen, W.-Y. Ke, L.-G. Pang, X.-N. Wang, {Deep learning assisted jet tomography for the study of Mach cones in QGP}, Eur. Phys. J. C 83~(7) (2023) 652.
\newblock \href {http://arxiv.org/abs/2206.02393} {\path{arXiv:2206.02393}}, \href {https://doi.org/10.1140/epjc/s10052-023-11807-1} {\path{doi:10.1140/epjc/s10052-023-11807-1}}.

\bibitem{Lee:2022kdn}
K.~Lee, J.~Mulligan, M.~P\l{}osko\'n, F.~Ringer, F.~Yuan, {Machine learning-based jet and event classification at the Electron-Ion Collider with applications to hadron structure and spin physics}, JHEP 03 (2023) 085.
\newblock \href {http://arxiv.org/abs/2210.06450} {\path{arXiv:2210.06450}}, \href {https://doi.org/10.1007/JHEP03(2023)085} {\path{doi:10.1007/JHEP03(2023)085}}.

\bibitem{Arratia2020CharmJA}
M.~Arratia, Y.~Furletova, T.~Hobbs, F.~I. Olness, S.~J. Sekula, Charm jets as a probe for strangeness at the future electron-ion collider, Physical Review D 103 (2020).

\bibitem{Arrington:2021yeb}
J.~Arrington, et~al., {EIC Physics from An All-Silicon Tracking Detector} (2 2021).
\newblock \href {http://arxiv.org/abs/2102.08337} {\path{arXiv:2102.08337}}.

\bibitem{Caletti:2021ysv}
S.~Caletti, O.~Fedkevych, S.~Marzani, D.~Reichelt, {Tagging the initial-state gluon}, Eur. Phys. J. C 81~(9) (2021) 844.
\newblock \href {http://arxiv.org/abs/2108.10024} {\path{arXiv:2108.10024}}, \href {https://doi.org/10.1140/epjc/s10052-021-09648-x} {\path{doi:10.1140/epjc/s10052-021-09648-x}}.

\bibitem{Baumgardt:1975qv}
H.~G. Baumgardt, J.~U. Schott, Y.~Sakamoto, E.~Schopper, H.~Stoecker, J.~Hofmann, W.~Scheid, W.~Greiner, {Shock Waves and MACH Cones in Fast Nucleus-Nucleus Collisions}, Z. Phys. A 273 (1975) 359--371.
\newblock \href {https://doi.org/10.1007/BF01435578} {\path{doi:10.1007/BF01435578}}.

\bibitem{Rischke:1990jy}
D.~H. Rischke, H.~Stoecker, W.~Greiner, {Flow in Conical Shock Waves: A Signal for the Deconfinement Transition?}, Phys. Rev. D 42 (1990) 2283--2292.
\newblock \href {https://doi.org/10.1103/PhysRevD.42.2283} {\path{doi:10.1103/PhysRevD.42.2283}}.

\bibitem{Casalderrey-Solana:2004fdk}
J.~Casalderrey-Solana, E.~V. Shuryak, D.~Teaney, {Conical flow induced by quenched QCD jets}, J. Phys. Conf. Ser. 27 (2005) 22--31.
\newblock \href {http://arxiv.org/abs/hep-ph/0411315} {\path{arXiv:hep-ph/0411315}}, \href {https://doi.org/10.1088/1742-6596/27/1/003} {\path{doi:10.1088/1742-6596/27/1/003}}.

\bibitem{Satarov:2005mv}
L.~M. Satarov, H.~Stoecker, I.~N. Mishustin, {Mach shocks induced by partonic jets in expanding quark-gluon plasma}, Phys. Lett. B 627 (2005) 64--70.
\newblock \href {http://arxiv.org/abs/hep-ph/0505245} {\path{arXiv:hep-ph/0505245}}, \href {https://doi.org/10.1016/j.physletb.2005.08.102} {\path{doi:10.1016/j.physletb.2005.08.102}}.

\bibitem{Dremin:2005an}
I.~M. Dremin, {Ring-like events: Cerenkov gluons or mach waves?}, Nucl. Phys. A 767 (2006) 233--247.
\newblock \href {http://arxiv.org/abs/hep-ph/0507167} {\path{arXiv:hep-ph/0507167}}, \href {https://doi.org/10.1016/j.nuclphysa.2005.12.015} {\path{doi:10.1016/j.nuclphysa.2005.12.015}}.

\bibitem{Koch:2005sx}
V.~Koch, A.~Majumder, X.-N. Wang, {Cerenkov radiation from jets in heavy-ion collisions}, Phys. Rev. Lett. 96 (2006) 172302.
\newblock \href {http://arxiv.org/abs/nucl-th/0507063} {\path{arXiv:nucl-th/0507063}}, \href {https://doi.org/10.1103/PhysRevLett.96.172302} {\path{doi:10.1103/PhysRevLett.96.172302}}.

\bibitem{Ma:2006mz}
G.~L. Ma, et~al., {Time evolution of Mach-like structure in a partonic transport model} (10 2006).
\newblock \href {http://arxiv.org/abs/nucl-th/0610088} {\path{arXiv:nucl-th/0610088}}.

\bibitem{Gubser:2007ga}
S.~S. Gubser, S.~S. Pufu, A.~Yarom, {Sonic booms and diffusion wakes generated by a heavy quark in thermal AdS/CFT}, Phys. Rev. Lett. 100 (2008) 012301.
\newblock \href {http://arxiv.org/abs/0706.4307} {\path{arXiv:0706.4307}}, \href {https://doi.org/10.1103/PhysRevLett.100.012301} {\path{doi:10.1103/PhysRevLett.100.012301}}.

\bibitem{Betz:2008js}
B.~Betz, M.~Gyulassy, D.~H. Rischke, H.~Stocker, G.~Torrieri, {Jet Propagation and Mach Cones in (3+1)d Ideal Hydrodynamics}, J. Phys. G 35 (2008) 104106.
\newblock \href {http://arxiv.org/abs/0804.4408} {\path{arXiv:0804.4408}}, \href {https://doi.org/10.1088/0954-3899/35/10/104106} {\path{doi:10.1088/0954-3899/35/10/104106}}.

\bibitem{Neufeld:2008dx}
R.~B. Neufeld, {Mach cones in the quark-gluon plasma: Viscosity, speed of sound, and effects of finite source structure}, Phys. Rev. C 79 (2009) 054909.
\newblock \href {http://arxiv.org/abs/0807.2996} {\path{arXiv:0807.2996}}, \href {https://doi.org/10.1103/PhysRevC.79.054909} {\path{doi:10.1103/PhysRevC.79.054909}}.

\bibitem{Torrieri:2008aqg}
G.~Torrieri, B.~Betz, J.~Noronha, M.~Gyulassy, {Mach cones in heavy ion collisions}, Acta Phys. Polon. B 39 (2008) 3281--3308.
\newblock \href {http://arxiv.org/abs/0901.0230} {\path{arXiv:0901.0230}}.

\bibitem{Qin:2009uh}
G.~Y. Qin, A.~Majumder, H.~Song, U.~Heinz, {Energy and momentum deposited into a QCD medium by a jet shower}, Phys. Rev. Lett. 103 (2009) 152303.
\newblock \href {http://arxiv.org/abs/0903.2255} {\path{arXiv:0903.2255}}, \href {https://doi.org/10.1103/PhysRevLett.103.152303} {\path{doi:10.1103/PhysRevLett.103.152303}}.

\bibitem{Roy:2009cc}
V.~Roy, A.~K. Chaudhuri, {Equation of state dependence of Mach cone like structures in Au+Au collisions}, J. Phys. G 37 (2010) 035105.
\newblock \href {http://arxiv.org/abs/0905.2807} {\path{arXiv:0905.2807}}, \href {https://doi.org/10.1088/0954-3899/37/3/035105} {\path{doi:10.1088/0954-3899/37/3/035105}}.

\bibitem{Tachibana:2015qxa}
Y.~Tachibana, T.~Hirano, {Interplay between Mach cone and radial expansion and its signal in \ensuremath{\gamma}-jet events}, Phys. Rev. C 93~(5) (2016) 054907.
\newblock \href {http://arxiv.org/abs/1510.06966} {\path{arXiv:1510.06966}}, \href {https://doi.org/10.1103/PhysRevC.93.054907} {\path{doi:10.1103/PhysRevC.93.054907}}.

\bibitem{Fukushima:2008xe}
K.~Fukushima, D.~E. Kharzeev, H.~J. Warringa, {The Chiral Magnetic Effect}, Phys. Rev. D 78 (2008) 074033.
\newblock \href {http://arxiv.org/abs/0808.3382} {\path{arXiv:0808.3382}}, \href {https://doi.org/10.1103/PhysRevD.78.074033} {\path{doi:10.1103/PhysRevD.78.074033}}.

\bibitem{Kharzeev:2013ffa}
D.~E. Kharzeev, {The Chiral Magnetic Effect and Anomaly-Induced Transport}, Prog. Part. Nucl. Phys. 75 (2014) 133--151.
\newblock \href {http://arxiv.org/abs/1312.3348} {\path{arXiv:1312.3348}}, \href {https://doi.org/10.1016/j.ppnp.2014.01.002} {\path{doi:10.1016/j.ppnp.2014.01.002}}.

\bibitem{Kharzeev:2015znc}
D.~E. Kharzeev, J.~Liao, S.~A. Voloshin, G.~Wang, {Chiral magnetic and vortical effects in high-energy nuclear collisions\textemdash{}A status report}, Prog. Part. Nucl. Phys. 88 (2016) 1--28.
\newblock \href {http://arxiv.org/abs/1511.04050} {\path{arXiv:1511.04050}}, \href {https://doi.org/10.1016/j.ppnp.2016.01.001} {\path{doi:10.1016/j.ppnp.2016.01.001}}.

\bibitem{Zhao:2019hta}
J.~Zhao, F.~Wang, {Experimental searches for the chiral magnetic effect in heavy-ion collisions}, Prog. Part. Nucl. Phys. 107 (2019) 200--236.
\newblock \href {http://arxiv.org/abs/1906.11413} {\path{arXiv:1906.11413}}, \href {https://doi.org/10.1016/j.ppnp.2019.05.001} {\path{doi:10.1016/j.ppnp.2019.05.001}}.

\bibitem{Li:2020dwr}
W.~Li, G.~Wang, {Chiral Magnetic Effects in Nuclear Collisions}, Ann. Rev. Nucl. Part. Sci. 70 (2020) 293--321.
\newblock \href {http://arxiv.org/abs/2002.10397} {\path{arXiv:2002.10397}}, \href {https://doi.org/10.1146/annurev-nucl-030220-065203} {\path{doi:10.1146/annurev-nucl-030220-065203}}.

\bibitem{Zhao:2021yjo}
Y.-S. Zhao, L.~Wang, K.~Zhou, X.-G. Huang, {Detecting the chiral magnetic effect via deep learning}, Phys. Rev. C 106~(5) (2022) L051901.
\newblock \href {http://arxiv.org/abs/2105.13761} {\path{arXiv:2105.13761}}, \href {https://doi.org/10.1103/PhysRevC.106.L051901} {\path{doi:10.1103/PhysRevC.106.L051901}}.

\bibitem{Lin:2004en}
Z.-W. Lin, C.~M. Ko, B.-A. Li, B.~Zhang, S.~Pal, {A Multi-phase transport model for relativistic heavy ion collisions}, Phys. Rev. C 72 (2005) 064901.
\newblock \href {http://arxiv.org/abs/nucl-th/0411110} {\path{arXiv:nucl-th/0411110}}, \href {https://doi.org/10.1103/PhysRevC.72.064901} {\path{doi:10.1103/PhysRevC.72.064901}}.

\bibitem{Ma:2011uma}
G.-L. Ma, B.~Zhang, {Effects of final state interactions on charge separation in relativistic heavy ion collisions}, Phys. Lett. B 700 (2011) 39--43.
\newblock \href {http://arxiv.org/abs/1101.1701} {\path{arXiv:1101.1701}}, \href {https://doi.org/10.1016/j.physletb.2011.04.057} {\path{doi:10.1016/j.physletb.2011.04.057}}.

\bibitem{Jin:2018fwq}
X.~Jin, J.~Chen, Z.~Lin, G.~Ma, Y.~Ma, S.~Zhang, {Explore the QCD phase transition phenomena from a multiphase transport model}, Sci. China Phys. Mech. Astron. 62~(1) (2019) 11012.
\newblock \href {https://doi.org/10.1007/s11433-018-9272-4} {\path{doi:10.1007/s11433-018-9272-4}}.

\bibitem{Shi:2017cpu}
S.~Shi, Y.~Jiang, E.~Lilleskov, J.~Liao, {Anomalous Chiral Transport in Heavy Ion Collisions from Anomalous-Viscous Fluid Dynamics}, Annals Phys. 394 (2018) 50--72.
\newblock \href {http://arxiv.org/abs/1711.02496} {\path{arXiv:1711.02496}}, \href {https://doi.org/10.1016/j.aop.2018.04.026} {\path{doi:10.1016/j.aop.2018.04.026}}.

\bibitem{Shi:2018sah}
S.~Shi, H.~Zhang, D.~Hou, J.~Liao, {Chiral Magnetic Effect in Isobaric Collisions from Anomalous-Viscous Fluid Dynamics (AVFD)}, Nucl. Phys. A 982 (2019) 539--542.
\newblock \href {http://arxiv.org/abs/1807.05604} {\path{arXiv:1807.05604}}, \href {https://doi.org/10.1016/j.nuclphysa.2018.10.007} {\path{doi:10.1016/j.nuclphysa.2018.10.007}}.

\bibitem{Shi:2019wzi}
S.~Shi, H.~Zhang, D.~Hou, J.~Liao, {Signatures of Chiral Magnetic Effect in the Collisions of Isobars}, Phys. Rev. Lett. 125 (2020) 242301.
\newblock \href {http://arxiv.org/abs/1910.14010} {\path{arXiv:1910.14010}}, \href {https://doi.org/10.1103/PhysRevLett.125.242301} {\path{doi:10.1103/PhysRevLett.125.242301}}.

\bibitem{Liu:2019jxg}
Z.~Liu, W.~Zhao, H.~Song, {Principal Component Analysis of collective flow in Relativistic Heavy-Ion Collisions}, Eur. Phys. J. C 79~(10) (2019) 870.
\newblock \href {http://arxiv.org/abs/1903.09833} {\path{arXiv:1903.09833}}, \href {https://doi.org/10.1140/epjc/s10052-019-7379-y} {\path{doi:10.1140/epjc/s10052-019-7379-y}}.

\bibitem{Mallick:2022alr}
N.~Mallick, S.~Prasad, A.~N. Mishra, R.~Sahoo, G.~G. Barnaf\"oldi, {Estimating elliptic flow coefficient in heavy ion collisions using deep learning}, Phys. Rev. D 105~(11) (2022) 114022.
\newblock \href {http://arxiv.org/abs/2203.01246} {\path{arXiv:2203.01246}}, \href {https://doi.org/10.1103/PhysRevD.105.114022} {\path{doi:10.1103/PhysRevD.105.114022}}.

\bibitem{Mallick:2023vgi}
N.~Mallick, S.~Prasad, A.~N. Mishra, R.~Sahoo, G.~G. Barnaf\"oldi, {Deep learning predicted elliptic flow of identified particles in heavy-ion collisions at the RHIC and LHC energies}, Phys. Rev. D 107~(9) (2023) 094001.
\newblock \href {http://arxiv.org/abs/2301.10426} {\path{arXiv:2301.10426}}, \href {https://doi.org/10.1103/PhysRevD.107.094001} {\path{doi:10.1103/PhysRevD.107.094001}}.

\bibitem{higdon2015bayesian}
D.~Higdon, J.~D. McDonnell, N.~Schunck, J.~Sarich, S.~M. Wild, A bayesian approach for parameter estimation and prediction using a computationally intensive model, Journal of Physics G: Nuclear and Particle Physics 42~(3) (2015) 034009.

\bibitem{higdon2008computer}
D.~Higdon, J.~Gattiker, B.~Williams, M.~Rightley, Computer model calibration using high-dimensional output, Journal of the American Statistical Association 103~(482) (2008) 570--583.

\bibitem{Liyanage:2022byj}
D.~Liyanage, Y.~Ji, D.~Everett, M.~Heffernan, U.~Heinz, S.~Mak, J.-F. Paquet, {Efficient emulation of relativistic heavy ion collisions with transfer learning}, Phys. Rev. C 105~(3) (2022) 034910.
\newblock \href {http://arxiv.org/abs/2201.07302} {\path{arXiv:2201.07302}}, \href {https://doi.org/10.1103/PhysRevC.105.034910} {\path{doi:10.1103/PhysRevC.105.034910}}.

\bibitem{Huang:2018fzn}
H.~Huang, B.~Xiao, Z.~Liu, Z.~Wu, Y.~Mu, H.~Song, {Applications of deep learning to relativistic hydrodynamics}, Phys. Rev. Res. 3~(2) (2021) 023256.
\newblock \href {http://arxiv.org/abs/1801.03334} {\path{arXiv:1801.03334}}, \href {https://doi.org/10.1103/PhysRevResearch.3.023256} {\path{doi:10.1103/PhysRevResearch.3.023256}}.

\bibitem{Taradiy:2021pxd}
K.~Taradiy, K.~Zhou, J.~Steinheimer, R.~V. Poberezhnyuk, V.~Vovchenko, H.~Stoecker, {Machine learning based approach to fluid dynamics} (6 2021).
\newblock \href {http://arxiv.org/abs/2106.02841} {\path{arXiv:2106.02841}}.

\bibitem{2021AcMSn..37.1727C}
S.~{Cai}, Z.~{Mao}, Z.~{Wang}, M.~{Yin}, G.~E. {Karniadakis}, {Physics-informed neural networks (PINNs) for fluid mechanics: a review}, Acta Mechanica Sinica 37~(12) (2021) 1727--1738.
\newblock \href {http://arxiv.org/abs/2105.09506} {\path{arXiv:2105.09506}}, \href {https://doi.org/10.1007/s10409-021-01148-1} {\path{doi:10.1007/s10409-021-01148-1}}.

\bibitem{RAO2020207}
C.~Rao, H.~Sun, Y.~Liu, \href{https://www.sciencedirect.com/science/article/pii/S2095034920300350}{Physics-informed deep learning for incompressible laminar flows}, Theoretical and Applied Mechanics Letters 10~(3) (2020) 207--212.
\newblock \href {https://doi.org/https://doi.org/10.1016/j.taml.2020.01.039} {\path{doi:https://doi.org/10.1016/j.taml.2020.01.039}}.
\newline\urlprefix\url{https://www.sciencedirect.com/science/article/pii/S2095034920300350}

\bibitem{BAI2022114740}
J.~Bai, Y.~Zhou, Y.~Ma, H.~Jeong, H.~Zhan, C.~Rathnayaka, E.~Sauret, Y.~Gu, \href{https://www.sciencedirect.com/science/article/pii/S0045782522000962}{A general neural particle method for hydrodynamics modeling}, Computer Methods in Applied Mechanics and Engineering 393 (2022) 114740.
\newblock \href {https://doi.org/https://doi.org/10.1016/j.cma.2022.114740} {\path{doi:https://doi.org/10.1016/j.cma.2022.114740}}.
\newline\urlprefix\url{https://www.sciencedirect.com/science/article/pii/S0045782522000962}

\bibitem{Habashy:2021qku}
D.~M. Habashy, M.~Y. El-Bakry, W.~Scheinast, M.~Hanafy, {Entropy per Rapidity in Pb-Pb Central Collisions using Thermal and Artificial Neural Network (ANN) Models at LHC Energies}, Chin. Phys. C 46~(7) (2022) 073103.
\newblock \href {http://arxiv.org/abs/2110.15026} {\path{arXiv:2110.15026}}, \href {https://doi.org/10.1088/1674-1137/ac5f9d} {\path{doi:10.1088/1674-1137/ac5f9d}}.

\bibitem{Rahman:2022tfq}
R.~M.~A. Rahman, M.~Y. El-Bakry, D.~M. Habashy, A.~N. Tawfik, M.~Hanafy, {Particle ratios with in Hadron Resonance Gas (HRG) and Artificial Neural Network (ANN) models} (1 2022).
\newblock \href {http://arxiv.org/abs/2201.04444} {\path{arXiv:2201.04444}}.

\bibitem{Habashy:2021poi}
D.~M. Habashy, M.~Y. El-Bakry, A.~N. Tawfik, R.~M. Abdel~Rahman, M.~Hanafy, {Particles multiplicity based on rapidity in Landau and artificial neural network (ANN) models}, Int. J. Mod. Phys. A 37~(02) (2022) 2250002.
\newblock \href {http://arxiv.org/abs/2109.07191} {\path{arXiv:2109.07191}}, \href {https://doi.org/10.1142/S0217751X22500026} {\path{doi:10.1142/S0217751X22500026}}.

\bibitem{Yousefnia:2021cup}
K.~V. Yousefnia, A.~Kotibhaskar, R.~Bhalerao, J.-Y. Ollitrault, {Bayesian approach to long-range correlations and multiplicity fluctuations in nucleus-nucleus collisions}, Phys. Rev. C 105~(1) (2022) 014907.
\newblock \href {http://arxiv.org/abs/2108.03471} {\path{arXiv:2108.03471}}, \href {https://doi.org/10.1103/PhysRevC.105.014907} {\path{doi:10.1103/PhysRevC.105.014907}}.

\bibitem{Fischer:2018sdj}
C.~S. Fischer, {QCD at finite temperature and chemical potential from Dyson\textendash{}Schwinger equations}, Prog. Part. Nucl. Phys. 105 (2019) 1--60.
\newblock \href {http://arxiv.org/abs/1810.12938} {\path{arXiv:1810.12938}}, \href {https://doi.org/10.1016/j.ppnp.2019.01.002} {\path{doi:10.1016/j.ppnp.2019.01.002}}.

\bibitem{Wilson:1974sk}
K.~G. Wilson, {Confinement of Quarks}, Phys. Rev. D 10 (1974) 2445--2459.
\newblock \href {https://doi.org/10.1103/PhysRevD.10.2445} {\path{doi:10.1103/PhysRevD.10.2445}}.

\bibitem{Duane:1987de}
S.~Duane, A.~D. Kennedy, B.~J. Pendleton, D.~Roweth, {Hybrid Monte Carlo}, Phys. Lett. B 195 (1987) 216--222.
\newblock \href {https://doi.org/10.1016/0370-2693(87)91197-X} {\path{doi:10.1016/0370-2693(87)91197-X}}.

\bibitem{Knechtli:2017sna}
F.~Knechtli, M.~G\"unther, M.~Peardon, {Lattice Quantum Chromodynamics: Practical Essentials}, SpringerBriefs in Physics, Springer, 2017.
\newblock \href {https://doi.org/10.1007/978-94-024-0999-4} {\path{doi:10.1007/978-94-024-0999-4}}.

\bibitem{Fukushima:2010bq}
K.~Fukushima, T.~Hatsuda, {The phase diagram of dense QCD}, Rept. Prog. Phys. 74 (2011) 014001.
\newblock \href {http://arxiv.org/abs/1005.4814} {\path{arXiv:1005.4814}}, \href {https://doi.org/10.1088/0034-4885/74/1/014001} {\path{doi:10.1088/0034-4885/74/1/014001}}.

\bibitem{Philipsen:2012nu}
O.~Philipsen, {The QCD equation of state from the lattice}, Prog. Part. Nucl. Phys. 70 (2013) 55--107.
\newblock \href {http://arxiv.org/abs/1207.5999} {\path{arXiv:1207.5999}}, \href {https://doi.org/10.1016/j.ppnp.2012.09.003} {\path{doi:10.1016/j.ppnp.2012.09.003}}.

\bibitem{Guenther:2020jwe}
J.~N. Guenther, {Overview of the QCD phase diagram: Recent progress from the lattice}, Eur. Phys. J. A 57~(4) (2021) 136.
\newblock \href {http://arxiv.org/abs/2010.15503} {\path{arXiv:2010.15503}}, \href {https://doi.org/10.1140/epja/s10050-021-00354-6} {\path{doi:10.1140/epja/s10050-021-00354-6}}.

\bibitem{Ratti:2018ksb}
C.~Ratti, {Lattice QCD and heavy ion collisions: a review of recent progress}, Rept. Prog. Phys. 81~(8) (2018) 084301.
\newblock \href {http://arxiv.org/abs/1804.07810} {\path{arXiv:1804.07810}}, \href {https://doi.org/10.1088/1361-6633/aabb97} {\path{doi:10.1088/1361-6633/aabb97}}.

\bibitem{Wolff:1989wq}
U.~Wolff, {CRITICAL SLOWING DOWN}, Nucl. Phys. B Proc. Suppl. 17 (1990) 93--102.
\newblock \href {https://doi.org/10.1016/0920-5632(90)90224-I} {\path{doi:10.1016/0920-5632(90)90224-I}}.

\bibitem{Boyda:2022nmh}
D.~Boyda, et~al., {Applications of Machine Learning to Lattice Quantum Field Theory}, in: {Snowmass 2021}, 2022.
\newblock \href {http://arxiv.org/abs/2202.05838} {\path{arXiv:2202.05838}}.

\bibitem{krauth2006statistical}
W.~Krauth, Statistical mechanics: algorithms and computations, Vol.~13, OUP Oxford, 2006.

\bibitem{Metropolis:1953am}
N.~Metropolis, A.~W. Rosenbluth, M.~N. Rosenbluth, A.~H. Teller, E.~Teller, {Equation of state calculations by fast computing machines}, J. Chem. Phys. 21 (1953) 1087--1092.
\newblock \href {https://doi.org/10.1063/1.1699114} {\path{doi:10.1063/1.1699114}}.

\bibitem{Hastings:1970mcs}
W.~K. Hastings, \href{https://doi.org/10.1093/biomet/57.1.97}{{Monte Carlo sampling methods using Markov chains and their applications}}, Biometrika 57~(1) (1970) 97--109.
\newblock \href {https://doi.org/10.1093/biomet/57.1.97} {\path{doi:10.1093/biomet/57.1.97}}.
\newline\urlprefix\url{https://doi.org/10.1093/biomet/57.1.97}

\bibitem{DelDebbio:2004xh}
L.~Del~Debbio, G.~M. Manca, E.~Vicari, {Critical slowing down of topological modes}, Phys. Lett. B 594 (2004) 315--323.
\newblock \href {http://arxiv.org/abs/hep-lat/0403001} {\path{arXiv:hep-lat/0403001}}, \href {https://doi.org/10.1016/j.physletb.2004.05.038} {\path{doi:10.1016/j.physletb.2004.05.038}}.

\bibitem{Schaefer:2010hu}
S.~Schaefer, R.~Sommer, F.~Virotta, {Critical slowing down and error analysis in lattice QCD simulations}, Nucl. Phys. B 845 (2011) 93--119.
\newblock \href {http://arxiv.org/abs/1009.5228} {\path{arXiv:1009.5228}}, \href {https://doi.org/10.1016/j.nuclphysb.2010.11.020} {\path{doi:10.1016/j.nuclphysb.2010.11.020}}.

\bibitem{Troyer:2004ge}
M.~Troyer, U.-J. Wiese, {Computational complexity and fundamental limitations to fermionic quantum Monte Carlo simulations}, Phys. Rev. Lett. 94 (2005) 170201.
\newblock \href {http://arxiv.org/abs/cond-mat/0408370} {\path{arXiv:cond-mat/0408370}}, \href {https://doi.org/10.1103/PhysRevLett.94.170201} {\path{doi:10.1103/PhysRevLett.94.170201}}.

\bibitem{Aarts:2015tyj}
G.~Aarts, {Introductory lectures on lattice QCD at nonzero baryon number}, J. Phys. Conf. Ser. 706~(2) (2016) 022004.
\newblock \href {http://arxiv.org/abs/1512.05145} {\path{arXiv:1512.05145}}, \href {https://doi.org/10.1088/1742-6596/706/2/022004} {\path{doi:10.1088/1742-6596/706/2/022004}}.

\bibitem{Berger:2019odf}
C.~E. Berger, L.~Rammelm\"uller, A.~C. Loheac, F.~Ehmann, J.~Braun, J.~E. Drut, {Complex Langevin and other approaches to the sign problem in quantum many-body physics}, Phys. Rept. 892 (2021) 1--54.
\newblock \href {http://arxiv.org/abs/1907.10183} {\path{arXiv:1907.10183}}, \href {https://doi.org/10.1016/j.physrep.2020.09.002} {\path{doi:10.1016/j.physrep.2020.09.002}}.

\bibitem{Alexandru:2020wrj}
A.~Alexandru, G.~Basar, P.~F. Bedaque, N.~C. Warrington, {Complex paths around the sign problem}, Rev. Mod. Phys. 94~(1) (2022) 015006.
\newblock \href {http://arxiv.org/abs/2007.05436} {\path{arXiv:2007.05436}}, \href {https://doi.org/10.1103/RevModPhys.94.015006} {\path{doi:10.1103/RevModPhys.94.015006}}.

\bibitem{Hasenbusch:2017fsd}
M.~Hasenbusch, {Fighting topological freezing in the two-dimensional CP$^{N-1}$ model}, EPJ Web Conf. 175 (2018) 02004.
\newblock \href {http://arxiv.org/abs/1709.09460} {\path{arXiv:1709.09460}}, \href {https://doi.org/10.1051/epjconf/201817502004} {\path{doi:10.1051/epjconf/201817502004}}.

\bibitem{2017arXiv171004987L}
Z.~{Liu}, S.~P. {Rodrigues}, W.~{Cai}, {Simulating the Ising Model with a Deep Convolutional Generative Adversarial Network}, arXiv e-prints (2017) arXiv:1710.04987\href {http://arxiv.org/abs/1710.04987} {\path{arXiv:1710.04987}}, \href {https://doi.org/10.48550/arXiv.1710.04987} {\path{doi:10.48550/arXiv.1710.04987}}.

\bibitem{Zhou:2018ill}
K.~Zhou, G.~Endr\H{o}di, L.-G. Pang, H.~St\"ocker, {Regressive and generative neural networks for scalar field theory}, Phys. Rev. D 100~(1) (2019) 011501.
\newblock \href {http://arxiv.org/abs/1810.12879} {\path{arXiv:1810.12879}}, \href {https://doi.org/10.1103/PhysRevD.100.011501} {\path{doi:10.1103/PhysRevD.100.011501}}.

\bibitem{Zhou:2021vza}
K.~Zhou, G.~Endr\H{o}di, L.-G. Pang, H.~St\"ocker, {Neural Network Study for 1+1d-Complex Scalar Field Theory}, Nucl. Phys. A 1005 (2021) 121847.
\newblock \href {https://doi.org/10.1016/j.nuclphysa.2020.121847} {\path{doi:10.1016/j.nuclphysa.2020.121847}}.

\bibitem{Zhou:2020yna}
K.~Zhou, G.~Endrodi, L.-G. Pang, H.~St\"ocker, {Generative Model Study for 1+1d-Complex Scalar Field Theory}, PoS AISIS2019 (2020) 007.
\newblock \href {https://doi.org/10.22323/1.372.0007} {\path{doi:10.22323/1.372.0007}}.

\bibitem{Singha:2021nht}
A.~Singha, D.~Chakrabarti, V.~Arora, {Generative learning for the problem of critical slowing down in lattice Gross-Neveu model}, SciPost Phys. Core 5 (2022) 052.
\newblock \href {http://arxiv.org/abs/2111.00574} {\path{arXiv:2111.00574}}, \href {https://doi.org/10.21468/SciPostPhysCore.5.4.052} {\path{doi:10.21468/SciPostPhysCore.5.4.052}}.

\bibitem{2021ScPP...11...43S}
J.~{Singh}, M.~{Scheurer}, V.~{Arora}, {Conditional generative models for sampling and phase transition indication in spin systems}, SciPost Physics 11~(2) (2021) 043.
\newblock \href {http://arxiv.org/abs/2006.11868} {\path{arXiv:2006.11868}}, \href {https://doi.org/10.21468/SciPostPhys.11.2.043} {\path{doi:10.21468/SciPostPhys.11.2.043}}.

\bibitem{Pawlowski:2018qxs}
J.~M. Pawlowski, J.~M. Urban, {Reducing Autocorrelation Times in Lattice Simulations with Generative Adversarial Networks}, Mach. Learn. Sci. Tech. 1 (2020) 045011.
\newblock \href {http://arxiv.org/abs/1811.03533} {\path{arXiv:1811.03533}}, \href {https://doi.org/10.1088/2632-2153/abae73} {\path{doi:10.1088/2632-2153/abae73}}.

\bibitem{2017PhRvB..95d1101L}
J.~{Liu}, Y.~{Qi}, Z.~Y. {Meng}, L.~{Fu}, {Self-learning Monte Carlo method}, Physical Review B 95~(4) (2017) 041101.
\newblock \href {http://arxiv.org/abs/1610.03137} {\path{arXiv:1610.03137}}, \href {https://doi.org/10.1103/PhysRevB.95.041101} {\path{doi:10.1103/PhysRevB.95.041101}}.

\bibitem{PhysRevB.102.041124}
Y.~Nagai, M.~Okumura, K.~Kobayashi, M.~Shiga, \href{https://link.aps.org/doi/10.1103/PhysRevB.102.041124}{Self-learning hybrid monte carlo: A first-principles approach}, Phys. Rev. B 102 (2020) 041124.
\newblock \href {https://doi.org/10.1103/PhysRevB.102.041124} {\path{doi:10.1103/PhysRevB.102.041124}}.
\newline\urlprefix\url{https://link.aps.org/doi/10.1103/PhysRevB.102.041124}

\bibitem{10.1162/089976602760128018}
G.~E. Hinton, \href{https://doi.org/10.1162/089976602760128018}{{Training Products of Experts by Minimizing Contrastive Divergence}}, Neural Computation 14~(8) (2002) 1771--1800.
\newblock \href {http://arxiv.org/abs/https://direct.mit.edu/neco/article-pdf/14/8/1771/815447/089976602760128018.pdf} {\path{arXiv:https://direct.mit.edu/neco/article-pdf/14/8/1771/815447/089976602760128018.pdf}}, \href {https://doi.org/10.1162/089976602760128018} {\path{doi:10.1162/089976602760128018}}.
\newline\urlprefix\url{https://doi.org/10.1162/089976602760128018}

\bibitem{PhysRevB.95.035105}
L.~Huang, L.~Wang, \href{https://link.aps.org/doi/10.1103/PhysRevB.95.035105}{Accelerated monte carlo simulations with restricted boltzmann machines}, Phys. Rev. B 95 (2017) 035105.
\newblock \href {https://doi.org/10.1103/PhysRevB.95.035105} {\path{doi:10.1103/PhysRevB.95.035105}}.
\newline\urlprefix\url{https://link.aps.org/doi/10.1103/PhysRevB.95.035105}

\bibitem{2016PhRvB..94p5134T}
G.~{Torlai}, R.~G. {Melko}, {Learning thermodynamics with Boltzmann machines}, Phys. Rev. B 94~(16) (2016) 165134.
\newblock \href {http://arxiv.org/abs/1606.02718} {\path{arXiv:1606.02718}}, \href {https://doi.org/10.1103/PhysRevB.94.165134} {\path{doi:10.1103/PhysRevB.94.165134}}.

\bibitem{Wang:2017mzw}
L.~Wang, {Exploring cluster Monte Carlo updates with Boltzmann machines}, Phys. Rev. E 96~(5) (2017) 051301.
\newblock \href {http://arxiv.org/abs/1702.08586} {\path{arXiv:1702.08586}}, \href {https://doi.org/10.1103/PhysRevE.96.051301} {\path{doi:10.1103/PhysRevE.96.051301}}.

\bibitem{2019PhRvE.100d3301P}
S.~{Pilati}, E.~M. {Inack}, P.~{Pieri}, {Self-learning projective quantum Monte Carlo simulations guided by restricted Boltzmann machines}, Phys. Rev. E 100~(4) (2019) 043301.
\newblock \href {http://arxiv.org/abs/1907.00907} {\path{arXiv:1907.00907}}, \href {https://doi.org/10.1103/PhysRevE.100.043301} {\path{doi:10.1103/PhysRevE.100.043301}}.

\bibitem{Liu:2016zmg}
J.~Liu, H.~Shen, Y.~Qi, Z.~Y. Meng, L.~Fu, {Self-learning Monte Carlo method and cumulative update in fermion systems}, Phys. Rev. B 95~(24) (2017) 241104.
\newblock \href {http://arxiv.org/abs/1611.09364} {\path{arXiv:1611.09364}}, \href {https://doi.org/10.1103/PhysRevB.95.241104} {\path{doi:10.1103/PhysRevB.95.241104}}.

\bibitem{2017PhRvB..96d1119X}
X.~Y. Xu, Y.~Qi, J.~Liu, L.~Fu, Z.~Y. Meng, {Self-learning quantum Monte Carlo method in interacting fermion systems}, Phys. Rev. B 96~(4) (2017) 041119.
\newblock \href {http://arxiv.org/abs/1612.03804} {\path{arXiv:1612.03804}}, \href {https://doi.org/10.1103/PhysRevB.96.041119} {\path{doi:10.1103/PhysRevB.96.041119}}.

\bibitem{2017PhRvB..96p1102N}
Y.~{Nagai}, H.~{Shen}, Y.~{Qi}, J.~{Liu}, L.~{Fu}, {Self-learning Monte Carlo method: Continuous-time algorithm}, Physical Review B 96~(16) (2017) 161102.
\newblock \href {http://arxiv.org/abs/1705.06724} {\path{arXiv:1705.06724}}, \href {https://doi.org/10.1103/PhysRevB.96.161102} {\path{doi:10.1103/PhysRevB.96.161102}}.

\bibitem{2018PhRvB..97t5140S}
H.~{Shen}, J.~{Liu}, L.~{Fu}, {Self-learning Monte Carlo with deep neural networks}, Physical Review B 97~(20) (2018) 205140.
\newblock \href {http://arxiv.org/abs/1801.01127} {\path{arXiv:1801.01127}}, \href {https://doi.org/10.1103/PhysRevB.97.205140} {\path{doi:10.1103/PhysRevB.97.205140}}.

\bibitem{Nagai:2018sav}
Y.~Nagai, M.~Okumura, A.~Tanaka, {Self-learning Monte Carlo method with Behler-Parrinello neural networks}, Phys. Rev. B 101~(11) (2020) 115111.
\newblock \href {http://arxiv.org/abs/1807.04955} {\path{arXiv:1807.04955}}, \href {https://doi.org/10.1103/PhysRevB.101.115111} {\path{doi:10.1103/PhysRevB.101.115111}}.

\bibitem{2019PhRvB.100d5153S}
T.~{Song}, H.~{Lee}, {Accelerated continuous time quantum Monte Carlo method with machine learning}, Phys. Rev. B 100~(4) (2019) 045153.
\newblock \href {http://arxiv.org/abs/1901.01501} {\path{arXiv:1901.01501}}, \href {https://doi.org/10.1103/PhysRevB.100.045153} {\path{doi:10.1103/PhysRevB.100.045153}}.

\bibitem{Nagai:2020jar}
Y.~Nagai, A.~Tanaka, A.~Tomiya, {Self-learning Monte Carlo for non-Abelian gauge theory with dynamical fermions}, Phys. Rev. D 107~(5) (2023) 054501.
\newblock \href {http://arxiv.org/abs/2010.11900} {\path{arXiv:2010.11900}}, \href {https://doi.org/10.1103/PhysRevD.107.054501} {\path{doi:10.1103/PhysRevD.107.054501}}.

\bibitem{Tomiya:2021ywc}
Y.~Nagai, A.~Tomiya, {Gauge covariant neural network for 4 dimensional non-abelian gauge theory} (3 2021).
\newblock \href {http://arxiv.org/abs/2103.11965} {\path{arXiv:2103.11965}}.

\bibitem{Shanahan:2018vcv}
P.~E. Shanahan, A.~Trewartha, W.~Detmold, {Machine learning action parameters in lattice quantum chromodynamics}, Phys. Rev. D 97~(9) (2018) 094506.
\newblock \href {http://arxiv.org/abs/1801.05784} {\path{arXiv:1801.05784}}, \href {https://doi.org/10.1103/PhysRevD.97.094506} {\path{doi:10.1103/PhysRevD.97.094506}}.

\bibitem{Peierls:1938zz}
R.~Peierls, {On a Minimum Property of the Free Energy}, Phys. Rev. 54 (1938) 918--919.
\newblock \href {https://doi.org/10.1103/PhysRev.54.918} {\path{doi:10.1103/PhysRev.54.918}}.

\bibitem{10.1063/1.1704383}
K.~Symanzik, \href{https://doi.org/10.1063/1.1704383}{Proof and refinements of an inequality of feynman}, Journal of Mathematical Physics 6~(7) (1965) 1155--1156.
\newblock \href {http://arxiv.org/abs/https://doi.org/10.1063/1.1704383} {\path{arXiv:https://doi.org/10.1063/1.1704383}}, \href {https://doi.org/10.1063/1.1704383} {\path{doi:10.1063/1.1704383}}.
\newline\urlprefix\url{https://doi.org/10.1063/1.1704383}

\bibitem{PhysRevLett.22.631}
J.~L. Lebowitz, E.~H. Lieb, \href{https://link.aps.org/doi/10.1103/PhysRevLett.22.631}{Existence of thermodynamics for real matter with coulomb forces}, Phys. Rev. Lett. 22 (1969) 631--634.
\newblock \href {https://doi.org/10.1103/PhysRevLett.22.631} {\path{doi:10.1103/PhysRevLett.22.631}}.
\newline\urlprefix\url{https://link.aps.org/doi/10.1103/PhysRevLett.22.631}

\bibitem{2019PhRvL.122h0602W}
D.~{Wu}, L.~{Wang}, P.~{Zhang}, {Solving Statistical Mechanics Using Variational Autoregressive Networks}, Physical Review Letter 122~(8) (2019) 080602.
\newblock \href {http://arxiv.org/abs/1809.10606} {\path{arXiv:1809.10606}}, \href {https://doi.org/10.1103/PhysRevLett.122.080602} {\path{doi:10.1103/PhysRevLett.122.080602}}.

\bibitem{Liu:2021bst}
J.-G. Liu, L.~Mao, P.~Zhang, L.~Wang, {Solving quantum statistical mechanics with variational autoregressive networks and quantum circuits}, Mach. Learn. Sci. Tech. 2~(2) (2021) 025011.
\newblock \href {https://doi.org/10.1088/2632-2153/aba19d} {\path{doi:10.1088/2632-2153/aba19d}}.

\bibitem{Tomiya:2022chr}
A.~Tomiya, {Schwinger model at finite temperature and density with beta VQE} (5 2022).
\newblock \href {http://arxiv.org/abs/2205.08860} {\path{arXiv:2205.08860}}.

\bibitem{Wang:2020hji}
L.~Wang, Y.~Jiang, L.~He, K.~Zhou, {Continuous-Mixture Autoregressive Networks Learning the Kosterlitz-Thouless Transition}, Chin. Phys. Lett. 39~(12) (2022) 120502.
\newblock \href {http://arxiv.org/abs/2005.04857} {\path{arXiv:2005.04857}}, \href {https://doi.org/10.1088/0256-307X/39/12/120502} {\path{doi:10.1088/0256-307X/39/12/120502}}.

\bibitem{Wu:2021tfb}
D.~Wu, R.~Rossi, G.~Carleo, {Unbiased Monte Carlo cluster updates with autoregressive neural networks}, Phys. Rev. Res. 3~(4) (2021) L042024.
\newblock \href {http://arxiv.org/abs/2105.05650} {\path{arXiv:2105.05650}}, \href {https://doi.org/10.1103/PhysRevResearch.3.L042024} {\path{doi:10.1103/PhysRevResearch.3.L042024}}.

\bibitem{2018PhRvL.121z0601L}
S.-H. {Li}, L.~{Wang}, {Neural Network Renormalization Group}, Phys. Rev. Lett. 121~(26) (2018) 260601.
\newblock \href {http://arxiv.org/abs/1802.02840} {\path{arXiv:1802.02840}}, \href {https://doi.org/10.1103/PhysRevLett.121.260601} {\path{doi:10.1103/PhysRevLett.121.260601}}.

\bibitem{Nicoli:2020njz}
K.~A. Nicoli, C.~J. Anders, L.~Funcke, T.~Hartung, K.~Jansen, P.~Kessel, S.~Nakajima, P.~Stornati, {Estimation of Thermodynamic Observables in Lattice Field Theories with Deep Generative Models}, Phys. Rev. Lett. 126~(3) (2021) 032001.
\newblock \href {http://arxiv.org/abs/2007.07115} {\path{arXiv:2007.07115}}, \href {https://doi.org/10.1103/PhysRevLett.126.032001} {\path{doi:10.1103/PhysRevLett.126.032001}}.

\bibitem{Muller:2019nis}
T.~M\"{u}ller, B.~Mcwilliams, F.~Rousselle, M.~Gross, J.~Nov\'{a}k, \href{https://doi.org/10.1145/3341156}{Neural importance sampling}, ACM Trans. Graph. 38~(5) (oct 2019).
\newblock \href {https://doi.org/10.1145/3341156} {\path{doi:10.1145/3341156}}.
\newline\urlprefix\url{https://doi.org/10.1145/3341156}

\bibitem{Singha:2023cql}
A.~Singha, D.~Chakrabarti, V.~Arora, {Conditional normalizing flow for Markov chain Monte~Carlo sampling in the critical region of lattice field theory}, Phys. Rev. D 107~(1) (2023) 014512.
\newblock \href {https://doi.org/10.1103/PhysRevD.107.014512} {\path{doi:10.1103/PhysRevD.107.014512}}.

\bibitem{DelDebbio:2021qwf}
L.~Del~Debbio, J.~M. Rossney, M.~Wilson, {Efficient modeling of trivializing maps for lattice \ensuremath{\phi}4 theory using normalizing flows: A first look at scalability}, Phys. Rev. D 104~(9) (2021) 094507.
\newblock \href {http://arxiv.org/abs/2105.12481} {\path{arXiv:2105.12481}}, \href {https://doi.org/10.1103/PhysRevD.104.094507} {\path{doi:10.1103/PhysRevD.104.094507}}.

\bibitem{Hu:2019nea}
H.-Y. Hu, S.-H. Li, L.~Wang, Y.-Z. You, {Machine Learning Holographic Mapping by Neural Network Renormalization Group}, Phys. Rev. Res. 2~(2) (2020) 023369.
\newblock \href {http://arxiv.org/abs/1903.00804} {\path{arXiv:1903.00804}}, \href {https://doi.org/10.1103/PhysRevResearch.2.023369} {\path{doi:10.1103/PhysRevResearch.2.023369}}.

\bibitem{Foreman:2021ixr}
S.~Foreman, X.-Y. Jin, J.~C. Osborn, {Deep Learning Hamiltonian Monte Carlo}, in: {9th International Conference on Learning Representations}, 2021.
\newblock \href {http://arxiv.org/abs/2105.03418} {\path{arXiv:2105.03418}}.

\bibitem{Boyda:2020hsi}
D.~Boyda, G.~Kanwar, S.~Racani\`ere, D.~J. Rezende, M.~S. Albergo, K.~Cranmer, D.~C. Hackett, P.~E. Shanahan, {Sampling using $SU(N)$ gauge equivariant flows}, Phys. Rev. D 103~(7) (2021) 074504.
\newblock \href {http://arxiv.org/abs/2008.05456} {\path{arXiv:2008.05456}}, \href {https://doi.org/10.1103/PhysRevD.103.074504} {\path{doi:10.1103/PhysRevD.103.074504}}.

\bibitem{Albergo:2021bna}
M.~S. Albergo, G.~Kanwar, S.~Racani\`ere, D.~J. Rezende, J.~M. Urban, D.~Boyda, K.~Cranmer, D.~C. Hackett, P.~E. Shanahan, {Flow-based sampling for fermionic lattice field theories}, Phys. Rev. D 104~(11) (2021) 114507.
\newblock \href {http://arxiv.org/abs/2106.05934} {\path{arXiv:2106.05934}}, \href {https://doi.org/10.1103/PhysRevD.104.114507} {\path{doi:10.1103/PhysRevD.104.114507}}.

\bibitem{Abbott:2022zhs}
R.~Abbott, et~al., {Gauge-equivariant flow models for sampling in lattice field theories with pseudofermions}, Phys. Rev. D 106~(7) (2022) 074506.
\newblock \href {http://arxiv.org/abs/2207.08945} {\path{arXiv:2207.08945}}, \href {https://doi.org/10.1103/PhysRevD.106.074506} {\path{doi:10.1103/PhysRevD.106.074506}}.

\bibitem{Luscher:2009eq}
M.~Luscher, {Trivializing maps, the Wilson flow and the HMC algorithm}, Commun. Math. Phys. 293 (2010) 899--919.
\newblock \href {http://arxiv.org/abs/0907.5491} {\path{arXiv:0907.5491}}, \href {https://doi.org/10.1007/s00220-009-0953-7} {\path{doi:10.1007/s00220-009-0953-7}}.

\bibitem{Foreman:2021ljl}
S.~Foreman, T.~Izubuchi, L.~Jin, X.-Y. Jin, J.~C. Osborn, A.~Tomiya, {HMC with Normalizing Flows}, PoS LATTICE2021 (2022) 073.
\newblock \href {http://arxiv.org/abs/2112.01586} {\path{arXiv:2112.01586}}, \href {https://doi.org/10.22323/1.396.0073} {\path{doi:10.22323/1.396.0073}}.

\bibitem{Albandea:2023wgd}
D.~Albandea, L.~Del~Debbio, P.~Hern\'andez, R.~Kenway, J.~Marsh~Rossney, A.~Ramos, {Learning trivializing flows}, Eur. Phys. J. C 83~(7) (2023) 676.
\newblock \href {http://arxiv.org/abs/2302.08408} {\path{arXiv:2302.08408}}, \href {https://doi.org/10.1140/epjc/s10052-023-11838-8} {\path{doi:10.1140/epjc/s10052-023-11838-8}}.

\bibitem{Gerdes:2022eve}
M.~Gerdes, P.~de~Haan, C.~Rainone, R.~Bondesan, M.~C.~N. Cheng, {Learning Lattice Quantum Field Theories with Equivariant Continuous Flows} (7 2022).
\newblock \href {http://arxiv.org/abs/2207.00283} {\path{arXiv:2207.00283}}.

\bibitem{2018arXiv180607366C}
R.~T.~Q. {Chen}, Y.~{Rubanova}, J.~{Bettencourt}, D.~{Duvenaud}, {Neural Ordinary Differential Equations}, arXiv e-prints (2018) arXiv:1806.07366\href {http://arxiv.org/abs/1806.07366} {\path{arXiv:1806.07366}}, \href {https://doi.org/10.48550/arXiv.1806.07366} {\path{doi:10.48550/arXiv.1806.07366}}.

\bibitem{Hackett:2021idh}
D.~C. Hackett, C.-C. Hsieh, M.~S. Albergo, D.~Boyda, J.-W. Chen, K.-F. Chen, K.~Cranmer, G.~Kanwar, P.~E. Shanahan, {Flow-based sampling for multimodal distributions in lattice field theory} (7 2021).
\newblock \href {http://arxiv.org/abs/2107.00734} {\path{arXiv:2107.00734}}.

\bibitem{Chen:2022ytr}
S.~Chen, O.~Savchuk, S.~Zheng, B.~Chen, H.~Stoecker, L.~Wang, K.~Zhou, {Fourier-flow model generating Feynman paths}, Phys. Rev. D 107~(5) (2023) 056001.
\newblock \href {http://arxiv.org/abs/2211.03470} {\path{arXiv:2211.03470}}, \href {https://doi.org/10.1103/PhysRevD.107.056001} {\path{doi:10.1103/PhysRevD.107.056001}}.

\bibitem{Komijani:2023fzy}
J.~Komijani, M.~K. Marinkovic, {Generative models for scalar field theories: how to deal with poor scaling?}, PoS LATTICE2022 (2023) 019.
\newblock \href {http://arxiv.org/abs/2301.01504} {\path{arXiv:2301.01504}}, \href {https://doi.org/10.22323/1.430.0019} {\path{doi:10.22323/1.430.0019}}.

\bibitem{Abbott:2022hkm}
R.~Abbott, et~al., {Sampling QCD field configurations with gauge-equivariant flow models}, PoS LATTICE2022 (2023) 036.
\newblock \href {http://arxiv.org/abs/2208.03832} {\path{arXiv:2208.03832}}, \href {https://doi.org/10.22323/1.430.0036} {\path{doi:10.22323/1.430.0036}}.

\bibitem{Abbott:2022zsh}
R.~Abbott, et~al., {Aspects of scaling and scalability for flow-based sampling of lattice QCD} (11 2022).
\newblock \href {http://arxiv.org/abs/2211.07541} {\path{arXiv:2211.07541}}.

\bibitem{2017NatPh..13..431C}
J.~{Carrasquilla}, R.~G. {Melko}, {Machine learning phases of matter}, Nature Physics 13~(5) (2017) 431--434.
\newblock \href {http://arxiv.org/abs/1605.01735} {\path{arXiv:1605.01735}}, \href {https://doi.org/10.1038/nphys4035} {\path{doi:10.1038/nphys4035}}.

\bibitem{2017NatSR...7.8823B}
P.~{Broecker}, J.~{Carrasquilla}, R.~G. {Melko}, S.~{Trebst}, {Machine learning quantum phases of matter beyond the fermion sign problem}, Scientific Reports 7 (2017) 8823.
\newblock \href {http://arxiv.org/abs/1608.07848} {\path{arXiv:1608.07848}}, \href {https://doi.org/10.1038/s41598-017-09098-0} {\path{doi:10.1038/s41598-017-09098-0}}.

\bibitem{Tanaka:2016rtu}
A.~Tanaka, A.~Tomiya, {Detection of phase transition via convolutional neural network}, J. Phys. Soc. Jap. 86~(6) (2017) 063001.
\newblock \href {http://arxiv.org/abs/1609.09087} {\path{arXiv:1609.09087}}, \href {https://doi.org/10.7566/JPSJ.86.063001} {\path{doi:10.7566/JPSJ.86.063001}}.

\bibitem{Li:2021yet}
X.~Li, R.~Guo, Y.~Zhou, K.~Liu, J.~Zhao, F.~Long, Y.~Wu, Z.~Li, {Machine learning phase transitions of the three-dimensional Ising universality class*}, Chin. Phys. C 47~(3) (2023) 034101.
\newblock \href {http://arxiv.org/abs/2112.13987} {\path{arXiv:2112.13987}}, \href {https://doi.org/10.1088/1674-1137/aca5f5} {\path{doi:10.1088/1674-1137/aca5f5}}.

\bibitem{2016PhRvB..94s5105W}
L.~{Wang}, {Discovering phase transitions with unsupervised learning}, Physical Review B 94~(19) (2016) 195105.
\newblock \href {http://arxiv.org/abs/1606.00318} {\path{arXiv:1606.00318}}, \href {https://doi.org/10.1103/PhysRevB.94.195105} {\path{doi:10.1103/PhysRevB.94.195105}}.

\bibitem{2017PhRvE..95f2122H}
W.~{Hu}, R.~R.~P. {Singh}, R.~T. {Scalettar}, {Discovering phases, phase transitions, and crossovers through unsupervised machine learning: A critical examination}, Physical Review E 95~(6) (2017) 062122.
\newblock \href {http://arxiv.org/abs/1704.00080} {\path{arXiv:1704.00080}}, \href {https://doi.org/10.1103/PhysRevE.95.062122} {\path{doi:10.1103/PhysRevE.95.062122}}.

\bibitem{2017NatPh..13..435V}
E.~P.~L. {van Nieuwenburg}, Y.-H. {Liu}, S.~D. {Huber}, {Learning phase transitions by confusion}, Nature Physics 13~(5) (2017) 435--439.
\newblock \href {http://arxiv.org/abs/1610.02048} {\path{arXiv:1610.02048}}, \href {https://doi.org/10.1038/nphys4037} {\path{doi:10.1038/nphys4037}}.

\bibitem{2018PhRvB..97d5207B}
M.~J.~S. {Beach}, A.~{Golubeva}, R.~G. {Melko}, {Machine learning vortices at the Kosterlitz-Thouless transition}, Physical Review B 97~(4) (2018) 045207.
\newblock \href {http://arxiv.org/abs/1710.09842} {\path{arXiv:1710.09842}}, \href {https://doi.org/10.1103/PhysRevB.97.045207} {\path{doi:10.1103/PhysRevB.97.045207}}.

\bibitem{2017PhRvE..96b2140W}
S.~J. {Wetzel}, {Unsupervised learning of phase transitions: From principal component analysis to variational autoencoders}, Physical Review E 96~(2) (2017) 022140.
\newblock \href {http://arxiv.org/abs/1703.02435} {\path{arXiv:1703.02435}}, \href {https://doi.org/10.1103/PhysRevE.96.022140} {\path{doi:10.1103/PhysRevE.96.022140}}.

\bibitem{2020NatSR..1013047W}
N.~{Walker}, K.-M. {Tam}, M.~{Jarrell}, {Deep learning on the 2-dimensional Ising model to extract the crossover region with a variational autoencoder}, Scientific Reports 10 (2020) 13047.
\newblock \href {http://arxiv.org/abs/2005.13742} {\path{arXiv:2005.13742}}, \href {https://doi.org/10.1038/s41598-020-69848-5} {\path{doi:10.1038/s41598-020-69848-5}}.

\bibitem{2020PhRvL.125q0603K}
K.~{Kottmann}, P.~{Huembeli}, M.~{Lewenstein}, A.~{Ac{\'\i}n}, {Unsupervised Phase Discovery with Deep Anomaly Detection}, Physical Review Letter 125~(17) (2020) 170603.
\newblock \href {http://arxiv.org/abs/2003.09905} {\path{arXiv:2003.09905}}, \href {https://doi.org/10.1103/PhysRevLett.125.170603} {\path{doi:10.1103/PhysRevLett.125.170603}}.

\bibitem{Contessi:2021mrn}
D.~Contessi, E.~Ricci, A.~Recati, M.~Rizzi, {Detection of Berezinskii-Kosterlitz-Thouless transition via Generative Adversarial Networks}, SciPost Phys. 12~(3) (2022) 107.
\newblock \href {http://arxiv.org/abs/2110.05383} {\path{arXiv:2110.05383}}, \href {https://doi.org/10.21468/SciPostPhys.12.3.107} {\path{doi:10.21468/SciPostPhys.12.3.107}}.

\bibitem{2017PhRvB..96t5146P}
P.~{Ponte}, R.~G. {Melko}, {Kernel methods for interpretable machine learning of order parameters}, Phys. Rev. B 96~(20) (2017) 205146.
\newblock \href {http://arxiv.org/abs/1704.05848} {\path{arXiv:1704.05848}}, \href {https://doi.org/10.1103/PhysRevB.96.205146} {\path{doi:10.1103/PhysRevB.96.205146}}.

\bibitem{2019PhRvB..99j4410L}
K.~{Liu}, J.~{Greitemann}, L.~{Pollet}, {Learning multiple order parameters with interpretable machines}, Phys. Rev. B 99~(10) (2019) 104410.
\newblock \href {http://arxiv.org/abs/1810.05538} {\path{arXiv:1810.05538}}, \href {https://doi.org/10.1103/PhysRevB.99.104410} {\path{doi:10.1103/PhysRevB.99.104410}}.

\bibitem{2019PhRvB..99f0404G}
J.~{Greitemann}, K.~{Liu}, L.~{Pollet}, {Probing hidden spin order with interpretable machine learning}, Phys. Rev. B 99~(6) (2019) 060404.
\newblock \href {http://arxiv.org/abs/1804.08557} {\path{arXiv:1804.08557}}, \href {https://doi.org/10.1103/PhysRevB.99.060404} {\path{doi:10.1103/PhysRevB.99.060404}}.

\bibitem{Wetzel:2017ooo}
S.~J. Wetzel, M.~Scherzer, {Machine Learning of Explicit Order Parameters: From the Ising Model to SU(2) Lattice Gauge Theory}, Phys. Rev. B 96~(18) (2017) 184410.
\newblock \href {http://arxiv.org/abs/1705.05582} {\path{arXiv:1705.05582}}, \href {https://doi.org/10.1103/PhysRevB.96.184410} {\path{doi:10.1103/PhysRevB.96.184410}}.

\bibitem{Blucher:2020mjt}
S.~Bl\"ucher, L.~Kades, J.~M. Pawlowski, N.~Strodthoff, J.~M. Urban, {Towards novel insights in lattice field theory with explainable machine learning}, Phys. Rev. D 101~(9) (2020) 094507.
\newblock \href {http://arxiv.org/abs/2003.01504} {\path{arXiv:2003.01504}}, \href {https://doi.org/10.1103/PhysRevD.101.094507} {\path{doi:10.1103/PhysRevD.101.094507}}.

\bibitem{Bachtis:2020dmf}
D.~Bachtis, G.~Aarts, B.~Lucini, {Extending machine learning classification capabilities with histogram reweighting}, Phys. Rev. E 102~(3) (2020) 033303.
\newblock \href {http://arxiv.org/abs/2004.14341} {\path{arXiv:2004.14341}}, \href {https://doi.org/10.1103/PhysRevE.102.033303} {\path{doi:10.1103/PhysRevE.102.033303}}.

\bibitem{Bachtis:2020ajb}
D.~Bachtis, G.~Aarts, B.~Lucini, {Mapping distinct phase transitions to a neural network}, Phys. Rev. E 102~(5) (2020) 053306.
\newblock \href {http://arxiv.org/abs/2007.00355} {\path{arXiv:2007.00355}}, \href {https://doi.org/10.1103/PhysRevE.102.053306} {\path{doi:10.1103/PhysRevE.102.053306}}.

\bibitem{Boyda:2020nfh}
D.~L. Boyda, M.~N. Chernodub, N.~V. Gerasimeniuk, V.~A. Goy, S.~D. Liubimov, A.~V. Molochkov, {Finding the deconfinement temperature in lattice Yang-Mills theories from outside the scaling window with machine learning}, Phys. Rev. D 103~(1) (2021) 014509.
\newblock \href {http://arxiv.org/abs/2009.10971} {\path{arXiv:2009.10971}}, \href {https://doi.org/10.1103/PhysRevD.103.014509} {\path{doi:10.1103/PhysRevD.103.014509}}.

\bibitem{Bulusu:2021rqz}
S.~Bulusu, M.~Favoni, A.~Ipp, D.~I. M\"uller, D.~Schuh, {Generalization capabilities of translationally equivariant neural networks}, Phys. Rev. D 104~(7) (2021) 074504.
\newblock \href {http://arxiv.org/abs/2103.14686} {\path{arXiv:2103.14686}}, \href {https://doi.org/10.1103/PhysRevD.104.074504} {\path{doi:10.1103/PhysRevD.104.074504}}.

\bibitem{Yoon:2018krb}
B.~Yoon, T.~Bhattacharya, R.~Gupta, {Machine Learning Estimators for Lattice QCD Observables}, Phys. Rev. D 100~(1) (2019) 014504.
\newblock \href {http://arxiv.org/abs/1807.05971} {\path{arXiv:1807.05971}}, \href {https://doi.org/10.1103/PhysRevD.100.014504} {\path{doi:10.1103/PhysRevD.100.014504}}.

\bibitem{2016arXiv160207576C}
T.~S. {Cohen}, M.~{Welling}, {Group Equivariant Convolutional Networks}, arXiv e-prints (2016) arXiv:1602.07576\href {http://arxiv.org/abs/1602.07576} {\path{arXiv:1602.07576}}, \href {https://doi.org/10.48550/arXiv.1602.07576} {\path{doi:10.48550/arXiv.1602.07576}}.

\bibitem{2019arXiv190204615C}
T.~S. {Cohen}, M.~{Weiler}, B.~{Kicanaoglu}, M.~{Welling}, {Gauge Equivariant Convolutional Networks and the Icosahedral CNN}, arXiv e-prints (2019) arXiv:1902.04615\href {http://arxiv.org/abs/1902.04615} {\path{arXiv:1902.04615}}, \href {https://doi.org/10.48550/arXiv.1902.04615} {\path{doi:10.48550/arXiv.1902.04615}}.

\bibitem{Bogatskiy:2022hub}
A.~Bogatskiy, et~al., {Symmetry Group Equivariant Architectures for Physics}, in: {Snowmass 2021}, 2022.
\newblock \href {http://arxiv.org/abs/2203.06153} {\path{arXiv:2203.06153}}.

\bibitem{Favoni:2020reg}
M.~Favoni, A.~Ipp, D.~I. M\"uller, D.~Schuh, {Lattice Gauge Equivariant Convolutional Neural Networks}, Phys. Rev. Lett. 128~(3) (2022) 032003.
\newblock \href {http://arxiv.org/abs/2012.12901} {\path{arXiv:2012.12901}}, \href {https://doi.org/10.1103/PhysRevLett.128.032003} {\path{doi:10.1103/PhysRevLett.128.032003}}.

\bibitem{Aronsson:2023rli}
J.~Aronsson, D.~I. M\"uller, D.~Schuh, {Geometrical aspects of lattice gauge equivariant convolutional neural networks} (3 2023).
\newblock \href {http://arxiv.org/abs/2303.11448} {\path{arXiv:2303.11448}}.

\bibitem{2018PhRvB..97h5104C}
J.~{Chen}, S.~{Cheng}, H.~{Xie}, L.~{Wang}, T.~{Xiang}, {Equivalence of restricted Boltzmann machines and tensor network states}, Phys. Rev. B 97~(8) (2018) 085104.
\newblock \href {http://arxiv.org/abs/1701.04831} {\path{arXiv:1701.04831}}, \href {https://doi.org/10.1103/PhysRevB.97.085104} {\path{doi:10.1103/PhysRevB.97.085104}}.

\bibitem{2018PhRvL.121p7204C}
K.~{Choo}, G.~{Carleo}, N.~{Regnault}, T.~{Neupert}, {Symmetries and Many-Body Excitations with Neural-Network Quantum States}, Physical Review Letter 121~(16) (2018) 167204.
\newblock \href {http://arxiv.org/abs/1807.03325} {\path{arXiv:1807.03325}}, \href {https://doi.org/10.1103/PhysRevLett.121.167204} {\path{doi:10.1103/PhysRevLett.121.167204}}.

\bibitem{Noormandipour:2020dqp}
M.~Noormandipour, Y.~Sun, B.~Haghighat, {Restricted Boltzmann machine representation for the groundstate and excited states of Kitaev Honeycomb model}, Mach. Learn. Sci. Tech. 3~(1) (2022) 015010.
\newblock \href {http://arxiv.org/abs/2003.07280} {\path{arXiv:2003.07280}}, \href {https://doi.org/10.1088/2632-2153/ac3ddf} {\path{doi:10.1088/2632-2153/ac3ddf}}.

\bibitem{2020PhRvL.124b0503S}
O.~{Sharir}, Y.~{Levine}, N.~{Wies}, G.~{Carleo}, A.~{Shashua}, {Deep Autoregressive Models for the Efficient Variational Simulation of Many-Body Quantum Systems}, Phys. Rev. Lett. 124~(2) (2020) 020503.
\newblock \href {http://arxiv.org/abs/1902.04057} {\path{arXiv:1902.04057}}, \href {https://doi.org/10.1103/PhysRevLett.124.020503} {\path{doi:10.1103/PhysRevLett.124.020503}}.

\bibitem{Carrasquilla:2021zlj}
J.~Carrasquilla, G.~Torlai, {How To Use Neural Networks To Investigate Quantum Many-Body Physics}, PRX Quantum 2~(4) (2021) 040201.
\newblock \href {https://doi.org/10.1103/PRXQuantum.2.040201} {\path{doi:10.1103/PRXQuantum.2.040201}}.

\bibitem{2021arXiv210111099C}
J.~{Carrasquilla}, G.~{Torlai}, {Neural networks in quantum many-body physics: a hands-on tutorial}, arXiv e-prints (2021) arXiv:2101.11099\href {http://arxiv.org/abs/2101.11099} {\path{arXiv:2101.11099}}, \href {https://doi.org/10.48550/arXiv.2101.11099} {\path{doi:10.48550/arXiv.2101.11099}}.

\bibitem{Asakawa:2000tr}
M.~Asakawa, T.~Hatsuda, Y.~Nakahara, {Maximum entropy analysis of the spectral functions in lattice QCD}, Prog. Part. Nucl. Phys. 46 (2001) 459--508.
\newblock \href {http://arxiv.org/abs/hep-lat/0011040} {\path{arXiv:hep-lat/0011040}}, \href {https://doi.org/10.1016/S0146-6410(01)00150-8} {\path{doi:10.1016/S0146-6410(01)00150-8}}.

\bibitem{Rothkopf:2022fyo}
A.~Rothkopf, {Inverse problems, real-time dynamics and lattice simulations}, EPJ Web Conf. 274 (2022) 01004.
\newblock \href {http://arxiv.org/abs/2211.10680} {\path{arXiv:2211.10680}}, \href {https://doi.org/10.1051/epjconf/202227401004} {\path{doi:10.1051/epjconf/202227401004}}.

\bibitem{Rothkopf:2019ipj}
A.~Rothkopf, {Heavy Quarkonium in Extreme Conditions}, Phys. Rept. 858 (2020) 1--117.
\newblock \href {http://arxiv.org/abs/1912.02253} {\path{arXiv:1912.02253}}, \href {https://doi.org/10.1016/j.physrep.2020.02.006} {\path{doi:10.1016/j.physrep.2020.02.006}}.

\bibitem{Zhao:2020jqu}
J.~Zhao, K.~Zhou, S.~Chen, P.~Zhuang, {Heavy flavors under extreme conditions in high energy nuclear collisions}, Prog. Part. Nucl. Phys. 114 (2020) 103801.
\newblock \href {http://arxiv.org/abs/2005.08277} {\path{arXiv:2005.08277}}, \href {https://doi.org/10.1016/j.ppnp.2020.103801} {\path{doi:10.1016/j.ppnp.2020.103801}}.

\bibitem{Ji:2020ect}
X.~Ji, Y.-S. Liu, Y.~Liu, J.-H. Zhang, Y.~Zhao, {Large-momentum effective theory}, Rev. Mod. Phys. 93~(3) (2021) 035005.
\newblock \href {http://arxiv.org/abs/2004.03543} {\path{arXiv:2004.03543}}, \href {https://doi.org/10.1103/RevModPhys.93.035005} {\path{doi:10.1103/RevModPhys.93.035005}}.

\bibitem{Constantinou:2020pek}
M.~Constantinou, {The x-dependence of hadronic parton distributions: A review on the progress of lattice QCD}, Eur. Phys. J. A 57~(2) (2021) 77.
\newblock \href {http://arxiv.org/abs/2010.02445} {\path{arXiv:2010.02445}}, \href {https://doi.org/10.1140/epja/s10050-021-00353-7} {\path{doi:10.1140/epja/s10050-021-00353-7}}.

\bibitem{Candido:2023nnb}
A.~Candido, L.~Del~Debbio, T.~Giani, G.~Petrillo, {Inverse Problems in PDF determinations}, PoS LATTICE2022 (2023) 098.
\newblock \href {http://arxiv.org/abs/2302.14731} {\path{arXiv:2302.14731}}, \href {https://doi.org/10.22323/1.430.0098} {\path{doi:10.22323/1.430.0098}}.

\bibitem{Shi:2022yqw}
S.~Shi, L.~Wang, K.~Zhou, {Rethinking the ill-posedness of the spectral function reconstruction \textemdash{} Why is it fundamentally hard and how Artificial Neural Networks can help}, Comput. Phys. Commun. 282 (2023) 108547.
\newblock \href {http://arxiv.org/abs/2201.02564} {\path{arXiv:2201.02564}}, \href {https://doi.org/10.1016/j.cpc.2022.108547} {\path{doi:10.1016/j.cpc.2022.108547}}.

\bibitem{J_G_McWhirter_1978}
J.~G. McWhirter, E.~R. Pike, \href{https://dx.doi.org/10.1088/0305-4470/11/9/007}{On the numerical inversion of the laplace transform and similar fredholm integral equations of the first kind}, Journal of Physics A: Mathematical and General 11~(9) (1978) 1729.
\newblock \href {https://doi.org/10.1088/0305-4470/11/9/007} {\path{doi:10.1088/0305-4470/11/9/007}}.
\newline\urlprefix\url{https://dx.doi.org/10.1088/0305-4470/11/9/007}

\bibitem{Wang:2021jou}
L.~Wang, S.~Shi, K.~Zhou, {Reconstructing spectral functions via automatic differentiation}, Phys. Rev. D 106~(5) (2022) L051502.
\newblock \href {http://arxiv.org/abs/2111.14760} {\path{arXiv:2111.14760}}, \href {https://doi.org/10.1103/PhysRevD.106.L051502} {\path{doi:10.1103/PhysRevD.106.L051502}}.

\bibitem{Tikhonov1943OnTS}
A.~N. Tikhonov, On the stability of inverse problems, Proceedings of the USSR Academy of Sciences 39 (1943) 195--198.

\bibitem{tikhonov1995numerical}
A.~N. Tikhonov, A.~Goncharsky, V.~V. Stepanov, A.~G. Yagola, Numerical methods for the solution of ill-posed problems, Vol. 328, Springer Science \& Business Media, 1995.

\bibitem{Otsuki:2017sma}
J.~Otsuki, M.~Ohzeki, H.~Shinaoka, K.~Yoshimi, \href{https://link.aps.org/doi/10.1103/PhysRevE.95.061302}{Sparse modeling approach to analytical continuation of imaginary-time quantum monte carlo data}, Phys. Rev. E 95 (2017) 061302.
\newblock \href {https://doi.org/10.1103/PhysRevE.95.061302} {\path{doi:10.1103/PhysRevE.95.061302}}.
\newline\urlprefix\url{https://link.aps.org/doi/10.1103/PhysRevE.95.061302}

\bibitem{Itou:2020azb}
E.~Itou, Y.~Nagai, {Sparse modeling approach to obtaining the shear viscosity from smeared correlation functions}, JHEP 07 (2020) 007.
\newblock \href {http://arxiv.org/abs/2004.02426} {\path{arXiv:2004.02426}}, \href {https://doi.org/10.1007/JHEP07(2020)007} {\path{doi:10.1007/JHEP07(2020)007}}.

\bibitem{Jarrell:1996rrw}
M.~Jarrell, J.~E. Gubernatis, {Bayesian inference and the analytic continuation of imaginary-time quantum Monte Carlo data}, Phys. Rept. 269 (1996) 133--195.
\newblock \href {https://doi.org/10.1016/0370-1573(95)00074-7} {\path{doi:10.1016/0370-1573(95)00074-7}}.

\bibitem{Burnier:2013nla}
Y.~Burnier, A.~Rothkopf, {Bayesian Approach to Spectral Function Reconstruction for Euclidean Quantum Field Theories}, Phys. Rev. Lett. 111 (2013) 182003.
\newblock \href {http://arxiv.org/abs/1307.6106} {\path{arXiv:1307.6106}}, \href {https://doi.org/10.1103/PhysRevLett.111.182003} {\path{doi:10.1103/PhysRevLett.111.182003}}.

\bibitem{2016arXiv161204895A}
L.-F. {Arsenault}, R.~{Neuberg}, L.~A. {Hannah}, A.~J. {Millis}, {Projected Regression Methods for Inverting Fredholm Integrals: Formalism and Application to Analytical Continuation}, arXiv e-prints (2016) arXiv:1612.04895\href {http://arxiv.org/abs/1612.04895} {\path{arXiv:1612.04895}}, \href {https://doi.org/10.48550/arXiv.1612.04895} {\path{doi:10.48550/arXiv.1612.04895}}.

\bibitem{2018PhRvB..98x5101Y}
H.~{Yoon}, J.-H. {Sim}, M.~J. {Han}, {Analytic continuation via domain knowledge free machine learning}, Physical Review B 98~(24) (2018) 245101.
\newblock \href {http://arxiv.org/abs/1806.03841} {\path{arXiv:1806.03841}}, \href {https://doi.org/10.1103/PhysRevB.98.245101} {\path{doi:10.1103/PhysRevB.98.245101}}.

\bibitem{PhysRevLett.124.056401}
R.~Fournier, L.~Wang, O.~V. Yazyev, Q.~Wu, \href{https://link.aps.org/doi/10.1103/PhysRevLett.124.056401}{Artificial neural network approach to the analytic continuation problem}, Phys. Rev. Lett. 124 (2020) 056401.
\newblock \href {https://doi.org/10.1103/PhysRevLett.124.056401} {\path{doi:10.1103/PhysRevLett.124.056401}}.
\newline\urlprefix\url{https://link.aps.org/doi/10.1103/PhysRevLett.124.056401}

\bibitem{Kades:2019wtd}
L.~Kades, J.~M. Pawlowski, A.~Rothkopf, M.~Scherzer, J.~M. Urban, S.~J. Wetzel, N.~Wink, F.~P.~G. Ziegler, {Spectral Reconstruction with Deep Neural Networks}, Phys. Rev. D 102~(9) (2020) 096001.
\newblock \href {http://arxiv.org/abs/1905.04305} {\path{arXiv:1905.04305}}, \href {https://doi.org/10.1103/PhysRevD.102.096001} {\path{doi:10.1103/PhysRevD.102.096001}}.

\bibitem{Chen:2021giw}
S.~Y. Chen, H.~T. Ding, F.~Y. Liu, G.~Papp, C.~B. Yang, {Machine learning spectral functions in lattice QCD} (10 2021).
\newblock \href {http://arxiv.org/abs/2110.13521} {\path{arXiv:2110.13521}}.

\bibitem{Zhou:2021bvw}
M.~Zhou, F.~Gao, J.~Chao, Y.-X. Liu, H.~Song, {Application of radial basis functions neutral networks in spectral functions}, Phys. Rev. D 104~(7) (2021) 076011.
\newblock \href {http://arxiv.org/abs/2106.08168} {\path{arXiv:2106.08168}}, \href {https://doi.org/10.1103/PhysRevD.104.076011} {\path{doi:10.1103/PhysRevD.104.076011}}.

\bibitem{Horak:2021syv}
J.~Horak, J.~M. Pawlowski, J.~Rodr\'\i{}guez-Quintero, J.~Turnwald, J.~M. Urban, N.~Wink, S.~Zafeiropoulos, {Reconstructing QCD spectral functions with Gaussian processes}, Phys. Rev. D 105~(3) (2022) 036014.
\newblock \href {http://arxiv.org/abs/2107.13464} {\path{arXiv:2107.13464}}, \href {https://doi.org/10.1103/PhysRevD.105.036014} {\path{doi:10.1103/PhysRevD.105.036014}}.

\bibitem{Wang:2021cqw}
L.~Wang, S.~Shi, K.~Zhou, {Automatic differentiation approach for reconstructing spectral functions with neural networks}, in: {35th Conference on Neural Information Processing Systems}, 2021.
\newblock \href {http://arxiv.org/abs/2112.06206} {\path{arXiv:2112.06206}}.

\bibitem{Chen:2012gg}
B.~Chen, K.~Zhou, P.~Zhuang, {Mean Field Effect on $J/\psi$ Production in Heavy Ion Collisions}, Phys. Rev. C 86 (2012) 034906.
\newblock \href {http://arxiv.org/abs/1202.3523} {\path{arXiv:1202.3523}}, \href {https://doi.org/10.1103/PhysRevC.86.034906} {\path{doi:10.1103/PhysRevC.86.034906}}.

\bibitem{Zhao:2010nk}
X.~Zhao, R.~Rapp, {Charmonium in Medium: From Correlators to Experiment}, Phys. Rev. C 82 (2010) 064905.
\newblock \href {http://arxiv.org/abs/1008.5328} {\path{arXiv:1008.5328}}, \href {https://doi.org/10.1103/PhysRevC.82.064905} {\path{doi:10.1103/PhysRevC.82.064905}}.

\bibitem{Zhou:2014kka}
K.~Zhou, N.~Xu, Z.~Xu, P.~Zhuang, {Medium effects on charmonium production at ultrarelativistic energies available at the CERN Large Hadron Collider}, Phys. Rev. C 89~(5) (2014) 054911.
\newblock \href {http://arxiv.org/abs/1401.5845} {\path{arXiv:1401.5845}}, \href {https://doi.org/10.1103/PhysRevC.89.054911} {\path{doi:10.1103/PhysRevC.89.054911}}.

\bibitem{CMS:2011all}
S.~Chatrchyan, et~al., {Indications of suppression of excited $\Upsilon$ states in PbPb collisions at $\sqrt{S_{NN}}$ = 2.76 TeV}, Phys. Rev. Lett. 107 (2011) 052302.
\newblock \href {http://arxiv.org/abs/1105.4894} {\path{arXiv:1105.4894}}, \href {https://doi.org/10.1103/PhysRevLett.107.052302} {\path{doi:10.1103/PhysRevLett.107.052302}}.

\bibitem{CMS:2012gvv}
S.~Chatrchyan, et~al., {Observation of Sequential Upsilon Suppression in PbPb Collisions}, Phys. Rev. Lett. 109 (2012) 222301, [Erratum: Phys.Rev.Lett. 120, 199903 (2018)].
\newblock \href {http://arxiv.org/abs/1208.2826} {\path{arXiv:1208.2826}}, \href {https://doi.org/10.1103/PhysRevLett.109.222301} {\path{doi:10.1103/PhysRevLett.109.222301}}.

\bibitem{Laine:2006ns}
M.~Laine, O.~Philipsen, P.~Romatschke, M.~Tassler, {Real-time static potential in hot QCD}, JHEP 03 (2007) 054.
\newblock \href {http://arxiv.org/abs/hep-ph/0611300} {\path{arXiv:hep-ph/0611300}}, \href {https://doi.org/10.1088/1126-6708/2007/03/054} {\path{doi:10.1088/1126-6708/2007/03/054}}.

\bibitem{Beraudo:2007ky}
A.~Beraudo, J.~P. Blaizot, C.~Ratti, {Real and imaginary-time Q anti-Q correlators in a thermal medium}, Nucl. Phys. A 806 (2008) 312--338.
\newblock \href {http://arxiv.org/abs/0712.4394} {\path{arXiv:0712.4394}}, \href {https://doi.org/10.1016/j.nuclphysa.2008.03.001} {\path{doi:10.1016/j.nuclphysa.2008.03.001}}.

\bibitem{Brambilla:2008cx}
N.~Brambilla, J.~Ghiglieri, A.~Vairo, P.~Petreczky, {Static quark-antiquark pairs at finite temperature}, Phys. Rev. D 78 (2008) 014017.
\newblock \href {http://arxiv.org/abs/0804.0993} {\path{arXiv:0804.0993}}, \href {https://doi.org/10.1103/PhysRevD.78.014017} {\path{doi:10.1103/PhysRevD.78.014017}}.

\bibitem{Brambilla:2010vq}
N.~Brambilla, M.~A. Escobedo, J.~Ghiglieri, J.~Soto, A.~Vairo, {Heavy Quarkonium in a weakly-coupled quark-gluon plasma below the melting temperature}, JHEP 09 (2010) 038.
\newblock \href {http://arxiv.org/abs/1007.4156} {\path{arXiv:1007.4156}}, \href {https://doi.org/10.1007/JHEP09(2010)038} {\path{doi:10.1007/JHEP09(2010)038}}.

\bibitem{Rothkopf:2011db}
A.~Rothkopf, T.~Hatsuda, S.~Sasaki, {Complex Heavy-Quark Potential at Finite Temperature from Lattice QCD}, Phys. Rev. Lett. 108 (2012) 162001.
\newblock \href {http://arxiv.org/abs/1108.1579} {\path{arXiv:1108.1579}}, \href {https://doi.org/10.1103/PhysRevLett.108.162001} {\path{doi:10.1103/PhysRevLett.108.162001}}.

\bibitem{Burnier:2014ssa}
Y.~Burnier, O.~Kaczmarek, A.~Rothkopf, {Static quark-antiquark potential in the quark-gluon plasma from lattice QCD}, Phys. Rev. Lett. 114~(8) (2015) 082001.
\newblock \href {http://arxiv.org/abs/1410.2546} {\path{arXiv:1410.2546}}, \href {https://doi.org/10.1103/PhysRevLett.114.082001} {\path{doi:10.1103/PhysRevLett.114.082001}}.

\bibitem{Burnier:2015tda}
Y.~Burnier, O.~Kaczmarek, A.~Rothkopf, {Quarkonium at finite temperature: Towards realistic phenomenology from first principles}, JHEP 12 (2015) 101.
\newblock \href {http://arxiv.org/abs/1509.07366} {\path{arXiv:1509.07366}}, \href {https://doi.org/10.1007/JHEP12(2015)101} {\path{doi:10.1007/JHEP12(2015)101}}.

\bibitem{Larsen:2019bwy}
R.~Larsen, S.~Meinel, S.~Mukherjee, P.~Petreczky, {Thermal broadening of bottomonia: Lattice nonrelativistic QCD with extended operators}, Phys. Rev. D 100~(7) (2019) 074506.
\newblock \href {http://arxiv.org/abs/1908.08437} {\path{arXiv:1908.08437}}, \href {https://doi.org/10.1103/PhysRevD.100.074506} {\path{doi:10.1103/PhysRevD.100.074506}}.

\bibitem{Larsen:2019zqv}
R.~Larsen, S.~Meinel, S.~Mukherjee, P.~Petreczky, {Excited bottomonia in quark-gluon plasma from lattice QCD}, Phys. Lett. B 800 (2020) 135119.
\newblock \href {http://arxiv.org/abs/1910.07374} {\path{arXiv:1910.07374}}, \href {https://doi.org/10.1016/j.physletb.2019.135119} {\path{doi:10.1016/j.physletb.2019.135119}}.

\bibitem{Larsen:2020rjk}
R.~Larsen, S.~Meinel, S.~Mukherjee, P.~Petreczky, {Bethe-Salpeter amplitudes of Upsilons}, Phys. Rev. D 102 (2020) 114508.
\newblock \href {http://arxiv.org/abs/2008.00100} {\path{arXiv:2008.00100}}, \href {https://doi.org/10.1103/PhysRevD.102.114508} {\path{doi:10.1103/PhysRevD.102.114508}}.

\bibitem{Shi:2021qri}
S.~Shi, K.~Zhou, J.~Zhao, S.~Mukherjee, P.~Zhuang, {Heavy quark potential in the quark-gluon plasma: Deep neural network meets lattice quantum chromodynamics}, Phys. Rev. D 105~(1) (2022) 014017.
\newblock \href {http://arxiv.org/abs/2105.07862} {\path{arXiv:2105.07862}}, \href {https://doi.org/10.1103/PhysRevD.105.014017} {\path{doi:10.1103/PhysRevD.105.014017}}.

\bibitem{Lin:2017snn}
H.-W. Lin, et~al., {Parton distributions and lattice QCD calculations: a community white paper}, Prog. Part. Nucl. Phys. 100 (2018) 107--160.
\newblock \href {http://arxiv.org/abs/1711.07916} {\path{arXiv:1711.07916}}, \href {https://doi.org/10.1016/j.ppnp.2018.01.007} {\path{doi:10.1016/j.ppnp.2018.01.007}}.

\bibitem{Forte:2020yip}
S.~Forte, S.~Carrazza, {Parton distribution functions} (8 2020).
\newblock \href {http://arxiv.org/abs/2008.12305} {\path{arXiv:2008.12305}}.

\bibitem{Forte:2002fg}
S.~Forte, L.~Garrido, J.~I. Latorre, A.~Piccione, {Neural network parametrization of deep inelastic structure functions}, JHEP 05 (2002) 062.
\newblock \href {http://arxiv.org/abs/hep-ph/0204232} {\path{arXiv:hep-ph/0204232}}, \href {https://doi.org/10.1088/1126-6708/2002/05/062} {\path{doi:10.1088/1126-6708/2002/05/062}}.

\bibitem{Ball:2009mk}
R.~D. Ball, L.~Del~Debbio, S.~Forte, A.~Guffanti, J.~I. Latorre, A.~Piccione, J.~Rojo, M.~Ubiali, {Precision determination of electroweak parameters and the strange content of the proton from neutrino deep-inelastic scattering}, Nucl. Phys. B 823 (2009) 195--233.
\newblock \href {http://arxiv.org/abs/0906.1958} {\path{arXiv:0906.1958}}, \href {https://doi.org/10.1016/j.nuclphysb.2009.08.003} {\path{doi:10.1016/j.nuclphysb.2009.08.003}}.

\bibitem{Ball:2010de}
R.~D. Ball, L.~Del~Debbio, S.~Forte, A.~Guffanti, J.~I. Latorre, J.~Rojo, M.~Ubiali, {A first unbiased global NLO determination of parton distributions and their uncertainties}, Nucl. Phys. B 838 (2010) 136--206.
\newblock \href {http://arxiv.org/abs/1002.4407} {\path{arXiv:1002.4407}}, \href {https://doi.org/10.1016/j.nuclphysb.2010.05.008} {\path{doi:10.1016/j.nuclphysb.2010.05.008}}.

\bibitem{Ball:2012cx}
R.~D. Ball, et~al., {Parton distributions with LHC data}, Nucl. Phys. B 867 (2013) 244--289.
\newblock \href {http://arxiv.org/abs/1207.1303} {\path{arXiv:1207.1303}}, \href {https://doi.org/10.1016/j.nuclphysb.2012.10.003} {\path{doi:10.1016/j.nuclphysb.2012.10.003}}.

\bibitem{NNPDF:2014otw}
R.~D. Ball, et~al., {Parton distributions for the LHC Run II}, JHEP 04 (2015) 040.
\newblock \href {http://arxiv.org/abs/1410.8849} {\path{arXiv:1410.8849}}, \href {https://doi.org/10.1007/JHEP04(2015)040} {\path{doi:10.1007/JHEP04(2015)040}}.

\bibitem{Bertone:2017tyb}
V.~Bertone, S.~Carrazza, N.~P. Hartland, E.~R. Nocera, J.~Rojo, {A determination of the fragmentation functions of pions, kaons, and protons with faithful uncertainties}, Eur. Phys. J. C 77~(8) (2017) 516.
\newblock \href {http://arxiv.org/abs/1706.07049} {\path{arXiv:1706.07049}}, \href {https://doi.org/10.1140/epjc/s10052-017-5088-y} {\path{doi:10.1140/epjc/s10052-017-5088-y}}.

\bibitem{AbdulKhalek:2019mzd}
R.~Abdul~Khalek, J.~J. Ethier, J.~Rojo, {Nuclear parton distributions from lepton-nucleus scattering and the impact of an electron-ion collider}, Eur. Phys. J. C 79~(6) (2019) 471.
\newblock \href {http://arxiv.org/abs/1904.00018} {\path{arXiv:1904.00018}}, \href {https://doi.org/10.1140/epjc/s10052-019-6983-1} {\path{doi:10.1140/epjc/s10052-019-6983-1}}.

\bibitem{AbdulKhalek:2020yuc}
R.~Abdul~Khalek, J.~J. Ethier, J.~Rojo, G.~van Weelden, {nNNPDF2.0: quark flavor separation in nuclei from LHC data}, JHEP 09 (2020) 183.
\newblock \href {http://arxiv.org/abs/2006.14629} {\path{arXiv:2006.14629}}, \href {https://doi.org/10.1007/JHEP09(2020)183} {\path{doi:10.1007/JHEP09(2020)183}}.

\bibitem{AbdulKhalek:2022fyi}
R.~Abdul~Khalek, R.~Gauld, T.~Giani, E.~R. Nocera, T.~R. Rabemananjara, J.~Rojo, {nNNPDF3.0: evidence for a modified partonic structure in heavy nuclei}, Eur. Phys. J. C 82~(6) (2022) 507.
\newblock \href {http://arxiv.org/abs/2201.12363} {\path{arXiv:2201.12363}}, \href {https://doi.org/10.1140/epjc/s10052-022-10417-7} {\path{doi:10.1140/epjc/s10052-022-10417-7}}.

\bibitem{Karpie:2019eiq}
J.~Karpie, K.~Orginos, A.~Rothkopf, S.~Zafeiropoulos, {Reconstructing parton distribution functions from Ioffe time data: from Bayesian methods to Neural Networks}, JHEP 04 (2019) 057.
\newblock \href {http://arxiv.org/abs/1901.05408} {\path{arXiv:1901.05408}}, \href {https://doi.org/10.1007/JHEP04(2019)057} {\path{doi:10.1007/JHEP04(2019)057}}.

\bibitem{Gao:2022iex}
X.~Gao, A.~D. Hanlon, N.~Karthik, S.~Mukherjee, P.~Petreczky, P.~Scior, S.~Shi, S.~Syritsyn, Y.~Zhao, K.~Zhou, {Continuum-extrapolated NNLO valence PDF of the pion at the physical point}, Phys. Rev. D 106~(11) (2022) 114510.
\newblock \href {http://arxiv.org/abs/2208.02297} {\path{arXiv:2208.02297}}, \href {https://doi.org/10.1103/PhysRevD.106.114510} {\path{doi:10.1103/PhysRevD.106.114510}}.

\bibitem{Forte:2002us}
S.~Forte, J.~I. Latorre, L.~Magnea, A.~Piccione, {Determination of alpha(s) from scaling violations of truncated moments of structure functions}, Nucl. Phys. B 643 (2002) 477--500.
\newblock \href {http://arxiv.org/abs/hep-ph/0205286} {\path{arXiv:hep-ph/0205286}}, \href {https://doi.org/10.1016/S0550-3213(02)00688-0} {\path{doi:10.1016/S0550-3213(02)00688-0}}.

\bibitem{Zhang:2019qiq}
R.~Zhang, Z.~Fan, R.~Li, H.-W. Lin, B.~Yoon, {Machine-learning prediction for quasiparton distribution function matrix elements}, Phys. Rev. D 101~(3) (2020) 034516.
\newblock \href {http://arxiv.org/abs/1909.10990} {\path{arXiv:1909.10990}}, \href {https://doi.org/10.1103/PhysRevD.101.034516} {\path{doi:10.1103/PhysRevD.101.034516}}.

\bibitem{DelDebbio:2020rgv}
L.~Del~Debbio, T.~Giani, J.~Karpie, K.~Orginos, A.~Radyushkin, S.~Zafeiropoulos, {Neural-network analysis of Parton Distribution Functions from Ioffe-time pseudodistributions}, JHEP 02 (2021) 138.
\newblock \href {http://arxiv.org/abs/2010.03996} {\path{arXiv:2010.03996}}, \href {https://doi.org/10.1007/JHEP02(2021)138} {\path{doi:10.1007/JHEP02(2021)138}}.

\bibitem{DelDebbio:2021whr}
L.~Del~Debbio, T.~Giani, M.~Wilson, {Bayesian approach to inverse problems: an application to NNPDF closure testing}, Eur. Phys. J. C 82~(4) (2022) 330.
\newblock \href {http://arxiv.org/abs/2111.05787} {\path{arXiv:2111.05787}}, \href {https://doi.org/10.1140/epjc/s10052-022-10297-x} {\path{doi:10.1140/epjc/s10052-022-10297-x}}.

\bibitem{Ferrenberg:1988yz}
A.~M. Ferrenberg, R.~H. Swendsen, {New Monte Carlo Technique for Studying Phase Transitions}, Phys. Rev. Lett. 61 (1988) 2635--2638.
\newblock \href {https://doi.org/10.1103/PhysRevLett.61.2635} {\path{doi:10.1103/PhysRevLett.61.2635}}.

\bibitem{Wang:2000fzi}
F.~Wang, D.~P. Landau, {Efficient, Multiple-Range Random Walk Algorithm to Calculate the Density of States}, Phys. Rev. Lett. 86~(10) (2001) 2050.
\newblock \href {http://arxiv.org/abs/cond-mat/0011174} {\path{arXiv:cond-mat/0011174}}, \href {https://doi.org/10.1103/PhysRevLett.86.2050} {\path{doi:10.1103/PhysRevLett.86.2050}}.

\bibitem{Allton:2002zi}
C.~R. Allton, S.~Ejiri, S.~J. Hands, O.~Kaczmarek, F.~Karsch, E.~Laermann, C.~Schmidt, L.~Scorzato, {The QCD thermal phase transition in the presence of a small chemical potential}, Phys. Rev. D 66 (2002) 074507.
\newblock \href {http://arxiv.org/abs/hep-lat/0204010} {\path{arXiv:hep-lat/0204010}}, \href {https://doi.org/10.1103/PhysRevD.66.074507} {\path{doi:10.1103/PhysRevD.66.074507}}.

\bibitem{Borsanyi:2015axp}
S.~Bors\'anyi, {Fluctuations at finite temperature and density}, PoS LATTICE2015 (2016) 015.
\newblock \href {http://arxiv.org/abs/1511.06541} {\path{arXiv:1511.06541}}, \href {https://doi.org/10.22323/1.251.0015} {\path{doi:10.22323/1.251.0015}}.

\bibitem{deForcrand:2002hgr}
P.~de~Forcrand, O.~Philipsen, {The QCD phase diagram for small densities from imaginary chemical potential}, Nucl. Phys. B 642 (2002) 290--306.
\newblock \href {http://arxiv.org/abs/hep-lat/0205016} {\path{arXiv:hep-lat/0205016}}, \href {https://doi.org/10.1016/S0550-3213(02)00626-0} {\path{doi:10.1016/S0550-3213(02)00626-0}}.

\bibitem{deForcrand:2009zkb}
P.~de~Forcrand, {Simulating QCD at finite density}, PoS LAT2009 (2009) 010.
\newblock \href {http://arxiv.org/abs/1005.0539} {\path{arXiv:1005.0539}}, \href {https://doi.org/10.22323/1.091.0010} {\path{doi:10.22323/1.091.0010}}.

\bibitem{Rossi:1984cv}
P.~Rossi, U.~Wolff, {Lattice {QCD} With Fermions at Strong Coupling: A Dimer System}, Nucl. Phys. B 248 (1984) 105--122.
\newblock \href {https://doi.org/10.1016/0550-3213(84)90589-3} {\path{doi:10.1016/0550-3213(84)90589-3}}.

\bibitem{Parisi:1980ys}
G.~Parisi, Y.-s. Wu, {Perturbation Theory Without Gauge Fixing}, Sci. Sin. 24 (1981) 483.

\bibitem{Aarts:2013uxa}
G.~Aarts, L.~Bongiovanni, E.~Seiler, D.~Sexty, I.-O. Stamatescu, {Controlling complex Langevin dynamics at finite density}, Eur. Phys. J. A 49 (2013) 89.
\newblock \href {http://arxiv.org/abs/1303.6425} {\path{arXiv:1303.6425}}, \href {https://doi.org/10.1140/epja/i2013-13089-4} {\path{doi:10.1140/epja/i2013-13089-4}}.

\bibitem{Attanasio:2020spv}
F.~Attanasio, B.~J\"ager, F.~P.~G. Ziegler, {Complex Langevin simulations and the QCD phase diagram: Recent developments}, Eur. Phys. J. A 56~(10) (2020) 251.
\newblock \href {http://arxiv.org/abs/2006.00476} {\path{arXiv:2006.00476}}, \href {https://doi.org/10.1140/epja/s10050-020-00256-z} {\path{doi:10.1140/epja/s10050-020-00256-z}}.

\bibitem{1997CPL...270..382R}
N.~{Rom}, D.~M. {Charutz}, D.~{Neuhauser}, {Shifted-contour auxiliary-field Monte Carlo: circumventing the sign difficulty for electronic-structure calculations}, Chemical Physics Letters 270~(3) (1997) 382--386.
\newblock \href {https://doi.org/10.1016/S0009-2614(97)00370-9} {\path{doi:10.1016/S0009-2614(97)00370-9}}.

\bibitem{Cristoforetti:2012su}
M.~Cristoforetti, F.~Di~Renzo, L.~Scorzato, {New approach to the sign problem in quantum field theories: High density QCD on a Lefschetz thimble}, Phys. Rev. D 86 (2012) 074506.
\newblock \href {http://arxiv.org/abs/1205.3996} {\path{arXiv:1205.3996}}, \href {https://doi.org/10.1103/PhysRevD.86.074506} {\path{doi:10.1103/PhysRevD.86.074506}}.

\bibitem{Cristoforetti:2013wha}
M.~Cristoforetti, F.~Di~Renzo, A.~Mukherjee, L.~Scorzato, {Monte Carlo simulations on the Lefschetz thimble: Taming the sign problem}, Phys. Rev. D 88~(5) (2013) 051501.
\newblock \href {http://arxiv.org/abs/1303.7204} {\path{arXiv:1303.7204}}, \href {https://doi.org/10.1103/PhysRevD.88.051501} {\path{doi:10.1103/PhysRevD.88.051501}}.

\bibitem{Alexandru:2015sua}
A.~Alexandru, G.~Basar, P.~F. Bedaque, G.~W. Ridgway, N.~C. Warrington, {Sign problem and Monte Carlo calculations beyond Lefschetz thimbles}, JHEP 05 (2016) 053.
\newblock \href {http://arxiv.org/abs/1512.08764} {\path{arXiv:1512.08764}}, \href {https://doi.org/10.1007/JHEP05(2016)053} {\path{doi:10.1007/JHEP05(2016)053}}.

\bibitem{Nishimura:2017vav}
J.~Nishimura, S.~Shimasaki, {Combining the complex Langevin method and the generalized Lefschetz-thimble method}, JHEP 06 (2017) 023.
\newblock \href {http://arxiv.org/abs/1703.09409} {\path{arXiv:1703.09409}}, \href {https://doi.org/10.1007/JHEP06(2017)023} {\path{doi:10.1007/JHEP06(2017)023}}.

\bibitem{Alexandru:2017czx}
A.~Alexandru, P.~F. Bedaque, H.~Lamm, S.~Lawrence, {Deep Learning Beyond Lefschetz Thimbles}, Phys. Rev. D 96~(9) (2017) 094505.
\newblock \href {http://arxiv.org/abs/1709.01971} {\path{arXiv:1709.01971}}, \href {https://doi.org/10.1103/PhysRevD.96.094505} {\path{doi:10.1103/PhysRevD.96.094505}}.

\bibitem{Wynen:2020uzx}
J.-L. Wynen, E.~Berkowitz, S.~Krieg, T.~Luu, J.~Ostmeyer, {Machine learning to alleviate Hubbard-model sign problems}, Phys. Rev. B 103~(12) (2021) 125153.
\newblock \href {http://arxiv.org/abs/2006.11221} {\path{arXiv:2006.11221}}, \href {https://doi.org/10.1103/PhysRevB.103.125153} {\path{doi:10.1103/PhysRevB.103.125153}}.

\bibitem{Rodekamp:2022xpf}
M.~Rodekamp, E.~Berkowitz, C.~G\"antgen, S.~Krieg, T.~Luu, J.~Ostmeyer, {Mitigating the Hubbard sign problem with complex-valued neural networks}, Phys. Rev. B 106~(12) (2022) 125139.
\newblock \href {http://arxiv.org/abs/2203.00390} {\path{arXiv:2203.00390}}, \href {https://doi.org/10.1103/PhysRevB.106.125139} {\path{doi:10.1103/PhysRevB.106.125139}}.

\bibitem{Rodekamp:2022ylw}
M.~Rodekamp, C.~G\"antgen, {Mitigating the Hubbard Sign Problem. A Novel Application of Machine Learning}, PoS LATTICE2022 (2023) 032.
\newblock \href {http://arxiv.org/abs/2211.09584} {\path{arXiv:2211.09584}}, \href {https://doi.org/10.22323/1.430.0032} {\path{doi:10.22323/1.430.0032}}.

\bibitem{Lawrence:2021izu}
S.~Lawrence, Y.~Yamauchi, {Normalizing Flows and the Real-Time Sign Problem}, Phys. Rev. D 103~(11) (2021) 114509.
\newblock \href {http://arxiv.org/abs/2101.05755} {\path{arXiv:2101.05755}}, \href {https://doi.org/10.1103/PhysRevD.103.114509} {\path{doi:10.1103/PhysRevD.103.114509}}.

\bibitem{Yamauchi:2021kpo}
Y.~Yamauchi, S.~Lawrence, {Normalizing flows for the real-time sign problem}, PoS LATTICE2021 (2022) 621.
\newblock \href {http://arxiv.org/abs/2112.15035} {\path{arXiv:2112.15035}}, \href {https://doi.org/10.22323/1.396.0621} {\path{doi:10.22323/1.396.0621}}.

\bibitem{Lawrence:2022afv}
S.~Lawrence, H.~Oh, Y.~Yamauchi, {Lattice scalar field theory at complex coupling}, Phys. Rev. D 106~(11) (2022) 114503.
\newblock \href {http://arxiv.org/abs/2205.12303} {\path{arXiv:2205.12303}}, \href {https://doi.org/10.1103/PhysRevD.106.114503} {\path{doi:10.1103/PhysRevD.106.114503}}.

\bibitem{Pawlowski:2022rdn}
J.~M. Pawlowski, J.~M. Urban, {Flow-based density of states for complex actions}, Phys. Rev. D 108~(5) (2023) 054511.
\newblock \href {http://arxiv.org/abs/2203.01243} {\path{arXiv:2203.01243}}, \href {https://doi.org/10.1103/PhysRevD.108.054511} {\path{doi:10.1103/PhysRevD.108.054511}}.

\bibitem{Wan:2020lff}
Z.-Q. Wan, S.-X. Zhang, H.~Yao, {Mitigating the fermion sign problem by automatic differentiation}, Phys. Rev. B 106~(24) (2022) L241109.
\newblock \href {http://arxiv.org/abs/2010.01141} {\path{arXiv:2010.01141}}, \href {https://doi.org/10.1103/PhysRevB.106.L241109} {\path{doi:10.1103/PhysRevB.106.L241109}}.

\bibitem{Mori:2017pne}
Y.~Mori, K.~Kashiwa, A.~Ohnishi, {Toward solving the sign problem with path optimization method}, Phys. Rev. D 96~(11) (2017) 111501.
\newblock \href {http://arxiv.org/abs/1705.05605} {\path{arXiv:1705.05605}}, \href {https://doi.org/10.1103/PhysRevD.96.111501} {\path{doi:10.1103/PhysRevD.96.111501}}.

\bibitem{Mori:2017nwj}
Y.~Mori, K.~Kashiwa, A.~Ohnishi, {Application of a neural network to the sign problem via the path optimization method}, PTEP 2018~(2) (2018) 023B04.
\newblock \href {http://arxiv.org/abs/1709.03208} {\path{arXiv:1709.03208}}, \href {https://doi.org/10.1093/ptep/ptx191} {\path{doi:10.1093/ptep/ptx191}}.

\bibitem{Kashiwa:2019lkv}
K.~Kashiwa, Y.~Mori, A.~Ohnishi, {Application of the path optimization method to the sign problem in an effective model of QCD with a repulsive vector-type interaction}, Phys. Rev. D 99~(11) (2019) 114005.
\newblock \href {http://arxiv.org/abs/1903.03679} {\path{arXiv:1903.03679}}, \href {https://doi.org/10.1103/PhysRevD.99.114005} {\path{doi:10.1103/PhysRevD.99.114005}}.

\bibitem{Kashiwa:2018vxr}
K.~Kashiwa, Y.~Mori, A.~Ohnishi, {Controlling the model sign problem via the path optimization method: Monte Carlo approach to a QCD effective model with Polyakov loop}, Phys. Rev. D 99~(1) (2019) 014033.
\newblock \href {http://arxiv.org/abs/1805.08940} {\path{arXiv:1805.08940}}, \href {https://doi.org/10.1103/PhysRevD.99.014033} {\path{doi:10.1103/PhysRevD.99.014033}}.

\bibitem{Bursa:2018ykf}
F.~Bursa, M.~Kroyter, {A simple approach towards the sign problem using path optimisation}, JHEP 12 (2018) 054.
\newblock \href {http://arxiv.org/abs/1805.04941} {\path{arXiv:1805.04941}}, \href {https://doi.org/10.1007/JHEP12(2018)054} {\path{doi:10.1007/JHEP12(2018)054}}.

\bibitem{Mori:2019tux}
Y.~Mori, K.~Kashiwa, A.~Ohnishi, {Path optimization in $0+1$D QCD at finite density}, PTEP 2019~(11) (2019) 113B01.
\newblock \href {http://arxiv.org/abs/1904.11140} {\path{arXiv:1904.11140}}, \href {https://doi.org/10.1093/ptep/ptz111} {\path{doi:10.1093/ptep/ptz111}}.

\bibitem{Detmold:2021ulb}
W.~Detmold, G.~Kanwar, H.~Lamm, M.~L. Wagman, N.~C. Warrington, {Path integral contour deformations for observables in $SU(N)$ gauge theory}, Phys. Rev. D 103~(9) (2021) 094517.
\newblock \href {http://arxiv.org/abs/2101.12668} {\path{arXiv:2101.12668}}, \href {https://doi.org/10.1103/PhysRevD.103.094517} {\path{doi:10.1103/PhysRevD.103.094517}}.

\bibitem{Alexandru:2018fqp}
A.~Alexandru, P.~F. Bedaque, H.~Lamm, S.~Lawrence, {Finite-Density Monte Carlo Calculations on Sign-Optimized Manifolds}, Phys. Rev. D 97~(9) (2018) 094510.
\newblock \href {http://arxiv.org/abs/1804.00697} {\path{arXiv:1804.00697}}, \href {https://doi.org/10.1103/PhysRevD.97.094510} {\path{doi:10.1103/PhysRevD.97.094510}}.

\bibitem{Alexandru:2018ddf}
A.~Alexandru, P.~F. Bedaque, H.~Lamm, S.~Lawrence, N.~C. Warrington, {Fermions at Finite Density in 2+1 Dimensions with Sign-Optimized Manifolds}, Phys. Rev. Lett. 121~(19) (2018) 191602.
\newblock \href {http://arxiv.org/abs/1808.09799} {\path{arXiv:1808.09799}}, \href {https://doi.org/10.1103/PhysRevLett.121.191602} {\path{doi:10.1103/PhysRevLett.121.191602}}.

\bibitem{Namekawa:2021nzu}
Y.~Namekawa, K.~Kashiwa, A.~Ohnishi, H.~Takase, {Gauge invariant input to neural network for path optimization method}, Phys. Rev. D 105~(3) (2022) 034502.
\newblock \href {http://arxiv.org/abs/2109.11710} {\path{arXiv:2109.11710}}, \href {https://doi.org/10.1103/PhysRevD.105.034502} {\path{doi:10.1103/PhysRevD.105.034502}}.

\bibitem{Namekawa:2022liz}
Y.~Namekawa, K.~Kashiwa, H.~Matsuda, A.~Ohnishi, H.~Takase, {Improving efficiency of the path optimization method for a gauge theory}, Phys. Rev. D 107~(3) (2023) 034509.
\newblock \href {http://arxiv.org/abs/2210.05402} {\path{arXiv:2210.05402}}, \href {https://doi.org/10.1103/PhysRevD.107.034509} {\path{doi:10.1103/PhysRevD.107.034509}}.

\bibitem{Dexheimer:2020zzs}
V.~Dexheimer, J.~Noronha, J.~Noronha-Hostler, C.~Ratti, N.~Yunes, {Future physics perspectives on the equation of state from heavy ion collisions to neutron stars}, J. Phys. G 48~(7) (2021) 073001.
\newblock \href {http://arxiv.org/abs/2010.08834} {\path{arXiv:2010.08834}}, \href {https://doi.org/10.1088/1361-6471/abe104} {\path{doi:10.1088/1361-6471/abe104}}.

\bibitem{Watts:2016uzu}
A.~L. Watts, et~al., {Colloquium : Measuring the neutron star equation of state using x-ray timing}, Rev. Mod. Phys. 88~(2) (2016) 021001.
\newblock \href {http://arxiv.org/abs/1602.01081} {\path{arXiv:1602.01081}}, \href {https://doi.org/10.1103/RevModPhys.88.021001} {\path{doi:10.1103/RevModPhys.88.021001}}.

\bibitem{Ozel:2016oaf}
F.~\"Ozel, P.~Freire, {Masses, Radii, and the Equation of State of Neutron Stars}, Ann. Rev. Astron. Astrophys. 54 (2016) 401--440.
\newblock \href {http://arxiv.org/abs/1603.02698} {\path{arXiv:1603.02698}}, \href {https://doi.org/10.1146/annurev-astro-081915-023322} {\path{doi:10.1146/annurev-astro-081915-023322}}.

\bibitem{Baym:2017whm}
G.~Baym, T.~Hatsuda, T.~Kojo, P.~D. Powell, Y.~Song, T.~Takatsuka, {From hadrons to quarks in neutron stars: a review}, Rept. Prog. Phys. 81~(5) (2018) 056902.
\newblock \href {http://arxiv.org/abs/1707.04966} {\path{arXiv:1707.04966}}, \href {https://doi.org/10.1088/1361-6633/aaae14} {\path{doi:10.1088/1361-6633/aaae14}}.

\bibitem{Baiotti:2019sew}
L.~Baiotti, {Gravitational waves from neutron star mergers and their relation to the nuclear equation of state}, Prog. Part. Nucl. Phys. 109 (2019) 103714.
\newblock \href {http://arxiv.org/abs/1907.08534} {\path{arXiv:1907.08534}}, \href {https://doi.org/10.1016/j.ppnp.2019.103714} {\path{doi:10.1016/j.ppnp.2019.103714}}.

\bibitem{Kojo:2020krb}
T.~Kojo, {QCD equations of state and speed of sound in neutron stars}, AAPPS Bull. 31~(1) (2021) 11.
\newblock \href {http://arxiv.org/abs/2011.10940} {\path{arXiv:2011.10940}}, \href {https://doi.org/10.1007/s43673-021-00011-6} {\path{doi:10.1007/s43673-021-00011-6}}.

\bibitem{Lattimer:2021emm}
J.~M. Lattimer, {Neutron Stars and the Nuclear Matter Equation of State}, Ann. Rev. Nucl. Part. Sci. 71 (2021) 433--464.
\newblock \href {https://doi.org/10.1146/annurev-nucl-102419-124827} {\path{doi:10.1146/annurev-nucl-102419-124827}}.

\bibitem{Fukushima:2013rx}
K.~Fukushima, C.~Sasaki, {The phase diagram of nuclear and quark matter at high baryon density}, Prog. Part. Nucl. Phys. 72 (2013) 99--154.
\newblock \href {http://arxiv.org/abs/1301.6377} {\path{arXiv:1301.6377}}, \href {https://doi.org/10.1016/j.ppnp.2013.05.003} {\path{doi:10.1016/j.ppnp.2013.05.003}}.

\bibitem{Burgio:2021vgk}
G.~F. Burgio, H.~J. Schulze, I.~Vidana, J.~B. Wei, {Neutron stars and the nuclear equation of state}, Prog. Part. Nucl. Phys. 120 (2021) 103879.
\newblock \href {http://arxiv.org/abs/2105.03747} {\path{arXiv:2105.03747}}, \href {https://doi.org/10.1016/j.ppnp.2021.103879} {\path{doi:10.1016/j.ppnp.2021.103879}}.

\bibitem{Drischler:2021kxf}
C.~Drischler, J.~W. Holt, C.~Wellenhofer, {Chiral Effective Field Theory and the High-Density Nuclear Equation of State}, Ann. Rev. Nucl. Part. Sci. 71 (2021) 403--432.
\newblock \href {http://arxiv.org/abs/2101.01709} {\path{arXiv:2101.01709}}, \href {https://doi.org/10.1146/annurev-nucl-102419-041903} {\path{doi:10.1146/annurev-nucl-102419-041903}}.

\bibitem{Ghiglieri:2020dpq}
J.~Ghiglieri, A.~Kurkela, M.~Strickland, A.~Vuorinen, {Perturbative Thermal QCD: Formalism and Applications}, Phys. Rept. 880 (2020) 1--73.
\newblock \href {http://arxiv.org/abs/2002.10188} {\path{arXiv:2002.10188}}, \href {https://doi.org/10.1016/j.physrep.2020.07.004} {\path{doi:10.1016/j.physrep.2020.07.004}}.

\bibitem{Miller:2013tca}
M.~C. Miller, {Astrophysical Constraints on Dense Matter in Neutron Stars}, Astrophys. Space Sci. Libr. 461 (2020) 1--51.
\newblock \href {http://arxiv.org/abs/1312.0029} {\path{arXiv:1312.0029}}, \href {https://doi.org/10.1007/978-3-662-62110-3_1} {\path{doi:10.1007/978-3-662-62110-3_1}}.

\bibitem{Miller:2016pom}
M.~C. Miller, F.~K. Lamb, {Observational Constraints on Neutron Star Masses and Radii}, Eur. Phys. J. A 52~(3) (2016) 63.
\newblock \href {http://arxiv.org/abs/1604.03894} {\path{arXiv:1604.03894}}, \href {https://doi.org/10.1140/epja/i2016-16063-8} {\path{doi:10.1140/epja/i2016-16063-8}}.

\bibitem{10.1117/12.2056811}
Z.~Arzoumanian, K.~C. Gendreau, C.~L. Baker, T.~Cazeau, P.~Hestnes, J.~W. Kellogg, S.~J. Kenyon, R.~P. Kozon, K.-C. Liu, S.~S. Manthripragada, C.~B. Markwardt, A.~L. Mitchell, J.~W. Mitchell, C.~A. Monroe, T.~Okajima, S.~E. Pollard, D.~F. Powers, B.~J. Savadkin, L.~B. Winternitz, P.~T. Chen, M.~R. Wright, R.~Foster, G.~Prigozhin, R.~Remillard, J.~Doty, \href{https://doi.org/10.1117/12.2056811}{{The neutron star interior composition explorer (NICER): mission definition}}, in: T.~Takahashi, J.-W.~A. den Herder, M.~Bautz (Eds.), Space Telescopes and Instrumentation 2014: Ultraviolet to Gamma Ray, Vol. 9144, International Society for Optics and Photonics, SPIE, 2014, p. 914420.
\newblock \href {https://doi.org/10.1117/12.2056811} {\path{doi:10.1117/12.2056811}}.
\newline\urlprefix\url{https://doi.org/10.1117/12.2056811}

\bibitem{gendreau2017searching}
K.~Gendreau, Z.~Arzoumanian, Searching for a pulse, Nature Astronomy 1~(12) (2017) 895--895.

\bibitem{Yunes:2022ldq}
N.~Yunes, M.~C. Miller, K.~Yagi, {Gravitational-wave and X-ray probes of the neutron star equation of state}, Nature Rev. Phys. 4~(4) (2022) 237--246.
\newblock \href {http://arxiv.org/abs/2202.04117} {\path{arXiv:2202.04117}}, \href {https://doi.org/10.1038/s42254-022-00420-y} {\path{doi:10.1038/s42254-022-00420-y}}.

\bibitem{Yagi:2013bca}
K.~Yagi, N.~Yunes, {I-Love-Q}, Science 341 (2013) 365--368.
\newblock \href {http://arxiv.org/abs/1302.4499} {\path{arXiv:1302.4499}}, \href {https://doi.org/10.1126/science.1236462} {\path{doi:10.1126/science.1236462}}.

\bibitem{LIGOScientific:2018mvr}
B.~P. Abbott, et~al., {GWTC-1: A Gravitational-Wave Transient Catalog of Compact Binary Mergers Observed by LIGO and Virgo during the First and Second Observing Runs}, Phys. Rev. X 9~(3) (2019) 031040.
\newblock \href {http://arxiv.org/abs/1811.12907} {\path{arXiv:1811.12907}}, \href {https://doi.org/10.1103/PhysRevX.9.031040} {\path{doi:10.1103/PhysRevX.9.031040}}.

\bibitem{LIGOScientific:2020ibl}
R.~Abbott, et~al., {GWTC-2: Compact Binary Coalescences Observed by LIGO and Virgo During the First Half of the Third Observing Run}, Phys. Rev. X 11 (2021) 021053.
\newblock \href {http://arxiv.org/abs/2010.14527} {\path{arXiv:2010.14527}}, \href {https://doi.org/10.1103/PhysRevX.11.021053} {\path{doi:10.1103/PhysRevX.11.021053}}.

\bibitem{LIGOScientific:2017vwq}
B.~P. Abbott, et~al., {GW170817: Observation of Gravitational Waves from a Binary Neutron Star Inspiral}, Phys. Rev. Lett. 119~(16) (2017) 161101.
\newblock \href {http://arxiv.org/abs/1710.05832} {\path{arXiv:1710.05832}}, \href {https://doi.org/10.1103/PhysRevLett.119.161101} {\path{doi:10.1103/PhysRevLett.119.161101}}.

\bibitem{LIGOScientific:2018cki}
B.~P. Abbott, et~al., {GW170817: Measurements of neutron star radii and equation of state}, Phys. Rev. Lett. 121~(16) (2018) 161101.
\newblock \href {http://arxiv.org/abs/1805.11581} {\path{arXiv:1805.11581}}, \href {https://doi.org/10.1103/PhysRevLett.121.161101} {\path{doi:10.1103/PhysRevLett.121.161101}}.

\bibitem{LIGOScientific:2020aai}
B.~P. Abbott, et~al., {GW190425: Observation of a Compact Binary Coalescence with Total Mass $\sim 3.4 M_{\odot}$}, Astrophys. J. Lett. 892~(1) (2020) L3.
\newblock \href {http://arxiv.org/abs/2001.01761} {\path{arXiv:2001.01761}}, \href {https://doi.org/10.3847/2041-8213/ab75f5} {\path{doi:10.3847/2041-8213/ab75f5}}.

\bibitem{LIGOScientific:2021qlt}
R.~Abbott, et~al., {Observation of Gravitational Waves from Two Neutron Star\textendash{}Black Hole Coalescences}, Astrophys. J. Lett. 915~(1) (2021) L5.
\newblock \href {http://arxiv.org/abs/2106.15163} {\path{arXiv:2106.15163}}, \href {https://doi.org/10.3847/2041-8213/ac082e} {\path{doi:10.3847/2041-8213/ac082e}}.

\bibitem{Annala:2021gom}
E.~Annala, T.~Gorda, E.~Katerini, A.~Kurkela, J.~N\"attil\"a, V.~Paschalidis, A.~Vuorinen, {Multimessenger Constraints for Ultradense Matter}, Phys. Rev. X 12~(1) (2022) 011058.
\newblock \href {http://arxiv.org/abs/2105.05132} {\path{arXiv:2105.05132}}, \href {https://doi.org/10.1103/PhysRevX.12.011058} {\path{doi:10.1103/PhysRevX.12.011058}}.

\bibitem{2020WDMKD..10.1349F}
C.~J. {Fluke}, C.~{Jacobs}, {Surveying the reach and maturity of machine learning and artificial intelligence in astronomy}, WIREs Data Mining and Knowledge Discovery 10~(2) (2020) e1349.
\newblock \href {http://arxiv.org/abs/1912.02934} {\path{arXiv:1912.02934}}, \href {https://doi.org/10.1002/widm.1349} {\path{doi:10.1002/widm.1349}}.

\bibitem{Schaffner-Bielich:2020psc}
J.~Schaffner-Bielich, {Compact Star Physics}, Cambridge University Press, 2020.
\newblock \href {https://doi.org/10.1017/9781316848357} {\path{doi:10.1017/9781316848357}}.

\bibitem{Tolman:1939jz}
R.~C. Tolman, {Static solutions of Einstein's field equations for spheres of fluid}, Phys. Rev. 55 (1939) 364--373.
\newblock \href {https://doi.org/10.1103/PhysRev.55.364} {\path{doi:10.1103/PhysRev.55.364}}.

\bibitem{Oppenheimer:1939ne}
J.~R. Oppenheimer, G.~M. Volkoff, {On massive neutron cores}, Phys. Rev. 55 (1939) 374--381.
\newblock \href {https://doi.org/10.1103/PhysRev.55.374} {\path{doi:10.1103/PhysRev.55.374}}.

\bibitem{Krastev:2021reh}
P.~G. Krastev, {Translating Neutron Star Observations to Nuclear Symmetry Energy via Deep Neural Networks}, Galaxies 10~(1) (2022) 16.
\newblock \href {http://arxiv.org/abs/2112.04089} {\path{arXiv:2112.04089}}, \href {https://doi.org/10.3390/galaxies10010016} {\path{doi:10.3390/galaxies10010016}}.

\bibitem{Han:2021kjx}
M.-Z. Han, J.-L. Jiang, S.-P. Tang, Y.-Z. Fan, {Bayesian Nonparametric Inference of the Neutron Star Equation of State via a Neural Network}, Astrophys. J. 919~(1) (2021) 11.
\newblock \href {http://arxiv.org/abs/2103.05408} {\path{arXiv:2103.05408}}, \href {https://doi.org/10.3847/1538-4357/ac11f8} {\path{doi:10.3847/1538-4357/ac11f8}}.

\bibitem{Demorest:2010bx}
P.~Demorest, T.~Pennucci, S.~Ransom, M.~Roberts, J.~Hessels, {Shapiro Delay Measurement of A Two Solar Mass Neutron Star}, Nature 467 (2010) 1081--1083.
\newblock \href {http://arxiv.org/abs/1010.5788} {\path{arXiv:1010.5788}}, \href {https://doi.org/10.1038/nature09466} {\path{doi:10.1038/nature09466}}.

\bibitem{Lobato:2022ajs}
R.~V. Lobato, E.~V. Chimanski, C.~A. Bertulani, {Unsupervised machine learning correlations in EoS of neutron stars}, PoS XVHadronPhysics (2022) 062.
\newblock \href {http://arxiv.org/abs/2202.13940} {\path{arXiv:2202.13940}}, \href {https://doi.org/10.22323/1.408.0062} {\path{doi:10.22323/1.408.0062}}.

\bibitem{Lindblom:2010bb}
L.~Lindblom, {Spectral Representations of Neutron-Star Equations of State}, Phys. Rev. D 82 (2010) 103011.
\newblock \href {http://arxiv.org/abs/1009.0738} {\path{arXiv:1009.0738}}, \href {https://doi.org/10.1103/PhysRevD.82.103011} {\path{doi:10.1103/PhysRevD.82.103011}}.

\bibitem{Lindblom:2022mkr}
L.~Lindblom, {Improved spectral representations of neutron-star equations of state}, Phys. Rev. D 105~(6) (2022) 063031.
\newblock \href {http://arxiv.org/abs/2202.12285} {\path{arXiv:2202.12285}}, \href {https://doi.org/10.1103/PhysRevD.105.063031} {\path{doi:10.1103/PhysRevD.105.063031}}.

\bibitem{Read:2008iy}
J.~S. Read, B.~D. Lackey, B.~J. Owen, J.~L. Friedman, {Constraints on a phenomenologically parameterized neutron-star equation of state}, Phys. Rev. D 79 (2009) 124032.
\newblock \href {http://arxiv.org/abs/0812.2163} {\path{arXiv:0812.2163}}, \href {https://doi.org/10.1103/PhysRevD.79.124032} {\path{doi:10.1103/PhysRevD.79.124032}}.

\bibitem{Ozel:2009da}
F.~Ozel, D.~Psaltis, {Reconstructing the Neutron-Star Equation of State from Astrophysical Measurements}, Phys. Rev. D 80 (2009) 103003.
\newblock \href {http://arxiv.org/abs/0905.1959} {\path{arXiv:0905.1959}}, \href {https://doi.org/10.1103/PhysRevD.80.103003} {\path{doi:10.1103/PhysRevD.80.103003}}.

\bibitem{Steiner:2010fz}
A.~W. Steiner, J.~M. Lattimer, E.~F. Brown, {The Equation of State from Observed Masses and Radii of Neutron Stars}, Astrophys. J. 722 (2010) 33--54.
\newblock \href {http://arxiv.org/abs/1005.0811} {\path{arXiv:1005.0811}}, \href {https://doi.org/10.1088/0004-637X/722/1/33} {\path{doi:10.1088/0004-637X/722/1/33}}.

\bibitem{Steiner:2012xt}
A.~W. Steiner, J.~M. Lattimer, E.~F. Brown, {The Neutron Star Mass-Radius Relation and the Equation of State of Dense Matter}, Astrophys. J. Lett. 765 (2013) L5.
\newblock \href {http://arxiv.org/abs/1205.6871} {\path{arXiv:1205.6871}}, \href {https://doi.org/10.1088/2041-8205/765/1/L5} {\path{doi:10.1088/2041-8205/765/1/L5}}.

\bibitem{Raithel:2016bux}
C.~A. Raithel, F.~Ozel, D.~Psaltis, {From Neutron Star Observables to the Equation of State: An Optimal Parametrization}, Astrophys. J. 831~(1) (2016) 44.
\newblock \href {http://arxiv.org/abs/1605.03591} {\path{arXiv:1605.03591}}, \href {https://doi.org/10.3847/0004-637X/831/1/44} {\path{doi:10.3847/0004-637X/831/1/44}}.

\bibitem{Raithel:2017ity}
C.~A. Raithel, F.~\"Ozel, D.~Psaltis, {From Neutron Star Observables to the Equation of State. II. Bayesian Inference of Equation of State Pressures}, Astrophys. J. 844~(2) (2017) 156.
\newblock \href {http://arxiv.org/abs/1704.00737} {\path{arXiv:1704.00737}}, \href {https://doi.org/10.3847/1538-4357/aa7a5a} {\path{doi:10.3847/1538-4357/aa7a5a}}.

\bibitem{Landry:2018prl}
P.~Landry, R.~Essick, {Nonparametric inference of the neutron star equation of state from gravitational wave observations}, Phys. Rev. D 99~(8) (2019) 084049.
\newblock \href {http://arxiv.org/abs/1811.12529} {\path{arXiv:1811.12529}}, \href {https://doi.org/10.1103/PhysRevD.99.084049} {\path{doi:10.1103/PhysRevD.99.084049}}.

\bibitem{Soma:2022qnv}
S.~Soma, L.~Wang, S.~Shi, H.~St\"ocker, K.~Zhou, {Neural network reconstruction of the dense matter equation of state from neutron star observables}, JCAP 08 (2022) 071.
\newblock \href {http://arxiv.org/abs/2201.01756} {\path{arXiv:2201.01756}}, \href {https://doi.org/10.1088/1475-7516/2022/08/071} {\path{doi:10.1088/1475-7516/2022/08/071}}.

\bibitem{Soma:2022vbb}
S.~Soma, L.~Wang, S.~Shi, H.~St\"ocker, K.~Zhou, {Reconstructing the neutron star equation of state from observational data via automatic differentiation}, Phys. Rev. D 107~(8) (2023) 083028.
\newblock \href {http://arxiv.org/abs/2209.08883} {\path{arXiv:2209.08883}}, \href {https://doi.org/10.1103/PhysRevD.107.083028} {\path{doi:10.1103/PhysRevD.107.083028}}.

\bibitem{Han:2022sxt}
M.-Z. Han, S.-P. Tang, Y.-Z. Fan, {Nonparametric Representation of Neutron Star Equation of State Using Variational Autoencoder}, Astrophys. J. 950~(2) (2023) 77.
\newblock \href {http://arxiv.org/abs/2205.03855} {\path{arXiv:2205.03855}}, \href {https://doi.org/10.3847/1538-4357/acd050} {\path{doi:10.3847/1538-4357/acd050}}.

\bibitem{Margueron:2017eqc}
J.~Margueron, R.~Hoffmann~Casali, F.~Gulminelli, {Equation of state for dense nucleonic matter from metamodeling. I. Foundational aspects}, Phys. Rev. C 97~(2) (2018) 025805.
\newblock \href {http://arxiv.org/abs/1708.06894} {\path{arXiv:1708.06894}}, \href {https://doi.org/10.1103/PhysRevC.97.025805} {\path{doi:10.1103/PhysRevC.97.025805}}.

\bibitem{Margueron:2017lup}
J.~Margueron, R.~Hoffmann~Casali, F.~Gulminelli, {Equation of state for dense nucleonic matter from metamodeling. II. Predictions for neutron star properties}, Phys. Rev. C 97~(2) (2018) 025806.
\newblock \href {http://arxiv.org/abs/1708.06895} {\path{arXiv:1708.06895}}, \href {https://doi.org/10.1103/PhysRevC.97.025806} {\path{doi:10.1103/PhysRevC.97.025806}}.

\bibitem{Xie:2019sqb}
W.-J. Xie, B.-A. Li, {Bayesian Inference of High-density Nuclear Symmetry Energy from Radii of Canonical Neutron Stars}, Astrophys. J. 883 (2019) 174.
\newblock \href {http://arxiv.org/abs/1907.10741} {\path{arXiv:1907.10741}}, \href {https://doi.org/10.3847/1538-4357/ab3f37} {\path{doi:10.3847/1538-4357/ab3f37}}.

\bibitem{Brandes:2022nxa}
L.~Brandes, W.~Weise, N.~Kaiser, {Inference of the sound speed and related properties of neutron stars}, Phys. Rev. D 107~(1) (2023) 014011.
\newblock \href {http://arxiv.org/abs/2208.03026} {\path{arXiv:2208.03026}}, \href {https://doi.org/10.1103/PhysRevD.107.014011} {\path{doi:10.1103/PhysRevD.107.014011}}.

\bibitem{Altiparmak:2022bke}
S.~Altiparmak, C.~Ecker, L.~Rezzolla, {On the Sound Speed in Neutron Stars}, Astrophys. J. Lett. 939~(2) (2022) L34.
\newblock \href {http://arxiv.org/abs/2203.14974} {\path{arXiv:2203.14974}}, \href {https://doi.org/10.3847/2041-8213/ac9b2a} {\path{doi:10.3847/2041-8213/ac9b2a}}.

\bibitem{Ecker:2022xxj}
C.~Ecker, L.~Rezzolla, {A General, Scale-independent Description of the Sound Speed in Neutron Stars}, Astrophys. J. Lett. 939~(2) (2022) L35.
\newblock \href {http://arxiv.org/abs/2207.04417} {\path{arXiv:2207.04417}}, \href {https://doi.org/10.3847/2041-8213/ac8674} {\path{doi:10.3847/2041-8213/ac8674}}.

\bibitem{Jiang:2022tps}
J.-L. Jiang, C.~Ecker, L.~Rezzolla, {Bayesian Analysis of Neutron-star Properties with Parameterized Equations of State: The Role of the Likelihood Functions}, Astrophys. J. 949~(1) (2023) 11.
\newblock \href {http://arxiv.org/abs/2211.00018} {\path{arXiv:2211.00018}}, \href {https://doi.org/10.3847/1538-4357/acc4be} {\path{doi:10.3847/1538-4357/acc4be}}.

\bibitem{d2003bayesian}
G.~D'Agostini, Bayesian inference in processing experimental data: principles and basic applications, Reports on Progress in Physics 66~(9) (2003) 1383.

\bibitem{berkson1980minimum}
J.~Berkson, Minimum chi-square, not maximum likelihood!, The Annals of Statistics 8~(3) (1980) 457--487.

\bibitem{Bogdanov:2022faf}
S.~Bogdanov, et~al., {Snowmass 2021 Cosmic Frontier White Paper: The Dense Matter Equation of State and QCD Phase Transitions}, in: {Snowmass 2021}, 2022.
\newblock \href {http://arxiv.org/abs/2209.07412} {\path{arXiv:2209.07412}}.

\bibitem{2013Sci...340..448A}
J.~Antoniadis, et~al., {A Massive Pulsar in a Compact Relativistic Binary}, Science 340 (2013) 6131.
\newblock \href {http://arxiv.org/abs/1304.6875} {\path{arXiv:1304.6875}}, \href {https://doi.org/10.1126/science.1233232} {\path{doi:10.1126/science.1233232}}.

\bibitem{NANOGrav:2019jur}
H.~T. Cromartie, et~al., {Relativistic Shapiro delay measurements of an extremely massive millisecond pulsar}, Nature Astron. 4~(1) (2019) 72--76.
\newblock \href {http://arxiv.org/abs/1904.06759} {\path{arXiv:1904.06759}}, \href {https://doi.org/10.1038/s41550-019-0880-2} {\path{doi:10.1038/s41550-019-0880-2}}.

\bibitem{Landry:2018jyg}
P.~Landry, B.~Kumar, {Constraints on the moment of inertia of PSR J0737-3039A from GW170817}, Astrophys. J. Lett. 868~(2) (2018) L22.
\newblock \href {http://arxiv.org/abs/1807.04727} {\path{arXiv:1807.04727}}, \href {https://doi.org/10.3847/2041-8213/aaee76} {\path{doi:10.3847/2041-8213/aaee76}}.

\bibitem{Miao:2021gmf}
Z.~Miao, A.~Li, Z.-G. Dai, {On the moment of inertia of PSR J0737-3039 A from LIGO/Virgo and NICER}, Mon. Not. Roy. Astron. Soc. 515~(4) (2022) 5071--5080.
\newblock \href {http://arxiv.org/abs/2107.07979} {\path{arXiv:2107.07979}}, \href {https://doi.org/10.1093/mnras/stac2015} {\path{doi:10.1093/mnras/stac2015}}.

\bibitem{Miller:2019cac}
M.~C. Miller, et~al., {PSR J0030+0451 Mass and Radius from $NICER$ Data and Implications for the Properties of Neutron Star Matter}, Astrophys. J. Lett. 887~(1) (2019) L24.
\newblock \href {http://arxiv.org/abs/1912.05705} {\path{arXiv:1912.05705}}, \href {https://doi.org/10.3847/2041-8213/ab50c5} {\path{doi:10.3847/2041-8213/ab50c5}}.

\bibitem{Riley:2019yda}
T.~E. Riley, et~al., {A $NICER$ View of PSR J0030+0451: Millisecond Pulsar Parameter Estimation}, Astrophys. J. Lett. 887~(1) (2019) L21.
\newblock \href {http://arxiv.org/abs/1912.05702} {\path{arXiv:1912.05702}}, \href {https://doi.org/10.3847/2041-8213/ab481c} {\path{doi:10.3847/2041-8213/ab481c}}.

\bibitem{Miller:2021qha}
M.~C. Miller, et~al., {The Radius of PSR J0740+6620 from NICER and XMM-Newton Data}, Astrophys. J. Lett. 918~(2) (2021) L28.
\newblock \href {http://arxiv.org/abs/2105.06979} {\path{arXiv:2105.06979}}, \href {https://doi.org/10.3847/2041-8213/ac089b} {\path{doi:10.3847/2041-8213/ac089b}}.

\bibitem{Riley:2021pdl}
T.~E. Riley, et~al., {A NICER View of the Massive Pulsar PSR J0740+6620 Informed by Radio Timing and XMM-Newton Spectroscopy}, Astrophys. J. Lett. 918~(2) (2021) L27.
\newblock \href {http://arxiv.org/abs/2105.06980} {\path{arXiv:2105.06980}}, \href {https://doi.org/10.3847/2041-8213/ac0a81} {\path{doi:10.3847/2041-8213/ac0a81}}.

\bibitem{Essick:2019ldf}
R.~Essick, P.~Landry, D.~E. Holz, {Nonparametric Inference of Neutron Star Composition, Equation of State, and Maximum Mass with GW170817}, Phys. Rev. D 101~(6) (2020) 063007.
\newblock \href {http://arxiv.org/abs/1910.09740} {\path{arXiv:1910.09740}}, \href {https://doi.org/10.1103/PhysRevD.101.063007} {\path{doi:10.1103/PhysRevD.101.063007}}.

\bibitem{1992ApJ...398..569L}
L.~{Lindblom}, {Determining the Nuclear Equation of State from Neutron-Star Masses and Radii}, Astrophysical Journal 398 (1992) 569.
\newblock \href {https://doi.org/10.1086/171882} {\path{doi:10.1086/171882}}.

\bibitem{PANG:2020lda}
L.~PANG, K.~ZHOU, X.~WANG, {Deep Learning for Nuclear Physics}, Nucl. Phys. Rev. 37~(3) (2020) 720--726.
\newblock \href {https://doi.org/10.11804/NuclPhysRev.37.2019CNPC41} {\path{doi:10.11804/NuclPhysRev.37.2019CNPC41}}.

\bibitem{Fujimoto:2021zas}
Y.~Fujimoto, K.~Fukushima, K.~Murase, {Extensive Studies of the Neutron Star Equation of State from the Deep Learning Inference with the Observational Data Augmentation}, JHEP 03 (2021) 273.
\newblock \href {http://arxiv.org/abs/2101.08156} {\path{arXiv:2101.08156}}, \href {https://doi.org/10.1007/JHEP03(2021)273} {\path{doi:10.1007/JHEP03(2021)273}}.

\bibitem{Fujimoto:2017cdo}
Y.~Fujimoto, K.~Fukushima, K.~Murase, {Methodology study of machine learning for the neutron star equation of state}, Phys. Rev. D 98~(2) (2018) 023019.
\newblock \href {http://arxiv.org/abs/1711.06748} {\path{arXiv:1711.06748}}, \href {https://doi.org/10.1103/PhysRevD.98.023019} {\path{doi:10.1103/PhysRevD.98.023019}}.

\bibitem{Fujimoto:2019hxv}
Y.~Fujimoto, K.~Fukushima, K.~Murase, {Mapping neutron star data to the equation of state using the deep neural network}, Phys. Rev. D 101~(5) (2020) 054016.
\newblock \href {http://arxiv.org/abs/1903.03400} {\path{arXiv:1903.03400}}, \href {https://doi.org/10.1103/PhysRevD.101.054016} {\path{doi:10.1103/PhysRevD.101.054016}}.

\bibitem{Morawski:2020izm}
F.~Morawski, M.~Bejger, {Neural network reconstruction of the dense matter equation of state derived from the parameters of neutron stars}, Astron. Astrophys. 642 (2020) A78.
\newblock \href {http://arxiv.org/abs/2006.07194} {\path{arXiv:2006.07194}}, \href {https://doi.org/10.1051/0004-6361/202038130} {\path{doi:10.1051/0004-6361/202038130}}.

\bibitem{Morawski:2022aud}
F.~Morawski, M.~Bejger, {Detecting dense-matter phase transition signatures in neutron star mass-radius measurements as data anomalies using normalizing flows}, Phys. Rev. C 106~(6) (2022) 065802.
\newblock \href {http://arxiv.org/abs/2212.05480} {\path{arXiv:2212.05480}}, \href {https://doi.org/10.1103/PhysRevC.106.065802} {\path{doi:10.1103/PhysRevC.106.065802}}.

\bibitem{Traversi:2020dho}
S.~Traversi, P.~Char, {Structure of Quark Star: A Comparative Analysis of Bayesian Inference and Neural Network Based Modeling}, Astrophys. J. 905~(1) (2020) 9.
\newblock \href {http://arxiv.org/abs/2007.10239} {\path{arXiv:2007.10239}}, \href {https://doi.org/10.3847/1538-4357/abbfb4} {\path{doi:10.3847/1538-4357/abbfb4}}.

\bibitem{Nattila:2017wtj}
J.~N\"attil\"a, M.~C. Miller, A.~W. Steiner, J.~J.~E. Kajava, V.~F. Suleimanov, J.~Poutanen, {Neutron star mass and radius measurements from atmospheric model fits to X-ray burst cooling tail spectra}, Astron. Astrophys. 608 (2017) A31.
\newblock \href {http://arxiv.org/abs/1709.09120} {\path{arXiv:1709.09120}}, \href {https://doi.org/10.1051/0004-6361/201731082} {\path{doi:10.1051/0004-6361/201731082}}.

\bibitem{Bogdanov:2019ixe}
S.~Bogdanov, et~al., {Constraining the Neutron Star Mass\textendash{}Radius Relation and Dense Matter Equation of State with $NICER$. I. The Millisecond Pulsar X-Ray Data Set}, Astrophys. J. Lett. 887~(1) (2019) L25.
\newblock \href {http://arxiv.org/abs/1912.05706} {\path{arXiv:1912.05706}}, \href {https://doi.org/10.3847/2041-8213/ab53eb} {\path{doi:10.3847/2041-8213/ab53eb}}.

\bibitem{Flanagan:2007ix}
E.~E. Flanagan, T.~Hinderer, {Constraining neutron star tidal Love numbers with gravitational wave detectors}, Phys. Rev. D 77 (2008) 021502.
\newblock \href {http://arxiv.org/abs/0709.1915} {\path{arXiv:0709.1915}}, \href {https://doi.org/10.1103/PhysRevD.77.021502} {\path{doi:10.1103/PhysRevD.77.021502}}.

\bibitem{Hinderer:2007mb}
T.~Hinderer, {Tidal Love numbers of neutron stars}, Astrophys. J. 677 (2008) 1216--1220.
\newblock \href {http://arxiv.org/abs/0711.2420} {\path{arXiv:0711.2420}}, \href {https://doi.org/10.1086/533487} {\path{doi:10.1086/533487}}.

\bibitem{Landry:2020vaw}
P.~Landry, R.~Essick, K.~Chatziioannou, {Nonparametric constraints on neutron star matter with existing and upcoming gravitational wave and pulsar observations}, Phys. Rev. D 101~(12) (2020) 123007.
\newblock \href {http://arxiv.org/abs/2003.04880} {\path{arXiv:2003.04880}}, \href {https://doi.org/10.1103/PhysRevD.101.123007} {\path{doi:10.1103/PhysRevD.101.123007}}.

\bibitem{Miao:2021nuq}
Z.~Miao, J.-L. Jiang, A.~Li, L.-W. Chen, {Bayesian Inference of Strange Star Equation of State Using the GW170817 and GW190425 Data}, Astrophys. J. Lett. 917~(2) (2021) L22.
\newblock \href {http://arxiv.org/abs/2107.13997} {\path{arXiv:2107.13997}}, \href {https://doi.org/10.3847/2041-8213/ac194d} {\path{doi:10.3847/2041-8213/ac194d}}.

\bibitem{Golomb:2021tll}
J.~Golomb, C.~Talbot, {Hierarchical Inference of Binary Neutron Star Mass Distribution and Equation of State with Gravitational Waves}, Astrophys. J. 926~(1) (2022) 79.
\newblock \href {http://arxiv.org/abs/2106.15745} {\path{arXiv:2106.15745}}, \href {https://doi.org/10.3847/1538-4357/ac43bc} {\path{doi:10.3847/1538-4357/ac43bc}}.

\bibitem{Chimanski:2022wzi}
E.~V. Chimanski, R.~V. Lobato, A.~R. Goncalves, C.~A. Bertulani, {Bayesian Exploration of Phenomenological EoS of Neutron/Hybrid Stars with Recent Observations}, Particles 6~(1) (2023) 198--216.
\newblock \href {http://arxiv.org/abs/2205.01174} {\path{arXiv:2205.01174}}, \href {https://doi.org/10.3390/particles6010011} {\path{doi:10.3390/particles6010011}}.

\bibitem{De:2018uhw}
S.~De, D.~Finstad, J.~M. Lattimer, D.~A. Brown, E.~Berger, C.~M. Biwer, {Tidal Deformabilities and Radii of Neutron Stars from the Observation of GW170817}, Phys. Rev. Lett. 121~(9) (2018) 091102, [Erratum: Phys.Rev.Lett. 121, 259902 (2018)].
\newblock \href {http://arxiv.org/abs/1804.08583} {\path{arXiv:1804.08583}}, \href {https://doi.org/10.1103/PhysRevLett.121.091102} {\path{doi:10.1103/PhysRevLett.121.091102}}.

\bibitem{LIGOScientific:2018hze}
B.~P. Abbott, et~al., {Properties of the binary neutron star merger GW170817}, Phys. Rev. X 9~(1) (2019) 011001.
\newblock \href {http://arxiv.org/abs/1805.11579} {\path{arXiv:1805.11579}}, \href {https://doi.org/10.1103/PhysRevX.9.011001} {\path{doi:10.1103/PhysRevX.9.011001}}.

\bibitem{Ferreira:2019bny}
M.~a. Ferreira, C.~Provid\^encia, {Unveiling the nuclear matter EoS from neutron star properties: a supervised machine learning approach}, JCAP 07 (2021) 011.
\newblock \href {http://arxiv.org/abs/1910.05554} {\path{arXiv:1910.05554}}, \href {https://doi.org/10.1088/1475-7516/2021/07/011} {\path{doi:10.1088/1475-7516/2021/07/011}}.

\bibitem{PhysRevD.100.103009}
F.~Hernandez~Vivanco, R.~Smith, E.~Thrane, P.~D. Lasky, C.~Talbot, V.~Raymond, \href{https://link.aps.org/doi/10.1103/PhysRevD.100.103009}{Measuring the neutron star equation of state with gravitational waves: The first forty binary neutron star merger observations}, Phys. Rev. D 100 (2019) 103009.
\newblock \href {https://doi.org/10.1103/PhysRevD.100.103009} {\path{doi:10.1103/PhysRevD.100.103009}}.
\newline\urlprefix\url{https://link.aps.org/doi/10.1103/PhysRevD.100.103009}

\bibitem{HernandezVivanco:2020cyp}
F.~Hernandez~Vivanco, R.~Smith, E.~Thrane, P.~D. Lasky, {A scalable random forest regressor for combining neutron-star equation of state measurements: A case study with GW170817 and GW190425}, Mon. Not. Roy. Astron. Soc. 499~(4) (2020) 5972--5977.
\newblock \href {http://arxiv.org/abs/2008.05627} {\path{arXiv:2008.05627}}, \href {https://doi.org/10.1093/mnras/staa3243} {\path{doi:10.1093/mnras/staa3243}}.

\bibitem{Goncalves:2022smd}
G.~Gon\c{c}alves, M.~Ferreira, J.~a. Aveiro, A.~Onofre, F.~F. Freitas, C.~Provid\^encia, J.~A. Font, {Machine-Learning Love: classifying the equation of state of neutron stars with Transformers} (10 2022).
\newblock \href {http://arxiv.org/abs/2210.08382} {\path{arXiv:2210.08382}}.

\bibitem{Farrell:2022lfd}
D.~Farrell, P.~Baldi, J.~Ott, A.~Ghosh, A.~W. Steiner, A.~Kavitkar, L.~Lindblom, D.~Whiteson, F.~Weber, {Deducing neutron star equation of state parameters directly from telescope spectra with uncertainty-aware machine learning}, JCAP 02 (2023) 016.
\newblock \href {http://arxiv.org/abs/2209.02817} {\path{arXiv:2209.02817}}, \href {https://doi.org/10.1088/1475-7516/2023/02/016} {\path{doi:10.1088/1475-7516/2023/02/016}}.

\bibitem{DAvanzo:2015kdp}
P.~D'Avanzo, {Short gamma-ray bursts: A review}, JHEAp 7 (2015) 73--80.
\newblock \href {https://doi.org/10.1016/j.jheap.2015.07.002} {\path{doi:10.1016/j.jheap.2015.07.002}}.

\bibitem{Cowperthwaite:2017dyu}
P.~S. Cowperthwaite, et~al., {The Electromagnetic Counterpart of the Binary Neutron Star Merger LIGO/Virgo GW170817. II. UV, Optical, and Near-infrared Light Curves and Comparison to Kilonova Models}, Astrophys. J. Lett. 848~(2) (2017) L17.
\newblock \href {http://arxiv.org/abs/1710.05840} {\path{arXiv:1710.05840}}, \href {https://doi.org/10.3847/2041-8213/aa8fc7} {\path{doi:10.3847/2041-8213/aa8fc7}}.

\bibitem{Huang:2022mqp}
Y.-J. Huang, L.~Baiotti, T.~Kojo, K.~Takami, H.~Sotani, H.~Togashi, T.~Hatsuda, S.~Nagataki, Y.-Z. Fan, {Merger and Postmerger of Binary Neutron Stars with a Quark-Hadron Crossover Equation of State}, Phys. Rev. Lett. 129~(18) (2022) 181101.
\newblock \href {http://arxiv.org/abs/2203.04528} {\path{arXiv:2203.04528}}, \href {https://doi.org/10.1103/PhysRevLett.129.181101} {\path{doi:10.1103/PhysRevLett.129.181101}}.

\bibitem{Fujimoto:2022xhv}
Y.~Fujimoto, K.~Fukushima, K.~Hotokezaka, K.~Kyutoku, {Gravitational Wave Signal for Quark Matter with Realistic Phase Transition}, Phys. Rev. Lett. 130~(9) (2023) 091404.
\newblock \href {http://arxiv.org/abs/2205.03882} {\path{arXiv:2205.03882}}, \href {https://doi.org/10.1103/PhysRevLett.130.091404} {\path{doi:10.1103/PhysRevLett.130.091404}}.

\bibitem{Baldo:2016jhp}
M.~Baldo, G.~F. Burgio, {The nuclear symmetry energy}, Prog. Part. Nucl. Phys. 91 (2016) 203--258.
\newblock \href {http://arxiv.org/abs/1606.08838} {\path{arXiv:1606.08838}}, \href {https://doi.org/10.1016/j.ppnp.2016.06.006} {\path{doi:10.1016/j.ppnp.2016.06.006}}.

\bibitem{Li:2021thg}
B.-A. Li, B.-J. Cai, W.-J. Xie, N.-B. Zhang, {Progress in Constraining Nuclear Symmetry Energy Using Neutron Star Observables Since GW170817}, Universe 7~(6) (2021) 182.
\newblock \href {http://arxiv.org/abs/2105.04629} {\path{arXiv:2105.04629}}, \href {https://doi.org/10.3390/universe7060182} {\path{doi:10.3390/universe7060182}}.

\bibitem{Zhang:2018vrx}
N.-B. Zhang, B.-A. Li, J.~Xu, {Combined Constraints on the Equation of State of Dense Neutron-rich Matter from Terrestrial Nuclear Experiments and Observations of Neutron Stars}, Astrophys. J. 859~(2) (2018) 90.
\newblock \href {http://arxiv.org/abs/1801.06855} {\path{arXiv:1801.06855}}, \href {https://doi.org/10.3847/1538-4357/aac027} {\path{doi:10.3847/1538-4357/aac027}}.

\bibitem{Li:2019xxz}
B.-A. Li, P.~G. Krastev, D.-H. Wen, N.-B. Zhang, {Towards Understanding Astrophysical Effects of Nuclear Symmetry Energy}, Eur. Phys. J. A 55~(7) (2019) 117.
\newblock \href {http://arxiv.org/abs/1905.13175} {\path{arXiv:1905.13175}}, \href {https://doi.org/10.1140/epja/i2019-12780-8} {\path{doi:10.1140/epja/i2019-12780-8}}.

\bibitem{Anil:2020lch}
M.~U. Anil, K.~Banerjee, T.~Malik, C.~Provid\^encia, {The neutron star outer crust equation of state: a machine learning approach}, JCAP 01~(01) (2022) 045.
\newblock \href {http://arxiv.org/abs/2004.14196} {\path{arXiv:2004.14196}}, \href {https://doi.org/10.1088/1475-7516/2022/01/045} {\path{doi:10.1088/1475-7516/2022/01/045}}.

\bibitem{Xie:2020rwg}
W.-J. Xie, B.-A. Li, {Bayesian inference of the dense-matter equation of state encapsulating a first-order hadron-quark phase transition from observables of canonical neutron stars}, Phys. Rev. C 103~(3) (2021) 035802.
\newblock \href {http://arxiv.org/abs/2009.13653} {\path{arXiv:2009.13653}}, \href {https://doi.org/10.1103/PhysRevC.103.035802} {\path{doi:10.1103/PhysRevC.103.035802}}.

\bibitem{Ferreira:2022nwh}
M.~Ferreira, V.~Carvalho, C.~Provid\^encia, {Extracting nuclear matter properties from the neutron star matter equation of state using deep neural networks}, Phys. Rev. D 106~(10) (2022) 103023.
\newblock \href {http://arxiv.org/abs/2209.09085} {\path{arXiv:2209.09085}}, \href {https://doi.org/10.1103/PhysRevD.106.103023} {\path{doi:10.1103/PhysRevD.106.103023}}.

\bibitem{Mondal:2021vzt}
C.~Mondal, F.~Gulminelli, {Can we decipher the composition of the core of a neutron star?}, Phys. Rev. D 105~(8) (2022) 083016.
\newblock \href {http://arxiv.org/abs/2111.04520} {\path{arXiv:2111.04520}}, \href {https://doi.org/10.1103/PhysRevD.105.083016} {\path{doi:10.1103/PhysRevD.105.083016}}.

\bibitem{Thete:2022eif}
A.~Thete, K.~Banerjee, T.~Malik, {Inferring the dense matter equation of state from neutron star observations via artificial neural networks} (8 2022).
\newblock \href {http://arxiv.org/abs/2208.13163} {\path{arXiv:2208.13163}}.

\bibitem{Serot:1984ey}
B.~D. Serot, J.~D. Walecka, {The Relativistic Nuclear Many Body Problem}, Adv. Nucl. Phys. 16 (1986) 1--327.

\bibitem{Oertel:2016bki}
M.~Oertel, M.~Hempel, T.~Kl\"ahn, S.~Typel, {Equations of state for supernovae and compact stars}, Rev. Mod. Phys. 89~(1) (2017) 015007.
\newblock \href {http://arxiv.org/abs/1610.03361} {\path{arXiv:1610.03361}}, \href {https://doi.org/10.1103/RevModPhys.89.015007} {\path{doi:10.1103/RevModPhys.89.015007}}.

\bibitem{Huth:2021bsp}
S.~Huth, et~al., {Constraining Neutron-Star Matter with Microscopic and Macroscopic Collisions}, Nature 606 (2022) 276--280.
\newblock \href {http://arxiv.org/abs/2107.06229} {\path{arXiv:2107.06229}}, \href {https://doi.org/10.1038/s41586-022-04750-w} {\path{doi:10.1038/s41586-022-04750-w}}.

\bibitem{Sorensen:2023zkk}
A.~Sorensen, et~al., {Dense Nuclear Matter Equation of State from Heavy-Ion Collisions} (1 2023).
\newblock \href {http://arxiv.org/abs/2301.13253} {\path{arXiv:2301.13253}}.

\bibitem{Demircik:2021zll}
T.~Demircik, C.~Ecker, M.~J\"arvinen, {Dense and Hot QCD at Strong Coupling}, Phys. Rev. X 12~(4) (2022) 041012.
\newblock \href {http://arxiv.org/abs/2112.12157} {\path{arXiv:2112.12157}}, \href {https://doi.org/10.1103/PhysRevX.12.041012} {\path{doi:10.1103/PhysRevX.12.041012}}.

\bibitem{Shirke:2022tta}
S.~Shirke, S.~Ghosh, D.~Chatterjee, {Constraining the Equation of State of Hybrid Stars Using Recent Information from Multidisciplinary Physics}, Astrophys. J. 944~(1) (2023) 7.
\newblock \href {http://arxiv.org/abs/2210.09077} {\path{arXiv:2210.09077}}, \href {https://doi.org/10.3847/1538-4357/acac31} {\path{doi:10.3847/1538-4357/acac31}}.

\bibitem{Sammarruca:2019ncy}
F.~Sammarruca, R.~Millerson, {Nuclear Forces in the Medium: Insight From the Equation of State}, Front. in Phys. 7 (2019) 213.
\newblock \href {https://doi.org/10.3389/fphy.2019.00213} {\path{doi:10.3389/fphy.2019.00213}}.

\bibitem{Drischler:2019xuo}
C.~Drischler, W.~Haxton, K.~McElvain, E.~Mereghetti, A.~Nicholson, P.~Vranas, A.~Walker-Loud, {Towards grounding nuclear physics in QCD}, Prog. Part. Nucl. Phys. 121 (2021) 103888.
\newblock \href {http://arxiv.org/abs/1910.07961} {\path{arXiv:1910.07961}}, \href {https://doi.org/10.1016/j.ppnp.2021.103888} {\path{doi:10.1016/j.ppnp.2021.103888}}.

\bibitem{Drischler:2020hwi}
C.~Drischler, R.~J. Furnstahl, J.~A. Melendez, D.~R. Phillips, {How Well Do We Know the Neutron-Matter Equation of State at the Densities Inside Neutron Stars? A Bayesian Approach with Correlated Uncertainties}, Phys. Rev. Lett. 125~(20) (2020) 202702.
\newblock \href {http://arxiv.org/abs/2004.07232} {\path{arXiv:2004.07232}}, \href {https://doi.org/10.1103/PhysRevLett.125.202702} {\path{doi:10.1103/PhysRevLett.125.202702}}.

\bibitem{Drischler:2020yad}
C.~Drischler, J.~A. Melendez, R.~J. Furnstahl, D.~R. Phillips, {Quantifying uncertainties and correlations in the nuclear-matter equation of state}, Phys. Rev. C 102~(5) (2020) 054315.
\newblock \href {http://arxiv.org/abs/2004.07805} {\path{arXiv:2004.07805}}, \href {https://doi.org/10.1103/PhysRevC.102.054315} {\path{doi:10.1103/PhysRevC.102.054315}}.

\bibitem{Malik:2022zol}
T.~Malik, M.~Ferreira, B.~K. Agrawal, C.~Provid\^encia, {Relativistic Description of Dense Matter Equation of State and Compatibility with Neutron Star Observables: A Bayesian Approach}, Astrophys. J. 930~(1) (2022) 17.
\newblock \href {http://arxiv.org/abs/2201.12552} {\path{arXiv:2201.12552}}, \href {https://doi.org/10.3847/1538-4357/ac5d3c} {\path{doi:10.3847/1538-4357/ac5d3c}}.

\bibitem{Zhu:2022ibs}
Z.~Zhu, A.~Li, T.~Liu, {A Bayesian Inference of a Relativistic Mean-field Model of Neutron Star Matter from Observations of NICER and GW170817/AT2017gfo}, Astrophys. J. 943~(2) (2023) 163.
\newblock \href {http://arxiv.org/abs/2211.02007} {\path{arXiv:2211.02007}}, \href {https://doi.org/10.3847/1538-4357/acac1f} {\path{doi:10.3847/1538-4357/acac1f}}.

\bibitem{Sun:2022yor}
X.~Sun, Z.~Miao, B.~Sun, A.~Li, {Astrophysical Implications on Hyperon Couplings and Hyperon Star Properties with Relativistic Equations of States}, Astrophys. J. 942~(1) (2023) 55.
\newblock \href {http://arxiv.org/abs/2205.10631} {\path{arXiv:2205.10631}}, \href {https://doi.org/10.3847/1538-4357/ac9d9a} {\path{doi:10.3847/1538-4357/ac9d9a}}.

\bibitem{Lovato:2022vgq}
A.~Lovato, et~al., {Long Range Plan: Dense matter theory for heavy-ion collisions and neutron stars} (11 2022).
\newblock \href {http://arxiv.org/abs/2211.02224} {\path{arXiv:2211.02224}}.

\bibitem{Kurkela:2009gj}
A.~Kurkela, P.~Romatschke, A.~Vuorinen, {Cold Quark Matter}, Phys. Rev. D 81 (2010) 105021.
\newblock \href {http://arxiv.org/abs/0912.1856} {\path{arXiv:0912.1856}}, \href {https://doi.org/10.1103/PhysRevD.81.105021} {\path{doi:10.1103/PhysRevD.81.105021}}.

\bibitem{Gorda:2022jvk}
T.~Gorda, O.~Komoltsev, A.~Kurkela, {Ab-initio QCD Calculations Impact the Inference of the Neutron-star-matter Equation of State}, Astrophys. J. 950~(2) (2023) 107.
\newblock \href {http://arxiv.org/abs/2204.11877} {\path{arXiv:2204.11877}}, \href {https://doi.org/10.3847/1538-4357/acce3a} {\path{doi:10.3847/1538-4357/acce3a}}.

\bibitem{Gorda:2022lsk}
T.~Gorda, K.~Hebeler, A.~Kurkela, A.~Schwenk, A.~Vuorinen, {Constraints on Strong Phase Transitions in Neutron Stars}, Astrophys. J. 955~(2) (2023) 100.
\newblock \href {http://arxiv.org/abs/2212.10576} {\path{arXiv:2212.10576}}, \href {https://doi.org/10.3847/1538-4357/aceefb} {\path{doi:10.3847/1538-4357/aceefb}}.

\bibitem{LIGOScientific:2019lzm}
R.~Abbott, et~al., {Open data from the first and second observing runs of Advanced LIGO and Advanced Virgo}, SoftwareX 13 (2021) 100658.
\newblock \href {http://arxiv.org/abs/1912.11716} {\path{arXiv:1912.11716}}, \href {https://doi.org/10.1016/j.softx.2021.100658} {\path{doi:10.1016/j.softx.2021.100658}}.

\bibitem{LIGOScientific:2023vdi}
R.~Abbott, et~al., {Open Data from the Third Observing Run of LIGO, Virgo, KAGRA, and GEO}, Astrophys. J. Suppl. 267~(2) (2023) 29.
\newblock \href {http://arxiv.org/abs/2302.03676} {\path{arXiv:2302.03676}}, \href {https://doi.org/10.3847/1538-4365/acdc9f} {\path{doi:10.3847/1538-4365/acdc9f}}.

\bibitem{Bogatskiy:2020}
A.~Bogatskiy, B.~Anderson, J.~T. Offermann, M.~Roussi, D.~W. Miller, R.~Kondor, Lorentz group equivariant neural network for particle physics, in: Proceedings of the 37th International Conference on Machine Learning, ICML'20, JMLR.org, 2020.

\bibitem{Onyisi:2022hdh}
P.~Onyisi, D.~Shen, J.~Thaler, {Comparing point cloud strategies for collider event classification}, Phys. Rev. D 108~(1) (2023) 012001.
\newblock \href {http://arxiv.org/abs/2212.10659} {\path{arXiv:2212.10659}}, \href {https://doi.org/10.1103/PhysRevD.108.012001} {\path{doi:10.1103/PhysRevD.108.012001}}.

\bibitem{Bachtis:2021eww}
D.~Bachtis, G.~Aarts, F.~Di~Renzo, B.~Lucini, {Inverse Renormalization Group in Quantum Field Theory}, Phys. Rev. Lett. 128~(8) (2022) 081603.
\newblock \href {http://arxiv.org/abs/2107.00466} {\path{arXiv:2107.00466}}, \href {https://doi.org/10.1103/PhysRevLett.128.081603} {\path{doi:10.1103/PhysRevLett.128.081603}}.

\bibitem{Shiina:2021pqe}
K.~Shiina, H.~Mori, Y.~Tomita, H.~K. Lee, Y.~Okabe, {Inverse renormalization group based on image super-resolution using deep convolutional networks}, Sci. Rep. 11~(1) (2021) 9617.
\newblock \href {http://arxiv.org/abs/2104.04482} {\path{arXiv:2104.04482}}, \href {https://doi.org/10.1038/s41598-021-88605-w} {\path{doi:10.1038/s41598-021-88605-w}}.

\bibitem{2016arXiv160307285D}
V.~{Dumoulin}, F.~{Visin}, {A guide to convolution arithmetic for deep learning}, arXiv e-prints (2016) arXiv:1603.07285\href {http://arxiv.org/abs/1603.07285} {\path{arXiv:1603.07285}}, \href {https://doi.org/10.48550/arXiv.1603.07285} {\path{doi:10.48550/arXiv.1603.07285}}.

\bibitem{Georges:1996zz}
A.~Georges, G.~Kotliar, W.~Krauth, M.~J. Rozenberg, {Dynamical mean-field theory of strongly correlated fermion systems and the limit of infinite dimensions}, Rev. Mod. Phys. 68 (1996) 13--125.
\newblock \href {https://doi.org/10.1103/RevModPhys.68.13} {\path{doi:10.1103/RevModPhys.68.13}}.

\bibitem{Hashimoto:2018ftp}
K.~Hashimoto, S.~Sugishita, A.~Tanaka, A.~Tomiya, {Deep learning and the AdS/CFT correspondence}, Phys. Rev. D 98~(4) (2018) 046019.
\newblock \href {http://arxiv.org/abs/1802.08313} {\path{arXiv:1802.08313}}, \href {https://doi.org/10.1103/PhysRevD.98.046019} {\path{doi:10.1103/PhysRevD.98.046019}}.

\bibitem{Hashimoto:2019bih}
K.~Hashimoto, {AdS/CFT correspondence as a deep Boltzmann machine}, Phys. Rev. D 99~(10) (2019) 106017.
\newblock \href {http://arxiv.org/abs/1903.04951} {\path{arXiv:1903.04951}}, \href {https://doi.org/10.1103/PhysRevD.99.106017} {\path{doi:10.1103/PhysRevD.99.106017}}.

\bibitem{Hashimoto:2021ihd}
K.~Hashimoto, K.~Ohashi, T.~Sumimoto, {Deriving the dilaton potential in improved holographic QCD from the meson spectrum}, Phys. Rev. D 105~(10) (2022) 106008.
\newblock \href {http://arxiv.org/abs/2108.08091} {\path{arXiv:2108.08091}}, \href {https://doi.org/10.1103/PhysRevD.105.106008} {\path{doi:10.1103/PhysRevD.105.106008}}.

\bibitem{Maldacena:1997re}
J.~M. Maldacena, {The Large N limit of superconformal field theories and supergravity}, Adv. Theor. Math. Phys. 2 (1998) 231--252.
\newblock \href {http://arxiv.org/abs/hep-th/9711200} {\path{arXiv:hep-th/9711200}}, \href {https://doi.org/10.4310/ATMP.1998.v2.n2.a1} {\path{doi:10.4310/ATMP.1998.v2.n2.a1}}.

\bibitem{Gubser:1998bc}
S.~S. Gubser, I.~R. Klebanov, A.~M. Polyakov, {Gauge theory correlators from noncritical string theory}, Phys. Lett. B 428 (1998) 105--114.
\newblock \href {http://arxiv.org/abs/hep-th/9802109} {\path{arXiv:hep-th/9802109}}, \href {https://doi.org/10.1016/S0370-2693(98)00377-3} {\path{doi:10.1016/S0370-2693(98)00377-3}}.

\bibitem{Witten:1998qj}
E.~Witten, {Anti-de Sitter space and holography}, Adv. Theor. Math. Phys. 2 (1998) 253--291.
\newblock \href {http://arxiv.org/abs/hep-th/9802150} {\path{arXiv:hep-th/9802150}}, \href {https://doi.org/10.4310/ATMP.1998.v2.n2.a2} {\path{doi:10.4310/ATMP.1998.v2.n2.a2}}.

\bibitem{mehta2014exact}
P.~Mehta, D.~J. Schwab, An exact mapping between the variational renormalization group and deep learning, arXiv preprint arXiv:1410.3831 (2014).

\bibitem{Hashimoto:2020jug}
K.~Hashimoto, H.-Y. Hu, Y.-Z. You, {Neural ordinary differential equation and holographic quantum chromodynamics}, Mach. Learn. Sci. Tech. 2~(3) (2021) 035011.
\newblock \href {http://arxiv.org/abs/2006.00712} {\path{arXiv:2006.00712}}, \href {https://doi.org/10.1088/2632-2153/abe527} {\path{doi:10.1088/2632-2153/abe527}}.

\bibitem{Zhou:2023tvv}
K.~Zhou, L.~Pang, S.~Shi, H.~Stoecker, {Deep Learning for inverse problems in nuclear physics}, PoS FAIRness2022 (2023) 064.
\newblock \href {https://doi.org/10.22323/1.419.0064} {\path{doi:10.22323/1.419.0064}}.

\bibitem{DelDebbio:2007ee}
L.~Del~Debbio, S.~Forte, J.~I. Latorre, A.~Piccione, J.~Rojo, {Neural network determination of parton distributions: The Nonsinglet case}, JHEP 03 (2007) 039.
\newblock \href {http://arxiv.org/abs/hep-ph/0701127} {\path{arXiv:hep-ph/0701127}}, \href {https://doi.org/10.1088/1126-6708/2007/03/039} {\path{doi:10.1088/1126-6708/2007/03/039}}.

\bibitem{Ball:2011gg}
R.~D. Ball, V.~Bertone, F.~Cerutti, L.~Del~Debbio, S.~Forte, A.~Guffanti, N.~P. Hartland, J.~I. Latorre, J.~Rojo, M.~Ubiali, {Reweighting and Unweighting of Parton Distributions and the LHC W lepton asymmetry data}, Nucl. Phys. B 855 (2012) 608--638.
\newblock \href {http://arxiv.org/abs/1108.1758} {\path{arXiv:1108.1758}}, \href {https://doi.org/10.1016/j.nuclphysb.2011.10.018} {\path{doi:10.1016/j.nuclphysb.2011.10.018}}.

\bibitem{Ball:2011uy}
R.~D. Ball, V.~Bertone, F.~Cerutti, L.~Del~Debbio, S.~Forte, A.~Guffanti, J.~I. Latorre, J.~Rojo, M.~Ubiali, {Unbiased global determination of parton distributions and their uncertainties at NNLO and at LO}, Nucl. Phys. B 855 (2012) 153--221.
\newblock \href {http://arxiv.org/abs/1107.2652} {\path{arXiv:1107.2652}}, \href {https://doi.org/10.1016/j.nuclphysb.2011.09.024} {\path{doi:10.1016/j.nuclphysb.2011.09.024}}.

\bibitem{Nocera:2014gqa}
E.~R. Nocera, R.~D. Ball, S.~Forte, G.~Ridolfi, J.~Rojo, {A first unbiased global determination of polarized PDFs and their uncertainties}, Nucl. Phys. B 887 (2014) 276--308.
\newblock \href {http://arxiv.org/abs/1406.5539} {\path{arXiv:1406.5539}}, \href {https://doi.org/10.1016/j.nuclphysb.2014.08.008} {\path{doi:10.1016/j.nuclphysb.2014.08.008}}.

\bibitem{Li:2022ozl}
F.-P. Li, H.-L. L\"u, L.-G. Pang, G.-Y. Qin, {Deep-learning quasi-particle masses from QCD equation of state}, Phys. Lett. B 844 (2023) 138088.
\newblock \href {http://arxiv.org/abs/2211.07994} {\path{arXiv:2211.07994}}, \href {https://doi.org/10.1016/j.physletb.2023.138088} {\path{doi:10.1016/j.physletb.2023.138088}}.

\bibitem{Lafferty:2019jpr}
D.~Lafferty, A.~Rothkopf, {Improved Gauss law model and in-medium heavy quarkonium at finite density and velocity}, Phys. Rev. D 101~(5) (2020) 056010.
\newblock \href {http://arxiv.org/abs/1906.00035} {\path{arXiv:1906.00035}}, \href {https://doi.org/10.1103/PhysRevD.101.056010} {\path{doi:10.1103/PhysRevD.101.056010}}.

\bibitem{Satz:2005hx}
H.~Satz, {Colour deconfinement and quarkonium binding}, J. Phys. G 32 (2006) R25.
\newblock \href {http://arxiv.org/abs/hep-ph/0512217} {\path{arXiv:hep-ph/0512217}}, \href {https://doi.org/10.1088/0954-3899/32/3/R01} {\path{doi:10.1088/0954-3899/32/3/R01}}.

\bibitem{Guo:2012hx}
X.~Guo, S.~Shi, P.~Zhuang, {Relativistic Correction to Charmonium Dissociation Temperature}, Phys. Lett. B 718 (2012) 143--146.
\newblock \href {http://arxiv.org/abs/1209.5873} {\path{arXiv:1209.5873}}, \href {https://doi.org/10.1016/j.physletb.2012.10.032} {\path{doi:10.1016/j.physletb.2012.10.032}}.

\bibitem{Narayan:1986wj}
R.~Narayan, R.~Nityananda, {Maximum entropy image restoration in astronomy}, Ann. Rev. Astron. Astrophys. 24 (1986) 127--170.
\newblock \href {https://doi.org/10.1146/annurev.aa.24.090186.001015} {\path{doi:10.1146/annurev.aa.24.090186.001015}}.

\bibitem{2020PhRvL.124e6401F}
R.~{Fournier}, L.~{Wang}, O.~V. {Yazyev}, Q.~{Wu}, {Artificial Neural Network Approach to the Analytic Continuation Problem}, Physical Review Letter 124~(5) (2020) 056401.
\newblock \href {https://doi.org/10.1103/PhysRevLett.124.056401} {\path{doi:10.1103/PhysRevLett.124.056401}}.

\bibitem{He:2023zin}
W.-B. He, Y.-G. Ma, L.-G. Pang, H.-C. Song, K.~Zhou, {High-energy nuclear physics meets machine learning}, Nucl. Sci. Tech. 34~(6) (2023) 88.
\newblock \href {http://arxiv.org/abs/2303.06752} {\path{arXiv:2303.06752}}, \href {https://doi.org/10.1007/s41365-023-01233-z} {\path{doi:10.1007/s41365-023-01233-z}}.

\bibitem{Calafiura:2022ges}
P.~Calafiura, D.~Rousseau, K.~Terao (Eds.), {Artificial Intelligence for High Energy Physics}, World Scientific, 2022.
\newblock \href {https://doi.org/10.1142/12200} {\path{doi:10.1142/12200}}.

\bibitem{Neubauer:2022zpn}
M.~S. Neubauer, A.~Roy, {Explainable AI for High Energy Physics}, in: {Snowmass 2021}, 2022.
\newblock \href {http://arxiv.org/abs/2206.06632} {\path{arXiv:2206.06632}}.

\bibitem{Cranmer:2019eaq}
K.~Cranmer, J.~Brehmer, G.~Louppe, {The frontier of simulation-based inference}, Proc. Nat. Acad. Sci. 117~(48) (2020) 30055--30062.
\newblock \href {http://arxiv.org/abs/1911.01429} {\path{arXiv:1911.01429}}, \href {https://doi.org/10.1073/pnas.1912789117} {\path{doi:10.1073/pnas.1912789117}}.

\bibitem{openai2023gpt4}
OpenAI, Gpt-4 technical report (2023).
\newblock \href {http://arxiv.org/abs/2303.08774} {\path{arXiv:2303.08774}}.

\bibitem{Alnuqaydan_2023}
A.~Alnuqaydan, S.~Gleyzer, H.~Prosper, \href{https://dx.doi.org/10.1088/2632-2153/acb2b2}{Symba: symbolic computation of squared amplitudes in high energy physics with machine learning}, Machine Learning: Science and Technology 4~(1) (2023) 015007.
\newblock \href {https://doi.org/10.1088/2632-2153/acb2b2} {\path{doi:10.1088/2632-2153/acb2b2}}.
\newline\urlprefix\url{https://dx.doi.org/10.1088/2632-2153/acb2b2}

\bibitem{yang2023leandojo}
K.~Yang, A.~M. Swope, A.~Gu, R.~Chalamala, P.~Song, S.~Yu, S.~Godil, R.~Prenger, A.~Anandkumar, Leandojo: Theorem proving with retrieval-augmented language models (2023).
\newblock \href {http://arxiv.org/abs/2306.15626} {\path{arXiv:2306.15626}}.

\end{thebibliography}
	
	\newpage
	\appendix
	\renewcommand*{\thesection}{\Alph{section}}
\end{document}